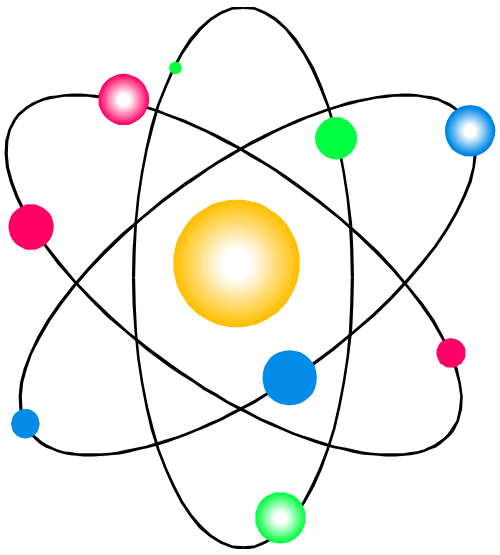

Дубовиченко С.Б.

# СВОЙСТВА ЛЕГКИХ АТОМНЫХ ЯДЕР В ПОТЕНЦИАЛЬНОЙ КЛАСТЕРНОЙ МОДЕЛИ

КАЗАХСКО - АМЕРИКАНСКИЙ УНИВЕРСИТЕТ
КАЗАХСКИЙ НАЦИОНАЛЬНЫЙ УНИВЕРСИТЕТ
им. АЛЬ-ФАРАБИ

***Дубовиченко С.Б.***

# Свойства легких атомных ядер в потенциальной кластерной модели

*Издание второе,*
*исправленное и дополненное*

Монография включает результаты научной работы автора в течение примерно 10 лет и посвящена теоретическим исследованиям структуры легких атомных ядер на основе потенциальной кластерной модели с запрещенными состояниями. Рассматриваются вопросы однозначного построения межкластерных потенциалов, содержащих запрещенные состояния и одновременно применимых в непрерывном и дискретном спектрах для легких ядерных систем с массой от 2 до 16. Изложен математический аппарат и некоторые методы расчетов, используемые в кластерной модели. Многие рассматриваемые здесь вопросы до сих пор не нашли отражения в монографической литературе.

Книга может представлять интерес для студентов старших курсов, стажеров, аспирантов и научных сотрудников, работающих в области теоретической ядерной физики.

# ОГЛАВЛЕНИЕ

# ПРЕДИСЛОВИЕ

Структура атомного ядра очень многообразна и порой обнаруживает, казалось бы, взаимоисключающие свойства. Например, в ядре могут реализоваться свойства независимого движения нуклонов, коллективные проявления степеней свободы, ассоциирование нуклонов в почти независимые группы - кластеры с характеристиками близкими к свойствам соответствующих свободных ядер. Ранее существовавшие представления о стабильно существующих в ядре кластерах заменились на понимание, что в процессе почти независимого движения нуклонов в ядре формируются и разрушаются виртуальные подсистемы - кластеры. Поэтому можно говорить лишь о вероятности существования того или иного кластерного канала.

Однако, если эта вероятность сравнительно велика, можно использовать одноканальную кластерную модель, которая во многих случаях оказывается хорошим приближением к реально существующей в ядре ситуации. Подобная модель позволяет сравнительно легко выполнять любые расчеты ядерных характеристик в процессах рассеяния и связанных состояниях, даже в тех системах, где методы решения задачи многих тел или очень громоздки в численном исполнении или вообще не приводят к конкретным количественным результатам.

Основная цель настоящей книги состоит в изучении именно таких простых двухкластерных моделей и выяснения их применимости для легких и легчайших атомных ядер с массовым числом от 2 до 16. В качестве межкластерных потенциалов взаимодействия выбран класс сравнительно новых потенциалов с запрещенными состояниями. Присутствие таких состояний позволяет эффективно учитывать принцип Паули без выполнения полной и явной антисимметризации волновых функций системы, что существенно упрощает всю вычислительную процедуру, не приводя, по-видимому, к заметному ухудшению результатов по сравнению с точными методами.

В последнее время получили большое распространение и интенсивно развивались различные варианты трехтельных моделей, применимых, например, для $^{6}$Li в трехкластерном np$^{4}$He канале, которые позволяют хорошо описать многие свойства этого ядра. Большие успехи достигнуты и в микроскопических моделях типа метода резонирующих групп, основывающихся на нуклон - нуклонных взаимодействиях с явным выделением кластерных каналов. Однако и двухкластерные потенциальные модели, использующие межкластерные силы с запрещенными состояниями, во многих слу-чаях, позволяют правильно описывать некоторые ядерные характеристики для самых различных легких и легчайших ядер и, по-видимому, не исчерпали еще полностью свои возможности.

Автор выражает большую признательность Неудачину В.Г, Кукулину В.И., Краснопольскому В.М, Померанцеву В.Н. (Научно - исследовательский институт ядерной физики МГУ, Москва), Strakovsky I.I. и Parke W.C. (Ядерный центр Университета Дж. Вашингтона, США), Узикову Ю.Н. (Объединенный институт ядерных исследований, Дубна), Дуйсебаеву А.Д., Буртебаеву Н. (Институт ядерной физики Национального ядерного центра РК, Алматы), Жусупову М.А. (Казахский национальный университет, Алматы) за исключительно полезные обсуждения отдельных, затронутых в книге вопросов, а так же Джазаирову - Кахраманову А.В. (Казахский национальный университет, Алматы) за большую помощь в написании некоторых программ, проведении численных расчетов и обсуждения полученных результатов.

Автор также выражает искреннюю благодарность Гарсону М., Де Ягеру К.В., Николенко Д.М. и Гилману Р. за экспериментальные данные по поляризациям дейтрона. Виринге Р. и Маклейдту Р. за волновые функции дейтрона для Аргонского и Боннского потенциалов.

# ВВЕДЕНИЕ

В простой кластерной модели считается, что атомное ядро состоит из двух бесструктурных фрагментов, свойства которых совпадают или близки к свойствам соответствующих ядер в свободном состоянии. Поэтому для многих характеристик кластеров, например, зарядового радиуса, кулоновского формфактора, квадрупольного и магнитного моментов, других характеристик связанных фрагментов принимаются характеристики не взаимодействующих легких ядер типа $^4$He, $^3$H и $^2$H и т.д. Классическим образцом кластерного объекта являются ядра $^6$Li и $^7$Li, в которых велика вероятность кластеризации в $^4$He$^2$H и $^4$He$^3$H каналах.

Полная волновая функция двухкластерной системы записывается в простом виде [1,2,3]

$$\Psi = A\left( \varphi_1(x_1)\varphi_2(x_2)\Psi_{JM}(\vec{R}) \right). \qquad (В.1)$$

Здесь A - оператор антисимметризации волновых функций по всем возможным перестановкам нуклонов между разными кластерами, если волновые функции кластеров, зависящие от своих внутренних координат $x_i$ , выбраны в правильном, антисимметризованном виде и $\Psi_{JM}$ - функция относительного движения, которая разделяется на радиальную $\Phi_L(R)$ и спин - угловую $Y^{LS}_{JM}$ функции [1,2]

$$\Psi_{JM}(\vec{R}) = \sum_L Y^{LS}_{JM}(\hat{R})\Phi_L(R). \qquad (В.2)$$

Спин - угловая часть волновой функции, определяемая в виде

$$Y^{LS}_{JM}(\hat{R}) = \sum_{m\sigma} (LmS\sigma|JM) Y_{Lm}(\hat{R})\chi_{s\sigma}(\sigma) \qquad (В.3)$$

связывает орбитальную $Y_{Lm}$ и спиновую $\chi_{s\sigma}$ компоненты волновой функции ядерной системы. Радиальная волновая функция относительного движения кластеров в ядре $\Phi_L(R)$ при заданном орбитальном моменте L зависит только от одной переменной R - радиус - вектора относительного движения фрагментов и является реше-нием радиального уравнения Шредингера [4]

$$u_L'(r)+(k^2+V(r))u_L(r)=0 \quad , \quad \Phi_L=\frac{u_L}{r}, \tag{B.4}$$

где V(r) - потенциал ядерного взаимодействия с учетом кулоновского и центробежного членов, $k^2=\hbar^2 E/2\mu$ - волновое число относительного движения фрагментов, $\mu$ - приведенная масса ядра в рассматриваемом кластерном канале, E - энергия относительного движения кластерной системы в центре масс.

В том случае, если ядерные ассоциации сильно обособлены, роль эффектов антисимметризации, т.е. обменных процессов между кластерами, оказывается малой и действием оператора A можно пренебречь. Однако сказать заранее, какова роль этих эффектов, достаточно сложно. Вообще говоря, в каждом конкретном случае, нужно рассматривать точную антисимметризованную волновую функцию системы, и только сравнивая ее с функцией без антисимметризации можно сделать на этот счет определенные выводы.

Процедура антисимметризации волновой функции обычно оказывается довольно сложной, поэтому часто используют приближенные способы учета принципа Паули [1,3]. В частности, в течение многих лет в потенциал межкластерного взаимодействия вводили отталкивающий кор, который не позволяет кластерам слиться в некоторую общую нуклонную систему, обеспечивая тем самым явное разделение ядра на два фрагмента. Использование потенциалов с кором приводило к вымиранию волновой функции относительного движения кластеров на малых расстояниях. В последствии появился другой класс ядерных, глубоких чисто притягивающих потенциалов, содержащих запрещенные состояния, благодаря которым обеспечивается выполнение принципа Паули [5]. Именно этот тип взаимодействий мы будем рассматривать далее, а поэтому остановимся более подробно на результатах, полученных с такими ядерными силами.

Около трех десятков лет назад в работах [5,6] было показано, что фазы упругого рассеяния легких кластерных систем могут быть описаны на основе глубоких чисто притягивающих потенциалов Вудс - саксоновского типа, которые содержат запрещенные связанные состояния. Структура запрещенных состояний определяется перестановочной симметрией волновых функций кластерной системы относительно нуклонных перестановок. Такой подход можно рассматривать, как определенную альтернативу концепции отталкивающего кора. Поведение фаз рассеяния при нулевой энергии для данного вида взаимодействий подчиняется обобщенной теореме Левинсона [6]:

$$\delta_L = \pi ( N_L + M_L ) , \qquad (В.5)$$

где $N_L$ и $M_L$ число запрещенных и разрешенных связанных состояний. Согласно выражению (В.5) фазы при больших энергиях стремятся к нулю, все время, оставаясь положительными. Радиальная волновая функция разрешенных состояний для потенциалов с запрещенными состояниями осциллирует на малых расстояниях, а не вымирает, как это было для взаимодействий с кором. Благодаря этому, в рассмотрение включается внутренняя структура ядра, которая определяется поведением волновой функции системы в области малых расстояний.

В дальнейшем [7] были получены центральные гауссовы потенциалы с запрещенными состояниями, параметры которых согласованы с фазами упругого рассеянии, а их использование в простых одноканальных $^4He^3H$ и $^4He^2H$ кластерных моделях позволяет вполне успешно описать некоторые характеристики основных состояний ядер $^6Li$, $^7Li$, вероятность кластеризации которых в этих каналах сравнительно высока. Различные оценки дают для $^4He^2H$ кластеризации величину 0.6-0.8 и около 0.9 для системы $^4He^3H$ [8,9].

Определенный успех одноканальной модели, основанной на таких потенциалах, обусловлен не только большой степенью кластеризации этих ядер, но и тем, что в каждом состоянии кластеров существует только одна разрешенная орбитальная схема Юнга [10], определяющая симметрию этого состояния. Тем самым, достигается некое "единое" описание непрерывного и дискретного спектра, и потенциалы, полученные на основе экспериментальных фаз рассеяния, вполне успешно используются для описания различных характеристик основного состояния ядер лития.

Для более легких кластерных систем вида $N^2H$, $^2H^2H$, $p^3H$, $n^3He$ и т.д. в состояниях рассеяния с минимальным спином уже возможно смешивание по орбитальным симметриям и ситуация оказывается более сложной. В состояниях с минимальным спином в непрерывном спектре таких систем разрешены две орбитальные симметрии с различными схемами Юнга, в то время, как основным связанным состояниям, по-видимому, соответствует только одна из этих схем [10,11].

Поэтому потенциалы, непосредственно полученные на основе экспериментальных фаз рассеяния, эффективно зависят от різлич-ных орбитальных схем и не могут в таком виде использоваться для описания характеристик основного состояния. Из таких взаимодействий, необходимо выделять чистую компоненту, применимую уже при анализе характеристик связанных состояний. Тогда результаты в основном будут зависеть от степени кластеризации ядра в рассматриваемый канал.

В работах [10,11] было показано, что для легчайших кластерных систем экспериментальные смешанные фазы могут быть представлены в виде полусуммы чистых фаз с определенными схемами Юнга. Обычно считают [10,11], что в качестве одной из чистых фаз канала с минимальным спином можно использовать аналогичную, чистую фазу другого спинового состояния или системы чистой по изоспину. В таком случае, по экспериментальным фазам легко найти чистую фазу максимальной симметрии канала с минимальным спином и по ней параметризовать чистые взаимодействия. В частности, в работах [10,11,12] получены такие $N^2H$, $N^3H$, $N^3He$, $^2H^2H$ и $^2H^3He$ чистые гауссовы взаимодействия и показано [11,12], что, в общем, удается правильно передать энергию связи ядер $^3H$, $^3He$ и $^4He$ в кластерных каналах, асимптотическую константу, зарядовый радиус и упругий кулоновский формфактор при малых переданных импульсах.

Отметим, что смешивание по орбитальным схемам Юнга в состояниях с минимальным спином характерно не только для большинства легких кластерных системах, но реализуется и в более тяжелых системах вида $N^6Li$, $N^7Li$ и $^2H^6Li$ [13].

Используя, полученные на основе фаз упругого рассеяния межкластерные взаимодействия с запрещенными состояниями, можно рассматривать многие ядерные характеристики. В том числе, сечения фотопроцессов, рассматривая их в кластерных потенциальных моделях, которые, несмотря на свою простоту, в ряде случаев, позволяют получить хорошие результаты [10-13,14, 15].

---

1. Неудачин В.Г., Смирнов Ю.Ф. - Нуклонные ассоциации в легких ядрах. М., Наука, 1969., 414с.

2. Вильдермут Л., Тан Я. - Единая теория ядра. М., Мир, 1980, 502с. (Wildermuth K., Tang Y.C. - A unified theory of the nucleus. Vieweg. Braunschweig. 1977.).

3. Немец О.Ф., Неудачин В.Г., Рудчик А.Т., Смирнов Ю.Ф., Чувильский Ю.М. - Нуклонные ассоциации в атомных ядрах и ядерные реакции многонуклонных передач. Киев, Наукова Думка, 1988, 488с.

4. Хюльтен Л., Сугавара М. - В кн.: Строение атомного ядра. М., ИЛ., 1959, с.9.

5. Neudatchin V.G., Kukulin V.I., Boyarkina A.N., Korennoy V.P. - Lett. Nuovo Cim., 1972, v.5, p.834; Neudatchin V.G., Kukulin V.I., Korotkikh V.L., Korennoy V.P. - Phys. Lett., 1971, v.34B, p.581; Kurdyumov I.V., Neudatchin V.G., Smirnov Y.F., Korennoy V.P. - Phys. Lett., 1972, v.40B, p.607.

6. Неудачин В.Г., Смирнов Ю.Ф. - Современные вопросы оптики и атомной физики. Киев, Киевский Гос. Университет, 1974, с.224; ЭЧАЯ, 1979, т.10, с.1236.

7. Дубовиченко С.Б., Джазаиров-Кахраманов А.В. - ЯФ, 1993, т.56, № 2, с.87; ЯФ, 1994, т.57, № 5, с.784.

8. Kukulin V.I., Krasnopol'sky V.M., Voronchev V.T., Sazonov P.B. - Nucl. Phys., 1984, v.A417, p.128; 1986, v.A453, p.365; Kukulin V.I., Voronchev V.T., Kaipov T.D., Eramzhyan R.A. - Nucl. Phys., 1990, v.A517, p.221.

9. Lehman D.R., Rajan M. - Phys. Rev., 1982, v.C25, p.2743; Lehman D.R. - Phys. Rev., 1982, v.C25, p.3146; Lehman D.R., Parke W.C., - Phys. Rev., 1983, v.C28, p.364.

10. Neudatchin V.G., Kukulin V.I., Pomerantsev V.N., Sakharuk A.A. - Phys. Rev., 1992, v.C45. p.1512; Неудачин В.Г., Сахарук А.А., Смирнов Ю.Ф. - ЭЧАЯ, 1993, т.23, с.480; Дубовиченко С.Б., Джазаиров - Кахраманов А.В. - ЭЧАЯ 1997, т.28, с.1529.

11. Искра В., Мазур А.И., Неудачин В.Г., Нечаев Ю.И., Смирнов Ю.Ф. - УФЖ, 1988, т.32, с.1141; Искра В., Мазур А.И., Неудачин В.Г., Смирнов Ю.Ф. - ЯФ, 1988, т.48, с.1674; Неудачин В.Г., Померанцев В.Н., Сахарук А.А. - ЯФ, 1990, т.52, с.738; Кукулин В.И., Неудачин В.Г., Померанцев В.Н., Сахарук А.А. - ЯФ, 1990, т.52, с.402; Дубовиченко С.Б., Неудачин В.Г., Смирнов Ю.Ф., Сахарук А.А.- Изв. АН СССР, сер. физ., 1990, т.54, с.911; Neudatchin V.G., Sakharuk A.A., Dubovichenko S.B. - Few Body Sys., 1995, v18, p.159.

12. Дубовиченко С.Б., Джазаиров-Кахраманов А.В. - ЯФ, 1990, т.51, № 6, с.1541; ЯФ, 1993, т.56, № 4, с.45.

13. Дубовиченко С.Б., Джазаиров-Кахраманов А.В. - ЯФ, 1992, т.55, № 11, с.2918; Дубовиченко С.Б., Джазаиров-Кахраманов А.В., Сахарук А.А.- ЯФ, 1993, т.56, № 8, с.90.

14. Дубовиченко С.Б., Джазаиров-Кахраманов А.В. - ЯФ, 1995, т.58, с.635; ЯФ, 1995, т.58, с.852.

15. Дубовиченко С.Б. Методы расчета ядерных характеристик. Алматы. Комплекс.2006. 311с.

# 1. МЕТОДЫ РАСЧЕТОВ ЯДЕРНЫХ ХАРАКТЕРИСТИК

В настоящей главе изложены методы расчетов различных ядерных характеристик в кластерной модели, как для связанных состояний, образующих ядро фрагментов, так и процессов их рассеяния. Рассмотрены способы расчета полных сечений фотоядерных реакций и методы, используемые в кластерных и нуклон - нуклонных системах с тензорными силами.

## 1.1 Векторные соотношения кластерной модели

Двухкластерная модель предполагает наличие только двух обособленных фрагментов - кластеров, между которыми перераспределены все нуклоны ядра. Первый кластер содержит $M_1$ нуклонов с зарядом $Z_1$, второй $M_2$ с зарядом $Z_2$. Векторная схема кластерной модели приведена на рисунке 1.1. Межкластерное расстояние R определяет относительное положение центров масс фрагментов. Радиусы $\rho_i$ и $\rho_j$ задают положение каждого нуклона относительно их центров масс в первом и втором кластерах соответственно. Радиусы $r_i$ и $r_j$ указывают положение каждого нуклона в обоих кластерах относительно общего центра масс ядра. Векторы $R_1$ и $R_2$ определяют положение центров масс кластеров относительно их общего центра масс. При таком определении радиус - векторов между ними существуют простые соотношения:

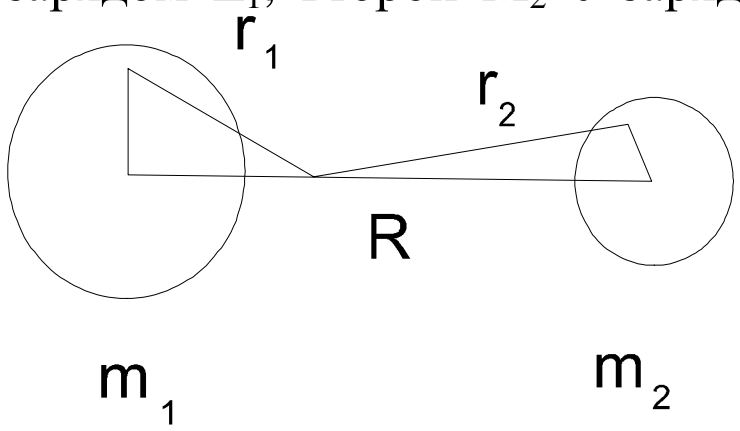

Рис. 1.1. Векторная схема кластерной модели.

$$\sum r_k = \sum r_i + \sum r_j = 0 \ , \qquad \sum \rho_i = \sum \rho_j = 0 \ ,$$

$$r_i = R_1 + \rho_i = RM_2/M + \rho_i \ , \quad r_j = R_2 + \rho_j = -RM_1/M + \rho_j \ , \quad R = R_1 - R_2 \ ,$$

$$R_1 = 1/M \sum_i r_i = \frac{M_2}{M} R \ , \qquad R_2 = 1/M \sum_j r_j = -\frac{M_1}{M} R \ , \qquad (1.1.1)$$

$$1 < i < M_1 \ , \quad M_1 + 1 < j < M \ , \quad 1 < k < M \ ,$$

$M = M_1 + M_2$ , $Z = Z_1 + Z_2$ , $\mu = M_1M_2/M$ .

Эти векторные соотношения будут использоваться в дальнейшем для вычисления различных ядерных характеристик в двухкластерной модели. Рассмотрим, например, вывод формулы для среднеквадратичного радиуса ядра, который определяется следующим образом

$$\langle r^2 \rangle = \langle \Psi | r^2 | \Psi \rangle \cdot$$

В кластерной модели квадрат радиус - вектора может быть представлен в виде

$$r^2 = 1/M \sum_k r_k^2 \cdot$$

Таким же образом определим зарядовые радиусы кластеров

$$\langle r^2 \rangle_{1,2} = \langle \Psi(1,2) | \frac{1}{M_{1,2}} \sum_n \rho_n^2 (1,2) | \Psi(1,2) \rangle ,$$

где 1,2 - первый или второй кластер, а индекс n определяет суммирование по i или j. Используя теперь выражения (1.1.1), устанавливающие связь между межкластерным расстоянием и векторами $r_k$

$$r^2 = 1/M \sum r_i^2 + 1/M \sum r_j^2 = 1/M \sum \rho_i^2 + 1/M \sum \rho_j^2 + \mu/M\, R^2 ,$$

для радиуса ядра в кластерной модели с волновыми функциями (В.2) получим окончательное выражение

$$R_r^2 = \frac{M_1}{M} \langle r^2 \rangle_1 + \frac{M_2}{M} \langle r^2 \rangle_2 + \frac{M_1 M_2}{M^2} I_2 , \qquad (1.1.2)$$

где

$$I_2 = \langle \Phi(R) | R^2 | \Phi(R) \rangle$$

матричный элемент по радиальным волновым функциям относительного движения кластеров от квадрата межкластерного расстояния. Таким образом, радиус ядра в кластерной модели может быть легко выражен через радиусы кластеров и эффективное межкластерное расстояние. Аналогичным образом можно использовать векторные соотношения кластерной модели при выводе формул для формфакторов, квадрупольных, магнитных моментов ядер, матричных элементов ядерных реакций, в частности процессов фоторазвала или радиационного захвата ассоциаций и т.д.

Рассмотрим далее методы вычисления полных сечений ядерных фотопроцессов, а также характеристик связанных состояний кластерных фрагментов в ядре для чисто центральных межкластерных потенциалов. Затем перейдем к учету тех эффектов, которые дают тензорные взаимодействия в двухчастичной системе, и приведем некоторые основные формулы для рассмотрения сечений рассеяния и реакций в супермультиплетном приближении, которое используется для анализа взаимодействий легчайших кластерных систем.

## 1.2 Процессы фоторазвала и радиационного захвата в кластерной модели

Одной из самых, пожалуй, интересных ядерных реакций является процесс ядерного фоторазвала или обратная ему реакция - радиационного захвата. Налетающая частица - фотон не вступает в сильные ядерные взаимодействия с ядром мишенью. Происходит только электромагнитное взаимодействие, операторы которого точно известны. Поэтому можно учитывать только ядерные взаимодействия связанных кластеров, что существенно упрощает рассмотрение по сравнению с трехтельной задачей, когда наряду с межкластерными силами надо включать и ядерное взаимодействие налетающей частицы. Общие методы расчета сечений подобных процессов подробно изложены в прекрасной монографии [1]. Поэтому далее будем исходить из уже известных определений дифференциальных сечений радиационных и фотоядерных процессов.

Для расчетов сечений радиационного захвата в длинноволновом приближении используем известное выражение [1,2]

$$\frac{d\sigma_c(N)}{d\Omega} = \frac{K\mu}{2\pi\,\hbar^2 q}\frac{1}{(2S_1+1)(2S_2+1)}\sum_{m_i,m_f,\lambda}|M_{J\lambda}(N)|^2\ , \qquad (1.2.1)$$

где N = E - электрические или M - магнитные переходы и

$$M_{J\lambda}(N)=\sum_J i^J\sqrt{2\pi\ (2J+1)}\,\frac{K^J}{(2J+1)!!}\left[\frac{J+1}{J}\right]^{1/2}\lambda\ \sum_m D^J_{m\lambda}\langle f\,|H_{Jm}(N)|\,i\rangle,$$

$H_{Jm}(E) = Q_{Jm}(L) + Q_{Jm}(S)$ ,

$H_{Jm}(M) = W_{Jm}(L) + W_{Jm}(S)$ ,

$$Q_{Jm}(L)=e\ \sum_i Z_i r_i^J Y_{Jm}(\Omega_i)\ ,$$

$$Q_{Jm}(S)=-\frac{e\,\hbar}{m_0c}K\left[\frac{J}{J+1}\right]^{1/2}\ \sum_i \mu_i\hat{S}_i r_i^J Y_{Jm}(\Omega_i)\ ,$$

$$W_{Jm}(L)=i\frac{e\,\hbar}{m_0c}\frac{1}{J+1}\ \sum_i\frac{Z_i}{M_i}\hat{L}_i\nabla_i(r_i^J Y_{Jm}(\Omega_i))\ ,$$

$$W_{Jm}(S)=i\frac{e\,\hbar}{m_0c}\ \sum_i \mu_i\hat{S}_i\nabla_i(r_i^J Y_{Jm}(\Omega_i))\ .$$

Здесь J - мультипольность, q - волновое число относительного движения кластеров, $D^J_{m\lambda}$ - функция Вигнера, μ - приведенная масса, $M_i$, $Z_i$, $\hat{S}_i$ и $\hat{L}_i$ - массы, заряды, спины и орбитальные моменты i - го кластера, $\mu_i$ - магнитные моменты кластеров, K - волновое число фотона, $m_0$ - масса нуклона. Знак оператора $Q_{Jm}(S)$ выбран отрицательным, как приведено в работе [3]. Интегрируя по углам и суммируя это выражение по λ, для полного сечения захвата получаем [3,4]

$$\sigma_c(J)=\frac{8\pi\ K^{2J+1}}{\hbar^2 q}\frac{\mu}{(2S_1+1)(2S_2+1)}\frac{J+1}{J[(2J+1)!!]^2}\sum_{m,m_i,m_f}\left|M_{Jm}\ (N)\right|^2,$$

$$M_{Jm}(N)=i^J\langle f\,|H_{Jm}(N)|\,i\rangle, \qquad (1.2.2)$$

где в кластерной модели электромагнитные операторы принимают простой вид

$$Q_{Jm}(L)=e\mu^J\left[\frac{Z_1}{M_1^J}+(-1)^J\frac{Z_2}{M_2^J}\right]R^J Y_{Jm}=A_J R^J Y_{Jm},$$

$$Q_{J\ m}(S) = -\frac{e\hbar}{m_0\ c} K\left[\frac{J}{J+1}\right]^{1/2}\left[\mu_1\hat{S}_1 \frac{M_2^J}{M^J} + (-1)^J \mu_2\hat{S}_2 \frac{M_1^J}{M^J}\right] R^J Y_{J\ m} =$$

$$= (B_{1J}\ \hat{S}_1 + B_{2J}\ \hat{S}_2) R^J Y_{J\ m},$$

$$W_{J\ m}(L) = i\frac{e\hbar}{m_0 c}\frac{\sqrt{J(2J+1)}}{J+1}\left[\frac{Z_1}{M_1}\frac{M_2^J}{M^J} + (-1)^{J-1}\ \frac{Z_2}{M_2}\frac{M_1^J}{M^J}\right] R^{J-1}\text{€} Y_{J\ m}^{J-1} =$$

$$= C_J\ R^{J-1}\text{€}\ Y_{J\ m}^{J-1},$$

$$W_{J\ m}(S) = i\frac{e\hbar}{m_0 c}\sqrt{J(2J+1)}\left[\mu_1\hat{S}_1 \frac{M_2^{J-1}}{M^{J-1}} + (-1)^{J-1}\ \mu_2\hat{S}_2 \frac{M_1^{J-1}}{M^{J-1}}\right] R^{J-1}\ Y_{J\ m}^{J-1} =$$

$$= (D_{1J}\ \hat{S}_1 + D_{2J}\ \hat{S}_2) R^{J-1}\ Y_{J\ m}^{J-1}.$$

Здесь R - межкластерное расстояние и M - масса ядра. Используем в дальнейшем волновые функции связанных состояний кластеров в обычной форме [2-4]

$$|f\rangle = \Psi_f = \sum_{L_f J_f} R_{L_f\ J_f}\ \Phi_{J_f m_f}^{L_f S},\ R_{LJ} = \frac{U_{LJ}}{r}. \tag{1.2.3}$$

Функцию рассеяния запишем в виде разложения по спин - угловым функциям [2]

$$|i\rangle = \Psi_i = \frac{1}{q}\sum_{L_i J_i} i^{L_i}\sqrt{4\pi(2L_i+1)}(L_i 0 S m_i \mid J_i m_i) e^{i\delta_{L_i J_i}} R_{L_i J_i}\ \Phi_{J_i m_i}^{L_i S}\ . \tag{1.2.4}$$

Здесь $R_{LJ}$ - радиальная волновая функция рассеяния, получаемая из решения уравнения Шредингера (В.4) с заданными межкластерными потенциалами, $\Phi_{Jm}^{LS}$ - спин - угловая функция начального i состояния системы, $\delta_{LJ}$ - фазы упругого рассеяния. Используя известные формулы для матричных элементов различных операторов, приведенные в [5], для полного сечения захвата можно получить окончательное выражение [4]

$$\sigma_c(J) = \frac{8\pi\ K^{2J+1}}{\hbar^2 q^3}\frac{\mu}{(2S_1+1)(2S_2+1)}\frac{J+1}{J[(2J+1)!!]^2}\sum_{\substack{L_i, L_f, \\ J_i, J_f}} \left|T_J\ (N)\right|^2, \tag{1.2.5}$$

где матричные элементы приобретают вид

$T_J(E) = A_J\, I_J\, P_J + (\, B_{1J}\, N_{1J} + B_{2J}\, N_{2J}\,)\, I_J$ ,

$T_J(M) = C_J\, I_{J-1}\, G_J + (\, D_{1J}\, N_{1J} + D_{2J}\, N_{2J}\,)\, I_{J-1}$ ,

$$P_J = \sqrt{4\pi}\left\langle J_f L_f S \left\| Y_J \right\| J_i L_i S\right\rangle = (-1)^{J+S+L_i+L_f}(L_i 0 J 0 \mid L_f 0)\times$$
$$\times\sqrt{(2J_i+1)(2J_f+1)(2J+1)(2L_i+1)}\begin{Bmatrix} L_i & S & J_i \\ J_f & J & L_f \end{Bmatrix},$$

$$G_J = \sqrt{4\pi}\left\langle J_f L_f S \left\| \hat{L} Y_J^k \right\| J_i L_i S\right\rangle = (-1)^{S+L_i+J_i}(L_i 0 k 0 \mid L_f 0)(2L_i+1)\times$$
$$\times\sqrt{L_i(2L_i+1)(2k+1)(2J_i+1)(2J_f+1)(2J+1)}\begin{Bmatrix} L_i & 1 & L_i \\ k & L_f & J \end{Bmatrix}\begin{Bmatrix} S & L_i & J_i \\ J & J_f & L_f \end{Bmatrix},$$

$$N_J = \sqrt{4\pi}\left\langle J_f L_f S \left\| \hat{S} Y_J^k \right\| J_i L_i S\right\rangle = (-1)^{k+1-J+L_i+L_f+2S-J_i-J_f}(L_i 0 k 0 \mid L_f 0)\times$$
$$\times\begin{Bmatrix} S & 1 & S \\ L_i & k & L_f \\ J_i & J & J_f \end{Bmatrix}\sqrt{S(S+1)(2S+1)(2k+1)(2L_i+1)(2J_i+1)(2J_f+1)},$$

а $I_J$ - радиальные интегралы от волновых функций вида

$I_J = < J_f L_f \mid R^J \mid J_i\, L_i >$ .

Сечение обратного процесса - фоторазвала можно получить из принципа детального равновесия [1,2]

$$\sigma_d(J) = \frac{q^2(2S_1+1)(2S_2+1)}{K^2\, 2(2J_0+1)}\sigma_c(J), \tag{1.2.6}$$

где $J_0$ - полный момент ядра в основном состоянии. В полученных выражениях аналитически вычисляются все величины, кроме радиальных интегралов, которые находятся численно по определен-ным из решения уравнения Шредингера волновым функциям связанных состояний и рассеяния. Асимптотика радиальной волновой функции рассеяния обычно представляется в виде суперпозиции кулоновских $F_L$ и $G_L$ функций на границе области ядерного взаимодействия при r = R

$$R_{LJ} \rightarrow N[\, F_L(qr)\, Cos(\delta_{LJ}) + G_L(qr)\, Sin(\delta_{LJ})\, ]\,, \qquad (1.2.7)$$

где $\delta_{LJ}$ - фазы рассеяния, N - нормировочная константа. Получив численную радиальную волновую функцию, из этого соотношения можно определить фазы рассеяния и нормировочную константу с данным орбитальным L и полным J моментами кластерной системы.

Приведенные выражения позволяют выполнять расчеты полных сечений ядерных фотопроцессов в кластерной модели ядра, когда известно межкластерное взаимодействие. Однако ядерные потенциалы, как правило, неизвестны и приходится использовать различные дополнительные методы и предположения для их определения. Одним из таких методов является анализ фаз упругого рассеяния кластеров, который позволяет определять приближенный вид межкластерных потенциалов.

Обычно считается, что если потенциалы способны правильно передать экспериментальные фазы рассеяния, то они могут быть использованы для рассмотрения ядерных характеристик связанных состояний кластеров в ядре. В дальнейшем результаты таких расчетов будут целиком зависеть от степени кластеризации ядра в рассматриваемый кластерный канал. Это предположение вытекает из общего принципа квантовой механики, который утверждает, что квантовая система должна иметь единый гамильтониан взаимодействия в дискретном и непрерывном спектре. Поэтому перейдем теперь к рассмотрению различных характеристик для связанных состояний, считая, что межкластерная волновая функция и потенциал взаимодействия в принципе известны или могут быть определены теми или иными методами.

## 1.3 Характеристики связанных состояний кластеров

К основным ядерным характеристикам связанных состояний можно отнести радиус ядра, выражения для которого были определены выше, кулоновские формфакторы, квадрупольный, октуполь-ный и магнитные моменты, вероятности электромагнитных переходов между разными уровнями ядра, энергию связи в кластерном канале и т.д. Отметим, что если некоторая модель ядра позволяет правильно описывать многие из этих характеристик, то можно, по-видимому, считать, что эта модель наиболее хорошо соответствует реальной ситуации, существующей в данном ядре. Кластерная модель позволяет правильно передать многие из этих характеристик для некоторых легких и даже легчайших ядер, если учитывать смешивание по орбитальным схемам Юнга. Этот факт, возможно, указывает на большую вероятность кластеризации таких ядер в данных каналах. Поэтому имеет смысл более подробно рассмотреть некоторые выражения для расчетов ядерных характеристик в связанном состоянии кластерной системы.

Используя формулы (1.2.2) можно вычислять некоторые статические электромагнитные характеристики ядер. В частности, для квадрупольного, магнитного, октупольного моментов и приведенной вероятности электрических и магнитных радиационных переходов можно написать [3]

$$Q=\left(\frac{16\pi}{5}\right)^{1/2}\left\langle J_0J_0\middle|Q_{20}(L)\middle|J_0J_0\right\rangle,$$

$$\mu=\left(\frac{4\pi}{3}\right)^{1/2}\left\langle J_0J_0\middle|W_{10}(L)+W_{10}(S)\middle|J_0J_0\right\rangle,$$

$$\Omega=\left(\frac{4\pi}{3}\right)^{1/2}\left\langle J_0J_0\middle|W_{30}(L)+W_{30}(S)\middle|J_0J_0\right\rangle, \qquad (1.3.1)$$

$$B(M1)=\frac{1}{2J_i+1}\left|\left\langle J_fL_f\middle\|W_1(L)+W_1(S)\middle\|J_iL_i\right\rangle\right|^2,$$

$$B(E2)=\frac{1}{2J_i+1}\left|\left\langle J_fL_f\middle\|Q_2(L)\middle\|J_iL_i\right\rangle\right|^2.$$

Отсюда, например, в двухкластерной модели для основного состояния $^7$Li с $J_0=3/2^-$ получаем [4,6]

$$Q=-\frac{2}{5}YI_2=-\frac{2\,34}{5\,49}I_2,$$

$$Y=(Z_1M_2^2+Z_2M_1^2)/M^2,$$

$$\frac{\mu}{\mu_0} = X + \mu_1 = \frac{17}{42} + \mu_1,$$

$$X = \frac{1}{M}\left(\frac{Z_1 M_2}{M_1} + \frac{Z_2 M_1}{M_2}\right), \qquad (1.3.2)$$

$$\frac{\Omega}{\mu} = \frac{3M_2^2}{5M^2} I_2 = \frac{48}{245} I_2,$$

$$B(M1) = \frac{1}{4\pi}(2\mu_1 - X)^2 I_0^2 = \frac{1}{4\pi}(3\mu_1 - \mu)^2 I_0^2,$$

$$B(E2) = \frac{1}{4\pi} Y^2 I_2^2.$$

Здесь первым кластером считается тритон или $^3$He, обладающие магнитным моментом и $\mu_0 = \frac{e\hbar}{2m_0c}$ - ядерный магнетон.

Магнитный радиус ядра $^7$Li в кластерной модели определяется через зарядовые и магнитные радиусы фрагментов в виде [6]

$$\mu R_m^2 = \frac{4}{21}\langle r_t^2\rangle + \frac{3}{14}\langle r_t^2\rangle + \mu_t\langle r_{tm}^2\rangle + \left(\frac{209}{3430} + \frac{432}{1225}\mu_t\right) I_2, \qquad (1.3.3)$$

где $\mu$ - магнитный момент ядра, $<r_i>$ - магнитные (m) и зарядовые радиусы кластеров, а интегралы $I_2$ - были определены выше в выражениях (1.2.5).

Импульсное распределение кластеров в ядре, определяемое, как Фурье - образ волновой функции относительного движения фрагментов и нормированное на единицу при переданном импульсе q=0 может быть записано [7]

$$P^2 = \sum_L P_L^2(q) \Big/ \sum_L P_L^2(0) \quad , \qquad P_L(q) = \int u_L j_L(qr) r dr \quad ,$$

где $j_L(x)$ - сферическая функция Бесселя.

Для расчетов продольных кулоновских формфакторов можно использовать, например, известное определение [7,8]

$$F(q) = 1/Z < \Psi_f | \sum (1/2 + t_{zk}) \exp(iqr_k) | \Psi_i > .$$

Аналогично запишем формфакторы кластеров

$$F_{1,2}(q) = 1/Z_{1,2} < \Phi_{1,2} \mid \sum (1/2+t_{zn}) \exp (iq\rho_n) \mid \Phi_{1,2} > ,$$

где $t_{zk}$ - проекция изоспина k - й частицы. Используя векторные соотношения (1.1.1) для формфактора двухкластерного ядра находим

$$F(q)=Z_1/Z\, F_1(q)<\Psi_f|\exp (iq_1R)|\Psi_i> + Z_2/Z\, F_2(q)<\Psi_f|\exp(iq_2R)|\Psi_i> .$$

Здесь $q_1 = - qM_2/M$ и $q_2 = qM_1/M$ , а $F_{1,2}$ (q) - собственные формфакторы ассоциаций в свободном состоянии. Разлагая плоские волны по функциям Бесселя и интегрируя по углам [5], квадрат кулоновского формфактора может быть представлен в виде [4]

$$F_J^2 = \frac{1}{Z^2} V_J^2 B_J, \tag{1.3.4}$$

$$B_J = (2J_f+1)(2J+1)(2L_i+1)(L_i 0 J 0 | L_f 0)^2 \begin{Bmatrix} L_i & S & J_i \\ J_f & J & L_f \end{Bmatrix}^2 ,$$

где $L_{i,f}$ и $J_{i,f}$ - орбитальные и полные моменты начального i и конечного f состояния ядра, J - мультипольность формфактора, S и Z - спин и заряд ядра, фигурная скобка - 6j символ Вигнера [5] и $V_J$ - структурный множитель, зависящий от характеристик фрагментов и их взаимного движения

$$V_J = Z_1 F_1 I_{2,J} + Z_2 F_2 I_{1,J} , \tag{1.3.5}$$

где $I_{k,J}$ - радиальные матричные элементы по функциям начального и конечного состояния от сферических функции Бесселя

$$I_{k,J} = < L_f J_f \mid j_J(g_k r) \mid L_i J_i > . \tag{1.3.6}$$

Здесь k=1 - й или 2 - й кластер, $g_k = (M_k / M)$ q , q - переданный импульс, $j_J(g_k r)$ - сферическая функция Бесселя. При вычислении неупругих формфакторов, когда конечное состояние лежит в непрерывном спектре, волновые функции рассеяния при резонансных энергиях необходимо нормировать на асимптотику вида [8]

$U_L = \exp(-\delta_L)\,[\,F_L \, Cos(\delta_L) + G_L \, Sin(\delta_L)]$ . (1.3.7)

Вероятность E2 - переходов и зарядовый радиус могут быть определены на основе кулоновских формфакторов [8,9] мультипольности CJ

$$B(E2) = \frac{225Z^2}{4\pi}\lim_{q\to 0}\left(\frac{F_{C2}^2(q)}{q^4}\right),$$

$$R_f^2 = 6\lim_{q\to 0}\left(\frac{1-F_{C0}(q)}{q^2}\right). \quad (1.3.8)$$

Волновые функции в матричных элементах (1.3.6) для основных и резонансных состояний представимы в виде разложения по гауссовому базису вида [9]

$$|LJ> = R_L = r^L \sum_i C_i \exp(-\alpha_i r^2) , \quad (1.3.9)$$

где $\alpha_i$ и $C_i$ - вариационные параметры и коэффициенты разложения, которые находятся вариационным методом для связанных состояний или аппроксимацией гауссойдами численных волновых функций резонансных уровней [9]. Вариационные параметры $\alpha_i$ могут быть, например, получены из квадратурной сетки вида [9]

$$\alpha_i = \alpha_0 \, tg^2 \{ \pi (2i - 1) / 4N \} .$$

Для определения спектра собственных энергий и волновых функций в таком вариационном методе решается обобщенная задача на собственные значения

$$\sum_i (H_{ij} - E\, L_{ij})C_i = 0 ,$$

где матрицы гамильтониана H и интегралов перекрывания L имеют вид [10]

$$H_{ij} = T_{ij} + V_{ij} + < i \,|\, Z_1 Z_2/r \,|\, j > + < i \,|\, \hbar^2 L(L+1)/2\mu r^2 \,|\, j > ,$$

$$T_{ij} = -\frac{\hbar^2}{2\mu}\frac{\sqrt{\pi}(2L-1)!!}{2^{L+1}\alpha_{ij}^{L+1/2}}\left\{L(2L+1) - L^2 - \frac{\alpha_i\alpha_j(2L+1)(2L+3)}{\alpha_{ij}^2}\right\},$$

$$V_{ij} = \int r^{2L+2}\, V(r)\exp(-\alpha_{ij} r^2)dr\ ,$$

$$L_{ij} = \frac{\sqrt{\pi}(2L+1)!!}{2^{L+2}\alpha_{ij}^{L+3/2}}\ , \qquad (1.3.10)$$

$$N_0 = [\ \sum C_i\, C_j\, L_{ij}\ ]^{-1/2},$$

$$< i\, |\, Z_1\ Z_2/r\, |\, j > = \frac{Z_1 Z_2\ L!}{2\alpha_{ij}^{L+1}}\ ,$$

$$< i\, |\, \hbar^2\, L(L+1)/2\mu r^2\, |\, j > = \frac{\sqrt{\pi}(2L-1)!!}{2^{L+1}\alpha_{ij}^{L+1/2}}\frac{L(L+1)\hbar^2}{2\mu}\ , \quad \alpha_{ij} = \alpha_i + \alpha_j.$$

В случае гауссова потенциала межкластерного взаимодействия

$$V(r)= V_0 \exp(-\beta r^2)$$

матричный элемент потенциала $V_{ij}$ определяется в аналитическом виде

$$V_{ij} = V_0 \frac{\sqrt{\pi}(2L+1)!!}{2^{L+2}(\alpha_{ij}+\beta)^{L+3/2}}\ .$$

Для приведенного выше вариационного разложения волновой функции матричные элементы формфактора также вычисляются аналитически и, например, для ядра $^6$Li имеют вид [10]

$$I_{k,0}(C0) = \frac{\sqrt{\pi}}{4}\sum_{i,j} C_i\, C_j\, W_{ij}\ /\alpha_{ij}^{3/2},$$

$$I_{k,2}(C2) = \frac{\sqrt{\pi}}{16}\sum_{i,j} C_i\, C_j g_k^2 W_{ij}\ /\alpha_{ij}^{7/2}. \qquad (1.3.11)$$

Аналогичные выражения можно получить для ядра $^7$Li

$$J_0 = 3/2; \qquad I_{k,0}(C0) = \frac{\sqrt{\pi}}{8}\sum_{i,j} C_i C_j W_{ij}\left(3 - \frac{g_k^2}{2\alpha_{ij}}\right)/\alpha_{ij}^{5/2},$$

$$J_0 = 3/2;1/2; \qquad I_{k,2}(C2) = \frac{\sqrt{\pi}}{16}\sum_{i,j} C_i C_j g_k^2 W_{ij}/\alpha_{ij}^{7/2}, \qquad (1.3.12)$$

$$J_0 = 7/2; \qquad I_{k,2}(C2) = \frac{\sqrt{\pi}}{32}\sum_{i,j} C_i C_j g_k^2 W_{ij}\left(7 - \frac{g_k^2}{2\alpha_{ij}}\right)/\alpha_{ij}^{9/2},$$

$$J_0 = 7/2; \qquad I_{k,4}(C4) = \frac{\sqrt{\pi}}{64}\sum_{i,j} C_i C_j g_k^4 W_{ij}/\alpha_{ij}^{11/2},$$

где

$$W_{ij} = \exp\left(-\frac{g_k^2}{4\alpha_{ij}}\right), \qquad \alpha_{ij} = \alpha_i + \alpha_j, \qquad g_k = \frac{M_k}{M}q \quad .$$

Используя разложение волновой функции по гауссойдам, для вероятности E2 переходов и квадрупольного момента имеем

$$B(E2) = \frac{1}{4\pi}W(M,Z)^2 R_0^4 B_2,$$

$$Q = -\frac{2}{5}W(M,Z)R_0^2 \quad , \qquad (1.3.13)$$

где

$$W(M,Z) = \frac{M_1^2 Z_2 + M_2^2 Z_1}{M^2},$$

$$R_0^2 = \frac{(2L+3)!!\sqrt{\pi}}{2^{L+3}}\sum_{i,j} C_i C_j \alpha_{ij}^{-(L+5/2)} ,$$

а величина $B_2$ была определена в выражении (1.3.4). При нахо-ждении волновых функций основных и резонансных состояний в двухкластерной системе, можно использовать не только вариационный [9], но и конечно - разностный [11] метод решения радиального уравнения Шредингера (В.4). Запишем еще раз уравнение Шредингера для волновой функции системы двух частиц в виде

$$\chi''_L + [\, k^2 - V(r)\,]\, \chi_L = 0 \,. \qquad (1.3.14)$$

Его решения для связанных состояний на бесконечности и в нуле подчиняются условиям

$$\chi_L(0) = \chi_L(\infty) = 0 \,.$$

Однако уравнение (1.3.14), на расстояниях больших, чем радиус действия ядерных сил, т.е. когда V(r)=0, имеет аналитическое решение и условие на бесконечности можно заменить на требование неразрывности логарифмической производной на границе области взаимодействия при r=R [11]

$$\chi'_L / \chi_L = W'_{\eta L}(2kR) / W_{\eta L}(2kR) = f(k,r) \,.$$

Здесь W(2kR) - функция Уиттекера для связанных состояний. В том случае, когда в потенциале не учитывается кулоновское взаимодействие, эта логарифмическая производная будет просто равна -k. На расстояниях порядка 10-20 Фм можно сшивать с асимптотикой, определяемой функцией Уиттекера, саму волновую функцию связанных состояний, определяя тем самым асимптотическую константу $C_0$

$$R_{LJ} = \frac{\sqrt{2k_0}}{r} C_0 W_{\eta L}(2k_0 r), \qquad (1.3.15)$$

где $\eta$ - кулоновский параметр и $k_0$ - волновое число, обусловленное энергией связи кластерной системы в ядре.

Уравнение Шредингера с тем или иным граничным условием образует краевую задачу типа Штурма - Лиувилля и при переходе к конечным разностям превращается в замкнутую систему линейных алгебраических уравнений [11]. Условие равенства нулю ее детерминанта позволяет определить энергию системы

$$D_N = \begin{pmatrix} \theta_1 & 1 & 0 & . & . & . & 0 \\ \alpha_2 & \theta_2 & 1 & 0 & . & . & 0 \\ 0 & \alpha_3 & \theta_3 & 1 & 0 & . & 0 \\ . & . & . & . & . & . & . \\ . & . & . & . & . & . & . \\ 0 & . & 0 & 0 & \alpha_{N-1} & \theta_{N-1} & 1 \\ 0 & . & 0 & 0 & 0 & \alpha_N & \theta_N \end{pmatrix} , \qquad (1.3.16)$$

где N - число уравнений, $h = \Delta r/N$ - шаг интегрирования, $\Delta r$ - интервал интегрирования, а

$\alpha_n = 1$ , $\alpha_N = 2$ , $\theta_n = k^2h^2 - 2 - V_nh^2$ ,

$\theta_N = k^2h^2 - 2 - V_nh^2 + 2hf(k,r_n)$ , $Z_n = 2kr_n$ ,

$r_n = nh$ , $n = 1,2 .....N$ , $f(k,r_n) = - k - 2k\eta/Z_n - 2k(L-\eta)/Z_n^2$ ,

и $V_n = V(r_n)$ - потенциал взаимодействия кластеров в точке $r_n$. Такая форма записи граничных условий f(k,r) позволяет приближенно учитывать кулоновские взаимодействия, т.е. эффекты, которые дает учет функции Уиттекера.

Вычисление $D_N$ проводится по рекуррентным формулам вида [11]

$$D_{-1} = 0 , \quad D_0 = 1 , \quad D_n = \theta_n D_{n-1} - \alpha_n D_{n-2} . \qquad (1.3.17)$$

Для нахождения формы волновых функций связанных состояний используется другой рекуррентный процесс

$$\chi_0 = 0 , \quad \chi_1 = \text{const.} , \quad \chi_n = \theta_{n-1} \chi_{n-1} + \alpha_{n-1} \chi_{n-2} . \qquad (1.3.18)$$

Тем самым, при заданной энергии системы удается найти детерминант и волновую функцию связанного состояния. Энергия, приводящая к нулю детерминанта, считается собственной энергией системы, а волновая функция (1.3.18) при этой энергии - собственной функцией задачи.

Определенные выше выражения, позволяют проводить расчеты многих свойств связанных состояний ядра в том случае, если межкластерные потенциалы имеют чисто центральный вид и не содер-жат слагаемых, зависящих от взаимной ориентации спинов частиц и их относительного расстояния. Подобные ядерные силы называются тензорными и приводят к смешиванию орбитальных состояний с различным L. В частности, в дейтроне такие силы приводят к появлению D компоненты в волновой функции системы. В следующем разделе мы перейдем к рассмотрению именно таких взаимодействий с изложением некоторых методов расчетов, применимые в данном случае.

## 1.4 Двухчастичная задача с тензорными силами. Процессы рассеяния

Включение тензорных сил в потенциал взаимодействия заметно расширяет круг вопросов, которые можно рассматривать на основе уравнения Шредингера. Первым следствием появления тензорных сил является смешивание различных орбитальных конфигураций, что в частности позволяет рассматривать квадрупольный момент дейтрона и некоторых других ядер. В расчетах появляется C2 компонента кулоновского формфактора, отсутствующая при чисто центральных взаимодействиях.

В волновых функциях относительного движения кластеров или нуклонов присутствует D компонента с весом от 2-3% до 7-8%, что позволяет более корректно описывать многие ядерные характеристики. Например, благодаря D компоненте волновой функции удается правильно передать поведение астрофизического S фактора в $^2H^2H$ взаимодействиях при сверхмалых энергиях, которые играют существенную роль для понимания процессов солнечного термоядерного синтеза легких ядер.

Для расчетов фаз и волновых функций рассеяния исходим из обычной системы уравнений Шредингера для тензорных потенциалов [12,13,14]

$$u''(r)+ [\, k^2 - V_c(r) - V_{cul}(r)]u(r) = \sqrt{8}\, V_t(r)w(r)\,, \qquad (1.4.1)$$

$$w''(r)+ [\, k^2 - V_c(r) - 6/r^2 - V_{cul}(r) + 2\, V_t(r)\,]w(r) = \sqrt{8}\, V_t(r)u(r)\,,$$

где $V_{cul}(r)= 2\mu/\hbar^2 Z_1Z_2/r$ - кулоновский потенциал, $\mu$ - приведенная масса ядра в рассматриваемом кластерном канале, константа $\hbar^2/M_N$ равна обычно 41.4686 или 41.47 МэВ Фм$^2$, $k^2 = 2\mu E/\hbar^2$ - волновое число относительного движения кластеров, $V_c$ - централь-ная часть потенциала, $V_t$ - тензорная часть потенциала взаимодействия, u и w - искомые функции рассеяния с орбитальными моментами 0 и 2.

Решением этой системы являются четыре волновые функции, получающиеся с начальными условиями типа [12]

$u_1(0)=0,\ u'_1(0)=1,\ w_1(0)=0,\ w'_1(0)=0$ ,

$u_2(0)=0,\ u'_2(0)=0,\ w_2(0)=0,\ w'_2(0)=1$ ,

которые образуют линейно независимые комбинации, представляемые в виде [12]

$u_\alpha = C_{1\alpha} u_1 + C_{2\alpha} u_2 \longrightarrow \mathrm{Cos}(\varepsilon)\,[F_0 \mathrm{Cos}(\delta_\alpha) + G_0 \mathrm{Sin}(\delta_\alpha)]$ ,

$w_\alpha = C_{1\alpha} w_1 + C_{2\alpha} w_2 \longrightarrow \mathrm{Sin}(\varepsilon)\,[F_2 \mathrm{Cos}(\delta_\alpha) + G_2 \mathrm{Sin}(\delta_\alpha)]$ ,

$u_\beta = C_{1\beta} u_1 + C_{2\beta} u_2 \longrightarrow -\mathrm{Sin}(\varepsilon)\,[F_0 \mathrm{Cos}(\delta_\beta) + G_0 \mathrm{Sin}(\delta_\beta)]$ ,

$w_\beta = C_{1\beta} w_1 + C_{2\beta} w_2 \longrightarrow \mathrm{Cos}(\varepsilon)\,[F_2 \mathrm{Cos}(\delta_\beta) + G_2 \mathrm{Sin}(\delta_\beta)]$ .

Здесь $F_L$ и $G_L$ - кулоновские функции рассеяния [13,14], $\delta_{\alpha,\beta}$ - фазы рассеяния, ε - параметр смешивания состояний с разными орбитальными моментами и полным спином ядра, равным 1. Если вынести в правой части Cos(δ), эти выражения преобразуются к форме

$u_{1\alpha} = C'_{1\alpha} u_1 + C'_{2\alpha} u_2 \longrightarrow \mathrm{Cos}(\varepsilon)\,[F_0 + G_0 \mathrm{tg}(\delta_\alpha)]$ ,

$w_{1\alpha} = C'_{1\alpha} w_1 + C'_{2\alpha} w_2 \longrightarrow \mathrm{Sin}(\varepsilon)\,[F_2 + G_2 \mathrm{tg}(\delta_\alpha)]$ ,

$u_{2\beta} = C'_{1\beta} u_1 + C'_{2\beta} u_2 \longrightarrow -\mathrm{Sin}(\varepsilon)\,[F_0 + G_0 \mathrm{tg}(\delta_\beta)]$ ,

$$w_{2\beta} = C'_{1\beta} w_1 + C'_{2\beta} w_2 \longrightarrow \mathrm{Cos}(\varepsilon)[F_2 + G_2 \mathrm{tg}(\delta_\beta)] , \qquad (1.4.2)$$

где C' = C/Cos(δ) и $u_{i\alpha} = u_\alpha / \mathrm{Cos}(\delta)$. В случае нуклон - нуклонной (np) задачи кулоновские функции заменяются на обычные функции Бесселя. Более компактно можно записать эти уравнения в матричном виде [15]

$$V = XC' \longrightarrow FU + GU\sigma , \qquad (1.4.3)$$

где

$$V = \begin{pmatrix} u_{1\alpha} & u_{2\beta} \\ w_{1\alpha} & w_{2\beta} \end{pmatrix}, \qquad X = \begin{pmatrix} u_1 & u_2 \\ w_1 & w_2 \end{pmatrix}, \qquad C' = \begin{pmatrix} C'_{1\alpha} & C'_{1\beta} \\ C'_{2\alpha} & C'_{2\beta} \end{pmatrix},$$

$$F = \begin{pmatrix} F_0 & 0 \\ 0 & F_2 \end{pmatrix}, \qquad G = \begin{pmatrix} G_0 & 0 \\ 0 & G_2 \end{pmatrix}, \qquad U = \begin{pmatrix} \mathrm{Cos}\varepsilon & -\mathrm{Sin}\varepsilon \\ \mathrm{Sin}\varepsilon & \mathrm{Cos}\varepsilon \end{pmatrix},$$

$$\sigma = \begin{pmatrix} \mathrm{tg}\delta_\alpha & 0 \\ 0 & \mathrm{tg}\delta_\beta \end{pmatrix}.$$

Аналогичное уравнение можно написать и для первых производных волновой функции

$$V' = X'C' \longrightarrow F'U + G'U\sigma .$$

Исключая из этих уравнений C', для К матрицы рассеяния, определяемой в виде $U\sigma U^{-1}$, окончательно будем иметь [15]

$$K = U\sigma U^{-1} = - [\, X(X')^{-1} G' - G]^{-1} [X(X')^{-1} F' - F] . \qquad (1.4.4)$$

Тем самым, К матрица рассеяния оказывается выраженной через кулоновские функции, численные решения исходных уравнений и их производные при некотором $r=R_0$.

Как известно, К матрица рассеяния в параметризации Блатта - Биденхарна выражается через фазы рассеяния и параметр смешивания следующим образом [12,13]

$$K = \begin{pmatrix} \mathrm{Cos}^2\varepsilon\,\mathrm{tg}\delta_\alpha + \mathrm{Sin}^2\varepsilon\,\mathrm{tg}\delta_\beta & \mathrm{Cos}\varepsilon\,\mathrm{Sin}\varepsilon(\mathrm{tg}\delta_\alpha - \mathrm{tg}\delta_\beta) \\ \mathrm{Cos}\varepsilon\,\mathrm{Sin}\varepsilon(\mathrm{tg}\delta_\alpha - \mathrm{tg}\delta_\beta) & \mathrm{Sin}^2\varepsilon\,\mathrm{tg}\delta_\alpha + \mathrm{Cos}^2\varepsilon\,\mathrm{tg}\delta_\beta \end{pmatrix}.$$

Тогда, приравнивая соответствующие матричные элементы, для К матрицы получим

$K_{12} = K_{21} = 1/2\ (tg\delta_\alpha - tg\delta_\beta)\ Sin(2\varepsilon)$ ,

$K_{11} + K_{22} = tg\delta_\alpha + tg\delta_\beta$ , $K_{11} - K_{22} = (tg\delta_\alpha - tg\delta_\beta)\ Cos(2\varepsilon)$ .

Откуда имеем

$tg(2\varepsilon) = 2K_{12}/(K_{11}-K_{22})$ , $tg\delta_\alpha = (A+B)/2$ , $tg\delta_\beta = (A-B)/2$ ,

$$A = K_{11}+K_{22}, \quad B = (K_{11}-K_{22})/Cos(2\varepsilon) . \qquad (1.4.5)$$

Здесь

$a = f\ (u_1w'_2 - u_2w'_1)$ , $b = f\ (u'_1u_2 - u_1u'_2)$ , $c = f\ (w_1w'_2 - w'_1w_2)$,

$$d = f\ (u'_1w_2 - u'_2w_1)\ , \quad f = (u'_1w'_2 - u'_2w'_1)^{-1} , \qquad (1.4.6)$$

$A = aG'_0 - G_0$ , $B = bG'_2$ , $E = cG'_0$ , $D = dG'_2 - G_2$ ,

$F = PD$ , $G = -PB$ , $N = -PE$ , $M = PA$ , $P = -(AD - BE)^{-1}$ ,

$R = aF'_0 - F_0$ , $S = bF'_2$ , $T = cF'_0$ , $Z = dF'_2 - F_2$ ,

$K_{11} = FR + GT$, $K_{12} = FS + GZ$ ,

$K_{21} = NS + MZ$ , $K_{22} = NR + MT$ .

Таким образом, получаются сравнительно простые выражения для определения фаз рассеяния и параметра смешивания. Для численных решений, производные и сами функции можно заменить на значения волновой функции в двух точках $R_1$ и $R_2$ при этом вид полученных выражений не изменяется. Надо только считать, что величины без штриха, например, находятся в первой точке, а со штрихом во второй.

По определенным фазам легко можно найти и коэффициенты линейных комбинаций C'

$C'=X^{-1}(FU+GU\sigma)$ .

Расписывая это матричное выражение, имеем

$C'_{1\alpha}=a(A+F)+b(E+H)$ , $C'_{2\alpha}=c(A+F)+d(E+H)$ ,

$C'_{1\beta}$=a(B+G)+b(D+K) , $C'_{2\beta}$=c(B+G)+d(D+K) , (1.4.7)

где

$a=fw_2$, $b=-fu_2$, $c=-fw_1$, $d=fu_1$, $f=(u_1w_2-u_2w_1)^{-1}$,

$A=F_0Cos(\varepsilon)$, $B=-F_0Sin(\varepsilon)$, $E=F_2Sin(\varepsilon)$, $D=F_2Cos(\varepsilon)$,

$F=G_0Cos(\varepsilon)tg(\delta_\alpha)$, $G=-G_0Sin(\varepsilon)tg(\delta_\beta)$,

$H=G_2Sin(\varepsilon)tg(\delta_\alpha)$, $K=G_2Cos(\varepsilon)tg(\delta_\beta)$ .

В результате, можно получить полный, нормированный на асимптотику вид волновой функции во всей области при $r<R_0$. Радиус сшивки $R_0$ обычно принимается равным 20-30 Фм. Для численного решения исходного уравнения можно использовать известный метод Рунге - Кутта с автоматическим выбором шага при заданной точности результатов по фазам рассеяния и параметру смешивания.

Полученные таким образом волновая функция рассеяния применяются в различных расчетах ядерных характеристик. В частности, при вычислениях эффективных радиусов и длин рассеяния в случае нуклон - нуклонной задачи, используются известные формулы, приведенные в работах [12,13].

Фазы рассеяния для нуклон - нуклонной задачи обычно представляют в параметризации Сака (Степпа), а не в используемом выше представлении Блатта - Биденхарна. Между этими представлениями фаз существует простая связь [12]

$$\theta_J^{J-1}+\theta_J^{J+1}=\delta_\alpha+\delta_\beta \;, \qquad tg(\theta_J^{J-1}-\theta_J^{J+1})=Cos(2\varepsilon)tg(\delta_\alpha-\delta_\beta) \;,$$

$$Sin(2\hat{\varepsilon})=Sin(2\varepsilon)Sin(\delta_\alpha-\delta_\beta) \;, \qquad (1.4.8)$$

где $\theta_J^{J+1}$, $\hat{\varepsilon}$ - фазы рассеяния и параметр смешивания в параметризации Сака.

Тензорные силы влияют не только на процессы рассеяния ядерных частиц, но и на их связанные состояния, где также появляется D компонента волновой функции. И хотя, как и в рассмотренном случае, учет тензорных сил несколько усложняет все расчетные формулы, переход к тензорным взаимодействиям несет много новой информации о структуре и свойствах ядра. Перейдем теперь к рассмотрению связанных состояний кластеров или других ядерных частиц для потенциалов с тензорной компонентой.

## 1.5. Двухчастичная задача с тензорными силами Связанные состояния

При рассмотрении связанных состояний двух частиц, исходим из обычных уравнений (1.4.1) для тензорных потенциалов, однако в данном случае граничные условия принимают вид

$$\chi_0 = C_1u_1 + C_2 u_2 = \exp(-kr) ,$$
$$\chi_2 = C_1w_1 + C_2w_2 = [1 + 3/kr + 3/(kr)^2]\exp(-kr) , \qquad (1.5.1)$$

или с учетом кулоновских сил

$$\chi_0 = C_1u_1 + C_2 u_2 = W_{-\gamma,0+1/2}(2kr) ,$$
$$\chi_2 = C_1w_1 + C_2w_2 = W_{-\gamma,2+1/2}(2kr) ,$$

где k - волновое число, определяемое энергией связи ядра в рассматриваемом канале, $\gamma$ - кулоновский параметр, $W_{-\gamma,L+1/2}(2kr)$ - функция Уиттекера. Волновые функции связанных состояний нормированы на единицу следующим образом

$$\int [\chi_0^2 + \chi_2^2]dr = 1 ,$$

а интеграл от квадрата волновой функции D состояния определяет ее вес, обычно выражаемый в процентах. Полная волновая функция связанной системы записывается в виде, приведенном во введении (В.2). Орбитальные состояния в тензорном потенциале смешиваются, так что сохраняется только полный момент системы.

Для нахождения численной волновой функции связанных состояний может быть использована комбинация численных и вариационных методов расчета. В частности, при некоторой заданной энергии связанного состояния численным методом находилась волновая функция системы (1.4.1). Затем она подставляется в исходную систему уравнений и вычисляется сумма невязок, которые находятся, как разница правой и левой частей обеих уравнений в каждой точке численной схемы. Изменяя энергию связанного состояния в некоторых пределах, проводится минимизация значений невязок каким - нибудь вариационным методом. Энергия, дающая минимум суммы невязок для обоих уравнений считается собственной энергией, а волновая функция, приводящая к минимуму - собственной функцией задачи на связанные состояния.

Изложенный метод будет использован далее для рассмотрения кластерной системы $^4He^2H$, когда в потенциале взаимодействия присутствует тензорная компонента, например, гауссового вида

$$V(r) = V_c(r) + V_t(r)\, S_{12}\,, \qquad S_{12} = [6(Sn)^2 - 2S^2]\,,$$

$$V_c(r) = -V_0 \exp(-\alpha r^2)\,, \qquad V_t(r) = -V_1 \exp(-\beta r^2)\,.$$

Здесь S - полный спин системы, n - единичный вектор, совпадающий по направлению с вектором межкластерного расстояния, $S_{12}$ - тензорный оператор. Под тензорным потенциалом в рассматриваемой системе следует понимать взаимодействие, оператор которого зависит от взаимной ориентации полного спина системы и межкластерного расстояния. Математическая форма записи такого оператора полностью совпадает с оператором двухнуклонной задачи, поэтому и потенциал, по аналогии, будем называть тензорным [12,13].

Рассмотрим теперь некоторые характеристики связанного состояния двух частиц при наличии тензорных сил. Квадрупольный момент системы заряженных частиц может быть представлен в виде (1.3.1). Используя векторные соотношения кластерной модели (1.1.1), можно получить выражение для квадрупольного момента ядра $^6Li$ с учетом момента дейтрона $Q_d$

$$Q = Q_d + Q_0\,, \qquad Q_0 = \sqrt{\frac{16\pi}{5}}\, C_{\alpha d} \sum_{LL'} I_{LL'} R_{LL'}\,,$$

где

$$C_{\alpha d} = \frac{Z_\alpha M_d^2 + Z_d M_\alpha^2}{M^2}\,, \qquad R_{LL'} = \langle \chi_L | r^2 | \chi_{L'} \rangle\,, \qquad \Phi_L = \chi_L / r$$

и

$$I_{LL'} = (-1)^{J+L'+S} \sqrt{\frac{5(2L+1)(2J+1)}{4\pi}}\,(L020|L'0)(JJ20|JJ)\begin{Bmatrix} L & S & J \\ J & 2 & L' \end{Bmatrix}\,.$$

Здесь $\chi_L$ - радиальные волновые функции связанных состояний, L и L' - могут принимать значения 0 и 2, Z и M - заряды и массы кластеров и ядра. В конечном итоге для величины $Q_0$ получаем выражение [16]

$$Q_0 = \frac{4\sqrt{2}}{15}\int r^2(\chi_0\chi_2 - \frac{1}{\sqrt{8}}\chi_2^2)dr \cdot \qquad (1.5.2)$$

Магнитный момент ядра, определенный в (1.3.1), для двухкластерной системе с тензорными силами, в случае, когда только один из кластеров имеет магнитный момент $\mu_d$ и спин 1, может быть представлен в виде [17]

$$\mu = \mu_d J + \frac{1}{2(J+1)}(B_{\alpha d} - \mu_d)(\hat{J} + \hat{L} - \hat{S})P_D \quad , \qquad \hat{A} = A(A+1) \quad , \quad (1.5.3)$$

где $P_D$ - величина примеси D состояния и

$$B_{\alpha d} = \frac{1}{M}\left(\frac{Z_\alpha M_d}{M_\alpha} + \frac{Z_d M_\alpha}{M_d}\right) \quad .$$

Магнитный момент дейтрона равен $0.857\mu_0$, а ядра $^6Li$ несколько меньше $0.822\,\mu_0$. Поэтому для получения, в рассматриваемой модели, правильного момента ядра $^6Li$ необходимо допустить примерно 6.5% примеси D состояния.

При расчетах кулоновских формфакторов используется выражение (1.3.4), а интегралы от радиальных функций связанных состояний представляются теперь в виде

$$I_{k,0} = \int(\chi_0^2 + \chi_2^2)j_0(g_k r)dr \quad , \qquad I_{k,2} = 2\int\chi_2(\chi_0 - \frac{1}{\sqrt{8}}\chi_2)j_2(g_k r)dr \quad . \qquad (1.5.4)$$

Здесь k=1 или 2 обозначает $^2H$ или $^4He$, $g_k=(M_k/M)q$ , J - мультипольность формфактора, равная 0 или 2, $j_J$ - сферическая функция Бесселя, q - переданный импульс. Формфактор $^4He$ кластера можно представить в виде параметризации [17,18]

$$F_\alpha=(1 - (aq^2)^n)\exp(-bq^2) \, , \qquad (1.5.5)$$

где a=0.09985 Фм$^2$, b=0.46376 Фм$^2$ и n=6. Для дейтрона может быть использована другая форма

$$F_d=\exp(-aq^2)+bq^2\exp(-cq^2) , \qquad (1.5.6)$$

с параметрами a=0.49029 Фм$^2$, b=0.01615 Фм$^2$ и c=0.16075 Фм$^2$. Для вычисления асимптотических констант $C_L^0$, $C_L^{W0}$ и $C_L^W$ используются выражения [19]:

$$\Phi_L = \frac{\sqrt{2k}}{r} C_L^0 A_L \exp(-kr), \qquad A_0 = 1, \qquad A_2 = [1+3/kr+3/(kr)^2],$$

$$(1.5.7)$$

$$\Phi_L = \frac{\sqrt{2k}}{r} C_L^{W0} A_L W^0_{-\gamma,L+1/2}(2kr), \qquad \Phi_L = \frac{\sqrt{2k}}{r} C_L^W W_{-\gamma,L+1/2}(2kr),$$

где k - волновое число, определяемое энергией связи ядра в рассматриваемом канале, $\gamma$ - кулоновский параметр, $W_{-\gamma,L+1/2}(2kr)$ - функция Уиттекера и $W^0_{-\gamma,L+1/2}(2k_0 r)=(2k_0 r)^{-\eta}\exp(-k_0 r)$ - ее асимптотика. Радиус ядра может быть вычислен через кулоновский формфактор согласно (1.3.8) и непосредственно в кластерной модели, где его можно представить в виде (1.1.2) с межкластерным расстоянием, учитывающим D компоненту волновой функции

$$I_2 = \int r^2(\chi_0^2+\chi_2^2)dr \quad . \qquad (1.5.8)$$

В случае нуклон - нуклонной задачи с тензорными силами несколько меняются формулы для формфакторов, которые учитывают магнитное рассеяние и принимают следующий вид [20]

$$\frac{d\sigma}{d\Omega} = \left(\frac{d\sigma_M}{d\Omega}\right)\left[A + B\operatorname{tg}^2\left(\frac{\theta}{2}\right)\right], \qquad (1.5.9)$$

$$A = G_0^2 + G_2^2 + \frac{2}{3}\eta(1+\eta)G_M^2, \qquad B = \frac{4}{3}\eta(1+\eta)^2 G_M^2 ,$$

$$G_0 = 2G_E C_E, \quad G_2 = 2G_E C_Q, \quad G_M = \frac{M_d}{M_P}(2G_{M0}C_S + G_E C_L),$$

$$2G_E = G_{Ep} + G_{En}, \qquad 2G_{M0} = G_{Mp} + G_{Mn},$$

$$\eta = \frac{(\hbar cq)^2}{4M_d^2} = 0.002767q^2,$$

$$C_E = \int \left(u^2 + w^2\right) j_0(x) dr, \qquad C_Q = 2\int w\left(u - \frac{w}{\sqrt{8}}\right) j_2(x) dr,$$

$$C_L = \frac{3}{2}\int w^2 \left(j_0(x) + j_2(x)\right) dr, \qquad x = \frac{qr}{2},$$

$$C_S = \int \left(u^2 - \frac{w^2}{2}\right) j_0(x) dr + \frac{1}{\sqrt{2}}\int w\left(u + \frac{w}{\sqrt{2}}\right) j_2(x) dr,$$

Здесь u(r) и w(r) - волновые функции связанного состояния, а $j_i$ - функции Бесселя i-го порядка. Для масс нуклонов можно использовать следующие значения $M_p$=938.28 МэВ и $M_n$=939.57 МэВ [21], масса дейтрона принимается равной 1875.63 МэВ. Зарядовый формфактор нейтрона можно считать равным нулю, а в качестве зарядового формфактора протона используется параметризация [22]

$$G_{Ep} = \frac{1}{\left(1 + 0.054844q^2\right)^2}.$$

Здесь переданный импульс q, который измеряется в $Фм^{-1}$. Магнитные формфакторы нуклонов находились на основе "масштабного закона" [22]

$$G_{Mp} = \mu_p G_{Ep}, \qquad G_{Mn} = \mu_n G_{Ep},$$

а в качестве магнитных моментов нуклонов использованы следующие величины

$$\mu_p = 2.7928\ \mu_0, \qquad \mu_n = -1.9131\ \mu_0.$$

Приведенные выражения определяют все основные характеристики двухчастичной или двухкластерной системы с тензорными силами.

## 1.6 Формализм супермультиплетной модели

В этом параграфе кратко остановимся на формализме супермультиплетного расщепления фаз в процессах рассеяния, более подробно изложенного в работах [17,23]. Рассмотрим его на примере рассеяния в ядерной $N^6Li$ системе. В рамках супермультиплетной модели обобщенного потенциального описания рассеяния парциальная амплитуда рассеяния в системе A + B представляется в виде [23]

$$T_L = \sum (t_A \tau_A t_B \tau_B | t\tau)(s_A \sigma_A s_B \sigma_B | s\sigma) <\{f_A\} s_A t_A, \{f_B\} s_B t_B | \{f\} st> T^{\{f\}}_L$$
$$(t'_A \tau'_A t'_B \tau'_B | t\tau)(s'_A \sigma'_A s'_B \sigma'_B | s\sigma) <\{f_A\} s'_A t'_A, \{f_B\} s'_B t'_B | \{f\} st> \; . \qquad (1.6.1)$$

Здесь S, t, {f} - спин, изоспин и спин - изоспиновая схема Юнга системы, $\sigma$ и $\tau$ - проекции спина и изоспина, $<\{f_a\} S_a t_a \{f_b\} S_b t_b | \{f\} ST>$ - изоскалярные множители коэффициентов Клебша - Гордана группы SU(4) [23], $T_L^{\{f\}}$- инвариантные, по отношению к преобразованиям группы SU(4), части амплитуды рассеяния, которые будем рассматривать далее, как обычные потенциальные амплитуды. Тогда сечение упругого рассеяния для неполяризованных частиц можно записать [23]

$$d\sigma/d\Omega = 1/4P_0^2 \sum B_L P_L(\cos\theta) \; , \qquad (1.6.2)$$

$$B_L = 1/[(2s_A+1)(2s_B+1)] \sum\sum (2L+1)(2L'+1)(L0L'0|l0)^2$$
$$\sum (t_A \tau_A t_B \tau_B | t\tau)(t'_A \tau'_A t'_B \tau'_B | t\tau)(t_A \tau_A t_B \tau_B | t'\tau')(t'_A \tau'_A t'_B \tau'_B | t'\tau')$$
$$\sum (2s+1) T^{\{f\}}_L [T^{\{f\}}_L]^* (<\{f_A\} s_A t_A \{f_B\} s_B t_B | \{f\} st>)^2$$
$$(<\{f_A\} s'_A t'_A \{f_B\} s'_B t'_B | \{f'\} s't'>)^2 \; . \qquad (1.6.3)$$

Учитывая конкретные значения изоскалярных множителей [17,23] и изоспиновых коэффициентов Клебша - Гордана, получим, что, например, в дублетных каналах $N^6Li$ системы парциальная амплитуда упругого рассеяния определяется суперпозицией двух потенциальных амплитуд с двумя различными перестановочными симметриями $\{f_1\} = \{421\}$ и $\{f_2\} = \{43\}$:

$$B_L = 1/3 \sum\sum (2L+1)(2L'+1)(L0L'0|l0)^2 (1/2T^{\{43\}}_L + 1/2T^{\{421\}}_{L'})$$

$(1/2\ [T^{\{43\}}{}_L\ ]^* + 1/2\ [T^{\{421\}}{}_{L'}\ ]^*)$ ,

где в дублетном канале

$T_{L,S=1/2} = 1/2\ T^{\{43\}}{}_L + 1/2\ T^{\{421\}}{}_L$ ,

в то время, как квартетные каналы непосредственно описываются потенциальными амплитудами

$$T_{L,S=3/2} = T_L{}^{\{421\}}\ . \tag{1.6.4}$$

Переходя к парциальной S матрице для упругого $N^6Li$ рассеяния

$$S_{LS}{}^{упр} = \eta_{LS} \exp(\ 2i\delta_{LS}\ ) = T_{LS} + 1, \tag{1.6.5}$$

$$T_L{}^{\{f\}} = \eta_{LS} \exp\ (\ 2i\delta_L{}^{\{f\}}\ ) -1\ \ ,$$

можно получить ряд интересных соотношений непосредственно для фазовых сдвигов $\delta_{LS}$ и коэффициентов неупругости $\eta_{LS}$. При этом будем полагать фазовые сдвиги $\delta$ действительными, то есть пренебрежем поглощением в каналах с фиксированной схемой Юнга {f}. В таком случае получаем

$$\delta_{L,S=1/2} = 1/2\ \delta_L{}^{\{43\}} + 1/2\ \delta_L{}^{\{421\}}\ , \tag{1.6.6}$$

$$\eta_{L,S=1/2} = |\ \mathrm{Cos}(\ \delta_L{}^{\{43\}} - \delta_L{}^{\{421\}}\ )\ |\ , \qquad \delta_{L,S=3/2} = \delta_L{}^{\{421\}}\ \ , \quad \eta_{L,S=3/2} = 1\ .$$

Отметим, что сделанное выше предположение не является принципиальным. При желании можно учесть возможную комплексность фазовых сдвигов, что приведет к заметному усложнению формул, но не изменит сути вопроса.

Из уравнения (1.6.6) вытекает, что матрица рассеяния в каналах с S=1/2 "не унитарна", что обусловлено связью упругого рассеяния с каналами перезарядки $p + {}^6Li \to n + {}^6Be$ и $n + {}^6Li \to p + {}^6He$, и неупругого рассеяния с возбуждением уровня TS=01 ядра $^6Li$. При этом происходит переворачивание спин - изоспина ядра $^6Li$ аналогично реакции $N+d \to N+d_s$, с образованием синглетного дейтрона $d_s$ с квантовыми числами TS=10 [17,23]. Из приведенных выше соотношений находим выражение для соответствующих элементов S матрицы:

$S_{L,S=1/2}^{пер}(p+^6Li \text{-->} n+^6Be)=(1/2\text{-}1/2 1 1|1/2\ 1/2)(1/2T_L^{\{421\}}-1/2T_L^{\{43\}})$,

$S_{L,S=1/2}^{пер}(n+^6Li \text{-->} p+^6He)=(1/2 1/2 1\text{-}1|1/2\ 1/2\ )(1/2\ T_L^{\{421\}}-1/2\ T_L^{\{43\}})$,

$S_{L,S=1/2}^{возб}(p+^6Li \text{-->} n+^6Li^*)=(1/2 1/2 1 0|1/2\ 1/2\ )(1/2\ T_L^{\{421\}}-1/2\ T_L^{\{43\}})$,

$S_{L,S=1/2}^{возб}(n+^6Li \text{-->} n+^6Li^*)=(1/2\text{-}1/2 1 0|1/2\text{-}1/2)(1/2T_L^{\{421\}}-1/2\ T_L^{\{43\}})$.

(1.6.7)

При учете этих неупругих каналов парциальная S матрица, как и полагается, оказывается унитарной

$$| S_{L,S=1/2}^{упр} |^2 + | S_{L,S=1/2}^{пер} |^2 + | S_{L,S=1/2}^{возб} |^2 = 1 , \qquad (1.6.8)$$

$$| S_{L,S=3./2}^{упр} |^2 = 1 .$$

На этой основе легко вычисляются дифференциальные сечения рассеяния соответствующих реакций. Так, например, сечение реакции $p+^6Li$ --> $n + ^6Be$ имеет вид [23]

$$d\sigma/d\Omega=1/4P_0^2 \frac{(2s_p+1)(2s_{Li}+1)}{(2s_n+1)(2s_{Be}+1)} (t_n\tau_n t_{Be}\tau_{Be}|1/2 1/2)^2 \sum (2L+1)P_L(\cos\theta)$$
$$|<\{1\}s_p t_p\{42\}s_{Li}t_{Li}|\{43\}1/2\ 1/2> <\{1\}s_n t_n\{42\}s_{Be}t_{Be}|\{43\}1/2 1/2>T^{\{43\}}{}_L+$$
$$+<\{1\}s_p t_p\{42\}s_{Li}t_{Li}|\{421\}1/2 1/2> <\{1\}s_n t_n\{42\}s_{Be}t_{Be}|\{421\}1/2 1/2>T^{\{421\}}{}_L|^2.$$

Установление непосредственной связи между упругим рассеянием и реакциями и, связанная с этим возможность простого вычисления сечений соответствующих неупругих процессов, представляются одним из несомненных достоинств используемого формализма [23].

## 1.7. Сечения упругого рассеяния ядерных частиц

В заключительном параграфе приведем некоторые основные выражения для расчетов сечений упругого рассеяния в ядерных системах с различным спином. Наиболее простые формулы получаются в случае рассеяния частиц со спином ноль, поскольку отсутствует спин - орбитальное расщепление фаз. Если частицы не тождественны, то сечение определяется наиболее просто, как квадрат модуля амплитуды рассеяния [24]

$$\frac{d\sigma(\theta)}{d\Omega} = \left| f(\theta) \right|^2 \quad , \qquad (1.7.1)$$

где сама амплитуда представляется в виде суммы кулоновской и ядерной амплитуд

$$f(\theta) = f_c(\theta) + f_N(\theta) \, , \qquad (1.7.2)$$

которые выражаются через ядерные $\delta_L$ и кулоновские $\sigma_L$ фазы рассеяния

$$f_c(\theta) = -\left( \frac{\eta}{2k \;\; Sin^2(\theta/2)} \right) \exp\{-i\eta \ln[Sin^2(\theta/2)] + 2i\sigma_0\} \quad ,$$

$$f_N(\theta) = \frac{1}{2ik} \sum_L (2L+1) \exp(2i\sigma_L)[\exp(2i\delta_L) - 1] P_L(Cos\theta) \quad . \qquad (1.7.3)$$

Здесь $P_L(x)$ - полином Лежандра

$$P_n(x) = \frac{1}{2^n n!} \frac{d^n}{dx^n} (x^2 - 1)^n \quad ,$$

$\eta = \dfrac{Z_1 Z_2 \mu}{k\hbar^2}$ - кулоновский параметр, $\mu$ - приведенная масса, k - волновое число относительного движения частиц - $k^2 = 2\mu E/\hbar^2$, E - энергия сталкивающихся частиц в центре масс и $S_L = \exp(2i\delta_L)$ - матрица рассеяния. В общем случае, фазы рассеяния являются комплексными величинами. Кулоновские фазы выражаются через Гамма - функцию

$$\sigma_L = \arg\{\Gamma(L+1+i\eta)\} \qquad (1.7.4)$$

и удовлетворяют рекуррентному процессу

$$\sigma_L = \sigma_{L+1} - Arctg\left( \frac{\eta}{L+1} \right) \quad . \qquad (1.7.5)$$

Откуда сразу можно получить следующее выражение для этих фаз

$$\alpha_L=\sigma_L - \sigma_{L-1} = \sum_{n=1}^{L} \mathrm{Arctg}\left(\frac{\eta}{n}\right), \qquad \alpha_0=0 . \tag{1.7.6}$$

Величина $\alpha_L$ используется в преобразованных выражениях (1.7.3), если вынести общий множитель $\exp(2i\sigma_0)$. Более подробно методы вычисления кулоновских функций и фаз рассеяния изложены в Приложении 1.

В случае рассеяния тождественных бозонов, например, ядер $^4$He, формула сечения (1.7.1) преобразуется к виду [25]

$$\frac{d\sigma(\theta)}{d\Omega} = \left|f(\theta)+f(\pi-\theta)\right|^2 , \tag{1.7.7}$$

что позволяет учитывать эффекты, которые дает симметризация волновых функций такой системы. В случае процессов рассеяния тождественных фермионов с полуцелым спином знак плюс в (1.7.7) заменяется на минус.

Рассмотрим теперь рассеяние в системе частиц с полным спином 1/2, т.е. одна частица имеет нулевой, а вторая полуцелый спин и учтем спин - орбитальное расщепление фаз. Такое рассеяние имеет место в ядерных системах N$^4$He, $^3$H$^4$He и т.д. Сечение рассеяния представляется в виде [24]

$$\frac{d\sigma(\theta)}{d\Omega} = \left|A(\theta)\right|^2 + \left|B(\theta)\right|^2 , \tag{1.7.8}$$

где

$$A(\theta) = f_c(\theta) + \frac{1}{2ik}\sum_{L=0}^{\infty}\{(L+1)S_L^+ + LS_L^- - (2L+1)\}\exp(2i\sigma_L)P_L(\mathrm{Cos}\theta),$$

$$B(\theta) = \frac{1}{2ik}\sum_{L=0}^{\infty}(S_L^+ - S_L^-)\exp(2i\sigma_L)P_L^1(\mathrm{Cos}\theta) \cdot \tag{1.7.9}$$

Здесь $S_L^{\pm} = \exp(2i\delta_L^{\pm})$ - матрица рассеяния, а знаки “±“ соответ-ствует полному моменту системы J=L±1/2, $P_n^m(x)$ - присоединенные полиномы Лежандра [26]

$$P_n^m(x) = (1-x^2)^{m/2}\frac{d^m P_n(x)}{dx^m} .$$

В случае системы частиц, когда одна из них имеет нулевой спин, а вторая равный 1, например, для $^2H^4He$ системы, формулы сечения с учетом только спин - орбитальных сил записываются в виде [24]

$$\frac{d\sigma(\theta)}{d\Omega} = 1/3\{|A|^2 + 2|B|^2 + |C|^2 + |D|^2 + |E|^2\} , \qquad (1.7.10)$$

где амплитуды рассеяния

$$A = f_c(q) + \frac{1}{2ik}\sum_{L=0}\{(L+1)\alpha_L^+ + L\alpha_L^-\}\exp(2i\sigma_L)P_L(Cos\theta) ,$$

$$B = f_c(\theta) + \frac{1}{4ik}\sum_{L=0}\{(L+2)\alpha_L^+ + (2L+1)\alpha_L^0 + (L-1)\alpha_L^-\}\exp(2i\sigma_L)P_L(Cos\theta),$$

$$C = \frac{1}{2ik\sqrt{2}}\sum_{L=1}\{\alpha_L^+ - \alpha_L^-\}\exp(2i\sigma_L)P_L^1(Cos\theta) , \qquad (1.7.11)$$

$$D = \frac{1}{2ik\sqrt{2}}\sum_{L=1}\frac{1}{L(L+1)}\{L(L+2)\alpha_L^+ - (2L+1)\alpha_L^0 -$$

$$-(L-1)(L+1)\alpha_L^-\}\exp(2i\sigma_L)P_L^1(Cos\theta),$$

$$E = \frac{1}{4ik}\sum_{L=2}\frac{1}{L(L+1)}\{L\alpha_L^+ - (2L+1)\alpha_L^0 + (L+1)\alpha_L^-\}\exp(2i\sigma_L)P_L^2(Cos\theta),$$

Здесь определена величина $\alpha_L = S_L - 1$ - для каждой спиновой переменной при J=L±1 и J=L. Существует и другая форма записи выражений для сечения, представленная через производные полиномов Лежандра [27].

При рассеянии нетождественных частиц с полуцелым спином, например, $N^3H$, $N^3He$ и т.д., с учетом тензорных взаимодействий, дифференциальное сечение рассеяния имеет более сложный вид, и в формулы для сечений входят, как фазы рассеяния, так и параметр смешивания состояний [28]

$$\frac{d\sigma(\theta)}{d\Omega} = \frac{1}{2k^2}\{|A|^2 + |B|^2 + |C|^2 + |D|^2 + |E|^2 + |F|^2 + |G|^2 + |H|^2\} \tag{1.7.12}$$

где амплитуды рассеяния записываются

$$A = f_c' + \frac{1}{4}\sum_{L=0}^{\infty} P_L(\text{Cos}\theta)\Big\{-\sqrt{L(L-1)}U^{L-1}_{L,1;L-2,1} + (L+2)U^{L+1}_{L,1;L,1} +$$

$$+(2L+1)U^{L}_{L,1;L,1} + (L-1)U^{L-1}_{L,1;L,1} - \sqrt{(L+1)(L+2)}\,U^{L+1}_{L,1;L+2,1}\Big\},$$

$$B = f_c' + \frac{1}{4}\sum_{L=0}^{\infty} P_L(\text{Cos}\theta)\Big\{\sqrt{L(L-1)}U^{L-1}_{L,1;L-2,1} + (L+1)U^{L+1}_{L,1;L,1} +$$

$$+(2L+1)U^{L}_{L,0;L,0} + LU^{L-1}_{L,1;L,1} + \sqrt{(L+1)(L+2)}\,U^{L+1}_{L,1;L+2,1}\Big\}, \tag{1.7.13}$$

$$C = \frac{1}{4}\sum_{L=0}^{\infty} P_L(\text{Cos}\theta)\Big\{\sqrt{L(L-1)}U^{L-1}_{L,1;L-2,1} + (L+1)U^{L+1}_{L,1;L,1} -$$

$$+(2L+1)U^{L}_{L,0;L,0} + LU^{L-1}_{L,1;L,1} + \sqrt{(L+1)(L+2)}\,U^{L+1}_{L,1;L+2,1}\Big\},$$

$$D = -\frac{1}{4}i\text{Sin}\theta\sum_{L=1}^{\infty} P_L'(\text{Cos}\theta)/\sqrt{L(L+1)}\Big\{-\sqrt{(L+1)(L-1)}U^{L-1}_{L,1;L-2,1} +$$

$$+\sqrt{L(L+1)}U^{L+1}_{L,1;L,1} - (2L+1)U^{L}_{L,1;L,0} - \sqrt{L(L+1)}U^{L-1}_{L,1;L,1} + \sqrt{L(L+2)}\,U^{L+1}_{L,1;L+2,1}\Big\}$$

$$E = -\frac{1}{4}i\text{Sin}\theta\sum_{L=1}^{\infty} P_L'(\text{Cos}\theta)/\sqrt{L(L+1)}\Big\{-\sqrt{(L+1)(L-1)}U^{L-1}_{L,1;L-2,1} + \sqrt{L(L+1)}U^{L+1}_{L,1;L,1}$$

$$F = -\frac{1}{4}\text{Sin}^2\theta\sum_{L=2}^{\infty} P_L''(\text{Cos}\theta)/\sqrt{(L-1)L(L+1)(L+2)}\Big\{-\sqrt{(L+1)(L+2)}U^{L-1}_{L,1;L-2,1} +$$

$$+\sqrt{\frac{L(L-1)(L+2)}{L+1}}U^{L+1}_{L,1;L,1} - (2L+1)\sqrt{\frac{(L-1)(L+2)}{L(L+1)}}U^{L}_{L,1;L,1} +$$

$$+\sqrt{\frac{(L-1)(L+1)(L+2)}{L}}U^{L-1}_{L,1;L,1} - \sqrt{L(L-1)}\,U^{L+1}_{L,1;L+2,1}\Big\},$$

$$G = -\frac{1}{4}i\text{Sin}\theta\sum_{L=1}^{\infty} P_L'(\text{Cos}\theta)/\sqrt{L(L+1)}\left\{\sqrt{(L-1)(L+1)}U^{L-1}_{L,1;L-2,1} + (L+2)\sqrt{\frac{L}{L+1}}U^{L+1}_{L,1;L,1} -\right.$$

$$-\frac{(2L+1)}{\sqrt{L(L+1)}}U^{L}_{L,1;L,1}-(L-1)\sqrt{\frac{(L+1)}{L}}U^{L-1}_{L,1;L,1}-\sqrt{L(L+2)}U^{L+1}_{L,1;L+2,1}-(2L+1)U^{L}_{L,0;L,1}\Big\},$$

$$H=-\frac{1}{4}i\mathrm{Sin}\theta\sum_{L=1}^{\infty}P^{'}_{L}(\mathrm{Cos}\theta)/\sqrt{L(L+1)}\Big\{\sqrt{(L-1)(L+1)}U^{L-1}_{L,1;L-2,1}+(L+2)\sqrt{\frac{L}{L+1}}U^{L+1}_{L,1;L,1}-$$

$$-\frac{(2L+1)}{\sqrt{L(L+1)}}U^{L}_{L,1;L,1}-(L-1)\sqrt{\frac{(L+1)}{L}}U^{L-1}_{L,1;L,1}-\sqrt{L(L+2)}U^{L+1}_{L,1;L+2,1}+(2L+1)U^{L}_{L,0;L,1}\Big\}$$

Здесь матрица рассеяния представляется в виде

$$U^{J}_{L,S;L',S'}=U^{J}_{L',S';L,S}=\exp[i(\alpha_L+\alpha_{L'})](\delta_{L,L'}\delta_{S,S'}-S^{J}_{L',S';L,S})\quad,$$

$$\alpha_0=0\;,\qquad \alpha_L=\sum_{k}^{L}\mathrm{arctg}(\eta/k)\;,\qquad\qquad (1.7.14)$$

где $\delta_{k,k}$ - дельта функция, штрихи у полиномов Лежандра обозначают производные, а кулоновская амплитуда записана в несколько другой форме

$$f^{'}_{c}(\theta)=\left(\frac{i\eta}{2\ \mathrm{Sin}^2(\theta/2)}\right)\exp\{-i\eta\ln[\mathrm{Sin}^2(\theta/2)]\}\quad.$$

В случае рассеяния частиц со спином 1/2 и 1, например, $p^2H$ $p^6Li$ и т.д., формулы для сечений упругого рассеяния приведены в работах [29].

Процессы рассеяния тождественных частиц с полуцелым спином, например, pp, $^3H^3H$, $^3He^3He$ и т.д. описаны в работе [30]. Во всех случаях, если оказываются открытыми неупругие процессы, фазы рассеяния становятся комплексными и мнимая часть учитывает переход сталкивающихся частиц в неупругий канал.

---

1. Айзенберг И., Грайнер В. - Механизмы возбуждения ядра. М. Атомиздат. 1973. 347с. (Eisenberg J.M., Greiner W. - Excitation mechanisms of the nucleus electromagnetic and wear interactions. North - Holland Publ. Comp. Amsterdam - London. 1970).

2. Tombrello T., Parker P.D. - Phys. Rev., 1963, v.131, p.2578.

3. Mertelmeir T., Hofmann H.M. - Nucl. Phys., 1986, v.A459, p.387.

4. Дубовиченко С.Б., Джазаиров - Кахраманов А.В. - ЯФ, 1995, т.58, с.635; ЯФ, 1995, т.58, с.852.

5. Варшалович Д.А., Москалев А.Н., Херсонский В.К. - Квантовая теория углового момента. Л. Наука. 1975. 436с.

6. Buck B., Baldock R.A., Rubio J.A. - J. Phys., 1985, v.11G, p.L11; Buck B., Merchant A.C. - J. Phys., 1988, v.14G, p.L211.

7. Ахиезер А.И., Ситенко А.Г., Тартаковский В.К. - Электродинамика ядер. Киев. Наукова Думка. 1989. 423с.

8. Bergstrom J.C. - Nucl. Phys., 1980, v.A341, p.13.

9. Kukulin V.I., Krasnopol'sky V.M., Voronchev V.T., Sazonov P.B. - Nucl. Phys., 1984, v.A417, p.128; 1986, v.A453, p.365; Kukulin V.I., Voronchev V.T., Kaipov T.D., Eramzhyan R.A. - Nucl. Phys., 1990, v.A517, p.221.

10. Дубовиченко С.Б., Джазаиров - Кахраманов А.В. - ЯФ, 1994, т.57, №5, с.784.

11. Марчук Г.И., Колесов В.Е. - Применение численных методов для расчета нейтронных сечений. М., Атомиздан, 1970, 304с.

12. Хюльтен Л., Сугавара М. - В кн. Строение атомного ядра. М., ИЛ., 1959, С.9. (In. Structure of atomic nuclei. Ed. Flugge S., Springer -Verlag. Berlin-Gottingen-Heidelberg. 1957).

13. Браун Д.Е., Джексон А.Д. - Нуклон - нуклонные взаимодействия. Москва, Атомиздат, 1979. 246С. (Brown G.E., Jackson A.D. The nucleon-nucleon interaction. North-Holland Pablishing Company. Amsterdam. 1976).

14. Reid R.V. - Ann. Phys. 1968. v.50. p.411.

15. Кукулин В.И., Краснопольский В.М., Померанцев В.Н., Сазонов П.Б. - ЯФ, 1986, т.43, с.559; Krasnopol'sky V.M., Kukulin V.I., Pomerantsev V.N., Sazonov P.B. - Phys. Lett., 1985, v.165B, p.7.

16. Merchant A.C. , Rowley N. - Phys. Lett., 1985, v.150B, p.35.

17. Дубовиченко С.Б., Джазаиров - Кахраманов А.В., Сахарук А.А. - ЯФ. 1993, т.56, № 8, с.90.

18. Дубовиченко С.Б., Джазаиров - Кахраманов А.В. - ЯФ, 1993, т.56, № 2, с.87.

19. Platner D. - In: Europ. Few Body Probl. Nucl. Part. Phys. Sesimbra., 1980, p.31; Platner G.R., Bornard M., Alder K. - Phys. Lett., 1976, v.61B, p.21; Bornard M., Platner G.R., Viollier R.D., Alder K. - Nucl. Phys., 1978, v.A294, p.492; Lim T. - Phys. Rev., 1976, v.C14, p.1243; Phys. Lett., 1975, v.56B, p.321; 1973, v.47B, p.397.

20. Benaksas D., Drickley D., Frerejacque D. - Phys. Rev., 1966, v.148, p.1327; McGurk N.J., Fiedeldey H. - Nucl. Phys., 1977, v.A281, p.310; Муфазанов В.М., Троицкий В.Е. - ЯФ, 1981, т.33, с.1461.

21. Ericson T.E. - Nucl. Phys., 1984, v.A416, p.281; Ericson T.E., Rosa - Costa M. - Nucl. Phys., 1983, v.A405, p.497; Ann. Rev. Nucl. Part. Sci., 1985, v.35, p.271.

22. Балдин А.М. - В кн. Электромагнитные взаимодействия и структура элементарных частиц. М., Мир, 1969, с.5.

23. Искра В., Мазур А.И., Неудачин В.Г., Нечаев Ю.И., Смирнов Ю.Ф. - УФЖ, 1988, т.32, с.1141; Искра В., Мазур А.И., Неудачин В.Г., Смирнов Ю.Ф. - ЯФ, 1988, т.48, с.1674; Неудачин В.Г., Померанцев В.Н., Сахарук А.А. - ЯФ, 1990, т.52, с.738; Кукулин В.И., Неудачин В.Г., Померанцев В.Н., Сахарук А.А. - ЯФ, 1990, т.52, с.402.

24. Ходгсон П.Е. - Оптическая модель упругого рассеяния. М., Атомиздат, 1966, 230с. (Hodgson P.E. - The optical vodel of elastic scattering. Clarendon press, Oxford, 1963).

25. Bussell J.L., Phillips Jr.G.C., Reich C.W. - Phys. rev., 1956, v.104, p.135; Nilson R., Jentschke W.K., Briggs G.R., Kerman R.O., Snyder J.N. - Phys. Rev., 1958, v.109, p.850; Tombrello T.A., Senhouse L.S. - Phys. Rev., 1963, v.129, p.2252; Darriulat P., Igo G., Pugh G. - Phys. Rev., 1965, v. 137, p.B315.

26. Янке Е., Емде Ф., Леш Ф. - Специальные функции. М., Наука, 1968, 344с. (Janke - Emde - Losch. - Tafeln hoherer funktionen., Stuttgard, 1960.)

27. Bruno M., Cannata F., D'Agostino M., Maroni C., Massa I. - Nuovo Cim., 1982, v.A68, p.35; Darriulat P., Garreta D., Tarrats A., Arvieux J. - Nucl. Phys., 1967, v.A94, p.653; Galonsky A., McEllistrem M.T. - Phys. Rev., 1955, v.98, p.590.

28. Tombrello T.A., Jones C.M., Phillips G.C., Weil J.L. - Nucl. Phys., 1962, v.39, p.541.

29. Arvieux J. - Nucl. Phys., 1967, v.A102, p.513; Van Oers W.T.H., Brockman K.W. - Nucl. Phys., 1967, v.A92, p.561; Jenny B., Gruebler W., Schmelzbach P.A., Konig V., Burgi H.R. - Nucl. Phys., 1980, v.A337, p.77.

30. Arndt R.A., Roper L.D., Bryan R.A., Clark R.B., VerWest B.J. - Phys. Rev., 1983, v.D28, p.97.

# 2. КЛАССИФИКАЦИЯ КЛАСТЕРНЫХ СОСТОЯНИЙ

Как уже говорилось во введении, системы кластеров типа $^2H^4He$, $^3H^4He$ и т.д. имеют одну определенную орбитальную схему Юнга, как в связанных состояниях, так и в состояниях рассеяния. В отличие от них в системах более легких кластеров $^2H^2H$, $N^3H$ и т.д. орбитальные состояния в канале с минимальным спином оказываются смешанными по схемам Юнга для состояний рассеяния, в то время, как основные состояния по - прежнему зависят только от одной орбитальной схемы. Именно поэтому потенциалы, извлекаемые из фаз рассеяния таких кластеров, будут эффективно зависеть от двух схем Юнга и не могут быть непосредственно применены для расчетов характеристик связанных состояний ядер $^3H$ и $^4He$ в кластерных моделях.

В предпоследнем параграфе первой главы было показано, что в таком случае фазы рассеяния могут быть представлены в виде полусуммы чистых по схемам Юнга фаз. Одна из таких фаз может быть взята из рассеяния в канале с максимальным спином, чистым по орбитальным схемам. В таком случае, имея смешанную экспериментальную фазу и одну из чистых фаз можно определить другую чистую фазу рассеяния, по которой параметризуется межкластерный потенциал, применимый уже для описания связанных состояний рассматриваемой кластерной системы.

Ниже, в табл.2.1, приведена классификация орбитальных состояний всех легких кластерных системах по схемам Юнга и показано, какие из этих систем являются чистыми, а какие смешаны в состояниях с минимальным спином. Смешанными оказываются состояния не только в легчайших кластерных ядрах, но и в более тяжелых $N^6Li$, $^2H^6Li$ и т.д. системах. Причем, во всех случаях, смешиваются по схемам Юнга только состояния с минимальным спином, а все другие спиновые состояния оказываются чистыми.

Рассмотрим, например, классификацию состояний в легчайшей $N^2H$ системе, чтобы пояснить на каком принципе строится полная волновая функция кластерного ядра и как проводится систематизация различных состояний в том виде, как она приводится в табл.2.1.

В случае $N^2H$ системы спин может принимать два значения 1/2 и 3/2, а изоспин одно - 1/2. Спиновая и изоспиновая волновые функции характеризуются определенными схемами Юнга, обозначаемыми символами $\{f\}_S$ и $\{f\}_T$, которые задают их симметрию относительно перестановок соответствующих координат нуклонов [1].

Так спиновая симметрия характеризуется схемами $\{3\}_S$ при спине S=3/2 и $\{21\}_S$ при S=1/2, а изоспиновая - $\{21\}_T$. Поскольку симметрия спин - изоспиновой волновой функции определяется прямым внутренним произведением $\{f\}_{ST} = \{f\}_S \otimes \{f\}_T$ [1], то в дублетном канале имеем следующие симметрии $\{f\}_{ST} = \{1^3\}+\{21\}+\{3\}$ [2]. В квартетном канале такое произведение дает только одну схему $\{21\}_{ST}$.

Симметрия полной волновой функции, с учетом ее орбитальной компоненты, определяется аналогично $\{f\} =\{f\}_L \otimes \{f\}_{ST}$. Полная волновая функция системы при антисимметризации не обращается тождественно в ноль, только если содержит антисимметричную компоненту $\{1^N\}$, что реализуется при перемножении сопряженных $\{f\}_L$ и $\{f\}_{ST}$. Поэтому схемы $\{f\}_L$, сопряженные к $\{f\}_{ST}$, считаются разрешенными в данном канале. Все остальные симметрии запрещены, так как приводят к нулевой полной волновой функции системы.

Возможные орбитальные схемы Юнга в системе $N=n_1+n_2$ частиц можно определить по теореме Литлвуда [1], как прямое внешнее произведение орбитальных схем каждой из подсистем, что в данном случае дает $\{f\}_L= \{2\}\times\{1\} = \{21\}_L+ \{3\}_L$. Орбитальная схема $\{2\}$ соответствует дейтрону в основном состоянии, а $\{1\}$ определяет нуклон.

Отсюда видно, что в квартетном канале разрешена только орбитальная волновая функция с симметрией $\{21\}_L$, а функция с $\{3\}_L$ оказывается запрещенной, так как произведение $\{21\}_{ST} \otimes \{3\}_L$ не приводит к антисимметричной компоненте волновой функции. В то же время, в дублетном канале имеем $\{3\}_{ST}\otimes\{1^3\}_L=\{1^3\}$ и $\{21\}_{ST}\otimes \{21\}_L \sim \{1^3\}$, и в обоих случаях получаем антисимметричную схему. Тем самым в дублетном канале оказываются разрешенными обе возможные орбитальные схемы Юнга $\{21\}_L$ и $\{3\}_L$ [3,4,5,6].

Именно этот результат приводит к понятию смешивания по орбитальным симметриям в состояниях с определенными ST = 1/2 1/2 при любых L. Поэтому дублетный потенциал взаимодействия, полученный из экспериментальных фаз рассеяния, эффективно зависит от обеих орбитальных схем, в то время, как основное состояние $^3$H или $^3$He соответствует чистой симметрии $\{3\}_L$ [3,4]. Значит, в N$^2$H системе эти потенциалы различны и из взаимодействий рассеяния $V^{\{3\}+\{21\}}$ надо выделять компоненту $V^{\{3\}}$, в принципе применимую для расчетов характеристик основных состояний [3-6].

В частности, при расчете фотопроцессов для описания связанных состояний нужно применять чистый по схемам Юнга потенци-ал, а для состояний рассеяния, например, для конечных состояний фоторазвала - потенциал, непосредственно получаемый из экспериментальных фаз.

Перейдем теперь к рассмотрению классификации орбитальных состояний в системе $^4He^{12}C$, где спин S и изоспин T равны нулю. Возможные орбитальные схемы Юнга определяются по теореме Литлвуда [1], что в данном случае дает {f} = {444} × {4} = {844} + {754} + {664} + {655} + {6442} + {5551} + {5542} + {5443} + {4444} + {7441}+{6541}. Здесь схемы {4} и {444} соответствуют ядрам $^4$He и $^{12}$C в основном состоянии. В соответствии с известными правилами [1], на основе этой теоремы, можно сделать вывод, что разрешенной схемой будет только {4444}, а все остальные орбитальные конфигурации запрещены. В частности, все возможные конфигурации, где в первой строке находится число больше четырех, не могут реализовываться, так как в S - оболочке не может быть больше четырех нуклонов.

Используя правило Элиота [1] можно определить орбитальные моменты, соответствующие различным схемам Юнга. Тогда получим, что момент L=0 может реализоваться для следующих орбитальных схем {4444}, {5551}, {664}, {844} и {6442}. Этот результат можно использовать для оценки числа связанных запрещенных состояний в потенциале основного состояния. Поскольку разрешена только симметрия {4444}, то остальные схемы будут запрещены и такой потенциал должен иметь четыре запрещенных состояния.

При рассмотрении систем $N^6Li$, $^2H^6Li$ и $N^7Li$ (табл.2.1), в качестве орбитальных схем основных состояний ядер $^6$Li и $^7$Li, принимаются схемы Юнга {6} и {42}, и {7} и {43} соответственно, что позволяет рассматривать полный набор возможных орбитальных симметрий.

*Таблица 2.1. Классификация разрешенных (РС) и запрещенных состояний (ЗС) в легких кластерных системах. L - орбитальный момент для связанных состояний, T и S - изоспин и спин.*

| Система | T | S | $\{f\}_S$ | $\{f\}_T$ | $\{f\}_{ST}$ | $\{f\}_L$ | L | $\{f\}_{рс}$ | $\{f\}_{зс}$ |
|---|---|---|---|---|---|---|---|---|---|
| $N^2H$ | 1/2 | 1/2 | {21} | {21} | {111}+{21}+{3} | {3}<br>{21} | 0<br>1 | {3}<br>{21} | -<br>- |
| | | 3/2 | {3} | {21} | {21} | {3}<br>{21} | 0<br>1 | -<br>{21} | {3}<br>- |
| $n^3He$<br>$p^3H$ | 0 | 0 | {22} | {22} | {4}+{22}+<br>+{1111} | {4}<br>{31} | 0<br>1 | {4}<br>- | -<br>{31} |
| | | 1 | {31} | {22} | {31}+{211} | {4}<br>{31} | 0<br>1 | -<br>{31} | {4}<br>- |
| $p^3He$<br>$n^3H$<br>$p^3H$<br>$n^3He$ | 1 | 0 | {22} | {31} | {31}+{211} | {4}<br>{31} | 0<br>1 | -<br>{31} | {4}<br>- |
| | | 1 | {31} | {31} | {4}+{31}+{22}+<br>+{211} | {4}<br>{31} | 0<br>1 | -<br>{31} | {4}<br>- |
| $^2H^2H$ | 0 | 0 | {22} | {22} | {4}+{22}+<br>+{1111} | {4}<br>{31}<br>{22} | 0<br>1<br>0,2 | {4}<br>-<br>{22} | -<br>{31}<br>- |
| | | 1 | {31} | {22} | {31}+{211} | {4}<br>{31}<br>{22} | 0<br>1<br>0,2 | -<br>{31} | {4}<br>-<br>{22} |
| | | 2 | {4} | {22} | {22} | {4}<br>{31}<br>{22} | 0<br>1<br>0,2 | -<br>-<br>{22} | {4}<br>{31}<br>- |
| $^2H^3He$<br>$^2H^3H$ | 1/2 | 1/2 | {32} | {32} | {5}+{41}+{32}+<br>+{311}+{221}+<br>+{2111} | {5}<br>{41}<br>{32} | 0<br>1<br>0,2 | -<br>{41}<br>{32} | {5}<br>-<br>- |
| | | 3/2 | {41} | {32} | {41}+{32}+{311}+<br>+{221} | {5}<br>{41}<br>{32} | 0<br>1<br>0,2 | -<br>-<br>{32} | {5}<br>{41}<br>- |
| $p^4He$<br>$n^4He$ | 1/2 | 1/2 | {32} | {32} | {5}+{41}+{32}+<br>+{311}+{221}+<br>+{2111} | {5}<br>{41} | 0<br>1 | -<br>{41} | {5}<br>- |
| $^4He^2H$ | 0 | 1 | {42} | {33} | {51}+{411}+{33}+<br>+{321}+{2211} | {6}<br>{51}<br>{42} | 0<br>1<br>0,2 | -<br>-<br>{42} | {6}<br>{51}<br>- |
| $^3He^3H$ | 0 | 0 | {33} | {33} | {42}+{222}+{6}+<br>+{3111} | {6}<br>{51}<br>{42}<br>{33} | 0<br>1<br>0,2<br>1,3 | -<br>-<br>-<br>{33} | {6}<br>{51}<br>{42}<br>- |
| | | 1 | {42} | {33} | {51}+{411}+{33}+<br>+{321}+{2211} | {6}<br>{51}<br>{42}<br>{33} | 0<br>1<br>0,2<br>1,3 | -<br>-<br>{42}<br>- | {6}<br>{51}<br>-<br>{33} |
| $^3H^3H$<br>$^3He^3He$<br>$^3He^3H$ | 1 | 0 | {33} | {42} | {51}+{411}+{33}+<br>+{321}+{2211} | {6}<br>{51}<br>{42}<br>{33} | 0<br>1<br>0,2<br>1,3 | -<br>-<br>{42}<br>- | {6}<br>{51}<br>-<br>{33} |
| | | 1 | {42} | {42} | {6}+2{42}+<br>+{51}+{411}+<br>+2{321}+<br>+{222}+{3111} | {6}<br>{51}<br>{42}<br>{33} | 0<br>1<br>0,2<br>1,3 | -<br>-<br>-<br>{33} | {6}<br>{51}<br>{42}<br>- |
| $^4He^3H$ | 1/2 | 1/2 | {43} | {43} | {7}+{61}+{52}+<br>+{511}+{43}+<br>+{421}+{4111}+<br>+{322}+{3211}+<br>+{2221}+{331}+ | {7}<br>{61}<br>{52}<br>{43} | 0<br>1<br>0,2<br>1,3 | -<br>-<br>-<br>{43} | {7}<br>{61}<br>{52}<br>- |
| n$^6$Li<br>p$^6$Li | 1/2 | 1/2 | {43} | {43} | {7}+{61}+{52}+<br>+{511}+<br>+{43}+{421}++{4111}+<br>+{322}+{3211}+<br>+{2221}+{331}+<br>+{61}+{52}+<br>+{511} | {7}<br>{61}<br>{52}<br>{43}<br>{421} | 0<br>1<br>0,2<br>1,3<br>1,2 | -<br>-<br>-<br>{43}<br>{421} | {7}<br>{61}<br>{52}<br>-<br>- |
| | | 3/2 | {52} | {43} | {43}+2{421}+<br>+{331}+{322}+<br>+{3211} | {7}<br>{61}<br>{52}<br>{43}<br>{421} | 0<br>1<br>0,2<br>1,3<br>1,2 | -<br>-<br>-<br>-<br>{421} | {7}<br>{61}<br>{52}<br>{43}<br>- |
| $^4$He$^4$He | 0 | 0 | {44} | {44} | {8}+{62}+{5111}+<br>+{44}+{422}+<br>+{3311}+{2222} | {8}<br>{71}<br>{62}<br>{53}<br>{44} | 0<br>1<br>0,2<br>1,3<br>0,2,4 | -<br>-<br>-<br>-<br>{44} | {8}<br>{71}<br>{62}<br>{53}<br>- |
| p$^7$Li<br>n$^7$Be | 0 | 1 | {53} | {44} | {71}+{611}+{53}+<br>+{521} + {431} +<br>+ {4211}+{332}+<br>+{3221} | {8}<br>{71}<br>{53}<br>{44}<br>{431} | 0<br>1<br>1,3<br>0,2,4<br>1,2,3 | -<br>-<br>-<br>-<br>{431} | {8}<br>{71}<br>{53}<br>{44}<br>- |
| | | 2 | {62} | {44} | {62}+{521}+<br>+{44}+{431}+<br>+{422}+<br>+{3311} | {8}<br>{71}<br>{53}<br>{44}<br>{431} | 0<br>1<br>1,3<br>0,2,4<br>1,2,3 | -<br>-<br>-<br>-<br>- | {8}<br>{71}<br>{53}<br>{44}<br>{431} |
| p$^7$Be<br>n$^7$Li<br>p$^7$Li<br>n$^7$Be | 1 | 1 | {53} | {53} | {8}+2{62}+{71}+<br>+{611}+{53}+{44}+<br>+2{521}+{5111}+<br>+{44}+{332}+<br>+2{431}+2{422}+<br>+{4211}+{3311}+{3221} | {8}<br>{71}<br>{53}<br>{44}<br>{431} | 0<br>1<br>1,3<br>0,2,4<br>1,2,3 | -<br>-<br>-<br>-<br>{431} | {8}<br>{71}<br>{53}<br>{44}<br>- |
| | | 2 | {62} | {53} | {71}+{62}+{611}+<br>+2{53}+2{521}+<br>+2{431}+{422}+ +{4211}<br>+{332} | {8}<br>{71}<br>{62}<br>{53}<br>{44}<br>{431} | 0<br>1<br>0,2<br>1,3<br>0,2,4<br>1,2,3 | -<br>-<br>-<br>-<br>-<br>- | {8}<br>{71}<br>{62}<br>{53}<br>{44}<br>{431} |
| $^2$H$^6$Li | 0 | 0 | {44} | {44} | {8}+{62}+{5111}+<br>+{44}+{422}+<br>+{3311}+{2222} | {8}<br>{71}<br>{62}<br>{53}<br>{521}<br>{44}<br>{431}<br>{422} | 0<br>1<br>0,2<br>1,3<br>1,2<br>0,2,4<br>1,2,3<br>0,2 | -<br>-<br>-<br>-<br>-<br>{44}<br>-<br>{422} | [8]<br>{71}<br>{62}<br>{53}<br>{521}<br>-<br>{431}<br>- |
| | | 1 | {53} | {44} | {71}+{611}+{53}+<br>+{521}+{431}+<br>+ {4211}+{332}+<br>+{3221} | {8}<br>{71}<br>{62}<br>{53}<br>{521}<br>{44}<br>{431}<br>{422} | 0<br>1<br>0,2<br>1,3<br>1,2<br>0,2,4<br>1,2,3<br>0,2 | -<br>-<br>-<br>-<br>-<br>-<br>{431}<br>- | {8}<br>{71}<br>{62}<br>{53}<br>{521}<br>{44}<br>-<br>{422} |
| | | 2 | {62} | {44} | {62}+{521}+<br>+{44}+{431}+<br>+{422}+{3311} | {8}<br>{71}<br>{62}<br>{53}<br>{521}<br>{44}<br>{431}<br>{422} | 0<br>1<br>0,2<br>1,3<br>1,2<br>0,2,4<br>1,2,3<br>0,2 | -<br>-<br>-<br>-<br>-<br>-<br>-<br>{422} | {8}<br>{71}<br>{62}<br>{53}<br>{521}<br>{44}<br>{431}<br>- |

1. Неудачин В.Г., Смирнов Ю.Ф. - Нуклонные ассоциации в легких ядрах. М., Наука, 1969., 414с.; Немец О.Ф., Неудачин В.Г., Рудчик А.Т., Смирнов Ю.Ф., Чувильский Ю.М. - Нуклонные ассоциации в атомных ядрах и ядерные реакции многонуклонных передач. Киев, Наукова Думка, 1988, 488с.

2. Itzykson C., Nauenberg M. - Rev. Mod. Phys., 1966, v.38, p.95.

3. Искра В., Мазур А.И., Неудачин В.Г., Нечаев Ю.И., Смирнов Ю.Ф. - УФЖ, 1988, т.32, с.1141; Искра В., Мазур А.И., Неудачин В.Г., Смирнов Ю.Ф. - ЯФ, 1988, т.48, с.1674; Неудачин В.Г., Померанцев В.Н., Сахарук А.А. - ЯФ, 1990, т.52, с.738; Кукулин В.И., Неудачин В.Г., Померанцев В.Н., Сахарук А.А. - ЯФ, 1990, т.52, с.402; Дубовиченко С.Б., Неудачин В.Г., Смирнов Ю.Ф., Сахарук А.А.- Изв. АН СССР, сер. физ., 1990, т.54, с.911; Neudatchin V.G., Sakharuk A.A., Dubovichenko S.B. - Few Body Sys., 1995, v.18, p.159.

4. Neudatchin V.G., Kukulin V.I., Pomerantsev V.N., Sakharuk A.A. - Phys. Rev., 1992, v.C45. p.1512; Неудачин В.Г., Сахарук А.А., Смирнов Ю.Ф. - ЭЧАЯ, 1993, т.23, с.480.

5. Дубовиченко С.Б., Джазаиров-Кахраманов А.В. - ЯФ, 1992, т.55, № 11, с.2918; Дубовиченко С.Б., Джазаиров-Кахраманов А.В., Сахарук А.А.- ЯФ, 1993, т.56, № 8, с.90.

6. Дубовиченко С.Б., Джазаиров-Кахраманов А.В. - ЯФ, 1990, т.51, № 6, с.1541; ЯФ, 1993, т.56, № 4, с.45; ЭЧАЯ, 1997, т.28, №6, с.1529.

# 3. ПОТЕНЦИАЛЫ МЕЖКЛАСТЕРНОГО ВЗАИМОДЕЙСТВИЯ

Рассмотрим вначале результаты, полученные для характеристик кластерных ядер в чистых по орбитальным симметриям системах $p^3He$, $n^3H$, $^4He^3H$, $^4He^3He$, $^4He^2H$, $^3He^3H$ и $^4He^{12}C$. Затем перейдем к потенциалам, смешанным по схемам Юнга, которые реализуются в системах $N^2H$, $p^3H$, $n^3He$, $^2H^2H$, $^2H^3He$, $^2H^3H$, $N^6Li$, $N^7Li$, $^2H^6Li$. Основное внимание будем уделять результатам, полученным в потенциальной кластерной модели для взаимодействий с запрещенными состояниями и сравнению их с результатами в других моделях, в частности, методе резонирующих групп (МРГ) и трехтельных моделях ядра.

Для расчетов волновых функций рассеяния и связанных состояний, на основе которых определялись характеристики кластерных систем, использовались центральные гауссовы потенциалы с запрещенными состояниями вида [1]

$$V(r) = V_0 \exp(-\alpha r^2) + V_c(r) \,,$$

$$V_c(r) = \begin{cases} \dfrac{Z_1 Z_2}{r} & r > R_c \\ Z_1 Z_2 \left(3 - \dfrac{r^2}{R_c^2}\right) \Big/ 2R_c & r < R_c \end{cases} \,. \qquad (3.1.1)$$

В некоторых случаях к ним добавлялось периферическое отталкивание экспоненциальной формы $V_1\exp(-\beta r)$, необходимое для правильного описания отрицательных значений D или F фаз при малых энергиях, если S или P фазы положительны. Потенциалы восстанавливались по фазам рассеяния кластеров так, чтобы одну парциальную волну в максимально широкой области энергий описывала одна гауссойда с определенными параметрами.

## 3.1. Кластерные каналы в ядрах $^6Li$ и $^7Li$

Как видно из таблицы 2.1, орбитальные состояния в системах $^4He^2H$, $^4He^3H$, $^4He^3He$ и $^3He^3H$ для ядер $^6Li$ и $^7Li$, $^7Be$ оказываются чистыми по схемам Юнга. Поэтому потенциалы, полученные на основе фаз рассеяния фрагментов, можно непосредственно использовать для рассмотрения характеристик связанных состояний этих ядер. Результаты будут зависеть только от степени кластеризации ядер в рассматриваемых кластерных каналах. А поскольку вероятность кластеризации таких ядер сравнительно высока, то и результаты расчетов должны в целом передавать экспериментальные данные. Параметры взаимодействий для чистых состояний в $^6$Li и $^7$Li, полученные в работах [1,2], приведены в табл.3.1. Взаимодействия в $^4$He$^3$H и $^4$He$^3$He системах отличаются только кулоновским членом. В $^3$He$^3$H системе при S=0 для D и F фаз используются те же потенциалы, что для S и P волн соответственно.

Для $^3$He$^3$He системы, из-за отсутствия экспериментальных данных, потенциалы строились исключительно по результатам вычисления фаз, полученных в методе резонирующих групп [3]. Параметры этих взаимодействий совпадают с потенциалами для $^3$He$^3$H системы при S=0. Поскольку $^3$He$^3$He это система тождественных частиц, здесь четные L соответствуют только нулевому спину, а нечетные - спину единица.

*Таблица 3.1. Параметры потенциалов в $^4$He$^3$H, $^4$He$^3$He, $^4$He$^2$H и $^3$He$^3$H системах [1,2]. Для $^4$He$^3$H и $^4$He$^3$He систем $\alpha$=0.15747 Фм$^{-2}$ и $R_c$=3.095 Фм. Для $^4$He$^2$H и $^3$He$^3$H - $R_c$ = 0 Фм.*

| $^7$Li ($^7$Be) | | $^6$Li ($^4$He$^2$H) | | | $^6$Li ($^3$He$^3$H) | | | |
|---|---|---|---|---|---|---|---|---|
| $L_J$ | $V_0$ (МэВ) | $L_J$ | $V_0$, (МэВ) | $\alpha$, (Фм$^{-2}$) | S=1 | | S=0 | |
| | | | | | $V_0$, (МэВ) | $\alpha$, (Фм$^{-2}$) | $V_0$, (МэВ) | $\alpha$, (Фм$^{-2}$) |
| S | -67.5 | S | -76.12 | 0.2 | -90 | 0.18 | -85.0 | 0.18 |
| $P_{1/2}$ | -81.92 | $P_0$ | -68.0 | 0.22 | -52.5 | 0.2 | | |
| $P_{3/2}$ | -83.83 | $P_1$ | -79.0 | 0.22 | -65.0 | 0.2 | -74.0 | 0.2 |
| $D_{3/2}$ | -66.0 | $P_2$ | -85.0 | 0.22 | -80.0 | 0.2 | | |
| $D_{5/2}$ | -69.0 | $D_1$ | -63.0 | 0.19 | -72.0 | 0.18 | | |
| $F_{5/2}$ | -75.9 | $D_2$ | -69.0 | 0.19 | -85.0 | 0.18 | | |
| $F_{7/2}$ | -84.8 | $D_3$ | -80.88 | 0.19 | -90.0 | 0.18 | | |

Качество описания фаз показано на рис.3.1, 3.2 и 3.3 вместе с экспериментальными данными из работ [4] для $^4$He$^2$H, [5,6] - $^4$He$^3$H и [7] - $^3$He$^3$H. На рис.3.3 крестиками показаны расчеты $^3$He$^3$H фаз, полученные в методе резонирующих групп [8].

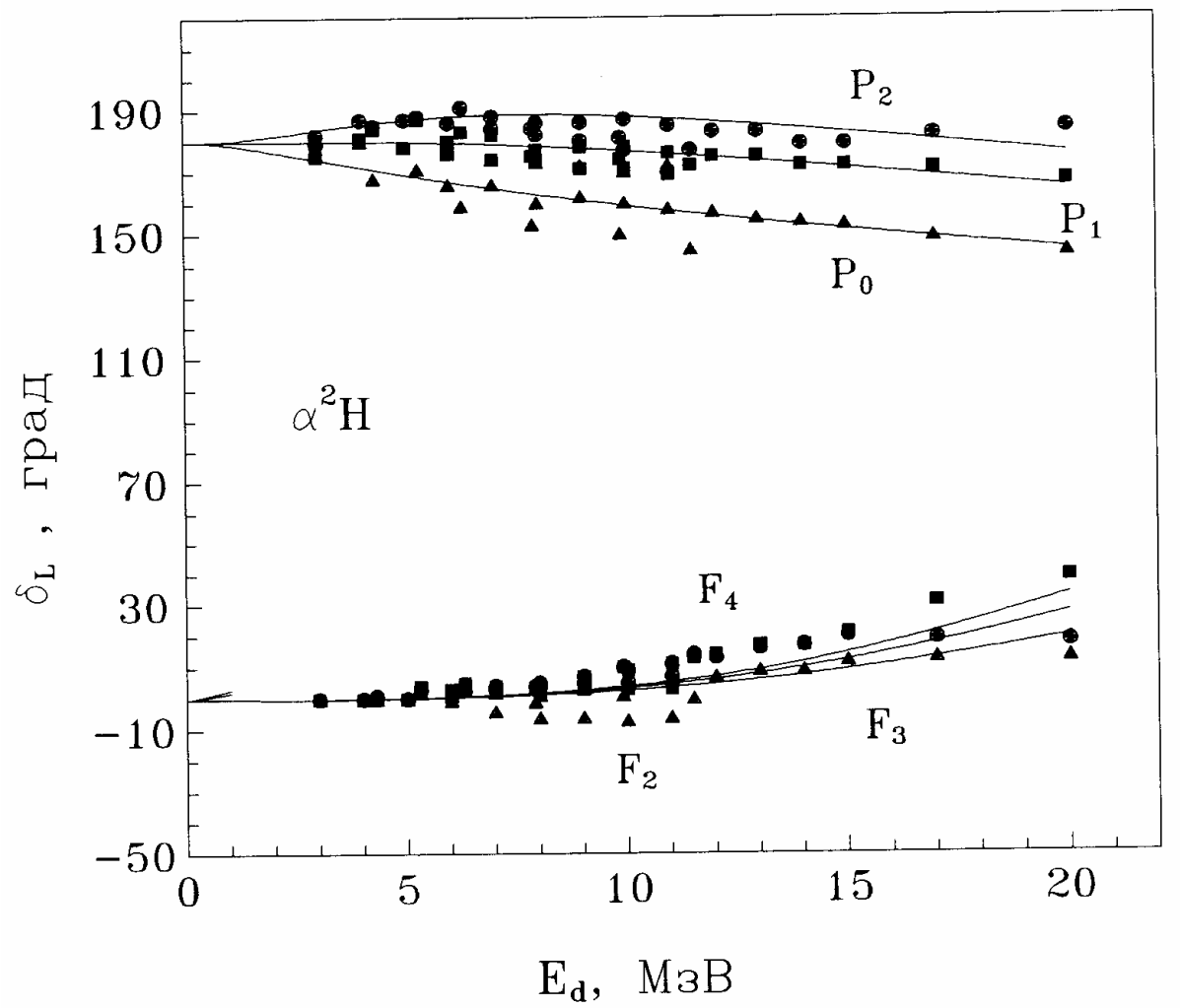

Рис.3.1а. Нечетные фазы упругого $^4He^2H$ рассеяния. Кривые - расчеты для потенциалов с параметрами из табл.3.1. Точки, треугольники, кружки и квадраты - экспериментальные данные работ [4].

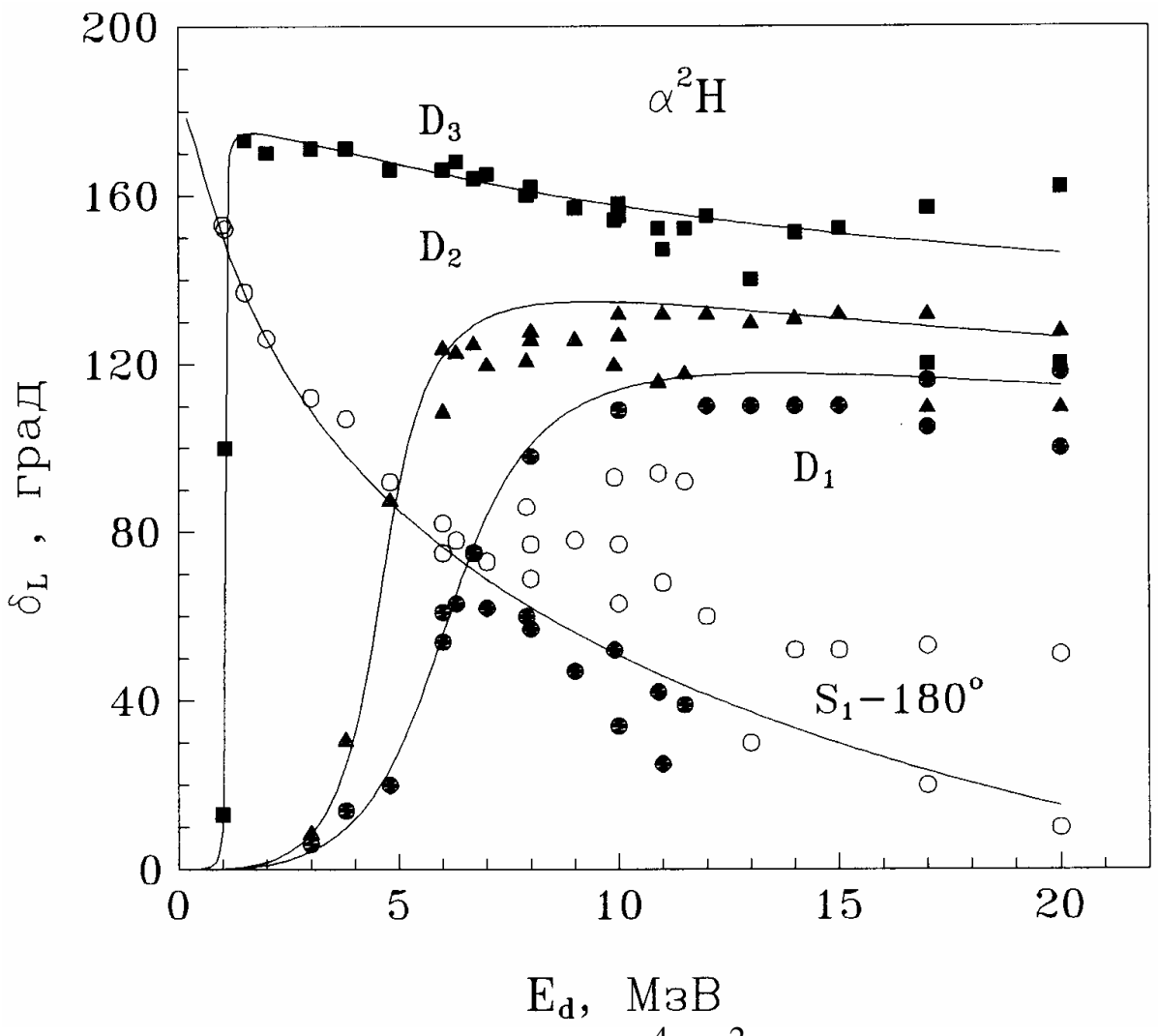

Рис.3.1б. Четные фазы упругого $^4He^2H$ рассеяния. Кривые - расчеты для потенциалов с параметрами из табл.3.1. Точки, треугольники, кружки и квадраты - данные из [4].

Приведенные в табл.3.1 триплетные $^3He^3H$ потенциалы, фазы которых показаны на рис.3.3а непрерывными линиями, неправильно передают энергию связанного состояния $^6Li$. Для получения правильной величины -15.8 МэВ для основного состояния необходимо увеличить глубину взаимодействия до 105 МэВ. Для описания энергии $D_3$ уровня требуется потенциал с глубиной 107.5 МэВ. Фазы этих потенциалов показаны на рис.3.3а штриховыми линиями.

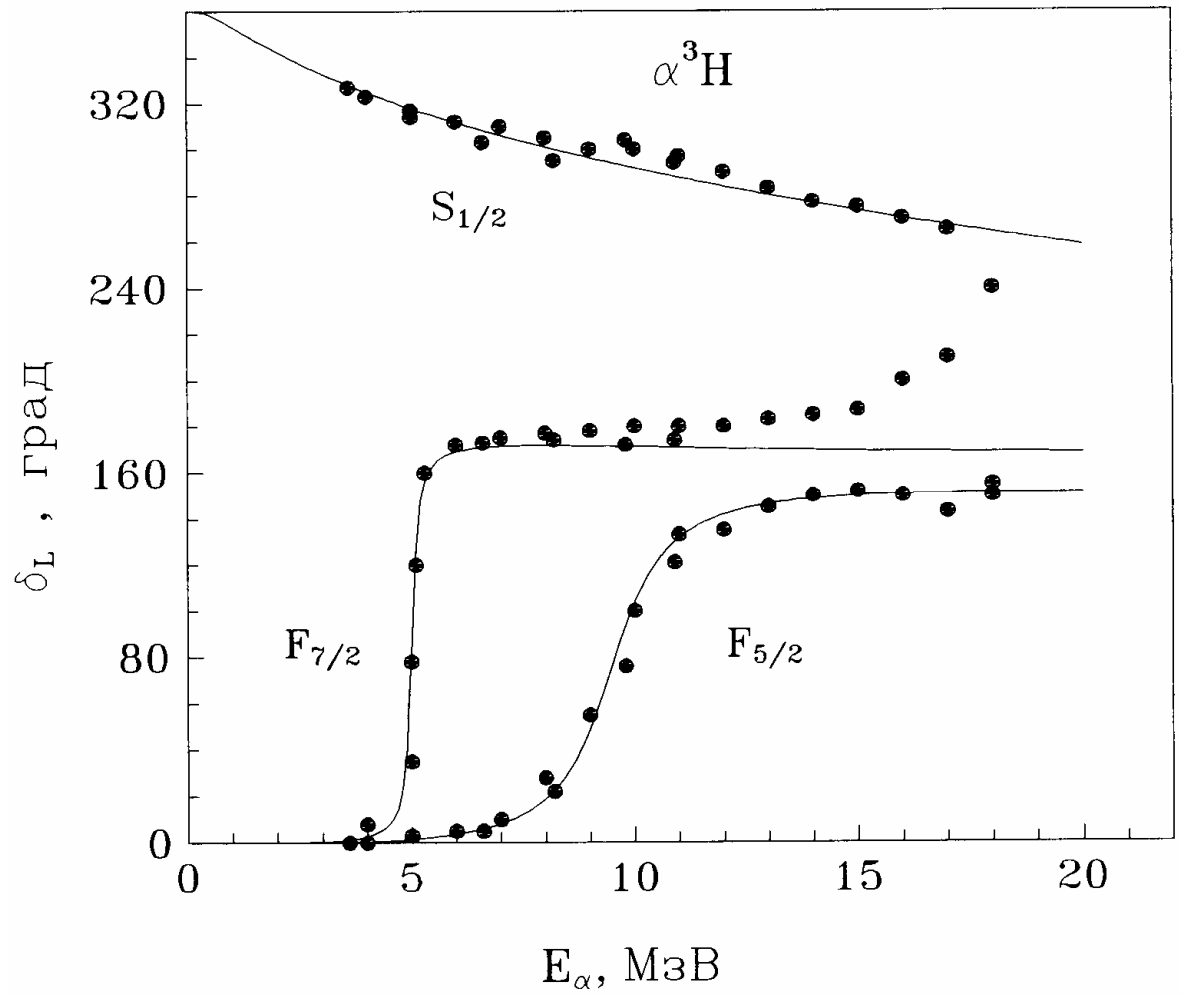

Рис.3.2а. Фазы упругого $^4He^3H$ рассеяния. Кривые - расчеты для потенциалов с параметрами из табл.3.1. Точки и треугольники - экспериментальные данные из работ [5,6].

Потенциал $^4He^2H$ в $D_3$ волне позволяет получить 90° фазы при энергии 0.71 МэВ в хорошем согласии с данными [9], где приведено значение 0.711 МэВ. Сечения $^4He^3H$ и $^4He^2H$ упругого рассеяния при 5.07 МэВ и 10.92 МэВ для первой системы и при 4 МэВ и 6 МэВ для второй показаны на рис.3.4. Как видно, они вполне передают экспериментальные данные работ [4,5]. На рис.3.5 показаны упругие сечения для $^3He^3H$ рассеяния при энергиях 5.79 МэВ, 19.91 МэВ вместе с экспериментальными данными работ [7,10].

В табл.3.2 даны энергии запрещенных состояний для потенциалов, параметры которых приведены выше. Полученные по фазам рассеяния межкластерные взаимодействия использовались далее для вычисления различных характеристик основных состояний $^6Li$, $^7Li$ и $^7Be$ , причем кластерам сопоставлялись определенные свойства соответствующих ядер в свободном состоянии.

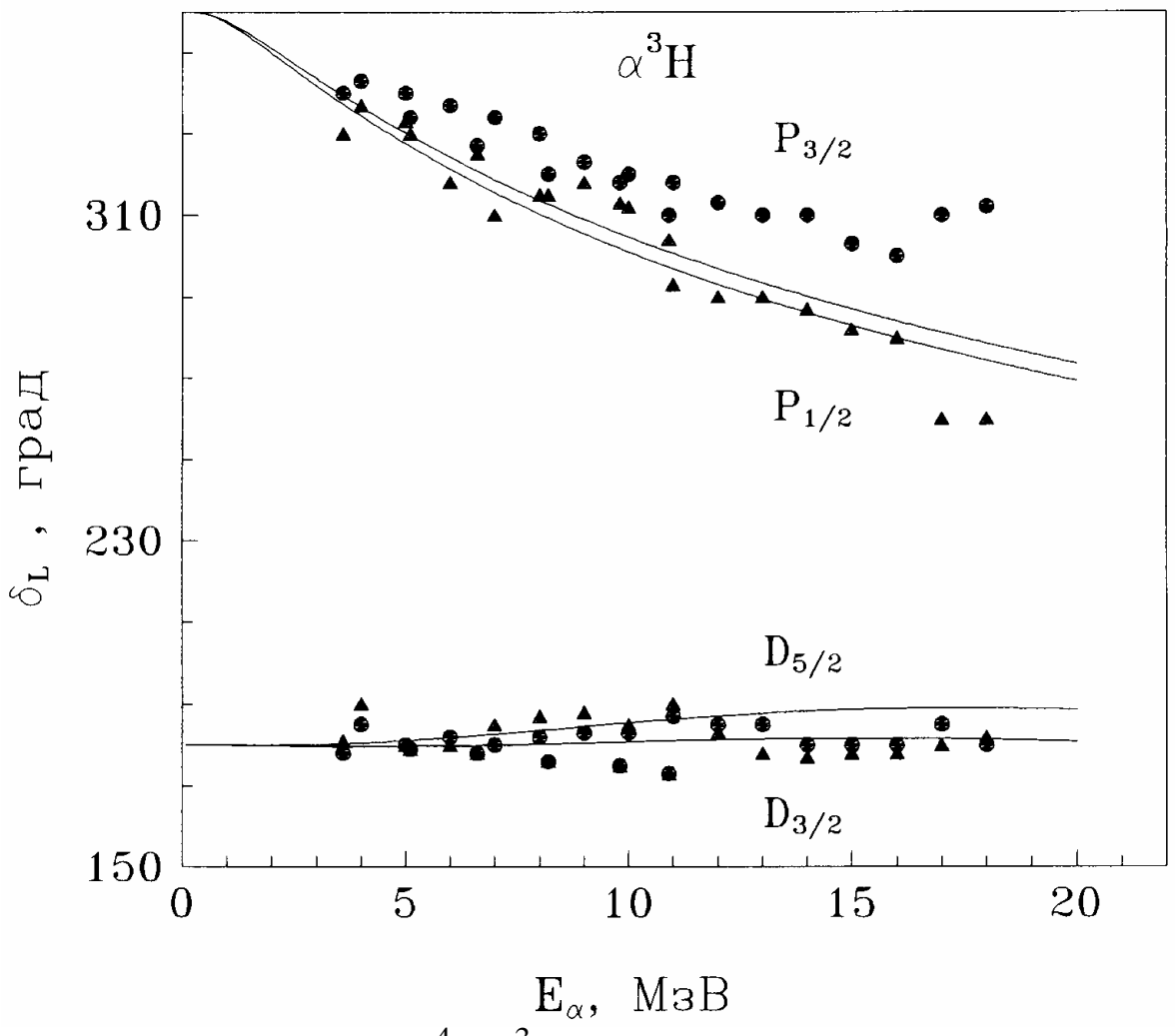

Рис.3.2б. Фазы упругого $^4He^3H$ рассеяния. Кривые - расчеты для потенциалов с параметрами из табл.3.1. Точки и треугольники - экспериментальные данные из работ [5,6].

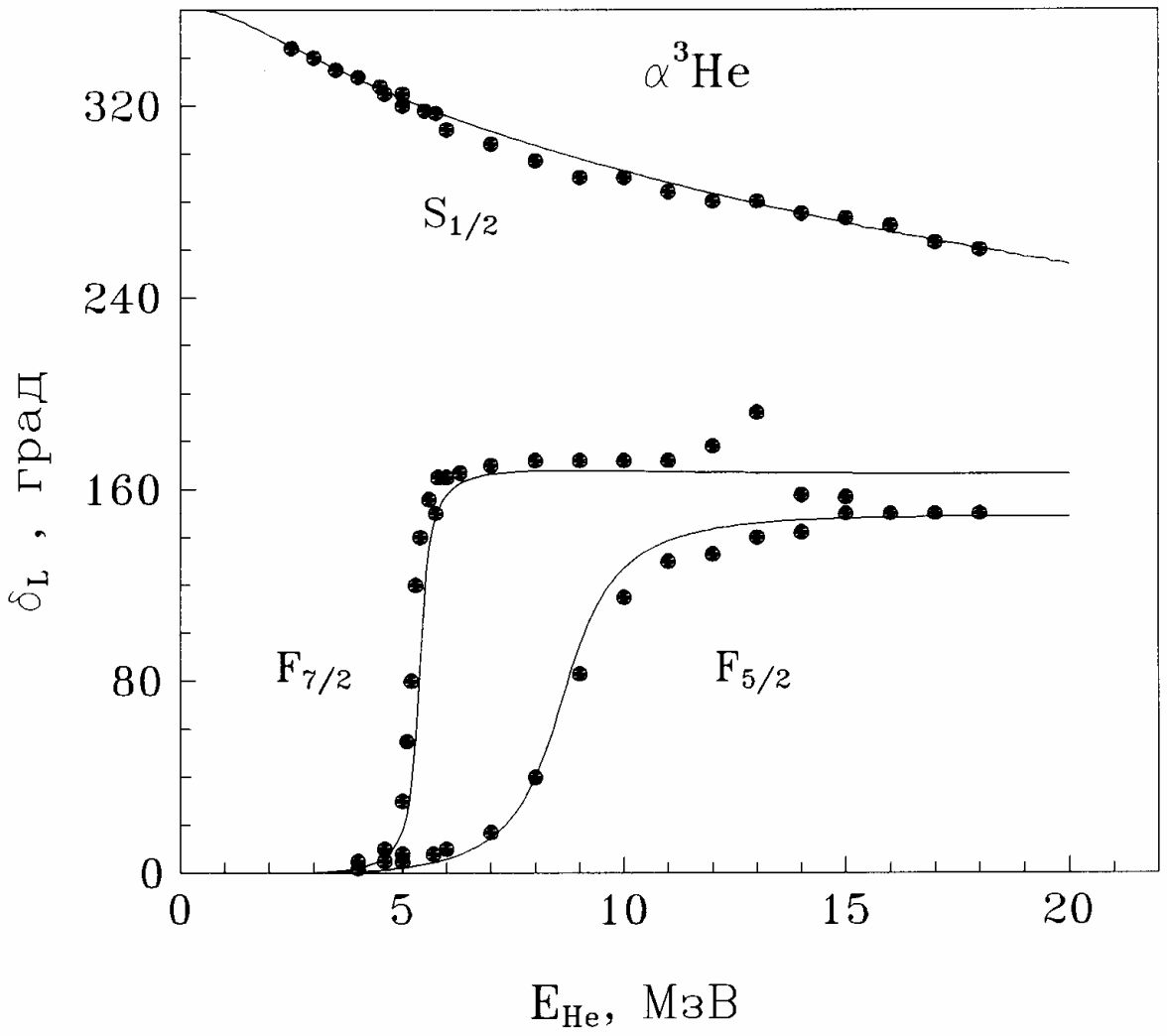

Рис.3.2в. Фазы упругого $^4He^3He$ рассеяния. Кривые - расчеты для потенциалов с параметрами из табл.3.1. Точки и треугольники - экспериментальные данные из работ [5,6].

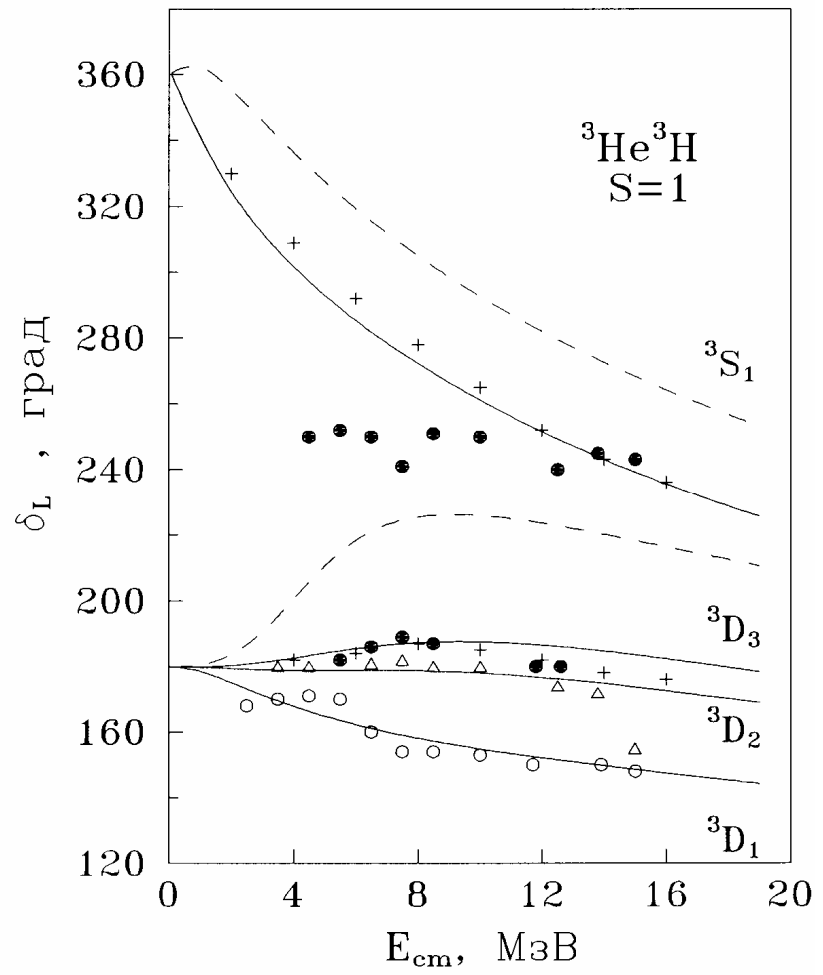

Рис.3.3а. Фазы упругого $^3He^3H$ рассеяния. Кривые - расчеты для потенциалов с параметрами из табл.3.1. Точки, треугольники и кружки - экспериментальные данные из работ [7]. Крестики - МРГ вычисления из [8].

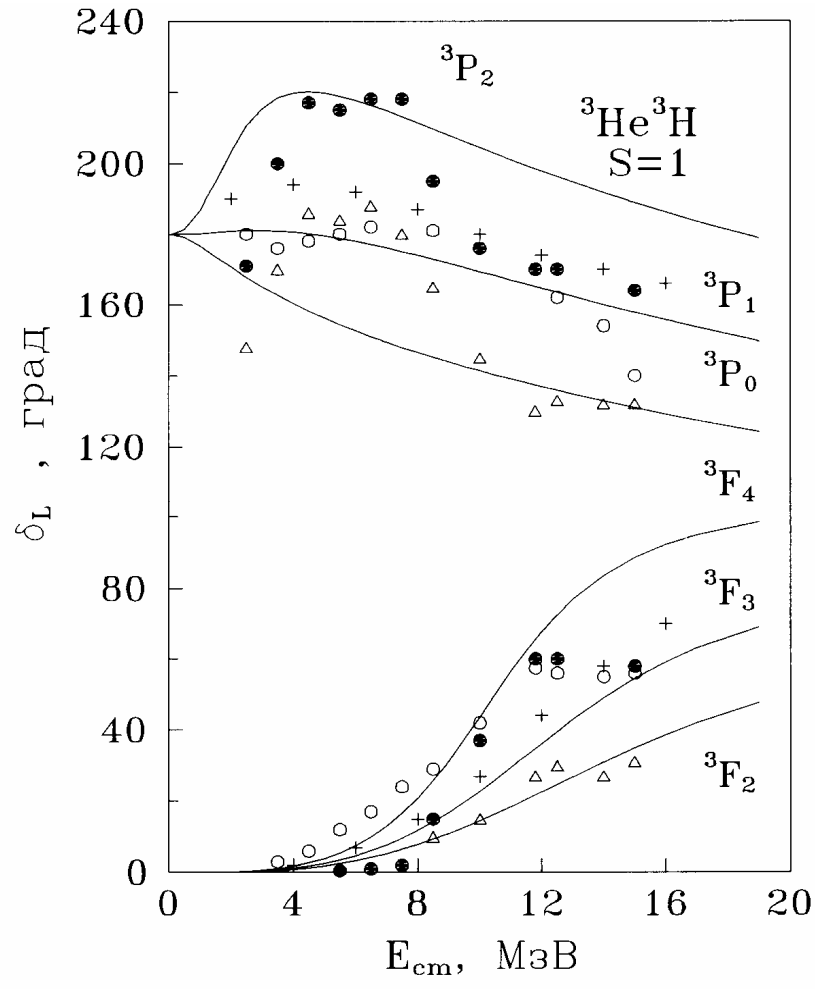

Рис.3.3б. Фазы упругого $^3He^3H$ рассеяния. Кривые - расчеты для потенциалов с параметрами из табл.3.1. Точки, треугольники и кружки - экспериментальные данные из работ [7]. Крестики - МРГ вычисления из [8].

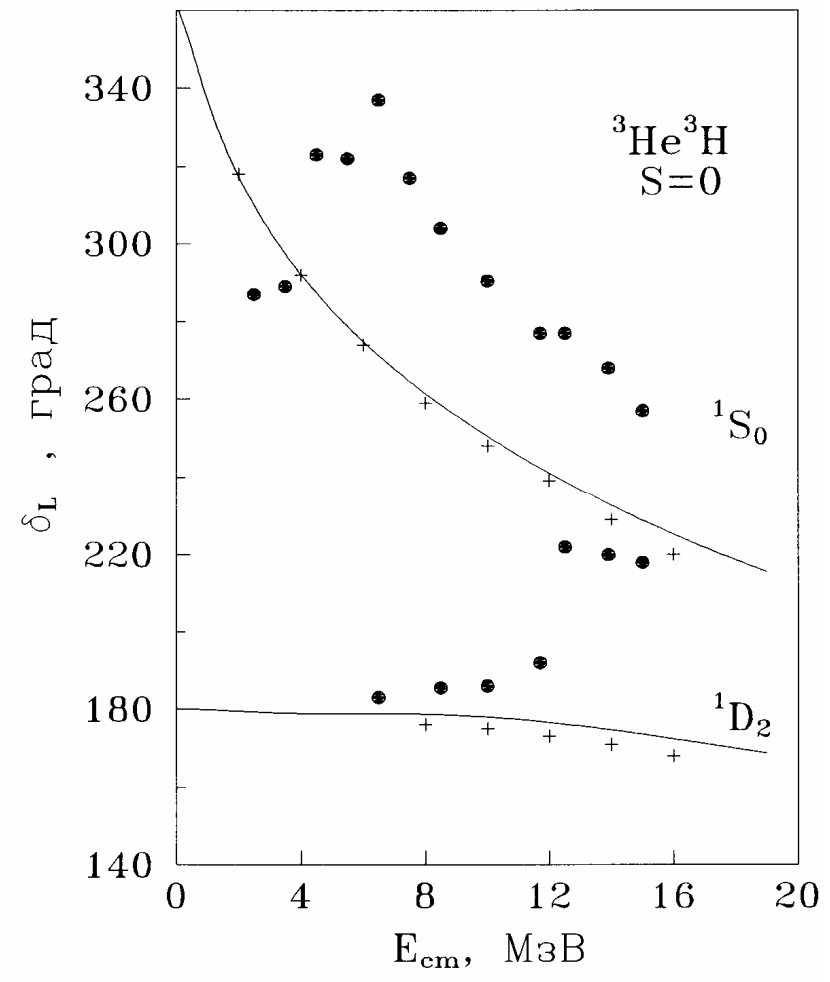

Рис.3.3в. Фазы упругого $^3He^3H$ рассеяния. Кривые - расчеты для потенциалов с параметрами из табл.3.1. Точки - экспериментальные данные из работ [7]. Крестики - результаты МРГ вычислений из работ [8].

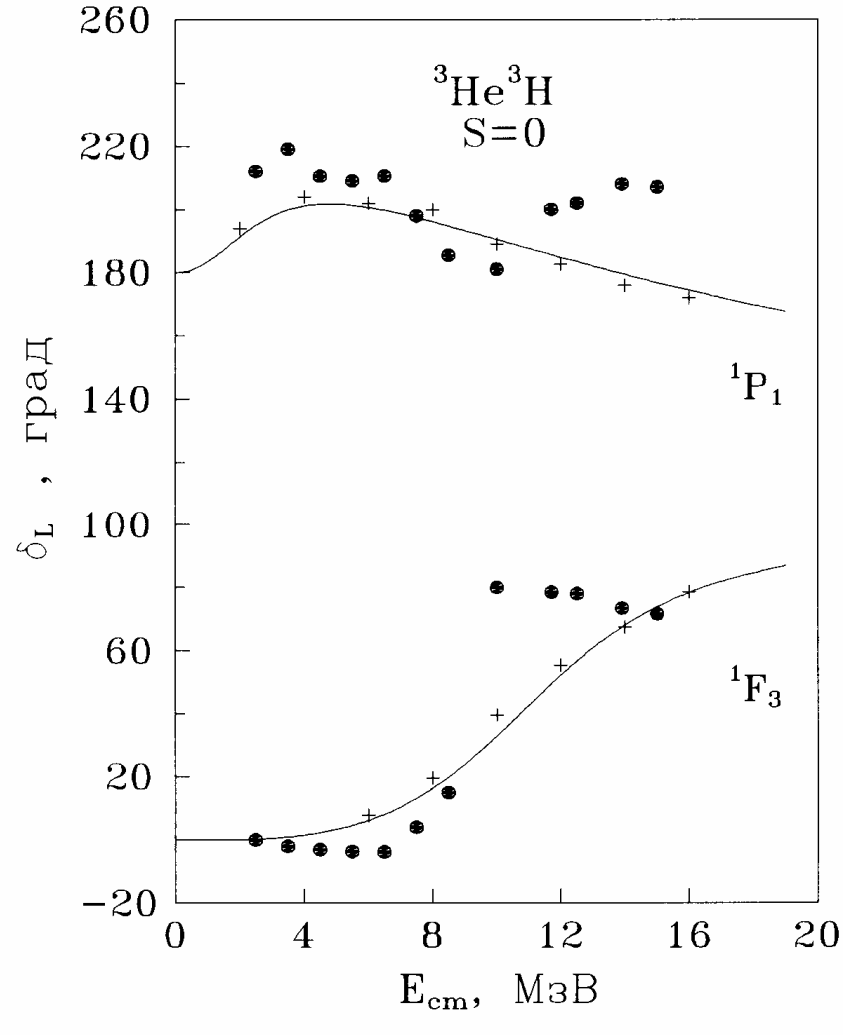

Рис.3.3г. Фазы упругого $^3He^3H$ рассеяния. Кривые - расчеты для потенциалов с параметрами из табл.3.1. Точки - экспериментальные данные из работ [7]. Крестики - МРГ результаты из работ [8].

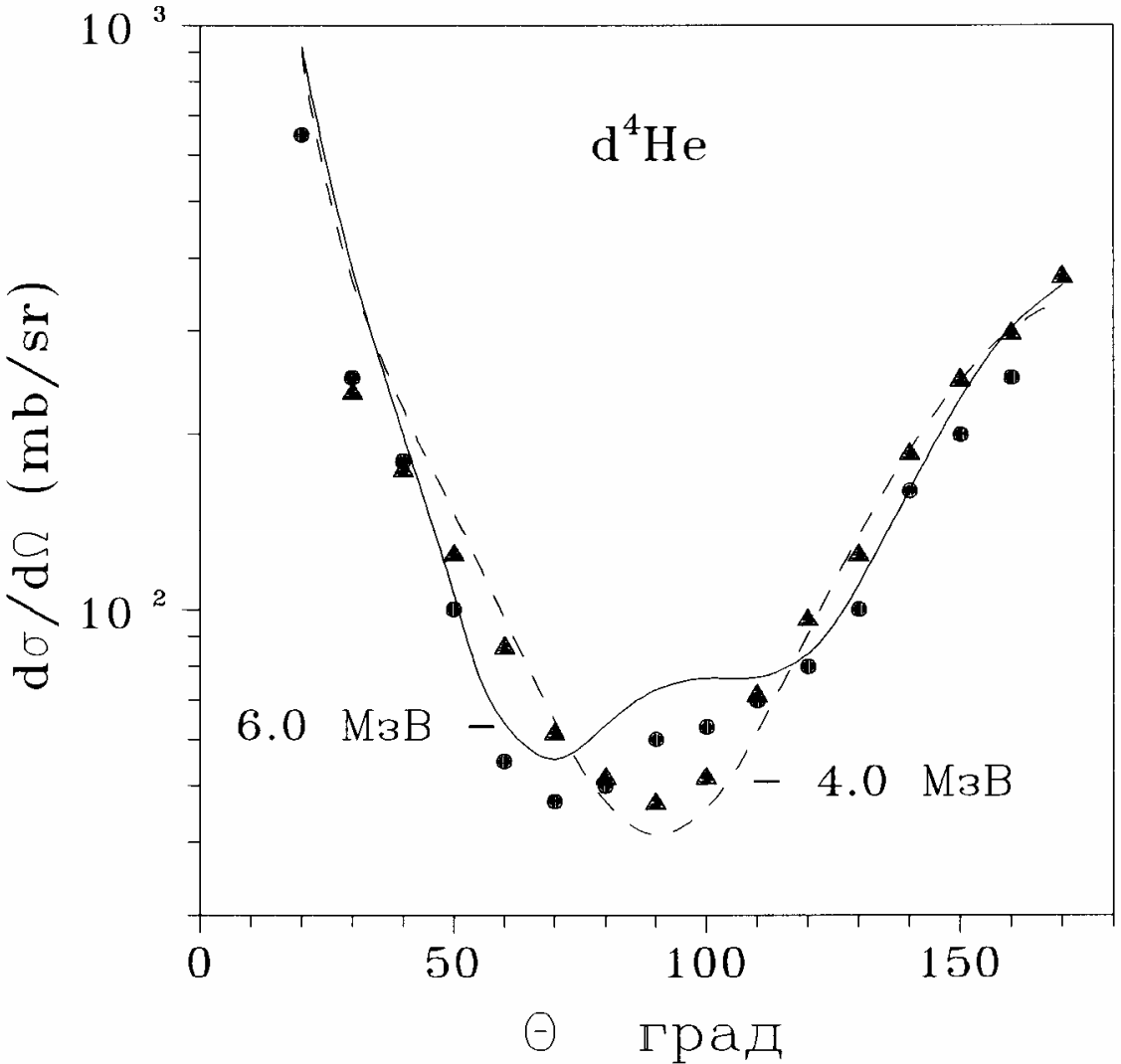

Рис.3.4а. Сечения упругого $^4He^2H$ рассеяния. Кривые - расчеты для потенциалов с параметрами из табл.3.1. Точки и треугольники - данные из работ [4,5].

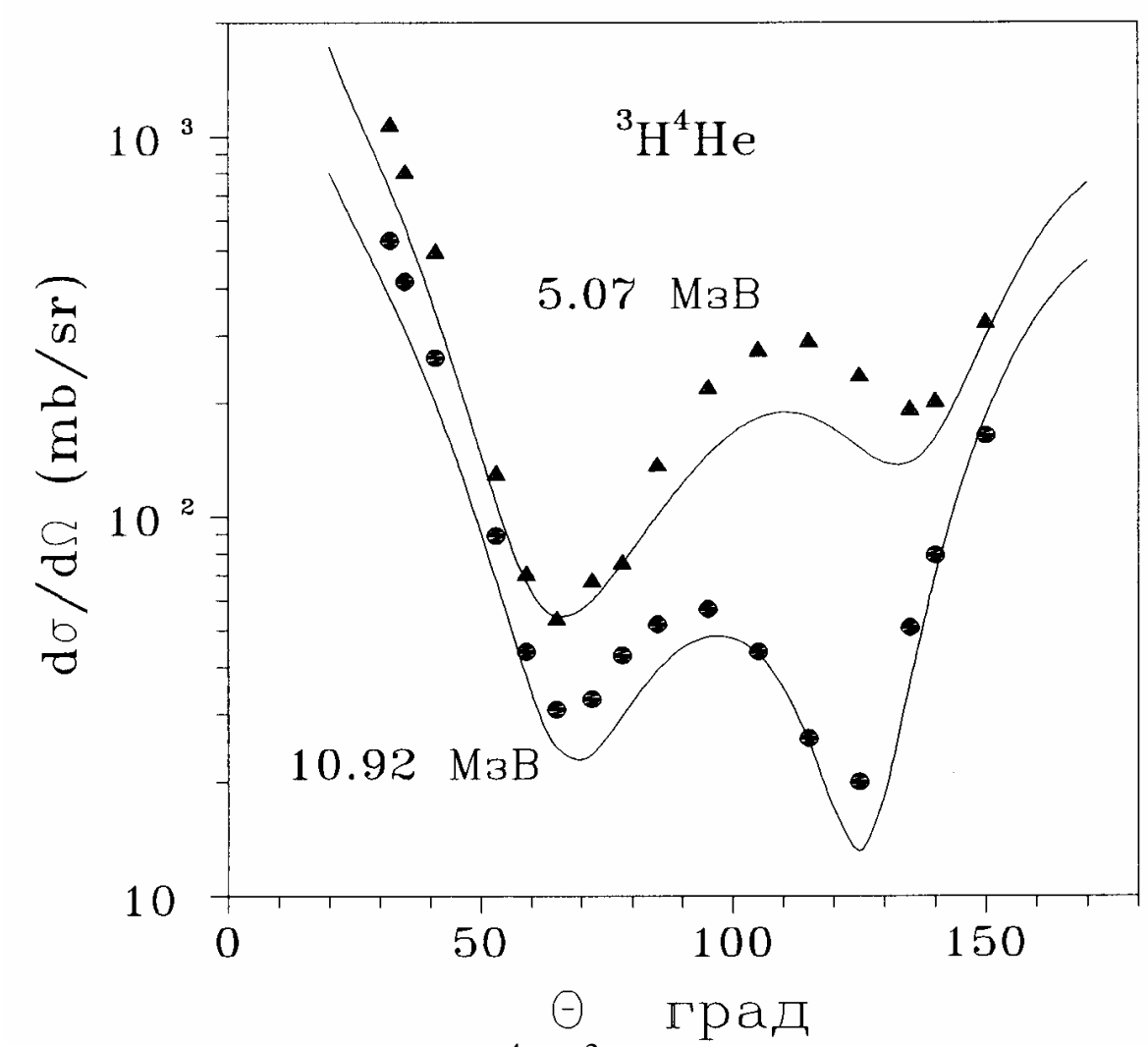

Рис.3.4б. Сечения упругого $^4He^3H$ рассеяния. Кривые - расчеты для потенциалов с параметрами из табл.3.1. Точки и треугольники - данные из работ [4,5].

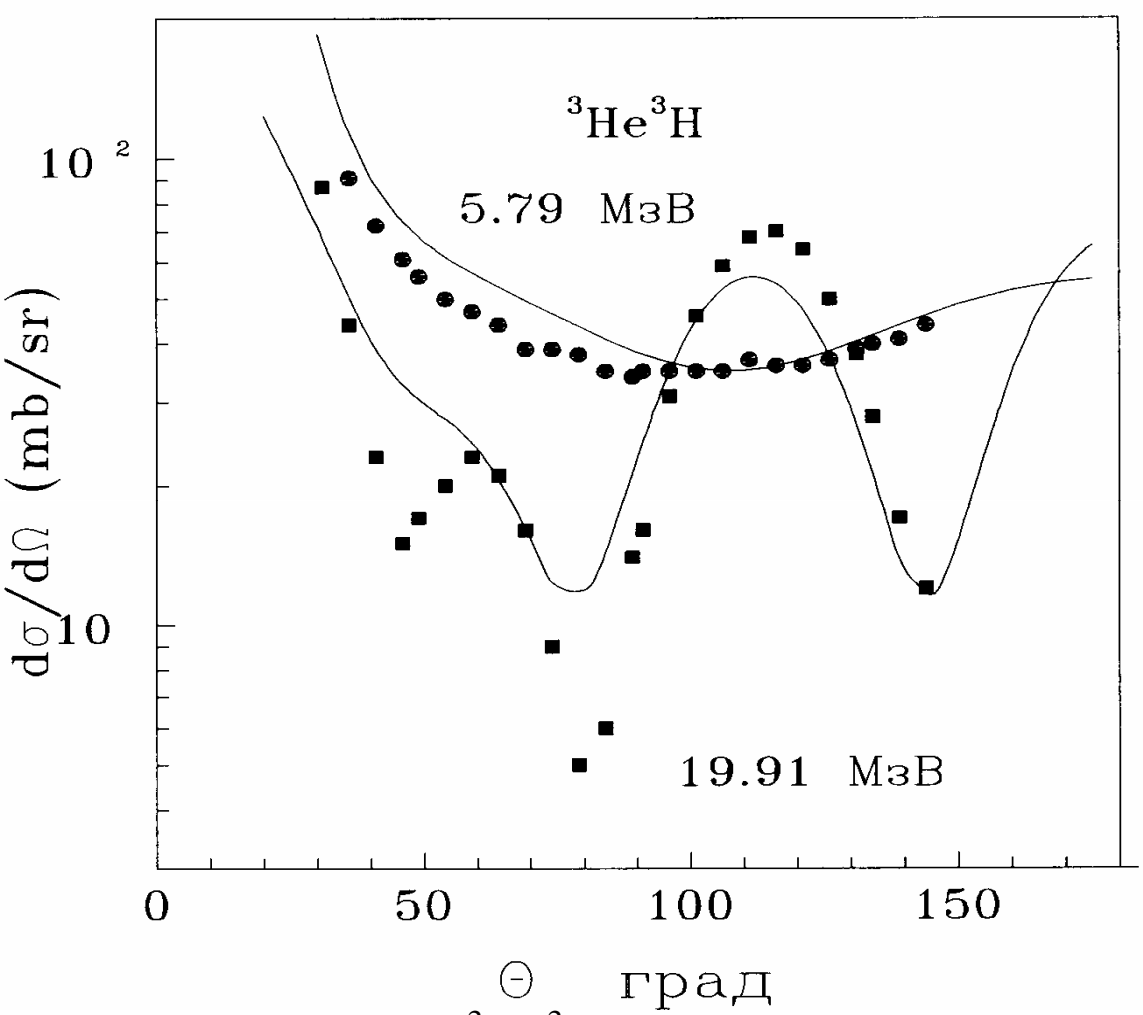

Рис.3.5. Сечения упругого $^3He^3H$ рассеяния. Кривые - расчеты для потенциалов с параметрами из табл.3.1. Точки и квадраты - экспериментальные данные из работ [7,10].

*Таблица 3.2. Энергии запрещенных состояний для ядер $^6Li$, $^7Li$ и $^7Be$ в кластерных каналах.*

| $L_J$ | $^7Li$ | $^7Be$ | $L_J$ | ($^4He\,^2H$) | S=1 ($^3H^3He$) | S=0 |
|---|---|---|---|---|---|---|
| S | -36.0 | -34.7 | S | -33.2 | -47.0 | -43.3 |
| | -7.4 | -6.3 | | | | -8.0 |
| $P_{1/2}$ | -27.5 | -26.0 | $P_0$ | -7.0 | -4.9 | |
| $P_{3/2}$ | -28.4 | -27.1 | $P_1$ | -11.7 | -9.8 | -14.1 |
| $D_{3/2}$ | -2.9 | -2.5 | $P_2$ | -14.5 | -17.1 | |
| $D_{5/2}$ | -4.1 | -3.7 | | | | |

В частности, для нахождения зарядового радиуса ядра использовались радиусы кластеров, приведенные в работах [11], значения которых даны в табл. 3.3. Магнитный радиус $^3H$ принимался равным 1.72(6) Фм [11]. Для радиуса протона использовалось значение 0.805(11) Фм.

Результаты расчетов зарядовых радиусов, энергии связи в кластерных каналах, магнитных, квадрупольных и октупольных моментов, вероятностей переходов $1/2^- \to 3/2^-$ в $^7Li$, перехода $1^+ \to 3^+$ в $^6Li$ и асимптотических констант приведены в табл. 3.4 вместе с экспе-риментальными данными из работ [12,13,14,15], МРГ расчетами [16,17] и расчетами в кластерной модели [18,19].

*Таблица 3.3. Экспериментальные данные по зарядовым радиусам легких ядер.*

| Ядро | Радиус, (Фм) |
|---|---|
| $^2$H | 1.9660(68); 1.950(3) |
| $^3$H | 1.70(5); 1.68(3) |
| $^3$He | 1.87(5); 1.844(45) |
| $^4$He | 1.63(4); 1.673(1) |

Следует отметить, что потенциалы, приведенные в табл. 3.1 для $^4$He$^3$H системы, мало отличаются, от полученных ранее в работах [18,19]. На рис.3.6 представлены импульсные распределения кластеров для ядра $^6$Li в $^4$He$^2$H, $^3$He$^3$H моделях и эксперимент [20]. На рис.3.7 показаны расчетные [2] и экспериментальные [9] спектры ядер $^6$Li и $^7$Li. В формуле (1.3.4) для формфактора ядра присутствуют не только матричные элементы, но и формфакторы кластеров, в качестве которых использовались формфакторы соответствующих ядер в свободном состоянии.

Параметризации этих формфакторов имеются в работах [2,12] и приведены в первой главе (1.5.5-1.5.6). Эти выражения позволяют правильно передать поведение формфакторов при импульсах до 20 Фм$^{-2}$, что вполне достаточно, так как формфакторы рассматриваемых ядер изменены только в области до 3,5-4,5 Фм$^{-1}$. Для формфактора протона использована обычная гауссова параметризация с $\alpha = 0.0864$ Фм$^2$ [21]. Качество описания экспериментальных данных [11] этими параметризациями показано на рис.3.8.

Кулоновские формфакторы ядер лития исследовались во многих работах, в том числе, методом резонирующих групп [22,23,24], на основе различных феноменологических подходов [25,26,27] и в кластерной модели [28]. В МРГ вычислениях упругих и неупругих формфакторов $^7$Li удается получить хорошее описание экспериментальных данных при малых переданных импульсах порядка 7-8 Фм$^{-2}$ [23]. Однако при 3-4 Фм$^{-1}$ в эксперименте [29] наблюдается второй максимум, который в МРГ расчетах отсутствует [22,23].

В случае ядра $^6$Li, существующие МРГ результаты, с учетом искажений или полной антисимметризации, в общем, правильно передают форму второго максимума упругого формфактора [24].

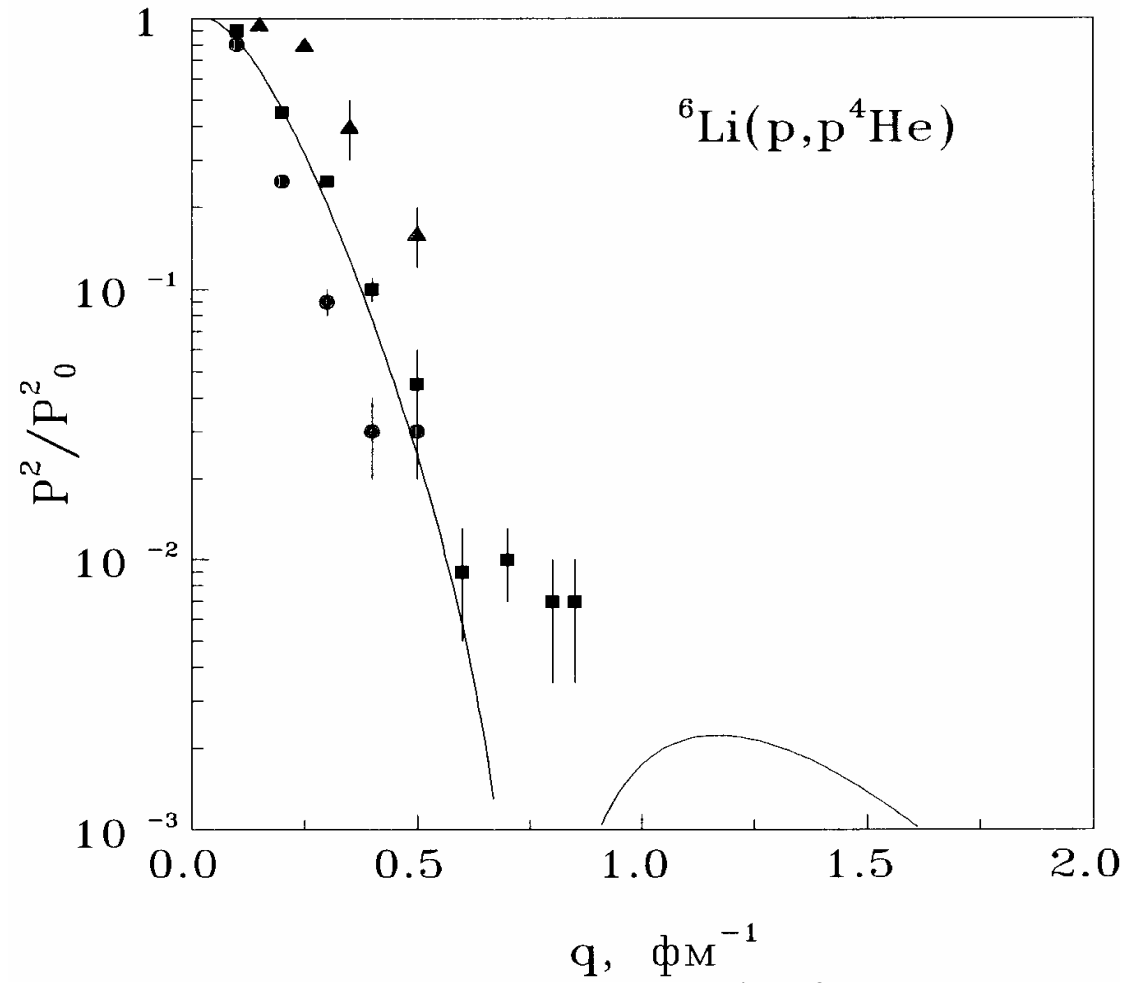

Рис.3.6а. Импульсные распределения $^4$He$^2$H кластеров в ядре $^6$Li. Кривые - расчеты [1,2] для потенциалов с параметрами из табл.3.1. Точки и квадраты - экспериментальные данные из работ [20].

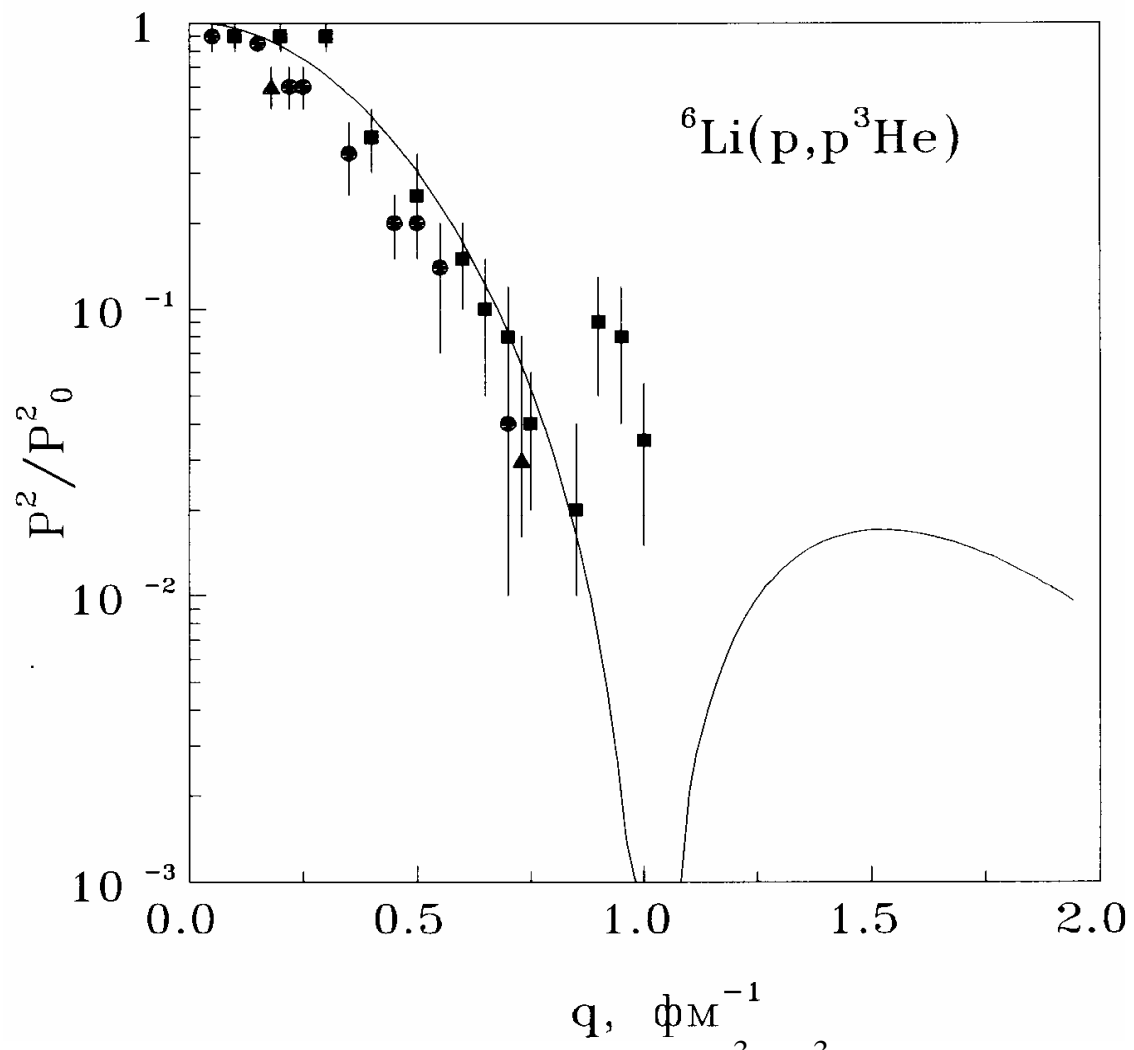

Рис.3.6б. Импульсные распределения $^3$He$^3$H кластеров в ядре $^6$Li. Кривые - расчеты [1,2] для потенциалов с параметрами из табл.3.1. Точки и квадраты - экспериментальные данные из работ [20].

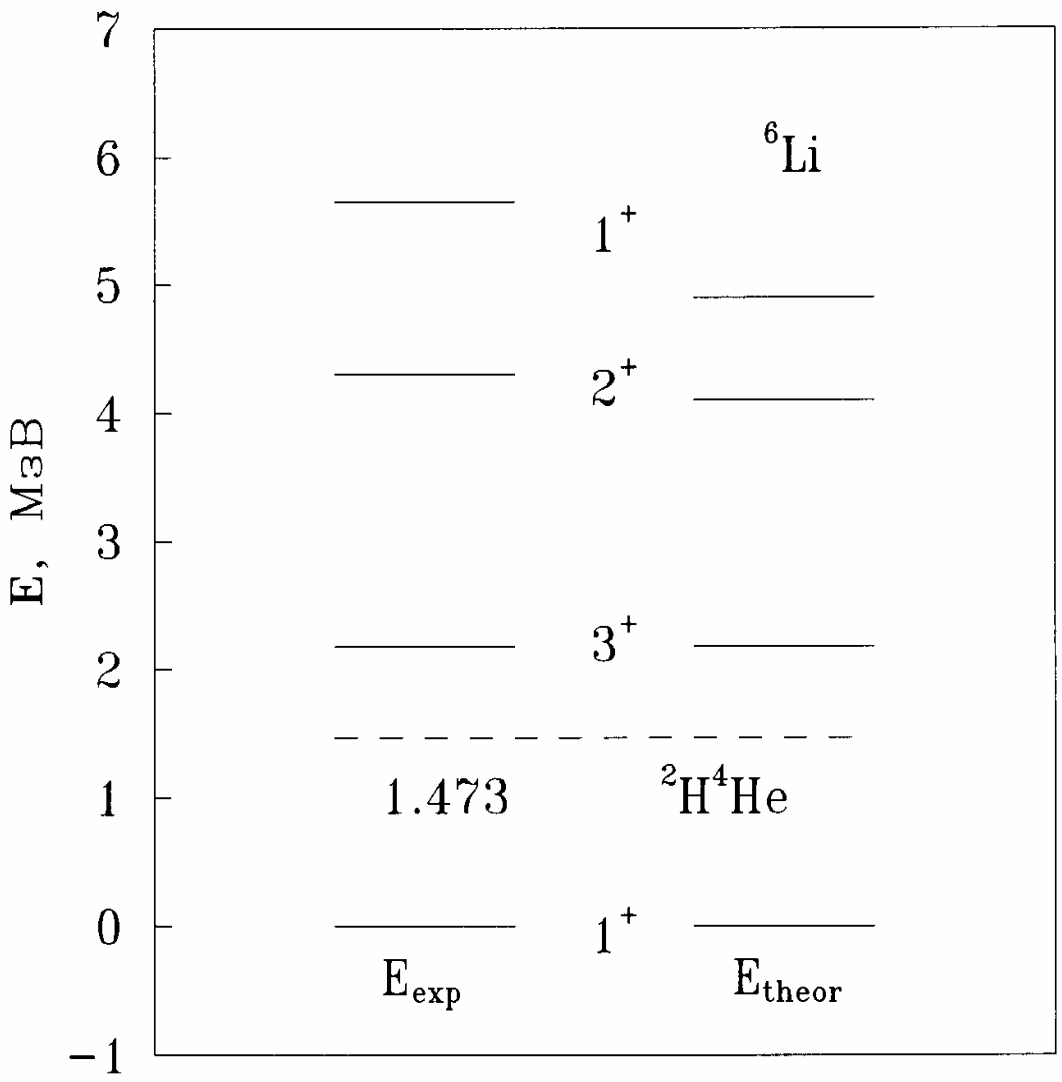

Рис.3.7а. Расчетные [1,2] и экспериментальные [9] спектры ядра $^{6}$Li. Параметры потенциалов из табл.3.1.

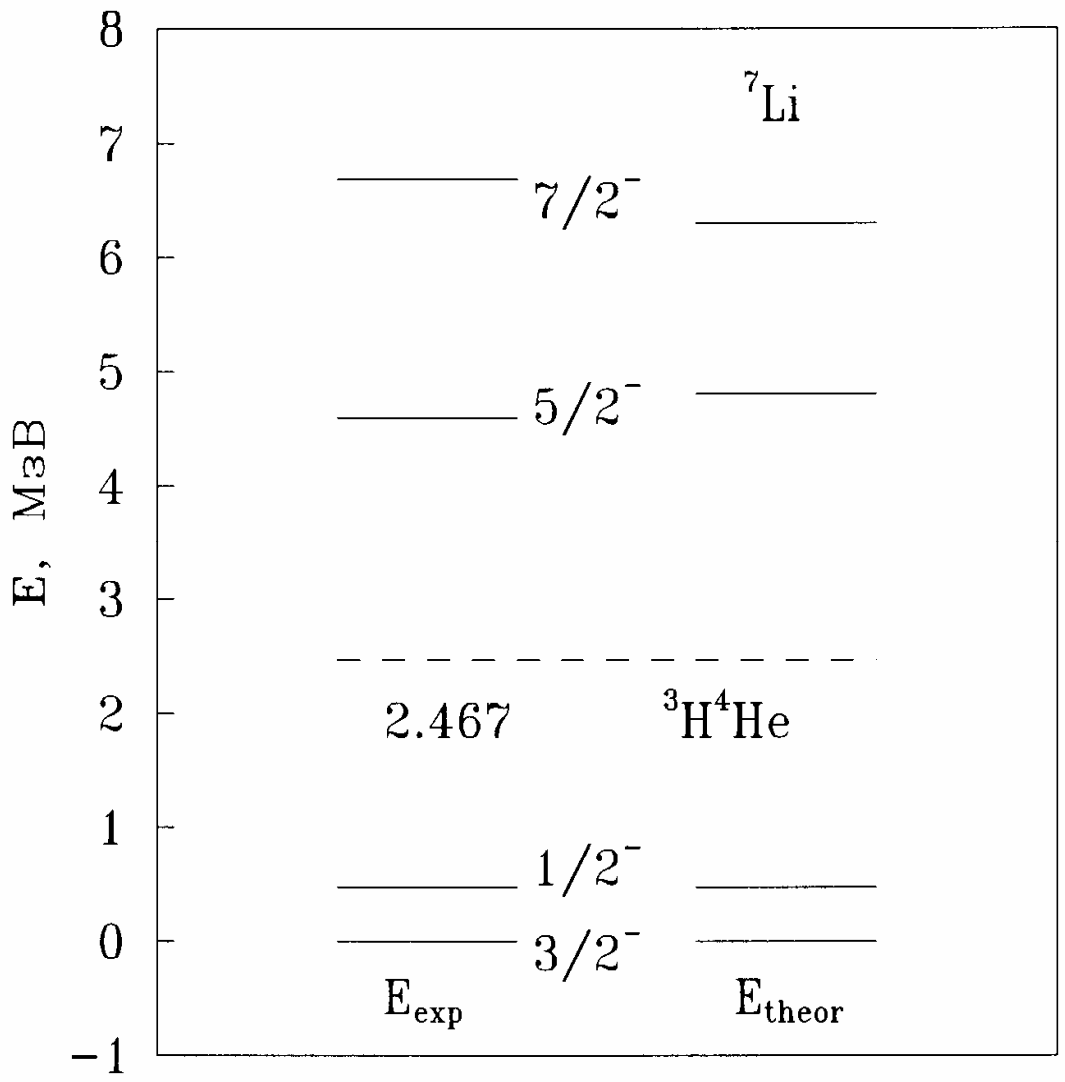

Рис.3.7б. Расчетные [1,2] и экспериментальные [9] спектры ядра $^{7}$Li. Параметры потенциалов из табл.3.1.

*Таблица 3.4. Результаты расчетов электромагнитных характеристик ядер $^7Li$, $^7Be$ и $^6Li$ в кластерных [1,2] моделях и сравнение их с МРГ результатами и результатами работ [18,19].*

| $^7$Li | | | | |
|---|---|---|---|---|
| Характеристики ядра | Расчет [1,2] | Экспер. [12-15] | МРГ [16,17] | Расчет [18, 19] |
| Q (мб) | -38.2 | -36.6(3); 40.6(3) | -34.2÷41.9 | -37.4 |
| $\mu$ ($\mu_0$) | 3.383 | 3.2564 | 2.79÷3.16 | 3.384 |
| $\Omega/\mu$ (Фм$^2$) | 2.70 | 2.4(5); 2.9(1) | 3.2÷3.6 | 2.48 |
| B(M1) ($\mu_0^2$) | 2.45 | 2.48(12) | 1.96÷2.17 | 2.45 |
| B(E2) ($e^2$ Фм$^4$) | 7.3 | 7.42(14) ÷8.3(6) | 5.4÷11.3 | 7.0 |
| $E_0$ (МэВ) | -2.47 | -2.467 | | |
| $E_1$ (МэВ) | -1.99 | -1.989 | | |
| $R_f$ (Фм) | 2.40 | 2.39(3) | | |
| $R_m$ (Фм) | 2.77 | 2.70(15); 2.98(5) | | |
| $C_0$ | 3.9 | - | | |
| $^7$Be | | | | |
| Характеристики ядра | Расчет [1,2] | Экспер. [12-15] | МРГ [16,17] | Расчет [18, 19] |
| Q (мб) | -59.3 | | -58.4 | |
| $\mu$ ($\mu_0$) | -1.532 | | -1.27 | -1.533 |
| $\Omega/\mu$ (Фм$^2$) | 2.85 | | 4.71 | |
| B(M1) ($\mu_0^2$) | 1.87 | 1.87(25) | 1.58 | 1.87 |
| B(E2) ($e^2$ Фм$^4$) | 17.5 | | 17.0 | |
| $E_0$ (МэВ) | -1.60 | -1.586 | | |
| $E_1$ (МэВ) | -1.14 | -1.157 | | |

| $^6$Li | | |
|---|---|---|
| Характеристики ядра | Расчет [1,2] | Эксперимент [12-15] |
| E(МэВ) | -1.472 | -1.4735 |
| $R_f$ (Фм) | 2.56 | 2.54(5) |
| $C_0$ | 3.22(2) | 2.9÷3.6 |
| B(E2) ($e^2$ Фм$^4$) | 25.1 | 25.6 |

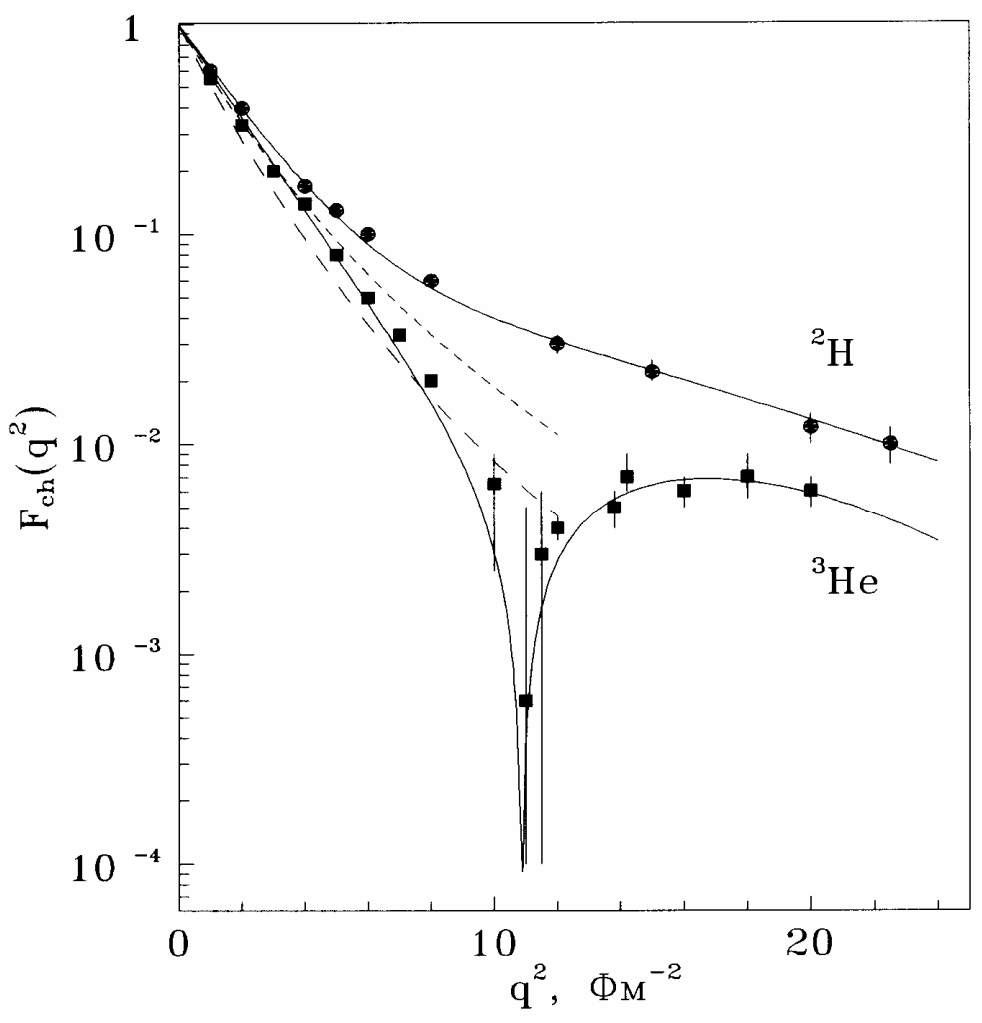

Рис.3.8а. Параметризация формфакторов легких ядер $^2$H и $^3$He. Непрерывные кривые - расчеты на основе формул (1.5.5, 1.5.6). Точечная и штриховые кривые - расчеты формфактора в p$^2$H кластерной модели. Точки и квадраты - эксперимент из [11].

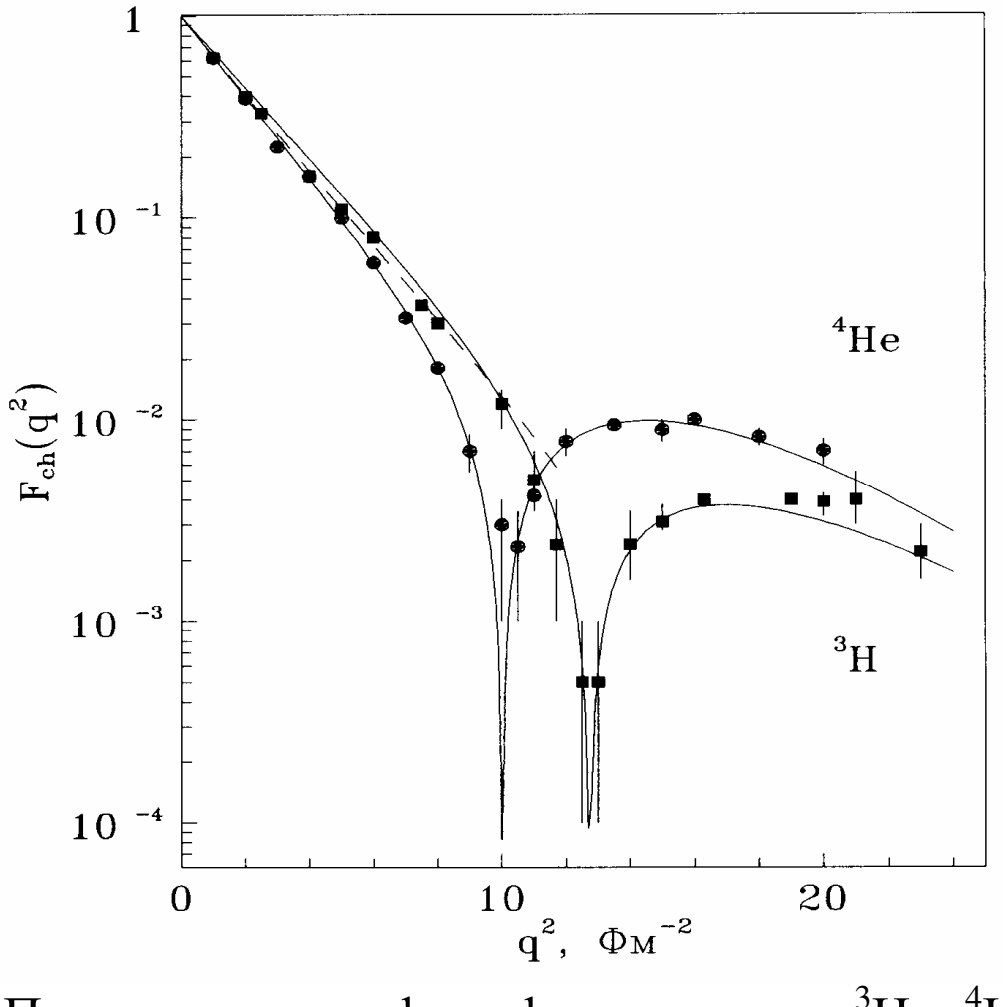

Рис.3.8б. Параметризация формфакторов ядер $^3$H и $^4$He. Непрерывные кривые - расчеты на основе формул (1.5.5, 1.5.6). Точки и квадраты - данные из работ [11]. Штриховая кривая - расчет формфактора $^4$He в p$^3$H кластерной модели.

Второй максимум упругого кулоновского формфактора $^6$Li удается передать и в рамках трехтельной модели ядра [30], где применялись потенциалы с запрещенными состояниями и выполнена антисимметризация волновой функции.

В работах [25-27] использовались феноменологические волновые функции ядра $^6$Li и получено хорошее описание упругого и неупругого $3^+$ формфакторов при малых переданных импульсах. В работе [2] вычисления формфакторов выполнялись на основе формул (1.3.11-1.3.12) в двухкластерной модели с запрещенными состояниями (коэффициенты $B_J$ в формуле для формфакторов (1.3.4) приведены в табл.3.5.) и получены параметры разложения вариационных волновых функций по гауссойдам (1.3.9), которые даны в табл.3.6 и 3.7.

*Таблица 3.5. Параметр $B_J$ для кулоновских формфакторов с переходом из основного состояния на уровень $L_f$, $J_f$.*

| Ядро | $L_f$ | $J_f$ | J | $B_J$ |
|---|---|---|---|---|
| $^6$Li | 0 | 1 | 0 | 1 |
| | 2 | 3 | 2 | 7/3 |
| $^7$Li | 1 | 3/2 | 0 | 1 |
| | 1 | 3/2 | 2 | 1 |
| | 1 | 1/2 | 2 | 1 |
| | 3 | 7/2 | 2 | 18/7 |
| | 3 | 7/2 | 4 | 10/7 |

Результаты расчета упругого формфактора $^6$Li [2] показаны на рис.3.9а непрерывной линией вместе с экспериментальными данными работ [29]. Штриховой линией приведены данные работ [23], полученные на основе МРГ вычислений, штрих - пунктиром даны результаты трехтельных расчетов, выполненных в [30]. Видно, что расчетная кривая в области 3-4 Фм$^{-1}$ идет несколько ниже экспериментальных данных, практически совпадая с трехтельными расчетами, и при средних импульсах не имеет явного минимума, что обусловлено плавным характером формфактора дейтрона, не имеющего минимума в этой области переданных импульсов.

В расчетах [2] никакой деформации дейтрона не проводилось, что, впрочем, оправдано только при 100% кластеризации ядра в $^4$He$^2$H канал. А так как вероятность кластеризации, по-видимому, не превышает 60%-80% [30], используемая модель без учета искажений является определенным приближением к реально существующей ситуации. Конечно, и в $^4He^2H$ модели можно ввести деформации дейтронного кластера. Однако, при этом появляется подгоночный параметр, характеризующий степень сжатия дейтрона.

*Таблица 3.6. Параметры $C_i$ и $\gamma_i$ в разложении волновых функций по гауссойдам для уровней $J=1^+$ в $^6Li$, $J=3/2^-$, и $J=1/2^-$ в $^7Li$.*

| i | $^6Li$ | |
|---|---|---|
| | j=1$^+$ | |
| | $\gamma_i$ | $C_i$ |
| 1 | 1.84375E-01 | 2.9320E-01 |
| 2 | 1.08665E-02 | 8.7096E-03 |
| 3 | 7.95585E-02 | 2.0572E-01 |
| 4 | 3.01026E-02 | 6.9455E-02 |
| 5 | 2.43701E-01 | -1.6228E-02 |
| 6 | 3.78324E-01 | -7.6704E-01 |
| 7 | 6.59311E-01 | -6.7846E-01 |
| 8 | 1.80306E+00 | 7.6962E+01 |
| 9 | 1.81111E+00 | -8.0035E+01 |
| 10 | 2.00011E+00 | 3.1766E+00 |

| $^7Li$ | | | | |
|---|---|---|---|---|
| | j=3/2$^-$ | | j=1/2$^-$ | |
| i | $\gamma_i$ | $C_i$ | $\gamma_i$ | $C_i$ |
| 1 | 2.00000E-01 | 1.3016E+00 | 4.10156E-01 | 8.9075E-01 |
| 2 | 2.56352E-02 | -2.8023E-03 | 3.57560E-02 | -1.1937E-02 |
| 3 | 6.64659E-02 | -3.6370E-02 | 1.05253E-01 | -1.2543E-01 |
| 4 | 1.65601E-01 | -6.9579E-01 | 2.52528E-01 | -2.8397E-01 |
| 5 | 2.43701E-01 | -1.2363E+00 | 7.46282E-01 | 2.8376E-01 |
| 6 | 3.79179E-01 | 1.0716E+00 | 9.75007E-01 | 2.5229E-01 |
| 7 | 6.40147E-01 | 1.8587E-01 | 1.43365E+00 | -3.7091E-01 |
| 8 | 9.03063E-01 | 4.5852E-01 | 2.15434E+00 | 4.8106E-01 |
| 9 | 1.52012E+00 | -3.5778E-01 | 2.68111E+00 | -4.1943E-01 |
| 10 | 2.10304E+00 | 1.8904E-01 | 3.20001E+00 | 1.3095E-01 |

*Таблица 3.7. Параметры $C_i$ и $\gamma_i$ в разложении волновых функций по гауссойдам для уровней $J=3^+$ в $^6Li$ и $J=7/2^-$ в $^7Li$.*

| i | $^6Li$ | |
|---|---|---|
| | $\gamma_i$ | $C_i$ |
| 1 | 9.466969E-03 | 1.07258E-04 |
| 2 | 3.609561E-02 | 1.59496E-03 |
| 3 | 6.945430E-02 | 2.74053E-03 |
| 4 | 1.111254E-01 | 1.21893E-02 |
| 5 | 1.655117E-01 | 2.10893E-02 |
| 6 | 2.416748E-01 | 5.91636E-02 |
| 7 | 3.599539E-01 | 1.11022E-01 |
| 8 | 5.759183E-01 | 7.26675E-02 |
| 9 | 1.108168E-00 | -3.47603E-03 |
| 10 | 4.225218E-00 | -3.87967E-03 |

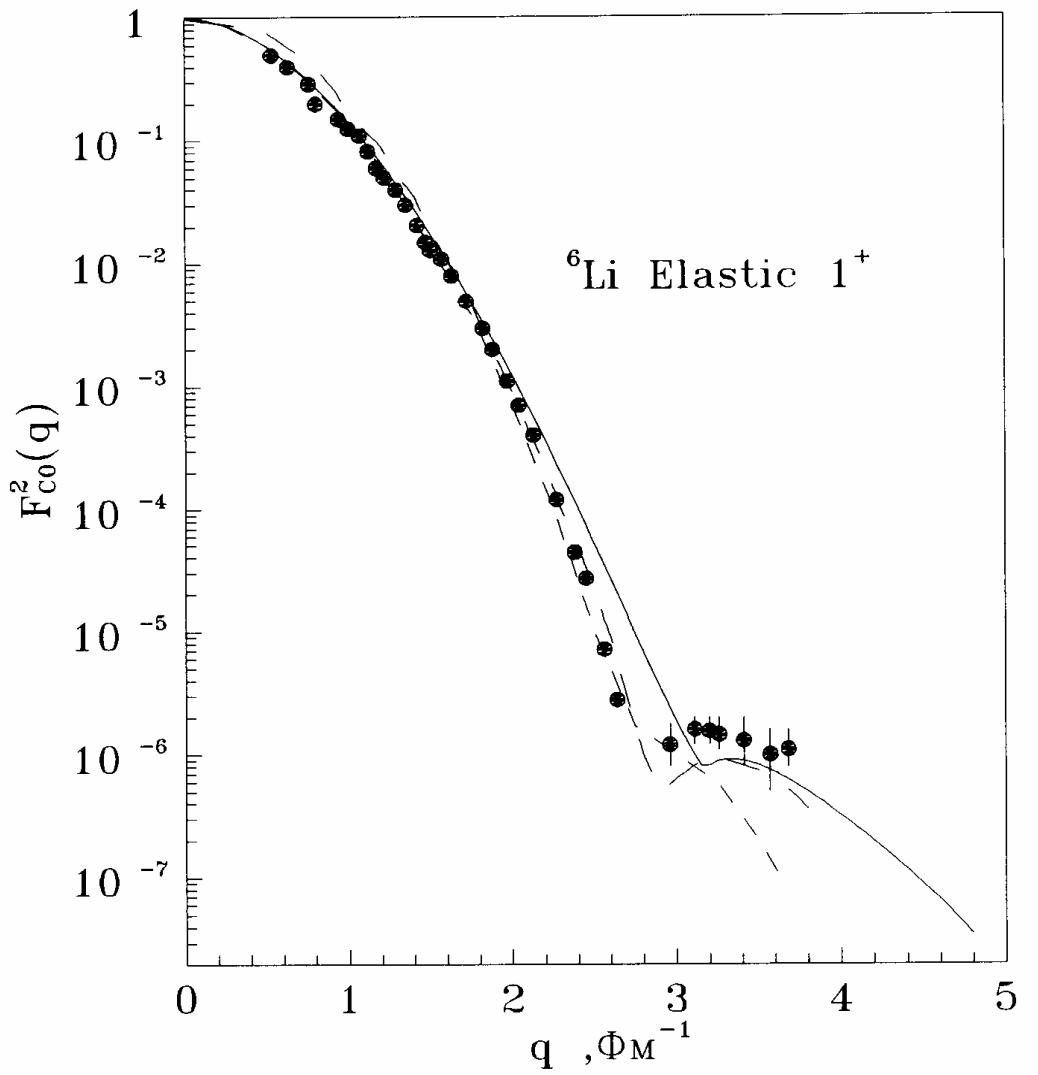

Рис.3.9а. Упругий кулоновские формфакторы ядра $^6$Li. Непрерывные кривые - расчеты в кластерной модели с потенциалами из табл.3.1. Штриховая кривая - МРГ расчеты из [23], штрих-пунктир - трехтельные расчеты [30]. Точки - экспериментальные данные из работ [29].

В тоже время, на основе расчетов без деформации кластеров можно судить о вероятности дейтронной кластеризации в ядре. При 100% кластеризации все результаты, полученные на основе простой двухкластерной модели должны хорошо согласовываться с экспериментом. А отклонение расчетных характеристик от экспериментальных данных будет свидетельствовать о меньшей степени кластеризации, когда кластеру уже нельзя полностью сопоставлять характеристики соответствующего ядра в свободном состоянии. Поэтому отклонение упругого формфактора от эксперимента, в первую очередь при средних импульсах, указывает на приближенность модели, которая не учитывает деформации. Меньшая величина формфактора при 3-4 $Фм^{-1}$ возможно обусловлена отсутствием D волны в волновой функции ядра, ввести которую можно, только учитывая тензорную компоненту в потенциале взаимодействия.

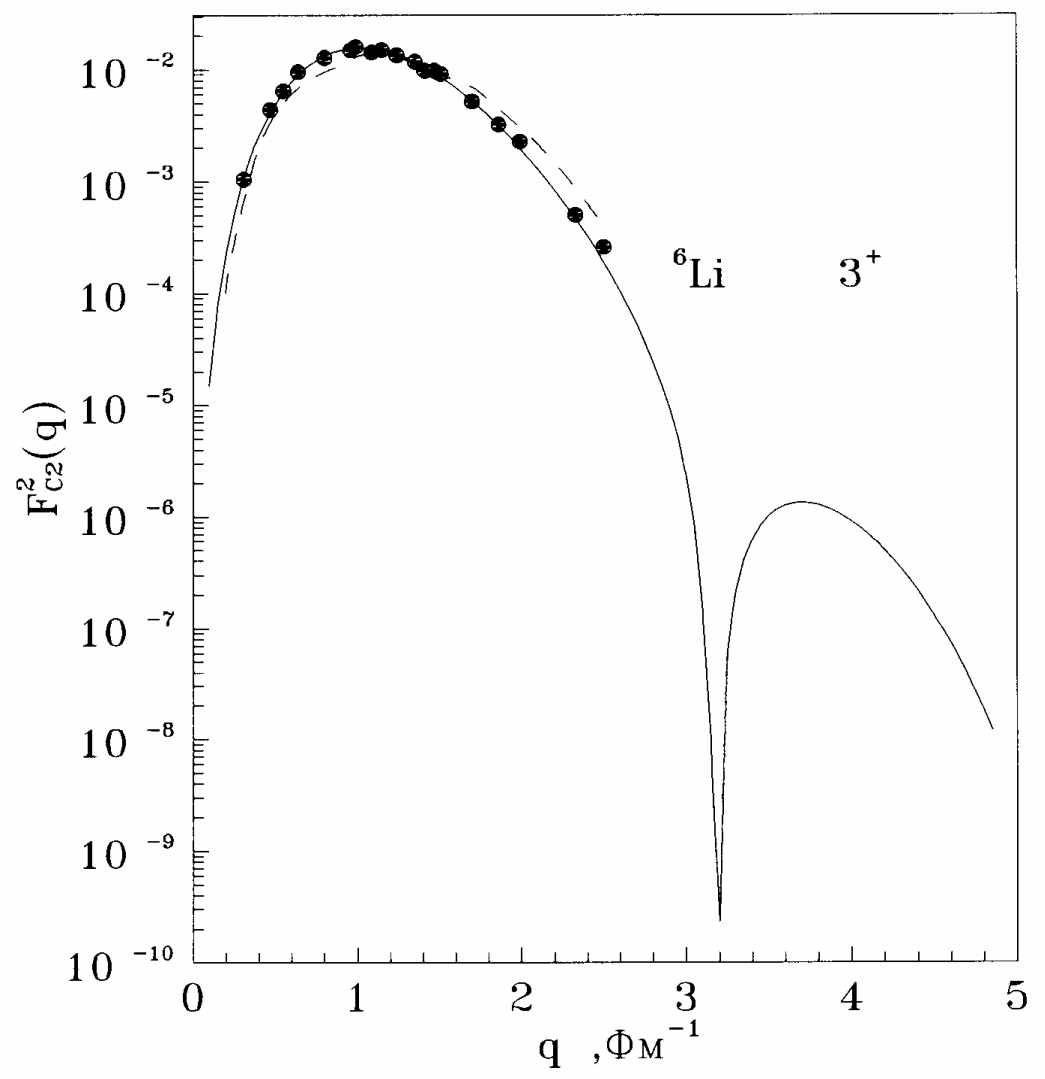

Рис.3.9б. Неупругий с переходом на 3+ уровень кулоновские формфакторы ядра $^6$Li. Непрерывные кривые - расчеты в кластерной модели с потенциалами из табл.3.1. Штриховая линия - трехтельные расчеты [30]. Точки - эксперимент [26,27].

На рис.3.9б непрерывной линией показан неупругий формфактор [2] с переходом на уровень $3^+$ в сравнении с результатами работ [30], для трехтельных волновых функций. Видно, что расчет практически совпадает с экспериментом [26,27], и имеет второй максимум. Надо отметить, что в отличие от вычислений упругого формфактора, неупругий формфактор оказывается очень чувствителен к глубине и форме потенциала. Изменение глубины на 0.03-0.05 МэВ практически не сказывается на поведении $3^+$ фазы рассеяния, но приводит к изменению формфактора в 2-3 раза. Для правильного описание этого формфактора пришлось несколько изменить потенциал $D_3$ волны и принять для его глубины величину 80.93 МэВ, что отличается от результатов, приведенных в табл.3.1 на 50 кэВ.

На рис.3.10а непрерывной линией показан вычисленный упругий кулоновский формфактор ядра $^7$Li [2], который имеет явный минимум при 3.2 $Фм^{-1}$ и подъем в области более высоких импульсов, практически лежащий в полосе экспериментальных ошибок [29]. Пунктиром здесь приведены МРГ вычисления работ [23], где расчетная кривая плавно спадает. На рис.3.10б и 3.10в даны неупругие кулоновские формфакторы с переходом на уровни $1/2^-$ и $7/2^-$ [2]. Пунктир - результаты работ [23]. Формфактор уровня $1/2^-$ описывается до второго максимума, где его величина заметно меньше эксперимента. Аналогичные результаты получены и для формфактора $7/2^-$. И здесь имеется второй максимум, полностью отсутствующий в МРГ вычислениях [22,23]. Надо отметить, что для расчетов использовались потенциалы из табл.3.1 без каких либо изменений параметров.

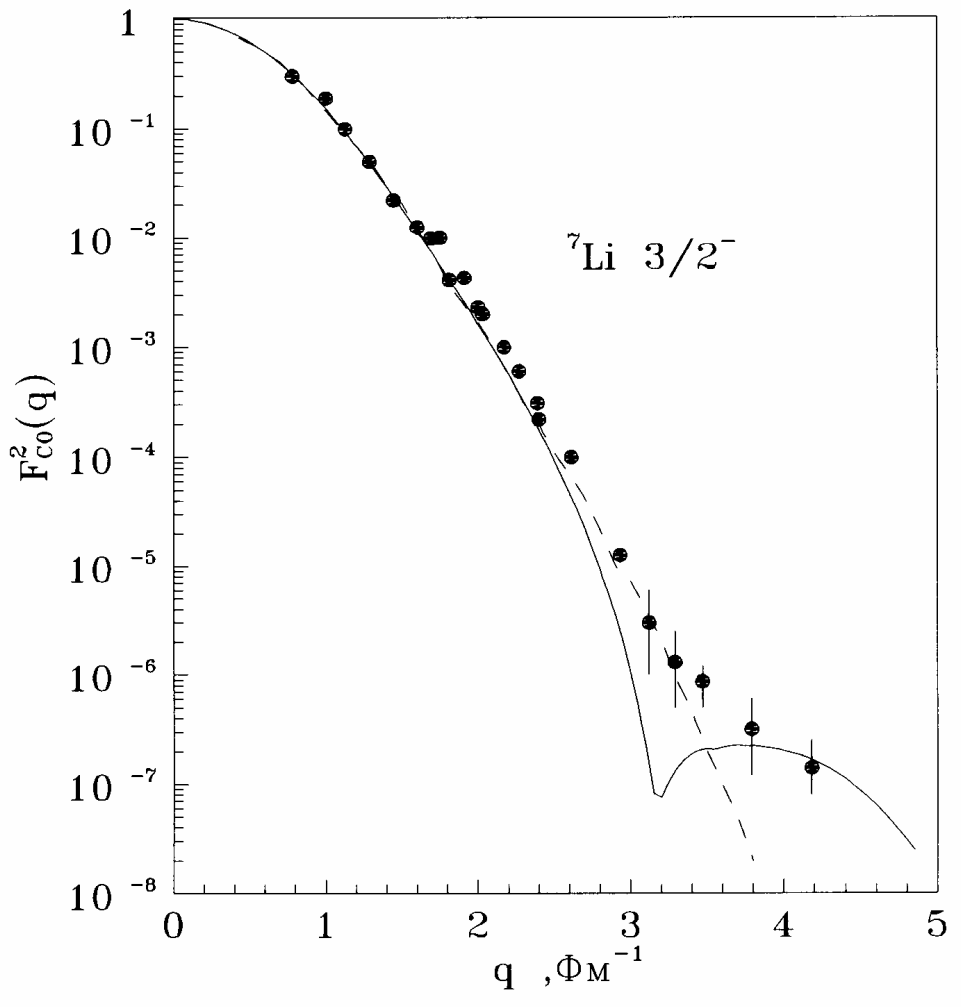

Рис.3.10а. Упругий кулоновский формфактор ядра $^7Li$. Непрерывная кривая - расчеты в кластерной модели с потенциалами из табл.3.1. Штриховая кривая - МРГ расчеты из [23]. Точки - экспериментальные данные [29].

На основе изложенных результатов можно считать, что расчеты упругих и неупругих кулоновских формфакторов ядер лития в двухкластерных моделях для потенциалов с запрещенными состояниями позволяют получить неплохое описание экспериментальных результатов даже при сравнительно больших переданных импульсах. Во всех случаях, в отличие от МРГ вычислений, в расчетах присутствует второй максимум формфактора, величина которого несколько занижена относительно экспериментальных данных.

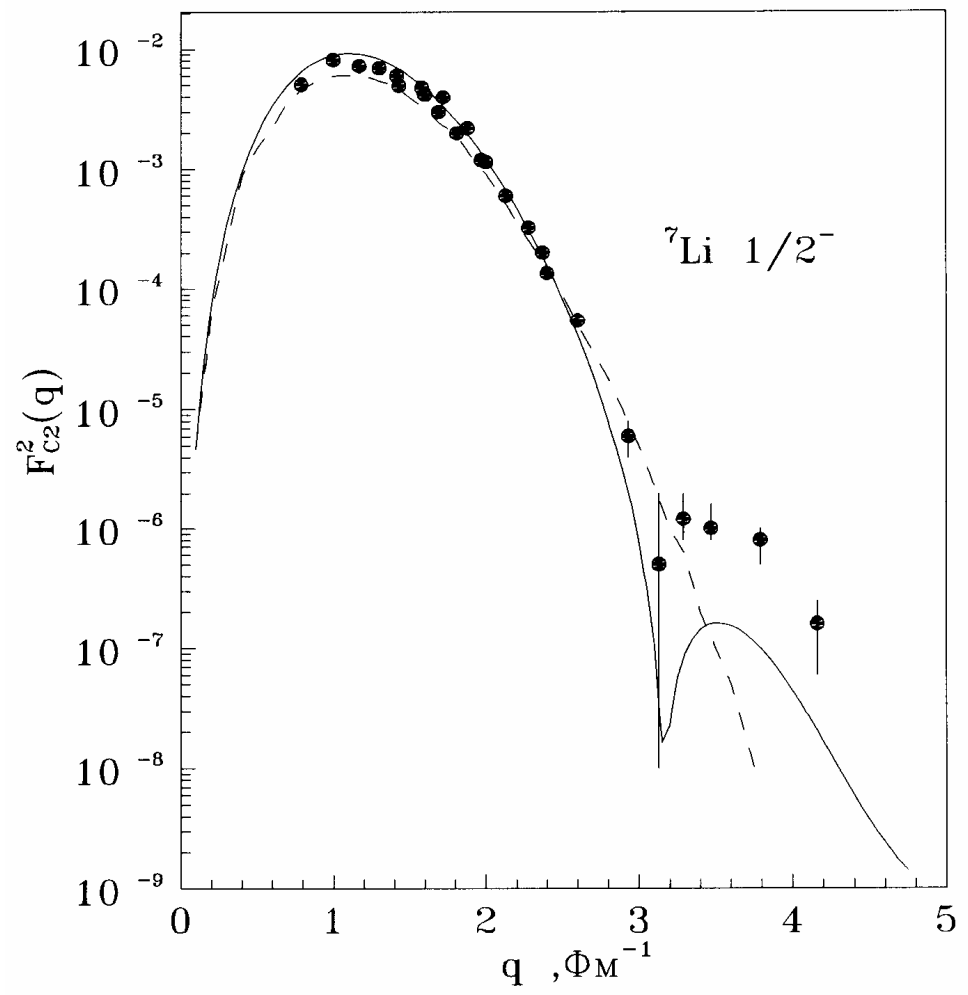

Рис.3.10б. Неупругий формфактор ядра $^7$Li для переходов на уровень 1/2$^-$. Непрерывная кривая - расчеты с потенциалами из табл.3.1. Штриховая кривая - МРГ расчеты из [23].

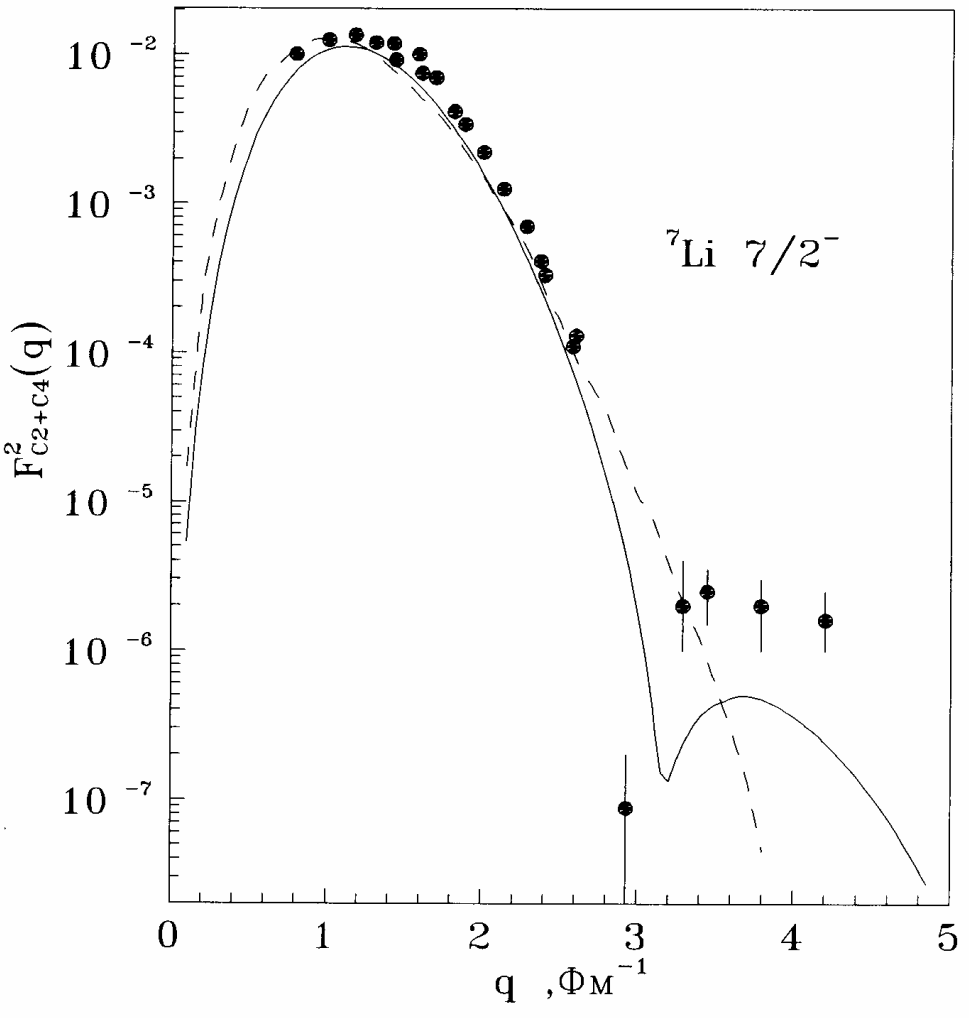

Рис.3.10в. Неупругий формфактор ядра $^7$Li для переходов на уровень 7/2$^-$. Непрерывная кривая - расчеты в кластерной модели с потенциалами из табл.3.1. Штриховая кривая - МРГ расчеты из [23]. Точки - экспериментальные данные [29].

Удается воспроизвести вероятности электромагнитных E2 и M1 переходов, магнитный, квадрупольный и октупольный моменты, магнитный и зарядовый радиус. Параметры потенциалов предварительно фиксированы по фазам упругого рассеяния кластеров и только при вычислении $3^+$ формфактора $^6$Li были несколько изменены для $D_3$ волны, что позволило хорошо передать поведение первого максимума.

## 3.2. Характеристики рассеяния в p$^3$He и n$^3$H системах

Кластерные системы p$^3$He и n$^3$H, чистые по изоспину с T=1, имеют только одну разрешенную орбитальную схему {31} с запрещенной в S волне конфигурацией {4} в триплетном и синглетном состояниях [31]. При низких энергиях (до 1 МэВ) фазовый анализ выполнен в работах [32], который приводит к заметной неоднозначности фаз - получено два различных набора решений. В области энергий 2-14 МэВ имеется две группы работ [33] и [34], где для синглетных и триплетных P фаз с J=1 так же получено два варианта результатов. В [33] синглетная (S=0) $^1P_1$ фаза при энергиях 6-12 МэВ идет на уровне 40-50 градусов, а $^3P_1$ для S=1 в области 20-30 градусов. В работе [34] синглетная фаза идет ниже - 20-30 градусов, а триплетная выше - 40-50 градусов. В работе [35] рассмотрены оба эти варианта, как решения A и C. Область энергий выше 18 МэВ рассмотрена в работах [36]. В табл. 3.8 приведены, полученные на основе фаз рассеяния параметры p$^3$He взаимодействий [31]. В четных волнах они совпадают с результатами работ [37,38]. Потенциал представлен двумя гауссойдами с притягивающей частью $V_0$ и периферическим отталкиванием $V_1$.

*Таблица 3.8. Потенциалы p$^3$He взаимодействия, чистые по орбитальным симметриям и энергии связанных состояний $E_{cc}$. В скобках даны энергии для n$^3$H системы.*

| $L_J$ | $V_0$, (МэВ) | $\alpha$, (Фм$^{-2}$) | $V_1$, (МэВ) | $\beta$, (Фм$^{-1}$) | $E_{cc}$, (МэВ) |
|---|---|---|---|---|---|
| S=0 | | | | | |
| Чет. | -110 | 0.37 | +45 | 0.67 | -9.0(-11.4) |
| Нечет | -14.0 | 0.1 | | | |
| S=1 | | | | | |
| Чет. | -43 | 0.26 | | | -3.6(5.3) |
| $P_0$ | -10 | 0.1 | | | |
| $P_1$ | -15 | 0.1 | | | |
| $P_2$ | -20 | 0.1 | | | |

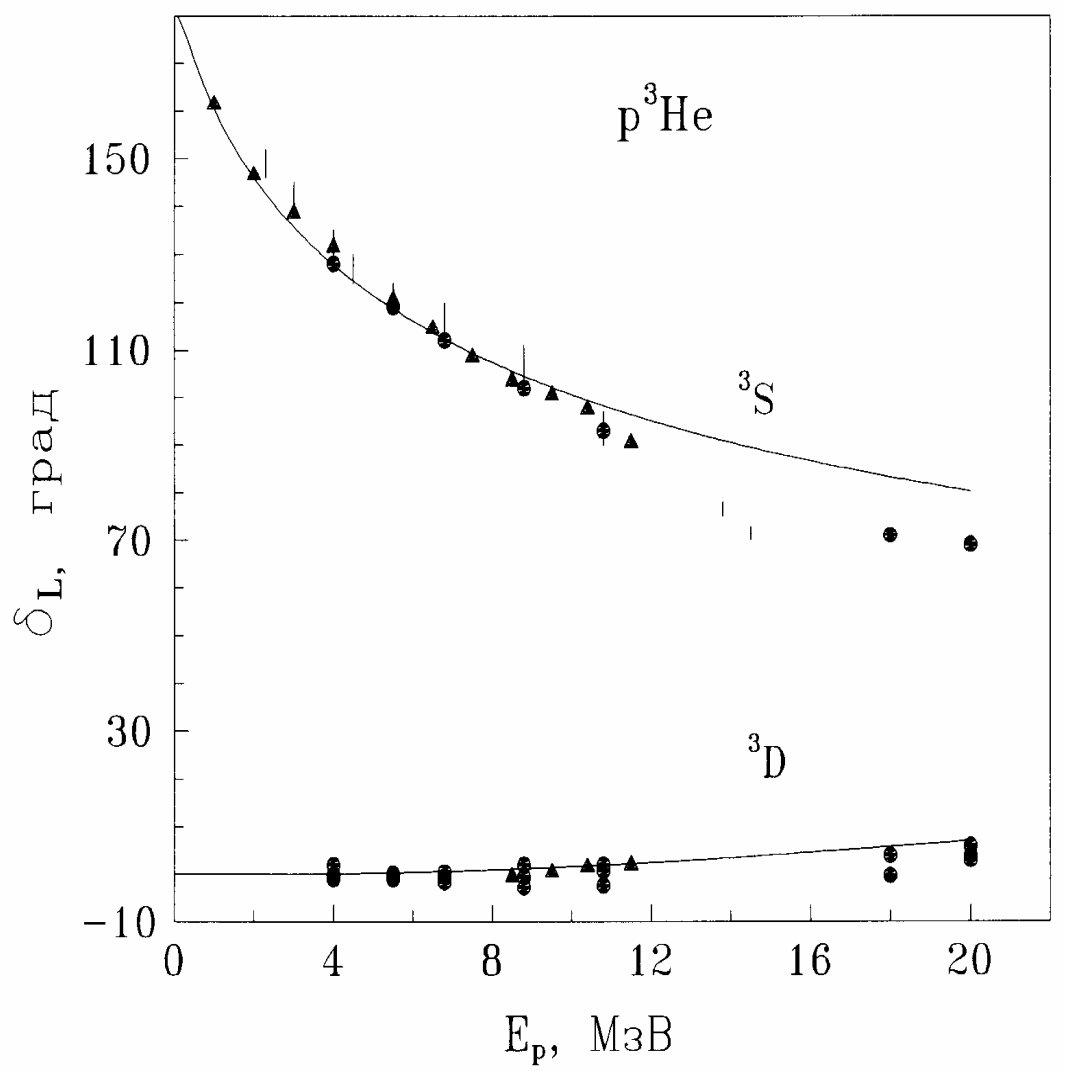

Рис.3.11а. Триплетные четные фазы упругого $p^3He$ рассеяния. Точки, квадраты и вертикальные линии - экспериментальные данные [32,33,35], треугольники из [34]. Кривые - результаты расчетов для различных потенциалов из табл.3.8.

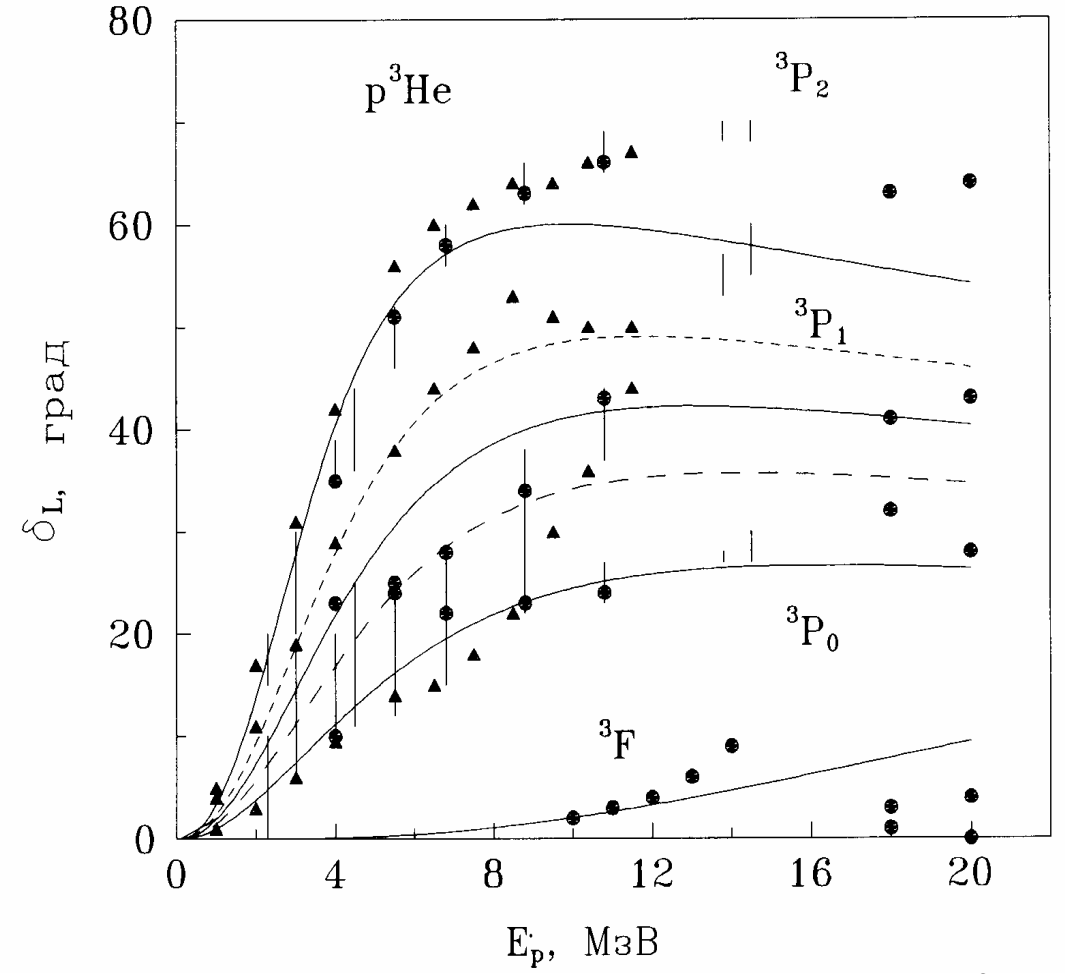

Рис.3.11б. Триплетные нечетные фазы упругого $p^3He$ рассеяния. Точки, квадраты и вертикальные линии - эксперимент из [32,33,35], треугольники из [34]. Кривые - результаты расчетов для различных потенциалов из табл.3.8.

На рис.3.11 и 3.12а непрерывными линиями показаны расчетные фазы для этих потенциалов и экспериментальные данные работ [32,33,35] - точки, квадратики и вертикальные линии и [34] - треугольники. На рис.3.12б показаны экспериментальные [32-35] и вычисленные с этими потенциалами сечения упругого $p^3He$ рассеяния при энергиях 6.82 МэВ и 10.77 МэВ. На рис. 3.13 представлены фазы $n^3H$ рассеяния и данные работы [39]. Крестиками показаны результаты МРГ вычислений из работ [40].

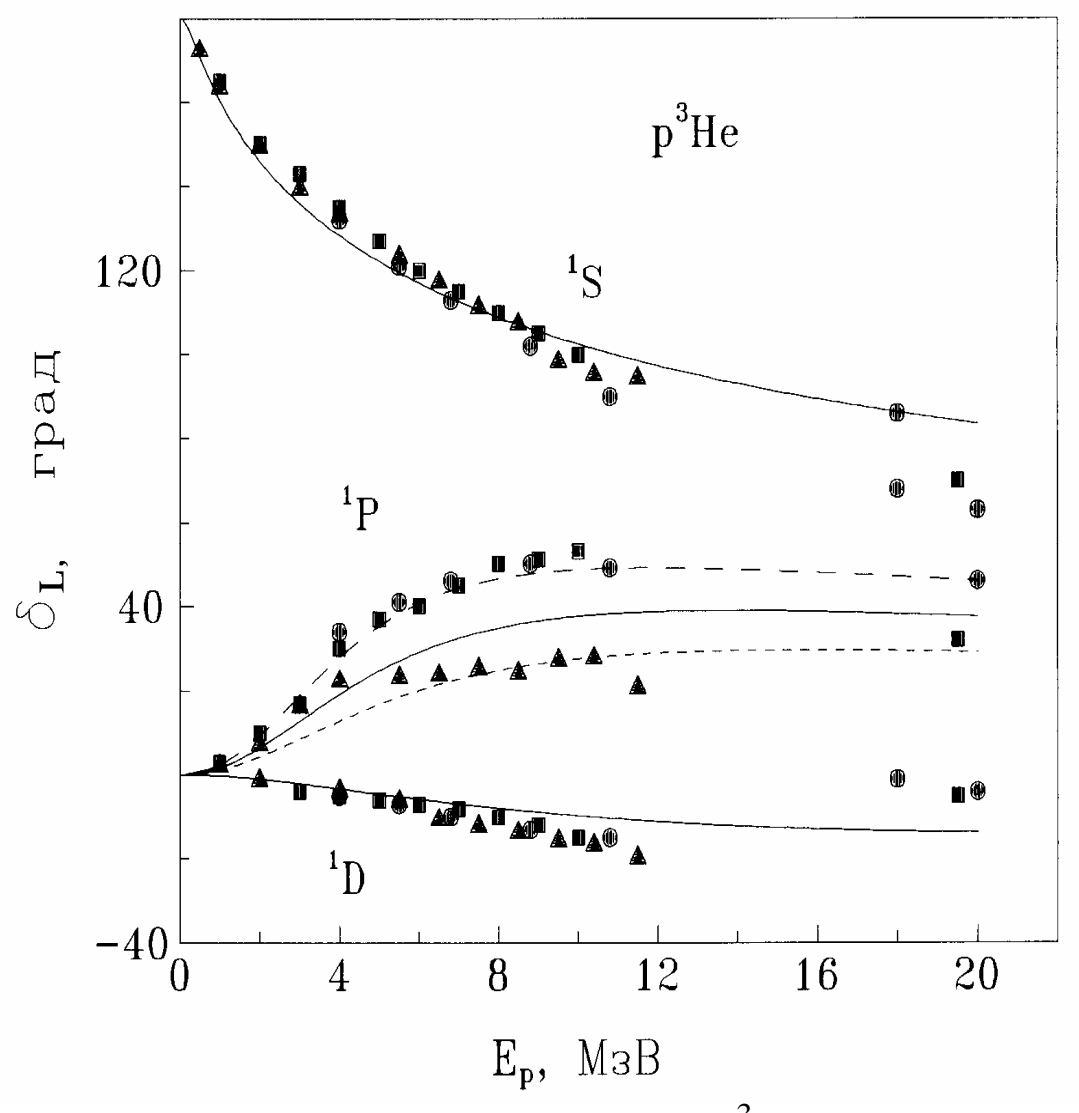

Рис.3.12а. Синглетные фазы упругого $p^3He$ рассеяния. Точки и квадраты - эксперимент из работ [32,33,35], треугольники из [34]. Кривые - расчеты для потенциалов из табл.3.8.

Поскольку имеется несколько различных вариантов фазовых анализов для синглетной $^1P_1$ и триплетной $^3P_1$ волн, параметры потенциала, приведенные в табл.3.8, подбирались так, чтобы получить определенный компромисс между разными фазовыми анализами.

В тоже время, для описания синглетных данных [34], показанных треугольниками [34], нужен потенциал с глубиной около -11 МэВ (точечная линия на рис.3.12а), а фазы работ [33] воспроизводятся более глубоким взаимодействием с $V_0$ = 17 МэВ (штриховая линия). Для воспроизведения триплетной $^3P_1$ волны [34] глубина потенциала должна быть 17 МэВ (точечная линия на рис.3.11б), а для данных работ [33] - 13 МэВ (штриховая линия).

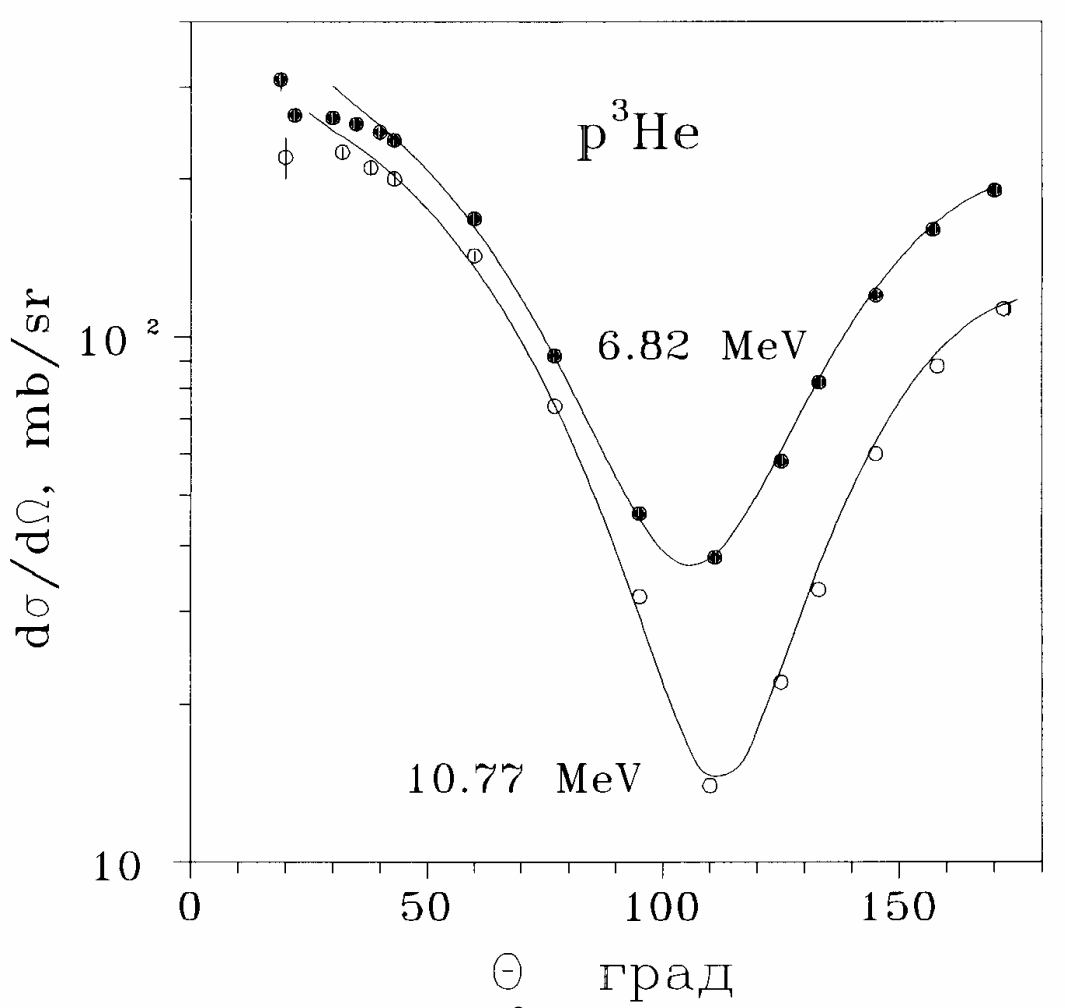

Рис.3.12б. Сечения упругого p$^3$He рассеяния. Точки и кружки - эксперимент из работ [32-35]. Непрерывные кривые - результаты расчетов для потенциалов с параметрами из табл.3.8.

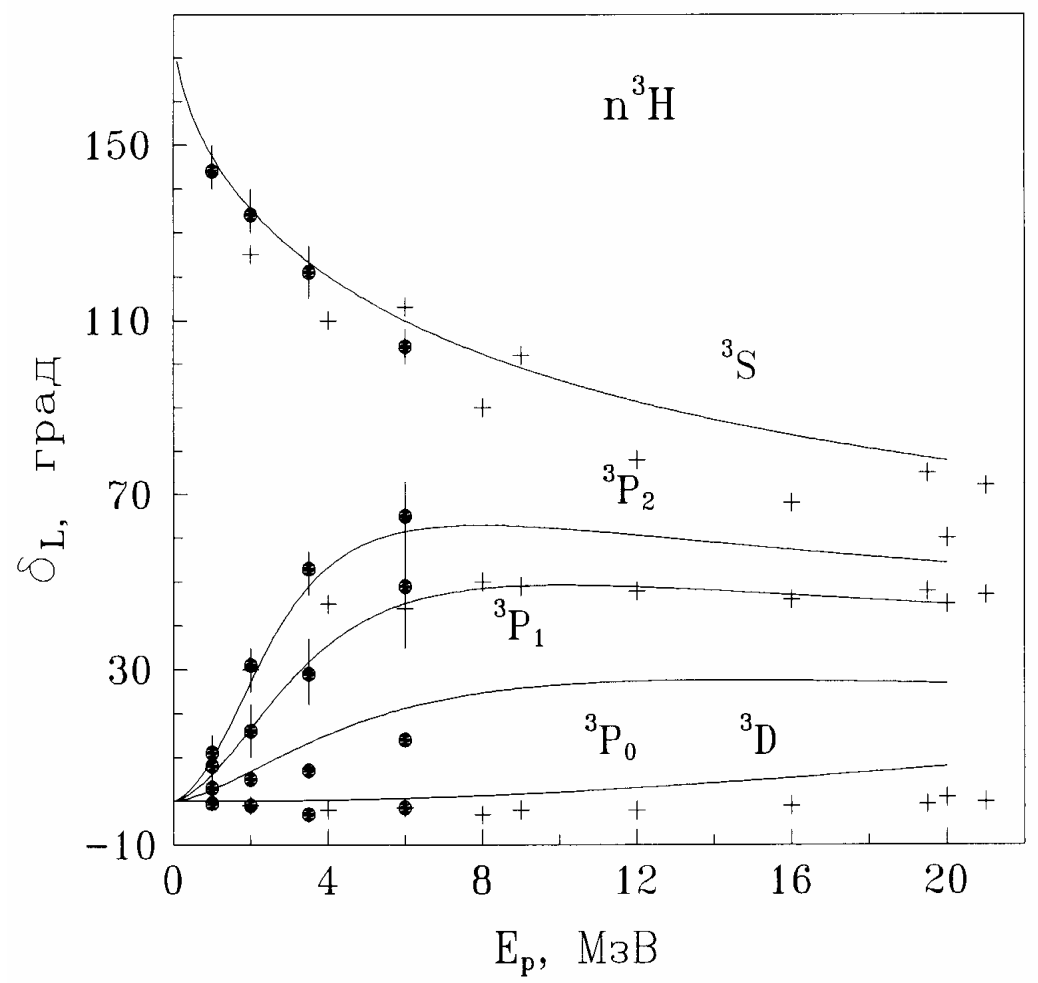

Рис.3.13а. Триплетные фазы упругого n$^3$H рассеяния. Точки с ошибками - эксперимент [39], крестики - МРГ вычисления из [40], непрерывные кривые - результаты расчетов для потенциалов с параметрами из табл.3.8.

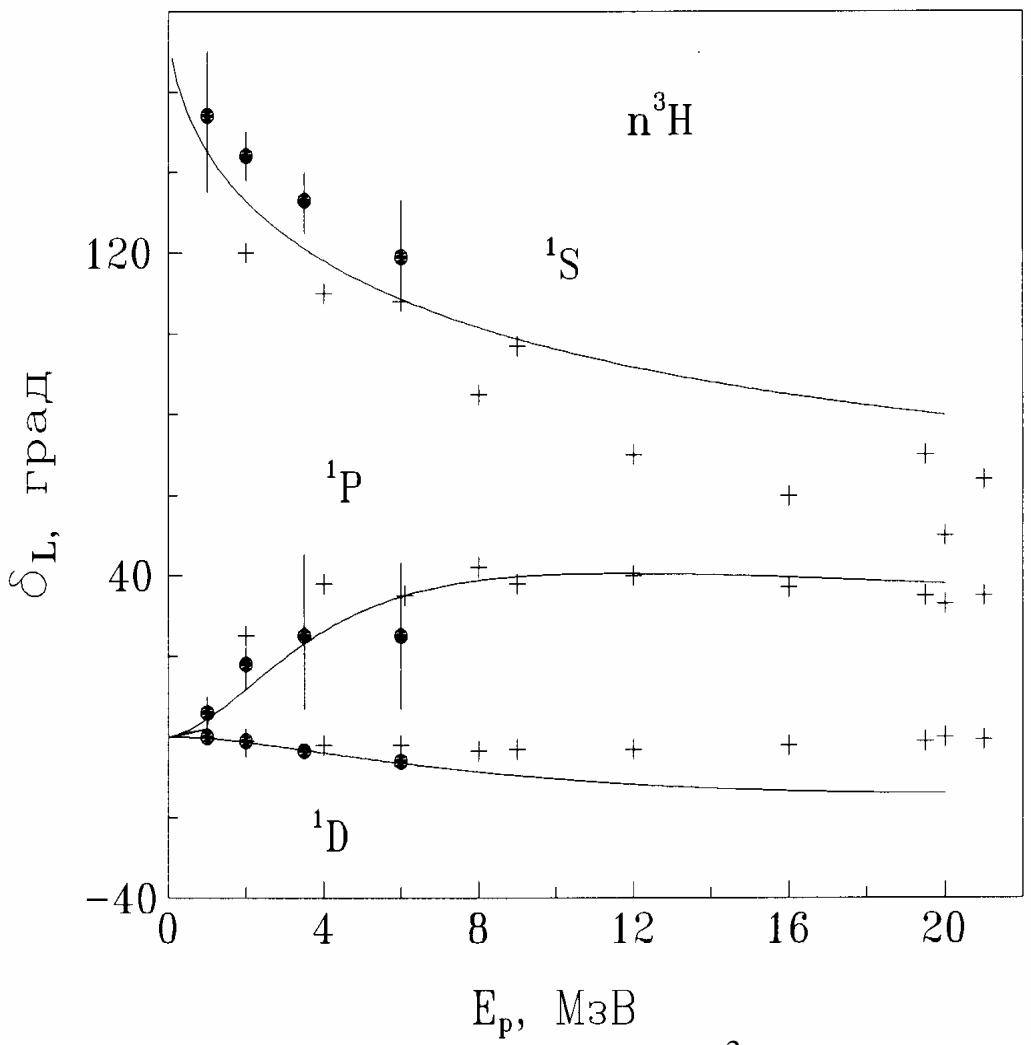

Рис.3.13б. Синглетные фазы упругого $n^3H$ рассеяния. Точки с ошибками - эксперимент [39], крестики - МРГ вычисления из [40], непрерывные кривые - результаты расчетов для потенциалов с параметрами из табл.3.8.

Энергии запрещенных состояний со схемой {4} приведены в табл.3.8. Разрешенные состояния симметрии {31} в Р волнах оказываются не связанными, так как на один квант возбуждения приходится около 10-15 МэВ энергии [41].

## 3.3. Кластерный канал $N^2H$ в ядрах $^3H$ и $^3He$

В этом и последующих параграфах перейдем к рассмотрению кластерных систем смешанных по орбитальным схемам Юнга в состояниях с минимальным спином.

Как видно из табл. 2.1, в квартетном канале $N^2H$ системы возможна только одна орбитальная схема {21}, а симметрия {3} запрещена. В то же время, дублетный канал совместим сразу с двумя схемами {21} и {3}. Поэтому, дублетные потенциалы, получаемые на основе фаз рассеяния, эффективно зависят от двух этих схем Юнга, в то время, как основному состоянию ядер $^3H$ и $^3He$ обычно сопоставляется только одна орбитальная симметрия {3}. Именно поэтому потенциал, полученный на основе фаз рассеяния, нельзя непосредст-венно использовать для описания характеристик связанных состояний.

Экспериментальные данные по фазовому анализу в $p^2H$ системе имеются в достаточно широкой энергетической области и результаты разных работ в целом согласуются между собой [42]. Используя эти данные, были получены потенциалы взаимодействия, параметры которых приведены в табл.3.9 [43].

На рис.3.14 показаны, вычисленные с этими потенциалами, и экспериментальные [42] дублетные $p^2H$ фазы при малых энергиях. Непрерывной линией на рис.14а даны результаты расчетов фаз для четных волн со вторым набором параметров для дублетных потенциалов из табл. 3.9. Фазы первого набора параметров изображены штриховой линией и приводят к несколько завышенной D фазе. Для P волны непрерывной линией показаны результаты для второго набора параметров, который имеет периферическое отталкивание, а точечной линией даны результаты для первого набора.

На рис.3.14б непрерывными линиями показаны результаты для квартетных потенциалов из табл.3.9, а штрихами приведены фазы взаимодействия из работ [38] с параметрами: $V_0$=57 МэВ, $\alpha$=0.37 Фм$^{-2}$, $V_1$=7.2 МэВ, $\beta$=0.36 Фм$^{-1}$ для четных волн и $V_0$=8.8 МэВ, $\alpha$ =0.06 Фм$^{-2}$ для нечетных.

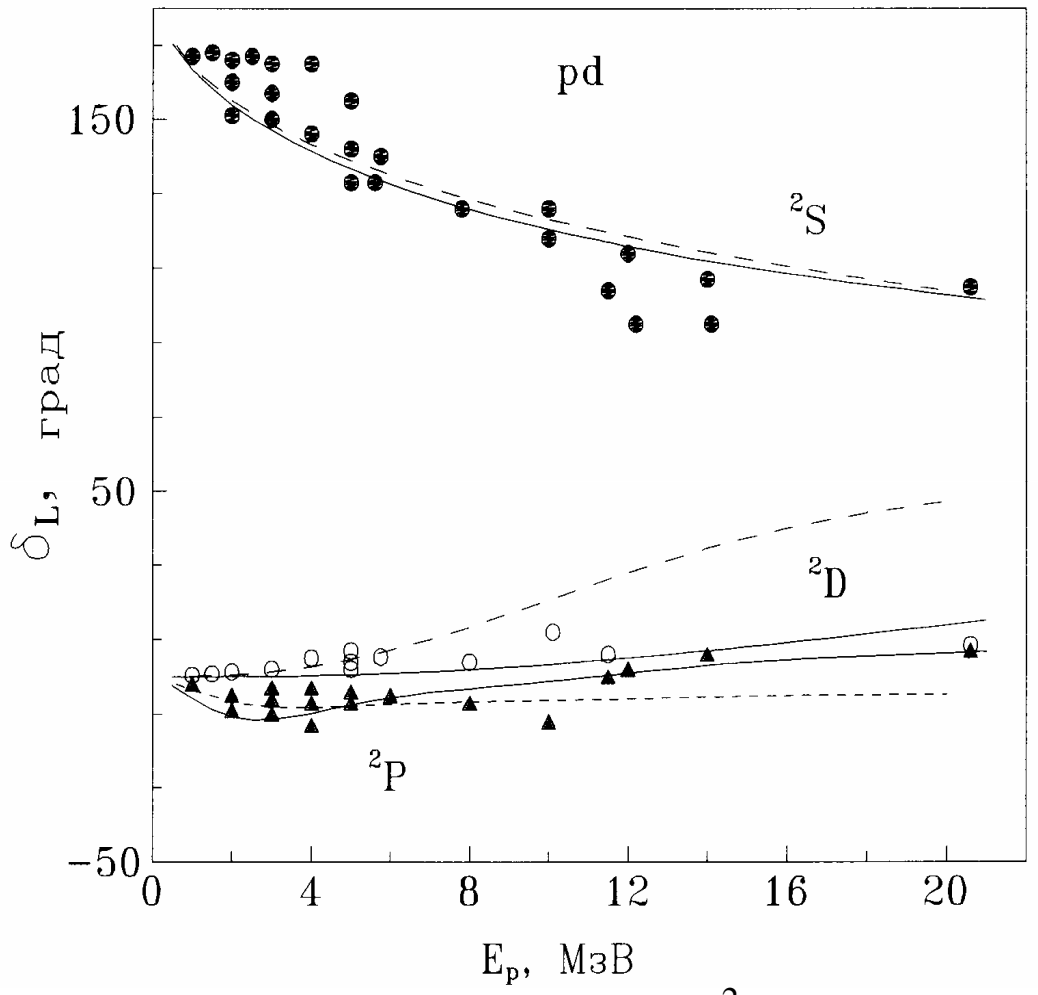

Рис.3.14а. Дублетные фазы упругого $p^2H$ рассеяния. Точки - экспериментальные данные [42], кривые - результаты расчетов для потенциалов с параметрами из табл.3.9.

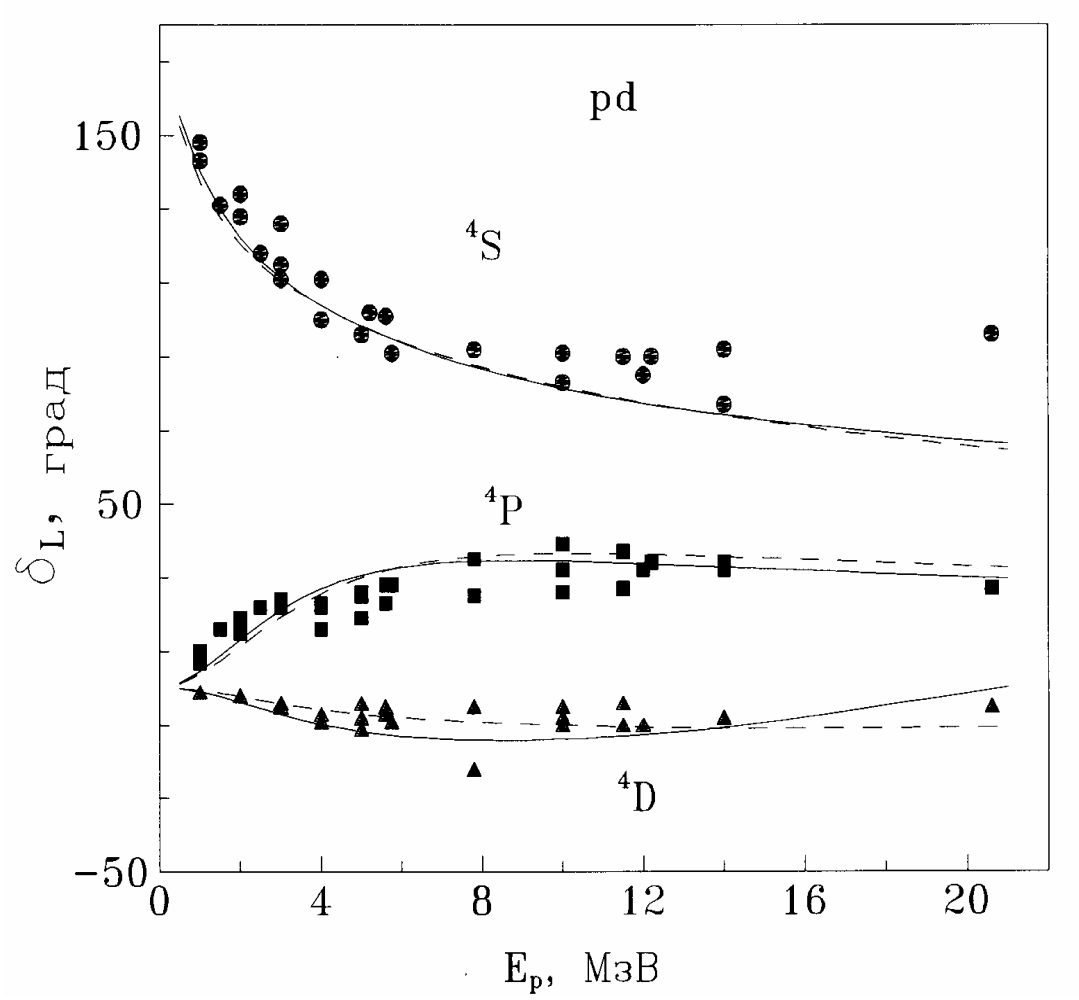

Рис.3.14б. Квартетные фазы упругого $p^2H$ рассеяния. Точки - экспериментальные данные [42], кривые - результаты расчетов для потенциалов с параметрами из табл.3.9.

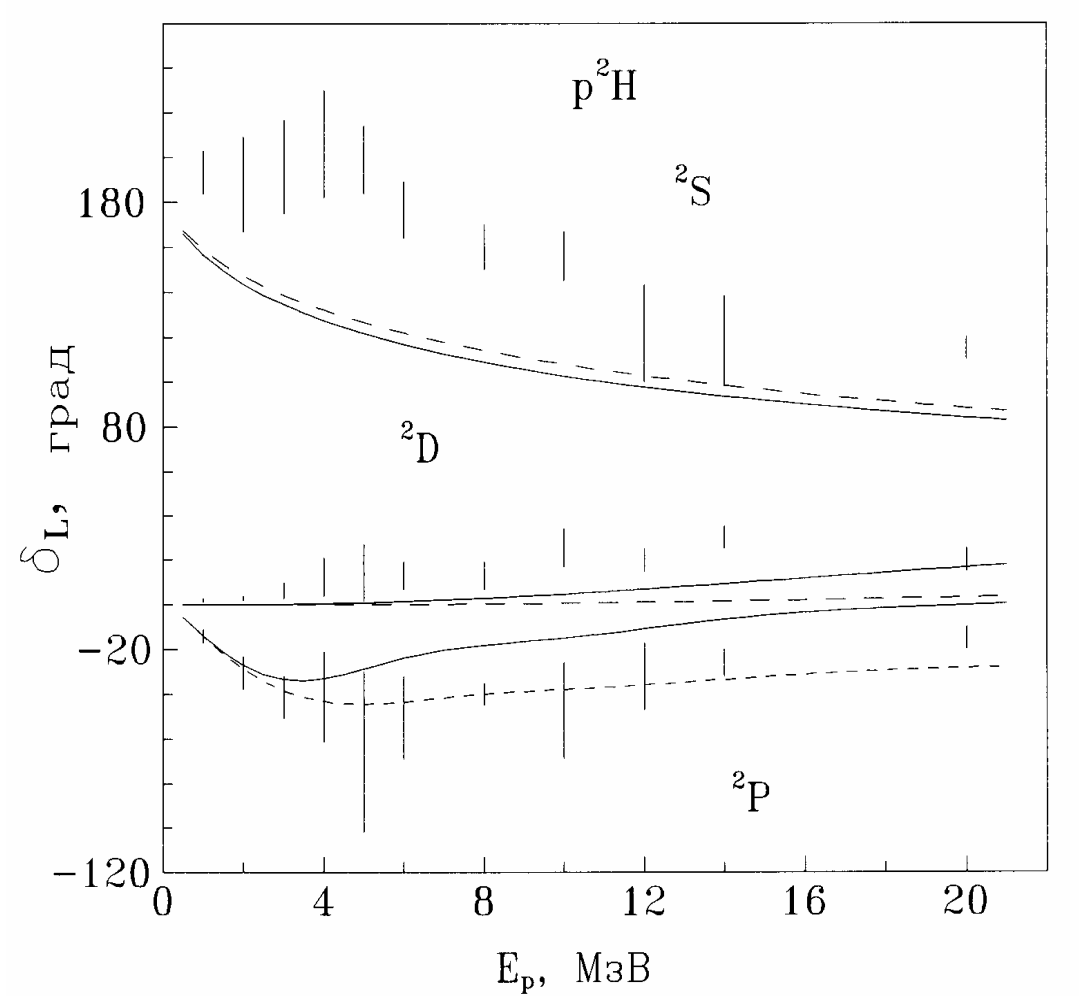

Рис.3.14в. Чистые по схемам Юнга фазы упругого $p^2H$ рассеяния. Вертикальные линии - полоса ошибок для чистых фаз со схемой {3}, кривые - результаты расчетов чистых фаз для потенциалов с параметрами из табл.3.10.

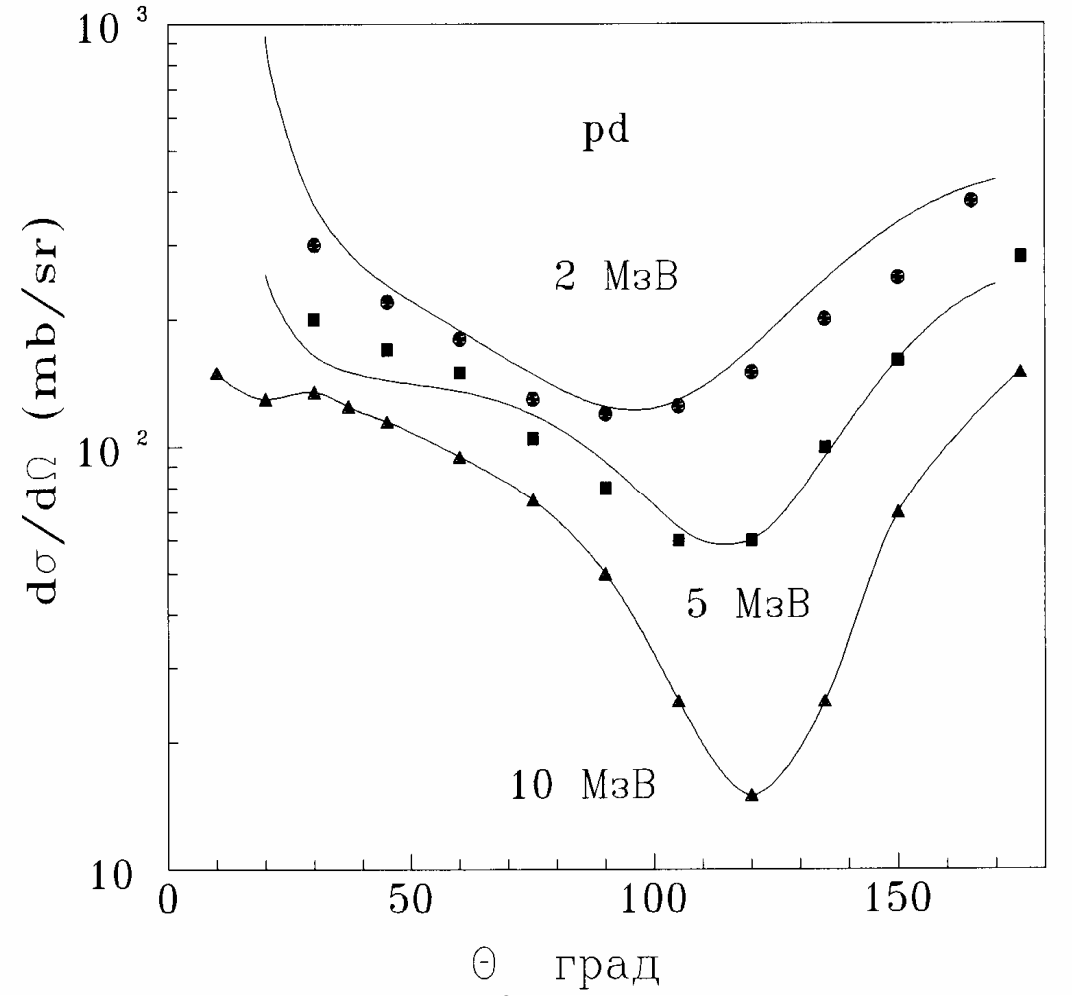

Рис.3.15а. Сечения упругого $p^2H$ рассеяния. Точки, треугольники и квадраты - экспериментальные данные [42,44], непрерывные кривые - результаты расчетов для потенциалов с параметрами из табл.3.9.

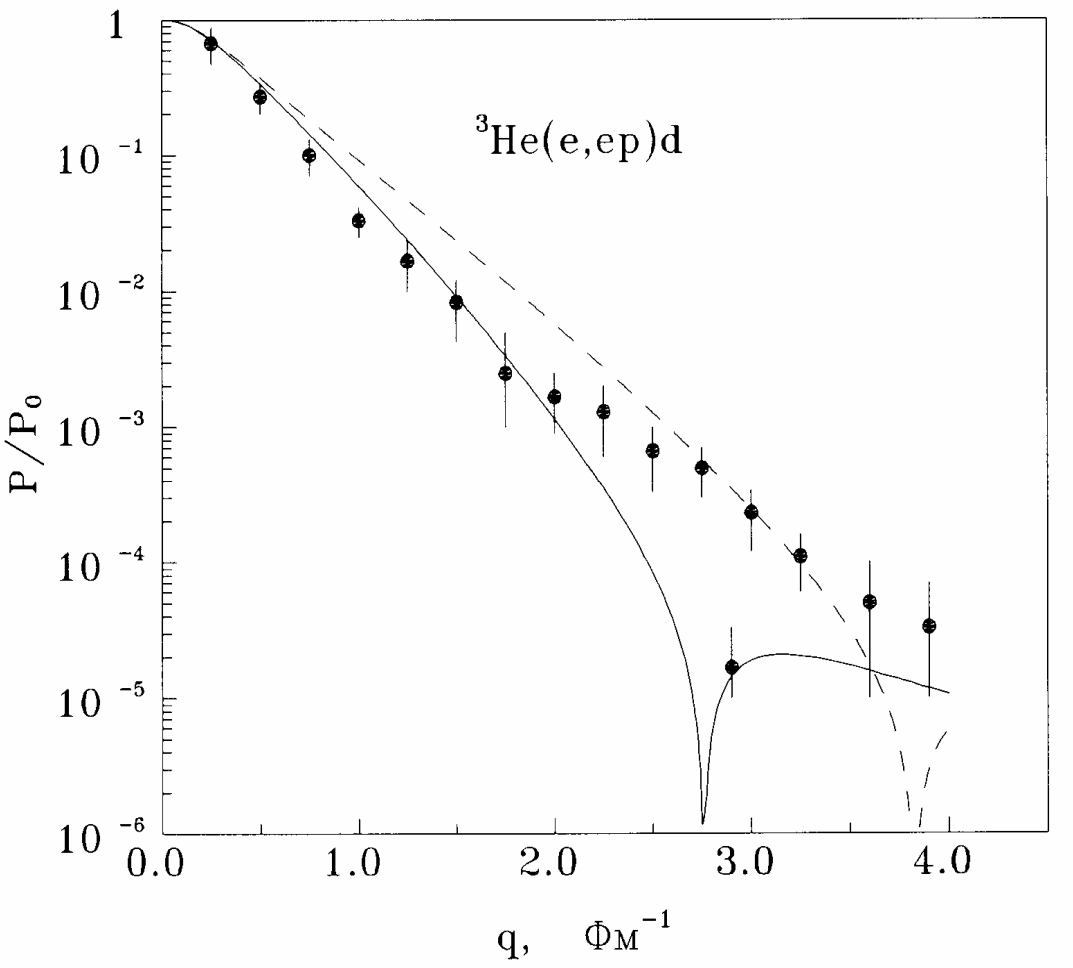

Рис.3.15б. Импульсные распределения $p^2H$ кластеров в ядре $^3He$. Точки - эксперимент из работы [45], кривые - расчеты для потенциалов основного состояния из табл.3.10.

Видно, что потенциалы без спин - орбитального расщепления вполне способны передать энергетическое поведение экспериментальных фаз. На рис.3.15а приведены вычисленные сечения упругого $p^2H$ рассеяния при 2, 5 и 10 МэВ, в сравнении с данными работ [42,44].

*Таблица 3.9. Потенциалы взаимодействия $p^2H$ и $^2H^3He$ кластерных систем, смешанные по схемам Юнга в дублетных каналах. Для обеих систем $R_c$ =0. Здесь $E_{cc}$ - энергии связанных состояний. В скобках приведены значения энергии для $n^2H$ и $^2H^3H$ систем.*

| Система | $L_J$ | $V_0$, (МэВ) | $\alpha$, ($Фм^{-2}$) | $V_1$, (МэВ) | $\beta$, ($Фм^{-1}$) | $E_{cc}$, (МэВ) |
|---|---|---|---|---|---|---|
| $p^2H$ | S=1/2 | | | | | |
| | Чет. | -35.0 | 0.1 | | | -9.3(-10.1) |
| | | -55.0 | 0.2 | | | -11.4(-12.3) |
| | Нечет. | +0.4 | 0.01 | | | |
| | | -10.0 | 0.16 | +0.6 | 0.1 | |
| | S=3/2 | | | | | |
| | Чет. | -41.9 | 0.13 | +13.7 | 0.36 | -4.2(-4.9) |
| | Нечет. | -7 | 0.05 | | | |
| $^2H^3He$ | S=1/2 | | | | | |
| | Чет. | -45.5 | 0.15 | | | -15.9(-17.0) |
| | Нечет. | -44 | 0.1 | | | -7.5(-8.1) |
| | S=3/2 | | | | | |
| | Чет. | -34.5 | 0.1 | | | -12.9(-13.8) |
| | Нечет. | -29 | 0.1 | | | -1.3(-1.8) |

В табл.3.9 приведены энергии связанных состояний для обоих спиновых состояний. В квартетном канале при L=0 имеется запрещенное состояние со схемой {3}, а разрешенное состояние с {21} для L=1 находится в непрерывном спектре, также как в случае $p^3He$ системы. Смешанный S потенциал дублетного канала приводит к связанному состоянию с энергией отличной от энергии связи ядра $^3He$.

Как говорилось выше, экспериментальные смешанные дублетные фазы могут быть представлены в виде полусуммы чистых фаз

$$\delta_L^{\{f_1\}+\{f_2\}} = \frac{1}{2}\delta_L^{\{f_1\}} + \frac{1}{2}\delta_L^{\{f_2\}} . \qquad (3.3.1)$$

В данном случае $\{f_1\}=\{3\}$ и $\{f_2\}=\{21\}$. Если допустить, что в качестве дублетных фаз с {21} могут быть использованы квартетные фазы той же симметрии {21}, то легко определить чистые дублетные фазы с {3}. На рис.3.14в вертикальными линиями, которые показывают полосу ошибок, приведены, полученные таким образом чистые $p^2H$ фазы. Параметры чистых взаимодействий даны в табл.3.10 вместе с энергиями связанных состояний.

*Таблица 3.10. Чистые по схемам Юнга потенциалы $p^2H$ и $^3He^2H$ взаимодействия в дублетных каналах. Для обеих систем $R_c$ =0. Здесь $E_{cc}$ - энергии связанных состояний. В скобках даны энергии для $n^2H$ и $^2H^3H$ систем.*

| Система | $L_J$ | $V_0$ , (МэВ) | $\alpha$ , (Фм$^{-2}$) | $V_1$ , (МэВ) | $\beta$ , (Фм$^{-1}$) | $E_{cc}$ , (МэВ) |
|---|---|---|---|---|---|---|
| $p^2H$ | Чет. | -34.75 | 0.15 | | | -5.49(-6.25) |
| | | -54.3 | 0.3 | | | -5.49(-6.40) |
| | Нечет. | +2.4 | 0.01 | | | |
| $^2H^3He$ | Чет. | -40.0 | 0.15 | +8.0 | 0.2 | -7.1(-8.0) |
| | $^2P_{3/2}$ | -75.5 | 0.15 | | | -16.4(-17.2) |
| | $^2P_{1/2}$ | -60.2 | 0.15 | | | -8.9(-9.6) |

В четных волнах фазы потенциала с первым набором параметров показаны на рис.3.14в непрерывной линией. Штриховой линией даны результаты для второго набора параметров, которые практически не отличаются, от приведенных в работах [38]. Фазы чисто отталкивающего P взаимодействия показаны точечной кривой, а непрерывной линией приведены фазы потенциала с периферическим отталкиванием из работы [38], для которого получены параметры:

$V_0$=-13.8 МэВ, $\alpha$=0.16 Фм$^{-2}$, $V_1$=+1.6 МэВ, $\beta$=0.09 Фм$^{-1}$.

Из рисунка видно, что вполне удается передать поведение чистой $p^2H$ фазы с L=1, в то время, как S фаза имеет характер близкий к резонансному. Поэтому не удалось найти взаимодействие способное описать ее форму, одновременно с характеристиками связанных состояний. Приведенные взаимодействия для основного состояния правильно описывают энергию связи ядер $^3He$ и $^3H$. Экспериментальные значения энергий равны -5.493 МэВ и -6.257 МэВ соответственно.

Оба чистых S потенциала $p^2H$ взаимодействия приводят примерно к одинаковым фазам рассеяния, но только второй из них без всяких деформаций дейтронного кластера позволяет относительно правильно передать кулоновский упругий формфактор при малых импульсах, как показано на рис.3.8а точечной линией.

Зарядовый радиус $^3$He, в случае этого взаимодействия - $R_f$=1.89 Фм (см. табл.3.3) и $R_r$=2.08 Фм, что немного больше экспериментальной величины, а асимптотическая константа - 1.8(1), при экспериментальных значениях находящихся в интервале 1.8-2.3 [13]. В случае n$^2$H системы второй потенциал несколько занижает энергию связи, но если уменьшить глубину до 53.9 МэВ, то энергия оказывается равна -6.25 МэВ, а радиусы $R_f$=1.89 Фм и $R_r$=2.01 Фм.

Для первого потенциала при описании радиуса и формфактора необходимо уже вводить деформации дейтрона, т.е. предполагать, что дейтронному кластеру внутри ядра $^3$He или $^3$H нельзя уже сопоставлять свойства свободного дейтрона. С точки зрения размеров дейтрона и ядер $^3$He и $^3$H это вполне обосновано, так как радиус дейтрона заметно больше радиусов этих ядер, как видно из табл.3.3. Поэтому для получения правильных радиусов и формфакторов рассматриваемых ядер необходимо сжать дейтронный кластер примерно на 30-40% [43]. Асимптотическая константа для этого потенциала в обоих каналах равна 2.3(1). Без учета деформаций для радиусов получены величины - $R_f$= 2.05 Фм и $R_r$=2.20 Фм в p$^2$H и $R_f$=1.95 Фм и $R_r$=2.12 Фм в n$^2$H. Формфактор без деформаций для этого потенциала показан на рис.3.8а штриховой линией.

На рис.3.15б непрерывной линией показаны импульсные распределения кластеров в ядре $^3$He, полученные с первым набором параметров потенциала основного состояния. Штриховой линией приведены результаты для второго потенциала. Экспериментальные данные из работы [45].

## 3.4. Рассеяние в кластерной системе $^2$H$^3$He

Перейдем теперь к рассмотрению $^2$H$^3$He системы. В дублетном канале фазы зависят от двух орбитальных схем {41} и {32}, как видно из табл.2.1. В квартетном канале разрешена только схема {32} и все состояния чистые.

Экспериментальные данные по фазам рассеяния имеются только в узкой области энергий 0-5 МэВ [46]. Поэтому для получения потенциала приходится использовать МРГ вычисления фаз, выполненные в работах [47]. Результаты расчета квартетных фаз [43] с потенциалами из табл.3.9 показаны на рис.3.16а непрерывными линиями. Здесь точки и квадраты - экспериментальные P и S фазы из [46], крестики и кружки - варианты МРГ вычислений из [47].

Штриховой линией показаны D фазы потенциала из работ [38] с параметрами $V_0$=-50 МэВ и $\alpha$=0.15 Фм$^{-2}$, приводящие к более правиль-ному описанию МРГ D фаз, при примерно таких же результатах для S фаз. В табл.3.9 приведены энергии запрещенных состояний для квартетных потенциалов со схемой Юнга {5} при орбитальном моменте L=0 и со схемой {41} при L=1. Разрешенное состояние со схемой {32} оказывается не связанным, так как имеет два кванта возбуждения.

Из рис.3.16б видно, что МРГ и экспериментальная дублетная S фаза существенно различаются между собой. Параметры взаимодействия, приведенные в табл.3.9, получены на основе экспериментальных данных, а его S фаза показана на рисунке точечной линией.

Для описания МРГ S фазы необходим потенциал с параметрами $V_0$=-25 МэВ и $\alpha$=0.15 Фм$^{-2}$, S и D фазы которого приведены непрерывными линиями. Дублетный P потенциал получен на основе МРГ вычислений и в целом согласуется с экспериментом. Сечения упругого рассеяния, вычисленные с этими взаимодействиями, приведены на рис.3.17а вместе с экспериментом работ [6,48].

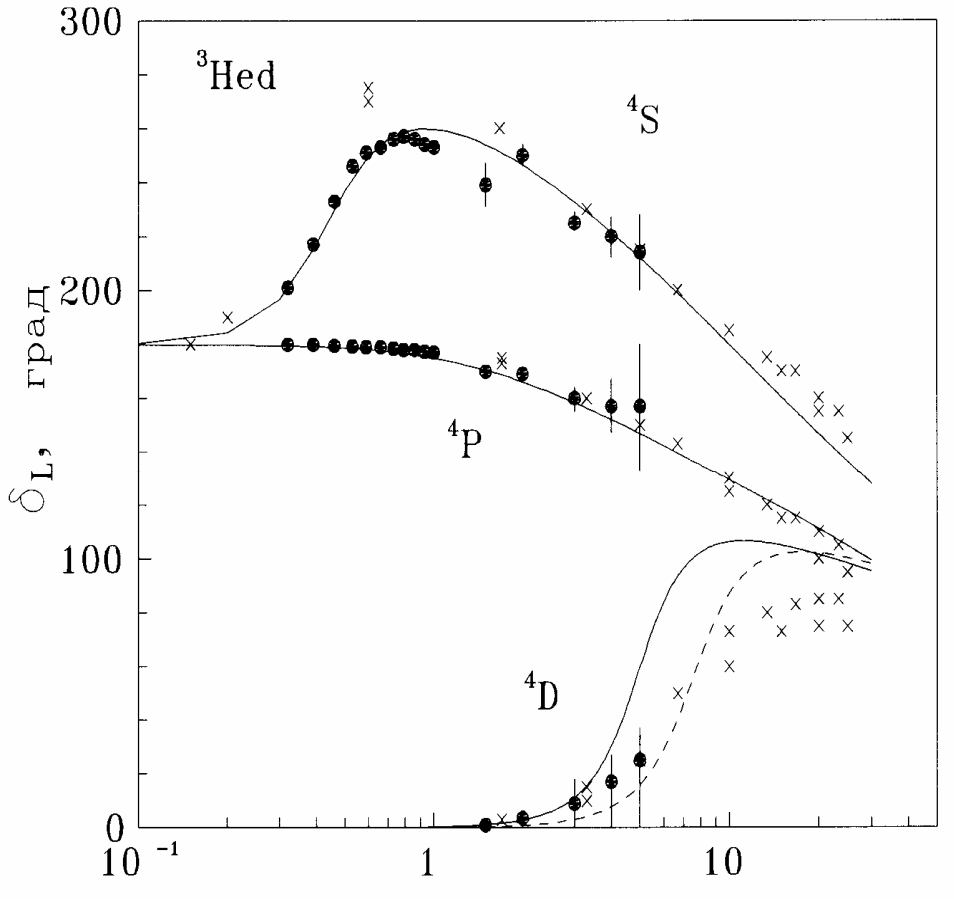

Рис.3.16а. Экспериментальные квартетные смешанные по орбитальным симметриям фазы упругого $^2$H$^3$He рассеяния. Точки - данные [46], крестики - различные варианты МРГ вычислений [47], кривые - результаты расчетов для потенциалов с параметрами из табл.3.9.

Из табл.3.9. видно, что разрешенное в дублетном канале P состояние имеет энергию, не согласующуюся со средней энергией связанных состояний для P уровней, спектр которых показан на рис.3.18а, так как потенциал зависит от двух орбитальных схем.

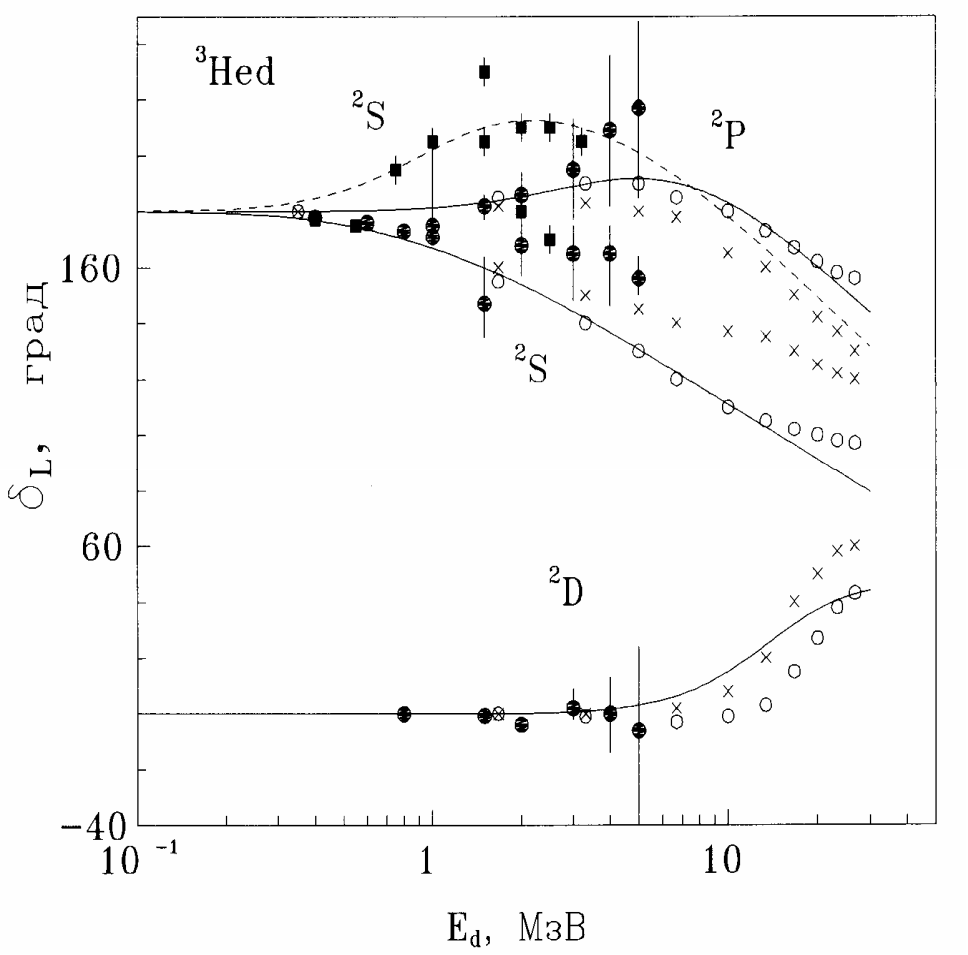

Рис.3.16б. Экспериментальные дублетные смешанные по орбитальным симметриям фазы упругого $^2H^3He$ рассеяния. Точки и квадраты - данные [46], крестики и кружки - различные варианты МРГ вычислений [47], кривые - результаты расчетов для потенциалов из табл.3.9.

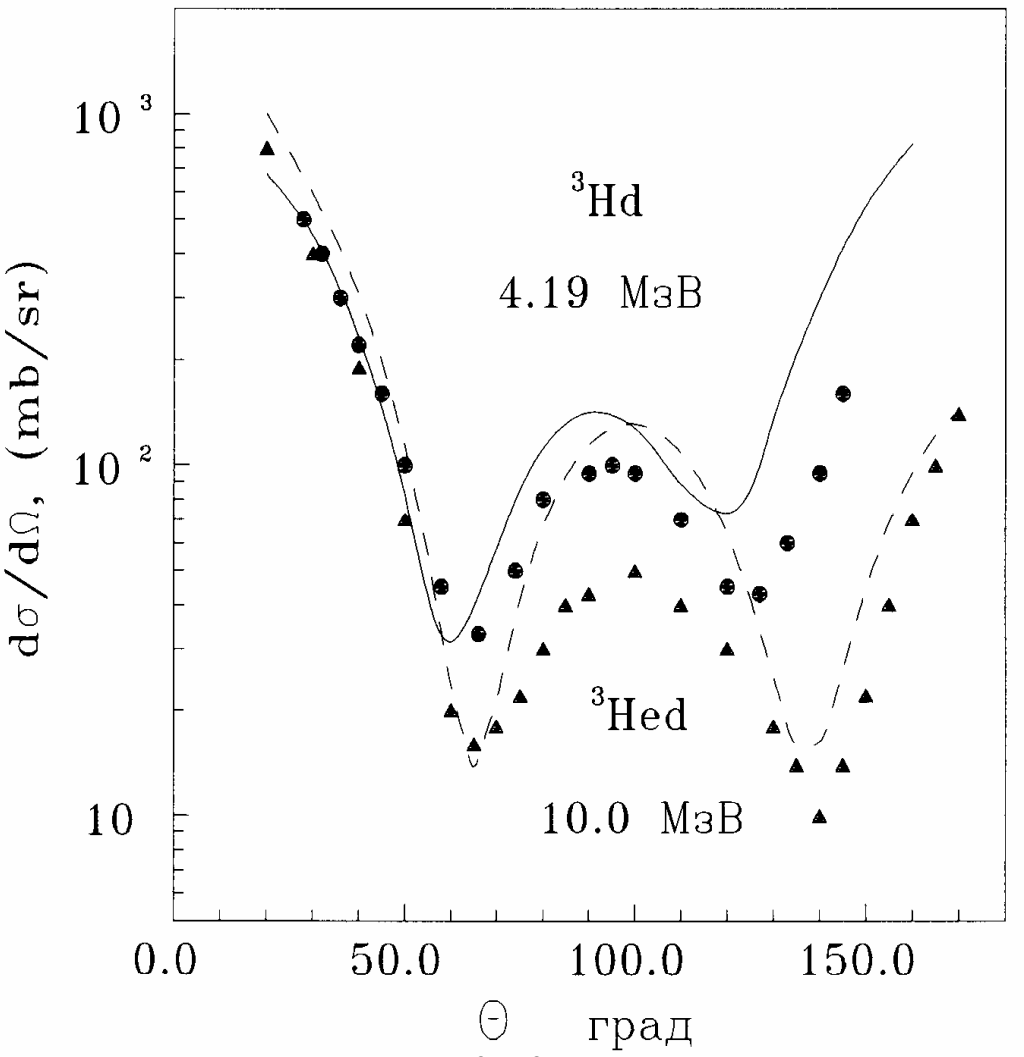

Рис.3.17а. Сечения упругого $^2H^3H$ рассеяния. Точки - экспериментальные данные [6,48], кривые - результаты расчетов для потенциалов с параметрами из табл.3.9.

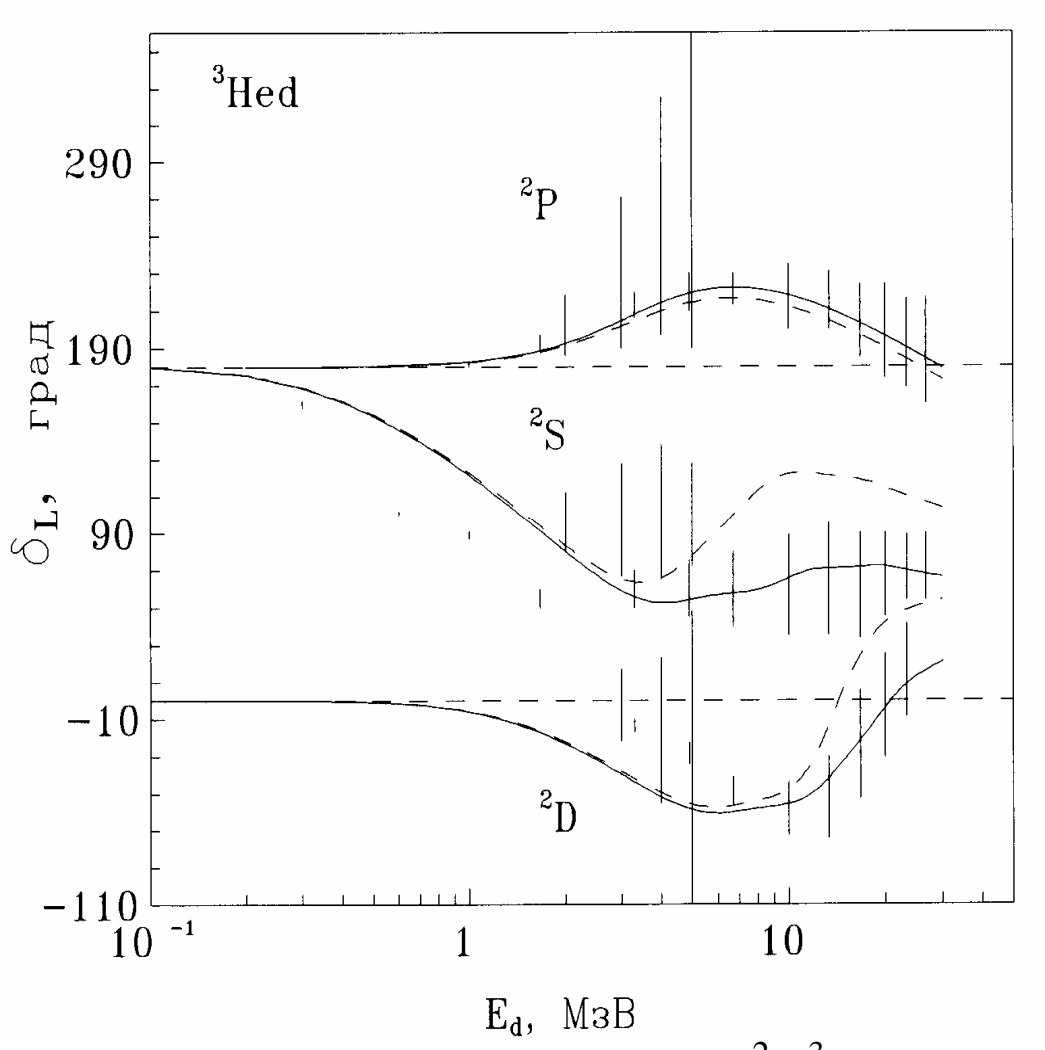

Рис.3.17б. Чистые по схемам Юнга фазы $^2H^3H$ рассеяния. Вертикальные линии - полоса ошибок для чистых фаз со схемой {41}, кривые - результаты расчетов чистых фаз для потенциалов с параметрами из табл.3.10.

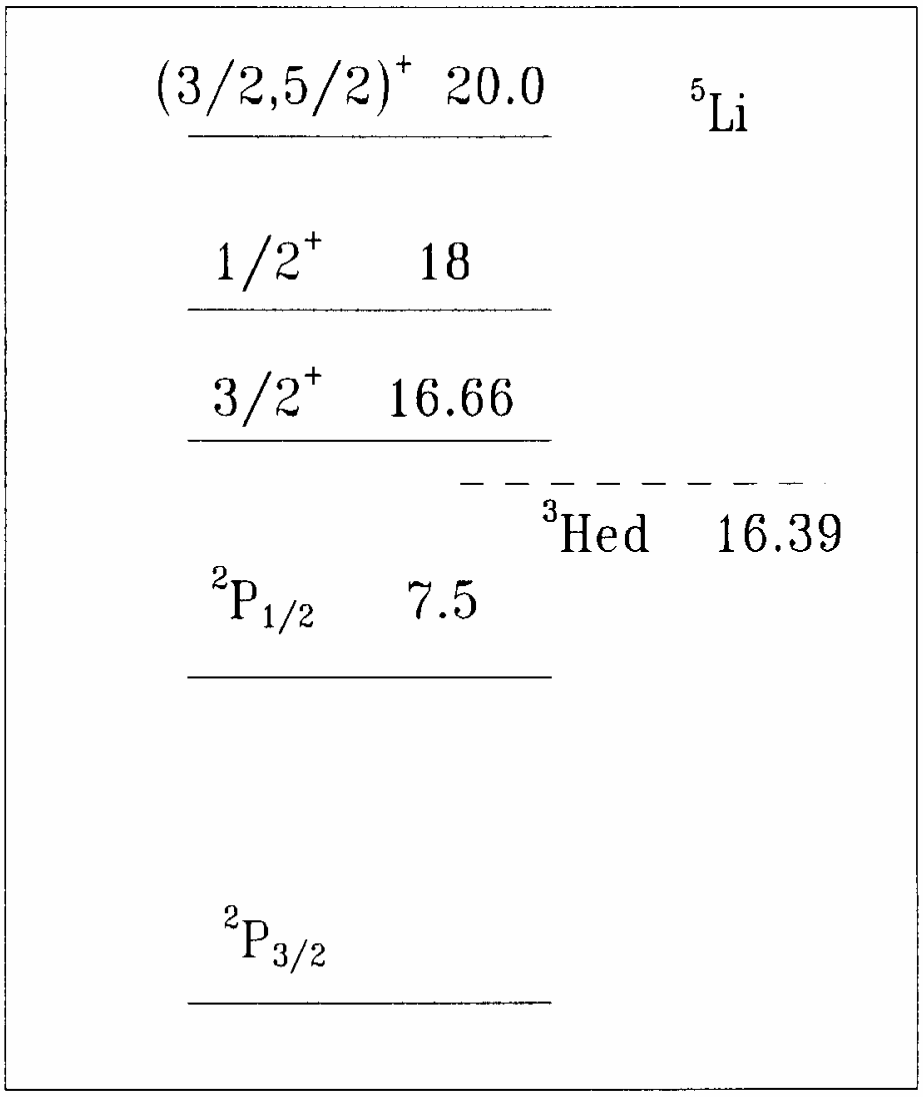

Рис.3.18а. Экспериментальные энергетические спектры ядра $^5Li$ [9].

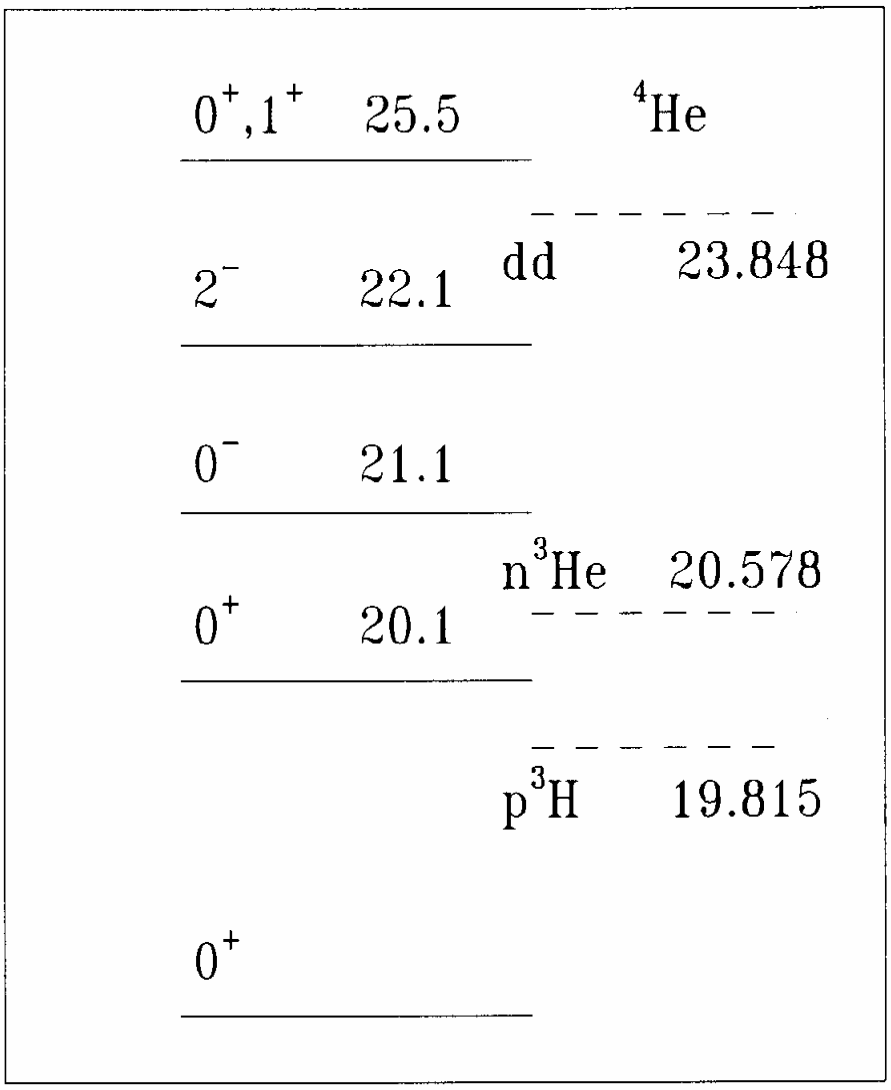

Рис.3.18б. Экспериментальные энергетические спектры ядра $^4$He [9].

На основе (3.3.1) получены чистые фазы рассеяния в дублетном канале, которые показаны на рис.3.17б вертикальными линиями. Параметры чистых потенциалов даны в табл.3.10, а на рисунке непрерывными линиями приведены результаты расчетов фаз с этими потенциалами для S, D и P волн. Штриховыми линиями показаны фазы потенциалов из работ [38] с параметрами: $V_0$=-57 МэВ, $\alpha$ = 0.16 Фм$^{-2}$, $V_1$=+8.4 МэВ, $\beta$=0.21 Фм$^{-1}$ для четных волн и $V_0$=-69 МэВ и $\alpha$ =0.14 Фм$^{-2}$ для нечетных. Чистые с {41} потенциалы дают правильную энергию P уровней ядра $^5$Li, а в S состоянии имеется запрещенный связанный уровень со схемой {5}.

## 3.5. Кластерная конфигурация $^2$H$^2$H ядра $^4$He

При рассеянии в $^2$H$^2$H системе в канале с S=2 разрешена орбитальная схема {22}, в триплетном канале схема {31}. Синглетный канал совместим с двумя симметриями {4} и {22}, как видно из табл.2.1.

Экспериментальные данные по фазам рассеяния в $^2$H$^2$H системе известны только в области 8-12 МэВ [49]. Поэтому приходится ис-пользовать результаты МРГ расчетов фаз [50]. На рис.3.19 и 3.20а крестиками показаны МРГ фазы, кружками результаты фазового анализа [49]. Из рис.3.19а видно большое различие между различными МРГ вычислениями в S волне синглетного канала.

Нижние результаты получены на основе многоканального варианта МРГ, верхние - одноканальные расчеты. Триплетные P фазы так же заметно отличаются в этих вариантах МРГ расчетов и лежат несколько ниже экспериментальных данных. В то же время, все D фазы и S фазы при S=2 не сильно отличаются между собой и в целом согласуются с имеющимся фазовым анализом экспериментальных данных по дифференциальным сечениям упругого рассеяния.

*Таблица 3.11. Смешанные в синглетном канале и чистые с {4} потенциалы взаимодействия $^2H^2H$ системы. Для всех случаев $R_c = 0$. $E_{cc}$ - энергии связанных состояний.*

| S, $L_J$ | $V_0$, (МэВ) | $\alpha$, ($Фм^{-2}$) | $E_{cc}$, (МэВ) |
|---|---|---|---|
| S=0 | -9.0 | 0.05 | -1.5 |
| | -108.0 | 0.125 | -17.1; -61.5 |
| S=1 | -51 | 0.16 | -2.0 |
| S=2 | -40 | 0.24 | -6.9 |
| $^1S$ {4} | -47.15 | 0.073 | -23.85; -3.73 |
| $^1D$ {4} | -35.0 | 0.11 | |

Триплетный потенциал $^2H^2H$ системы строился по экспериментальным данным [49], взаимодействия с L=0 и 2 по эксперименту и МРГ вычислениям. Параметры потенциалов даны в табл.3.11 [43]. Результаты расчета фаз показаны на рисунках непрерывными линиями.

В синглетном канале получено два варианта взаимодействий, первый из которых содержит одно связанное состояние и описывает многоканальные МРГ результаты, как показано на рис.3.19а штриховой линией. Второй, правильно воспроизводя D фазу, приводит к определенному компромиссу между различными МРГ вычислениями - непрерывная линия на рис.3.19а. Поскольку при S=0 разрешены две схемы, совместимые с L=0, а {22} допускает и L=2, то второй вариант синглетного взаимодействия, кажется более реальным, поскольку имеет два глубоких связанных уровня в S и одно в D состоянии, которое находится при энергии -13.3 МэВ. Триплетный потенциал имеет разрешенное состояние, согласующееся с уровнем $2^-$ при 22.1 МэВ, как показано на рис.3.18б.

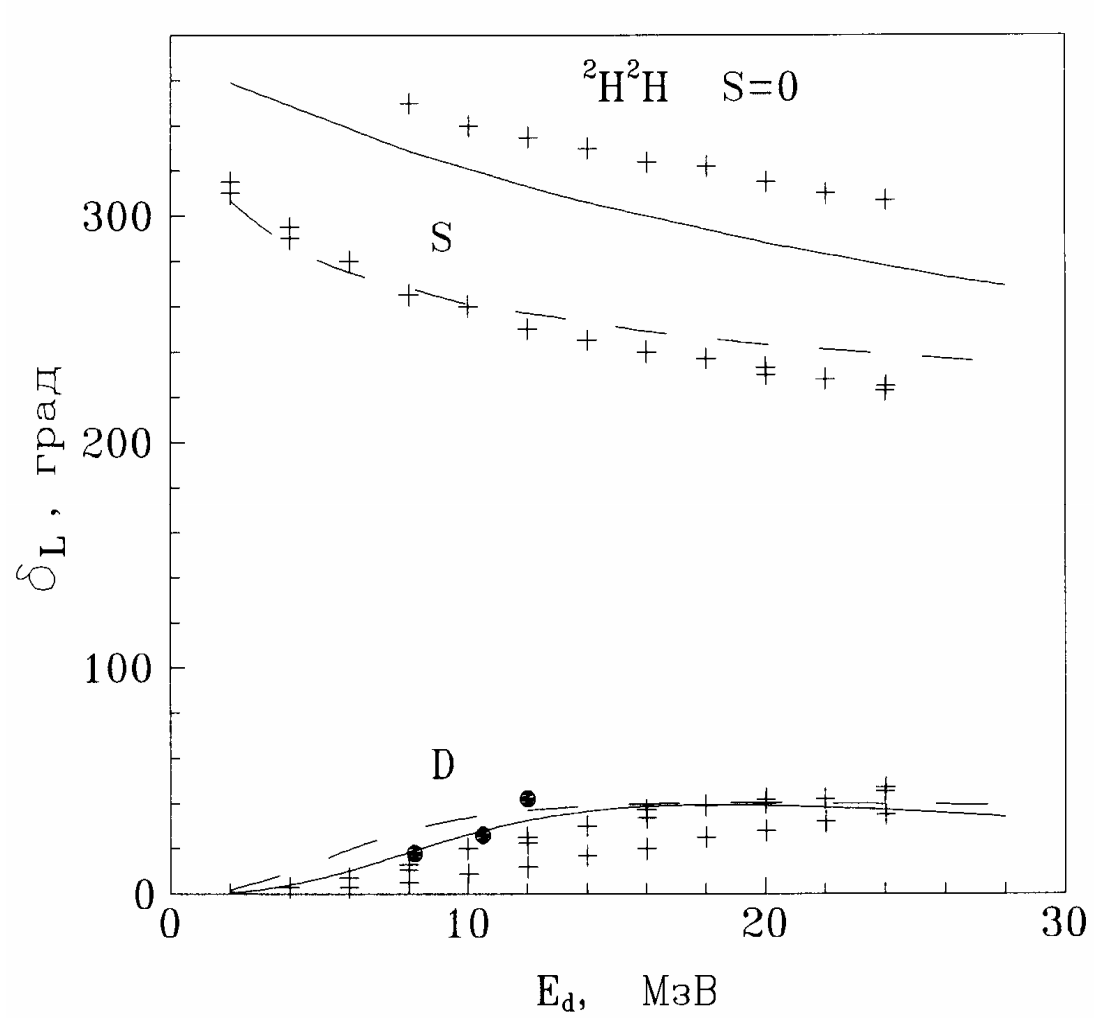

Рис.3.19а. Синглетные фазы упругого $^2H^2H$ рассеяния. Точки - экспериментальные данные [49], крестики - различные варианты МРГ вычислений [50], кривые - результаты расчетов для потенциалов с параметрами из табл.3.11.

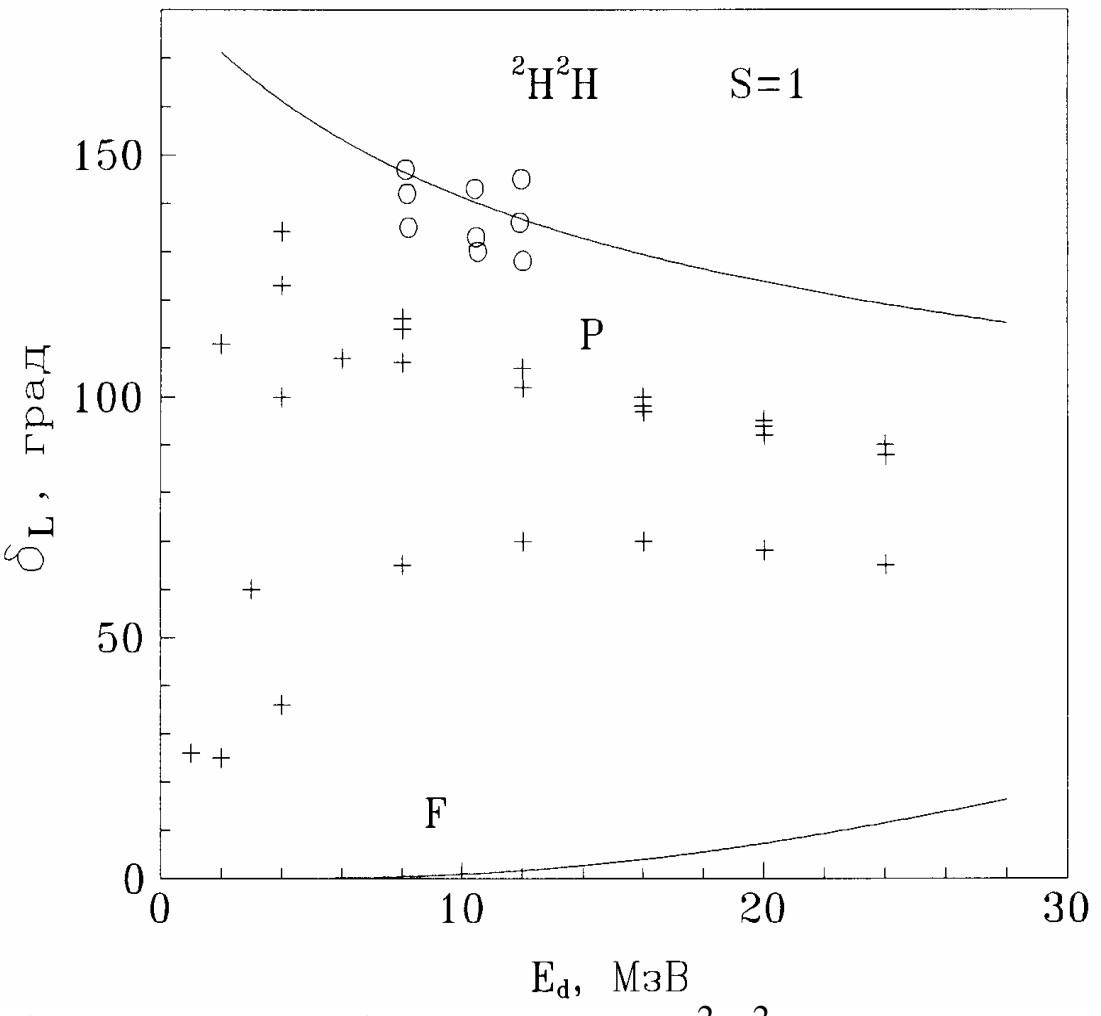

Рис.3.19б. Триплетные фазы упругого $^2H^2H$ рассеяния. Кружки - экспериментальные данные [49], крестики - различные варианты МРГ вычислений [50], кривые - результаты расчетов для потенциалов с параметрами из табл.3.11.

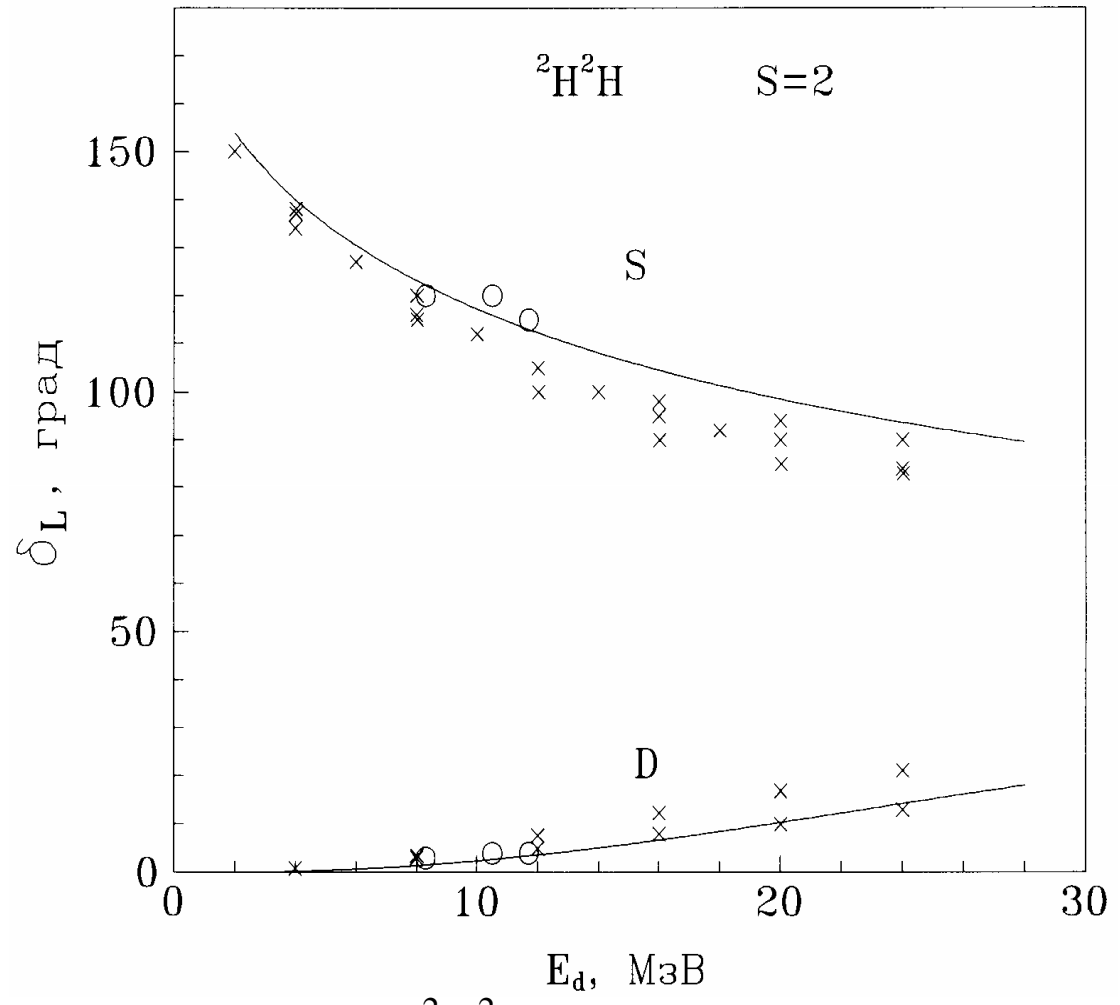

Рис.3.20а. Фазы упругого $^2H^2H$ рассеяния со спином 2. Кружки - экспериментальные данные [49], крестики - различные варианты МРГ вычислений [50], кривые - результаты расчетов для потенциалов с параметрами из табл.3.11.

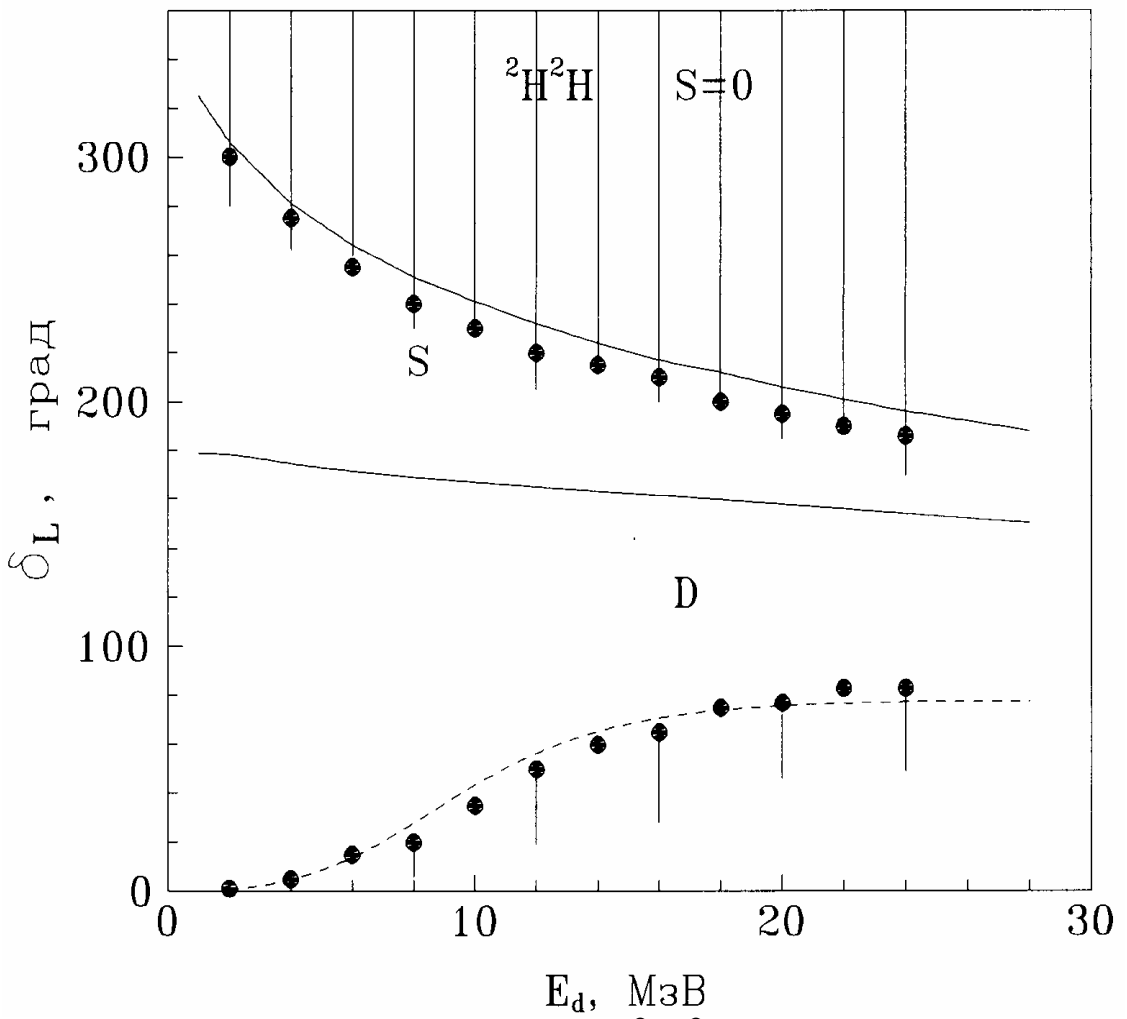

Рис.3.20б. Чистые фазы упругого $^2H^2H$ рассеяния. Вертикальные линии - полоса ошибок для фаз со схемой {4}, кривые - результаты расчетов чистых фаз для потенциалов с параметрами из табл.3.11.

Сечения упругого рассеяния при 13.9 МэВ, полученные с этими взаимодействиями, показаны на рис.3.21а непрерывной линией. Точечной линией приведен вклад синглетного канала, штриховой - триплетного и штрих - пунктирной - канала со спином 2.

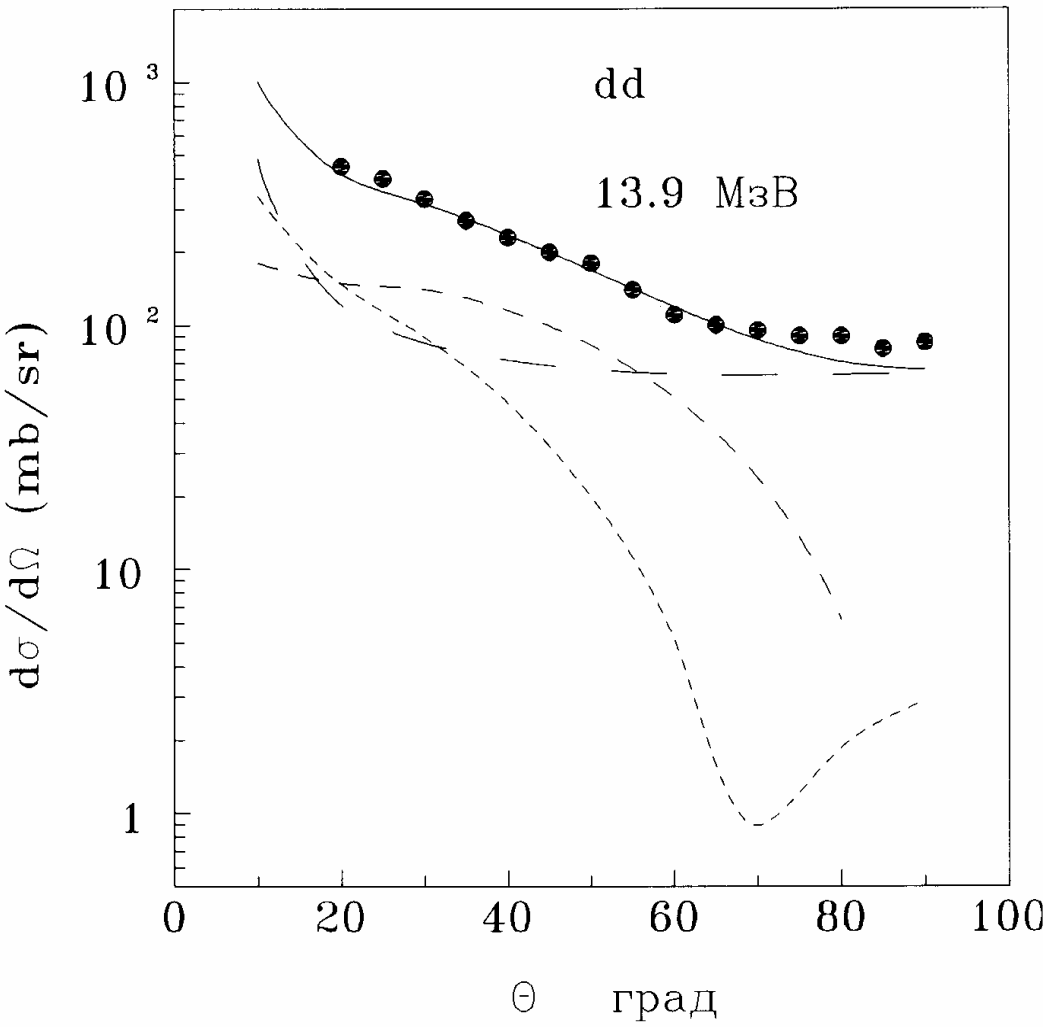

Рис.3.21а. Сечения упругого $^2H^2H$ рассеяния. Точки - экспериментальные данные [49], кривые - результаты расчетов для потенциалов с параметрами из табл.3.11.

Чистые фазы рассеяния в синглетном $^2H^2H$ канале получены на основе формул (3.3.1) для схем Юнга $\{f_1\}=\{4\}$ и $\{f_2\}=\{22\}$. При использовании только результатов многоканального МРГ, чистые фазы находятся вполне определенно и показаны точками на рис.3.20б.

При использовании обеих МРГ результатов, для чистых фаз получается полоса значений, приведенная на рисунке вертикальными линиями. Параметры чистых межкластерных потенциалов даны в табл.3.11. Приведенный S потенциал правильно воспроизводит положения связанных состояний спектра (см. рис.18б) при энергиях -28.848 МэВ и -3.748 МэВ относительно порога кластерного канала.

По энергиям этих уровней параметры взаимодействия фиксируются вполне однозначно. Однако, этот потенциал, хотя и позволяет описать S фазу, приводит к D фазе начинающейся со 180 градусов и содержит ненаблюдаемое в спектре связанное состояние при -1.5 МэВ. Эти результаты показаны на рис.3.20б непрерывными линиями. Поэтому для D волны приходится использовать потенциал с другими параметрами, не приводящий к связанному состоянию, но описывающий фазу, как показано на рис.3.20б точечной линией.

Чистый потенциал связанного состояния $^2H^2H$ системы приводит к правильному радиусу 1.64 Фм только при деформации дейтрона примерно на 40%, также как для $p^2H$ канала ядра $^3He$. Асимптотическая константа оказывается равна 13.7(1.3), что заметно больше данных [13], где приводятся значения в интервале 6-9.

## 3.6. Кластерные системы $p^3H$ и $n^3He$ в ядре $^4He$

Системы $p^3H$ и $n^3He$ оказываются смешанными по изоспину, так как при $T_3$=0 возможны значения T=0,1. Здесь и триплетные и синглетные фазы, а значит, и потенциалы эффективно зависят от двух значений изоспина. Следствием смешивания по изоспину является смешивание по схемам Юнга. В частности в синглетном состоянии разрешены две орбитальные схемы {31} и {4}, как видно из табл.2.1. Поэтому для получения чистых фаз с T=0, согласно (3.3.1) надо использовать фазы, чистой по изоспину $p^3He$ системы с T=1. В триплетном состоянии возможна только схема {31}, но оно так же смешано по изоспину [31,37,38].

*Таблица 3.12. Потенциалы взаимодействия в $p^3H$ системе, смешанные по схемам Юнга с $R_c$ =0. Здесь $E_{cc}$ - энергии связанных состояний. В скобках даны энергии для $n^3He$ системы.*

| $L_J$ | $V_0$ , (МэВ) | α, ($Фм^{-2}$) | $V_1$ , (МэВ) | β , ($Фм^{-1}$) | $E_{cc}$ , (МэВ) |
|---|---|---|---|---|---|
| S=0 | | | | | |
| Чет. | -50 | 0.2 | | | -10.4 (-11.4) |
| Нечет | +3 | 0.03 | | | |
| S=1 | | | | | |
| Чет. | -55 | 0.39 | +6.0 | 0.39 | -1.2 (-2.0) |
| $P_0$ | -11 | 0.1 | | | |
| $P_1$ | -17.5 | 0.1 | | | |
| $P_2$ | -22 | 0.1 | | | |

Имеется сравнительно много экспериментальных данных по $p^3H$ фазам рассеяния [51,52], причем в работах [52] приведены два набора фаз, позволяющих воспроизвести сечения рассеяния. Первый из них, в целом, согласуется с данными работ [51]. Параметры потенциалов, полученные на основе этих экспериментальных данных, даны в табл.3.12. [31]. На рис.3.22 и 3.23а показаны смешанные по изоспину и вычисленные фазы p$^3$H упругого рассеяния, а p$^3$H сечения упругого рассеяния приведены на рис.3.21б.

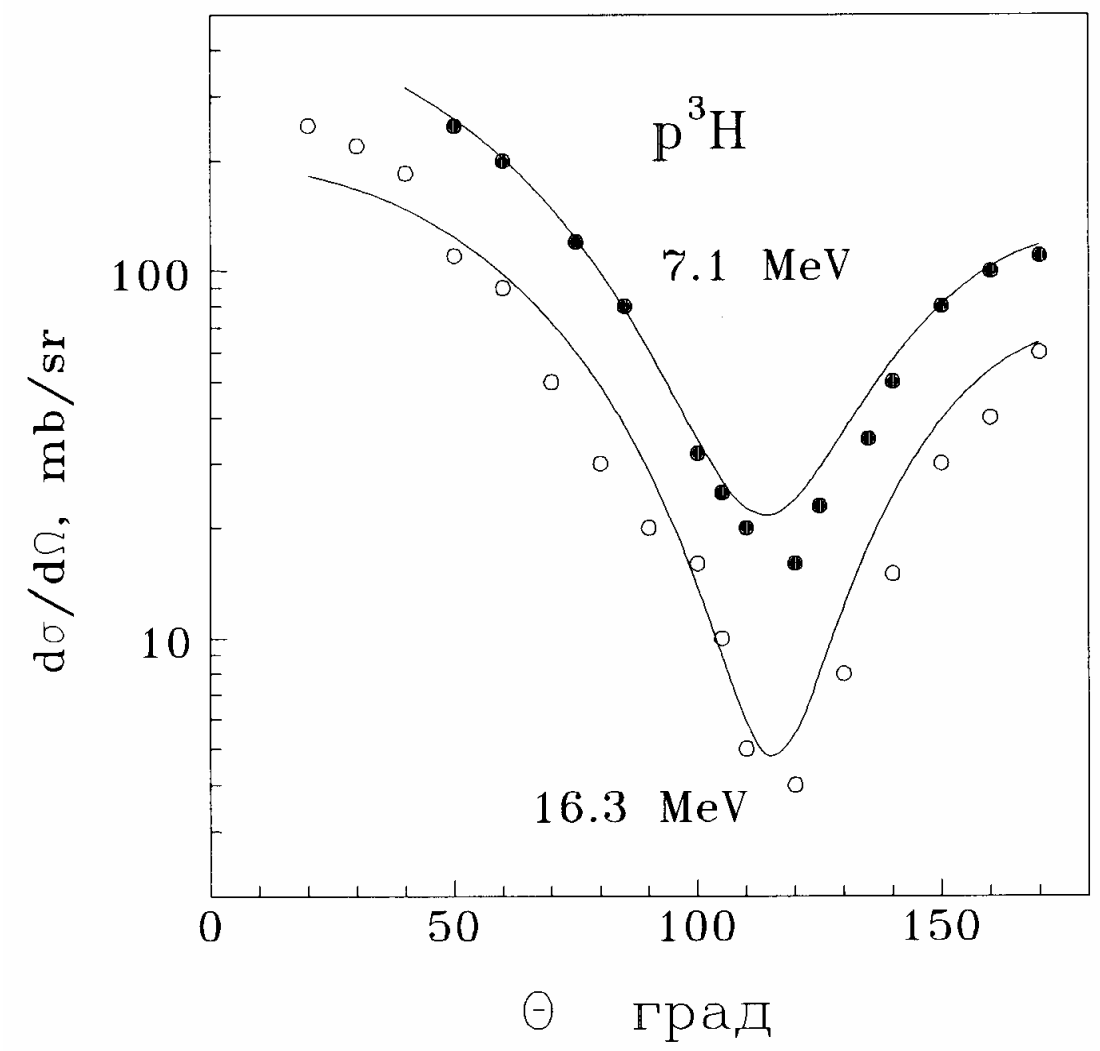

Рис.3.21б. Сечения упругого p$^3$H рассеяния. Точки и треугольники - экспериментальные данные [51, 52], кривые - результаты расчетов для потенциалов с параметрами из табл.3.12.

*Таблица 3.13. Чистые по схемам Юнга потенциалы взаимодействия p$^3$H кластерной системы. Для всех потенциалов $R_c = 0$. В скобках даны энергии для n$^3$He системы.*

| $L_J$ | $V_0$, (МэВ) | α, (Фм$^{-2}$) | $V_1$, (МэВ) | β, (Фм$^{-1}$) | $E_{cc}$, (МэВ) |
|---|---|---|---|---|---|
| S=0 | | | | | |
| Чет. | -63.1 | 0.17 | | | -19.82 |
| Нечет | +8 | 0.03 | | | |
| S=1 | | | | | |
| Чет. | -70.0 | 0.39 | +15 | 0.39 | -1.4(-2.4) |
| $P_0$ | -11 | 0.1 | | | |
| $P_1$ | -19 | 0.1 | | | |
| $P_2$ | -24 | 0.1 | | | |

Поскольку потенциалы смешанны по изоспину, энергия синглетного S взаимодействия не соответствует экспериментальной энергии связи ядра $^4$He (см. рис.3.18б). Разрешенные в обоих каналах состояния с {31} имеют один квант возбуждения и Р потенциалы не содержат связанных состояний.

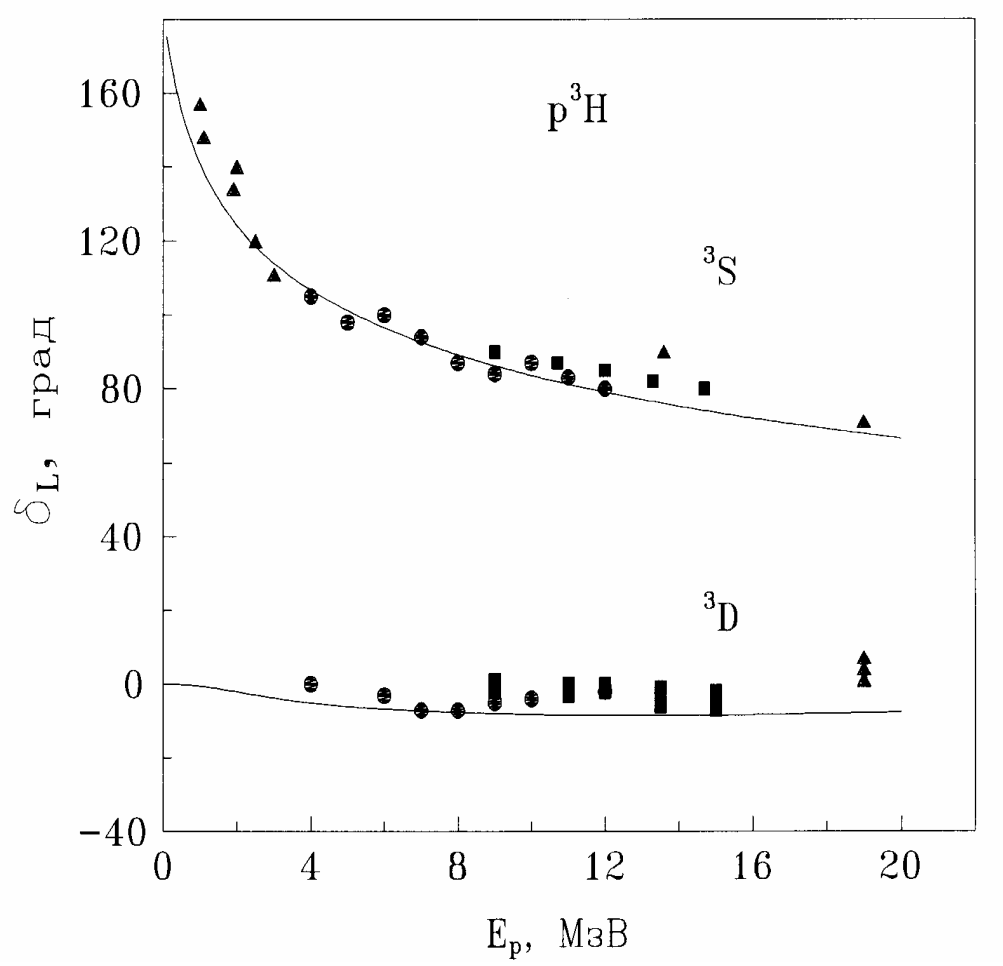

Рис.3.22а. Смешанные триплетные четные фазы упругого $p^3H$ рассеяния. Кружки, треугольники и квадраты - экспериментальные данные [51,52], кривые - результаты расчетов для потенциалов с параметрами из табл.3.12.

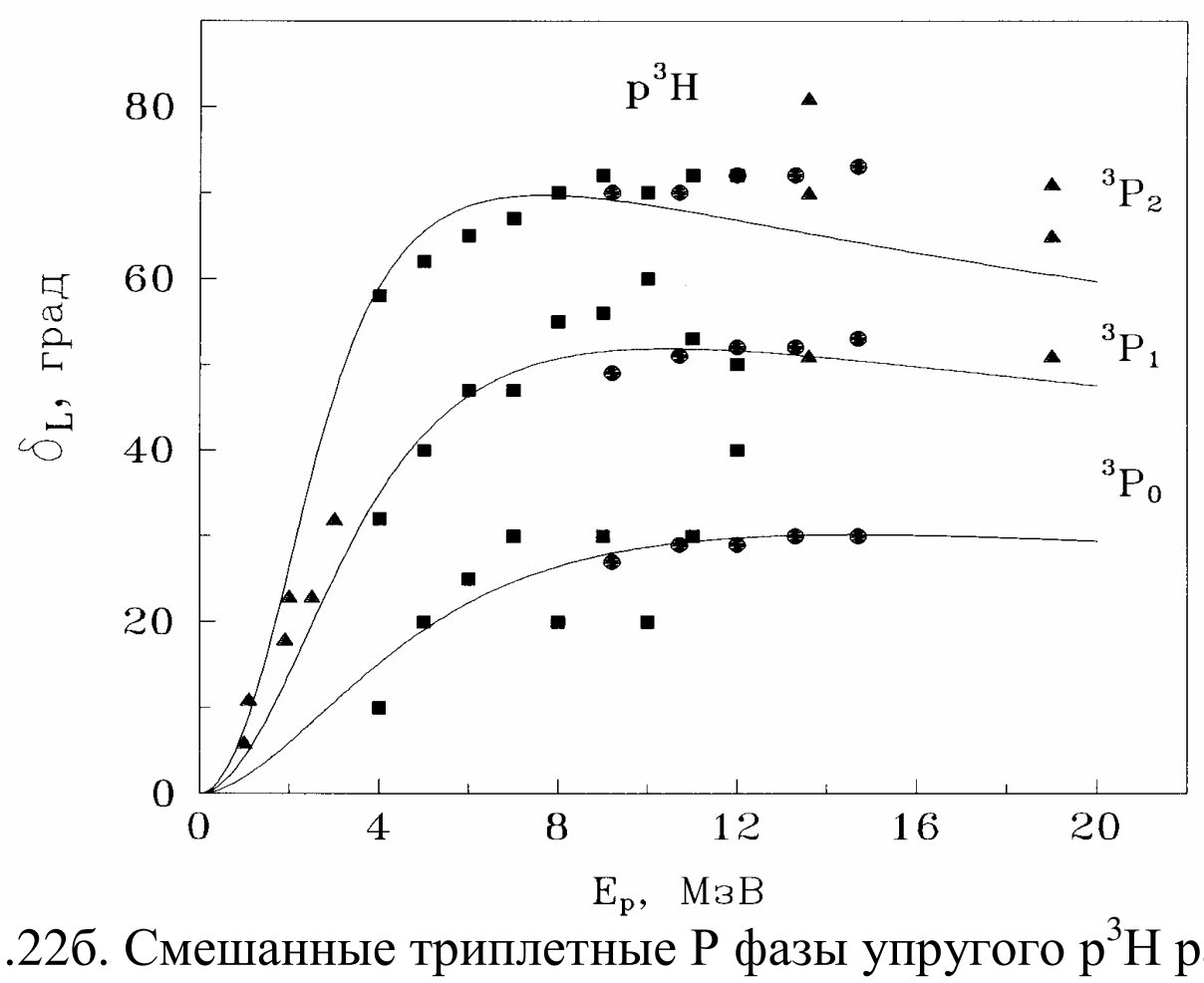

Рис.3.22б. Смешанные триплетные Р фазы упругого $p^3H$ рассеяния. Кружки, треугольники и квадраты - эксперимент из [51,52], кривые - результаты расчетов для потенциалов с параметрами из табл.3.12.

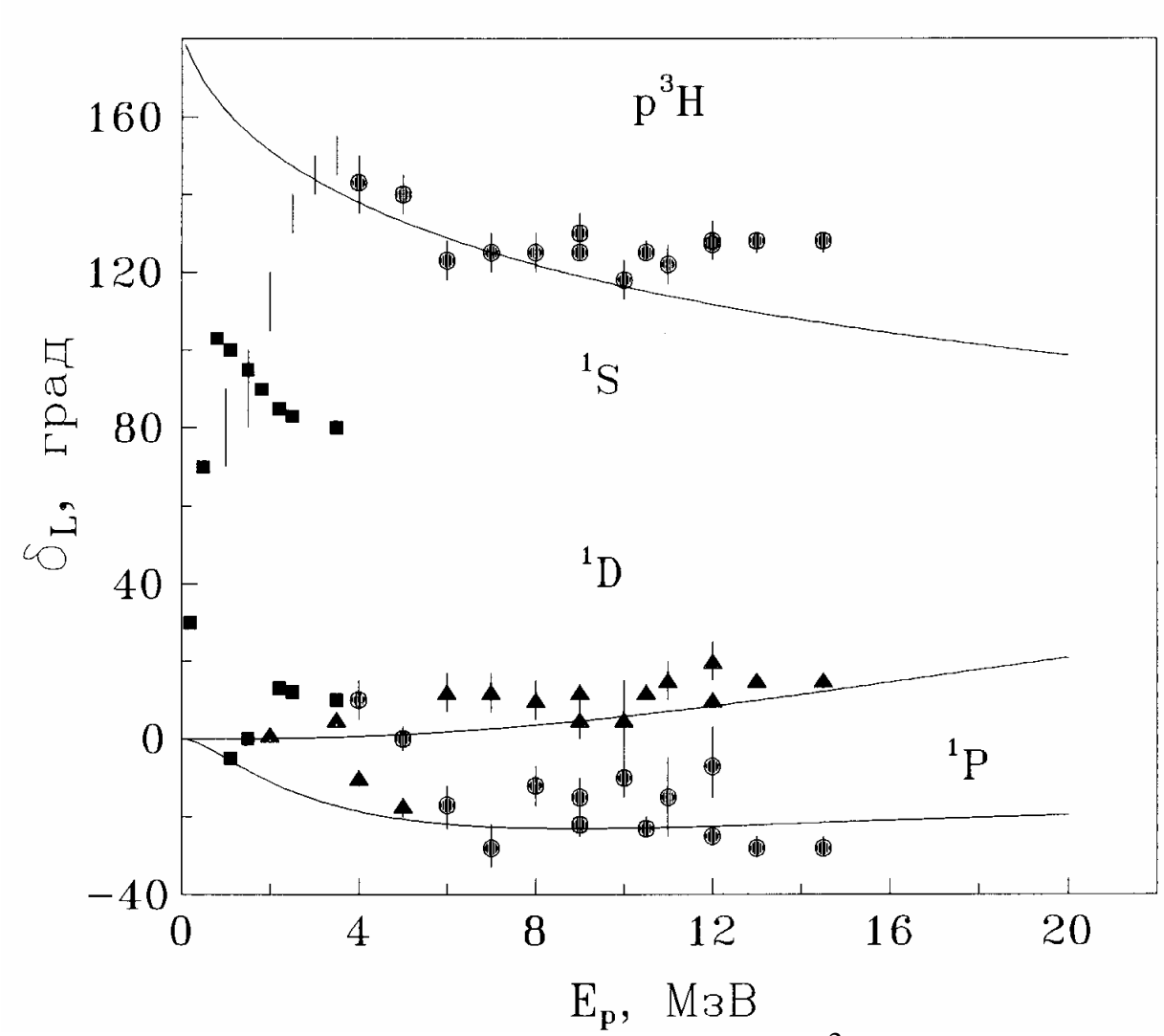

Рис.3.23а. Смешанные синглетные фазы p$^3$H рассеяния. Кружки, треугольники и квадраты - данные [51,52], кривые - расчеты для потенциалов из табл.3.12.

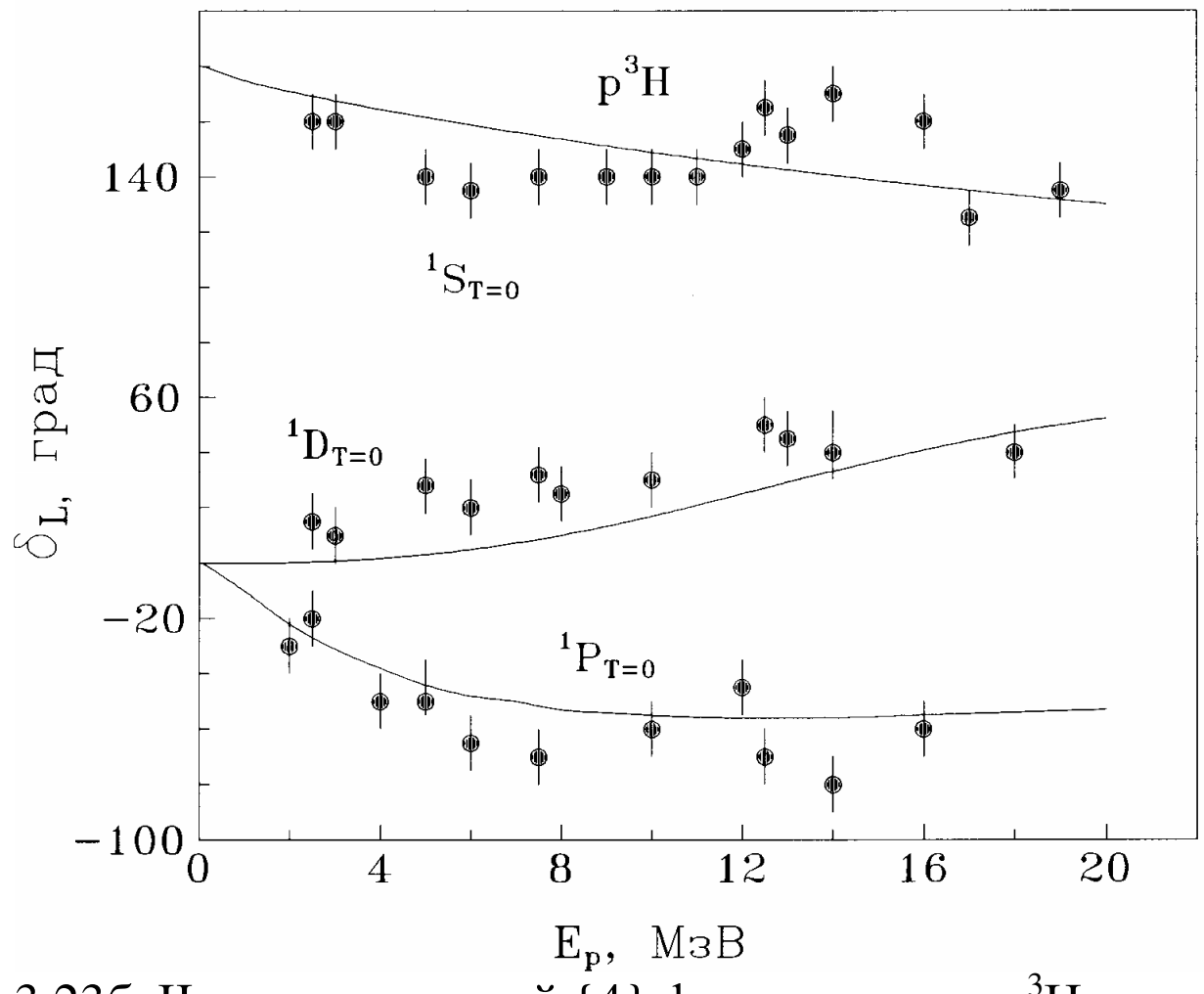

Рис.3.23б. Чистые со схемой {4} фазы упругого p$^3$H рассеяния - точки, кривые - результаты расчетов чистых фаз для потенциалов с параметрами из табл.3.13.

Определив на основе выражения (3.3.1) чистые фазы с {4}, когда в качестве чистой фазы с T=1 берутся фазы системы $p^3He$, находим чистые потенциалы взаимодействия при T=0, параметры которых приведенными табл.3.13 [31,37,38].

*Таблица 3.14. Характеристики основного состояния $^4He$ в $n^3He$ и $p^3H$ кластерных моделях. Экспериментальные данные из работ [11,13] и табл.3.3.*

| Канал | E , (МэВ) | $C_0$ , (Фм$^{-1/2}$) | $R_r$, (Фм) | $R_f$, (Фм) |
|---|---|---|---|---|
| $p^3H$ | -19.822 | 4.5(1) | 1.761 | 1.693 |
| Экспер. | -19.815 | 5.2(1); 4.2(2) | | 1.673(1) |
| $n^3He$ | -20.861 | 4.4(1) | 1.838 | 1.699 |
| Экспер. | -20.578 | 5.1(4) | | 1.673(1) |

Результаты вычислений чистых фаз для этих потенциалов при S=0 показаны на рис.3.23б непрерывными кривыми, а вычисленные характеристики ядра $^4He$ в $p^3H$ и $n^3He$ моделях даны в табл.3.14 [31,37,38], вместе с экспериментальными данными работ [11,13], приведенными в табл.3.3. На рис.3.8б штриховой линией показан расчетный формфактор ядра $^4He$ в $p^3H$ кластерной модели.

Видно, что полученные, для приведенных выше потенциалов, результаты в целом согласуются с экспериментальными данными по характеристикам связанных состояний и позволяют описать формфактор при малых переданных импульсах.

## 3.7. Процессы рассеяния в системе $N^4He$

Рассмотрим теперь кластерную систему $N^4He$ для ядер $^5Li$, $^5He$, которая не имеет связанных стабильных состояний. Однако потенциалы в таких кластерных каналах представляют не малый интерес с точки зрения использования их в трехтельных расчета, например, ядра $^6Li$ в $np^4He$ модели. Рассматриваемая система чистая по орбитальным симметриям, как видно из табл.2.1.

Спин и изоспин системы принимают только одно значение 1/2 со схемой Юнга {32}. Их произведение дает допустимые спин - изоспиновые симметрии - {5} + {41} + {32} + {311} + {221} + {2111}. Возможные орбитальные конфигурации определяются схемами {5}+{41}. Отсюда видно, что имеется только одна разрешенная орбитальная схема {41}, а вторая запрещена, так как отсутствует, соот-ветствующая ей, полностью антисимметричная спин - изоспиновая симметрия. Значит, в S волне должен быть запрещенный уровень, и соответствующая фаза будет начинаться со $180^0$. Разрешенное P состояние находится в непрерывном спектре и фазы идут от нуля.

Анализ экспериментальных $p^4$He и $n^4$He фаз рассеяния при энергиях до 20 МэВ [53,54] показывает, что параметры гауссового потенциала с точечным кулоном могут быть представлены в форме [43]

$$V_0 = V_1 + (-1)^L \Delta V_1 + (LS)\,[V_2 - (-1)^L \Delta V_2]\,,$$

$$\alpha = \alpha_1 + (-1)^L \Delta\alpha_1 + \overline{(LS)}\,[\,\alpha_2 - (-1)^L \Delta\alpha_2\,]\,, \qquad (3.7.1)$$

где $\overline{(LS)}$ - среднее значение спин - орбитального оператора. Однако, проще записать эти параметры в явном виде, как приведено в табл.3.15. Результаты расчетов фаз с этими потенциалами показаны непрерывными линиями на рис. 3.24а и 3.24б для $p^4$He системы и 3.24в, 3.24г для $n^4$He рассеяния. Видно, что удается правильно воспроизвести энергетическое поведение S, P и D фаз до 15-20 МэВ, а F и G фазы хорошо описываются до 40 МэВ. Пунктиром приведены результаты для потенциала Сака - Биденхарна, который неверно передает D фазы рассеяния [55].

*Таблица 3.15. Параметры гауссовых $N^4$He потенциалов с запрещенным в S волне состоянием.*

| Фазы, $L_J$ | $p^4$He | | $n^4$He | |
|---|---|---|---|---|
| | V, (МэВ) | α, (Фм$^{-2}$) | V, (МэВ) | α, (Фм$^{-2}$) |
| S | -73.4 | 0.428 | -66.35 | 0.404 |
| $P_{1/2}$ | -28.98 | 0.16 | -28.10 | 0.16 |
| $P_{3/2}$ | -56.86 | 0.20 | -53.75 | 0.19 |
| $D_{3/2}$ | -80.0 | 0.422 | -55.0 | 0.395 |
| $D_{5/2}$ | -105.0 | 0.432 | -95.0 | 0.41 |

Взаимодействие с приведенными выше параметрами дает правильное положение $P_{3/2}$ резонанса в ядрах $^5$Li - $^5$He при энергиях 2.2 и 1.4 МэВ с ширинами 1.8 и 0.7 МэВ соответственно. Эти результаты хорошо согласуется с данными [9], где для энергий приводится 2.0 и 0.9 МэВ, а для ширин 1.5 и 0.6 МэВ. Запрещенное состояние со схемой {5} в S волне найдено при энергии -9.1 МэВ в $n^4$He и -8.5 МэВ в $p^4$He системах.

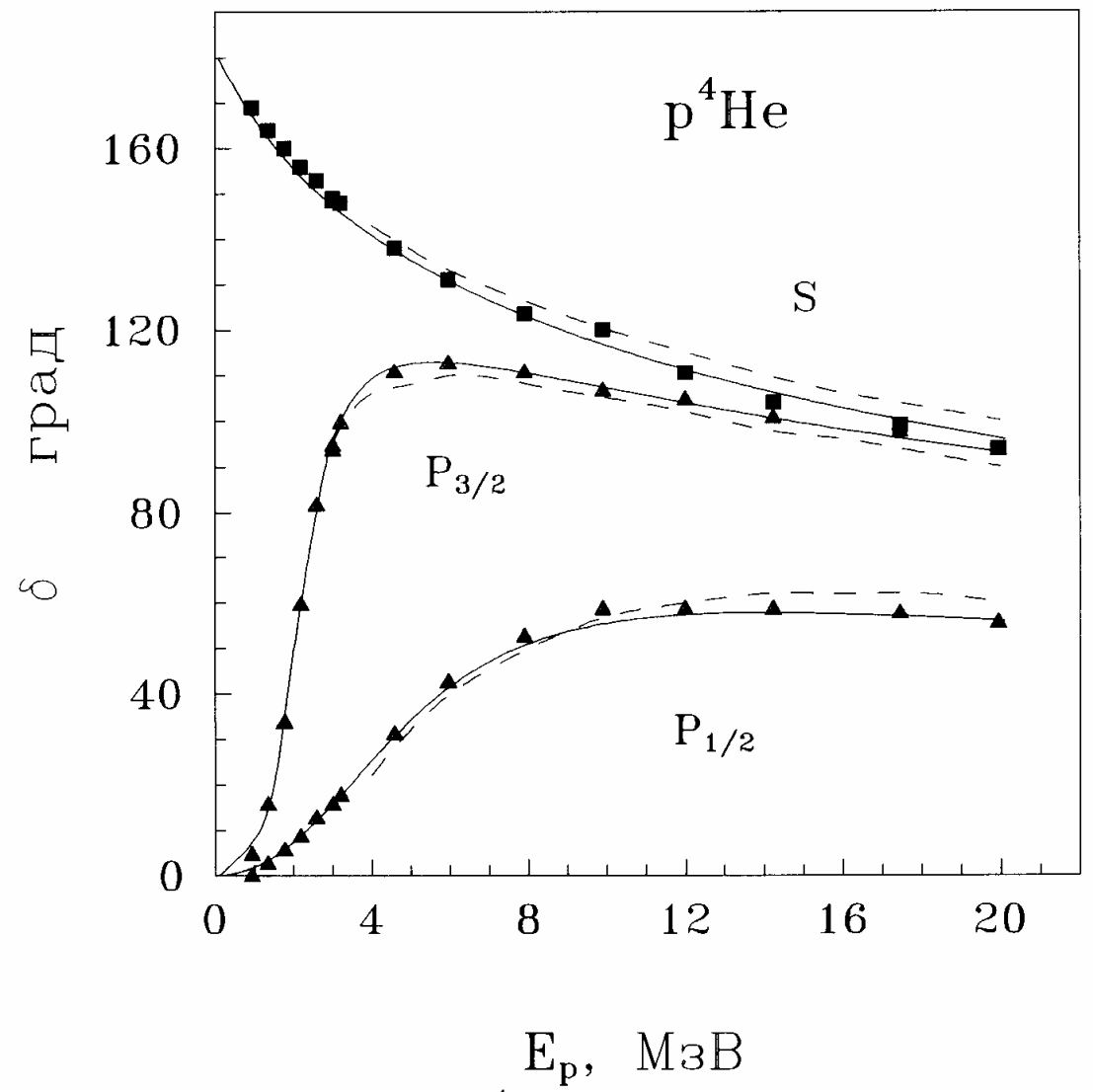

Рис.3.24а. Фазы упругого $p^4He$ рассеяния для потенциалов с запрещенными состояниями. Экспериментальные данные из работ [53,54].

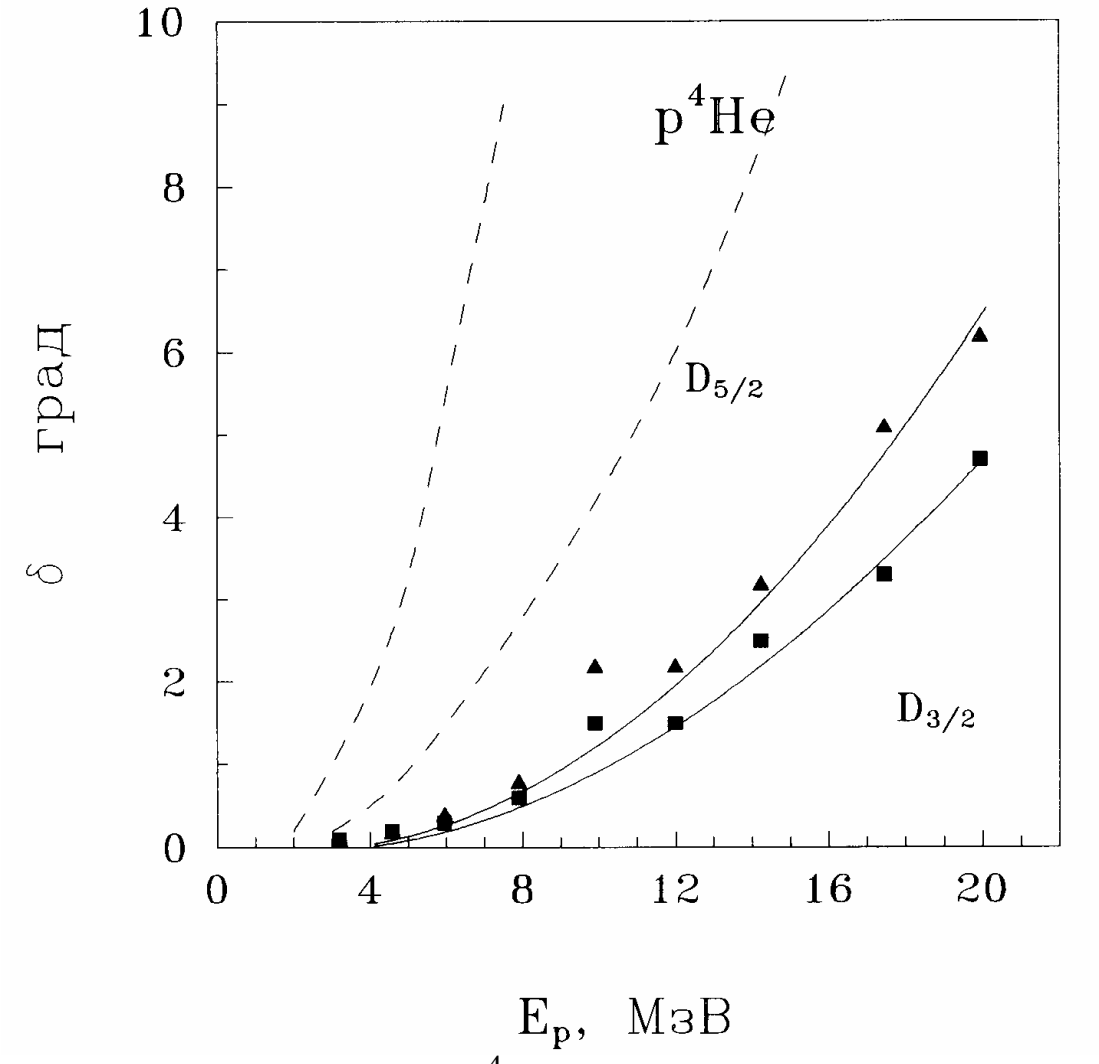

Рис.3.24б. Фазы упругого $p^4He$ рассеяния для потенциалов с запрещенными состояниями. Эксперимент из работ [53,54].

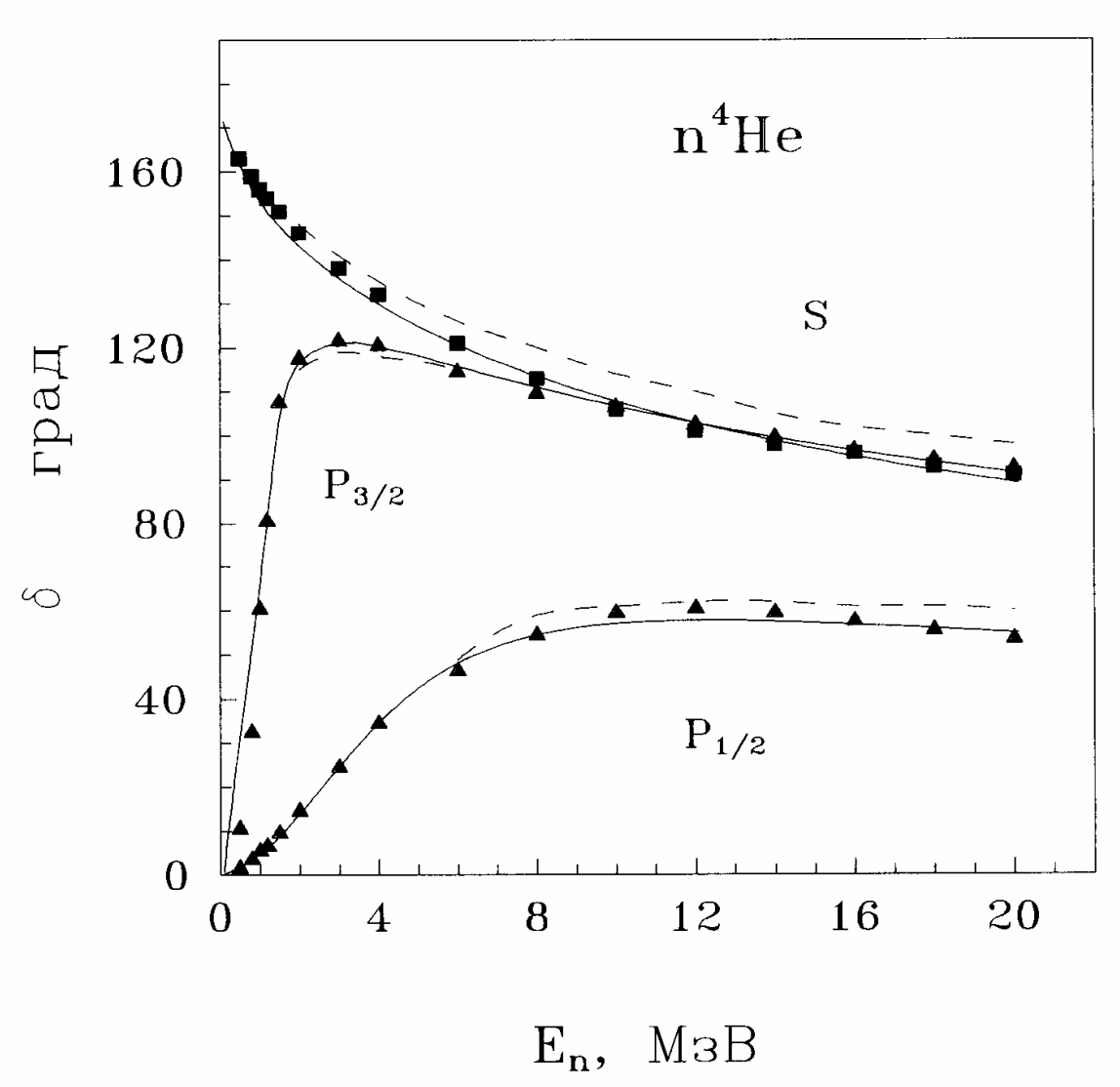

Рис.3.24в. Фазы упругого $n^4He$ рассеяния для потенциалов с запрещенными состояниями. Эксперимент из работ [53,54].

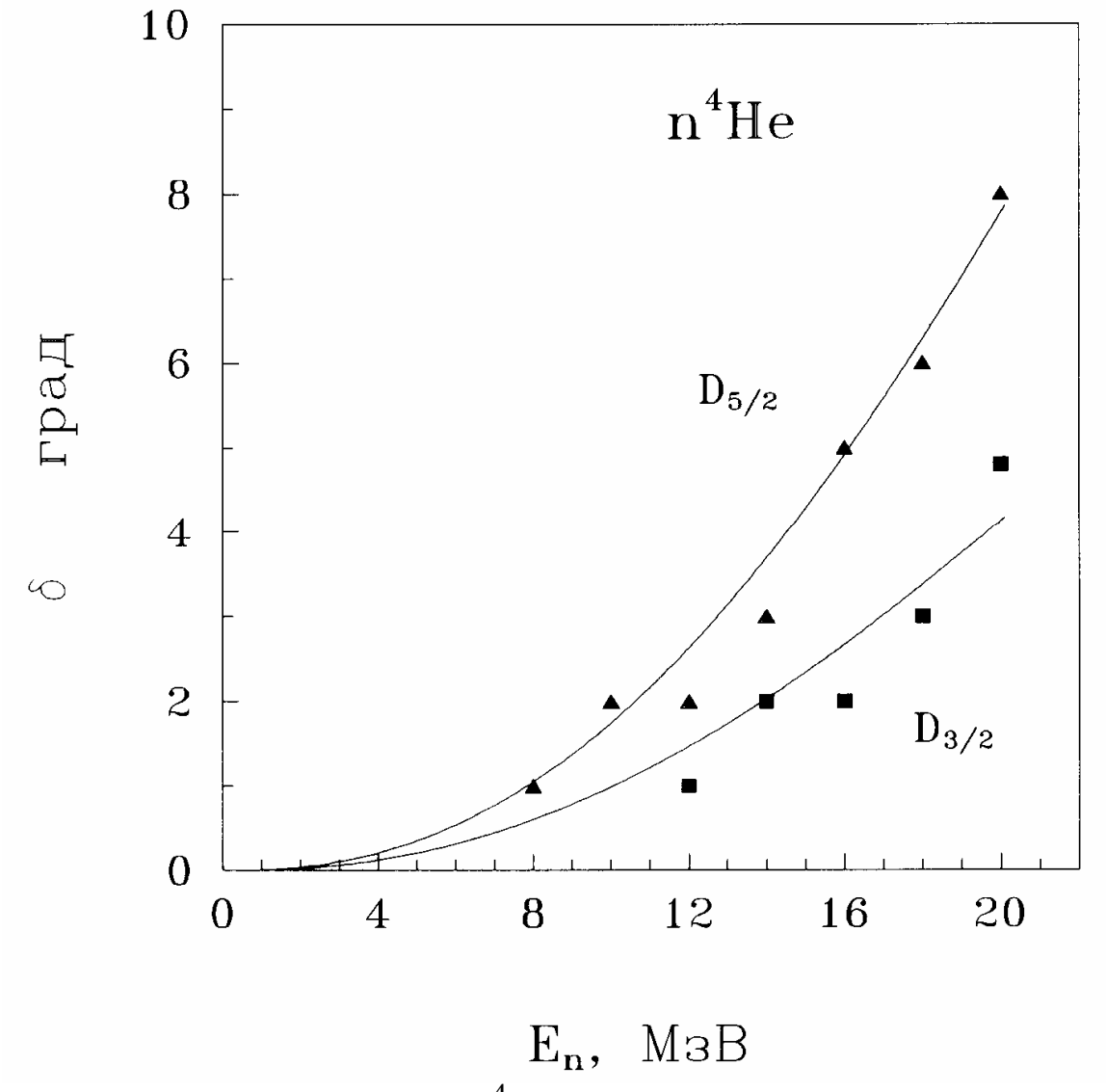

Рис.3.24г. Фазы упругого $n^4He$ рассеяния для потенциалов с запрещенными состояниями. Эксперимент из работ [53,54].

## 3.8. Характеристики ядра $^7$Li в трехтельной $^4$He$^2$Hn модели

В заключение этой главы рассмотрим трехтельную модель ядра $^7$Li и ее возможности по описанию некоторых характеристик связанного состояния $^4$He$^2$Hn кластеров.

В работах [55,56] были подробно рассмотрены возможности трехтельной модели ядра $^6$Li и показана ее способность правильно описывать почти все наблюдаемые характеристики, включая электромагнитные формфакторы, если выполнить антисимметризацию волновой функции [30]. Исключение составляет только квадрупольный момент ядра, который во всех расчетах получается положительным, в то время как экспериментальный измерения дают отрицательную величину. Учет антисимметризации волновой функции позволил существенно улучшить качество описания поперечных формфакторов [30], но мало изменил другие характеристики ядра. Этот результат может объяснить определенные успехи простых двухкластерных моделей легких ядер с запрещенными состояниями, в частности $^2$H$^4$He и $^3$H$^4$He моделей ядер $^6$Li и $^7$Li, в которых получается хорошее описание многих экспериментальных характеристик, но плохо воспроизводятся поперечные формфакторы при больших переданных импульсах [28].

Антисимметризация волновой функции, выполненная в [30], затрагивает, в основном, внутреннюю область ядра и заметно изменяет волновую функцию на малых расстояниях, которые определяют поведение высокоимпульсной компоненты формфакторов. Область больших расстояний при этом меняется мало, что приводит к не существенным изменениям других расчетных характеристик, зависящих в основном от поведения волновой функции периферийной области ядра. Поэтому, вполне можно предположить, что проведение антисимметризации волновой функции в двухкластерных моделях $^7$Li или $^6$Li с тензорными силами, которые позволяют передать квадрупольный момент ядра $^6$Li [57] может заметно улучшить описание поперечных формфакторов при больших переданных импульсах.

Ядро $^7$Li, не смотря на вполне успешное описание многих его характеристик на основе простой двухкластерной системы [1, 31, 37, 38], можно рассматривать в трехтельной n$^2$H$^4$He модели, которая имеет больше возможностей и, в частности, позволяет выделять различные двухчастичные каналы.

Будем считать, что в основании треугольника из трех частиц находятся $^2$Hn кластеры (частицы 23) с радиус - вектором относительного расстояния $r = r_{23}$ и орбитальным моментом относительного движения $\lambda$, которые находятся в дублетном спиновом состоянии. Ядро $^4$He (частица 1) находится в вершине треугольника и его положение относительно центра масс двухкластерной системы определяется вектором $R=R_{(23),1}$ и моментом l. Полный орбитальный момент системы $L=l+\lambda$ , равный 1 может быть получен из комбинации l=1 и $\lambda$=0, которая позволяет рассматривать систему $^2$Hn, как связанное состояние ядра трития [58].

В качестве парных межкластерных потенциалов выбирались взаимодействия гауссовой формы с отталкивающим кором, позволяющие правильно передавать соответствующие фазы рассеяния. В паре частиц (13) используется чистый по схемам Юнга n$^4$He потенциал для S - волны с параметрами, описывающими экспериментальную фазу [54], как показано на рис.3.25. В паре (12) использован $P_0$ - потенциал $^2$H$^4$He взаимодействия, параметры которого уточнялись по трехтельной энергии, поскольку $P_0$ - фазы (см. рис.3.26), полученные в разных работах [4] имеют большую неоднозначность.

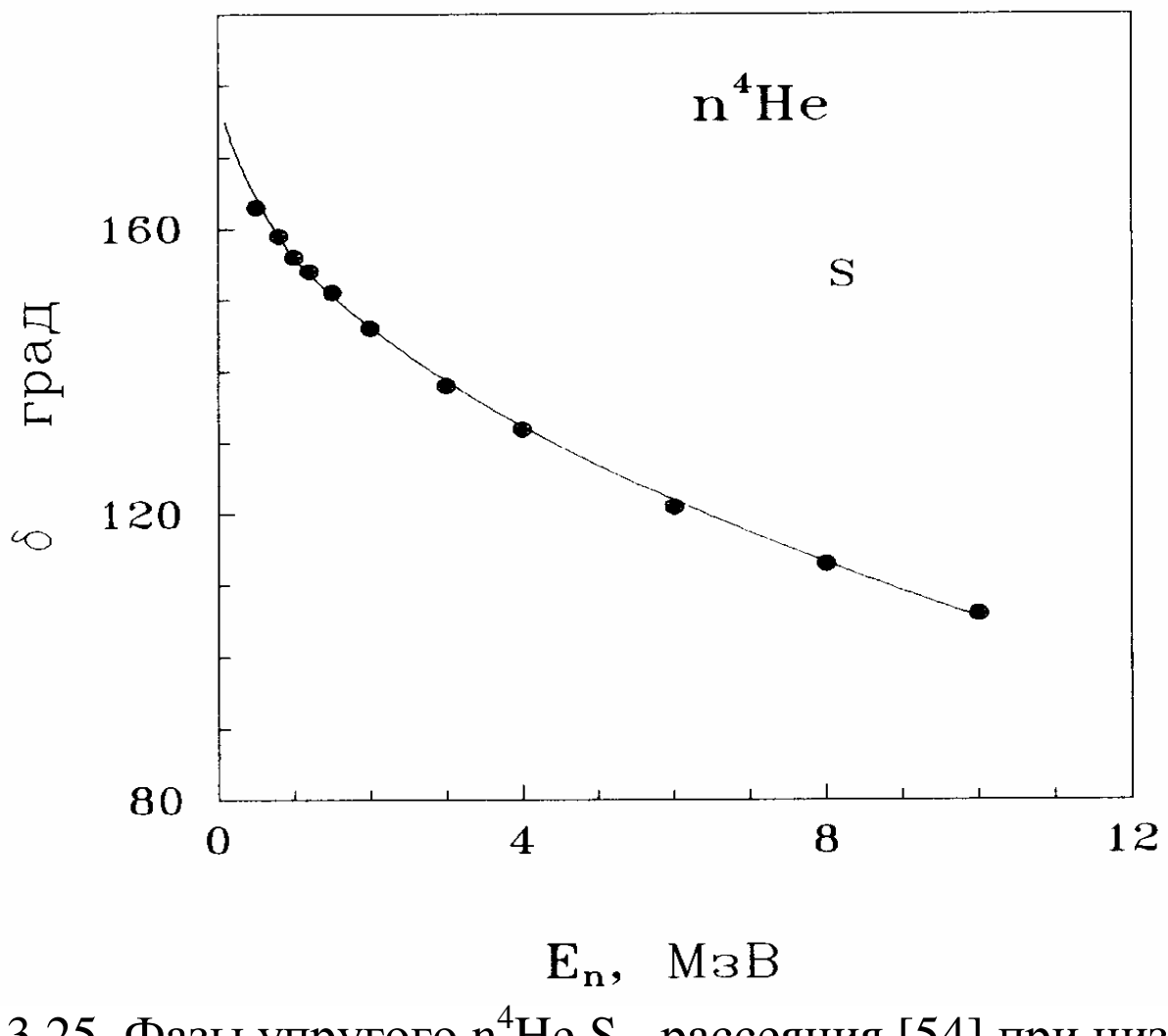

Рис.3.25. Фазы упругого n$^4$He S - рассеяния [54] при низких энергиях для потенциала из табл.3.16.

В паре частиц (23) взято n$^2$H чистое по орбитальным симметриям дублетное S - взаимодействие с отталкиванием, параметры которого фиксированы по характеристикам связанного состояния ядра трития, а фазы показаны на рис.3.27 в сравнении с извлеченными из эксперимента чистыми фазами [1,31,37,38,58].

В каждой паре частиц использован только один потенциал для определенной парциальной фазы взаимодействия. Это представляется вполне оправданным, если потенциалы для остальных парциальных волн в каждой паре вносят меньший вклад и приводят только к небольшим поправкам к расчетным характеристикам ядра.

В отличие от трехчастичной модели ядра $^6$Li [55], где потенциалы в NN и N$^4$He системах хорошо определены по экспериментальным фазам рассеяния, имеющим сравнительно малые ошибки, параметры $^2$H$^4$He взаимодействия имеют заметную неопределенность. Поэтому представляется интересным выяснить - можно ли в $^4$He$^2$Hn модели согласовать, в пределах имеющихся экспериментальных неоднозначностей по фазам, параметры двухкластерных потенциалов с энергией связи ядра $^7$Li.

Парные межкластерные взаимодействия имеют вид суммы двух гауссойд

$$V(r)=V_1\exp(-\gamma r^2) + V_2\exp(-\delta r^2) , \qquad (3.8.1)$$

а их параметры даны в табл.3.16.

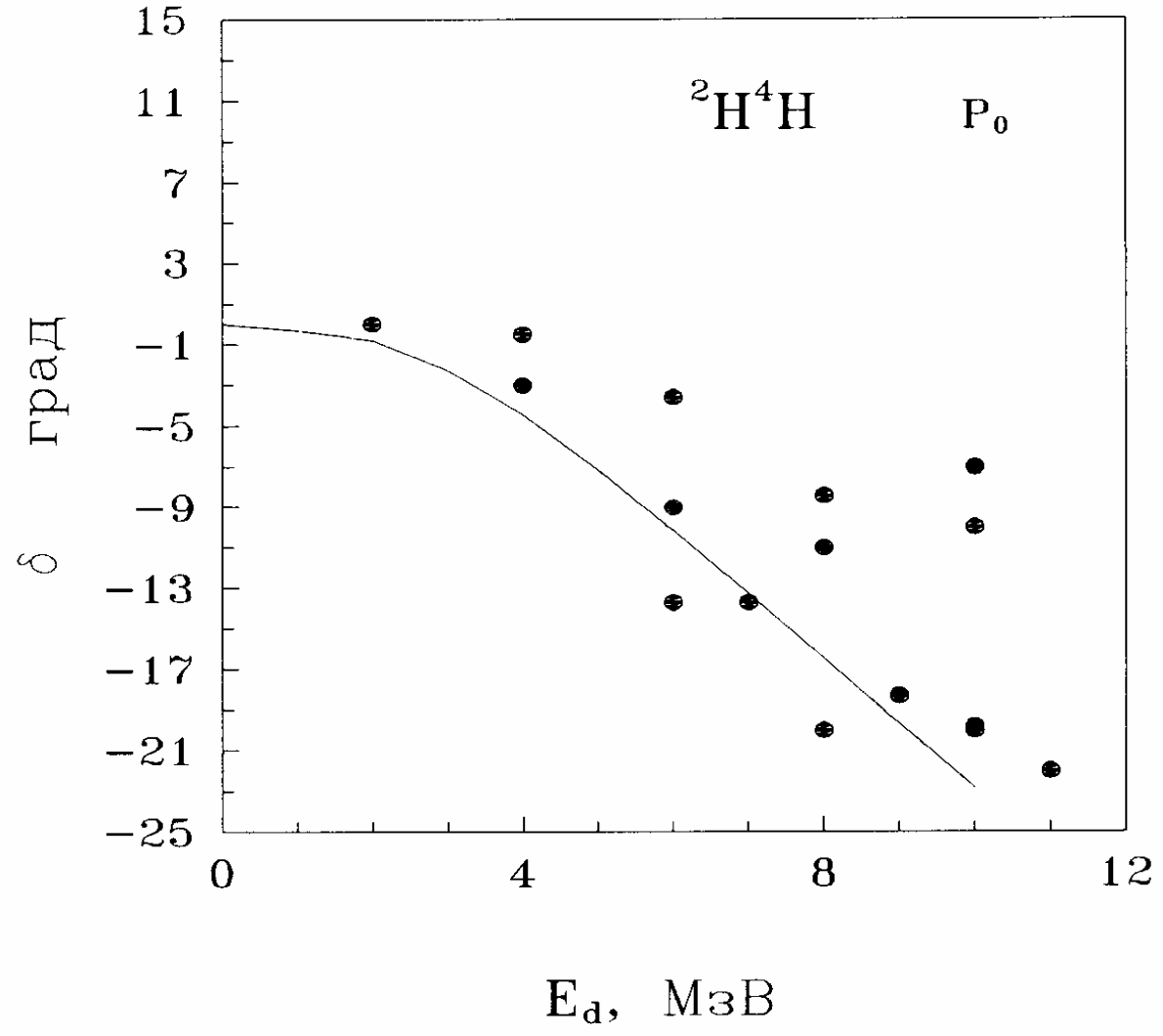

Рис.3.26. Фазы упругого $^4$He$^2$H $P_0$ - рассеяния [4] при низких энергиях для потенциала из табл.3.16.

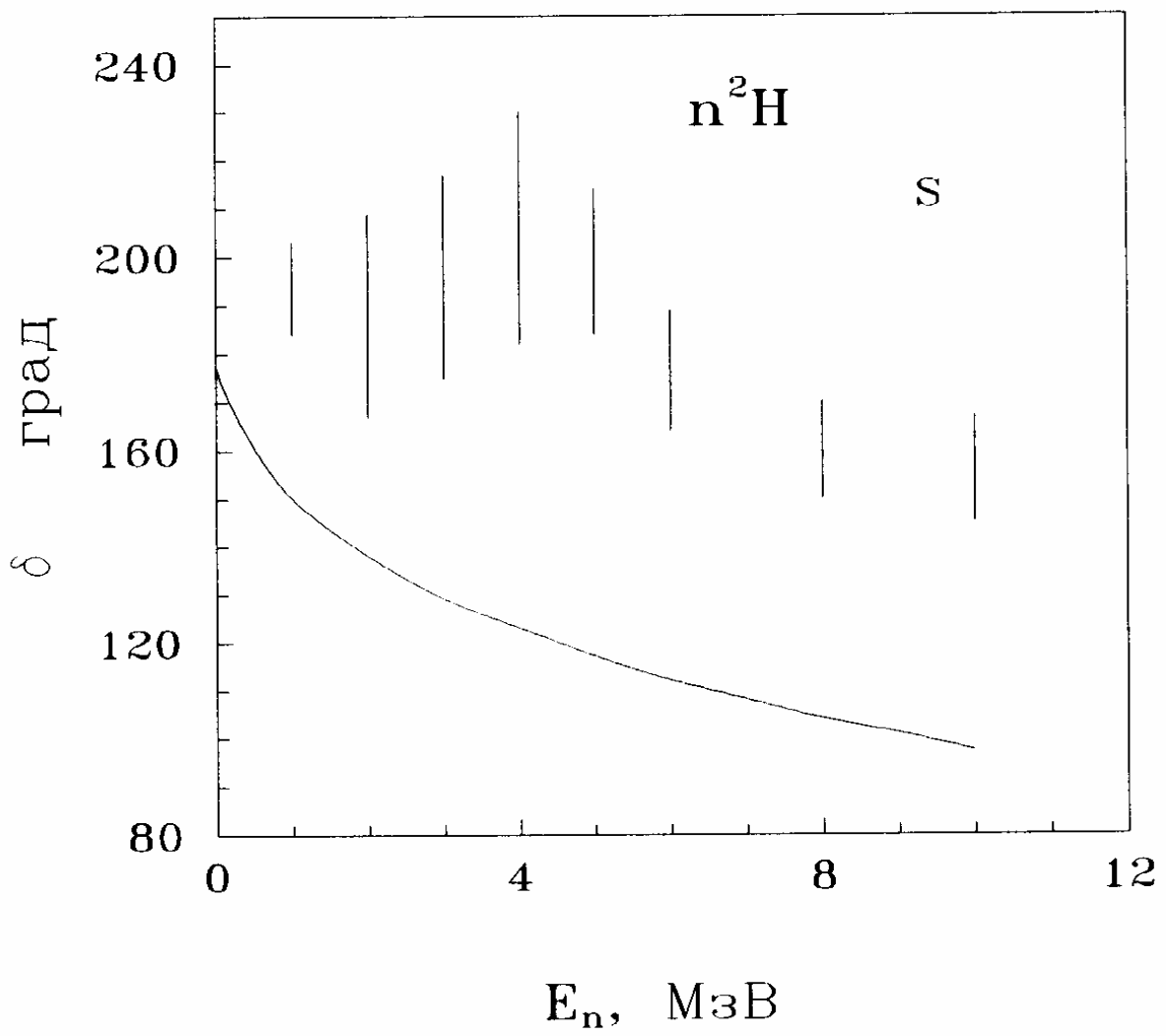

Рис.3.27. Чистые со схемой Юнга {3} дублетные фазы упругого $n^2H$ S - рассеяния [1,37,38] при низких энергиях для потенциала из Табл.3.16.

*Таблица 3.16. Параметры парных межкластерных потенциалов для различных относительных моментов $L_{1,2}$.*

| Система кластеров | $L_{1,2}$ | $V_1$, МэВ | $\gamma$, Фм$^{-2}$ | $V_2$, МэВ | $\delta$, Фм$^{-2}$ |
|---|---|---|---|---|---|
| $^2H^4He$ | 1 | -10.0 | 0.1 | +72.0 | 0.2 |
| $n^4He$ | 0 | -115.5 | 0.16 | +500 | 1.0 |
| $n^2H$ | 0 | -78.78 | 0.3 | +200 | 2 |

Потенциал основного состояния в $n^2H$ системе дает энергию связи -6.25 МэВ, асимптотическую константу $C_0$=2.0(1) в хорошем соответствии с экспериментом [11,13] и среднеквадратичный радиус 2.12 Фм, который несколько больше известной величины 1.70(3) Фм [11,12]. Здесь, как и в работах [1,37,38], не удается полностью согласовать $n^2H$ потенциал со всеми наблюдаемыми. Расчетная фаза лежит несколько ниже извлеченных из эксперимента чистых по схеме Юнга {3} фаз, а радиус ядра завышен.

Последнее вполне объяснимо, поскольку дейтрон имеет радиус больше, чем тритий и не может находиться внутри него без деформаций, т.е. дейтронному кластеру в такой системе нельзя полностью сопоставлять характеристики свободного дейтрона [1,37,38]. Для того чтобы получить правильный радиус трития необходимо деформировать дейтрон, уменьшив его радиус примерно на 27% и принять 1.42 Фм, что приводят к расчетному зарядовому радиусу 1.70 Фм в хорошем согласии с экспериментом.

Для нахождения энергии трехкластерной системы использовался неортогональный вариационный метод [55]. Полная трехтельная волновая функция имеет вид

$$\Psi(r,R) = \sum_{l,\lambda} \Phi_{l,\lambda}(r,R) Y_{LS}^{JM}(\hat{r},\hat{R}) \quad ,$$

где угловая часть записывается

$$Y_{LS}^{JM}(\hat{r},\hat{R}) = \sum_{M_S M_L} \langle LM_L SM_S | JM \rangle Y_{LM_L}(\hat{r},\hat{R}) \chi_{SM_S}(\sigma) \ .$$

Радиальная волновая функция представляется в форме разложения по гауссойдам, также как в трехтельной модели ядра $^6$Li [55]

$$\Phi_{l,\lambda}(r,R) = N r^{\lambda} R^{l} \sum_i C_i \exp(-\alpha_i r^2 - \beta_i R^2) = N \sum_i C_i \Phi_i \ . \qquad (3.8.2)$$

Исходное радиальное уравнения Шредингера системы трех частиц запишем в форме

$$( H - E ) \Phi_{l,\lambda} = 0 \ , \qquad (3.8.3)$$

где

$$H = T + V \ , \quad T = T_1 + T_2 = -\frac{\hbar^2}{2\mu}\Delta_r - \frac{\hbar^2}{2\mu_0}\Delta_R \quad , \quad V=V_{12}+V_{23}+V_{13} \ ,$$

$$\mu = \frac{m_2 m_3}{m_{23}}; \quad \mu_0 = \frac{m_1 m_{23}}{m}; \quad m_{23} = m_2 + m_3; \quad m = m_1 + m_2 + m_3$$

Подставляя разложение (3.8.2) в уравнение (3.8.3), домножая слева это уравнение на функцию $\Phi_j$ , и интегрируя по всем переменным, приводим (3.8.3) к матричному виду

$$\sum_i ( H_{ij} - E L_{ij}) C_i = 0 \qquad (3.8.4)$$

или

$$KC = 0 \ ,$$

где матрица K определяется в виде

$$K = H - E L \ .$$

В этих выражениях H - матрица гамильтониана, L - матрица интегралов перекрывания, которая при использовании ортогонального базиса переходит в единичную матрицу. Отметим, что матрица K не диагональна по энергии и вместо обычной задачи на собственные значения мы имеем обобщенный вариант этой задачи. Поскольку уравнение (3.8.4) однородное, оно будет иметь не тривиальные решения только тогда, когда детерминант матрица K равен нулю.

Условие равенства нулю ее детерминанта позволяет найти все собственные значения E системы (при заранее заданных параметрах $\alpha_i$ и $\beta_i$), а по ним все собственные вектора C, а значит, и саму радиальную функцию $\Phi_{l,\lambda}$ в выражении (3.8.2). При каждом значении вариационных параметров $\alpha_i$ и $\beta_i$ находим некоторую энергию системы, а затем, варьируя эти параметры, проводим поиск минимума этой энергии. Минимальная энергия и будет реальной энергией трехчастичной системы.

Матричные элементы гамильтониана системы и интегралов перекрывания, вычисленные по базисным функциям $\Phi_i$ имеют вид

$$T_{ij} = \frac{\pi}{16} N^2 \frac{(2l+1)!!(2\lambda+1)!!}{2^{l+\lambda}} \frac{\hbar^2}{m_N} \alpha_{ij}^{-\lambda-1/2} \beta_{ij}^{-l-1/2} G_{ij} \ ,$$

$$G_{ij} = \frac{B_{ij}(\alpha,\lambda)}{\mu\beta_{ij}} + \frac{B_{ij}(\beta,l)}{\mu_0\alpha_{ij}} \ , \qquad B_{ij}(\delta,\nu) = \frac{\nu^2}{2\nu+1} + \frac{\delta_i\delta_j}{\delta_{ij}^2}(2\nu+3) - \nu \ ,$$

$$L_{ij} = \frac{\pi}{16} N^2 \frac{(2l+1)!!(2\lambda+1)!!}{2^{l+\lambda}} \alpha_{ij}^{-\lambda-3/2} \beta_{ij}^{-l-3/2} \ ,$$

$$(V_{23})_{ij} = \frac{\pi}{16} N^2 \frac{(2l+1)!!(2\lambda+1)!!}{2^{l+\lambda}} V_{23}(r)(\alpha_{ij}+\gamma)^{-\lambda-3/2} \beta_{ij}^{-l-3/2} \ ,$$

$$[(V_{цб})_R]_{ij} = \frac{\pi}{16} N^2 l(l+1) \frac{(2l-1)!!(2\lambda+1)!!}{2^{l+\lambda}} \frac{\hbar^2}{\mu_0} \alpha_{ij}^{-\lambda-3/2} \beta_{ij}^{-l-1/2} \quad ,$$

$$[(V_{цб})_r]_{ij} = \frac{\pi}{16} N^2 \lambda(\lambda+1) \frac{(2l+1)!!(2\lambda-1)!!}{2^{l+\lambda}} \frac{\hbar^2}{\mu} \alpha_{ij}^{-\lambda-1/2} \beta_{ij}^{-l-3/2} \quad ,$$

$$[(V_k)_r]_{ij} = Z_2 Z_3 \frac{\pi}{16} N^2 \frac{2}{\sqrt{\pi}} \frac{(2l+1)!!}{2^l} \frac{\lambda!}{\alpha_{ij}^{\lambda+1} \beta_{ij}^{l+3/2}} \quad ,$$

$$N = \left( \sum_{ij} C_i C_j L_{ij} \right)^{-1/2} \quad , \qquad \alpha_{ij} = \alpha_i + \alpha_j \ , \qquad \beta_{ij} = \beta_i + \beta_j \ .$$

Далее, например, при значениях l=1 и λ=0 (набор моментов для $^4He^2Hn$ системы в ядре $^7Li$) имеем

$$(V_{12})_{ij} = \frac{\pi}{32} N^2 \frac{3}{A^{3/2}(\beta_{ij}+\gamma)} \left[ \frac{a^2\gamma^2}{A} + 1 \right] V_{12}(r_{12}) \quad ,$$

$$A = \alpha_{ij}\beta_{ij} + \gamma(\alpha_{ij} + a^2\beta_{ij}) \ , \qquad a = m_3/m_{23} \ .$$

В случае l=0 и λ=0 для этой части потенциала будем иметь

$$(V_{12})_{ij} = \frac{\pi}{16} N^2 \left[ \alpha_{ij}\beta_{ij} + \gamma(\alpha_{ij} + a^2\beta_{ij}) \right]^{-3/2} V_{12}(r_{12}) \quad .$$

Здесь величина γ является параметром ширины гауссового потенциала между частицами с номерами 1 и 2.

Среднеквадратичный массовый радиус ядра в такой модели представляется в виде

$$< r^2 >_m = m_1/m < r^2 >_{m1} + m_2/m < r^2 >_{m2} + m_3/m < r^2 >_{m3} + A/m \qquad (3.8.5)$$

$$A = \frac{\pi}{16} N^2 \frac{(2l+1)!!(2\lambda+1)!!}{2^{l+\lambda+1}} \sum_{i,j} C_i C_j \alpha_{ij}^{-\lambda-3/2} \beta_{ij}^{-l-3/2} \left( \frac{2\lambda+3}{\alpha_{ij}} \mu + \frac{2l+3}{\beta_{ij}} \mu_0 \right)$$

где

$$\mu = \frac{m_2 m_3}{m_{23}}; \quad \mu_0 = \frac{m_1 m_{23}}{m}; \quad m_{23} = m_2 + m_3; \quad m = m_1 + m_2 + m_3$$

Квадрупольный момент ядра с учетом момента дейтрона записывается [57,59]

$$Q = Q_d - \frac{2}{5} B , \tag{3.8.6}$$

$$B = \frac{\pi}{16} N^2 \frac{(2l+1)!!(2\lambda+1)!!}{2^{l+\lambda+1}} \sum_{i,j} C_i C_j \alpha_{ij}^{-\lambda-3/2} \beta_{ij}^{-l-3/2} \left( \frac{2l+3}{\beta_{ij}} C + \frac{2\lambda+3}{\alpha_{ij}} D \right) +$$

$$+ N^2 E \sum_{i,j} C_i C_j \frac{(\lambda+1)!(l+1)!}{2\alpha_{ij}^{\lambda+2} \beta_{ij}^{l+2}},$$

$$C = \frac{Z_1 m_{23}^2 + Z_{23} m_1^2}{m^2}, \quad D = \frac{Z_2 m_3^2 + Z_3 m_2^2}{m_{23}^2}, \quad E = \frac{m_1}{m m_{23}} (Z_3 m_2 - Z_2 m_3),$$

$$Z_{23} = Z_1 + Z_2 .$$

Среднеквадратичный зарядовый радиус имеет вид [55]

$$< r^2 >_z = Z_1 / Z < r^2 >_{z1} + Z_2 / Z < r^2 >_{z2} + Z_3 / Z < r^2 >_{z3} + B / Z . \tag{3.8.7}$$

В качестве зарядовых радиусов кластеров принимались величины $<r>_{mn}$ = 0.8 Фм, $<r>_{zn}$ = 0 Фм, $<r>_{md}$ = $<r>_{zd}$ = 1.96 Фм, $<r>_{m\alpha}$= $<r>_{z\alpha}$ = 1.67 Фм [12], а квадрупольный момент дейтрона 2.86 мб. [59]. Экспериментальное значение квадрупольного момента $^7$Li: -36.6(3) мб. [9]. Для нахождения волновой функции относительного движения и вероятности двухкластерной $^3$H$^4$He конфигурации использовалась волновая функция ядра $^3$H в n$^2$H модели в виде простого разложения по гауссойдам

$$\varphi(r) = N_0 \sum_j B_j \exp(-\chi_j r^2) . \tag{3.8.8}$$

Здесь параметры $\chi_j$ и коэффициенты разложения $B_j$ находились на основе n$^2$H потенциала основного состояния, приведенного в табл.3.16. При использовании волновых функций $^3$H вида (3.8.8), функция относительного движения двух кластеров в $^3$H$^4$He канале ядра $^7$Li при l=1 и λ=0 может быть представлена в виде [60]

$$X(R) = \int \Phi(r,R)\varphi(r)r^2 dr = \frac{\sqrt{\pi}}{4} R \sum_i C_i \exp(-\beta_i R^2) \sum_j B_j(\alpha_i + \chi_j)^{-3/2} \cdot \qquad (3.8.9)$$

Интеграл от квадрата модуля этой волновой функции дает вероятность двухкластерной конфигурации в трехтельной модели [60]

$$P = \frac{3\pi^{3/2}}{16 \cdot 8} \sum_{i,k} C_i C_k \beta_{ik}^{-5/2} \sum_{j,n} B_j B_n [(\alpha_i + \chi_j)(\alpha_k + \chi_n)]^{-3/2} . \qquad (3.8.10)$$

Получив волновую функцию двухчастичной системы (3.8.9), можно использовать ее для расчетов среднеквадратичных радиусов $^7$Li в обычной двухкластерной модели, которые записываются [1, 37, 38]

$$\langle r^2 \rangle_m = m_1/m \langle r^2 \rangle_{m1} + m_2/m \langle r^2 \rangle_{m2} + \frac{m_1 m_2}{m^2} R_{1,2}^2 \quad , \qquad (3.8.11)$$

$$\langle r^2 \rangle_z = Z_1/Z \langle r^2 \rangle_{z1} + Z_2/Z \langle r^2 \rangle_{z2} + \frac{(Z_1 m_2^2 + Z_2 m_1^2)}{Z m^2} R_{1,2}^2 \quad .$$

Здесь первая масса и заряд относятся, например, к $^4$He, вторая к $^3$H, m и Z - полная масса и заряд $^7$Li, а $R^2$ матричный элемент вида

$$R_{1,2}^2 = \langle X(R) | R^2 | X(R) \rangle \quad ,$$

который определяет относительное межкластерное расстояние для выделяемой пары частиц. При поиске энергии связи ядра в трехтельной модели начальные значения вариационных параметров $\alpha_i$ и $\beta_i$ находились из линейной сетки вида [55]

$$\alpha_i = i/10 \; , \qquad \beta_i = 2\alpha_i \quad .$$

Затем проводилось независимое варьирование каждого из них так, чтобы минимизировать энергию системы с точностью до $10^{-3}$ МэВ, т.е. параметры изменяются до тех пор, пока изменение энергии не станет меньше заданной величины. Для проверки метода расчета и компьютерной программы рассматривалась модельная задача для трех частиц, взаимодействующих в потенциале Афнана - Танга [61] с усреднением триплетных и синглетных состояний. Для энергии такой системы в [61] получено -7.74 МэВ, а в работах [62], где использовался неортогональный вариационный метод с изменением параметров волновой функции на основе тангенциальной сетки, найдено -7.76 МэВ. Здесь при независимом варьировании всех параметров и размерности базиса N=5 получено -7.83 МэВ, т.е. энергия изменилась примерно на 1% относительно результатов [61,62].

Результаты расчета вариационной энергии ядра $^7$Li, полученные изложенным методом с использованием потенциалов из табл.3.16, в зависимости от размерности вариационного базиса даны в табл.3.17. Экспериментальная трехтельная энергия ядра в этом канале составляет -8.725 МэВ [9]. Из таблицы видно, что при размерности N=7-9 энергия системы практически сходится, и дальнейшее увеличение базиса может привести, по - видимому, к изменению энергии на величину порядка 0.01 - 0.015 МэВ. Как уже говорилось, параметры взаимодействия в $^2$H$^4$He системе из-за различия разных экспериментальных данных имеют большую неоднозначность. Однако теперь становится ясно, что в пределах этой неопределенности можно найти параметры, позволяющие правильно воспроизвести энергию связи ядра $^7$Li.

С полученными волновыми функциями, для массового и зарядового среднеквадратичных радиусов, найдено 2.78 Фм и 2.51 Фм соответственно, что больше эксперимента [12], где для зарядового радиуса получены величины 2.39(3) и 2.35(10) Фм.

*Таблица 3.17. Сходимость трехтельной энергии ядра $^7$Li в зависимости от размерности базиса вариационных функций.*

| N | 3 | 5 | 7 | 9 |
|---|---|---|---|---|
| E($^7$Li), МэВ | -7.68 | -8.63 | -8.68 | -8.71 |

Однако здесь использовался n$^2$H потенциал, приводящий к завышенному радиусу трития, что повлияло и на радиус самого ядра $^7$Li. Тем самым дейтронный кластер нужно деформировать, как в ядре трития, так и в $^7$Li, поскольку в свободном состоянии дейтрон очень "рыхлая" система. Для того, чтобы получить правильный зарядовый радиус ядра 2.39 Фм необходимо уменьшить радиус дейтронного кластера, также как это было сделано выше для трития и принят его равным 1.42 Фм. Тем самым получаем, что дейтронный кластер одинаково деформирован, как в тритии, так и в $^7$Li, что хорошо согласуется с $^3$H$^4$He моделью этого ядра.

Второй причиной завышенного зарядового радиуса ядра без деформаций дейтрона может быть отсутствие учета в разных парах частиц потенциалов для других парциальных волн. Например, наряду с P волной в $^2H^4He$ системе можно учитывать S взаимодействие, а в $n^4He$ канале P волну. Но поскольку радиус завышен всего на 5-7%, то именно рассмотренные парные потенциалы дают основной вклад в структуру ядра, правильно объясняя и его квадрупольный момент, который оказывается равен -35.4 мб, что меньше экспериментальной величины всего на 3%. Отсюда видно, что учет дополнительных парциальных волн парных потенциалов, скорее всего, приведет только к небольшим поправкам для полученных величин.

Найденная вероятность двухчастичного $^3H^4He$ канала 98.1% вполне объясняет успешное использование простой двухкластерной модели, позволяющей получить хорошие результаты для многих характеристик ядра $^7Li$ [1,37,38]. Для двухчастичных радиусов на основе (3.8.11) найдено 2.68 Фм и 2.63 Фм соответственно, что несколько больше экспериментальной величины и результатов, получаемых в двухкластерной модели с запрещенными состояниями [1,38]. Кулоновская энергия ядра, которая представляется в виде среднего от кулоновского матричного элемента, оказалась равна 0.77 МэВ.

Для энергии связи ядра $^7Be$, если рассматривать его с теми же параметрами волновой функции, но учесть кулоновское взаимодействие между частицами (13) и (23), найдено -7.15 МэВ, что только на 1% отличается от экспериментальной величины -7.08 МэВ [9].

Таким образом, видно, что рассмотренная трехкластерная модель позволяет вполне разумно передать известные экспериментальные данные по некоторым характеристикам ядра $^7Li$ и приводит к большой вероятности $^3H^4He$ канала. Именно использованные парные взаимодействия дают наибольший вклад в рассмотренные характеристики, а учет деформаций дейтрона приводит к правильному зарядовому радиусу [63].

---

1. Дубовиченко С.Б., Джазаиров-Кахраманов А.В. - ЯФ, 1995, т.58, № 4, с.635; ЯФ, 1995, т.58, № 5, с.852; ЭЧАЯ, 1997, т.28, №6, с.1529.

2. Дубовиченко С.Б., Мажитов М. - Изв. АН КазССР, сер. физ.-мат., 1987, № 4, с.55; Дубовиченко С.Б., Джазаиров-Кахраманов А.В. - ЯФ, 1993, т.56, № 2, с.87; ЯФ, 1994, т.57, № 5, с.784.

3. Baye D., Hanck M. - J. Phys. G., 1981, v.7, p.1073; Tompson D., Tang Y.C. - Phys. Rev., 1967, v.159, p.806.

4. Schmelzbach P., Gruebler W., Konig V., Marmier P. - Nucl. Phys., 1972, v.A184, p.193; McIntair L., Haeberli W. - Nucl. Phys., 1967, v.A91, p.382; Bruno M., Cannata F., D'Agostino M., Maroni C., Massa I. - Nuovo Cim., 1982, v.A68, p.35; Jenny B., Gruebler W., Konig V., Schmelzbach P.A., Schweizer C. - Nucl. Phys., 1983, v.A397, p.61; Darriulat P., Garreta D., Tarrats A., Arvieux J. - Nucl. Phys., 1967, v.A94, p.653; Keller L., Haeberli W. - Nucl. Phys., 1970, v.A156, p.465; Барит И.Я., Бровкина Л.Н., Дулькова Л.С., Краснопольский В.М., Кузнецова Е.В., Кукулин В.И. - Препринт ИЯИ, Москва, 1987, № П-0513.

5. Barnard A.C., Jones C.M., Phillips G.C. - Nucl. Phys., 1964, v.50, p.629; Spiger R., Tombrello T.A. - Phys. Rev., 1967, v.163, p.964.

6. Ivanovich M., Young P.G., Ohlsen G.G. - Nucl. Phys., 1968, v.A110, p.441.

7. Vlastou J. et al. - Nucl. Phys., 1977, v.A292, p.29; Batten R., Glough D.L., England B.A., Harris R.G., Worledge D.H. - Nucl. Phys., 1970, v.A151, p.56.

8. Tompson D., Tang Y.C. - Nucl. Phys., 1968, v.A106, p.591.

9. Ajzenberg-Selove F. - Nucl. Phys., 1979, v.A320, p.1; Fiarman S., Meyerhof W.E. - Nucl. Phys., 1973, v.A206, p.1.

10. Bacher A.D, Spiger R.J., Tombrello T.A. - Nucl. Phys., 1968, v.A119, p.481.

11. Collard H., Hofstadter R., Hughes E.B., Johansson A., Yearian M.R. - Phys. Rev., 1965, v.138, p.B57; Juster F.P. et al. - Phys. Rev. Lett., 1985, v.55, p.2261; Beck D. et al. - Phys. Rev., 1984, v.C30, p.1403; Simon G. - Nucl. Phys., 1981, v.A364, p.285; Klasfeld S. et al. - Nucl. Phys., 1986, v.A456, p.373; Borie B. et al. - Nucl. Phys., 1977, v.A275, p.246; Phys. Rev., 1978, v.A18, p.324; Hand L. et al. - Rev. Mod. Phys., 1963, v.35, p.335; Frosh R.F., Mc Carthy J.S., Rand R.E., Yearian M.R. - Phys. Rev., 1965, v.160, p.874; McCarthy J.S., Sick I., Whitney R.R., Yearian M.R. - Phys. Rev. Lett., 1970, v.25, p.884; McGarthy J.S., Sick I., Whitney R.R. - Phys. Rev., 1977, v.C15, p.1396; Arnold R.G. et al. - Phys. Rev. Lett., 1978, v.40, p.1429; Dunn P.C., Kowalski S.B., Rad F.N., Sarget C.P., Turchinetz W.E., Goloskie R., Saylor D.P. - Phys. Rev., 1983, v.C27, p.71; Sick I. - Phys. Lett., 1982, v.B116, p.212.

12. Van Niftric G.J.C., Brockman K.W., Van Oers W.T.H. - Nucl. Phys., 1971, v.A174, p.173; Hausser O. et al. - Nucl. Phys., 1973, v.A212, p.613; Green S. et al. - Phys. Rev., 1971, v.A4, p.251; Sundholm D. et al. - Chem. Phys. Lett., 1984, v.112, p.1; Vermeer W. et al. - Austr. J. Phys., 1984, v.37, p.273; Phys. Lett., 1984, v.B138, p.365; Weller A. et al. - Phys. Rev. Lett., 1985, v.55, p.480; Rand R., Frosch R., Yearian

M.R. - Phys. Rev., 1966, v.144, p.859; Bamberger A. et al. - Nucl. Phys., 1972, v.194, p.193; De Vries H. et al. - Atom Data and Nucl. Data Tables., 1987, v.36, p.495; Suelzle L.R., Yearian M.R., Crannell H. - Phys. Rev., 1967, v.162, p.992.

13. Platner D. - In: Europ. Few Body Probl. Nucl. Part. Phys. Sesimbra., 1980, p.31; Platner G.R., Bornard M., Alder K. - Phys. Lett., 1976, v.61B, p.21; Bornard M., Platner G.R., Viollier R.D., Alder K. - Nucl. Phys., 1978, v.A294, p.492; Lim T. - Phys. Rev., 1976, v.C14, p.1243; Phys. Lett., 1975, v.56B, p.321; 1973, v.47B, p.397.

14. Вильдермут Л., Тан Я. - Единая теория ядра. М., Мир, 1980, 502с. (Wildermuth K., Tang Y.C. - A unified theory of the nucleus. Vieweg. Braunschweig. 1977).

15. Ланько Э.В., Домбровская Г.С., Шубный Ю.К. - Вероятности электромагнитных переходов атомных ядер, Л.: Наука, 1972, 701с.

16. Mertelmeir T., Hofmann H.M. - Nucl. Phys., 1986, v.A459, p.387.

17. Rao K.S., Sridhar R., Susila S. - Phys. Scr., 1981, v.24, p.925; Rao K.S., Sridhar R. - Phys. Scr., 1978, v.17, p.557; Liu Q., Kanada H., Tang Y.C. - Z. Phys., 1981, v.A303, p.253; Bouten M., Bouten M.C. - J. Phys. G., 1982, v.8, p.1641.

18. Buck B., Baldock R.A., Rubio J.A. - J. Phys., 1985, v.11G, p.L11.

19. Buck B., Merchant A.C. - J. Phys., 1988, v.14G, p.L211.

20. Roos P., Goldberg D.A., Chant N.S., Woody R. - Nucl. Phys., 1976, v.A257, p.317; Watson J. et al. - Nucl. Phys., 1971, v.A172, p.513; Alder J. et al. - Phys. Rev., 1972, v.C6, p.1.

21. Janssens T. et al. - Phys. Rev., 1966, v.142, p.922.

22. Kruppa A., Lovas R.G., Beck R., Dickmann F. - Phys. Lett., 1986, v.B179, p.317; Lovas R.G., Kruppa A., Beck R., Dickmann F. - Nucl. Phys., 1987, v.A474, p.451; Kruppa A., Beck R., Dickmann F. - Phys. Rev., 1987, v.C36, p.327; Sharma V.K., Nagarajan M.A. - J. Phys. G., 1984, v.10, p.1703; Kajino T., Matsuse T., Arima A. - Nucl. Phys., 1984, v.A413, p.323; v.A414, p.185; Kaneko T., Shirata M., Kanada H., Tang Y.C. - Phys. Rev., 1986, v.C43, p.771; Kanada H., Liu Q.K.K., Tang Y.C. - Phys. Rev., 1980, v.C22, p.813.

23. Unkelbach M., Hofman H. - Phys. Lett., 1991, v.B261, p.211; Few Body Systems., 1991, v.11, p.143.

24. Kanada H., Kaneko T., Tang Y.C. - Nucl. Phys., 1982, v.A389, p.285; Kanada H., Kaneko T., Nomoto M., Tang Y.C. - Progr. Theor. Phys., 1984, v.72, p.369.

25. Il-Tong Cheon - Phys. Lett., 1969, v.B30, p.81; 1971, v.B35, p.276; Phys. Rev., 1971, v.C3, p.1023.

26. Bergstrom J.C. - Nucl. Phys., 1980, v.A341, p.13.

27. Bergstrom J.C. - Nucl. Phys., 1976, v.A262, p.196; 1979, v.A327, p.458; Phys. Rev., 1982, v.C25, p.1156.

28. Walliser H., Fliesbach T. - Phys. Rev., 1985, v.C31, p.2242; Афанасьев В. Д., и др. // ЯФ, 1996, т.60, с.97.

29. Lichtenstadt J., Alster J., Moinester M.A., Dubach J., Hicks R.S., Peterson G.A. - Phys. Lett., 1989, v.B219, p.394; 1990, v.B244, p.173; Li G.C., Sick I., Whitney R.R., Yearian M.R. - Nucl. Phys., 1971, v.A162, p.583.

30. Кукулин В.И., Рыжих Г.Г., Чувильский Ю.М., Эрамжян Р.А. - Препр. ИЯИ АН СССР П-0685, 1990, 36с; Изв. АН СССР, сер. физ., 1989, т.53, с.121; Кукулин В.И. - Изв. АН КазССР, сер. физ.-мат., 1988, N2, с.44; Kukulin V.I. et al. - J. Phys. G., 1989, v.58, p.777; Eramzhyan R.A., Ryzhikh G.G., Kukulin V.I., Tchuvil'sky Yu.M. - Phys. Lett., 1989, v.B228, p.1.

31. Дубовиченко С.Б. - ЯФ, 1993, т.56, № 4, с.45.

32. Berg H., Arnold W., Huttel E., Krause H.H., Ulbricht J., Claunitzer G. - Nucl. Phys., 1980, v.A334, p.21; Kavanagh R.W., Parker P.D. - Phys. Rev., 1966, v.143, p.143.

33. McSherry D., Baker S.D. - Phys. Rev., 1970, v.1C, p.888; Szaloky G., Seiler F. - Nucl. Phys., 1978, v.A303, p.57; Morrow L., Haeberli W. - Nucl. Phys., 1969, v.A126, p.225.

34. Tombrello T. - Phys. Rev., 1965, v.138, p.40B.

35. Drigo L., Pisent G. - Nuovo Cim., 1967, v.LI B, p.419.

36. Morales J.R., Cahill T.A., Shadoan D.J., Willmes H. - Phys. Rev., 1975, v.11C, p.1905; Mudroch B.T., Hasell D.K., Sourkes A.M., Van Oers W.T.H., Verheijen P.J.T. - Phys. Rev., 1984, v.C29, p.2001.

37. Neudatchin V.G., Kukulin V.I., Pomerantsev V.N., Sakharuk A.A. - Phys. Rev., 1992, v.C45. p.1512; Phys. Lett., 1991, v.B255, p.482; Неудачин В.Г., Сахарук А.А., Смирнов Ю.Ф. - ЭЧАЯ, 1993, т.23, с.480.

38. Искра В., Мазур А.И., Неудачин В.Г., Нечаев Ю.И., Смирнов Ю.Ф. - УФЖ, 1988, т.32, с.1141; Искра В., Мазур А.И., Неудачин В.Г., Смирнов Ю.Ф. - ЯФ, 1988, т.48, с.1674; Неудачин В.Г., Померанцев В.Н., Сахарук А.А. - ЯФ, 1990, т.52, с.738; Кукулин В.И.,

Неудачин В.Г., Померанцев В.Н., Сахарук А.А. - ЯФ, 1990, т.52, с.402; Дубовиченко С.Б., Неудачин В.Г., Смирнов Ю.Ф., Сахарук А.А.- Изв. АН СССР, сер. физ., 1990, т.54, с.911; Neudatchin V.G., Sakharuk A.A., Dubovichenko S.B. - Few Body Sys., 1995, v18, p.159.

39. Tombrello T. - Phys. Rev., 1966, v.143, p.772.

40. Shen P.N., Tang Y.C., Kanada H., Kaneko T. - Phys. Rev., 1986, v.C33, p.1214; LeMere M. et al. - Phys. Rev., 1975, v.C12, p.1140.

41. Неудачин В.Г., Смирнов Ю.Ф. - Нуклонные ассоциации в легких ядрах. М.: Наука, 1969, 414с.

42. Schmelzbach P., Gruebler W., White R.E., Konig V., Risler R., Marmier P. - Nucl. Phys., 1972, v.A197, p.237; Arviex J. - Nucl. Phys., 1967, v.A102, p.513; Van Oers W.T.H., Brockman K.W. - Nucl. Phys., 1967, v.A92, p. 561; Chauvin J., Arvieux J. - Nucl. Phys., 1975, v.A247, p.347; Sloan J. - Nucl. Phys., 1971, v.A168, p.211; Huttel E., Arnold W., Baumgart H., Berg H., Clausnitzer G. - Nucl. Phys., 1983, v.A406, p.443.

43. Дубовиченко С.Б., Джазаиров-Кахраманов А.В. - ЯФ, 1990, т.51, № 6, с.1541.

44. Kocher D. et al. - Nucl. Phys., 1969, v.A132, p.455.

45. Sick I. et al. - Phys. Rev. Lett., 1980, v.45, p.871; Ciofi degli Atti C., Pace E., Salme G. - Lect. Not. Phys., 1981, v.137, p.115.

46. Jenny B., Gruebler W., Konig V., Schmelzbach P.A., Konig V., Burgi H.R. - Nucl. Phys., 1980, v.A337, p.77.

47. Kanada H., Kaneko T., Shen P.N., Tang Y.C. - Nucl. Phys., 1986, v.A457, p.93; Kanada H., Kaneko T., Tang Y.C. - Nucl. Phys., 1989, v.A504, p.529; Chwieroth F.S., Tang Y.C., Tompson D.R. - Phys. Rev., 1974, v.C9, p.56; Chwieroth F.S., Brown R.E., Tang Y.C., Tompson D.R. - Phys. Rev., 1973, v.C8, p.938; Shen P.N., Tang Y.C., Fujiwara Y., Kanada H. - Phys. Rev., 1985, v.C31, p.2001.

48. Brolley J. et al. - Phys. Rev., 1960, v.117, p.1307; King T. et al. - Nucl. Phys., 1972, v.A183, p.657.

49. Lien P. - Nucl. Phys., 1972, v.A178, p.375.

50. Kanada H., Kaneko T., Tang Y.C. - Phys. Rev., 1986, v.C34, p.22; Chwieroth F.S., Tang Y.C., Tompson D.R. - Nucl. Phys., 1972, v.A189, p.1; Thompson D. - Nucl. Phys., 1970, v.A143, p.304.

51. Frank R.M., Gammel J.L. - Phys. Rev., 1953, v.99, p.1406; Darves-Blane R. - Nucl. Phys., 1972, v.A191, p.353; Detch J. et al. - Phys. Rev., 1971, v.C4, p.52; Werntz C., Meyerhof W.E. - Nucl. Phys., 1968, v.A121, p.38; Meyerhof W.E., Mc Elearney J.N. - Nucl. Phys., 1965, v.74, p.533; Барит И.Я., Сергеев В.А. - ЯФ, 1971, т.13, с.1230.

52. Hardekopf R.A., Lisowski P.W., Rhea T.C., Walter R.L. - Nucl. Phys., 1972, v.A191, p.481; Kankovsky R., Fritz J.C., Kilian K., Neufert A., Fick D. - Nucl. Phys., 1976, v.A263, p.29.

53. Brown L., Trachslin W. - Nucl. Phys., 1967, v.A90, p.334; Brown L., Haeberli W., Trachslin W. - Nucl. Phys., 1967, v.A90, p.339; Hoop Jr., Barschall H.H. - Nucl. Phys. 1966, v.A83, p.65.

54. Ali S., Ahman A.A.Z., Ferdous N. - Int. Cent. Theor. Phys., Miramare - Trieste, 1984, № IC/84/195, 108p.

55. Kukulin V.I., Krasnopol'sky V.M., Voronchev V.T., Sazonov P.B. - Nucl. Phys., 1984, v.A417, p.128; 1986, v.A453, p.365; Kukulin V.I., Voronchev V.T., Kaipov T.D., Eramzhyan R.A. - Nucl. Phys., 1990, v.A517, p.221.

56. Lehman D.R. // Phys. Rev., 1982, v.C25, p.3146; Lehman D.R., Parke W.C. // Phys. Rev., 1983, v.C28, p.364; Eskandarian A., Lehman D.R., Parke W.C. // Phys. Rev., 1988, v.C38, p.2341.

57. Дубовиченко С.Б. // ЯФ, 1998, т.61, с.210; Kukulin V.I., Pomerantsev V.N., Cooper S.G., Dubovichenko S.B. // Phys. Rev., 1998, v.C57, p. 2462.

58. Кукулин В.И., Неудачин В.Г., Смирнов Ю.Ф. // ЭЧАЯ, 1979, т.10, с.1236.

59. Merchant A.C. , Rowley N. // Phys. Lett., 1985, v.150B, p.35.

60. Кукулин В.И., Краснопольский В.М., Миселхи М.А., Ворончев В.Т. // ЯФ, 1981, т.34, с.21.

61. Afnan I.R., Tang Y.C. // Phys. Rev., 1968, v.175, p.1337.

62. Krasnopolsky V.M., Kukulin V.I. // Czech. J. Phys., 1977, v.B27, p.290; J. Phys., 1977, v.G3, p.795; Дубовиченко С.Б., Жусупов М.А. // Изв. АН Каз. ССР., 1987, № 4, с.64.

63. Дубовиченко С.Б. // Изв. РАН сер. физ., 2000, т.64, с.2289.

# 4. ФОТОПРОЦЕССЫ НА ЛЕГКИХ ЯДРАХ

Перейдем теперь к рассмотрению ядерных фотопроцессов, включающих фоторазвал ядра в двухчастичный канал гамма квантом и радиационный захват двух кластеров с образование ядра в основном состоянии с испусканием гамма кванта. Все формулы, необходимые для расчетов полных сечений фотопроцессов приведены в первой главе. Классификация кластерных состояний легких ядер дана во второй главе, а потенциалы межкластерного взаимодействия с разделением по схемам Юнга там, где это необходимо, получены в третьей главе.

## 4.1. Фотопроцессы на ядрах $^6$Li и $^7$Li в кластерных моделях

Расчеты полных сечений фотопроцессов ранее выполнялись в феноменологической кластерной модели [1], аналогичной используемой здесь, и в методе резонирующих групп [2,3,4] для $^4$He$^2$H, $^4$He$^3$H и $^4$He$^3$He систем. Для взаимодействий с запрещенными состояниями были проведены расчеты полных сечений для $^4$He$^2$H кластерного канала ядра $^6$Li на основе трехтельных волновых функций основного состояния [5]. В двухкластерных моделях с запрещенными состояниями на основе гауссовых потенциалов, согласованных с фазами упругого рассеяния расчеты полных сечений фотопроцессов для ядер $^6$Li и $^7$Li были выполнены в работах [6].

Переходя к рассмотрению сечений фотопроцессов, заметим, что спиновая часть электрического оператора $Q_{Jm}(S)$ в (1.2.2), вообще говоря, дает сравнительно малый вклад в общее сечение процесса. Это позволяет рассматривать только орбитальный $Q_{Jm}(L)$ оператор EJ переходов. Однако, например, для E1 процессов в $^4$He$^2$H системе спиновое слагаемое может давать заметный вклад из-за малости сечения, обусловленного орбитальным оператором, т.к. в ядрах с N=Z переходы типа E1 с $\Delta T=0$ сильно подавлены.

В кластерной модели этот факт отражается множителем $(Z_1 / M_1 - Z_2 / M_2)^2$ в операторе $Q_{Jm}(L)$, который вообще равен нулю при этих условиях. Переход E1, обусловленный орбитальным оператором в таких ядрах может появиться только благодаря неточности условия M=2Z, так как, например, в $^4$He$^2$H системе ядра $^6$Li массы дейтрона и $^4$He не являются целыми числами и примерно равны $M_d$=2.013554 и $M_\alpha$=4.001506. Поэтому в $^4$He$^2$H канале ядра $^6$Li будем рассматривать E1 переходы с учетом обоих операторов - $Q_{Jm}(S)$ и $Q_{Jm}(L)$, E2 (только орбитальный член $Q_{Jm}(L)$) и M2 переходы, которые могут давать заметный вклад в полные сечения фотопроцессов. Сечения, обусловленные мультиполями более высокого порядка, оказываются на 2-3 порядка меньше.

### *Кластерные $^4He^3H$ и $^4He^3He$ каналы*

При расчете сечений в $^4He^3H$ и $^4He^3He$ системах рассматривались орбитальные E1, E2 и M1 переходы и вклад спиновой части электрического оператора $Q_{Jm}(S)$ для E1 процесса. Рассмотрим, например, E1 переход, когда радиационный захват может происходить, как на основное связанное состояние с $J=3/2^-$ и энергиями -2.47 МэВ в $^7Li$ и -1.59 МэВ в $^7Be$, так и на первое возбужденное с $J=1/2^-$ и энергиями -1.99 МэВ и -1.16 МэВ соответственно (см. спектр уровней ядер $^7Li$ и $^7Be$ в гл.3). Захват на основное состояние ядра происходит из состояний рассеяния S, $D_{3/2}$ и $D_{5/2}$, а на первое возбужденное состояние из S и $D_{3/2}$ волн, т.к. для начального $J_i$, конечного $J_f$ полных моментов системы и мультипольности перехода J должно выполняться условие

$$/J_i - J_f/ \leq J \leq J_i+J_f .$$

При E1, E2 и M1 переходах значения величин $P_J$, $N_J$ и $G_J$ из (1.2.5) приведены в табл.4.1, а параметры потенциалов, используемых при расчетах сечений, даны в табл.3.1.

Результаты расчета сечений EJ и M1 захвата для $^4He^3H$ и $^4He^3He$ каналов приведены на рис.4.1 [6]. Показаны также экспериментальные данные из [7,8] и расчеты для E1 из работ [7] (точечная кривая вверху на рис.4.1а). Штрих - пунктиром на рис.4.1б даны МРГ вычисления для E1 перехода из [4,9]. Из рисунков видно, что сечения E2 и M1 процессов на полтора - три порядка меньше, чем E1 и не дают существенного вклада в полные сечения.

Пик в E2 сечении для $^4He^3H$ захвата соответствует резонансу в $F_{7/2}$- фазе при энергии 2.16 МэВ в с.ц.м. Аналогичный пик наблюдается и в сечении $^4He^3He$ захвата при энергии 7.0 МэВ, что отвечает резонансу при 2.98 МэВ (с.ц.м.) относительно кластерных порогов. Вычисленные M1 сечения хорошо согласуются с МРГ результатами [4]. На рис.4.1а штрих - пунктиром с обозначением E1' приведено сечение, обусловленное спиновым членом $Q_{Jm}(S)$ электрического E1 оператора, которое в $^4He^3H$ системе заметно меньше орбитального слагаемого $Q_{Jm}(L)$ и его вкладом в полное E1 сечение действительно можно пренебречь. В расчетах для магнитных моментов кластеров использовались величины: $\mu_H = 2.9786\ \mu_0$ и $\mu_{He} = -2.1274\ \mu_0$.

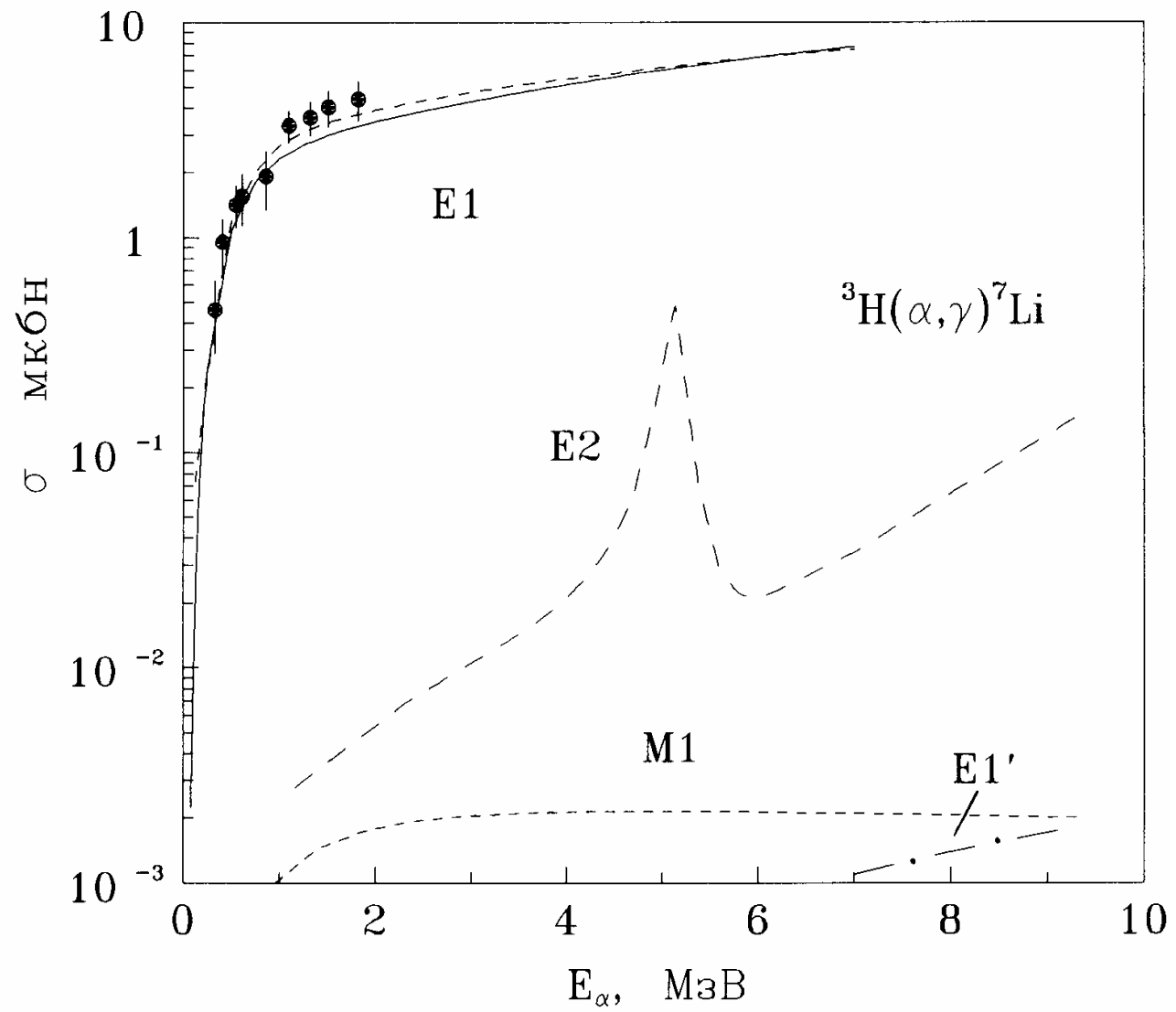

Рис.4.1а. Полные сечения процессов радиационного захвата в $^4He^3H$ системе с образованием ядра $^7Li$ в основном и первом возбужденном состояниях. Непрерывными линиями показаны расчетные E1 сечения для потенциалов из табл.3.1. Точечные линии внизу - сечение M1 процесса, штриховые линии - E2 сечение. Штрих - пунктир - E1 сечение, обусловленное спиновой частью электрического оператора. Точечная линия вверху - результаты расчетов E1 из работ [7]. Точки - экспериментальные данные из работ [7,8].

*Таблица 4.1. Значения коэффициентов $P_J$, $N_J$ и $G_J$ для (1.2.5) в $^4He^3H$ и $^4He^3He$ системах.*

| $(L_J)_i$ | $(L_J)_f$ | $P_J^2$(E1) | $(L_J)_i$ | $(L_J)_f$ | $P_J^2$(E2) | $(L_J)_i$ | $(L_J)_f$ | $N_J$(M1) | $G_J$(M1) |
|---|---|---|---|---|---|---|---|---|---|
| S | $P_{3/2}$ | 4 | $P_{3/2}$ | $P_{1/2}$ | 4 | $P_{1/2}$ | $P_{1/2}$ | $-\sqrt{1/6}$ | $2\sqrt{2/3}$ |
| $D_{3/2}$ | $P_{3/2}$ | 4/5 | $F_{5/2}$ | $P_{1/2}$ | 6 | $P_{3/2}$ | $P_{1/2}$ | $2\sqrt{1/3}$ | $-2\sqrt{1/3}$ |
| $D_{5/2}$ | $P_{3/2}$ | 36/5 | $P_{1/2}$ | $P_{3/2}$ | 4 | $P_{1/2}$ | $P_{3/2}$ | $-2\sqrt{1/3}$ | $2\sqrt{1/3}$ |
| S | $P_{1/2}$ | 2 | $P_{3/2}$ | $P_{3/2}$ | 4 | $P_{3/2}$ | $P_{3/2}$ | $15/2\sqrt{15}$ | $2\sqrt{5/2}$ |
| $D_{3/2}$ | $P_{1/2}$ | 4 | $F_{5/2}$ | $P_{3/2}$ | 12/7 | | | | |
| | | | $F_{7/2}$ | $P_{3/2}$ | 72/7 | | | | |

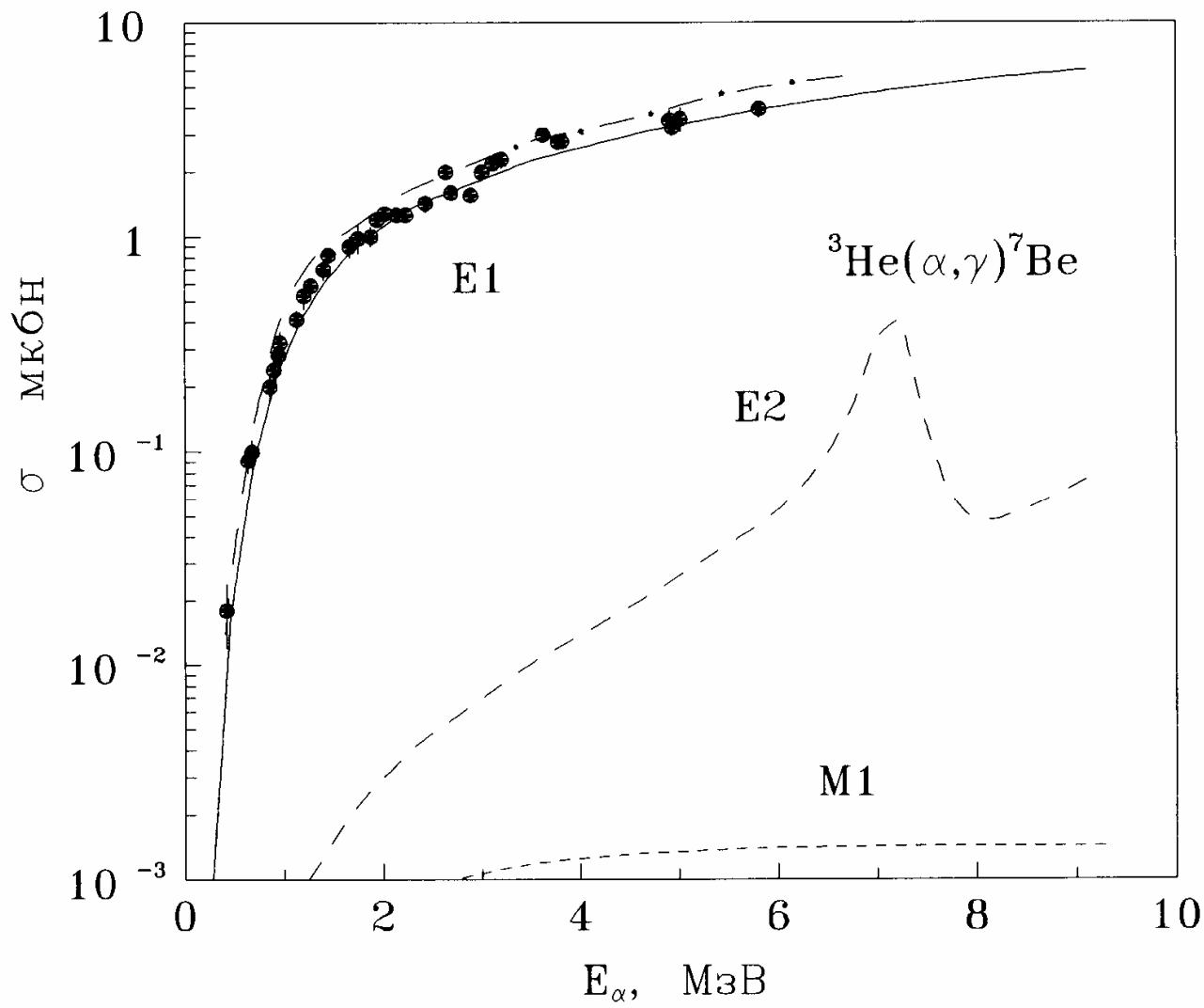

Рис.4.1б. Полные сечения процессов радиационного захвата в $^4$He$^3$He системе с образованием ядра $^7$Be в основном и первом возбужденном состояниях. Непрерывными линиями показаны расчетные E1 сечения для потенциалов из табл.3.1. Точечные линии внизу - сечение M1 процесса, штриховые линии - E2 сечения. Штрих - пунктир - результаты МРГ расчетов из [4,9].

При $^4$He$^3$H захвате был вычислен астрофизический S фактор [6], который при 20 кэВ оказался равен 0.087 кэВ бн. в сравнении с известной величиной 0.064(16) кэВ бн, приведенной в работах [7,8]. Отметим, что существуют и другие экспериментальные данные - 0.134(20) кэВ бн [10]. При $^4$He$^3$He захвате S фактор при 40 кэВ равен 0.47 кэВ бн [6], что согласуется с данными [7,11], лежащими в пределах 0.47-0.63 кэВ бн. В МРГ расчетах обычно получают величину от 0.5 до 0.6 кэВ бн [3,9]. В потенциальном подходе [12,13,14] вычисления дают 0.47(2) кэВ бн, а в работе [1] получено 0.56 кэВ бн.

На рис. 4.2 приведены астрофизические S факторы для $^4$He$^3$H и $^4$He$^3$He захвата в области энергий от 10 кэВ до 3 МэВ, полученные в [6] на основе E1 сечений, и результаты МРГ вычислений [2,3,9] (штриховые кривые). Эксперимент приведен в работах [7,8].

### *Кластерный $^4$He$^2$H канал*

Величина $P_J$ в (1.2.5) для $^4$He$^2$H системы в случае E2 переходов может быть представлена в виде

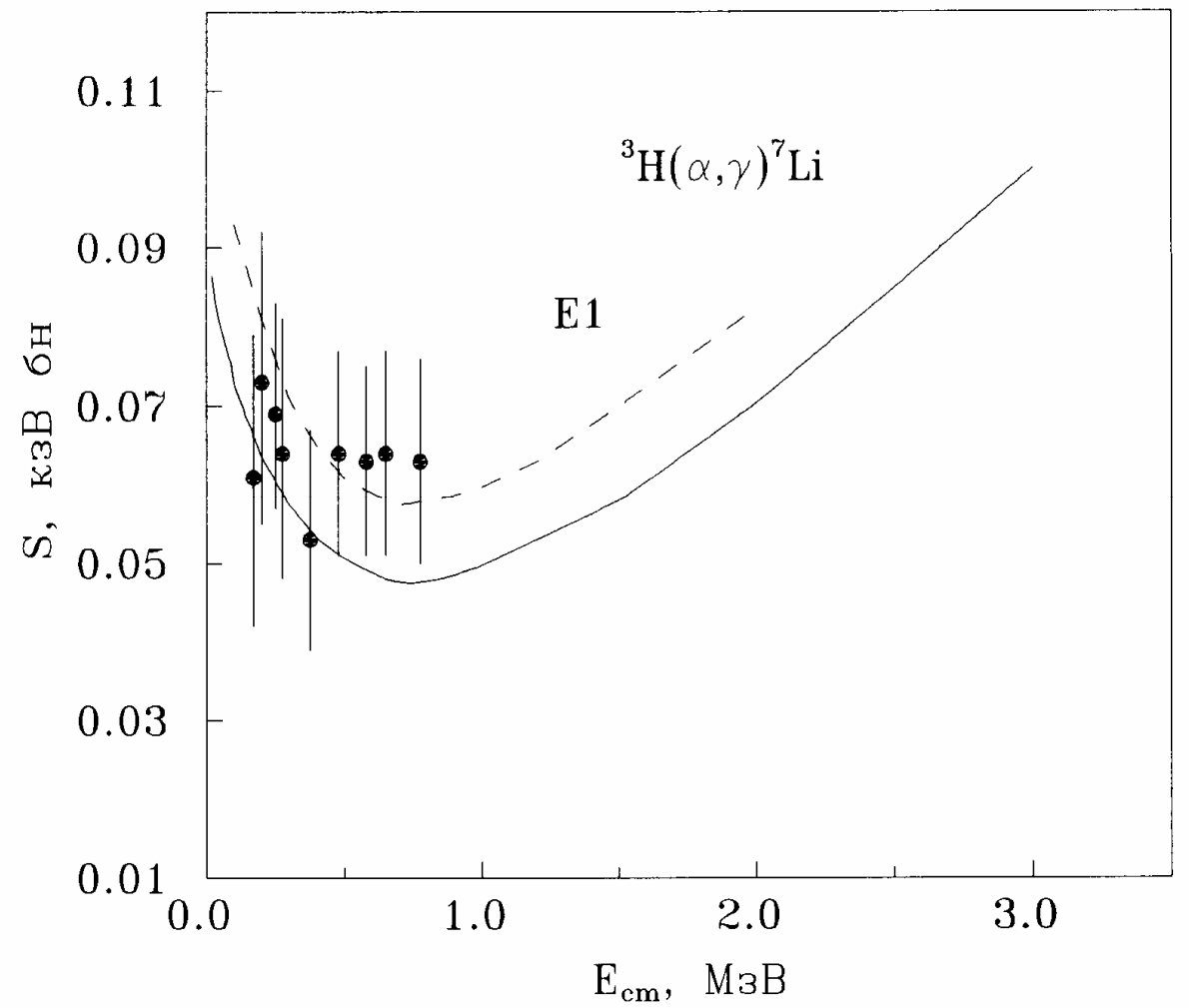

Рис.4.2а. Астрофизический S фактор при малых энергиях для $^4$He$^3$H захвата. Точки - экспериментальные данные из [7,8]. Непрерывные кривые - расчеты для потенциалов из табл.3.1. Штриховые линии - МРГ расчеты из [2,3,9].

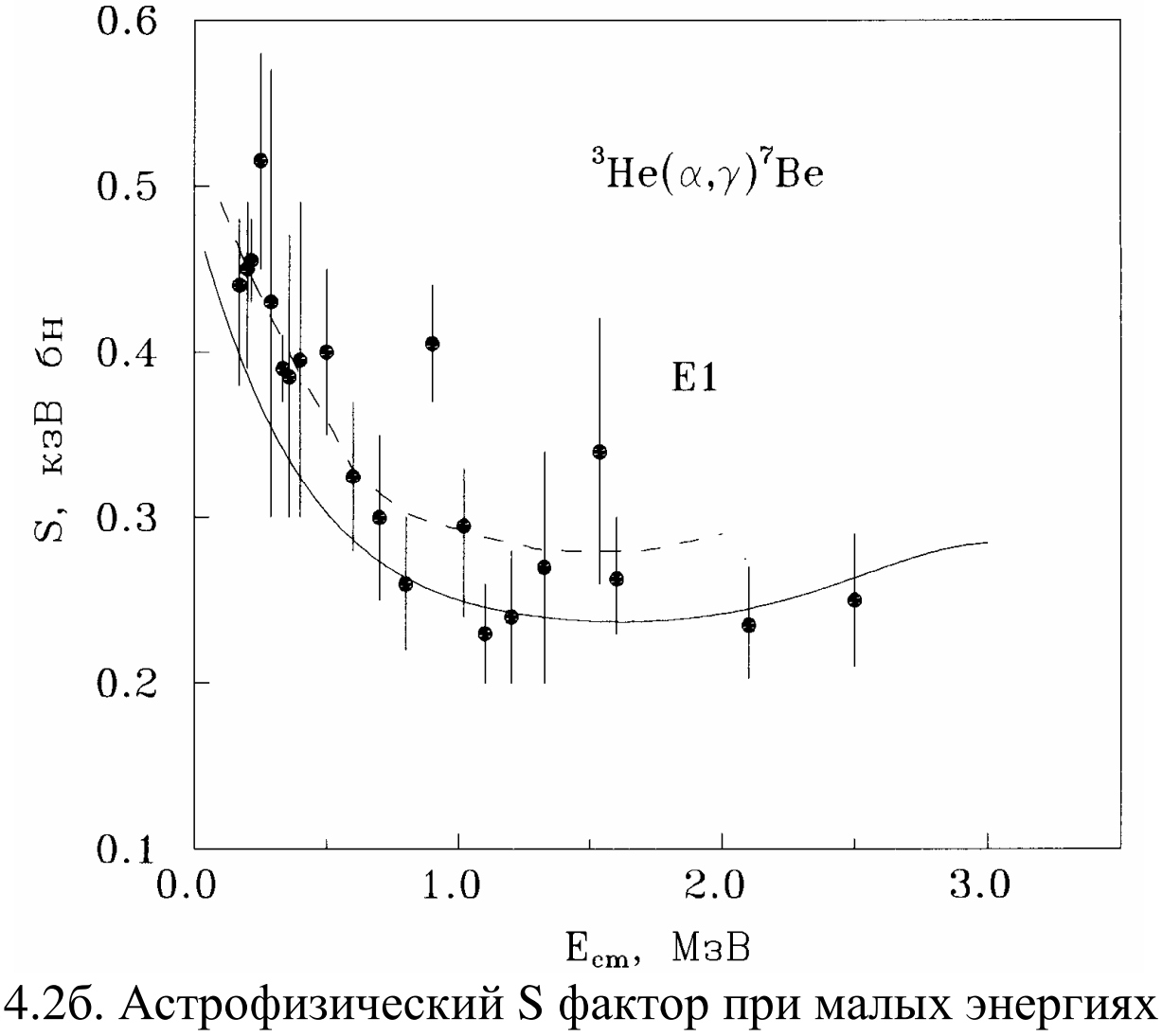

Рис.4.2б. Астрофизический S фактор при малых энергиях для $^4$He$^3$He захвата. Непрерывные кривые - расчеты для потенциалов из табл.3.1. Штриховые линии - МРГ расчеты из [2,3,9].

$$P_J^2 = 2J_i + 1 , \qquad (4.1.1)$$

если захват происходит на основное состояние из состояний рассеяния с L=2 и $J_i$=1,2,3. Параметры потенциалов используемых при расчетах приведены в табл.3.1.

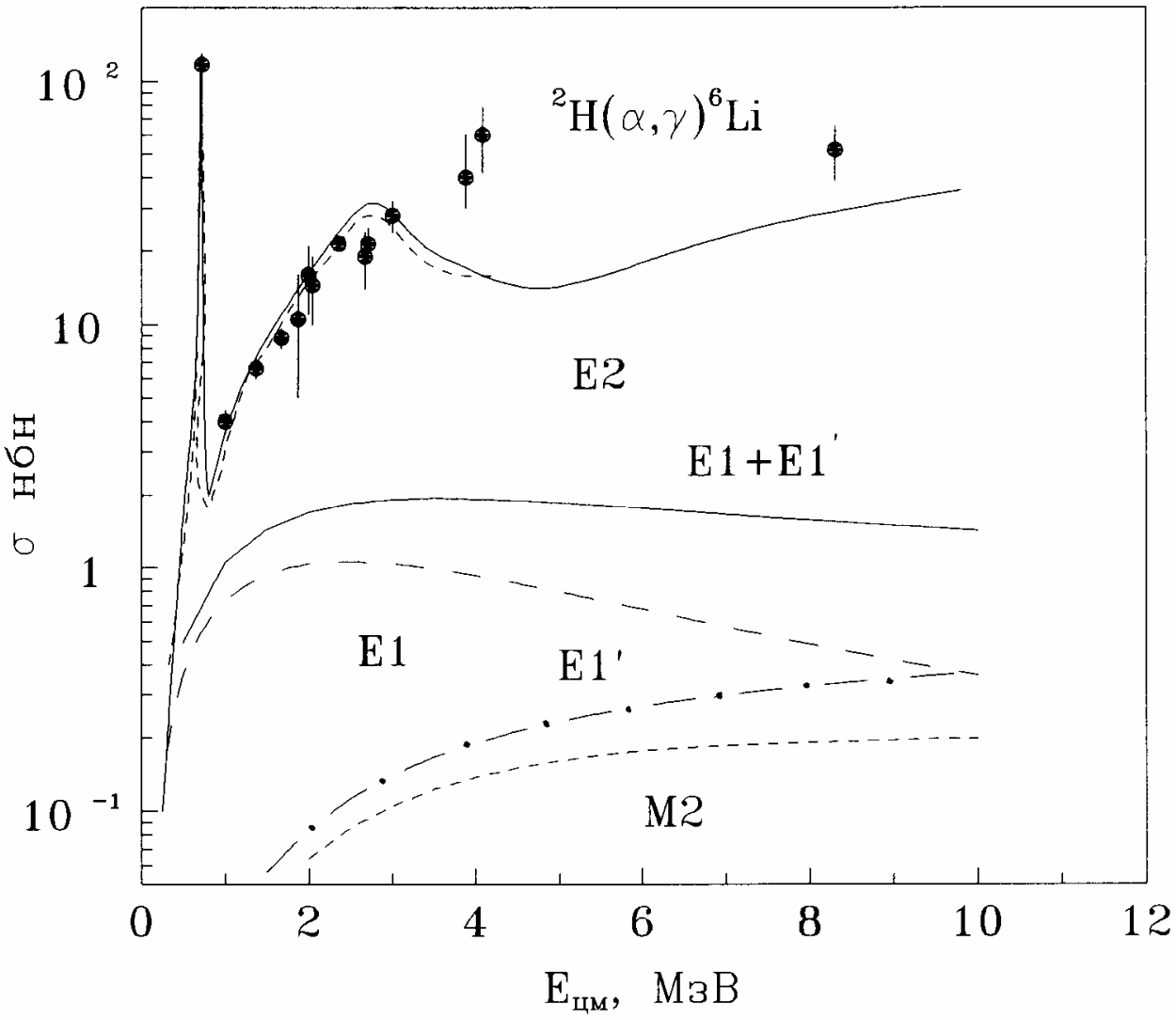

Рис.4.3а. Полные сечения процесса радиационного захвата в $^4$He$^2$H системе с образованием ядра $^6$Li в основном состоянии. Непрерывная линия вверху - расчетные E2 сечения для потенциалов из табл.3.1. Точечная линия внизу - сечения M2 процесса, штриховая линия - E1 сечения, обусловленное орбитальной частью электрического оператора, штрих-пунктир - E1 сечение, обусловленное спиновой частью электрического оператора, непрерывная линия внизу - E1 сечение, полученное с учетом орбитальной и спиновой частей электрического оператора. Точечная линия вверху - результаты расчетов E2 из работы [1]. Точки - данные из работ [15].

В сечении E2 захвата, показанном на рис.4.3 непрерывной линией, вместе с экспериментальными данными, приведенными в работах [15], кроме пика при 0.71 МэВ наблюдается и второй максимум при энергии около 2.5-3 МэВ, который соответствует резонансу в $D_2$ волне с энергией 2.84 МэВ (см. фазы рассеяния в гл.3). На

рис.4.3а точечной линией внизу показаны сечения M2 захвата из P состояний рассеяния на основное состояние ядра. Точечная линия вверху показывает результаты расчетов E2 сечений, выполненных в работе [1]. Значения коэффициентов $P_J$, $N_J$ и $G_J$ из формулы (1.2.5) даны в табл.4.2.

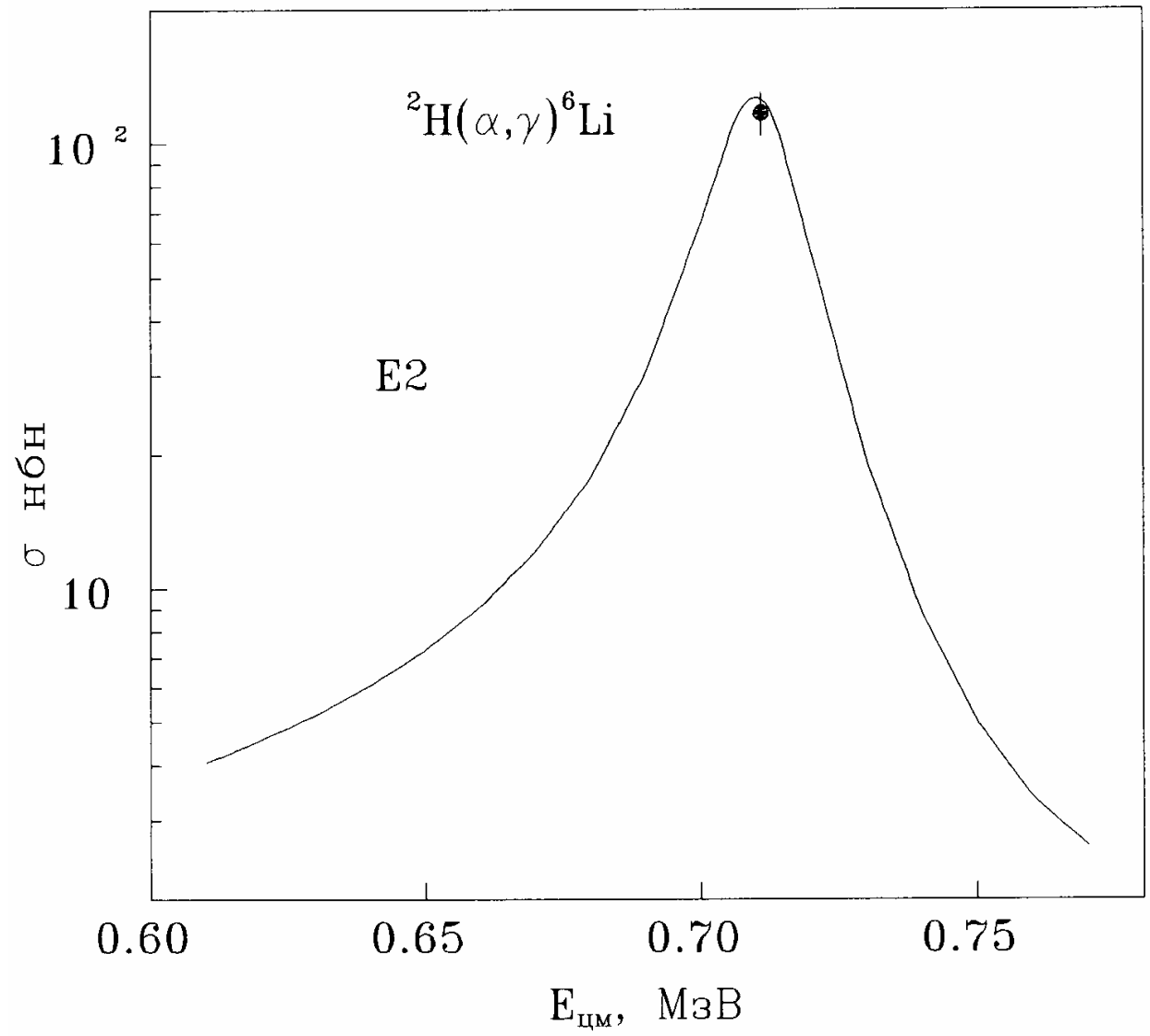

Рис.4.3б. Полные сечения процесса радиационного захвата в $^4He^2H$ системе с образованием ядра $^6Li$ в основном состоянии. Непрерывная линия - расчетные E2 сечения для потенциала из табл.3.1 в области $3^+$ резонанса. Точка - экспериментальные данные из работ [15].

*Таблица 4.2. Коэффициенты $P_J$, $N_J$ и $G_J$ для формул (1.2.5) в $^4He^2H$ и $^3He^3H$ кластерных системах ядра $^6Li$.*

| $^4He^2H$ | | | | | | $^3He^3H$ ($3^+$) | |
|---|---|---|---|---|---|---|---|
| $L_J$ | $N_J$(M2) | $G_J$(M2) | $L_J$ | $N_J$(E1) | $P_J$(E1) | $L_J$ | $P_J$(E1) |
| $P_1$ | $-\sqrt{5/2}$ | $\sqrt{10}$ | $P_0$ | $-\sqrt{2}$ | $-1$ | $^3P_2$ | 42/5 |
| $P_2$ | $-\sqrt{15/2}$ | $\sqrt{50/3}$ | $P_1$ | $-\sqrt{3/2}$ | $-\sqrt{3}$ | $^3F_2$ | 1/35 |
| | | | $P_2$ | $\sqrt{5/2}$ | $\sqrt{5}$ | $^3F_3$ | 1 |
| | | | | | | $^3F_4$ | 81/7 |

В работе [6] были выполнены и расчеты E1 захвата на основное состояние из континуума с $J_i$=0,1,2 и L=1 и параметрами потенциалов из табл.3.1 для орбитального (E1) и спинового (E1') членов в отдельности и их общего сечения с учетом интерференции (см. рис.4.3а непрерывная линия внизу). Это сечение заметно меньше чем E2 и практически не вносит вклада в полные сечения при энергии больше 0.5 МэВ. На рис.4.3б показаны $^4He^2H$ полные сечения захвата при малых энергиях, где хорошо видно положение и высота расчетного и экспериментального резонанса при 0.71 МэВ [6].

На рис.4.4а непрерывной линией показаны сечения фоторазвала ядра $^6Li$ в $^4He^2H$ канал [6] вместе с экспериментом [15,16] и расчетами (штриховая линия), полученными в работе [5] на основе трехтельных волновых функций основного состояния.

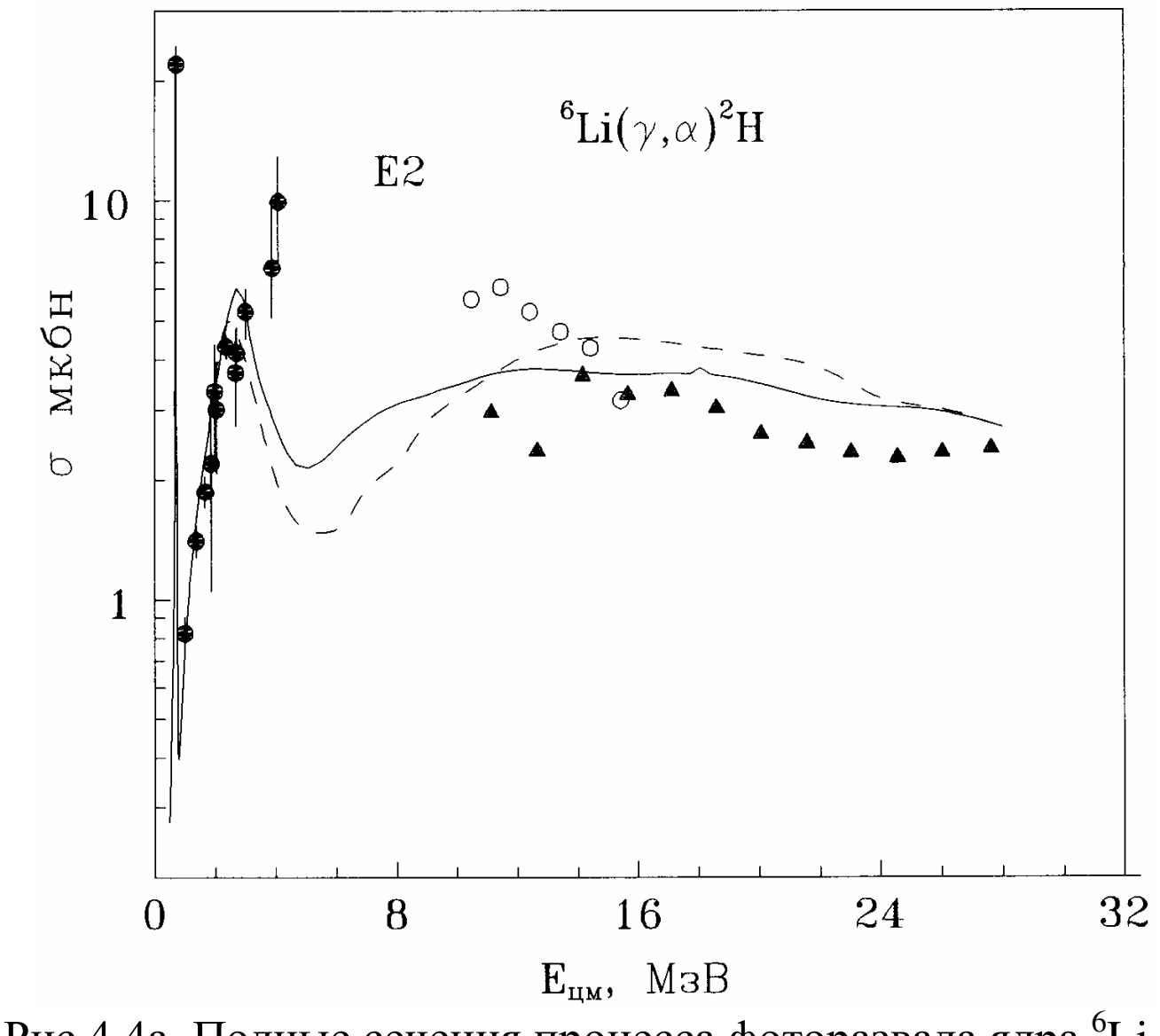

Рис.4.4а. Полные сечения процесса фоторазвала ядра $^6Li$ в $^4He^2H$ канал. Непрерывная линия - расчетные E2 сечения для потенциалов из табл.3.1. Штриховая линия - расчетные сечения из работы [5] с трехтельными волновыми функциями основного состояния.

На рис.4.4б даны астрофизические S факторы, полученные из расчетных E1 и E2 сечений [6]. Хорошо видно, что в области низких энергий преобладающим оказывается сечение E1 процесса. Линейная экстраполяция S факторов к нулю дает S(E2)=3 $10^{-7}$ кэВ бн и $S(E1)=1.2\ 10^{-6}$ кэВ бн, так что общий S фактор примерно равен $1.5\ 10^{-6}$ кэВ бн. Этот результат хорошо согласуются с вычислениями, выполненными в работе [17].

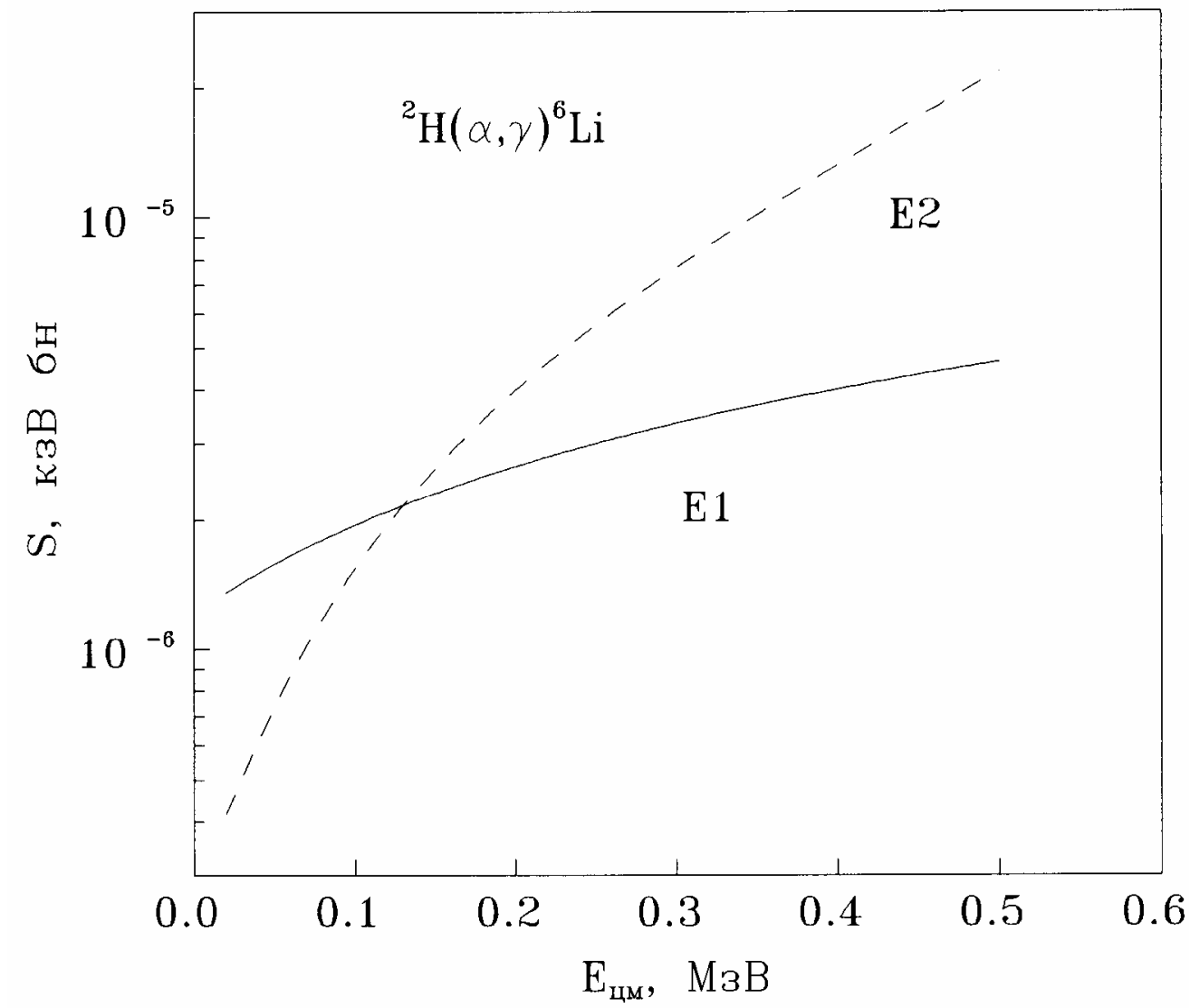

Рис.4.4б. Астрофизические S факторы для E1 и E2 процессов при фотозахвате в $^4He^2H$ канал [6].

### *Кластерный $^3He^3H$ канал*

В случае E1 захвата в $^3He^3H$ кластерном канале на основное состояние $^6Li$ величина $P_J$ представляется в виде (4.1.1) для начальных состояний с $J_i$=0,1,2 и L=1. В кластерной $^3He^3H$ модели можно рассмотреть и E1 переход на $3^+$ резонансное состояние. В этом случае величина $P_J$ приведена в табл.4.2. Параметры $^3He^3H$ потенциалов с учетом спин - орбитального расщепления даны в табл.3.1.

Результаты расчетов полного сечения захвата с S потенциалом, правильно передающим энергию связи ядра (см. гл.3) показаны на рис.4.5а непрерывной линией [6]. Экспериментальные данные из работы [18]. Видно, что использование такого потенциала и P взаимодействия, правильно передающего энергетический ход фаз рассеяния, позволяет хорошо описать экспериментальные данные. Отметим, что существуют и другие измерения сечений [19], заметно отличающиеся, от приведенных на рисунке.

В случае M1 переходов на основное состояние, рассматривался процесс, когда происходит изменение спинового состояния с синглетного на триплетное. В операторе перехода остается только спиновый член $W_{Jm}(S)$ с коэффициентом $-\sqrt{3/2}$ . Для E2 переходов на основное состояние из D волны с $J_i$ =1, 2, 3 $P_J$ находится из (4.1.1) . Результаты этих расчетов приведены на рис.4.5а точечной и штриховой кривыми [6].

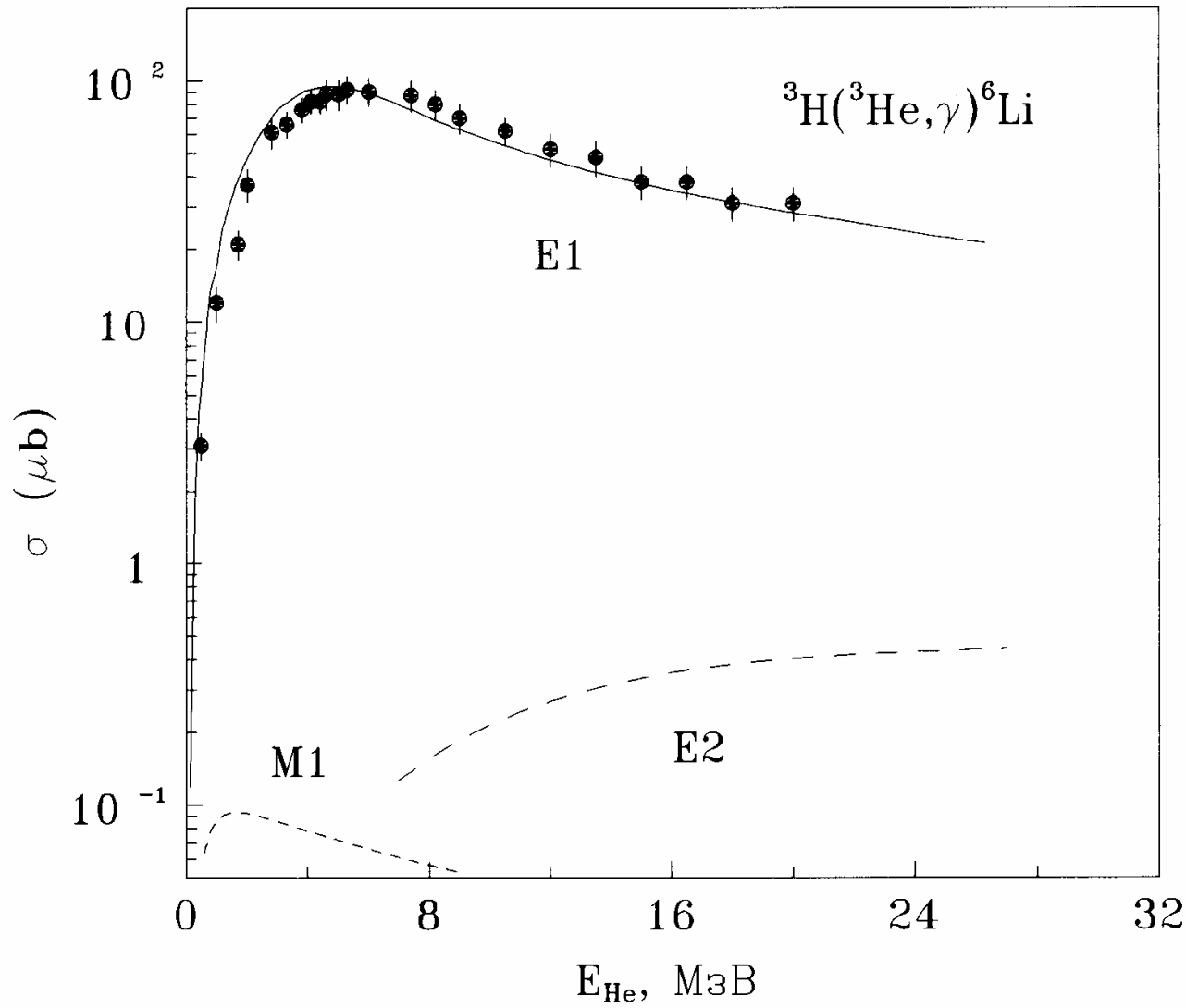

Рис.4.5а. Полные сечения процесса радиационного захвата в $^3He^3H$ канале с образованием ядра $^6Li$ в основном состоянии. Непрерывная линия - расчетные E1 сечения для потенциалов рассеяния из табл.3.1. Точечная линия - сечения M1 процесса, штриховая линия - E2 сечения. Точки - экспериментальные данные из работ [18].

Различия в полных экспериментальных сечениях для $^3He^3H$ развала больше, чем для радиационного захвата. На рис.4.5б показаны результаты измерений, выполненных в работах [20]. Непрерывной линией даны результаты, полученные на основе принципа детального равновесия из расчетных сечений захвата [6].

На рис.4.6а приведен астрофизический S фактор для $^3He^3H$ захвата при малых энергиях. Линейной экстраполяцией S фактора при нулевой энергии в E1 процессе получено 0.06 кэВ бн. На рис.4.6б показаны результаты расчета сечений захвата на уровень $3^+$ для потенциалов из табл.3.1 [6] вместе с данными и расчетами (точечная кривая), полученными в работе [21]. Здесь в качестве потенциала связанного $D_3$ состояния использовались взаимодействия с глубиной 105 МэВ (непрерывная линия) и 107.5 МэВ (штриховая линия), обсуждение которых дано в третьей главе.

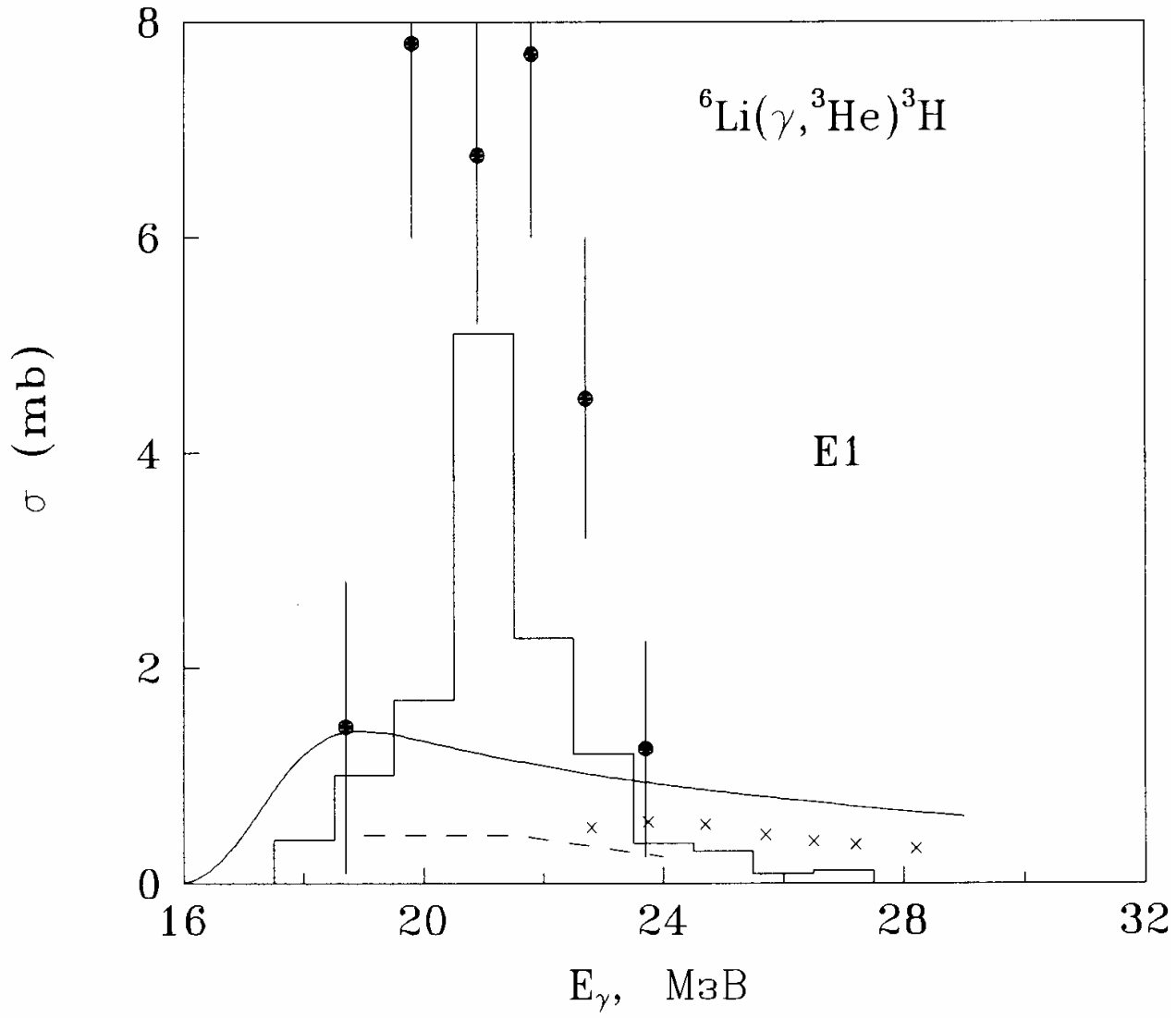

Рис.4.5б. Полные сечения процесса фоторазвала ядра $^6$Li. Непрерывная линия - сечение, полученное на основе принципа детального равновесия из расчетных сечения захвата [6]. Точки, гистограмма, штрихованная линия и крестики - экспериментальные данные из работ [20].

### *Кластерный $p^5He$ канал ядра $^6Li$*

В принципе можно рассмотреть $p^5$He и $n^5$Li каналы развала ядра $^6$Li. Однако здесь отсутствуют, как экспериментальные, так и МРГ фазы рассеяния и нет возможности построения взаимодействий на используемом ранее принципе. Тем не менее, можно попытаться найти некие качественные критерии, которые позволят, так или иначе, фиксировать параметры потенциалов необходимых для рассмотрения E1 процесса. В частности ясно, что S и P взаимодействия должны содержать запрещенные состояния, а S потенциал еще и уровень, соответствующий разрешенному основному состоянию ядра.

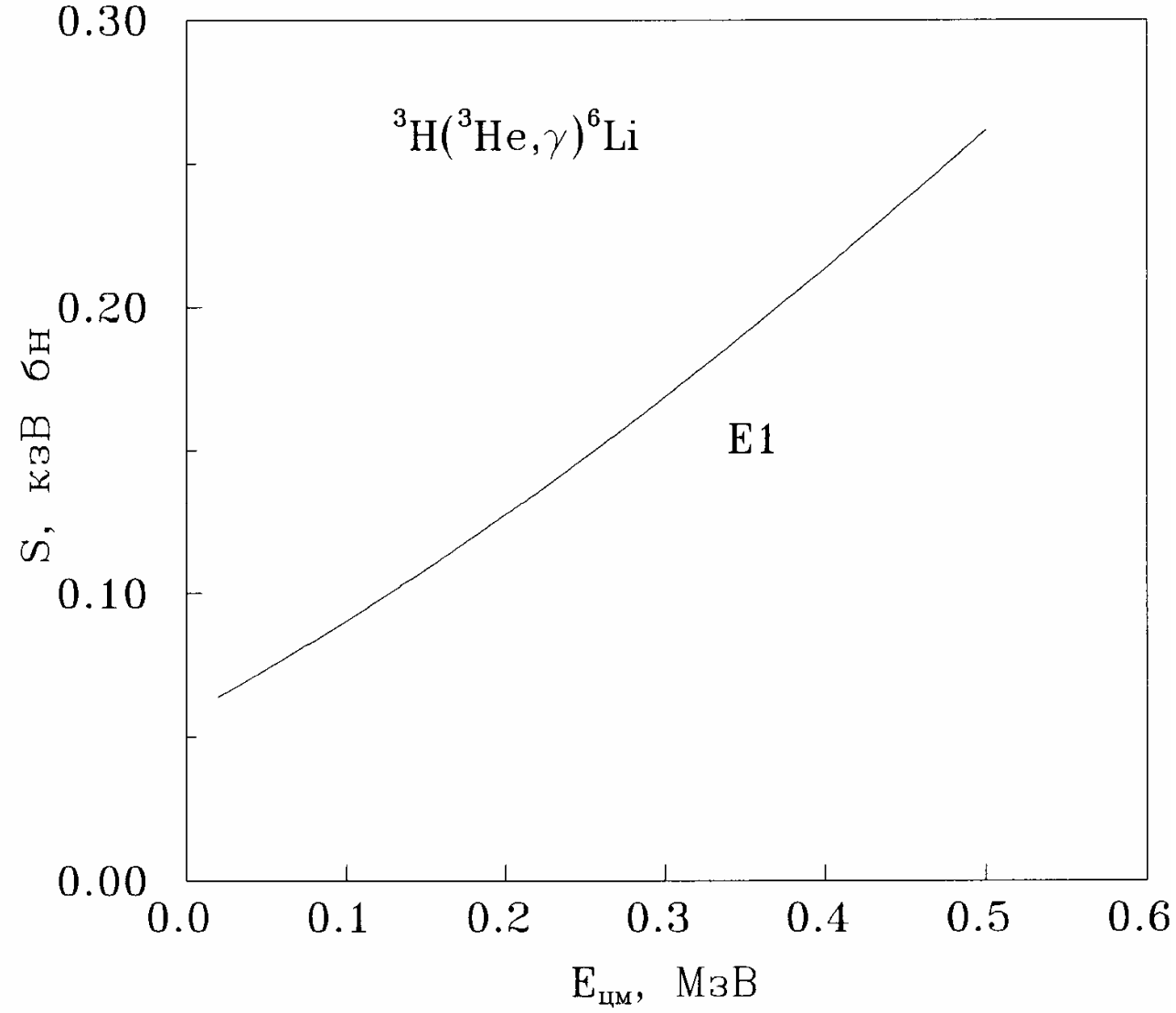

Рис.4.6а. Астрофизический S фактор для E1 процесса при $^3He^3H$ захвате [6].

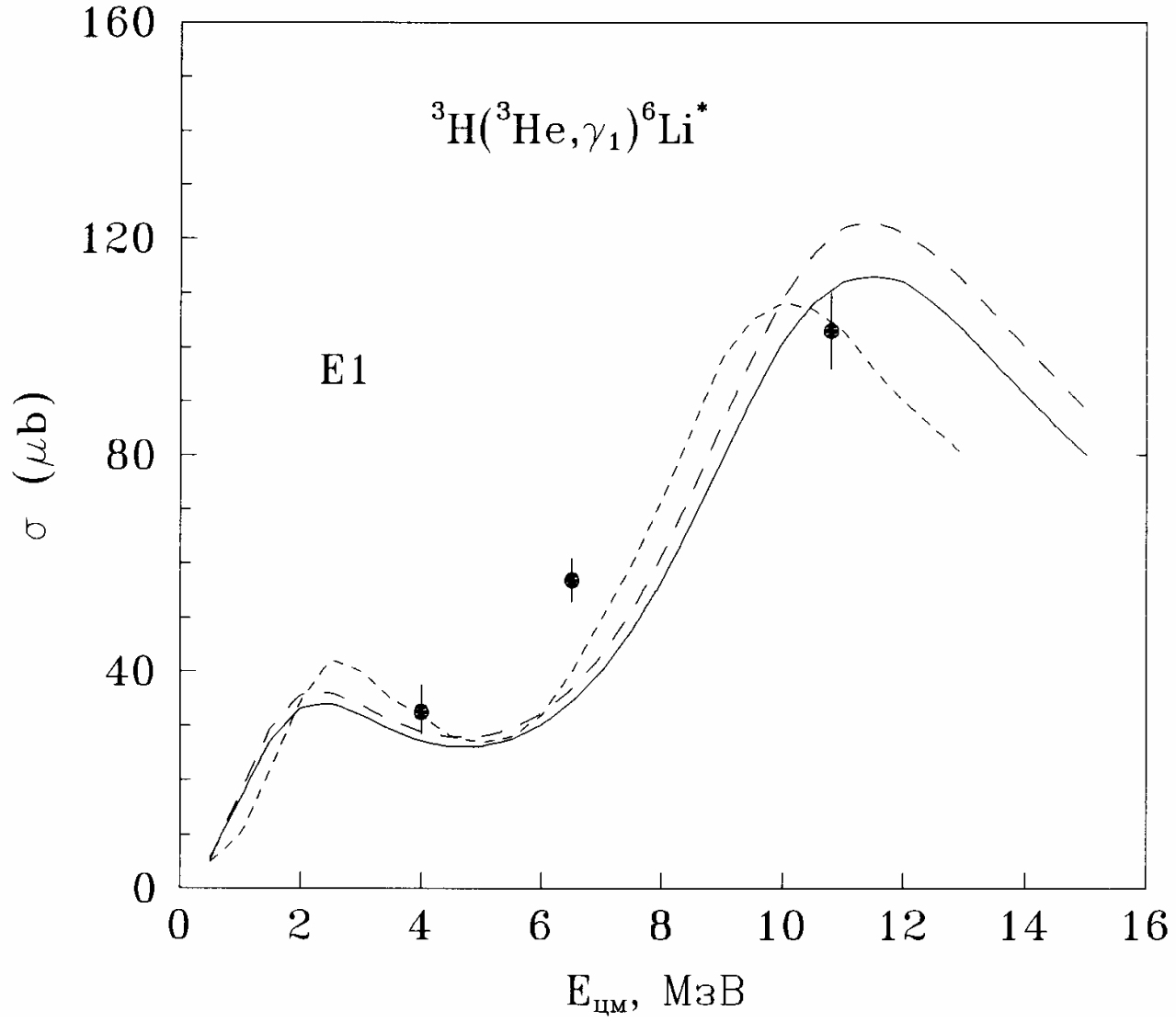

Рис.4.6б. Полные сечения процесса радиационного захвата в $^3He^3H$ канале с образованием ядра $^6Li$ в возбужденном $3^+$ состоянии. Точки - экспериментальные данные из работы [21].

Значит, потенциалы будут достаточно глубокими, что бы вместить запрещенные и разрешенные состояния при энергиях -4.59 МэВ для $p^5He$ и -5.66 МэВ для $n^5Li$ системы. Кроме того, для этих систем должен быть один потенциал, который приводит к правильным энергиям связи только при изменении кулоновского члена. Перечисленных выше условий вполне достаточно для построения S волнового потенциала основного состояния. В результате были найдены следующие параметры - $V_s$ = -128.5 МэВ и $\alpha$ = 0.2 $Фм^{-2}$ [6]. Потенциал приводит к энергиям разрешенных состояний -4.6 МэВ и -6.0 МэВ, а запрещенные состояния лежат при -59.6 МэВ и -62.5 МэВ.

В случае P потенциала ситуация более сложная, так как в спектре ядра отсутствуют связанные состояния или резонансы в P волне, по которым можно было бы определить параметры. По-видимому, единственное предположение, позволяющее конкретизировать потенциал, может основываться на схожести структуры запрещенных состояний в P волнах для $^4He^2H$ и $p^5He$ систем.

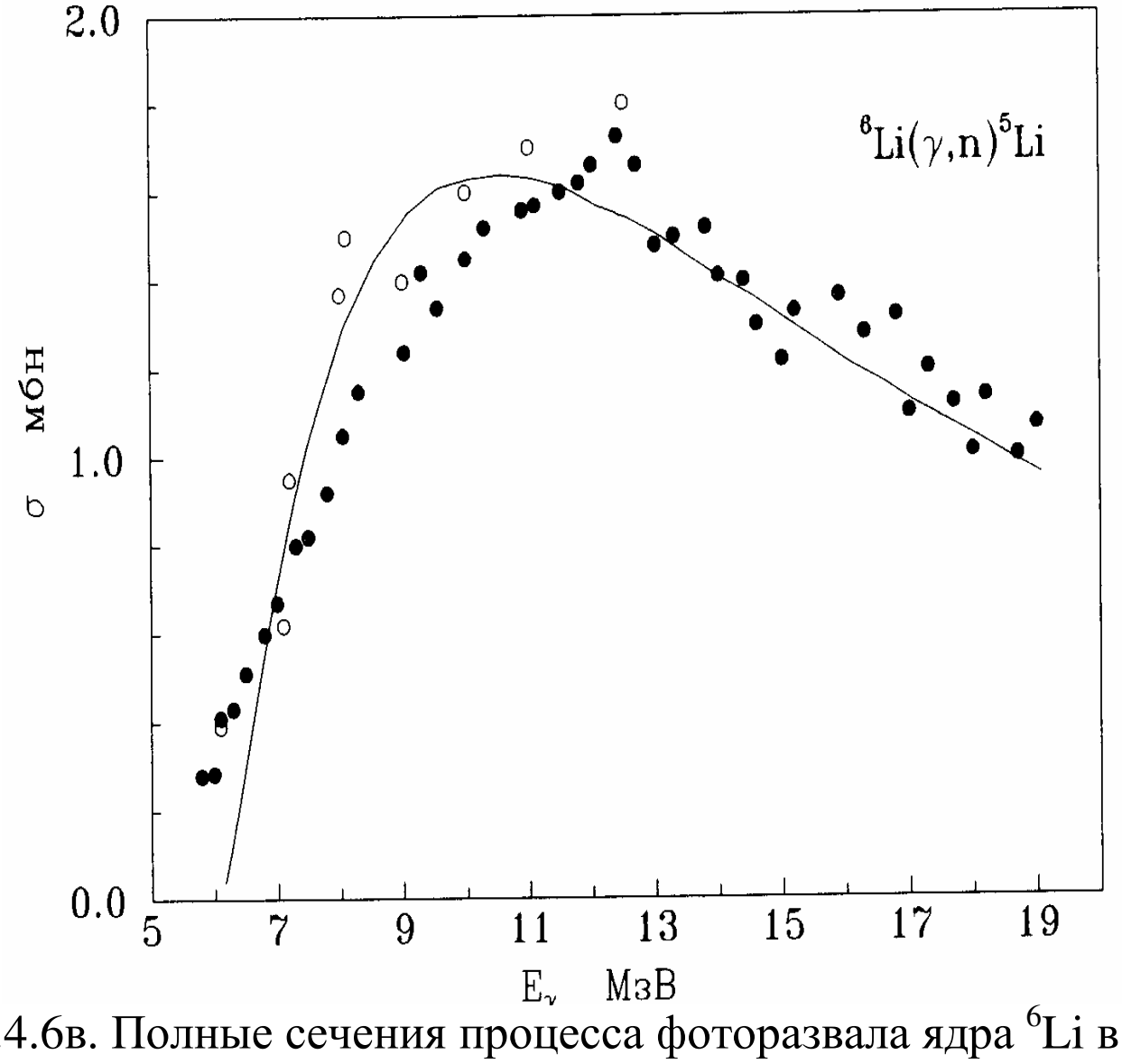

Рис.4.6в. Полные сечения процесса фоторазвала ядра $^6Li$ в $n^5Li$ канал. Точки и кружки - эксперимент из работ [16]. Кривая - расчеты с потенциалами, параметры которых приведены в тексте [6].

Тогда можно предположить, что запрещенные состояния долж-ны быть примерно при одинаковых энергиях. Усредненные энергии запрещенных состояний в $^4$He$^2$H системе приводят к величине 11-12 МэВ. Поэтому, оставляя геометрию P потенциала такой же, как в S волне можно найти его глубину отвечающую такому запрещенному состоянию. Тогда для глубины взаимодействия без спин-орбитального расщепления можно получить - $V_p$ = -100 МэВ. Энергия запрещенного состояния равна -11.5 МэВ.

Коэффициент $P_J$ для E1 развала находим из (4.1.1) с конечными $J_f$=0,1,2 и L=1. С этими потенциалами проведены расчеты полных сечений для E1 фоторазвала ядра $^6$Li в n$^5$Li канал. Результаты представлены на рис.4.6в вместе с экспериментальными данными [16]. Видно, что потенциал, основанный на таких чисто качественных рассуждениях, в принципе позволяет правильно передать величину сечения и его энергозависимость во всей известной области энергий. Отметим, что для потенциалов без запрещенных состояний вообще не удается воспроизвести экспериментальные данные при любых значениях параметров [6]. Эти результаты могут свидетельствовать о малой чувствительности полных сечений рассмотренного процесса к форме N$^5$Li потенциала, но эти взаимодействия обязательно должны содержать запрещенные состояния.

Таким образом, кластерная потенциальная модель для взаимодействий с запрещенными состояниями позволяет правильно описывать не только статические электромагнитные характеристики ядер лития, но и полные сечения фотопроцессов во всей рассмотренной области энергий, включая астрофизические S факторы при малых энергиях.

Большая вероятность кластеризации ядер лития в $^4$He$^2$H и $^4$He$^3$H каналы позволяет вполне успешно применять простую одноканальную кластерную модель. Одни и те же наборы потенциалов дают возможность описывать различные ядерные характеристики. Для получения согласия с экспериментом не требуется вводить какие-либо искажения характеристик кластеров. Предположение о том, что данным кластерам в ядре можно сопоставлять свойства соответствующих свободных частиц вполне оправдывается, что согласуется и с МРГ результатами, где искажения приводят только к незначительным изменениям характеристик ядер $^6$Li и $^7$Li.

Определенное исключение составляет только сечение $^3$He$^3$H развала, где имеются очень большие экспериментальные неоднозначности в сечениях, и не удается согласовать фазы рассеяния с энергией основного состояния.

Видно, что n$^5$Li потенциал, основанный на таких чисто качест-венных рассуждениях, в принципе позволяет правильно передать и величину сечения, и его энергозависимость во всей известной области энергий. Отметим, что для потенциалов без запрещенных состояний не удалось подобрать параметры взаимодействий, чтобы правильно описать фотосечения.

## 4.2. Фотопроцессы для ядер $^3$He, $^3$H и $^5$Li

Расчеты дифференциальных сечений фотопроцессов в N$^2$H, N$^3$H и $^2$H$^3$He системах для потенциалов с запрещенными состояниями и разделением по орбитальным схемам ранее выполнены в [22]. Полные сечения для таких кластерных систем для потенциалов с запрещенными состояниями и разделением по схемам Юнга рассматривались в [23].

### *Кластерный N$^2$H канал*

При расчете полных сечений учитывались E1 и E2 переходы, обусловленные орбитальной частью электрического оператора $Q_{Jm}(L)$. Магнитные сечения и сечения, зависящие от спиновой части электрического оператора, оказались сравнительно малыми. Электрические E1 переходы в N$^2$H системе возможны между основным чистым $^2$S состоянием и $^2$P состоянием рассеяния. Величина $P_J^2$ в (1.2.5) для фоторазвала может быть представлена в виде (4.1.1), если заменить $J_i$ на $J_f$, где $J_f$ - момент конечного состояния, который может принимать значения $1/2^-$ и $3/2^-$.

В случае E2 процессов переходы возможны между чистым основным состоянием и дублетной D волной рассеяния. И в этом случае $P_J^2$ представляется в виде (4.1.1) с $J_f = 3/2^+$ и $5/2^+$. Сечения фотопроцессов пропорциональны множителю $(Z_1/M^J{}_1+(-1)^J Z_2/M^J{}_2)^2$, который имеет одинаковую величину в n$^2$H и p$^2$H системах в случае E1 переходов и сильно отличается для E2 процессов. Поэтому E2 сечение оказывается заметным только в p$^2$H системе, что, впрочем, не объясняет разницу в величине экспериментальных сечениях для p$^2$H и n$^2$H фоторазвала (см. например [23]).

На рис.4.7 непрерывными линиями приведены результаты расчетов полных сечений процессов фоторазвала ядер $^3$H и $^3$He в n$^2$H и p$^2$H каналы с первым вариантом потенциала основного состояния и P взаимодействием с периферическим отталкиванием (см. табл. 3.9 и 3.10). Эксперимент приводится в работах [24,25].

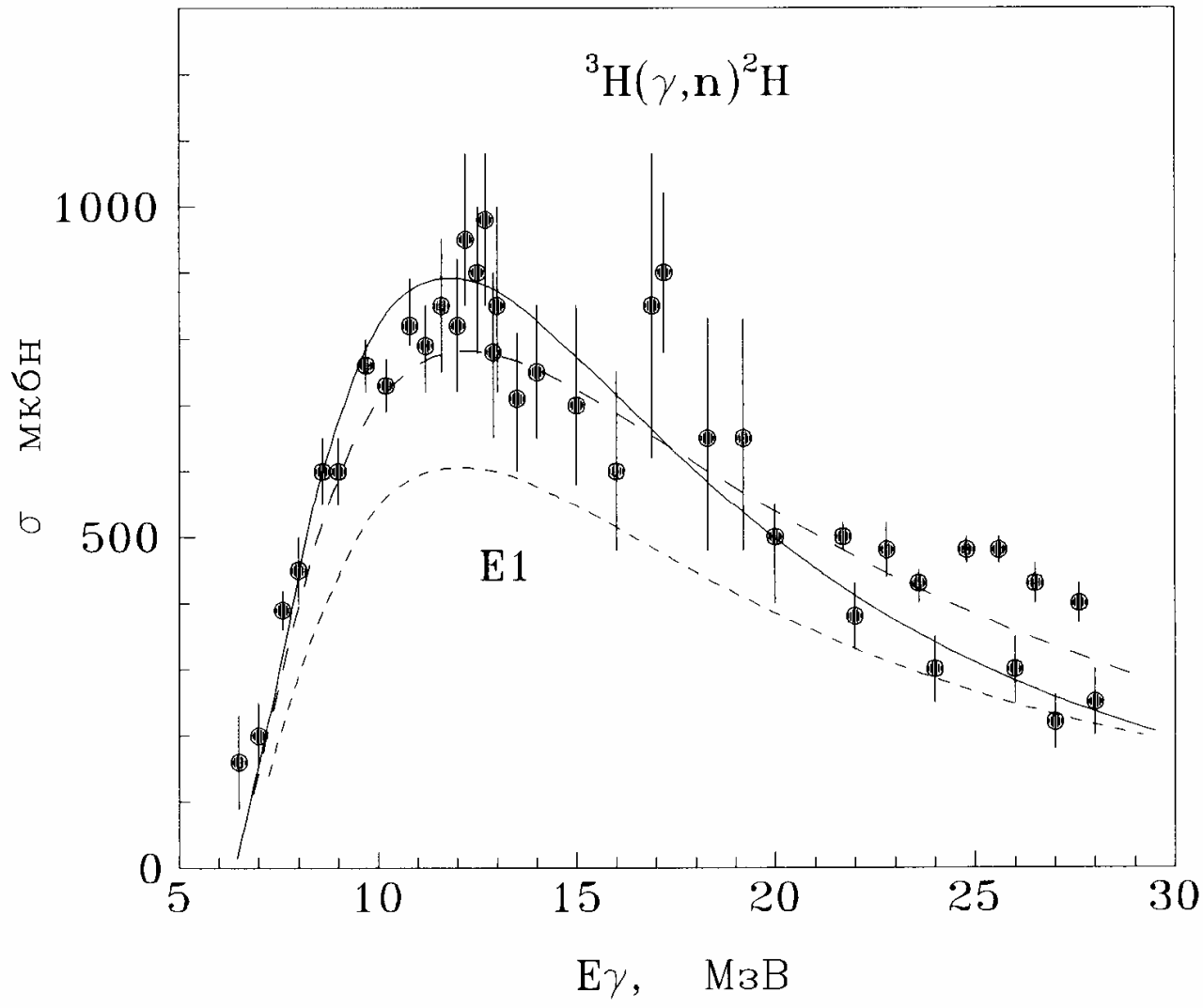

Рис.4.7а. Полные сечения фоторазвала ядра $^3$H в n$^2$H канал. Кривые - расчеты для потенциалов из табл.3.9, 3.10 [23].

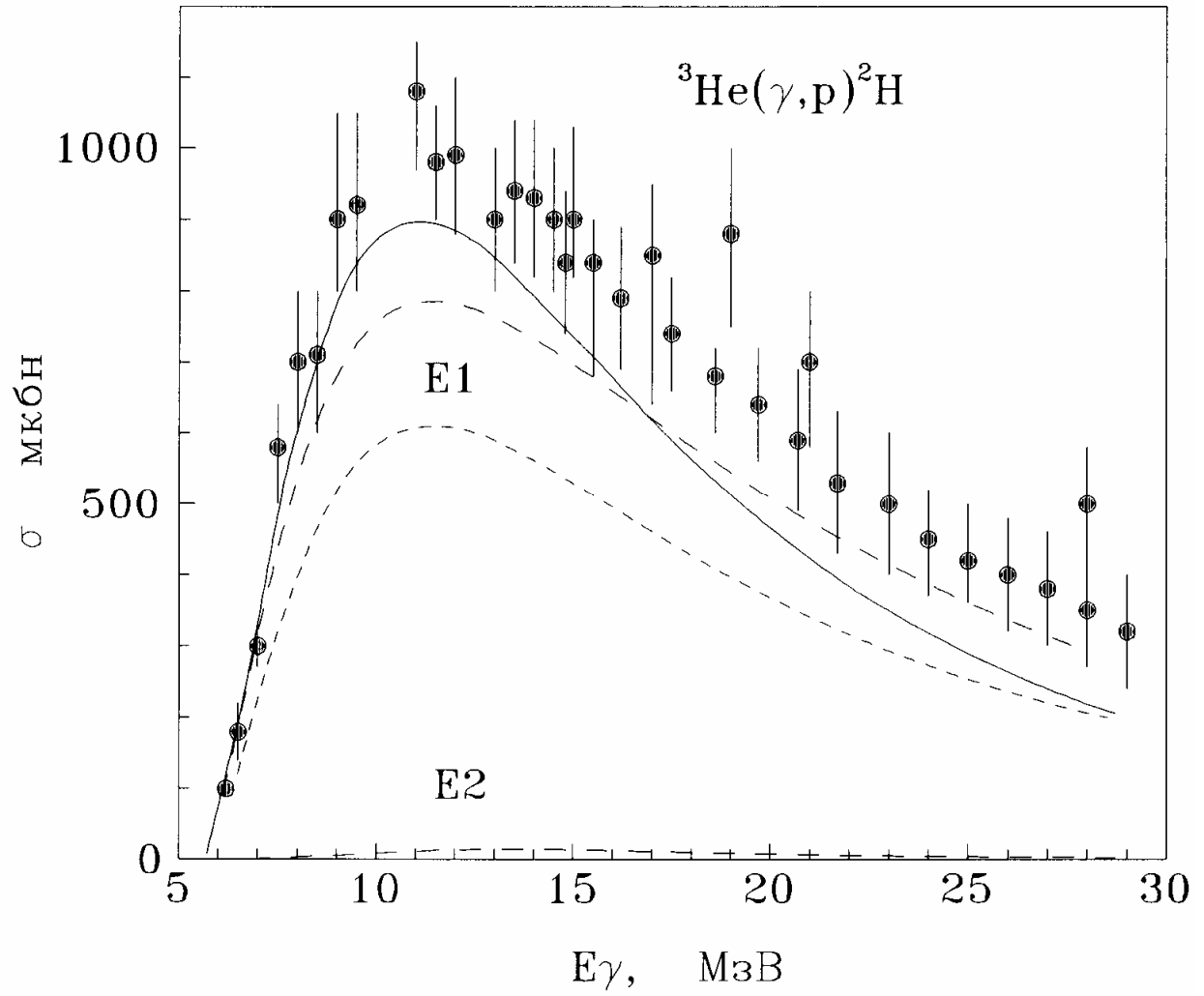

Рис.4.7б. Полные сечения фоторазвала ядер $^3$He в p$^2$H канал. Кривые - расчеты для потенциалов из табл.3.9, 3.10 [23]. Эксперимент из работ [24,25].

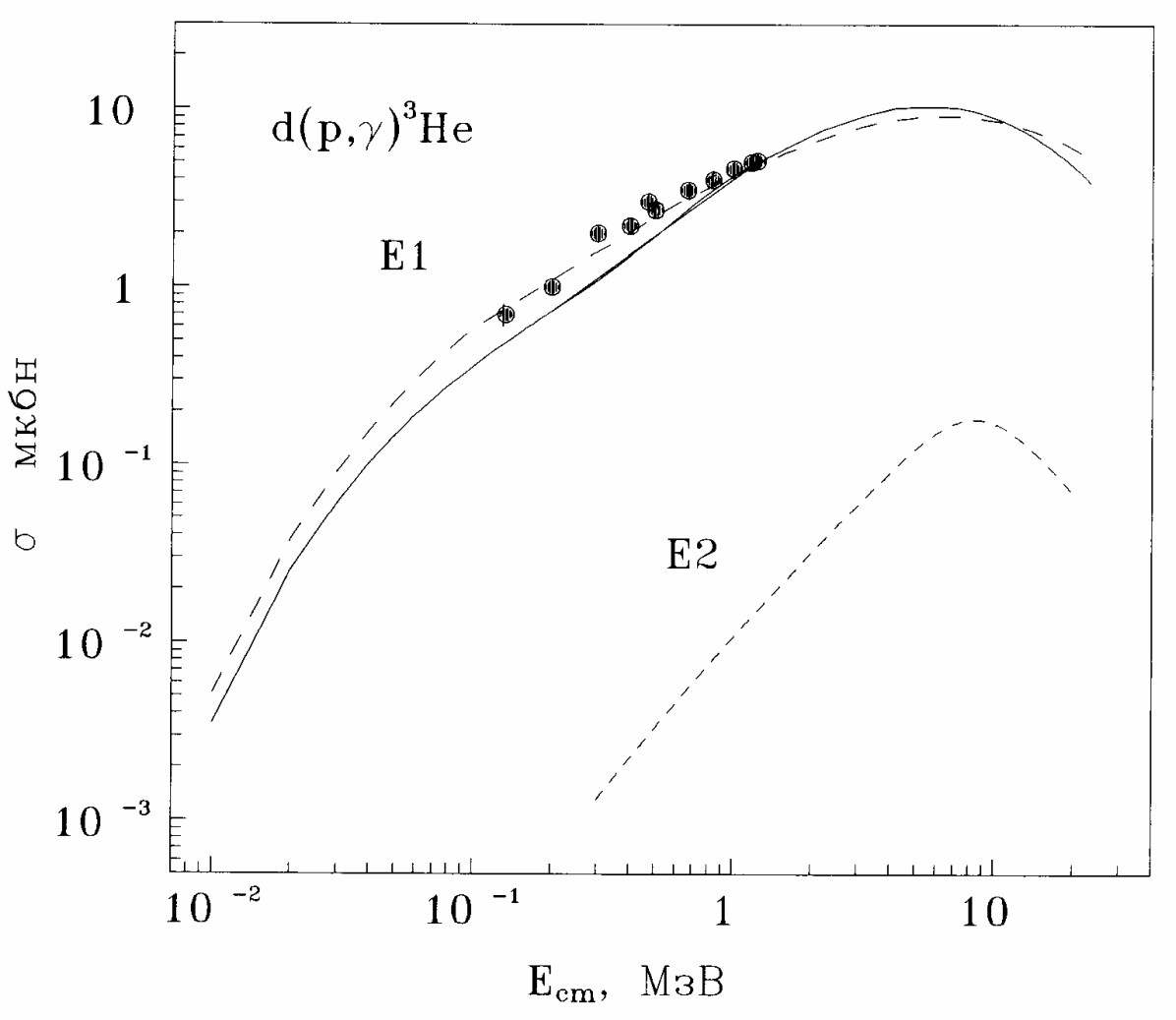

Рис.4.8а. Полные сечения p$^2$H захвата для ядра $^3$He. Кривые - расчеты для потенциалов из табл.3.9, 3.10 [23]. Эксперимент из работы [26].

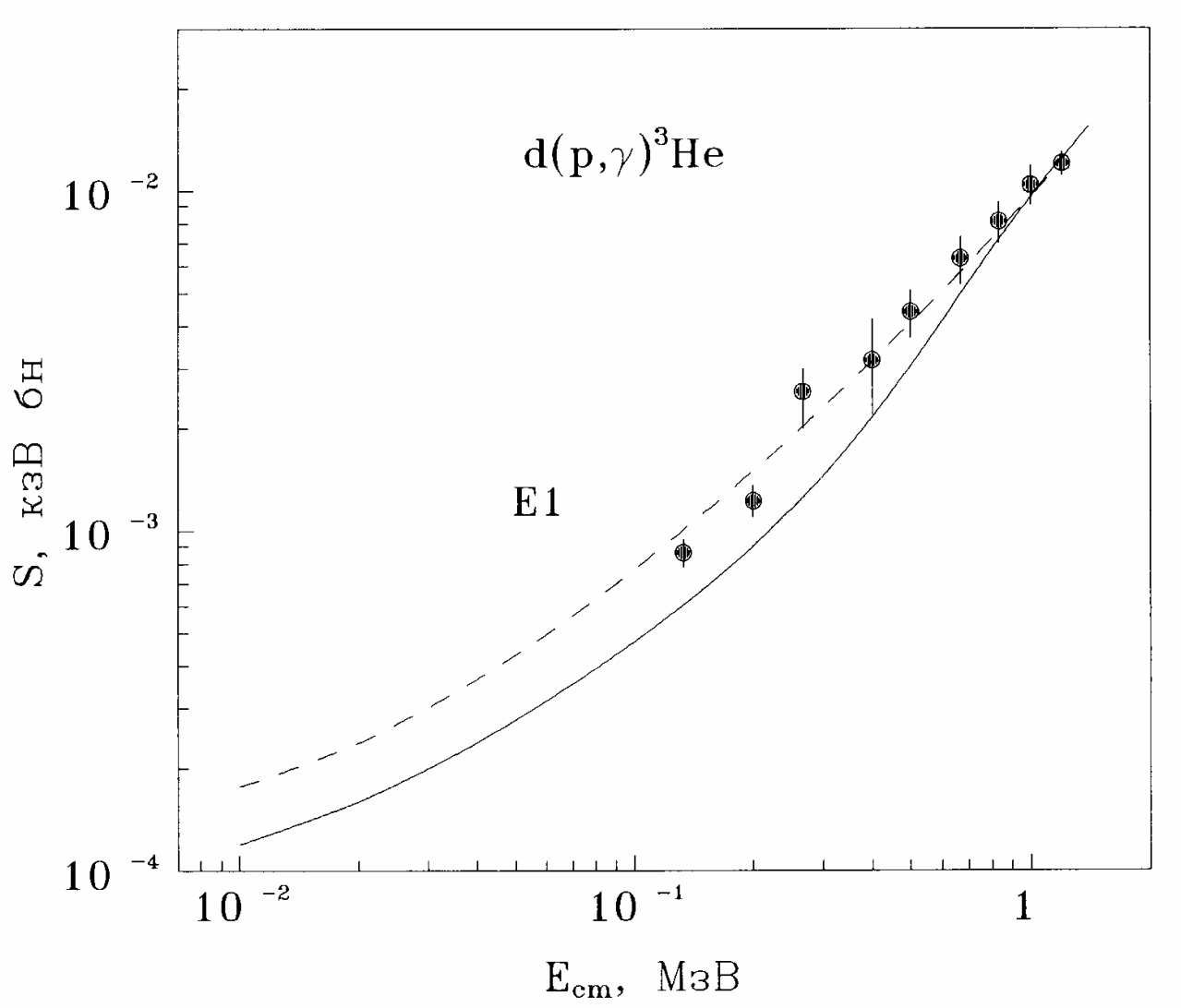

Рис.4.8б. Астрофизический S фактор для p$^2$H захвата. Кривые - расчеты для потенциалов из табл.3.9, 3.10 [23]. Эксперимент из работы [26].

Штриховой линией показаны результаты, полученные с тем же потенциалом основного состояния, но с чисто отталкивающим вариантом взаимодействия в P волне. Использование второго варианта потенциала основного состояния при любом P взаимодействии приводит к сечениям фоторазвала, которые в максимуме достигают только 600 мб, как показано на рис.4.7 точечными линиями. Штриховой линией внизу рис.4.7б приведено сечение E2 процесса.

На рис.4.8а показаны сечения радиационного захвата для $p^2H$ системы и экспериментальные данные, взятые из работ [26]. Точечной линией показан вклад E2 процесса, который на два порядка меньше, чем E1. На рис.4.8б приведен астрофизический S фактор обусловленный E1 переходом. Эксперимент получен пересчетом данных работы [26]. Обозначения кривых на рис.4.8 такие же, как рис.4.7.

Линейная экстраполяция S(E1) фактора к нулевой энергии дает для непрерывной линии величину $1.0(2)10^{-4}$ кэВ бн, для штриховой $1.6(2)\ 10^{-4}$ кэВ бн. Из рисунков видно, что сечения, показанные непрерывной линией, вполне описывают экспериментальные данные в максимуме, но при малых энергиях идут несколько ниже [23].

Существует достаточно много расчетов сечений этих процессов на основе различных модельных подходов. В частности, в некоторых вариантах метода гиперсферических функций удается хорошо передать полные сечения при невысоких энергиях [25]. Однако в таких подходах обычно не рассматривалась супермультиплетная симметрия волновой функции с разделением по схемам Юнга, позволяющая анализировать структуру межкластерных взаимодействий, определять наличие и положение разрешенных и запрещенных состояний, как это было сделано здесь и в работе [23].

### *Кластерный $^2H^3He$ канал*

Величина $P_J^2$ для E1 захвата в $^2H^3He$ системе может быть представлена в виде (4.1.1) с $J_f = 1/2^-$, $3/2^-$ если захват происходит из S волны в $^2P^{\{41\}}$ состояния конечного ядра, которые не стабильны. Энергия $P_{1/2}^{\{41\}}$ уровня около 9 МэВ с шириной порядка 5 МэВ [27] (см. рис.3.18а). При захвате из $^2D$ волны величина $P_J^2$ в два раза больше, чем при S захвате, если не учитывается спин - орбитальное расщепление [23].

На рис.4.9а представлены результаты расчетов сечения E1 радиационного захвата в $^2H^3He$ системе в области энергий 10 кэВ - 40

МэВ. Штрих - пунктирная линия показывает полное суммарное сечение E1 переходов из S и D волн рассеяния на уровень $P_{3/2}$, когда в качестве потенциала рассеяния используется взаимодействие из табл.3.9 для четных волн. Его S фаза показана на рис.3.16б точечной линией и описывает экспериментальные данные. Непрерывной линией приведено E1 сечение для S и D потенциалов рассеяния с параметрами V=-25 МэВ и $\alpha$=0.15 $Фм^{-2}$, приводящими к описанию МРГ S фазы, как показано на рис.3.16б непрерывной линией.

Штриховыми линиями показаны вклады из S и D волн для перехода на уровень $P_{3/2}$. Точечная линия полное сечение перехода на уровень $P_{1/2}$ для потенциалов, полученных из МРГ данных. Величина E2 сечения в области до 10 МэВ примерно на два порядка меньше. Два пика, наблюдаемые в сечениях захвата на оба $^2P$ уровня обусловлены резонансами в $^2S$ и $^2D$ волнах.

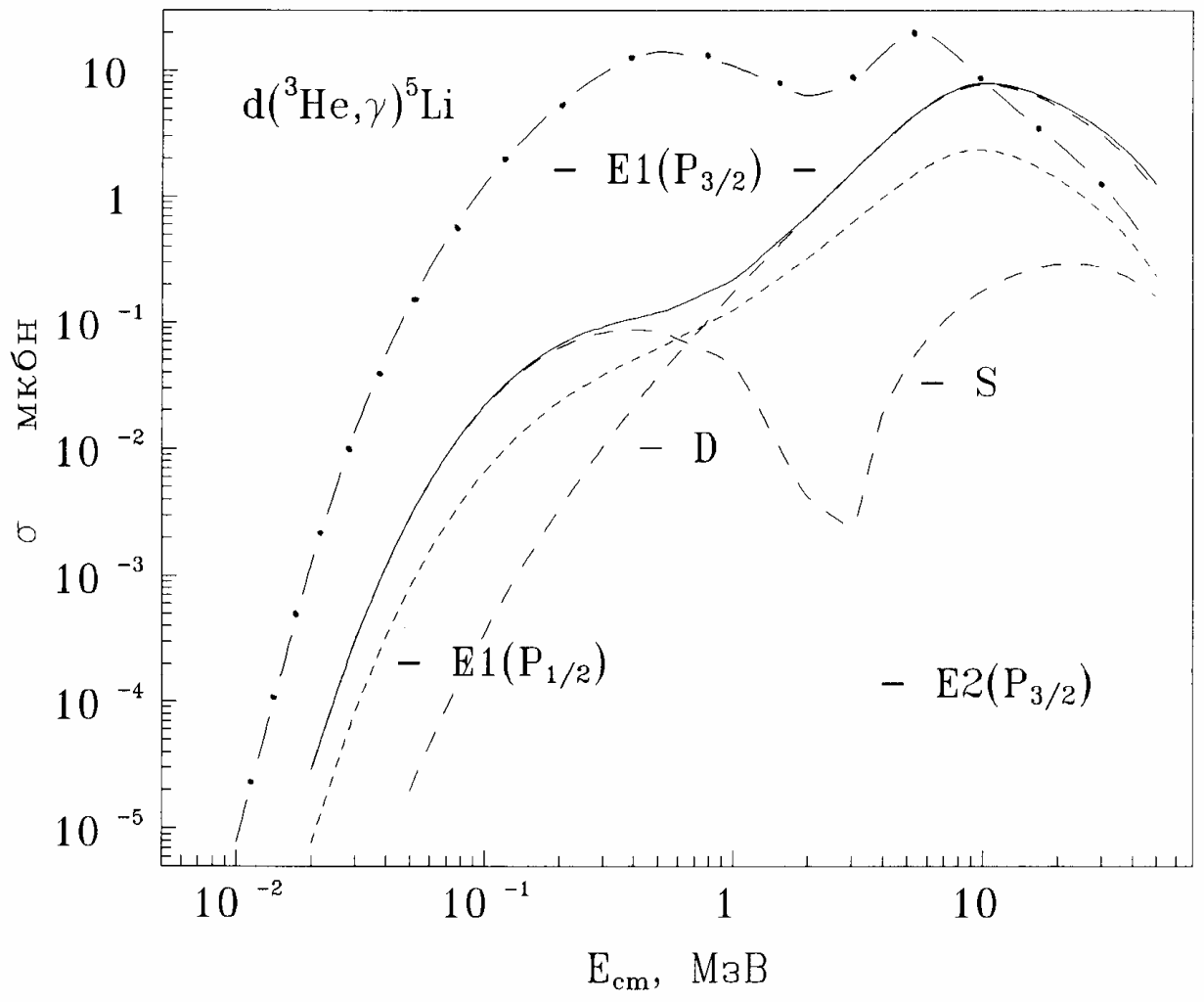

Рис.4.9а. Полные сечения $^3He^2H$ захвата для ядра $^5Li$. Кривые - расчеты для потенциалов из табл.3.9, 3.10 [23].

На рис. 4.9б приведены астрофизические S факторы для перехода на $P_{3/2}$ (непрерывная линия) и $P_{1/2}$ (штриховая линия), обусловленные E1 процессом, для потенциалов, описывающих МРГ результаты по фазам. Линейная экстраполяция S факторов к нулевой энергии дает S(3/2)=3(1) $10^{-3}$ кэВ бн и S(1/2)=7(1) $10^{-4}$ кэВ бн. Для потенциала процессов рассеяния, описывающего экспериментальную S фазу, астрофизический фактор показан штрих - пунктиром и при нулевой энергии равен 1.3(2)$10^{-1}$ кэВ бн [23].

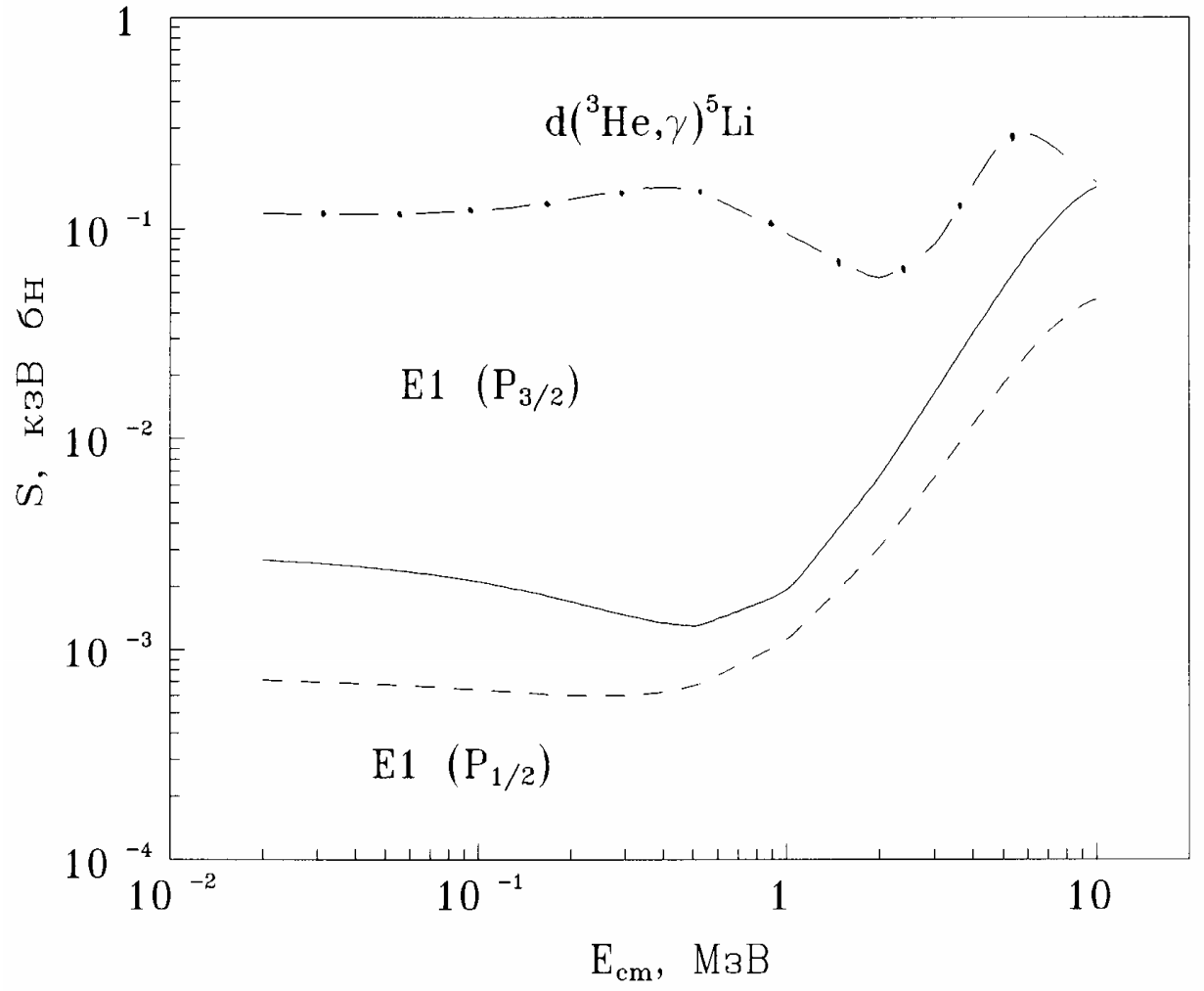

Рис.4.9б. Астрофизический S фактор для $^3He^2H$ захвата. Кривые - расчеты для потенциалов из табл.3.11, 3.12 [23].

Из приведенных результатов видно, что применение одноканальной потенциальной кластерной модели с запрещенными состояниями и разделением по схемам Юнга позволяет воспроизвести имеющиеся экспериментальные данные по полным сечениям фоторазвала и радиационного захвата в $N^2H$ системе при небольших энергиях на основе потенциала, согласованного с чистыми фазами, фазами рассеяния и характеристиками связанных состояний. Оказывается возможным получить вполне конкретные качественные результаты для системы $^2H^3He$.

## 4.3. Фотопроцессы в кластерном $^2H\,^2H$ канале ядра $^4He$

Процессы фоторазвала ядра $^4He$ рассматривались во многих работах (см., например, [28]), однако в них не использовались взаимодействия с запрещенными состояниями и разделение фаз и потенциалов по схемам Юнга. Такие взаимодействия применялись в [22] для расчетов дифференциальных сечений, а полные сечения рассматривались в работе [29]. Поскольку в $^2H^2H$ системе основной вклад дает E2 переход, который возможен между чистым основным S состоянием и синглетной D волной рассеяния, то можно уточнить параметры этих двух взаимодействий. В качестве чистого S потенциала будем использовать набор параметров, приведенный в табл.3.11, который позволяет точно воспроизвести энергии связанных состояний.

Для синглетного потенциала рассмотрим два варианта взаимодействий из табл.3.11, первое из которых описывает S и D фазы рассеяния, как показано на рис.3.19а и не содержит связанного состояния в D волне, а второе описывает D фазу, имеет запрещенное состояние и позволяет получить некий компромисс между разными МРГ S фазами (см. рис.3.19а непрерывная линия).

При расчете полных сечений рассматривался E2 переход, обусловленный только орбитальной частью электрического оператора $Q_{Jm}(L)$. Магнитные сечения и сечения, зависящие от спиновой части электрического оператора оказались сравнительно малыми, а E1 процессы вообще запрещены из - за наличия кластерного множителя, равного нулю в $^2H^2H$ системе. Величина $P_J$ представляется в виде (4.1.1), где $J_f$ может принимать при фоторазвале только одно значение, равное 2.

На рис.4.10а непрерывной линией приведены результаты расчетов полных сечений фоторазвала ядра $^4He$ в $^2H\,^2H$ канал для глубокого синглетного потенциала рассеяния, содержащего связанное состояние в D волне [29]. Эксперимент, полученный из пересчета сечений захвата до 10 МэВ, и из сечений электроразвала при более высоких энергиях приводится в работах [30,31] соответственно (см. рис 4.10а, точки и треугольники). Кружками показаны результаты измерений, выполненных в работе [32]. Видно, что только в серии работ [30] приводятся хорошо согласующиеся между собой данные, которые совпадают с результатами [32] при низких энергиях.

При более высоких энергиях различные измерения оказываются противоречивыми. Так в [31] появляются указания на присутствие второго максимума при 20 МэВ. В то же время в [32] при энергиях более 10 МэВ наблюдается явный спад сечения.

Расчетные сечения с мелким синглетным D взаимодействием оказываются на порядок больше экспериментальных результатов. Любые другие наборы параметров, позволяющие описать D фазу, и не приводящие к связанному состоянию в D волне, дают завышенные относительно эксперимента расчетные сечения. И если исходить из требования описания сечений фоторазвала, то приходится отдать предпочтение глубоким синглетным взаимодействиям рассеяния, имеющим запрещенное состояние в D волне.

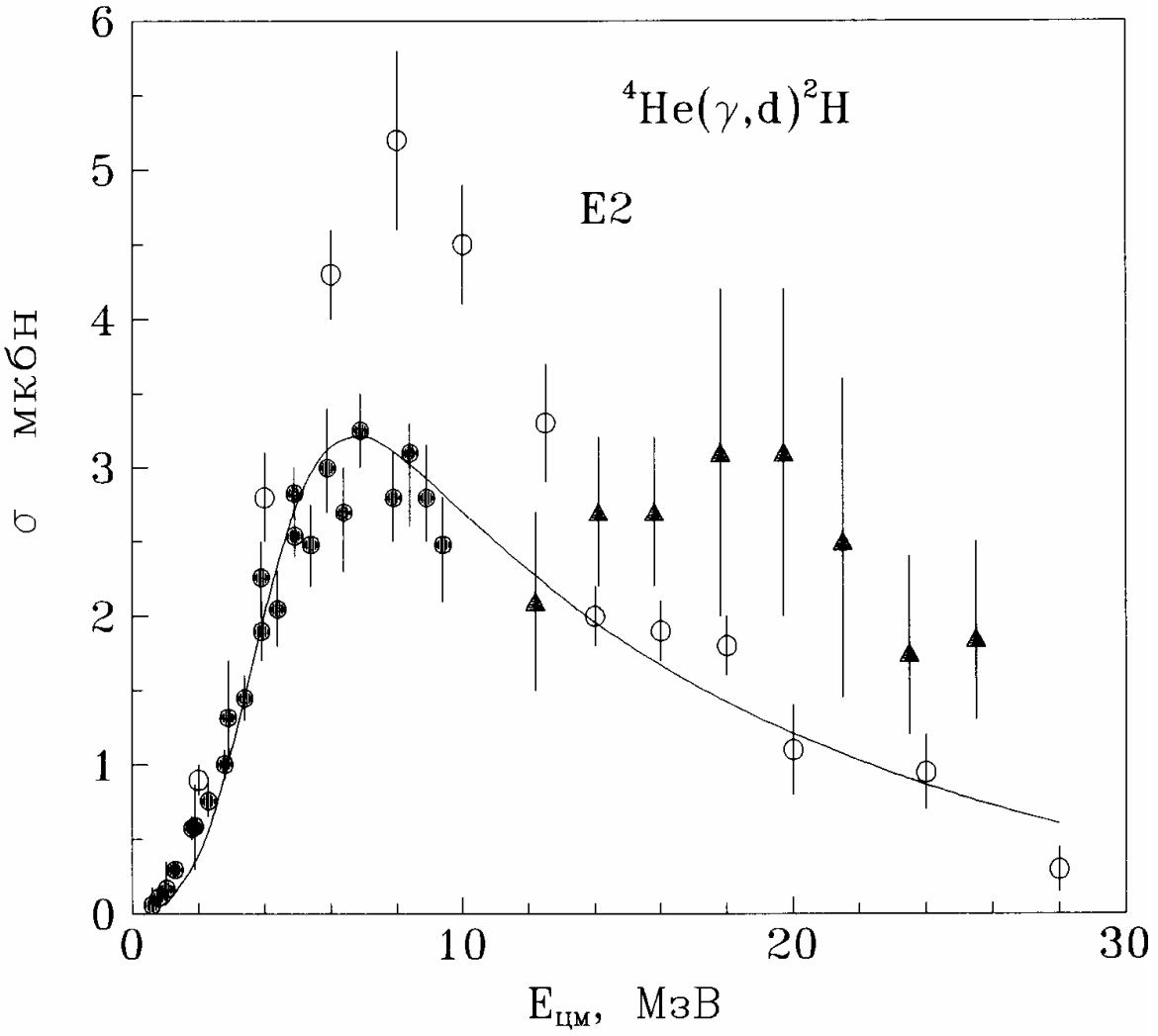

Рис.4.10а. Полные сечения фоторазвала ядра $^4$He в $^2$H$^2$H канал. Кривая - расчеты для глубокого D потенциала из табл.3.11 [29]. Точки, треугольники и кружки - эксперимент из работ [30-32].

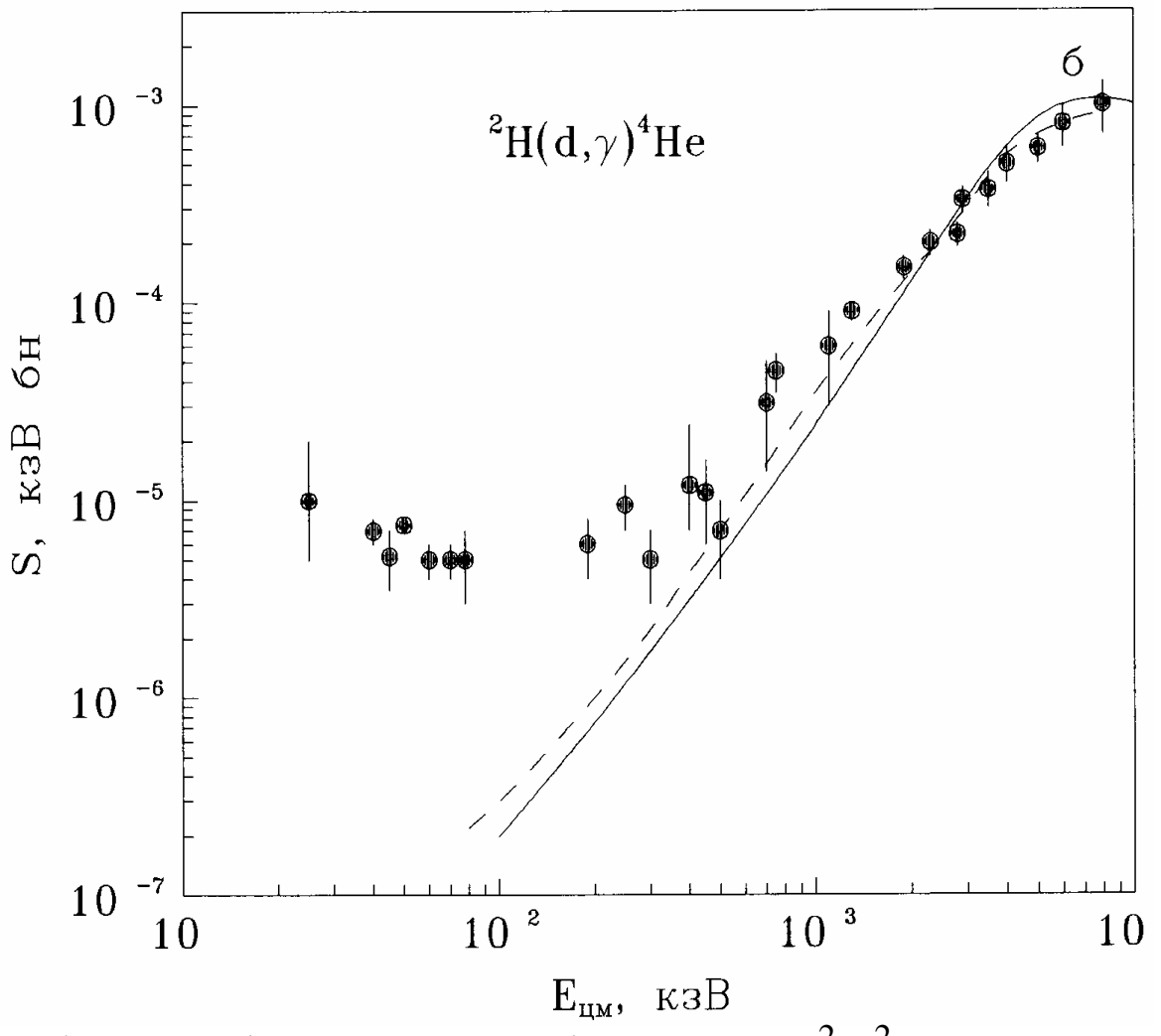

Рис.4.10б. Астрофизический S фактор для $^2$H$^2$H захвата. Непрерывная кривая - расчеты для потенциалов из табл.3.11 [29]. Штриховая кривая - результаты расчетов из работ [28]. Точки - эксперимент из [30].

Заметим, что если для чистого потенциала основного состояния использовать взаимодействие с другими параметрами [29], которые не точно передают энергии двух связанных уровней, то сечение процесса фоторазвала, сохраняя максимум при тех же энергиях, уменьшается в 3-5 раз.

На рис.4.10б показан вычисленный S фактор $^2H^2H$ захвата при энергиях от 100 кэВ до 10 МэВ (непрерывная линия) в сравнении с экспериментальными данными [30] и одним из вариантов расчетов, выполненных в [28] (штриховая линия). Видно, что расчетный S фактор описывает экспериментальные данные при энергиях до 1-2 МэВ, и имеет примерно такую же форму, как в работе [28]. При более низких энергиях сечение E2 переходов типа $^2S \rightarrow {}^2D$ оказывается слишком малым, чтобы объяснить экспериментальные данные. При энергиях ниже 1 МэВ основной вклад дают сечения E2 процесса с переходом из квинтетной S волны в D компоненту волновой функции основного состояния [28], которая отсутствует в используемой здесь модели.

Из сказанного видно, что кластерная модель с расщеплением по схемам Юнга позволяет описать определенные экспериментальные данные по полным сечениям фоторазвала и радиационного захвата в $^2H^2H$ канале при небольших энергиях. Астрофизический S фактор вполне согласуется с результатами других расчетов при не очень малых энергиях.

Оказывается возможным избавиться от неоднозначностей при выборе S потенциала основного состояния ядра и согласовать его параметры с чистыми фазами. Уточняется так же форма и структура синглетного D взаимодействия, позволяющего одновременно воспроизвести фазы упругого рассеяния и сечения фотопроцессов.

## 4.4. Фотопроцессы в кластерных $p^3H$ и $n^3He$ каналах ядра $^4He$

Расчеты дифференциальных сечений для $p^3H$ системы для потенциалов с запрещенными состояниями и разделением по схемам Юнга выполнялись в работах [22]. Полные сечения для таких же взаимодействий рассматривались нами в [33]. При E1 процессах возможны переходы между основным чистым с T=0 состоянием и синглетной P волной рассеяния. Если считать, что основной вклад в сечения дают процессы с изменением изоспина $\Delta T=1$ [34], то необходимо использовать P потенциал рассеяния из чистого по изоспину с T=1 синглетного состояния $p^3He$ системы.

В случае переходов без изменения изоспина с $\Delta T=0$ будем ис-пользовать потенциал чистой синглетной Р фазы с Т=0 для системы $p^3H$. В работе [33], для основного состояния с Т=0 приводятся три варианта взаимодействия, однако, здесь мы будем использовать только одно из них, параметры которого приведены в табл.3.13.

Параметры Р взаимодействия $p^3He$ приведены в табл.3.8, а фазы рассеяния показаны на рис.3.11 и 3.12. На рис.4.11а штриховой линией приведены результаты расчета полных сечений фоторазвала $^4He$ в $p^3H$ канал при переходах с $\Delta T=1$.

Эксперимент взят из работ [35,36] - кружки, [37] - треугольники и [38] - точки. Видно, что разброс различных экспериментальных данных достигает 20-30%, причем более поздние измерения [37,38] лежат заметно ниже результатов, полученных ранее в работах [35, 36].

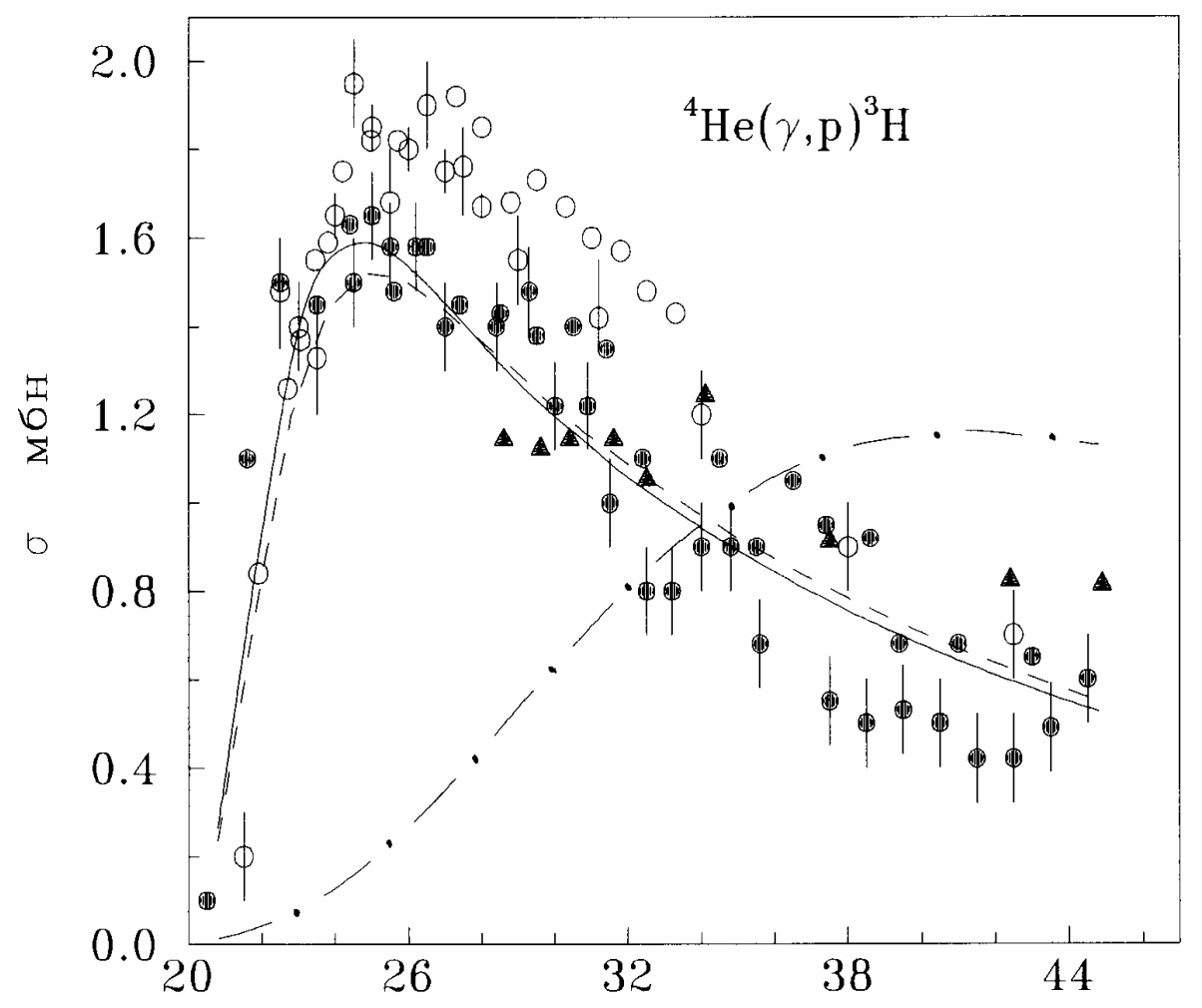

Рис.4.11а. Полные сечения фоторазвала ядра $^4He$ в $p^3H$ канал. Кривые - расчеты для потенциалов из табл.3.8, 3.13. Точки, треугольники и кружки - эксперимент из работ [35-40].

Поскольку Р потенциал выбирался, как некий компромисс между различными фазовыми анализами, его величина определена не однозначно. Поэтому можно несколько увеличить его глубину и принять $V_0$=-15 МэВ и $\alpha_0$=0.1 Фм$^{-2}$. Тогда в максимуме сечение увеличивается примерно на 0.1 мб. и несколько лучше согласуется с экспериментальными данными [37,38], как показано на рис.4.11а непрерывной линией. Фазы этого Р потенциала практически не отличаются, от показанных на рис.3.12а непрерывной линией.

Для рассмотрения переходов без изменения изоспина с ΔT=0 необходимо использовать отталкивающий Р волновой потенциал из $p^3H$ системы, приведенный в табл.3.12 (см. рис.3.23а). В этом случае, вообще не удается правильно передать даже форму экспериментальных сечений, как показано на рис.4.11а штрих - пунктирной линией.

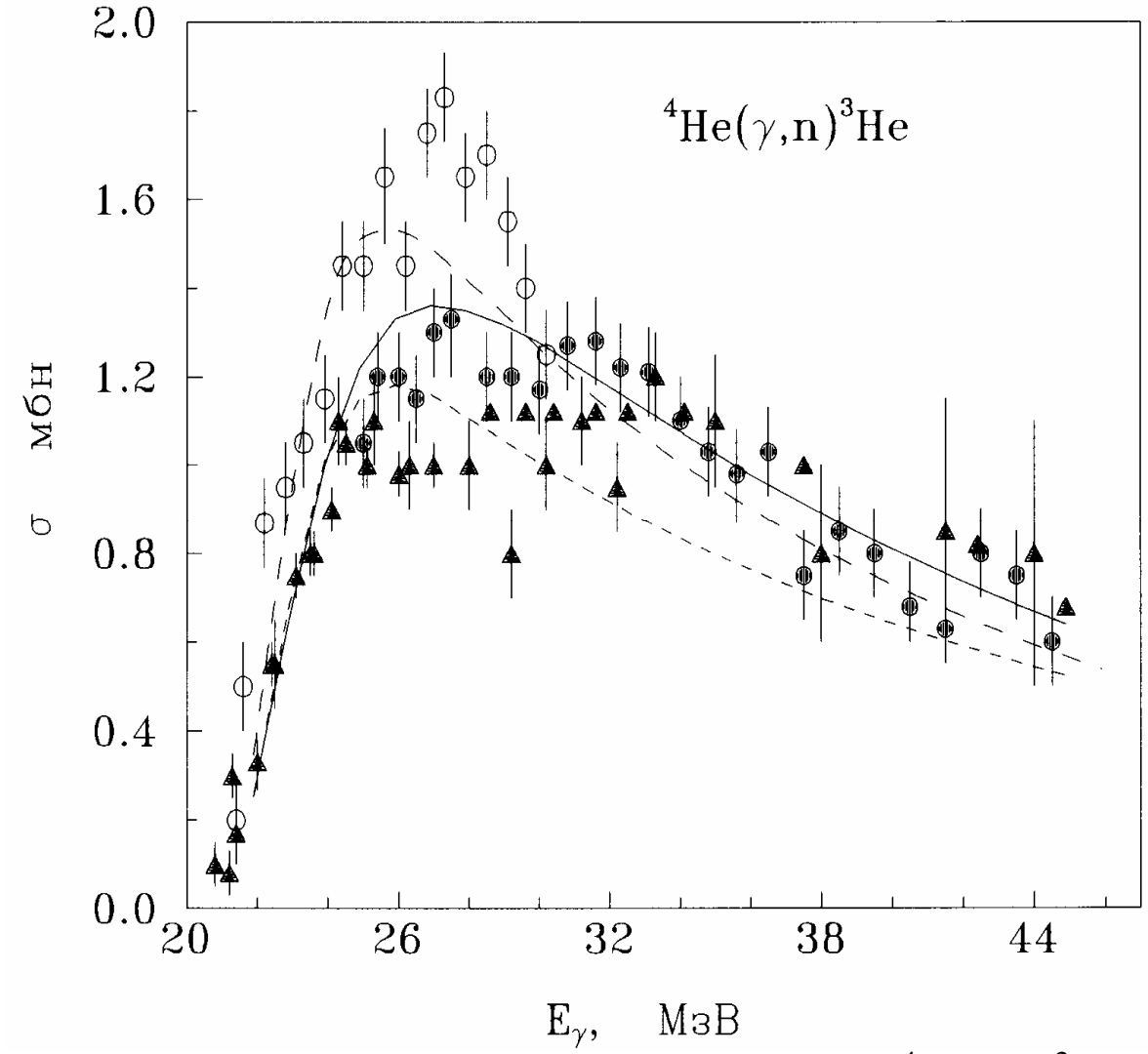

Рис.4.11б. Полные сечения фоторазвала ядра $^4$He в $n^3$He канал. Кривые - расчеты для потенциалов из табл.3.8, 3.13. Точки, треугольники и кружки - эксперимент из работ [38-40].

На рис.4.11б штриховой линией показаны результаты расчетов для реакции $^4$He(γ,n)$^3$He с ΔT=1 для Р взаимодействия из табл.3.8 вместе с экспериментальными данными работ [38] - точки, [39] - треугольники и [40] - кружки. И здесь более ранние измерения сечений [40] лежат заметно выше последних данных [38,39]. Если несколько уменьшить глубину в Р волне до -11 МэВ при той же геометрии, то удается описать данные работ [38,39]. Это сечение показано на рис.4.11б непрерывной линией. Фазы такого Р потенциала приведены на рис.3.12а точечной линией и согласуются с данными фазового анализа из работы [41].

Существующие расчеты сечений этих фотореакций в различных моделях и подходах (см., например, [35-40]), в общем, приводят к согласию с имеющимися экспериментальными данными. Однако, экспериментальные неоднозначности столь велики, что трудно отдать предпочтение тем или иным результатам. Неоднозначность фазовых анализов позволяет несколько менять параметры потенциалов и получать результаты, согласующиеся с теми или иными экспериментальными данными по сечениям фоторазвала.

На рис.4.12а даны расчетные сечения радиационного $p^3H$ захвата для ΔT=1, полученные с P волновым взаимодействием рассеяния при глубине -15 МэВ, которые приводит к описанию процесса фоторазвала. Экспериментальные данные работ [35-40]. Для этого же потенциала рассматривался и астрофизический S фактор при малых энергиях, приведенный на рис.4.12б.

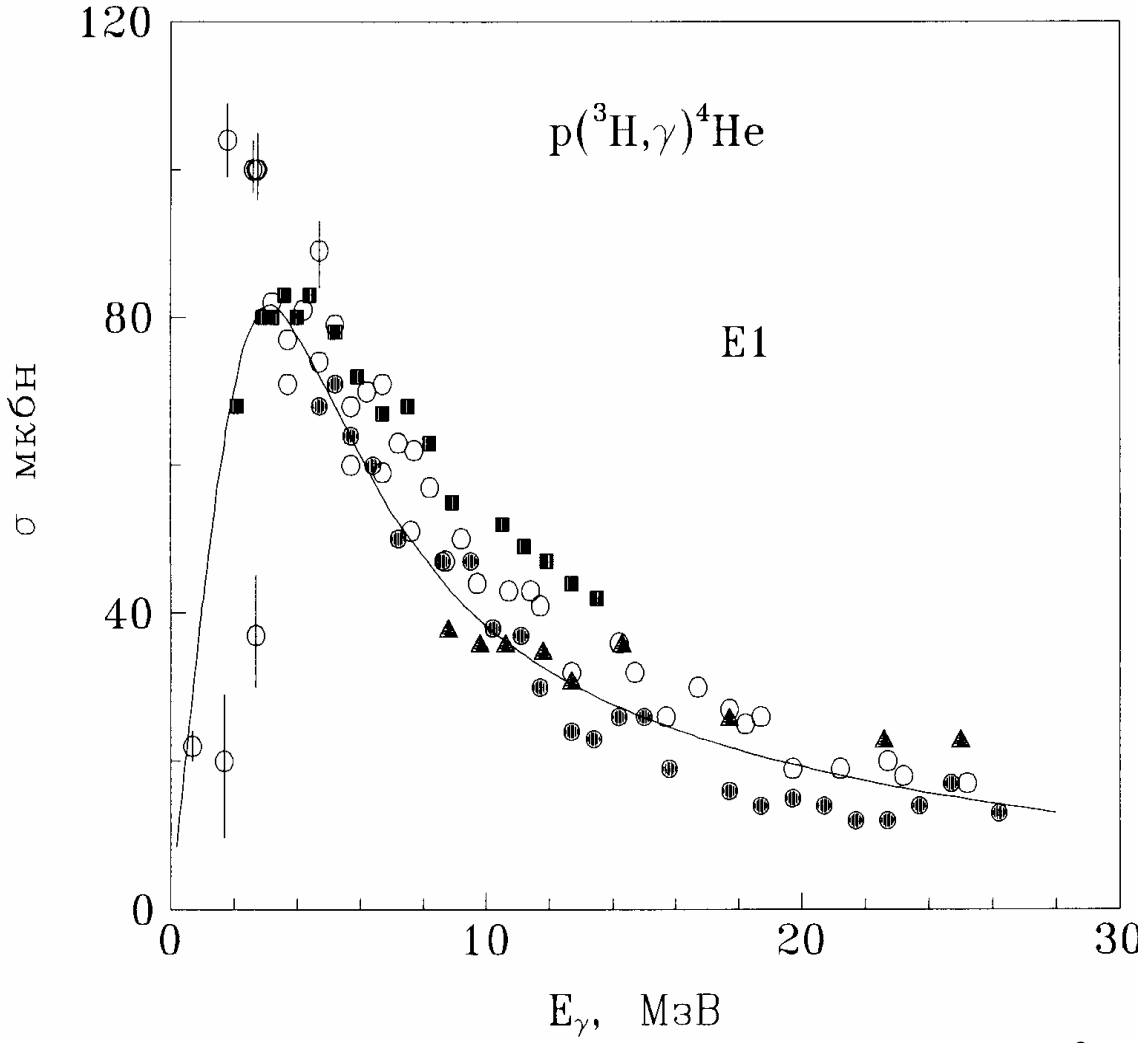

Рис.4.12а. Полные сечения радиационного захвата в $p^3H$ канал. Непрерывная кривая - расчеты для потенциалов из табл.3.8, 3.13. Точки, треугольники, квадраты и кружки - эксперимент [35-40].

Видно, что, не смотря на большие экспериментальные ошибки, в целом удается передать измеренные сечения [35-38] и при низких энергиях 0.7-3 МэВ. Линейная экстраполяция S фактора к нулевой энергии дает величину около $1.3(6)10^{-3}$ кэВ б.

Для тех же потенциалов рассматривался и E2 процесс фоторазвала ядра $^4He$ в $p^3H$ канал. Результаты расчетов полного сечения приведены на рис. 4.12в вместе с экспериментом [35-38]. Обозначения кривых такие же, как на рис. 4.11а.

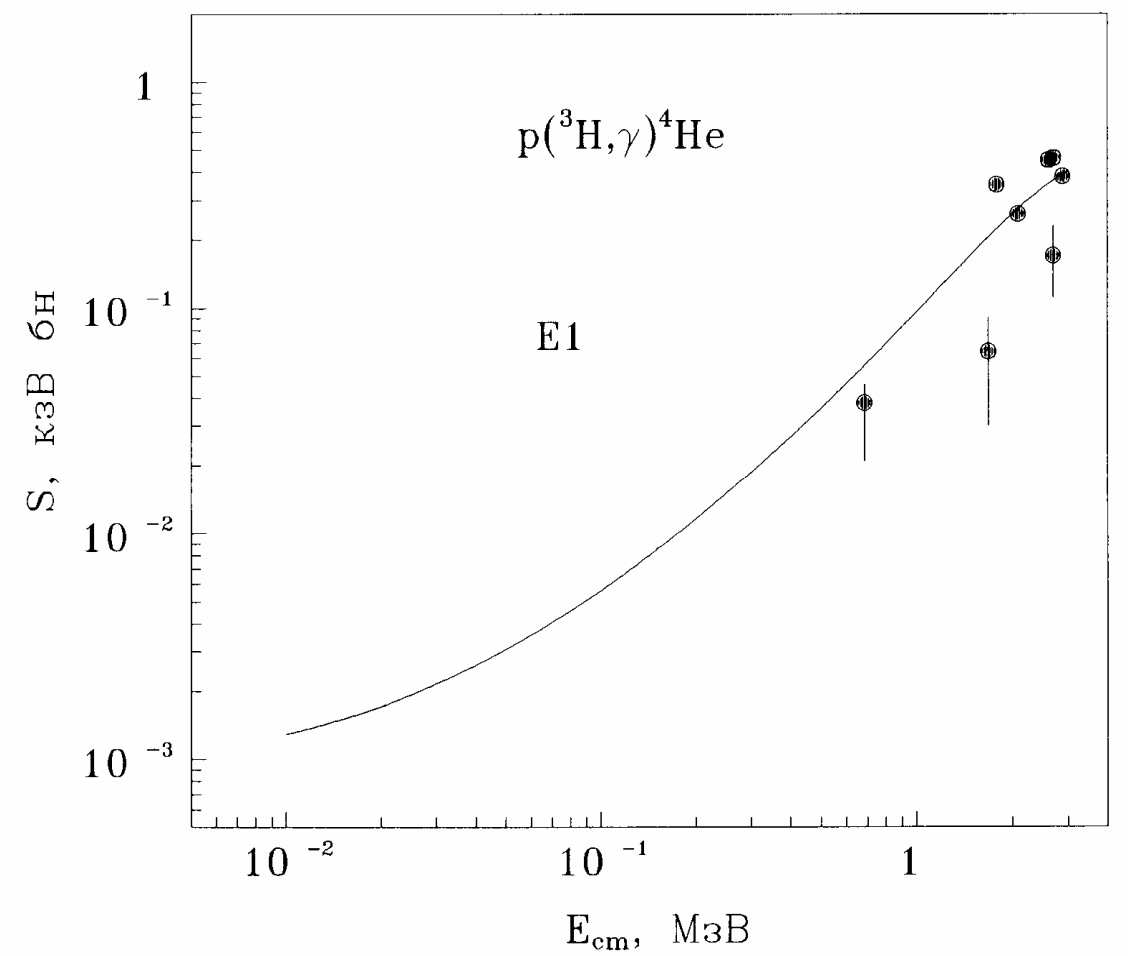

Рис.4.12б. Астрофизический S фактор для $p^3H$ захвата. Непрерывная кривая - расчеты для потенциалов из табл.3.8, 3.13. Экспериментальные данные из работ [35-40].

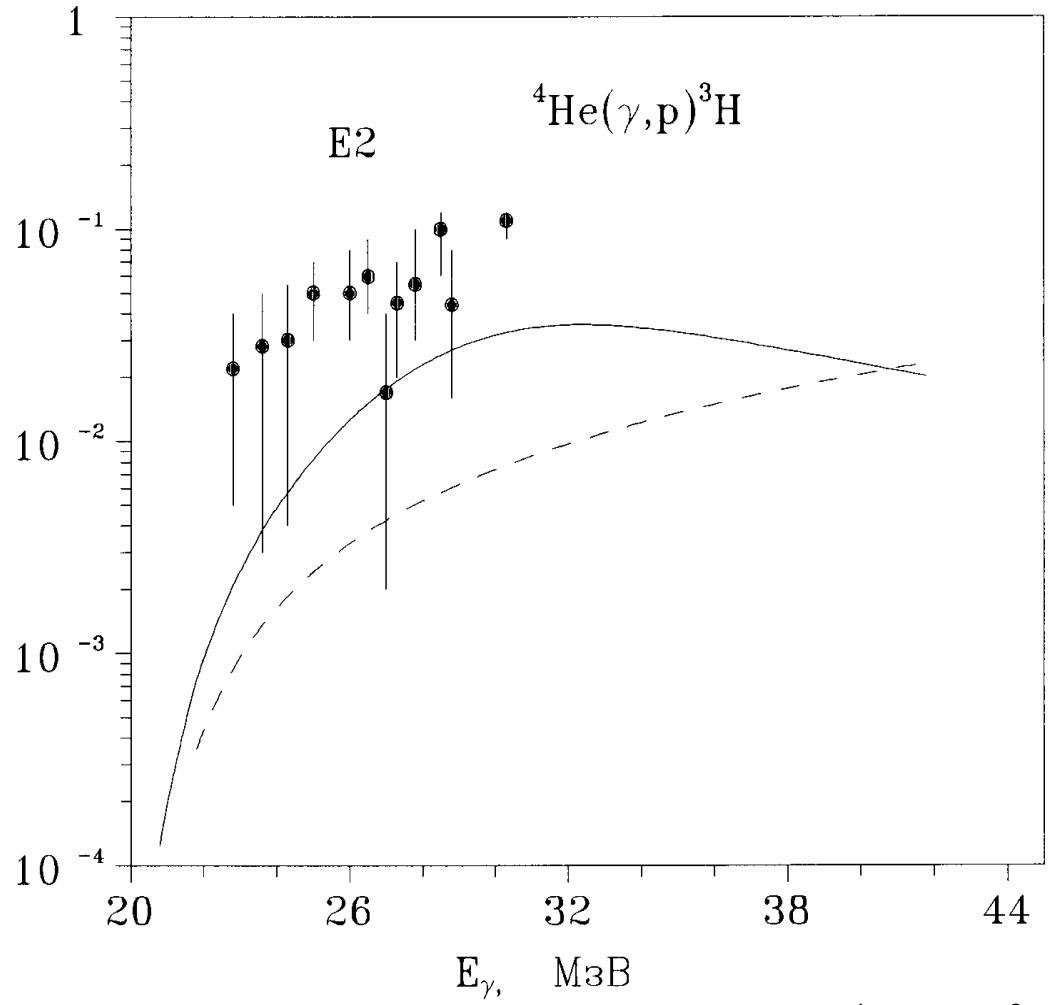

Рис.4.12в. Полные сечения фоторазвала ядра $^4He$ в $p^3H$ канал для E2 процесса. Кривые - расчеты для потенциалов из табл.3.8, 3.13. Экспериментальные данные из работ [35-40].

Тем самым, потенциальная кластерная модель вполне позволяет передать форму сечений фоторазвала $^4He$ при использовании E1

переходов с ΔT=1 для рассмотренных реакций в $p^3H$ и $n^3He$ каналах. Величина расчетных сечений находится в пределах неоднозначностей различных экспериментальных результатов. Варьируя глубину P взаимодействия в пределах 10-20% можно несколько улучшить согласие расчетов с теми или иными экспериментальными данными по фотосечениям. Фазы измененных таким образом P потенциалов находятся в интервале экспериментальных неоднозначностей различных фазовых анализов.

---

1 . Langanke K. - Nucl. Phys., 1986, v.A457, p.351.

2. Rao K.S., Sridhar R., Susila S. - Phys. Scr., 1981, v.24, p.925; Rao K.S., Sridhar R. - Phys. Scr., 1978, v.17, p.557; Liu Q., Kanada H., Tang Y.C. - Z. Phys., 1981, v.A303, p.253; Bouten M., Bouten M.C. - J. Phys. G., 1982, v.8, p.1641.

3. Liu Q.K.K., Kanada H., Tang Y.C. - Phys. Rev., 1981, v.C23, p.645; Walliser H., Kanada H., Tang Y.C. - Nucl. Phys., 1984, v.A419, p.133; Phys. Rev., 1983, v.C28, p.57.

4. Kajino T. - Nucl. Phys., 1986, v.A460, p.559.

5. Burkova N.A. et al. - Phys. Lett., 1990, v.B24, p.15.

6. Дубовиченко С.Б., Джазаиров-Кахраманов А.В. - ЯФ, 1995, т.58, № 4, с.635; ЯФ, 1995, т.58, № 5, с.852; ЭЧАЯ, 1997, т.28, №6, с.1529.

7. Parker P.D., Kavanagh R.W. - Phys. Rev., 1963, v.131, p.2582; Nagatani K. et al. - Nucl. Phys., 1969, v.A128, p.325; Osborn J.L., Barnes C.A., Kavanagh R.W., Kremer R.M., Mathews G.J., Zyskind J.L., Parker P.D., Howard A.J. - Phys. Rev. Lett., 1982, v.42, p.1664; Nucl. Phys., 1984, v.A419, p.115; Robertson R. et al. - Phys. Rev., 1983, v.C27, p.11; Krawinkel H. et al. - Z. Phys., 1982, v.304, p.307.

8. Griffiths G., Morrow R.A., Riley P.J., Warren J.B. - Can. J. Phys., 1961, v.39, p.1397.

9. Mertelmeir T., Hofmann H.M. - Nucl. Phys., 1986, v.A459, p.387.

10. Fowler W. et al. - Ann. Rev. Astr. Astrophys., 1975, v.13, p.69.

11. Alexander T. et al. - Nucl. Phys., 1984, v.A427, p.526.

12. Buck B., Baldock R.A., Rubio J.A. - J. Phys., 1985, v.11G, p.L11.

13. Buck B., Merchant A.C. - J. Phys., 1988, v.14G, p.L211.

14. Buck B., Friedrich H., Wheatley C. - Nucl. Phys., 1976, v.A275, p.246; Buck B. - In: Int. 4th Conf. on Clust. Aspects of Nucl. Struct. and Nucl. React., Chester, 1984.

15. Robertson R.G.H., Dyer P., Warner R.A., Melin R.C., Bowles T.J., Mc Donald A.B., Ball G.C., Davies W.G., Earle E.D. - Phys. Lett., 1981, v.47, p.1867.

16. Taneichi H., Ueno H., Shoda K., Kawazoe Y., Tsukamoto T. - Nucl. Phys., 1986, v.A448, p.315; Skopik D.M., Tomusiak E.L., Dressier E.T., Shin Y.M., Murphy J.J. - Phys. Rev., 1976, v.C14, p.789; Berman B.L. et al. - Phys. Rev. Lett.,1965, v.15, p.727.

17. Жусупов М.А., Кужевский Б.М., Маханов Б.Б. - Изв. АН КазССР, сер. физ.-мат., 1991, т.2, с.30.

18. Young A.M., Blatt S.L., Seyler R.G. - Phys. Rev. Lett., 1970, v.25, p.1764; Blatt S.L., Young A.M., Ling S.C., Moon K.J., Porterfield C.D. - Phys. Rev., 1968, v.176, p.1147.

19. Nusslin F., Werner H., Zimmerer J. - Z. Naturf., 1966, v.A21, p.1195.

20. Murakami A. - Nuovo Cim., 1968, v.B60, p.604; J. Phys. Soc. Jap., 1970, v.28, p.191; Bazhanov E.B., Komar A.P., Kulikov A.V., Makhnovsky E.D. - Nucl. Phys., 1965, v.65, p.191; Sherman N.K., Baglin J.E.E., Owens R.O. - Phys. Rev., 1968, v.169, p.771; Phys. Rev. Lett., 1966, v.17, p.31.

21. Mondragon A., Hernandez E. - Phys. Rev., 1990, v.C41, p.1975.

22. Neudatchin V.G., Kukulin V.I., Pomerantsev V.N., Sakharuk A.A. - Phys. Rev., 1992, v.C45. p.1512; Неудачин В.Г., Сахарук А.А., Смирнов Ю.Ф. - ЭЧАЯ, 1993, т.23, с.480.

23. Дубовиченко С.Б. - ЯФ, 1995, т.58, № 7, с.1253.

24. Fetisov V.N., Gorbunov A.N., Varfolomeev A.T. - Nucl. Phys., 1965, v.71, p.305; Stewart J.R., Morrison R.C., O'Connell J.S. - Phys. Rev., 1965, v.138, p.B372; Kundu S.K., Shin Y.M., Wait G.D. - Nucl. Phys., 1971, v.A171, p.384; Berman B.L., Koester Jr., Smith J.H. - Phys. Rev., 1964, v.133, p.B117.

25. Faul D.D., Berman B.L., Meyer P., Olson D.L. - Phys. Rev., 1981, v.C24, p.849; Skopik D.M., Beck D.H., Asai J., Murphy J.J. - Phys. Rev., 1981, v.C24, p.1791.

26. Griffiths G.M., Larson E.A., Robertson L.P. - Can. J. Phys., 1962, v.40, p.402.

27. Ajzenberg-Selove F. - Nucl. Phys., 1979, v.A320, p.1.

28. Weller H.R., Lehman D.R. - Ann. Rev. Nucl. Part., 1988, v.38, p.563; Bluge G., Assenbaum H.J., Langanke K. - Phys. Rev., 1987, v.C36, p.21; Assenbaum H.J., Langanke K. - Phys. Rev., 1987, v.C36, p.17; Wachter B., Mertelmeir T., Hofmann H.M. - Phys. Lett., 1988, v.B200, p.246; Phys. Rev., 1988, v.C38, p.1139.

29. Дубовиченко С.Б. - ЯФ, 1995, т.58, № 11, с.1973.

30. Zurmuhle R.W., Stephens W.E., Staub H.H. - Phys. Rev., 1963, v.132, p.751; Meyerhof W.E., Feldman W., Gilbert S., O'Connell W. - Nucl. Phys., 1969, v.A131, p.489; Wilkinson F.J., Cecil F.E. - Phys. Rev., 1985, v.C31, p.2036.

31. Skopik D.M., Dodge W.R. - Phys. Rev., 1972, v.C6, p.43.

32. Аркатов Ю.М., Вацет П.И., Волощук В.И., Гурьев В.Н., Золенко В.А., Прохорец И.М. - УФЖ, 1978, т.23, с.918.

33. Дубовиченко С.Б. - ЯФ, 1995, т.58, № 8, с.1377.

34. Gibson B.F. - Nucl. Phys., 1981, v.A353, p.85.

35. Meyerhof W.E., Suffert M., Feldman W. - Nucl. Phys., 1970, v.A148, p.211; Arkatov Yu. M. et al. - Sov. J. Nucl. Phys., 1970, v.10, p.639; 1971, v.12, p.123; 1976, v.21, p.475.

36. Gemmel D.S., Jones G.A. - Nucl. Phys., 1962, v.33, p.102; Gorbunov A. - Phys. Lett., 1968, v.B27, p.436.

37. Bernabei R. et al. - Phys. Rev., 1988, v.C38, p.1990.

38. Balestra F., Bollini E., Busso L., Garfagnini R., Guaraldo G., Piragino G., Scrimaglio R., Zanini A. - Nuovo Cim., 1977, v.A38, p.145.

39. Calarco J.R., Berman B.L., Donnelly T.W. - Phys. Rev., 1983, v.C27, p.1866; Berman B.L., Faul D.D., Meyer P., Olson D.L. - Phys. Rev., 1980, v.C22, p.2273.

40. Irish J.D., Johnson R.G., Berman B.L., Thomas B.J., Mc Neill K.G., Jury J.W. - Can. J. Phys., 1975, v.53, p.802.

41. Tombrello T. - Phys. Rev., 1965, v.138, p.40B.

# 5. ХАРАКТЕРИСТИКИ ТЯЖЕЛЫХ КЛАСТЕРНЫХ СИСТЕМ

Как видно из табл.2.1, практически все тяжелые кластерные системы типа $N^6Li$, $N^7Li$ и $^2H^6Li$ оказываются смешанными по орбитальным схемам Юнга. Исключение составляет только $^4He^4He$ канал ядра $^8Be$. Поэтому здесь применимы все методы разделения фаз и потенциалов, которые использовались ранее для легчайших кластеров в состояниях с минимальным спином. Перейдем к рассмотрению результатов, полученных в потенциальной кластерной модели для тяжелых кластерных систем с учетом смешивания орбитальных состояний.

## 5.1. Потенциалы упругого $N^6Li$ рассеяния

При построении эффективных потенциалов $p^6Li$ и $n^6Li$ взаимодействий, как и раньше, естественно опираться на экспериментальные данные по фазам упругого рассеяния и параметрам неупругости, получаемым в результате фазового анализа. К сожалению, экспериментальные исследования фазовых сдвигов $p^6Li$ рассеяния выполнены лишь в узком интервале энергий до 6 МэВ [1]. Поэтому наряду с экспериментальными, приходится использовать теоретические МРГ фазы, приведенные в [2].

Для системы $n^6Li$ экспериментальные данные по фазам вообще отсутствуют. В тоже время, в работах [3,4] приведены $n^6Li$ фазовые сдвиги, найденные в одноканальном МРГ. Дальнейшее развитие эти результаты получили в работах [5,6]. Так, в работе [5] рассмотрены три кластерных канала $n^6Li$, $n^6Li^*$ (TS=10) и $^3H^4He$, а в [6] к ним добавлен четвертый канал $^2H^5He$. Именно эти фазовые сдвиги и будут являться исходными данными для построения $n^6Li$ потенциалов.

Восстановление $p^6Li$ взаимодействия по имеющимся экспериментальным и МРГ фазам приводит к параметрам гауссового потенциала вида (3.1.1) с точечным кулоновским членом, которые могут быть представлены следующим образом [7]

$$V_0=V_1 - (-1)^L V_2 + (\mathbf{sl})V_3 , \quad \alpha=\alpha_1 - (-1)^L\alpha_2 + (sl)\alpha_3 . \qquad (5.1.1)$$

где (**sl**) - спин - орбитальный оператор и (sl) - его среднее значение. В различных спиновых каналах, без учета расщепления по схемам Юнга, т.е. для смешанных в дублетном канале потенциалов, в работе [7] было получено

$$
\begin{array}{llll}
S=1/2: & V_1 = -162.87 \text{ МэВ}, & V_2 = -122.87 \text{ МэВ}, & V_3 = -219.93 \text{ МэВ}, \\
& \alpha_1 = 0.3 \text{ Фм}^{-2}, & \alpha_2 = 0.1 \text{ Фм}^{-2}, & \alpha_3 = 0.2 \text{ Фм}^{-2}, \qquad (5.1.2) \\
S=3/2: & V_1 = -330.23 \text{ МэВ}, & V_2 = -270.29 \text{ МэВ}, & V_3 = -114.36 \text{ МэВ}, \\
& \alpha_1 = 0.46 \text{ Фм}^{-2}, & \alpha_2 = 0.26 \text{ Фм}^{-2}, & \alpha_3 = 0.13 \text{ Фм}^{-2}.
\end{array}
$$

На рис.5.1а-в представлены результаты вычислений $p^6Li$ фаз рассеяния с этими параметрами потенциалов. На рис.5.1а и 5.1б штриховыми линиями приведены средние $^2P$ и $^4P$ фазовые сдвиги, вычисленные на основе (5.1.2), но с выключенным спин - орбитальным взаимодействием, т.е. при условии $V_3=0$. Фазовые сдвиги $^2D$ и $^4D$ также получены без учета спин - орбитальной части взаимодействия с S волновыми потенциалами (5.1.2) при L=2 и $V_3=0$.

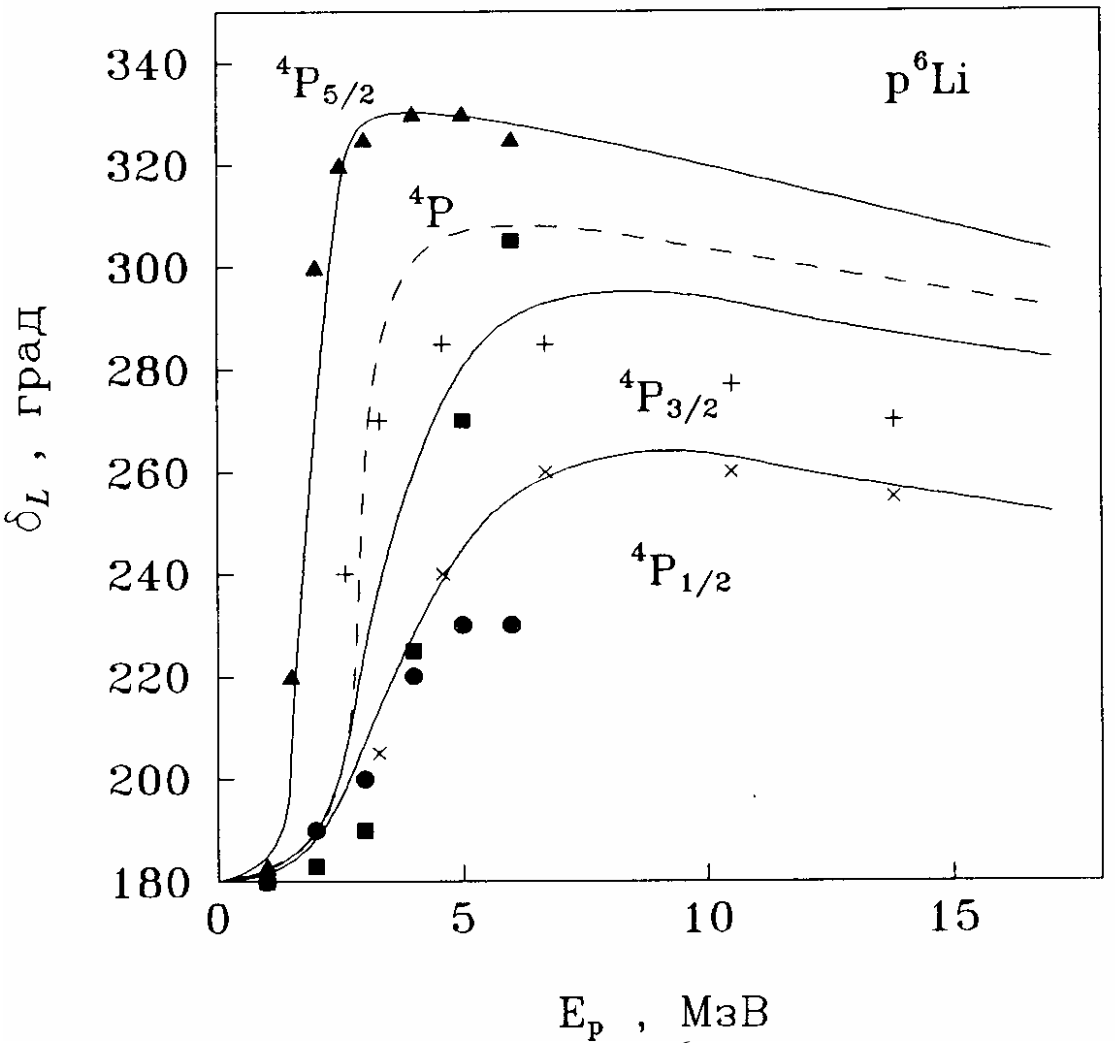

Рис.5.1а. Фазовые сдвиги упругого $p^6Li$ рассеяния в квартетных спиновых каналах для L=1 (чистая симметрия {f}={421}). Точки, треугольники и квадраты - экспериментальные данные [1], крестики - МРГ результаты, взятые из работы [2].

В работе [7] считается, что основному состоянию $^6Li$ соответствует орбитальная схема {42}. В результате волновая функция основного чистого P состояния оказывается безузловой, а запрещенный уровень имеется только в S волне. Исходя из результатов [7], будем считать, что в дублетных каналах $p(n)^6Li$ рассеяния реализуются две допустимые пространственные симметрии $\{f_1\}$ = {421} и $\{f_2\}$ = {43}, совместимые с нечетным орбитальным моментом.

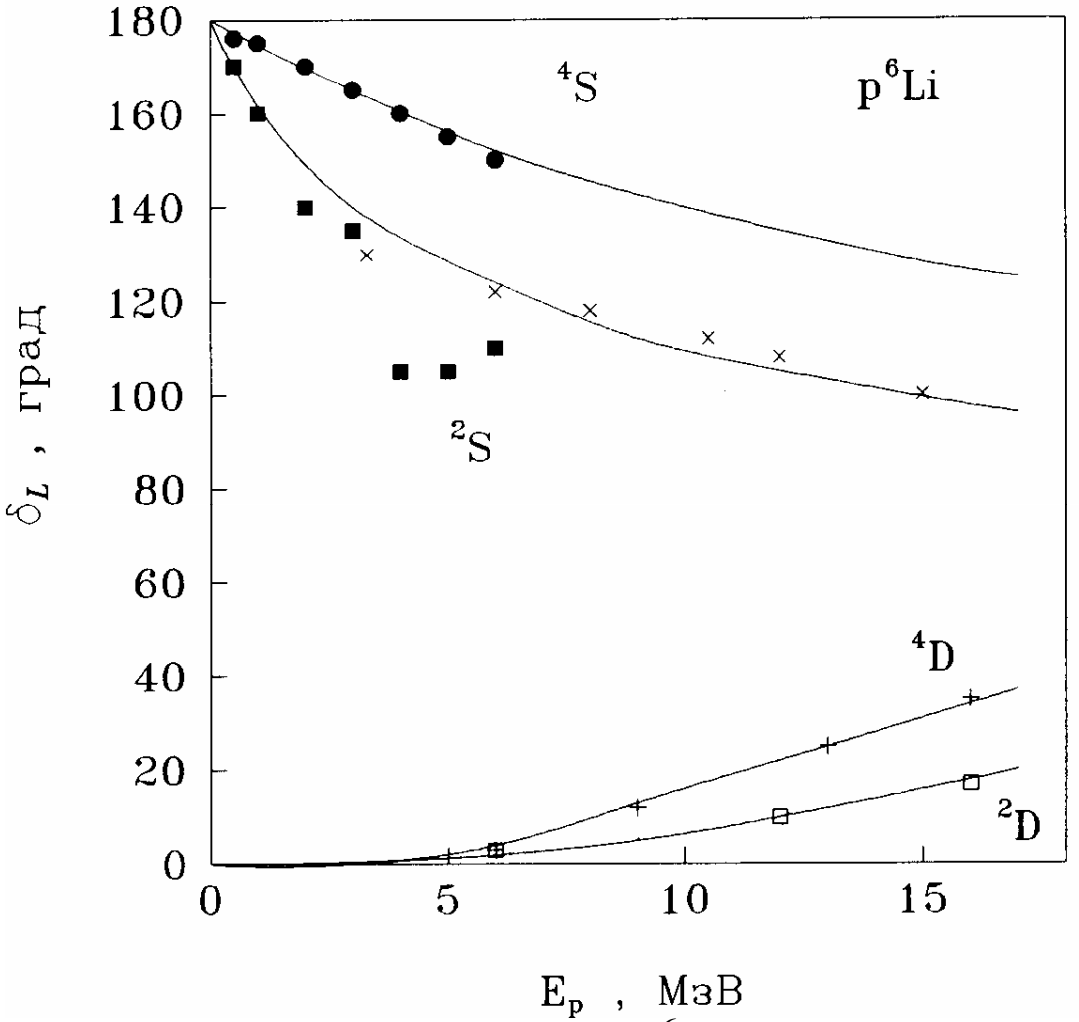

Рис.5.1б. Фазовые сдвиги упругого $p^6Li$ рассеяния в квартетных и дублетных спиновых каналах без спин - орбитального расщепления, отвечающие схеме Юнга {f}={421}. Точки и квадраты - экспериментальные данные [1], крестики и открытые квадраты - МРГ результаты, взятые из работы [2].

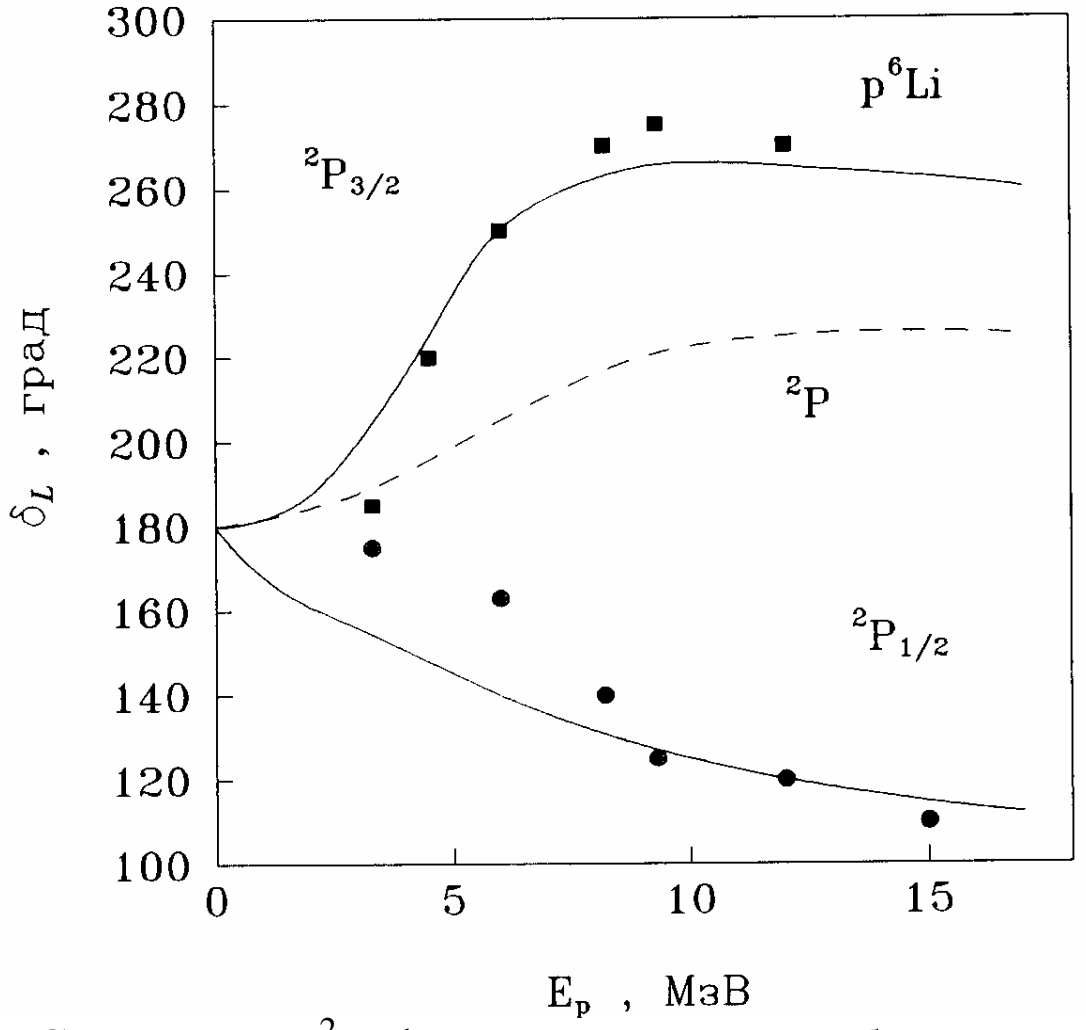

Рис.5.1в. Смешанные ${}^2P_J$ фазовые сдвиги в дублетных спиновых каналах для упругого $p^6Li$ рассеяния. Точки и квадраты - МРГ данные, взятые из работы [2].

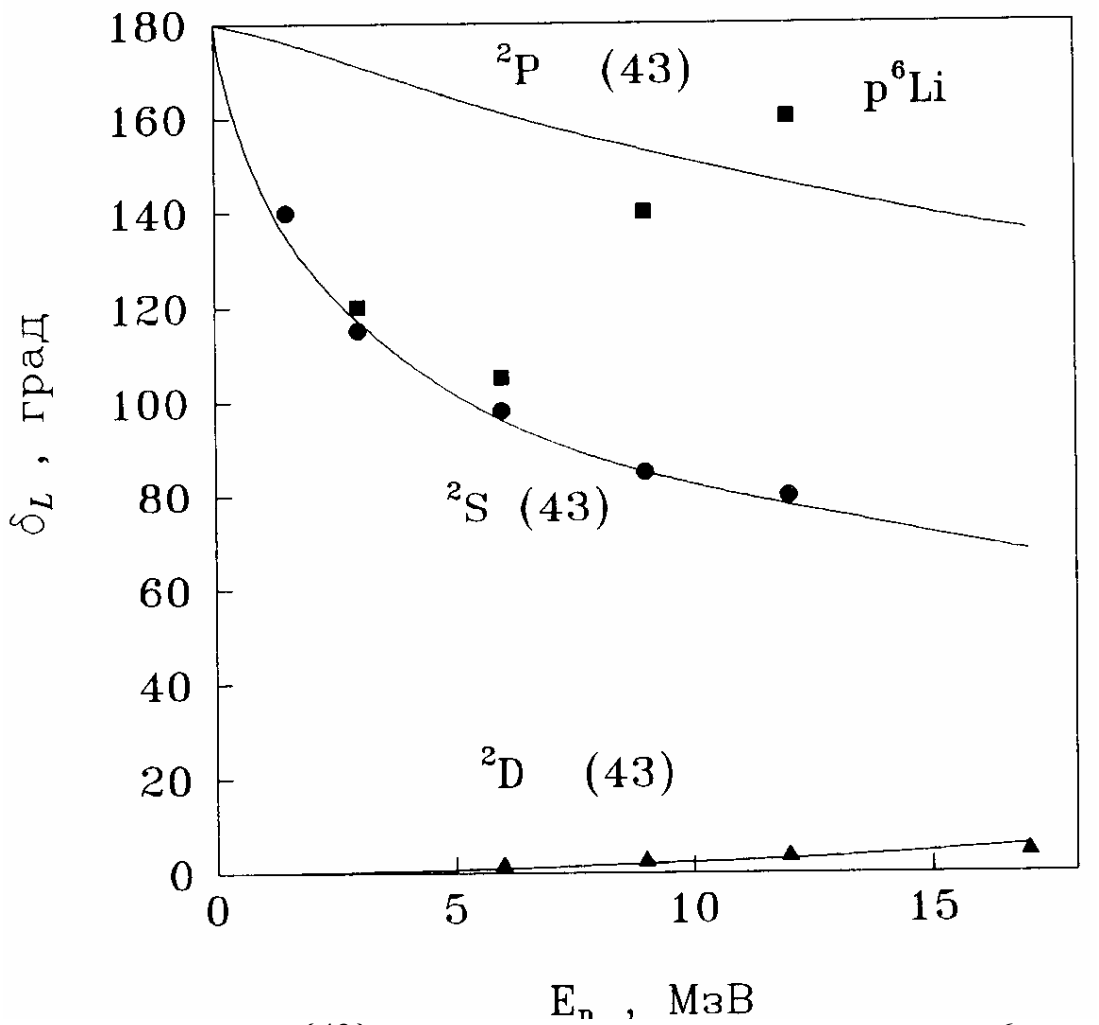

Рис.5.1г. Чистые $\delta_L^{\{43\}}$ фазовые сдвиги упругого $p^6$Li рассеяния. Точки, квадраты и треугольники, выделенные из экспериментальных и МРГ данных [1,2], чистые S, P и D фазы соответственно.

Поэтому для описания характеристик ядра $^7$Be необходимо определить чистый потенциал $p^6$Li взаимодействия, отвечающий схеме {f}={43}. На основе (3.3.1) из фазовых сдвигов $\delta_{L,S=3/2} = \delta_{\{421\}}$ и $\delta_{L,S=1/2} = \delta_{эксп.,S=1/2}$ можно найти чистые фазы $\delta^{\{43\}}$ с S=1/2, а значит и потенциалы взаимодействия. Чистая P фаза определяется из средних экспериментальных $^2$P и $^4$P фазовых сдвигов.

Потенциал, отвечающий фазовому сдвигу $\delta_0^{\{43\}}$, по полученным точкам для чистой фазы строится вполне однозначно и имеет запрещенное состояние со схемой {52}, а его параметры при одногауссойдной параметризации могут быть записаны в следующем виде

$$V_s=-33{,}16 \text{ МэВ}, \alpha_s=0{,}3 \text{ Фм}^{-2}. \qquad (5.1.3)$$

Получить чистый по орбитальным симметриям P волновой потенциал с {f}={43} гораздо сложнее, ввиду возможной сильной неупругости, делающей при $\eta_{L,S=1/2} << 1$ фазовый анализ существенно неустойчивым. Поэтому потенциал строился таким образом, чтобы в первую очередь описать канальную энергию - энергию связи ядра

$^7$Be, как системы p$^6$Li и его среднеквадратичный радиус. Тогда параметры чистого P$^{\{43\}}$ потенциала при одногауссойдной параметризации представляются в виде

$$V_p=-99{,}5\ \text{МэВ},\ \alpha_p=0{,}25\ \text{Фм}^{-2}. \qquad (5.1.4)$$

Результаты расчета фаз показаны на рис. 5.1г, вместе с извлеченными из эксперимента и МРГ результатов, чистыми фазами (точки, треугольники, квадраты) [7]. Удается хорошо передать S и D чистые фазы, а для правильного воспроизведения P волны нужен, по - видимому, потенциал с периферическим отталкиванием.

Восстановление потенциалов n$^6$Li взаимодействия по МРГ фазовым сдвигам [3-6] дает следующие параметры для потенциалов без спин - орбитальной части и разделения по схемам Юнга [7]

$$\begin{array}{lll} S=1/2 & V_1 = -114{,}7\ \text{МэВ}, & V_2 = -83{,}7\ \text{МэВ}, \\ & \alpha_1 = 0{,}225\ \text{Фм}^{-2}, & \alpha_2 = 0{,}075\ \text{Фм}^{-2}, \\ S=3/2 & V_1 = -222{,}7\ \text{МэВ}, & V_2 = -173{,}7\ \text{МэВ}, \\ & \alpha_1 = 0{,}35\ \text{Фм}^{-2}, & \alpha_2 = 0{,}15\ \text{Фм}^{-2}. \end{array} \qquad (5.1.5)$$

На рис.5.2а,б представлены фазовые сдвиги, отвечающие этим потенциалам, а точками, треугольниками и квадратами показаны исходные МРГ фазовые сдвиги из [3-6].

Отметим, что $^2$P МРГ фаза с этой системе не имеет резонансного характера (точки на рис.5.2б), как в случае p$^6$Li рассеяния, средняя $^2$P фаза которого показана на рис.5.1б. Поэтому для получения чистой $\delta_1^{\{43\}}$ фазы n$^6$Li системы будем использовать параметры p$^6$Li взаимодействия, учитывая изменение потенциала, вызванное различием кулоновских сил [7]. Отсюда и из $^4$P фазового сдвига (рис.5.2а) был получен чистый фазовый сдвиг $\delta_1^{\{43\}}$, аналогично тому, как это делалось при рассмотрении p$^6$Li рассеяния.

Параметры дублетного n$^6$Li взаимодействия, отвечающего {f} = {43}, имеют вид [7]

$$\begin{array}{ll} V_1 = -58{,}75\ \text{МэВ}, & V_2 = -40{,}75\ \text{МэВ}, \\ \alpha_1 = 0{,}715\ \text{Фм}^{-2}, & \alpha_2 = 0{,}075\ \text{Фм}^{-2}. \end{array} \qquad (5.1.6)$$

Результаты расчета чистых фаз рассеяния показаны на рис.5.2в непрерывными линиями, а сами фазы точками и квадратами. Из рисунков видно, что определенные таким образом потенциалы взаимодействия вполне приемлемо передают энергетический ход фаз рассеяния [7]. Некоторые характеристики ядер $^7$Li и $^7$Be в каналах p$^6$Li и n$^6$Li, рассчитанные на основе таких взаимодействий [7], приведены в табл.5.1 вместе с экспериментальными данными работ [1-6].

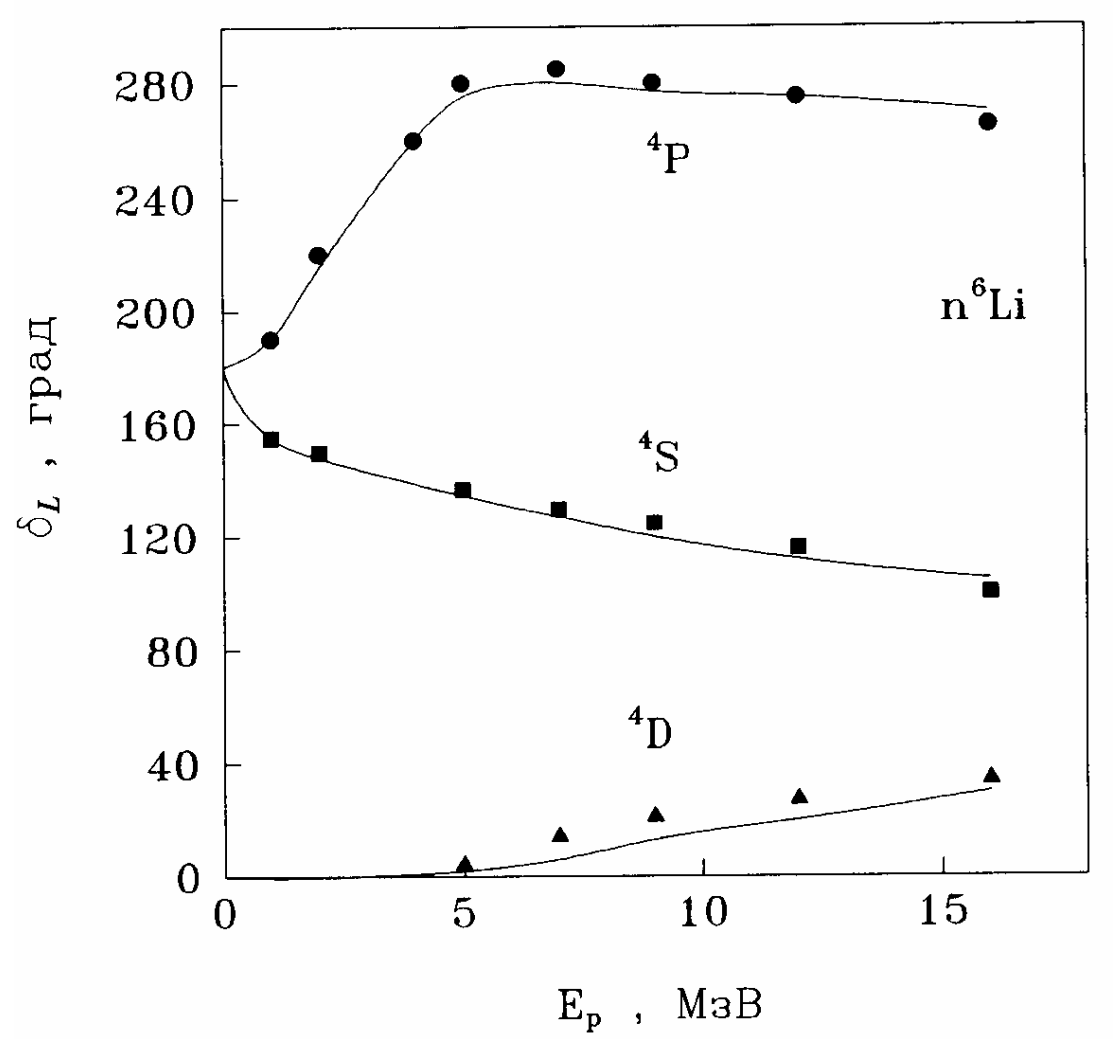

Рис.5.2а. Фазовые сдвиги упругого n$^6$Li рассеяния в квартетных спиновых каналах, отвечающие схеме Юнга {f}={421}. Точки, треугольники и квадраты - МРГ фазовые сдвиги, взятые из работ [3-6].

*Таблица 5.1. Характеристики ядра $^7$Li в кластерном n$^6$Li канале, полученные для чистых по орбитальным симметриям потенциалов [7], и эксперимент [1-6].*

| | $E_0$,(МэВ) | $R_r$, (Фм) | $R_f$, (Фм) | Q, (мб) | B(E2), ($e^2$ Фм$^4$) | $C_0$, $^7$Li | $C_0$, $^7$Be |
|---|---|---|---|---|---|---|---|
| Расчет | -7.254 | 2.51 | 2.31 | -0.154 | 0.0122 | 2.1(1) | 1.6 (1) |
| Эксп. | -7.467 | 2.405 | | -37(8) | 7.4(1) | 2-3 | --- |

Несмотря на то, что радиус и энергия связи передаются сравнительно хорошо, другие характеристики в канале n$^6$Li корректно описать не удается. Упругий кулоновский формфактор ядра $^7$Li при переданном импульсе выше 2 Фм$^{-2}$ резко спадает и не согласуется с экспериментальными данными [8]. В определенной мере, эти результаты могут служить подтверждением малой вероятности кластеризации основного состояния ядра в рассматриваемом канале [7].

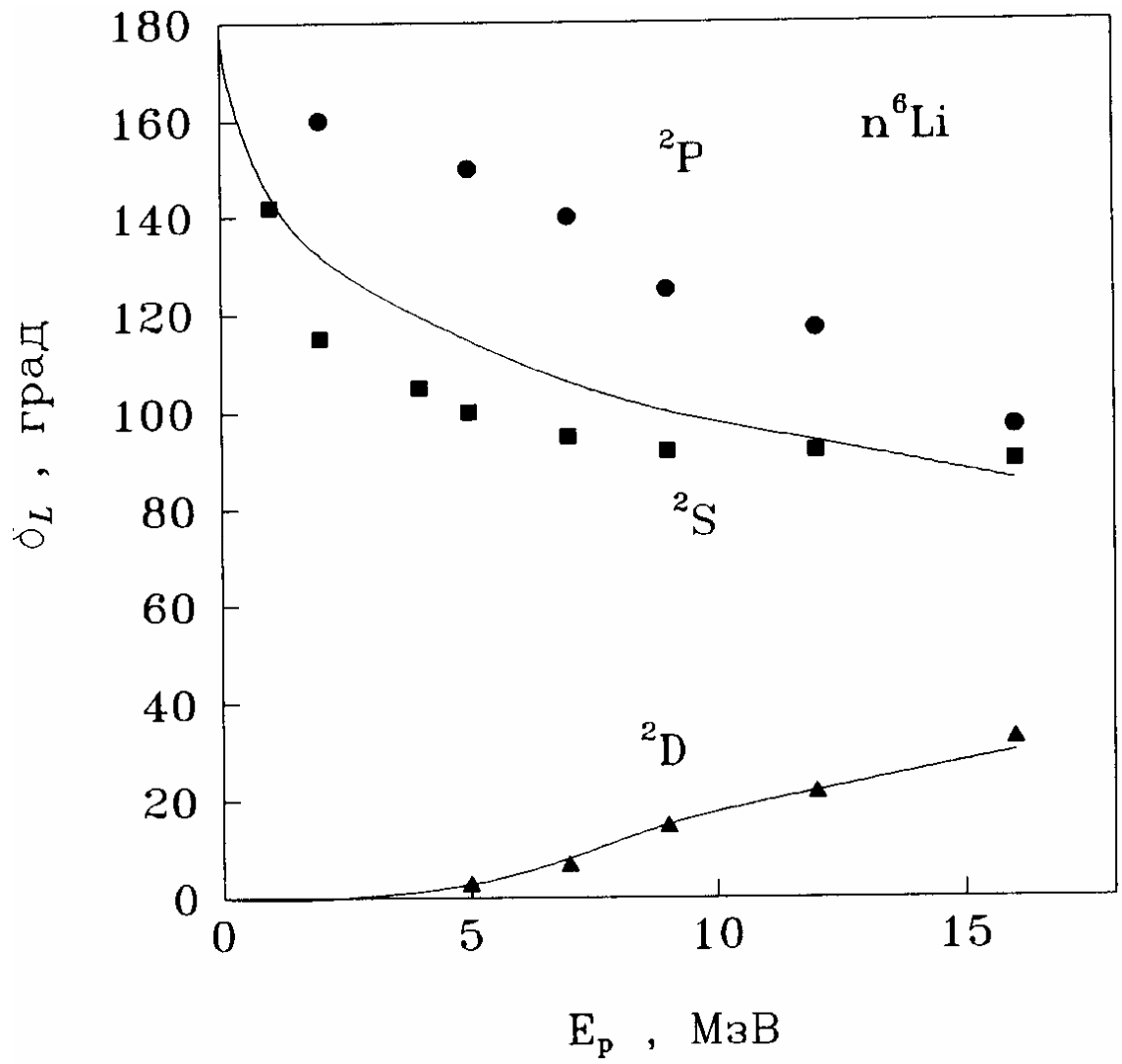

Рис.5.2б. Смешанные фазовые сдвиги упругого n$^6$Li рассеяния в дублетном спиновом канале. Точки, треугольники и квадраты - МРГ фазовые сдвиги, взятые из работ [3-6].

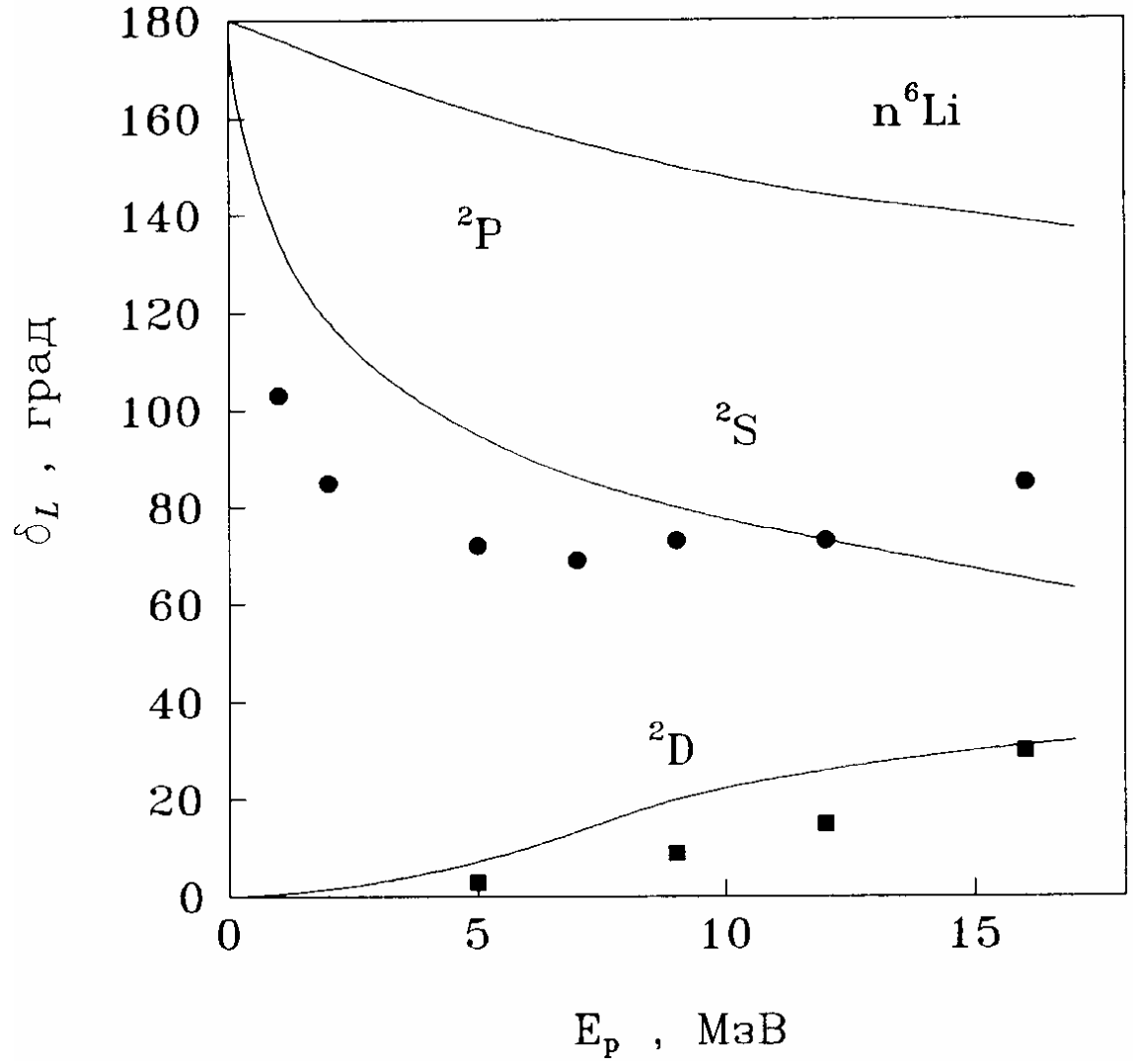

Рис.5.2в. Чистые $\delta_L^{\{43\}}$ фазовые сдвиги упругого n$^6$Li рассеяния. Точки и квадраты - чистые фазы, полученные из МРГ фазовых сдвигов, приведенных в работах [3-6].

С полученными потенциалами (5.1.1.)-(5.1.6) проведены расчеты сечений упругого n$^6$Li и p$^6$Li рассеяния при энергиях $E_n$=14,2 МэВ и $E_p$=25,9 МэВ соответственно, а результаты представлены на рис.5.3 [7]. Экспериментальные данные по сечениям взяты из работ [9,10].

Неупругости $\eta_{L,S}$ , рассчитанные по формулам (1.6.6) в предположении отсутствия поглощения в каналах с фиксированными симметриями {f}, не позволяют полностью передать поведение экспериментальных упругих сечении, поэтому в обоих спиновых каналах приходится вводить дополнительные неупругости, величина которых определялась из лучшего согласия вычисленных сечений с экспериментом.

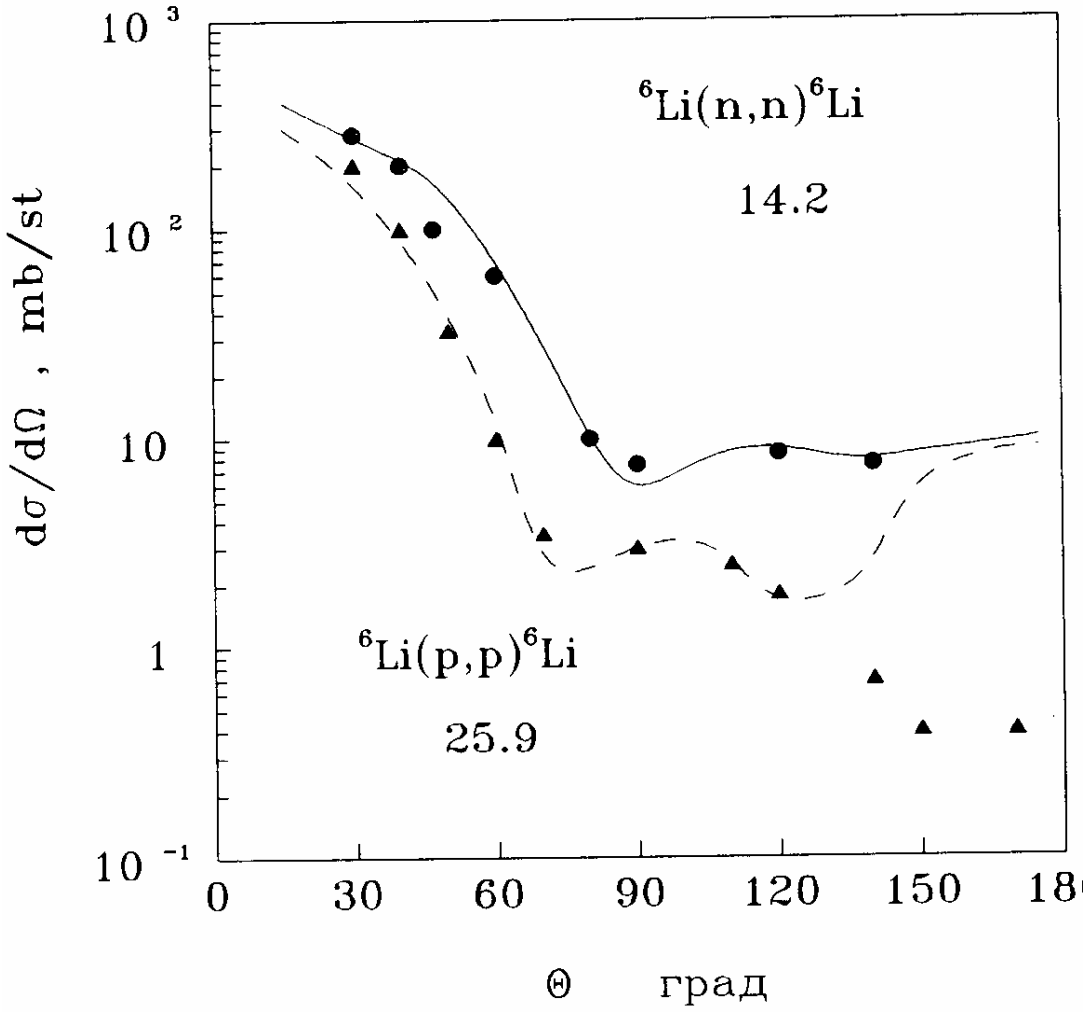

Рис.5.3. Дифференциальные сечения упругого n$^6$Li рассеяния при энергии 14.2 МэВ и 25,9 МэВ. Экспериментальные данные работ [9,10]. Кривые - расчеты с приведенными в тексте потенциалами.

На рис.5.4 представлены сечения процесса перезарядки $^6$Li(n,p)$^6$He при энергиях $E_n$=6,77 и $E_p$=14 МэВ, вычисленные с чистыми потенциалами взаимодействия $V^{\{f\}}$. Экспериментальные данные взяты из работ [11,12,13]. Сечения определяется согласно выражениям (1.6.2, 1.6.3) [7].

В заключение отметим, что далее, в этой главе будет приведена более полная классификация состояний в N$^6$Li кластерной системе, которая учитывает для основного состояния $^6$Li две возможные орбитальные схемы {6} и {42}.

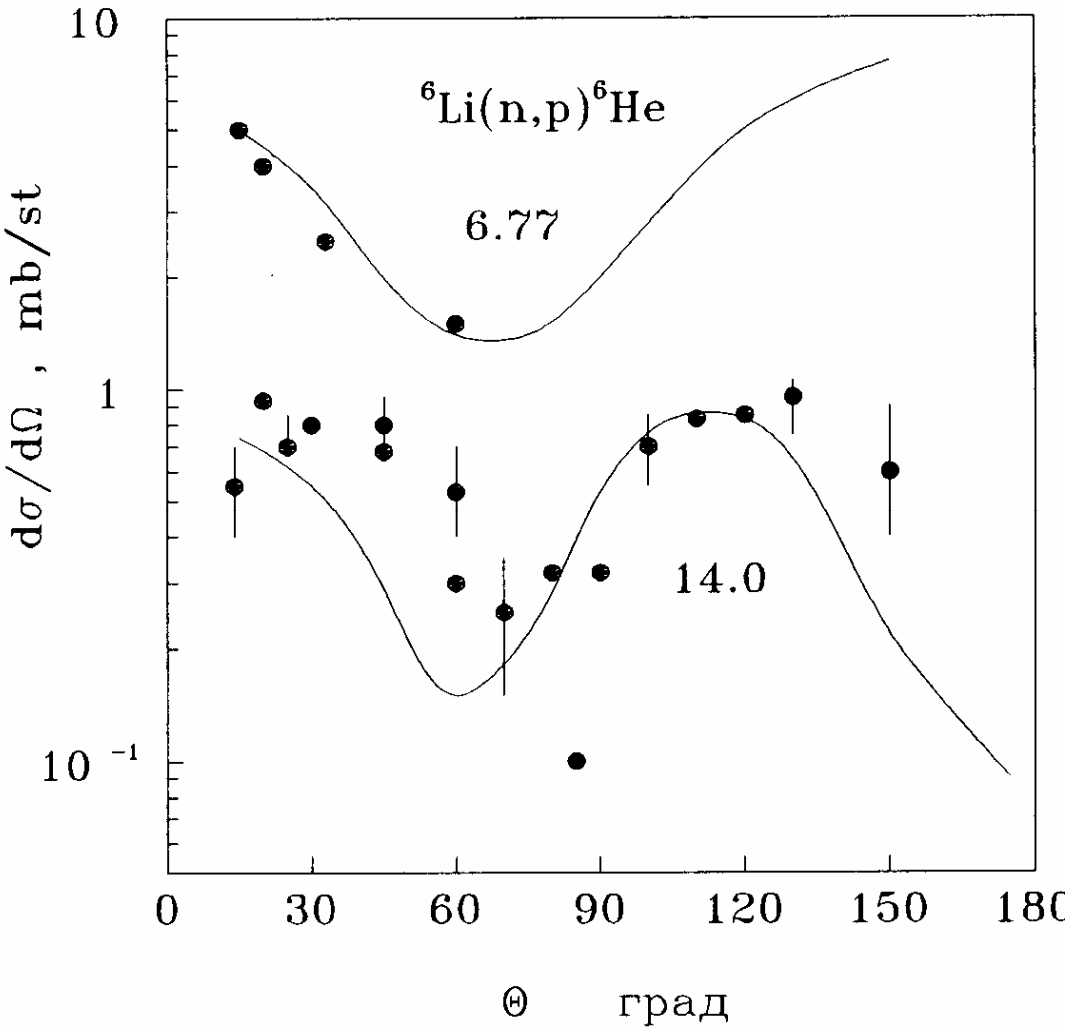

Рис.5.4. Дифференциальные сечения реакции перезарядки $^6$Li(n,p)$^6$He при энергиях 6.77 МэВ и 14 МэВ. Экспериментальные данные работ [11-13]. Кривые - расчеты с приведенными в тексте потенциалами.

Это приводит к запрещенному состоянию в чистой P волне и волновая функция основного состояния имеет узел, что позволяет более правильно описывать некоторые характеристики ядра $^7$Li и, в частности, полные сечения фотопроцессов.

## 5.2. Описание $^4$He$^4$He, $^2$H$^6$Li и N$^7$Li рассеяния

Рассмотрим теперь три кластерные системы, которые приводят к ядру $^8$Be в $^4$He$^4$He, $^2$H$^6$Li и p$^7$Li каналах. Первая их этих систем - чистая по орбитальным симметриям в любых состояниях, а две другие смешаны, и из экспериментальных фаз рассеяния необходимо выделять чистые по схемам Юнга состояния.

*Кластерная $^4$He$^4$He система*

В работе [14] приведен подробный обзор имеющихся экспериментальных результатов для $^4$He$^4$He системы, который охватывает энергетический диапазон до 150 МэВ. Кроме того, существует дос-таточно много феноменологических $^4He^4He$ потенциалов различной формы. Так, в работах [15,16] приведены взаимодействия Вудс - саксоновского и гауссова типа с отталкивающим кором. Однако, для этих потенциалов, при описании различных экспериментальных фаз рассеяния, приходится менять радиус кора в зависимости от орбитального момента. В работах [17,18] получены глубокие чисто притягивающие взаимодействия с запрещенными состояниями и используется обобщенная теорема Левинсона. В работе [19] получены глубокие гауссовы потенциалы со сферическим кулоновским членом, но не приводится структура запрещенных состояний.

Несмотря на имеющиеся и достаточно полные результаты, представляется интересным получить простой гауссовый потенциал $^4He^4He$ взаимодействия, позволяющий передать энергетический ход фаз рассеяния и основанный на идее запрещенных состояний. Используя классификацию, изложенную во второй главе, считаем, что основное состояние ядра $^8Be$ должно соответствовать 4S уровню со схемой Юнга {44}, а конфигурации 0S и 2S со схемами {8} и {62} являются запрещенными. Запрещенным также оказывается и уровень с конфигурацией 2D, совместимый с симметрией {62}.

В результате можно получить два потенциала с точечным и сферическим кулоновским членом [20]. В первом случае S фазу рассеяния хорошо описывает потенциал с параметрами $V_0$=-129 МэВ и $\alpha$ =0.225 Фм$^{-2}$. Однако D и G фазы оказываются заметно завышенными, что показано на рис.5.5 точечной линией. Для согласования этих парциальных волн необходимо уменьшить глубину до -126 МэВ (рис. 5.5, непрерывная линия), что приводит к L зависимому взаимодействию.

В то же время, введя кулоновский потенциал в виде сферы с радиусом 2,0 Фм, легко удается получить потенциал с $V_0$=-124.9 МэВ, $\alpha$=0.225 Фм$^{-2}$, хорошо передающий энергетический ход экспериментальных фаз рассеяния [14] при энергиях до 25 МэВ, как показано на рис. 5.5 штриховой линией.

Потенциал с такими параметрами, приводит к резонансу в S волне при 95 кэВ, а при нулевой энергии S фаза начинается с $2\pi$, что соответствует наличию двух запрещенных состояний при энергиях -74,4 и -26,5 МэВ. На рисунках, исключительно для наглядности, S фаза сдвинута вниз на $\pi$. Запрещенный уровень в D волне найден при энергии -22,6 МэВ [20].

Для расчета ширины резонансов $2^+$ и $4^+$ использовалось выражение

$\Gamma = 2(d\delta /dE)^{-1}$.

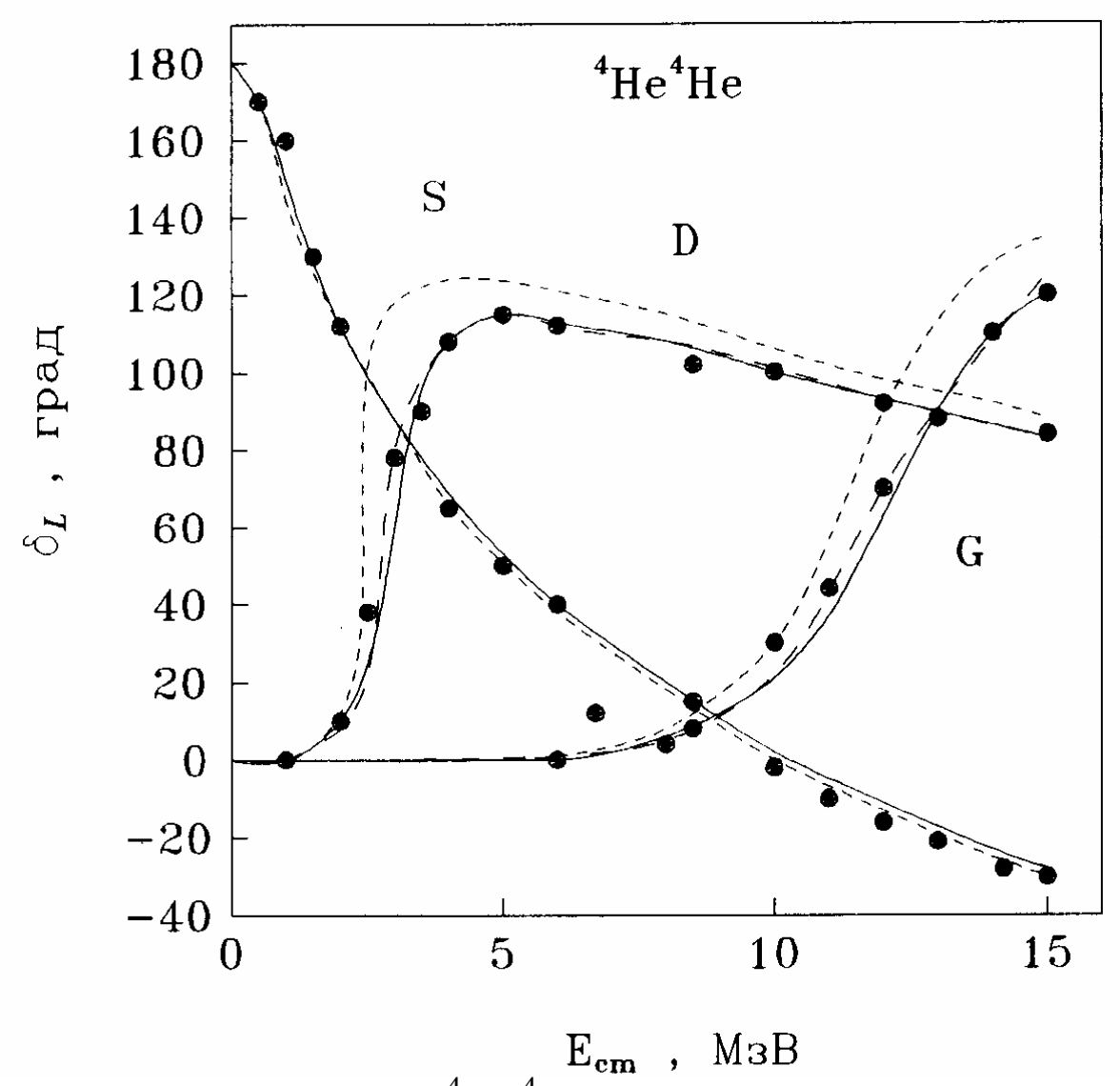

Рис.5.5. Фазы упругого $^4$He$^4$He рассеяния. Эксперимент из работы [14].

В результате получены величины 2,3 и 4,6 МэВ соответственно, что вполне согласуется с экспериментом [21], где приведены следующие данные - 1,6 и 3,5 МэВ [20]. На рис.5.6 показаны результаты расчета сечений упругого $^4$He$^4$He рассеяния при энергиях 11.88 и 24.11 МэВ, проведенного на основе L независимого потенциала.

### *Кластерная $^2H^6L$ система*

При рассмотрении системы $^2$H$^6$Li, существуют две возможности выбора орбитальных схем Юнга для основного состояния ядра $^6$Li, как было рассмотрено в [20]. Если симметрия этого состояния определяется схемой {6}, то в системе оказываются только запрещенные уровни {8}, {71} и {62}. Именно такой классификации соответствуют имеющиеся МРГ фазы, с одним исключением, когда D волна при S=2 идет не от π, а от нуля [22,23]. Другая возможность заключается в выборе симметрии {42}, т.е. учитывается наличие запрещенных состояний в самом ядре $^6$Li. Тогда получается более сложная классификация уровней, причем возможными оказываются симметрии

{62}, {53}, {44}, {521}, {422}, {431}.

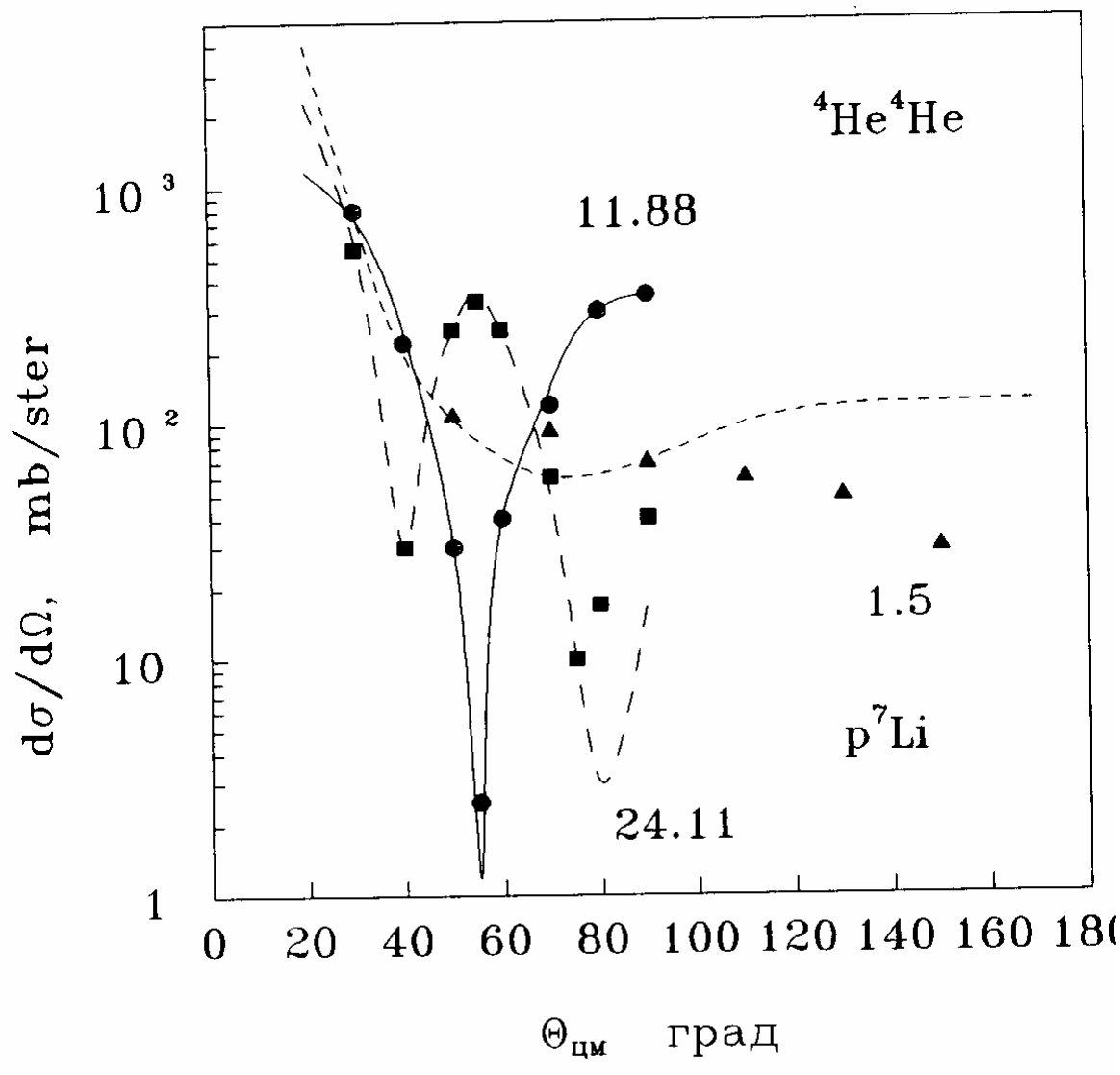

Рис.5.6. Сечение упругого $^4$He$^4$He при лабораторной энергии 11,88 и 24,11 МэВ и p$^7$Li рассеяния с $E_{с.ц.м.}$ = 1,5 МэВ. Экспериментальные данные взяты из работ [14] для $^4$He$^4$He системы и [24] для p$^7$Li канала.

В синглетном спиновом состоянии запрещен уровень с конфигурацией {62} при четных L, а состояния с {44} и {422} разрешены. При S=2 в четных волнах запрещены уже две конфигурации - {62} и {44}, при разрешенной симметрии {422}. Триплетное спиновое состояние характеризуется тремя запрещенными конфигурациями - {62}, {44}, {422}, и разрешенное состояние в четных волнах отсутствует. В нечетных парциальных волнах в спиновых состояниях 0 и 2 запрещены все три возможные симметрии {53}, {521} и {431}. А в триплетном разрешена схема {431} и запрещенных состояния только два - {53}, {521}.

Используя данные по МРГ фазам и приведенную выше классификацию орбитальных состояний в [20] были получены потенциалы взаимодействия, параметры которых приведены в табл.5.2, а результаты расчета фаз показаны на рис.5.7а-в. Причем, во всех нечетных состояниях, связанным считается только нижний уровень со схемой {53}, а остальные, запрещенные или разрешенные в разных спиновых каналах, находятся в непрерывном спектре.

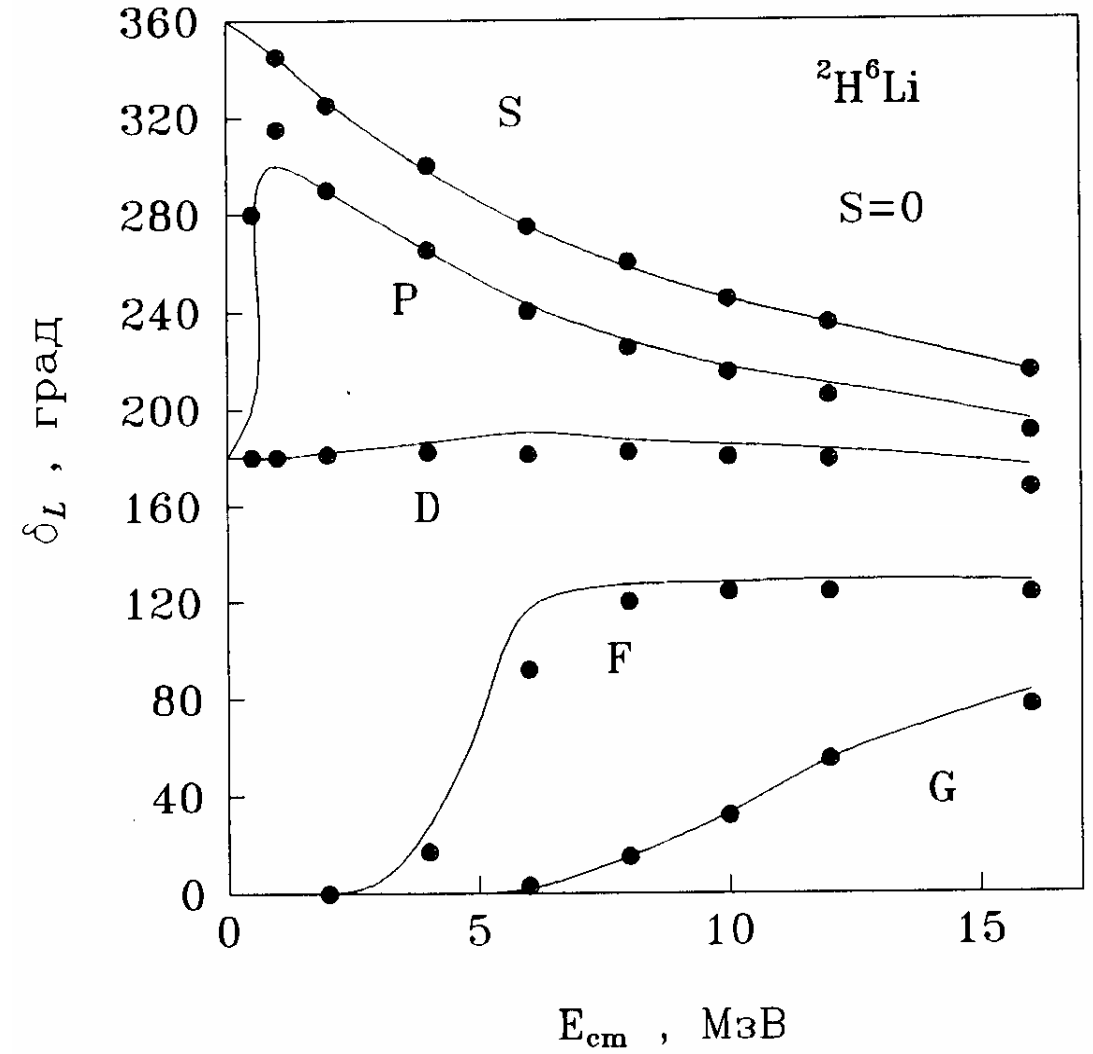

Рис.5.7а. Фазовые сдвиги для $^2H^6Li$ рассеяния при S=0. Точки - МРГ данные работ [22,23]. Линии - расчет с потенциалом, приведенным в табл.5.2.

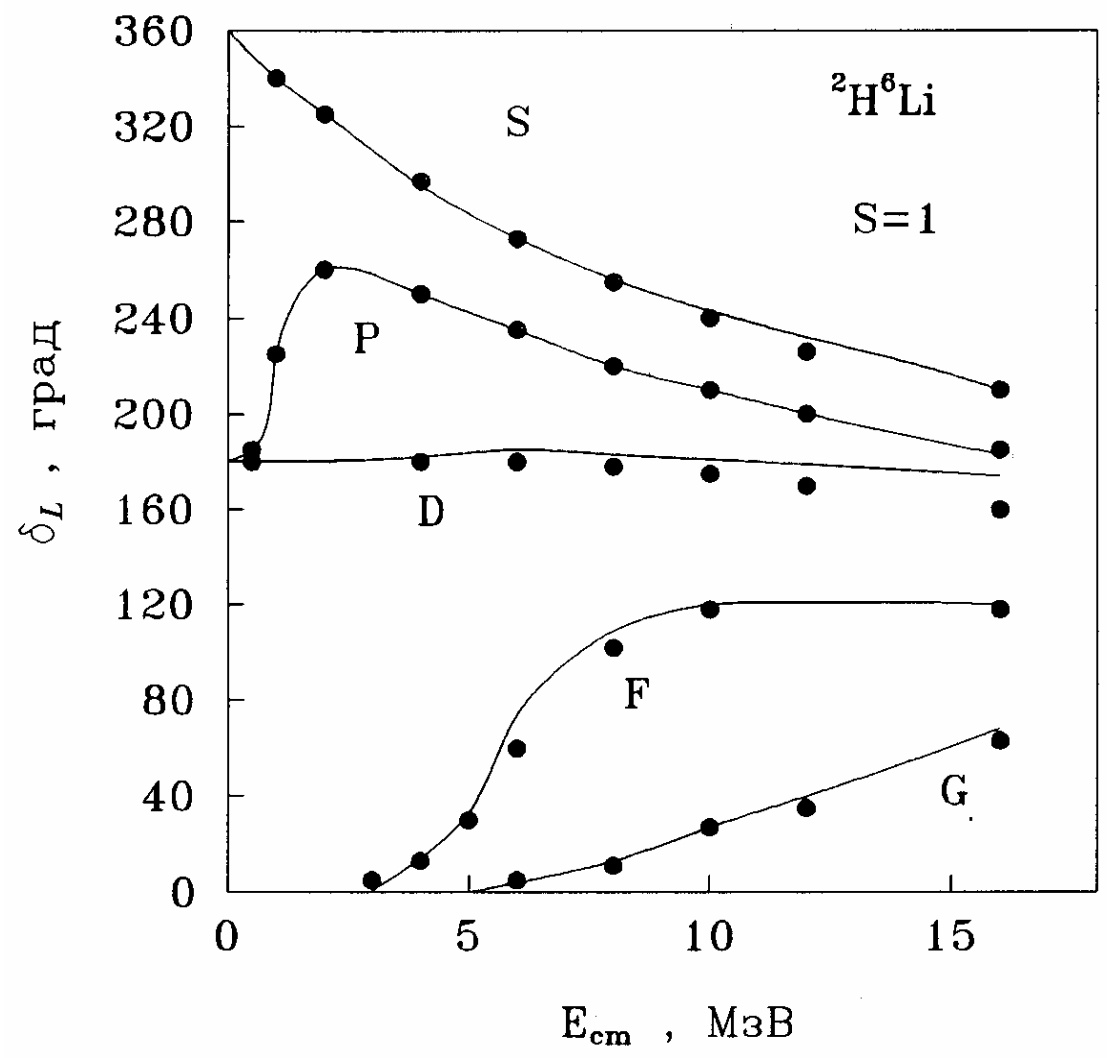

Рис.5.7б. Фазовые сдвиги для $^2H^6Li$ рассеяния при S=1. Точки - МРГ данные работ [22,23]. Линии - расчет с потенциалом, приведенным в табл.5.2.

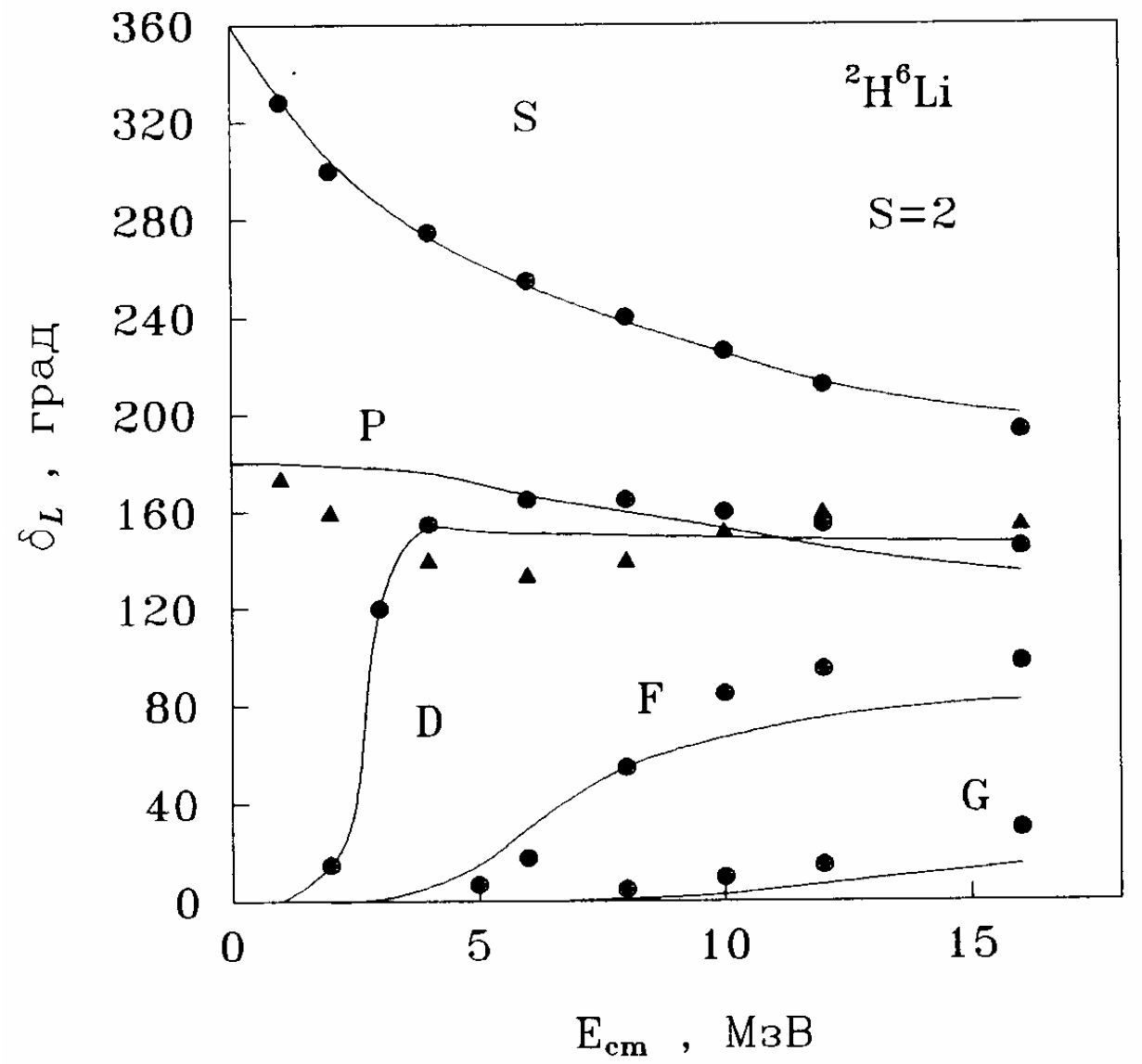

Рис.5.7в. Фазы рассеяния для $^2H^6Li$ рассеяния при S=2. Точки - МРГ данные работ [22,23]. Линии - расчет с потенциалом, приведенным в табл.5.2.

*Таблица 5.2. Параметры потенциалов $^2H^6Li$ взаимодействия и энергии связанных запрещенных (ЗС) и разрешенных (РС) уровней.*

| S; {f} | L | $V_0$, (МэВ) | α, (Фм$^{-2}$) | $E_{зс}$, (МэВ) | $E_{рс}$, (МэВ) |
|---|---|---|---|---|---|
| 0; {44}+{422} | 0 | -57.0 | 0.11 | -28.6 | -5.7 |
| | 1 | -55.0 | 0.11 | -14.3 | |
| 1; {431} | 0 | -60.0 | 0.12 | -29.7;-5.5 | |
| | 2 | --- | --- | -3.5 | |
| | 1 | -56.0 | 0.12 | -13.5 | |
| 2; {422} | 0 | -86.0 | 0.24 | -37.3;-1.9 | |
| | 1 | -33.0 | 0.1 | -1.0 | |
| 0; {44} | 0 | -59.0 | 0.1 | -31.0 | -8.0 |
| | 2 | --- | --- | -6.4 | |
| | 1 | -80.0 | 0.15 | -22.8 | |

В синглетном состоянии в четных волнах имеется два связан-ных уровня - запрещенный с {62} и разрешенный при {44}, а состояние с {422} считается не связанным. В триплетном и квартетном каналах связаны запрещенные уровни с {62} и {44}, а состояние симметрии {422} находится в непрерывном спектре. Поэтому надо иметь в виду, что на рис.5.7в D фаза должна начинаться со $180^0$.

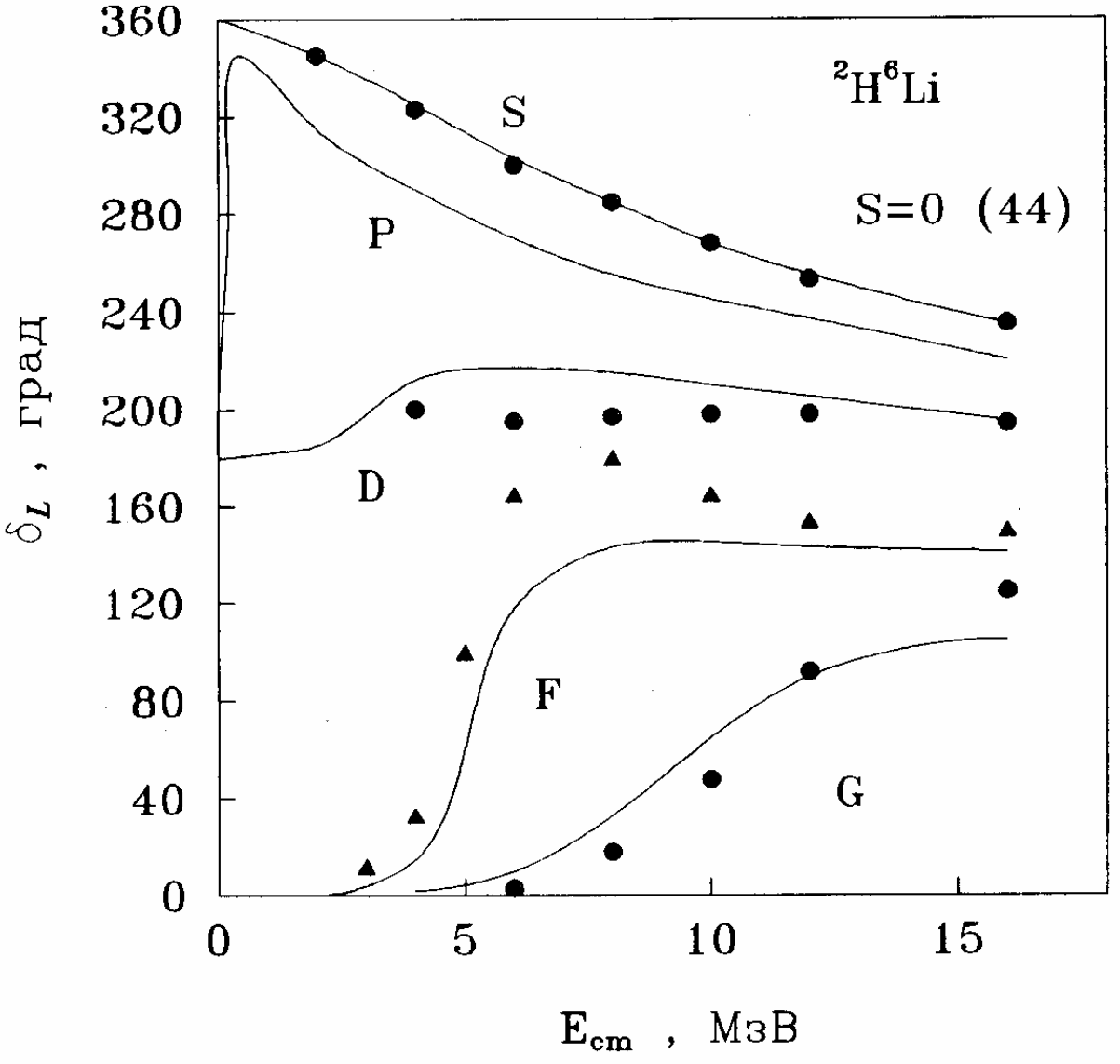

Рис.5.7г. Выделенные чистые фазы (S=0,{44}) для $^2H^6Li$ рассеяния. Точки - результаты для чистых фаз, полученные из МРГ данные работ [22,23]. Линии - расчет с потенциалом, приведенным в табл.5.2.

Разделение потенциалов на четные и нечетные, позволило хорошо воспроизвести МРГ результаты почти для всех энергий и спиновых состояниях. Несколько хуже передается P фаза при S= 2. Форма МРГ фазы такова, что для ее описания необходимо вводить периферическое отталкивание. Такая форма P волны определяет поведение чистой P фазы, которая при энергиях 0-6 МэВ расположена выше $360^0$ . Резонансная форма некоторых фаз предполагает наличие в спектре ядра $^8Be$ резонансных уровней. Однако, надежные данные по спектрам этого ядра выше порога $^2H^6Li$ канала практически отсутствуют [21].

Разделение смешанных по схемам Юнга состояний может быть проведено на основе T матрицы вида (1.6.1). Раньше было показано, что для всех легких систем полная T матрица, отвечающая экспери-ментальным результатам, представима в виде полусуммы чистых Т матриц. Отсюда может быть найдено и соотношение для экспериментальных, смешанных по симметриям, фаз рассеяния с чистыми фазами, зависящими только от одной схемы Юнга в виде полусуммы фаз (1.6.6). Причем, в качестве $T^{\{422\}}$ при S=0 можно принять $T^{\{422\}}$ для S=2 и таким образом по экспериментальным данным определить чистую фазу с {44}. Используя эти результаты в [20] было проведено разделение фаз рассеяния при S=0 и выделены чистые фазы с {44}, показанные на рис.5.7г, а спектр уровней полученных взаимодействий приведен в табл.5.2. Потенциал строился только исходя из возможности описания чистых фаз, и не согласован со спектром связанных уровней, так как $0^+$, $2^+$ и $4^+$ уровни вероятнее всего обусловлены $^4He^4He$ кластерным каналом.

В табл.2.1 приведена более общая классификация состояний по орбитальным схемам Юнга, чем изложенная выше, на основе результатов работы [20]. По - видимому, и здесь более правильно учитывать обе возможные орбитальные схемы {6} и {42} для основного состояния $^6Li$, также как в $N^6Li$ системе.

В таком случае, к возможным состояниям $^2H^6Li$ системы, которые использовались для полученных выше потенциалов, добавляются схемы {8} и {71}, первая из которых совместима с L=0, а вторая с L=1. Это, в свою очередь, изменяет число запрещенных связанных состояний во всех спиновых каналах. В частности, в синглетном чистом канале в S состоянии должно быть два связанных запрещенных и один разрешенный уровень со схемами {8}, {62} и {44} соответственно.

## *Кластерная $p^7L$ система*

Канал $p^7Li$ ядра $^8Be$ оказывается смешанным по изоспину с T=0,1, хотя в обоих случаях соответствует одной схеме Юнга {431}. Поэтому Т матрица представляется в виде (1.6.1)

$$T_L = 1/2\ T_{t=0} + 1/2\ T_{t=1}\ . \qquad (5.2.1)$$

Чтобы выделить чистое с T=0 состояние из смешанных экспериментальных данных с T=0,1 надо использовать Т матрицу систем $n^7Li$ или $p^7Be$, чистых по изоспину с T=1.

Надежные данные по этим системам отсутствуют, поэтому в [20] рассматривались только смешанные по изоспину потенциалы, выбираемые, как обычно, в простом гауссовом виде (3.1.1) с точечным кулоновским членом.

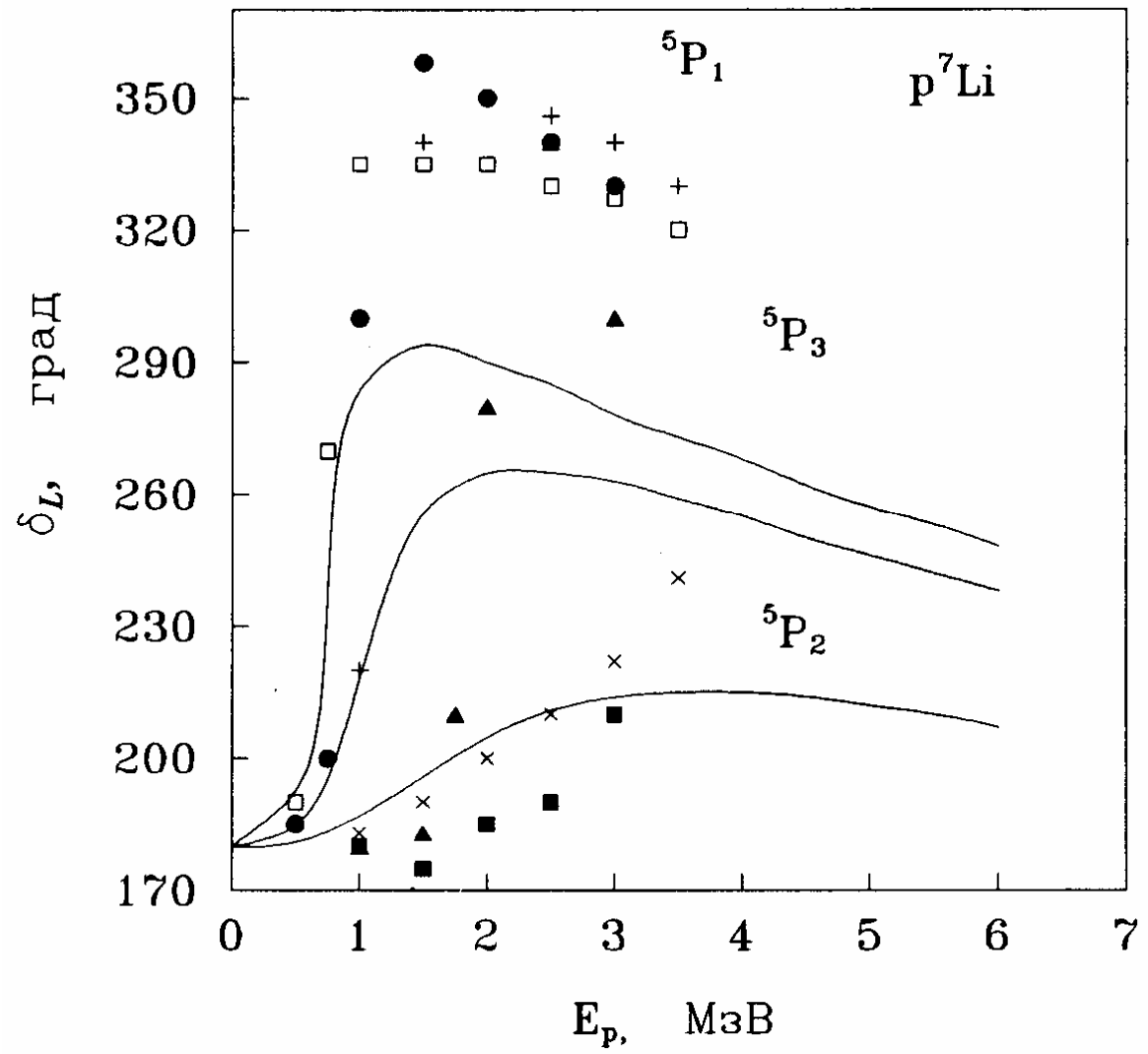

Рис.5.8а. Фазовые сдвиги для $p^7Li$ рассеяния при S=2. Точки, квадраты и треугольники - экспериментальные данные работы [24]. Крестики и открытые квадраты - МРГ данные работы [25].

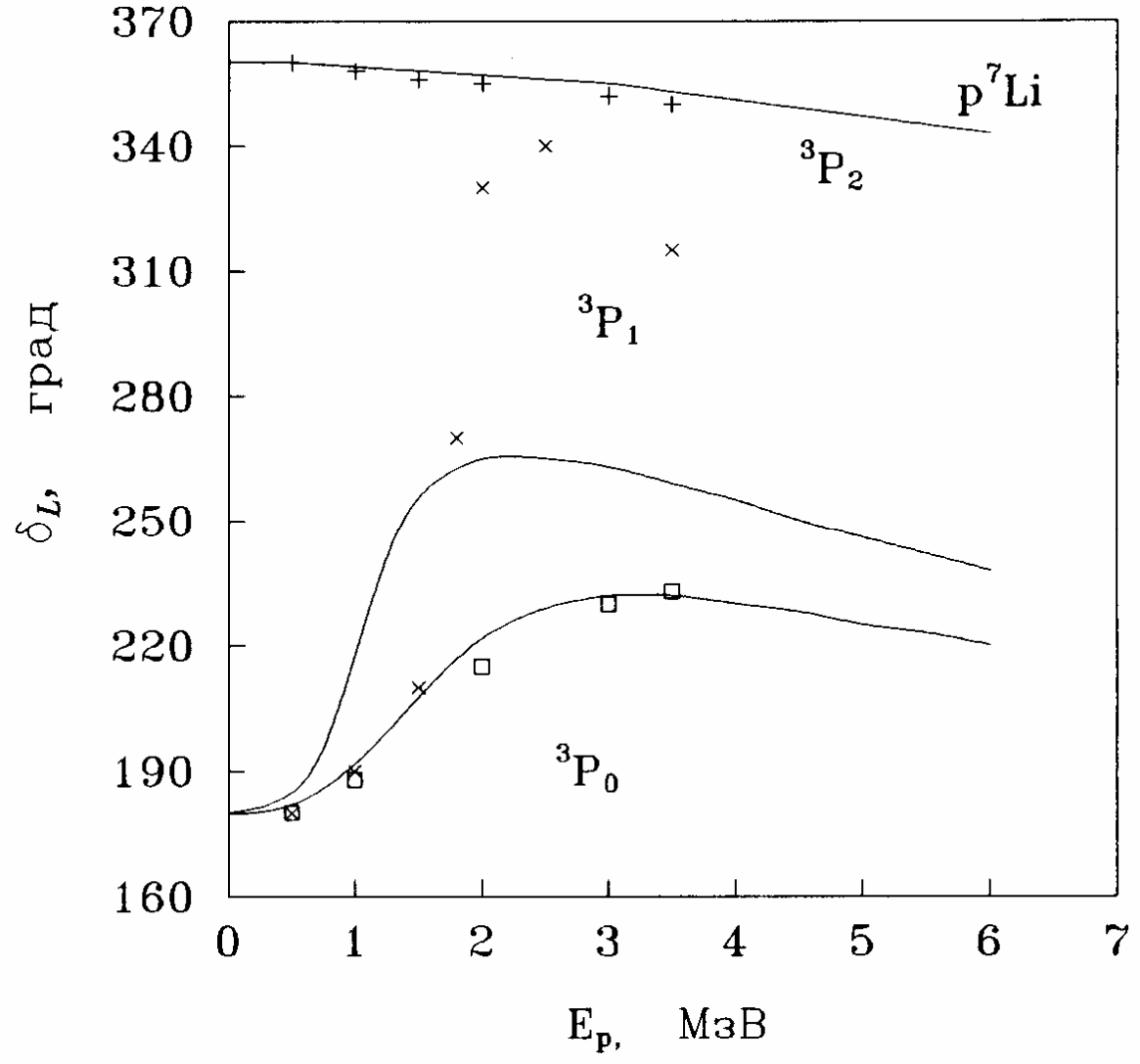

Рис.5.8б. Фазовые сдвиги для $p^7Li$ рассеяния при S=1. Крестики и открытые квадраты - МРГ данные работы [25].

В этой системе, как и в предыдущем случае, возможны два варианта схем Юнга для основного состояния ядра $^7Li$. В первом случае схема {7} приводит только к двум запрещенным состояниям в системе $p^7Li$ с конфигурациями {8} и {71}.

Во втором, когда для основного состояния $^7Li$ принимается схема {43}, в триплетном спиновом состоянии имеются запрещенные уровни {53} и {44} и разрешенный, с конфигурацией {431}. Значит, потенциал должен иметь запрещенное связанное {44} состояние в S волне и запрещенное и разрешенное связанные уровни в P волне с {53} и {431} соответственно. Причем сделанные выводы правильны для любого изоспинового состояния системы.

Экспериментальные фазы $p^7Li$ канала имеются только при энергиях до 2 МэВ [24], а МРГ вычисления проводились в области 0-3 МэВ [25]. Фазы, приведенные в этих работах, даны в обычном виде с учетом спин - орбитального расщепления, в отличие от результатов работ [26,27,28], где показаны фазы полного момента.

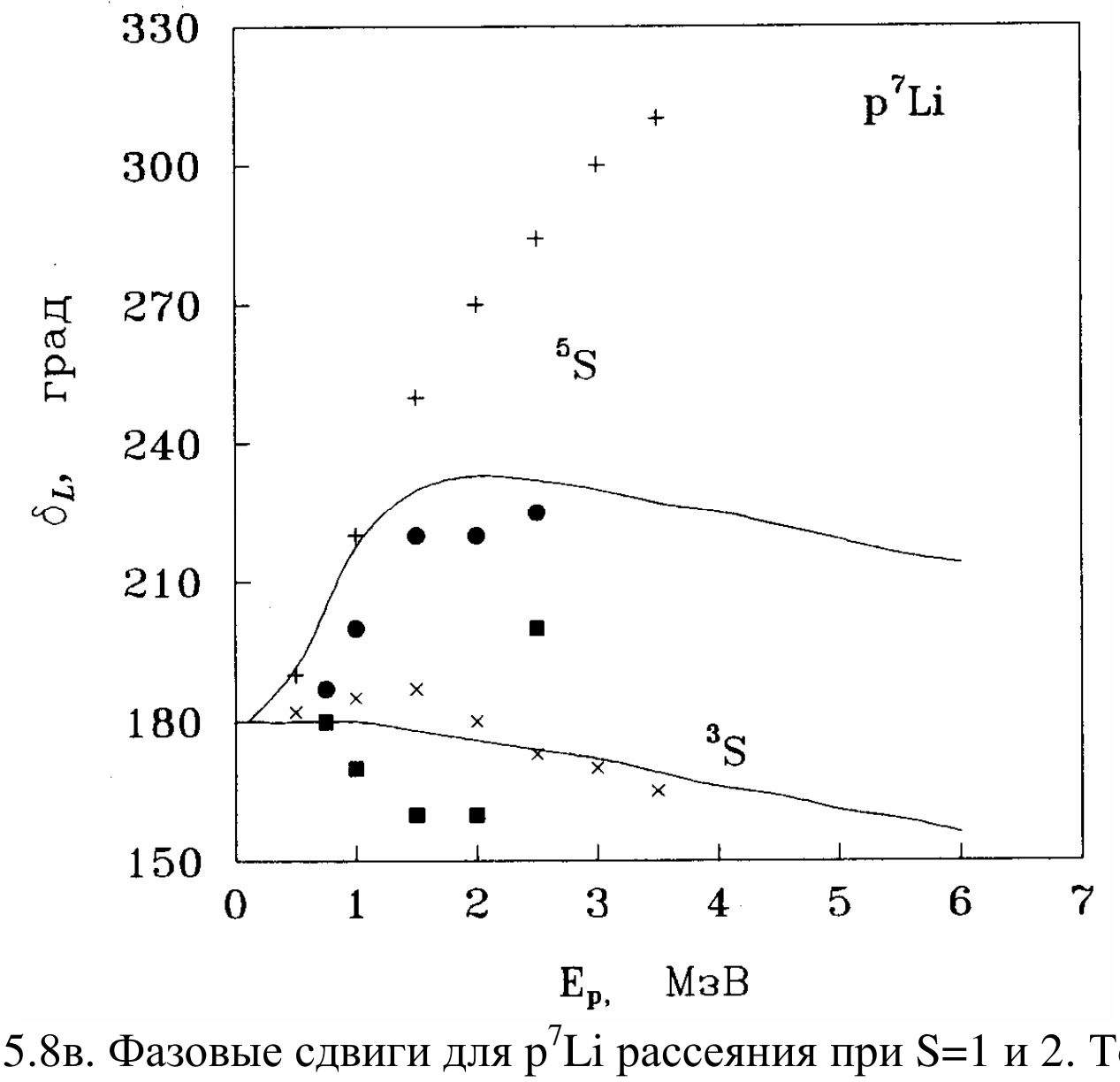

Рис.5.8в. Фазовые сдвиги для $p^7Li$ рассеяния при S=1 и 2. Точки и квадраты - экспериментальные данные работы [24]. Крестики - МРГ данные работы [25].

Экспериментальные и МРГ фазы, в общем, согласуются между собой. Исключение составляет только $^5P_2$ волна, экспериментальные данные которой показаны на рис.5.8 черными квадратами, а МРГ результаты открытыми квадратами.

Используя эти данные, были получены два вида потенциалов, параметры которых и структура уровней даны в табл.5.3, а результаты вычислений для второго варианта показаны на рис.5.8 [20]. Экспериментальные данные показаны точками, квадратами и треугольниками, остальные обозначения - МРГ данные.

*Таблица 5.3. Параметры потенциалов $p^7Li$ взаимодействия*

| | | Потенциал 1 | | | Потенциал 2 | | | |
|---|---|---|---|---|---|---|---|---|
| S | L | $V_0$, (МэВ) | α, ($Фм^{-2}$) | $E_{зс}$, (МэВ) | $V_0$, (МэВ) | α, ($Фм^{-2}$) | $E_{зс}$, (МэВ) | $E_{рс}$, (МэВ) |
| 1 | S | -35.0 | 0.1 | -3.66 | --- | --- | --- | --- |
| | $P_0$ | -72.0 | 0.1 | -16.1 | -155.0 | 0.1 | -67.6 | -20.3 |
| | $P_1$ | -80.0 | 0.1 | -20.4 | -163.0 | 0.1 | -73.1 | -23.7 |
| | $P_2$ | -50.0 | 0.1 | -5.6 | -130.0 | 0.1 | -50.9 | -10.7 |
| 2 | S | -88.0 | 0.2 | -32.4 | --- | --- | --- | --- |
| | $P_1$ | -85.0 | 0.1 | -23.2 | -168.0 | 0.1 | -76.5;-25.9 | --- |
| | $P_2$ | -87.0 | 0.1 | -24.3 | -150.0 | 0.1 | -64.1;-18.2 | --- |
| | $P_3$ | -80.0 | 0.1 | -20.4 | -163.0 | 0.1 | -73.1;-23.7 | --- |

Экспериментальные и МРГ фазы вполне согласуются со спектром надпороговых уровней $^8Be$ [21]. Так, резонанс в $^5S$ волне соответствует уровню $2^-$ при энергии 18,91 МэВ. Резонанс $^5P_3$ совпадает с уровнями $3^+$ при 19,24 (19,07) МэВ. Поведение $^5P_1$ фазы согласуется с уровнями $1^+$ при 17,64 и 18,5 МэВ, смешанными по изоспину. На рис.5.6 показаны дифференциальные сечения упругого $p^7Li$ рассеяния при энергии 1,5 МэВ, вычисленные с полученными потенциалами. Экспериментальные данные взяты из работ [24].

По - видимому, как и в предыдущих системах, здесь более правильно рассматривать обе возможные схемы {7} и {43} для основного состояния ядра $^7Li$. Тогда классификация уровней будет несколько иной, и число запрещенных состояний возрастет, и в каждой волне добавится лишний запрещенный связанный уровень. Такая более полная схема состояний приведена в табл.2.1.

## 5.3. Фотопроцессы на ядре $^7Li$ в $n^6Li$ канале

Продолжая изучение фотопроцессов для ядер лития, рассмотрим процесс прямого кластерного фоторазвала $^7Li$ в $n^6Li$ канал. Па-раметры таких взаимодействий предварительно были получены в [7], и приведены в первом параграфе данной главы, на основе расчетов фаз, выполненных в методе резонирующих групп [4-6]. Отличие канала $n^6Li$ от $^4He^3H$ состоит в том, что в данном случае, в состояниях с минимальным спином имеется смешивание по орбитальным схемам Юнга. В работе [7] было показано, что при S=1/2 разрешены две схемы {43} и {421} в то время, как в квартетном канале разрешена только симметрия {421}.

Рассмотрим более подробно классификацию состояний в системе $n^6Li$. Если ядру $^6Li$ сопоставлять только орбитальную схему {42}, то на основе теоремы Литтлвуда получаем набор возможных орбитальных схем ядра $^7Li$ в виде {52}, {43} и {421}, как это было сделано в работе [7]. Однако, для получения более полной структуры запрещенных и разрешенных состояний необходимо рассматривать не только {42}, но и орбитальную схему {6}. В этом случае возможными орбитальными состояниями будут {7} и {61}.

Полный набор возможных орбитальных схем составляют оба этих варианта. Разрешенными состояниями в дублетном канале будут состояния с {43} и {421}, а запрещенные имеют симметрии {7}, {61} и {52}. Схема {61} для связанного состояния совместима с орбитальным моментом 1, а {7} и {52} с нулевым, причем последняя возможна и при L=2. Таким образом, в S волне имеются два запрещенных состояния и по одному запрещено в P и D волнах, которые соответствуют симметриям {61} и {52} соответственно.

В квартетном канале запрещена так же и симметрия {43}, что приводит только к одному разрешенному состоянию с {421}. В такой ситуации, если по прежнему считать, что основному состоянию ядра $^7Li$ соответствует только разрешенная схема {43}, необходимо из экспериментальных фаз рассеяния в дублетном канале выделить чистую компоненту с {43}, которая в принципе уже применима для анализа характеристик основного состояния.

Эта классификация приводит к вполне определенному типу взаимодействий, структура которых полностью определена. Кроме того, разрешенное состояние в P волне должно приводить к правильной энергии связи в $n^6Li$ канале, соответствующей основному состоянию ядра $^7Li$ при -7.25 МэВ [21].

Используя тот факт, что смешанные фазы рассеяния представляются в виде полусуммы чистых фаз [7] и, принимая в качестве дублетной фазы с {421} квартетную фазу той же симметрии, легко можно получить чистые дублетные фазы с {43}. На основе этих фаз параметризуются чистые взаимодействия, соответствующие опре-деленным схемам Юнга.

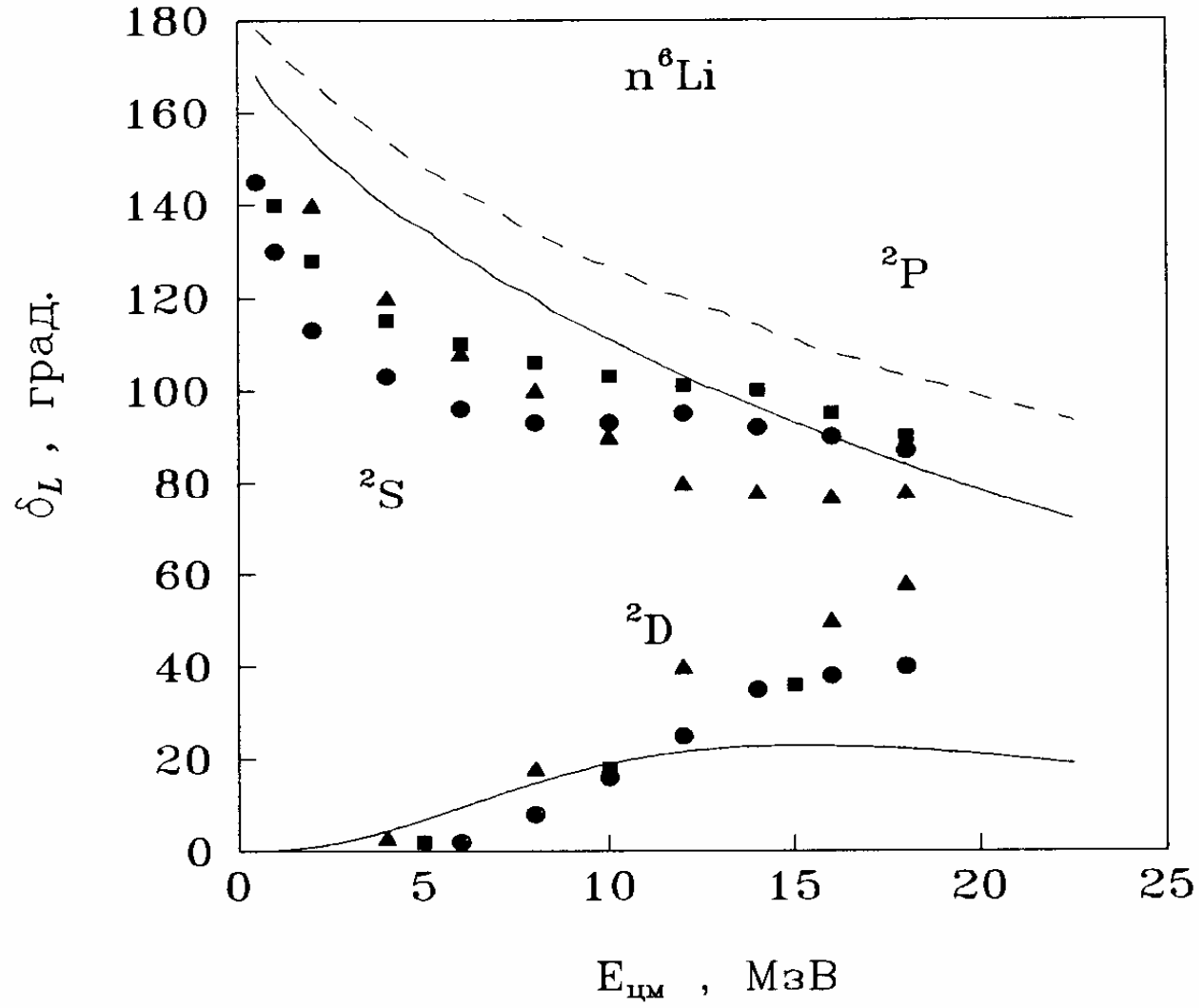

Рис.5.9. Смешанные дублетные фазы упругого n$^6$Li рассеяния. Линии - расчеты с полученным потенциалом. Точки, треугольники и кружки - МРГ вычисления из работ [4-6].

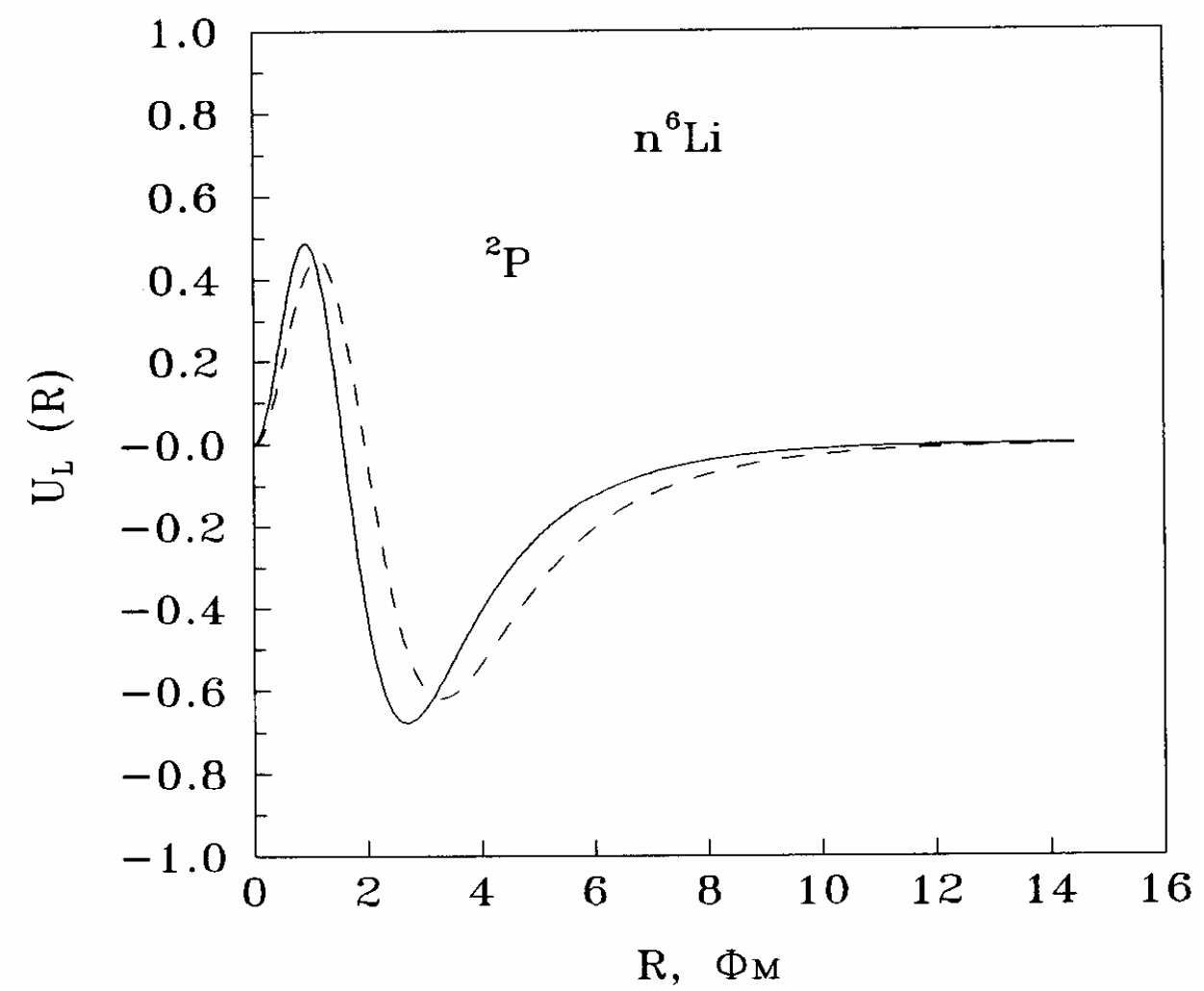

Рис.5.10. Волновые функции в n$^6$Li канале - непрерывная линия и в $^4$He$^3$H канале - штриховая линия.

Отметим, что учитывать возможную орбитальную схему {6} необходимо не только при рассмотрении $N^6Li$ системы, но так же в канале $^2H^6Li$ и схему {7} для $^7Li$ в $N^7Li$ системе. Эти изменения приводят к более полной, чем приведено в [7], классификации запрещенных состояний для ядер с A=7,8, которая была рассмотрена в работе [29] и дана в табл.2.1. Допустимые орбитальные симметрии в кластерных системах можно получить по теореме Литтлвуда [30] или на основе таблиц работы [31], где так же приведены схемы Юнга для спин - орбитальных частей волновых функций. Орбитальные моменты для связанных состояний определяются на основе правила Эллиота [30].

Потенциал взаимодействия, по - прежнему, будем выбирать в гауссовом виде с точечным кулоновским членом. Используя изложенные выше соображения, для четных дублетных состояний рассеяния найден потенциал с параметрами -140 МэВ и 0.15 Фм$^{-2}$, который имеет два запрещенных состояния в $^2S$ волне при энергиях -71.9 МэВ и -16.4 МэВ, а так же содержит запрещенное состояние в $^2D$ волне при -10.6 МэВ. Чистый потенциал для основного состояния в $^2P$ волне может быть представлен параметрами -252.6 МэВ и 0.25 Фм$^{-2}$. Он содержащий запрещенный уровень при -85.6 МэВ, а разрешенный совпадает с энергией основного состояния ядра $^7Li$ в $n^6Li$ канале при -7.25 МэВ [29]. Смешанные потенциалы рассеяния в целом передают дублетные четные МРГ фазы [4-6], как показано на рис.5.9 непрерывными линиями. Различные варианты МРГ фаз взяты из работ [4-6] и показаны точками, треугольниками и квадратами. Штриховой линией показана расчетная, чистая с {43} фаза для $^2P$ потенциала основного состояния.

На рис.5.10 непрерывной линией показана волновая функция относительного движения кластеров для $^7Li$ в $n^6Li$ канале. Штриховой линией приведена функция канала $^4He^3H$ для гауссового потенциала из работ [20]. Видно, что обе функции не сильно отличаются друг от друга. Обе имеют узловую структуру с узлом, примерно, в одной и той же области, а амплитуды их практически совпадают. Определенная схожесть волновых функций приводит к близким величинам и для эффективного межкластерного расстояния. В рассматриваемом канале имеем 3.09 Фм [29], а для $^4He^3H$ системы в [7] было получено 3.71 Фм.

Отметим, что используемые здесь потенциалы взаимодействия, полученные в [29], отличаются от приведенных в работе [7], т.к. в ней для получения структуры запрещенных и разрешенных состояний не учитывалась орбитальная схема Юнга {6} в ядре $^6Li$. Однако, на основе используемого здесь потенциала основного состояния, как и в работе [7] не удается описать квадрупольный момент и вероятность перехода на первый возбужденный уровень ядра $^7$Li. Эти величины хорошо воспроизводятся только в $^4$He$^3$H модели ядра [20].

При расчетах полных сечений фоторазвала рассматривался только E1 процесс с переходом из основного состояния на $^2$S и $^2$D состояния рассеяния. Параметры потенциалов приведены выше, а результаты расчета сечений представлены на рис.5.11 непрерывной линией. Пунктирами показаны вклады сечений, обусловленные переходами на $^2$S и $^2$D состояния рассеяния. Видно, что расчетное сечение практически верно передает величину и форму экспериментальных данных [32]. Основной вклад дает $^2$D волна, а сечение, обусловленное $^2$S волной заметно, только при малых энергиях.

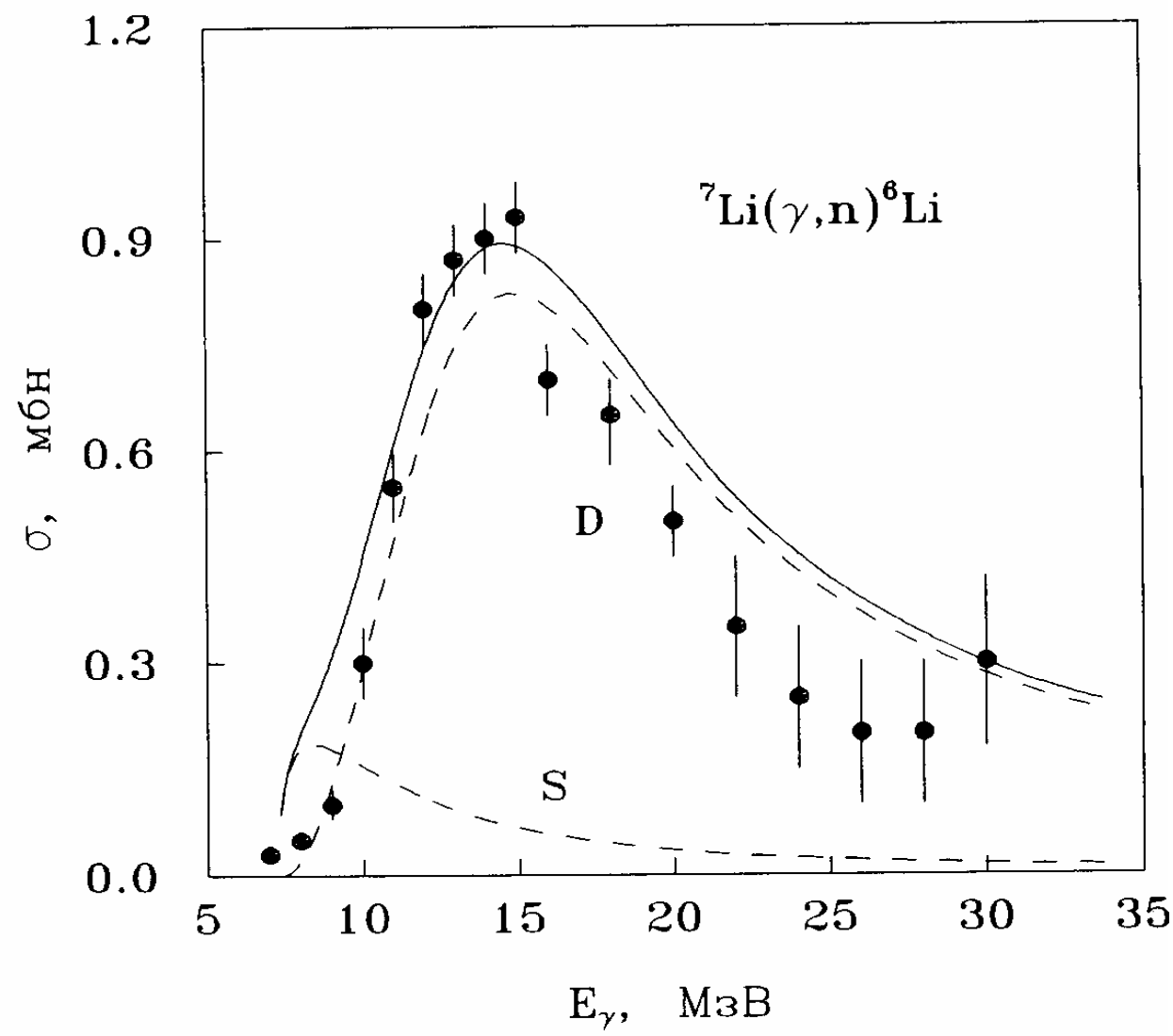

Рис.5.11. Полные сечения фоторазвала ядра $^7$Li в n$^6$Li канал. Непрерывная линия - расчет полных E1 сечений с полученным потенциалом. Штриховые линии - вклады фотопроцессов от $^2$S и $^2$D волн рассеяния. Точки - эксперимент из работы [32].

Отметим, что использование других потенциалов основного состояния или рассеяния, например, из работы [7] или каких - либо других, с другой структурой запрещенных уровней, не позволяет воспроизвести экспериментальные данные по фотосечениям. Результаты в этом случае или имеют совершенно другую форму или не описывают абсолютную величину сечений.

Таким образом, видно, что используемая модель с параметризованными здесь взаимодействиями, которые в целом согласованы с МРГ фазами и соответствуют приведенной выше структуре запрещенных и разрешенных состояний вполне способна передать известные экспериментальные данные по полным сечениям фоторазвала в рассматриваемый канал.

## 5.4. Кластерная система $^4He^{12}C$ в ядре $^{16}O$

В $^4He^{12}C$ системе не удается получить потенциалы способные одновременно правильно описывать фазы рассеяния и характеристики связанного состояния [33], как это было сделано ранее для более легких кластерных конфигураций. Поэтому приходится основываться на двух не связанных группах взаимодействий.

Для основного состояния потенциал строится исходя из требований описания таких характеристик, как энергия связи, зарядовый радиус, кулоновский формфактор при малых импульсах. Такой потенциал может иметь три или четыре запрещенных и одно разрешенное состояния при энергии связи ядра $^{16}O$ в $^4He^{12}C$ канале.

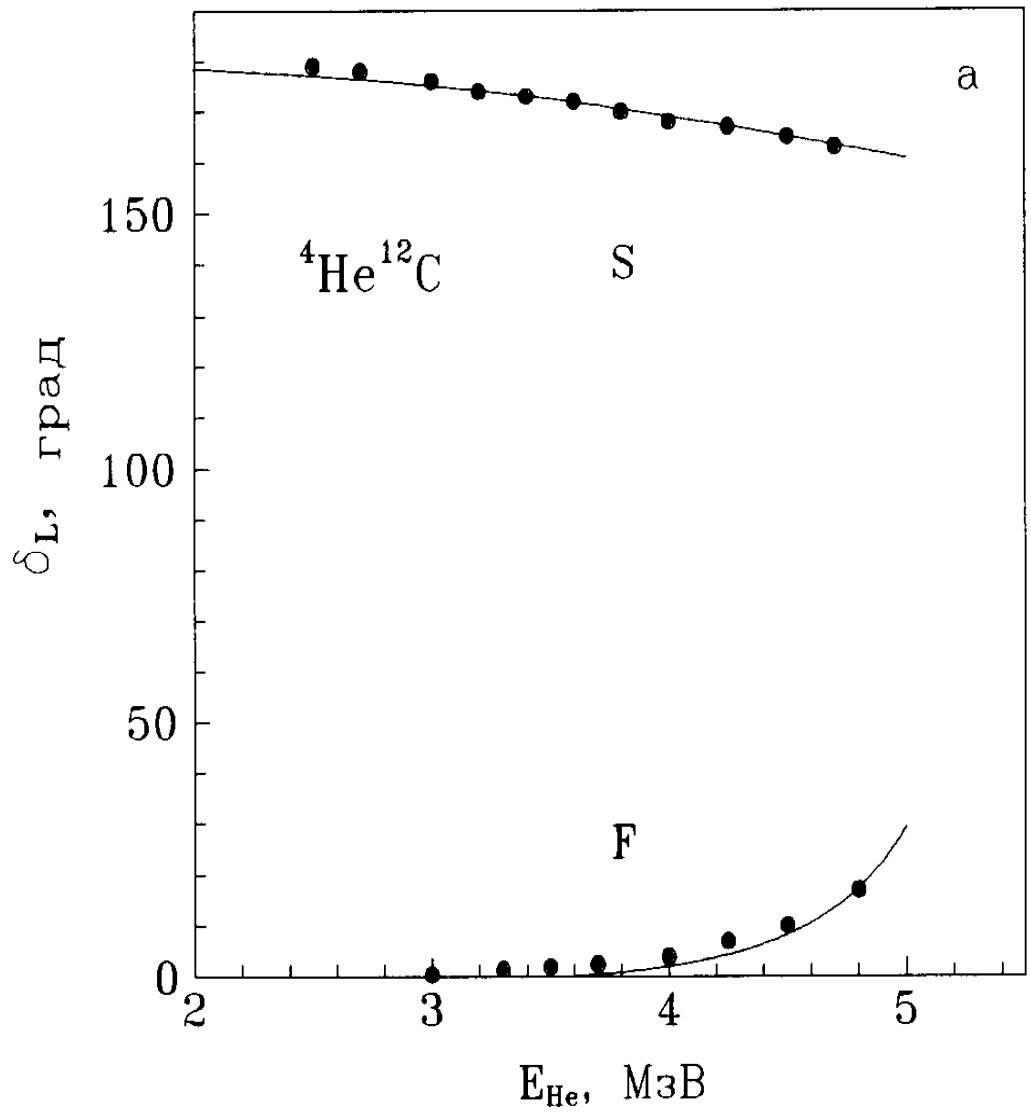

Рис.5.12а. Фазы упругого $^4He^{12}C$ рассеяния. Точки - экспериментальные данные [34], непрерывные кривые - результаты расчетов для потенциалов с параметрами из табл.5.4.

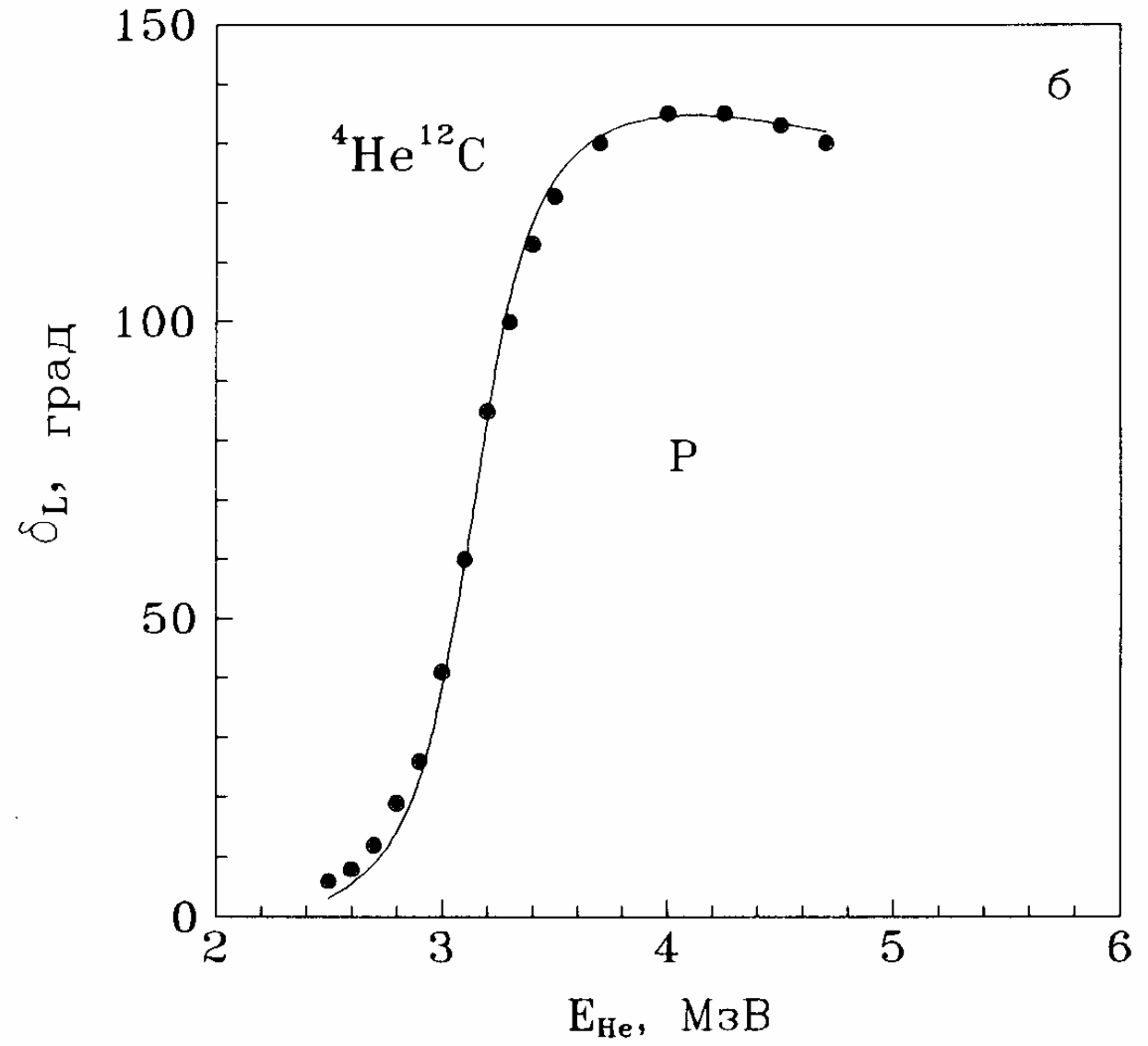

Рис.5.12б. Фазы упругого $^4He^{12}C$ рассеяния. Точки - экспериментальные данные [34], непрерывная кривая - результаты расчетов для потенциалов с параметрами из табл.5.4.

Потенциалы других связанных состояний строились исходя из требования описания ими приведенных вероятностей электромагнитных переходов между различными уровнями и энергий этих состояний. Параметры взаимодействий для различных уровней приведены в табл.5.4 под номерами 1-6 [33]. Уровни обозначены цифрой и буквой, которые определяют порядковый номер состояния и его орбитальный момент.

Для состояний рассеяния параметры взаимодействий подбирались так, что бы правильно воспроизвести энергетическое поведение экспериментальных фаз. Эти потенциалы приведены под номерами 7-13 и отличаются от взаимодействий для связанного состояния [33]. Результаты расчетов фаз для этих потенциалов вместе с экспериментальными данными работы [34] показаны на рис.5.12, 5.13. Потенциалы для P и G волн примерно совпадают с приведенными в работе [35].

Экспериментальный спектр ядра $^{16}O$ показан на рис.5.14а [36], а вычисленные энергии уровней приведены в табл.5.4. Для S и D фаз приводится два варианта взаимодействий, дающие одинаковое описание эксперимента, но содержащие разное число запрещенных состояний.

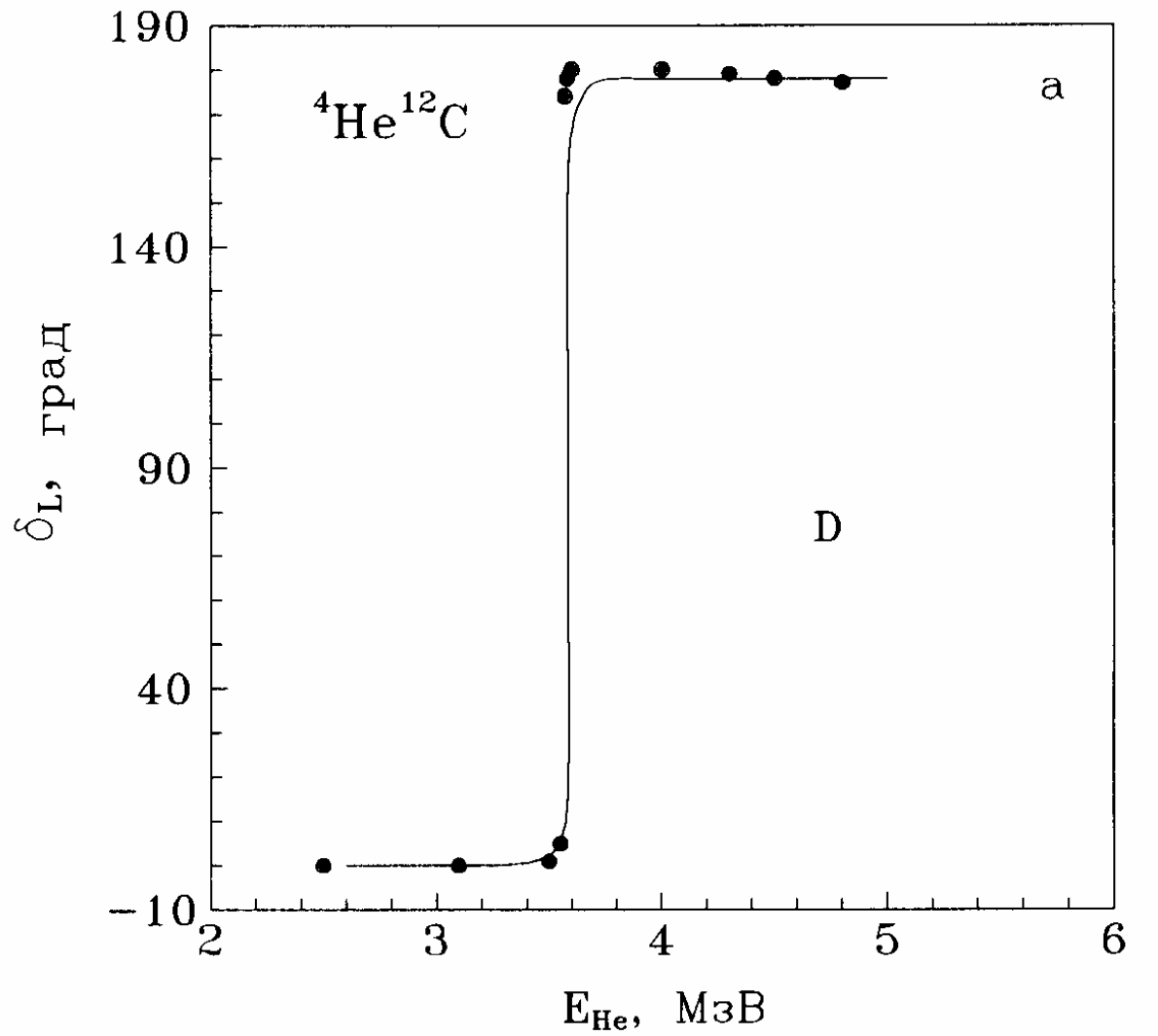

Рис.5.13а. Фазы упругого $^4He^{12}C$ рассеяния. Точки - экспериментальные данные [34], непрерывная кривая - результаты расчетов для потенциалов с параметрами из табл.5.4.

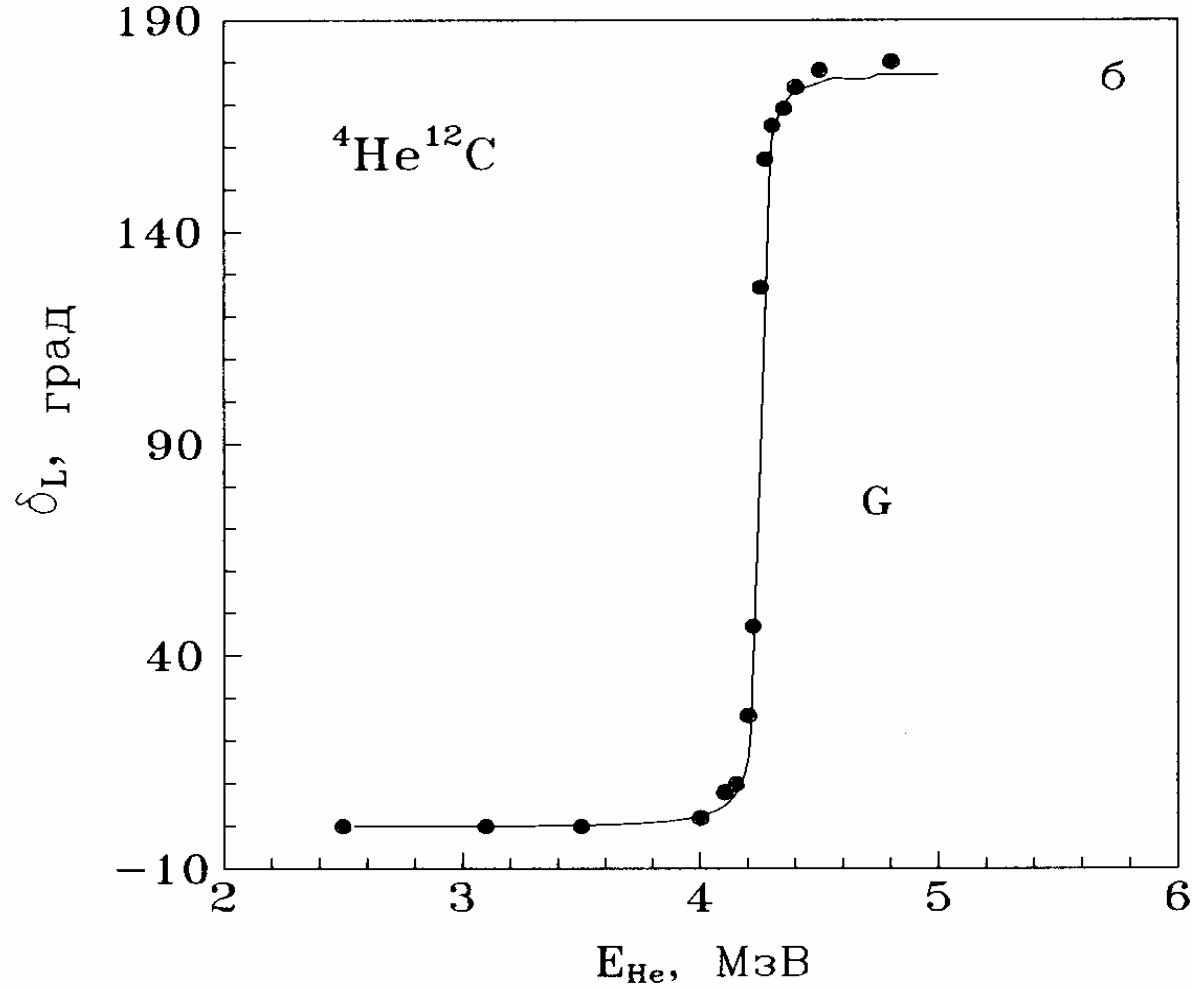

Рис.5.13б. Фазы упругого $^4He^{12}C$ рассеяния. Точки - экспериментальные данные [34], непрерывная кривая - результаты расчетов для потенциалов с параметрами из табл.5.4.

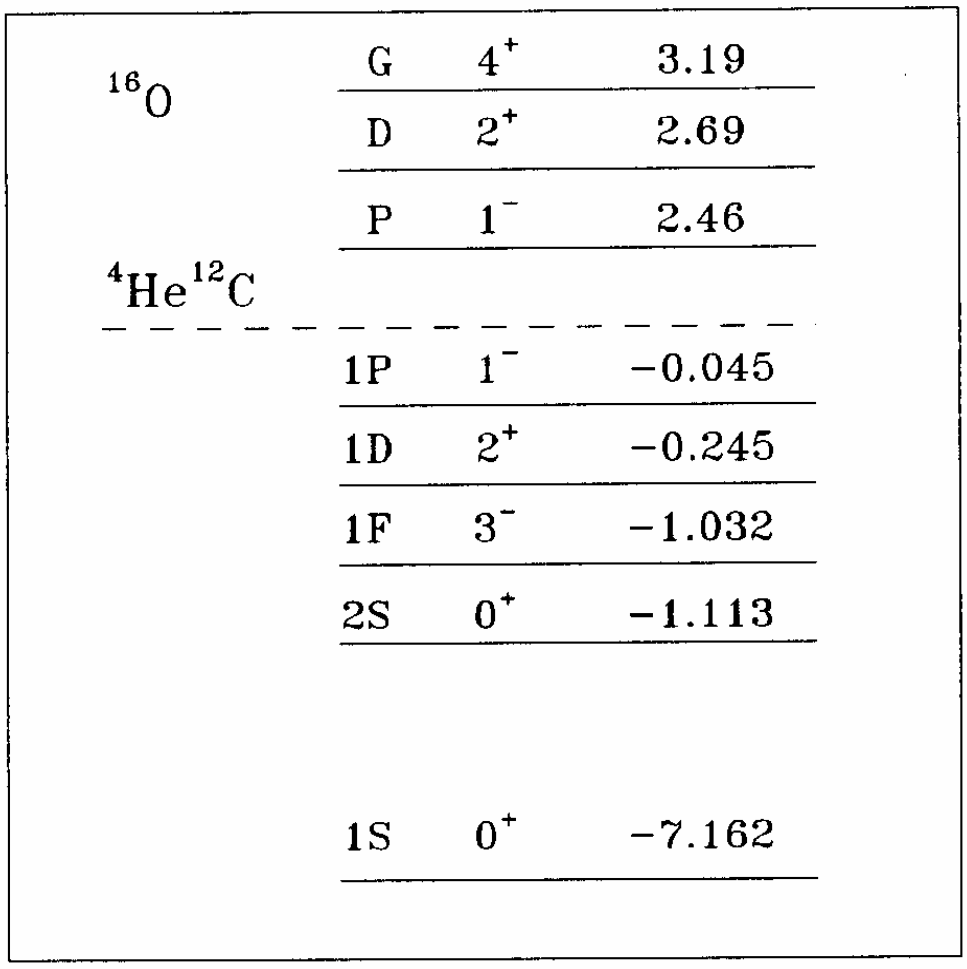

Рис.5.14а. Экспериментальный спектр энергетических уровней ядра $^{16}O$ [36].

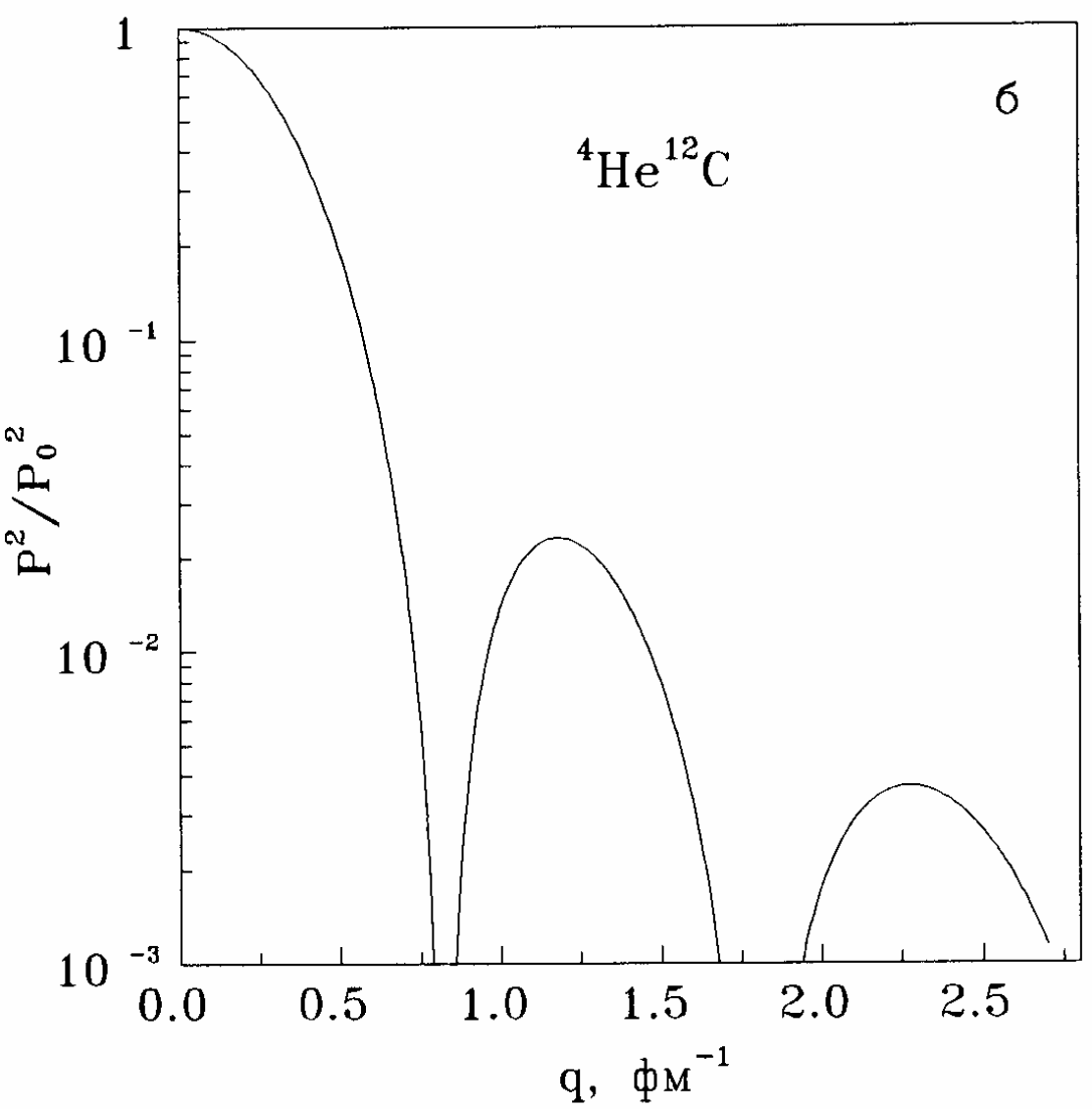

Рис.5.14б. Импульсные распределения $^4He^{12}C$ кластеров в ядре $^{16}O$.

Взаимодействия основного состояния, данные в табл.5.4 под номерами 1 и 2, приводят к зарядовому радиусу 2.66 Фм и 2.72 Фм соответственно при экспериментальной величине 2.710(15) Фм [36] и в целом описывает кулоновский формфактор при малых переданных импульсах.

*Таблица 5.4. Потенциалы взаимодействия в $^4He\,^{12}C$ системе. Теоретические $E_т$ и экспериментальные $E_э$ энергии уровней. $E_{cc}$ - энергии связанных запрещенных состояний. Здесь* $R_0^2 = 1/\alpha$ *в (3.1.1 и $R_c$ = 3.55 Фм.*

| N | L | $V_0$, (МэВ) | $R_0$, (Фм) | $E_т$, (МэВ) | $E_э$, (МэВ) | $E_{cc}$, (МэВ) |
|---|---|---|---|---|---|---|
| 1 | 1S | -176.8 | 2.3 | -7.17 | -7.162 | 35.9; 76.5; 126.6 |
| 2 | 1S | -256.65 | 2.3 | -7.16 | | 37.5; 80.7; 134.3; 197.0 |
| 3 | 2S | -97.78 | 3.0 | -1.11 | -1.113 | 16.0; 38.3; 66.2 |
| 4 | 1P | -104.1 | 2.5 | -0.048 | -0.045 | 19.2; 48.6 |
| 5 | 1D | -88.53 | 3.2 | -0.246 | -0.245 | 13.7; 33.5 |
| 6 | 1F | -191.42 | 1.9 | -1.03 | -1.032 | 38.3 |
| 7 | S | -90.0 | 2.3 | | | 20.7; 52.9 |
| 8 | S | -155.0 | 2.3 | | | 1.37; 25.2; 61.5; 107.7 |
| 9 | P | -145.0 | 2.5 | | | 13.6; 42.2; 79.7 |
| 10 | D | -254.8 | 1.3 | | | 57.0 |
| 11 | D | -434.9 | 1.3 | | | 61.8; 166.9 |
| 12 | F | -140.0 | 2.6 | | | 11.8; 39.4 |
| 13 | G | -111.15 | 2.8 | | | 13.6 |

Требования, предъявляемые к потенциалу основного состояния, позволяют вполне однозначно фиксировать его параметры. В частности, потенциал из работы [35] с параметрами $V_0$ = -110 МэВ и $R_0$ = 2.3 Фм, который содержит два запрещенных состояния, приводит к радиусу 2.6 Фм, а формфактор идет несколько выше экспериментальных данных.

Взаимодействия 1D и 2S, полученные в [33] не существенно отличаются, от приведенных в [35], а это ведет к подобным результатам для вероятности 1D $\to$ 2S перехода. Так, в работе [35], получено значение 63 $e^2$ $Фм^4$. Результаты расчетов [33] вероятностей переходов на различные уровни ядра $^{16}O$ приведены в табл.5.5 вместе с экспериментом [36]. Импульсные распределения кластеров в ядре $^{16}O$, полученные на основе волновых функций основного состояния показаны на рис.5.14б.

При вычислениях формфактора ядра $^{16}O$ использовалась пара-метризация формфактора $^{12}C$ (1.5.5) с параметрами: a=0.31 Фм$^2$, n=4.5, b=1.18 Фм$^2$. Результаты параметризации показаны на рис.5.15а вместе с экспериментальными данными работ [37,38]. Результаты расчета формфактора $^{16}O$ с первым вариантом потенциала основного состояния показаны на рис.5.15б непрерывной линией. Эксперимент по формфактору $^{16}O$ взят из работ [38]. Штриховой линией показан формфактор для потенциала основного состояния с параметрами из работы [35].

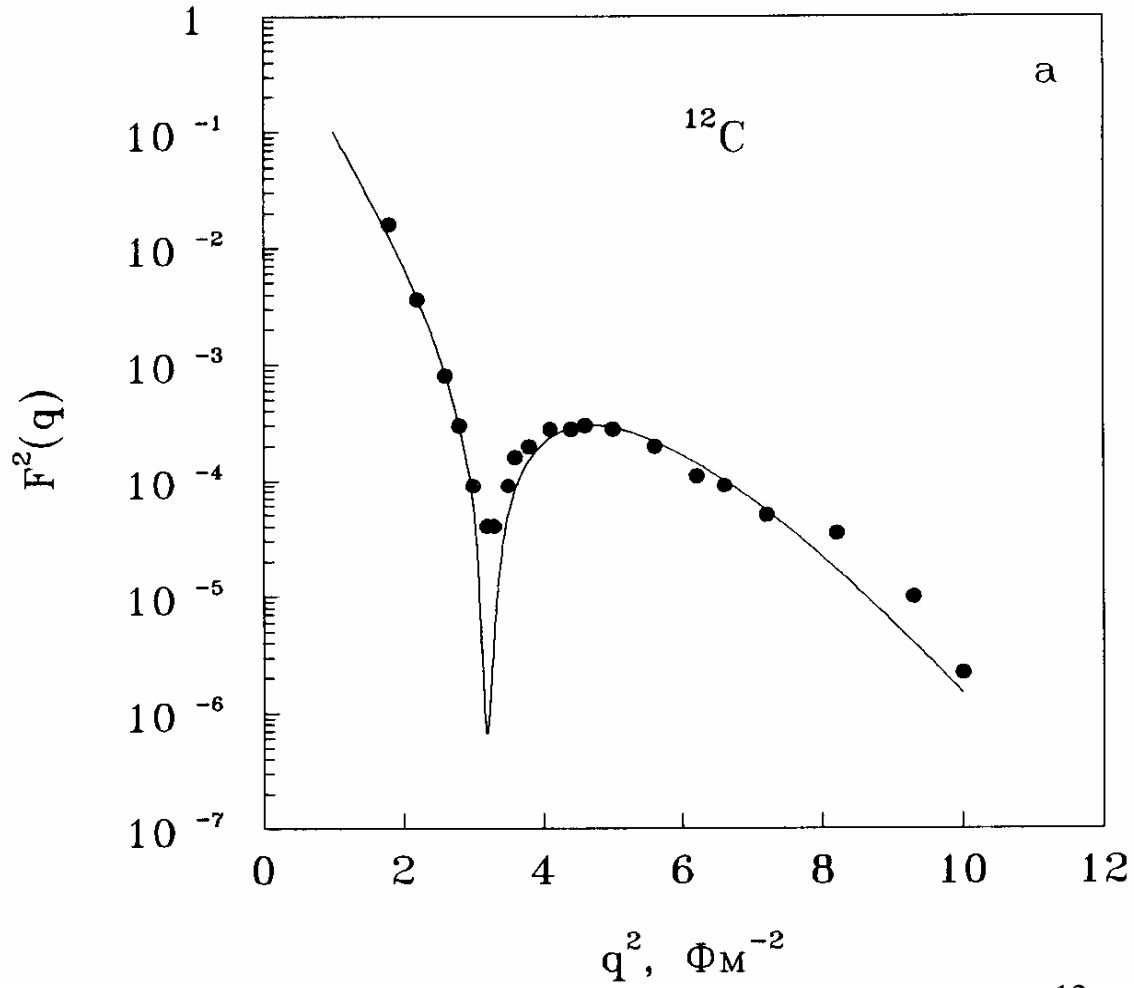

Рис.5.15а. Параметризация формфактора ядра $^{12}C$. Эксперимент из работы [37].

*Таблица 5.5. Теоретические и экспериментальные вероятности переходов в $^4He^{12}C$ системе.*

| Переход $L_i \to L_f$ | $B_т$, (e$^2$ Фм$^4$) Потен.1 | $B_т$, (e$^2$ Фм$^4$) Потен.2 | $B_э$, (e$^2$ Фм$^4$) |
|---|---|---|---|
| 1F → 1S | 15.0 | 11.8 | 13.9(1.2); 4.6(1.5) |
| 1F → 2S | 7.0 | 7.0 | - |
| 1D → 1S | 6.6 | 14.8 | 4.6 ÷ 7.9 |
| 1D → 2S | 67.1 | 67.1 | 63.0; 71(8) |

Таким образом, видно, что, несмотря на существенную простоту использованного здесь метода, он позволяет получить вполне приемлемые результаты при описании характеристик связанного состояния и рассеяния кластеров в ядре $^{16}O$.

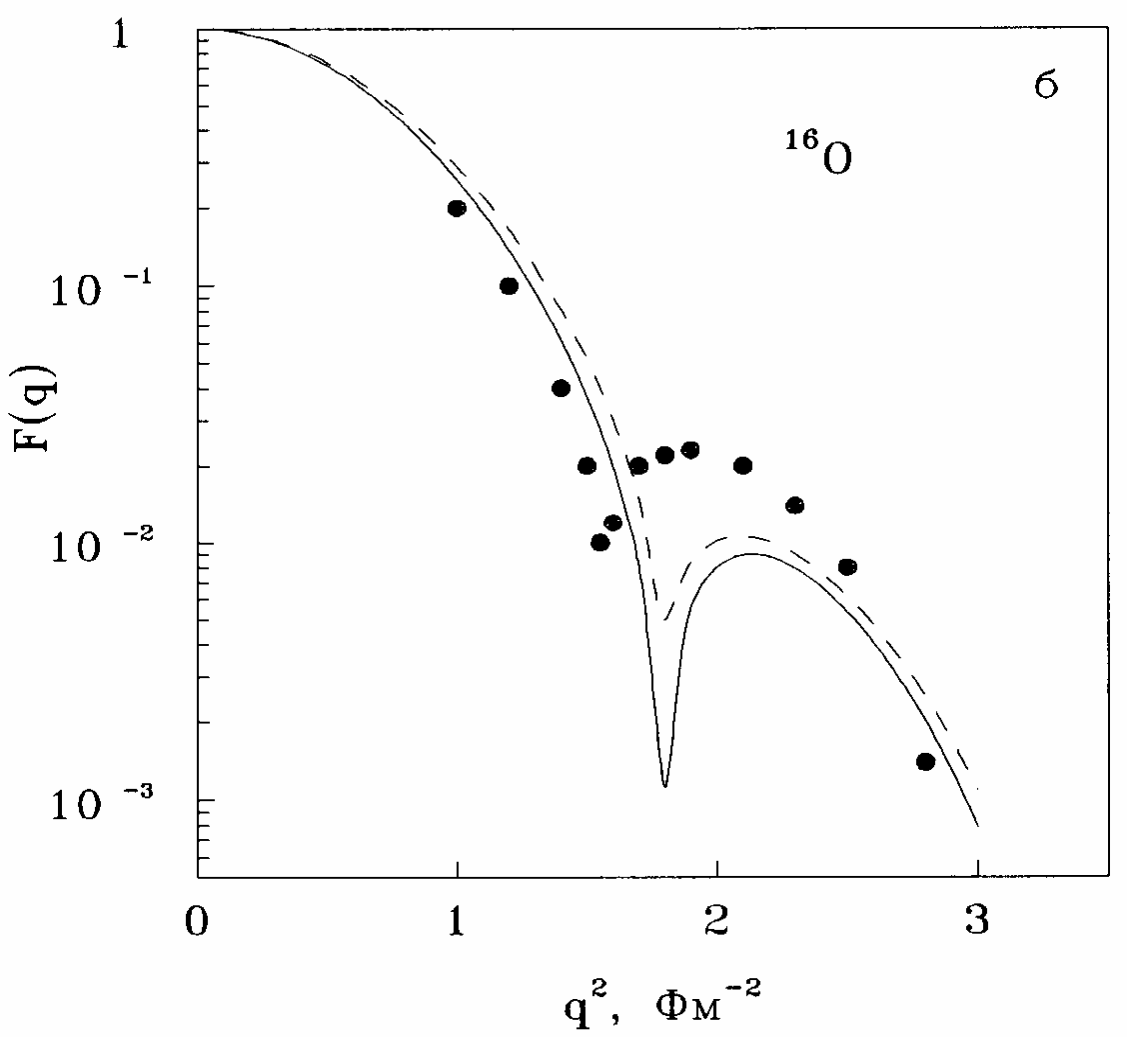

Рис.5.15б. Формфактор ядра $^{16}$O. Эксперимент из работы [38].

## 5.5. Фотопроцессы на ядре $^{16}$O в $^{4}$He$^{12}$C канале

В работах [35] рассматривались фотопроцессы в $^{4}$He $^{12}$C канале ядра $^{16}$O на основе двухкластерной модели с глубокими межкластерными взаимодействиями. Однако, отсутствовал анализ структуры запрещенных состояний в потенциалах, отвечающих различным парциальным волнам. Поэтому представляется интересным вернуться к рассмотрению этого процесса и заново проанализировать все возможные EJ переходы на различные уровни ядра $^{16}$O исключительно в рамках кластерной модели для потенциалов с запрещенными состояниями и подробным исследованием структуры запрещенных и разрешенных состояний, как было сделано в работе [33].

При рассмотрении процессов радиационного захвата учитывались только возможные E2 и E3 переходы. Дипольные переходы в такой модели оказываются запрещенными из-за множителя $( Z_1 / M_1 - Z_2 / M_2 )^2$, который практически равен нулю, даже с учетом отличия масс кластеров от целых чисел. Экспериментальные данные по сечениям захвата были получены в работе [39] для переходов на основное 1S состояние и связанный в $^{4}$He $^{12}$C канале 1D уровень ядра $^{16}$O.

Вычисленные сечения E4 переходов оказались на несколько порядков меньше, чем E3 и их вкладом можно пренебречь [33]. Не-которые из возможных EJ переходов с начального i на конечное f состояния и отвечающие им коэффициенты $P_J^2$ из (1.2.5) с мультипольностью J приведены в табл.5.6. Экспериментальные сечения для переходов на основное состояние имеют явно выраженный максимум при энергии около 2.4 МэВ, что соответствует $1^-$ резонансу при 2.46 МэВ, как видно на рис.5.16а. Такая форма сечения указывает на присутствие E1 процесса, который в данной модели отсутствует.

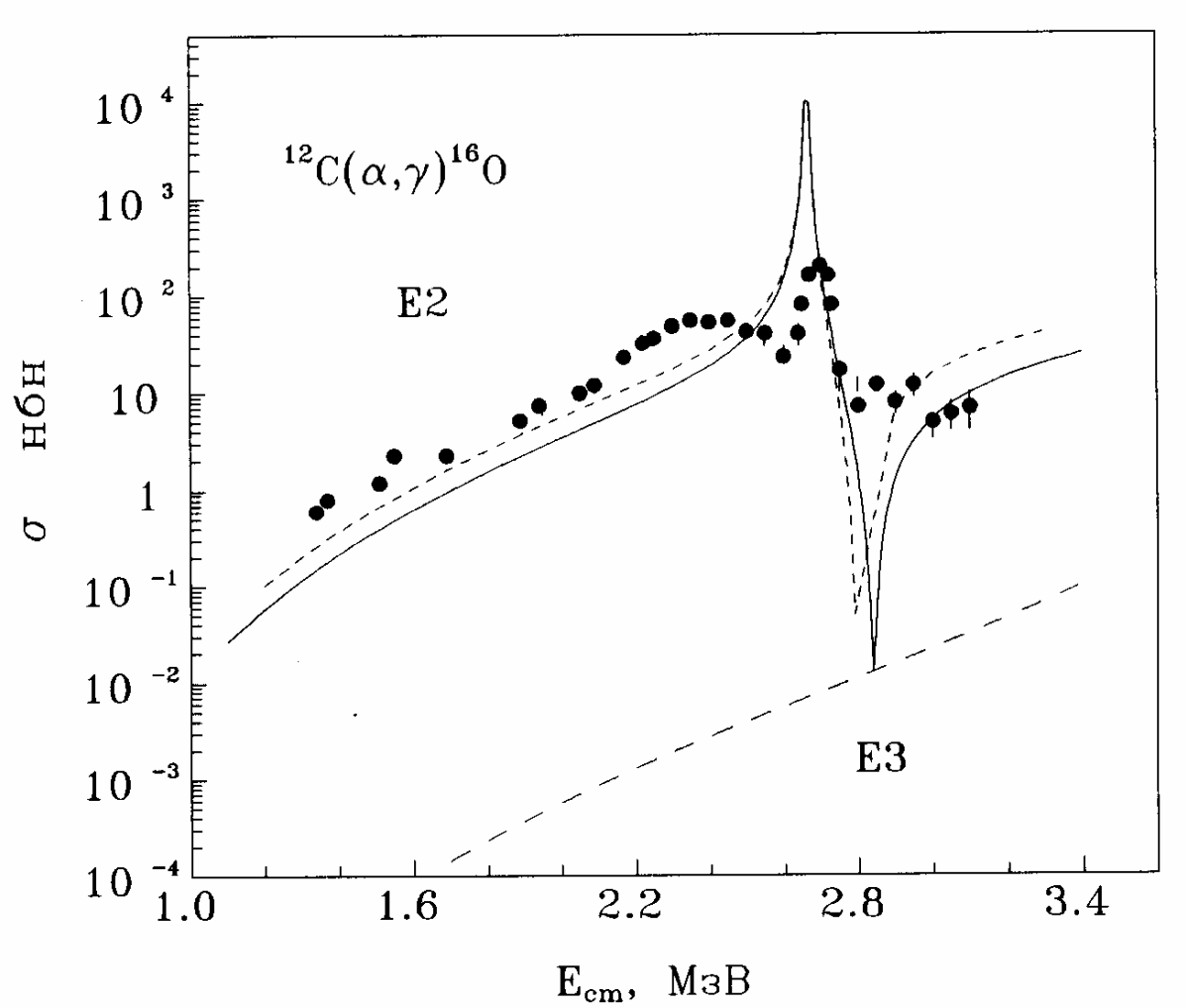

Рис.5.16а. Полные сечения $^4He^{12}C$ захвата на основное состояние ядра $^{16}O$. Кривые - расчеты для потенциалов из табл.5.4 [33]. Эксперимент из работы [39].

*Таблица 5.6. Значения коэффициентов $P_J^2$ для некоторых E J переходов в $^4He^{12}C$ системе.*

| $J_i$ | $J_f$ | J | $P_J^2$ | $J_i$ | $J_f$ | J | $P_J^2$ |
|---|---|---|---|---|---|---|---|
| D | 1S | 2 | 5 | S | 1F | 3 | 7 |
| D | 2S | 2 | 5 | F | 1S | 3 | 7 |
| S | 1D | 2 | 5 | F | 2S | 3 | 7 |
| D | 1D | 2 | 50/7 | | | | |
| G | 1D | 2 | 90/7 | G | 1S | 4 | 9 |
| P | 1F | 2 | 9 | G | 2S | 4 | 9 |
| F | 1F | 2 | 4 | G | 1D | 4 | 900/77 |
| P | 1P | 2 | 6 | | | | |
| F | 1P | 2 | 3 | | | | |

Однако, как видно на рис.5.16а, в сечении имеется и пик при 2.69 МэВ, отвечающий $2^+$ резонансу, который появляется благодаря возможному E2 переходу D→ 1S. На рис.5.16а показаны результаты расчета сечения этого процесса для потенциала основного состояния № 1 (непрерывная линия) и для № 2 (точечная линия).

Для D волны использовался первый вариант взаимодействия. Штриховой линией приведено расчетное сечение для E3 перехода из F волны рассеяния на основное состояние. Это сечение имеет нерезонансный характер, также как фаза рассеяния, показанная на рис.5.13а и плавно поднимается, оставаясь примерно на два порядка меньше, чем E2. Видно, что E2 процесс в принципе объясняет положение пика в сечении, хотя его величина заметно больше экспериментальных данных. Для более полной картины процесса необходимо рассматривать и E1 переход, учет которого может изменить суммарное сечение.

Расчетный астрофизический S фактор имеет примерно одинаковую величину в области 0.3-1.5 МэВ и при 0.3 МэВ равен 0.01 МэВ бн для первого потенциала основного состояния и 0.02 МэВ бн для второго, что заметно меньше экспериментальных величин, обзор которых приведен в работах [35].

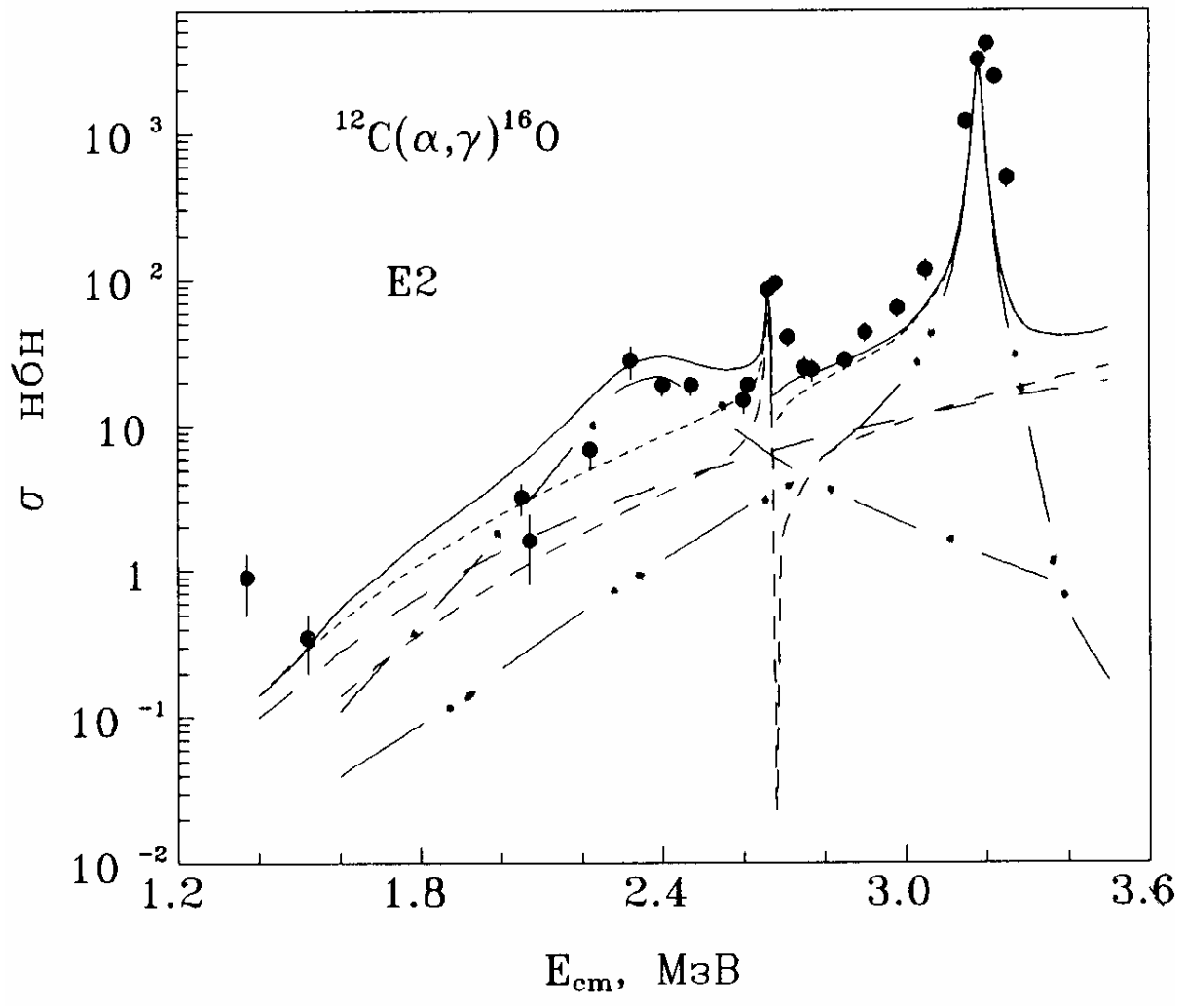

Рис. 5.16б. Полные сечения $^4He^{12}C$ захвата на $2^+$ состояние ядра $^{16}O$. Кривые - расчеты для потенциалов из табл.5.4 [33]. Эксперимент из работы [39].

При переходах на связанное (в этой модели) 1D состояние возможно несколько E2 процессов, в частности, это 1. G → 1D; 2. D → 1D и 3. S → 1D. На рис. 5.16б показаны вычисленные сечения, отвечающие этим процессам (1 - двойной штрих-пунктир, 2 - короткие штрихи и 3 - длинные штрихи). Суммарное сечение показано точечной линией. Параметры потенциалов даны в табл.5.4.

Видно, что рассмотренные переходы, в общем, позволяют объяснить экспериментальные данные, включая положение и высоту пиков при энергиях 2.69 МэВ и 3.19 МэВ, отвечающих D и G резонансам. Однако в экспериментальных сечениях наблюдается и небольшой пик при 2.46 МэВ явно обусловленный переходом из резонансной P волны рассеяния, который может быть объяснен вкладом E1 процесса.

Тем не менее, если допустить, что экспериментальное сечение включает и переход на 1P уровень при энергии -0.045 МэВ (см. рис.5.14а), которая только на 0.2 МэВ отличается от энергии 1D состояния, то можно рассматривать E2 переход типа P → 1P. Сечение такого процесса показано на рис.5.16б штрих - пунктиром. Непрерывной линией дано полное, с учетом этого перехода, сечение, которое практически полностью воспроизводит форму экспериментальных данных [33].

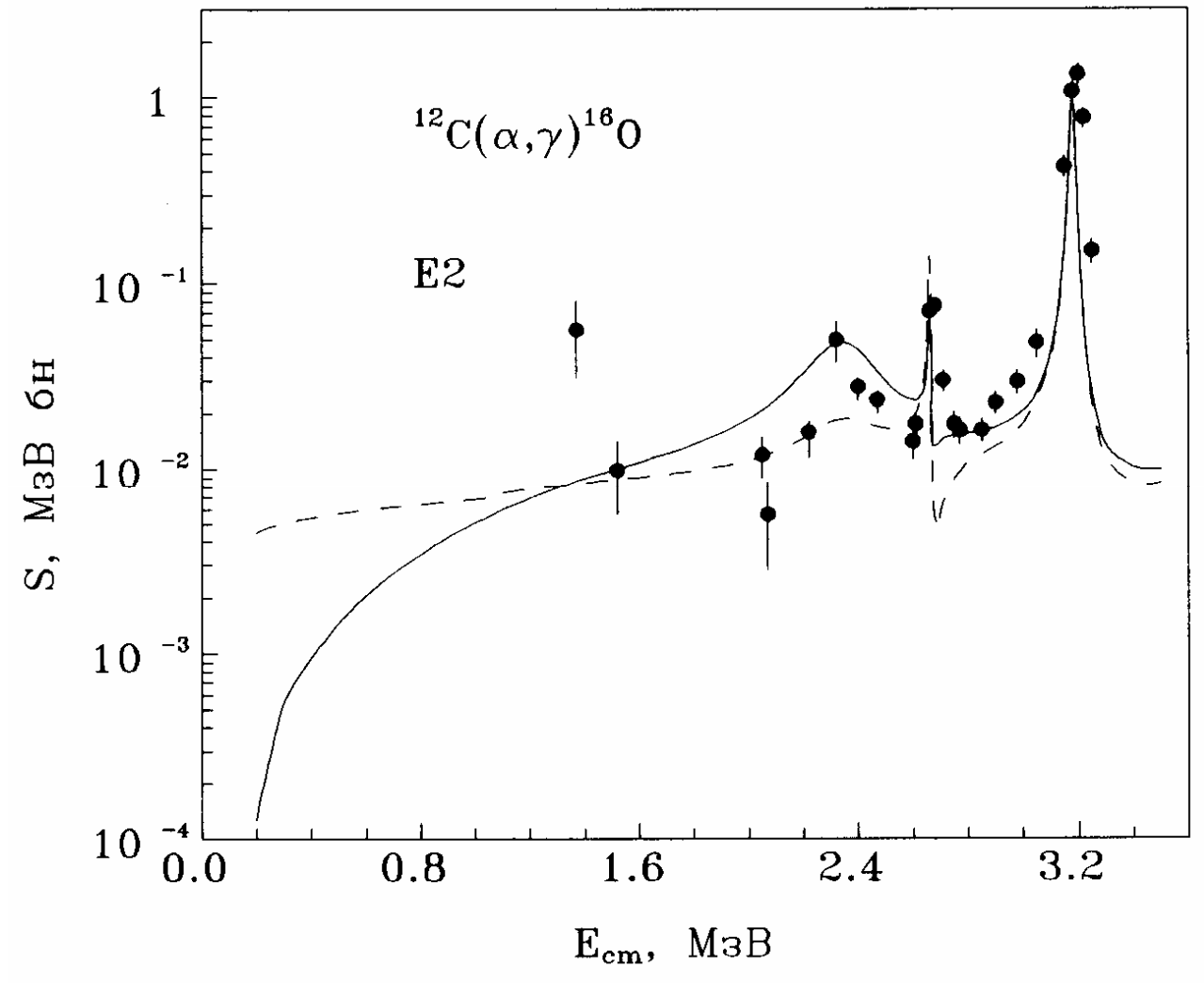

Рис.5.17а. Астрофизический S фактор $^4He^{12}C$ захвата на $2^+$ состояние ядра $^{16}O$. Кривые - расчеты для потенциалов из табл.5.4 [33]. Эксперимент из работы [39].

На рис.5.17а непрерывной линией приведен астрофизический S фактор переходов на 1D состояние, который при 300 кэВ равен примерно 0.001 МэВ бн и резко спадает с уменьшением энергии. Экспериментальные данные приведены в работах [39].

Интересно отметить, что если вместо переходов 2. D $\to$ 1D и 3. S $\to$ 1D рассматривать переходы 4. F $\to$ 1F; 5. D $\to$ 2S и 6. P $\to$ 1F, то, как показано на рис.5.17б вполне удается передать общий вид экспериментальных сечений. Здесь точечная линия обозначает переход 4, штриховая линия переход 5 и штрих-пунктир - 6.

Двойной штрих - пунктир по - прежнему обозначает сечение перехода номер 1. Непрерывная кривая представляет полной сечение с учетом всех этих процессов. Если на рис.5.16б основной вклад в сечения при малых энергиях дает процесс S $\to$ 1D, то здесь определяющим оказывается переход D $\to$ 2S, а вклад других сечений на порядок меньше. На рис.5.17а, штриховой линией показан соответствующий S фактор, который при низких энергиях идет несколько выше, чем S фактор для перехода на 1D уровень и при 300 кэВ дает величину 0.005 МэВ бн [33].

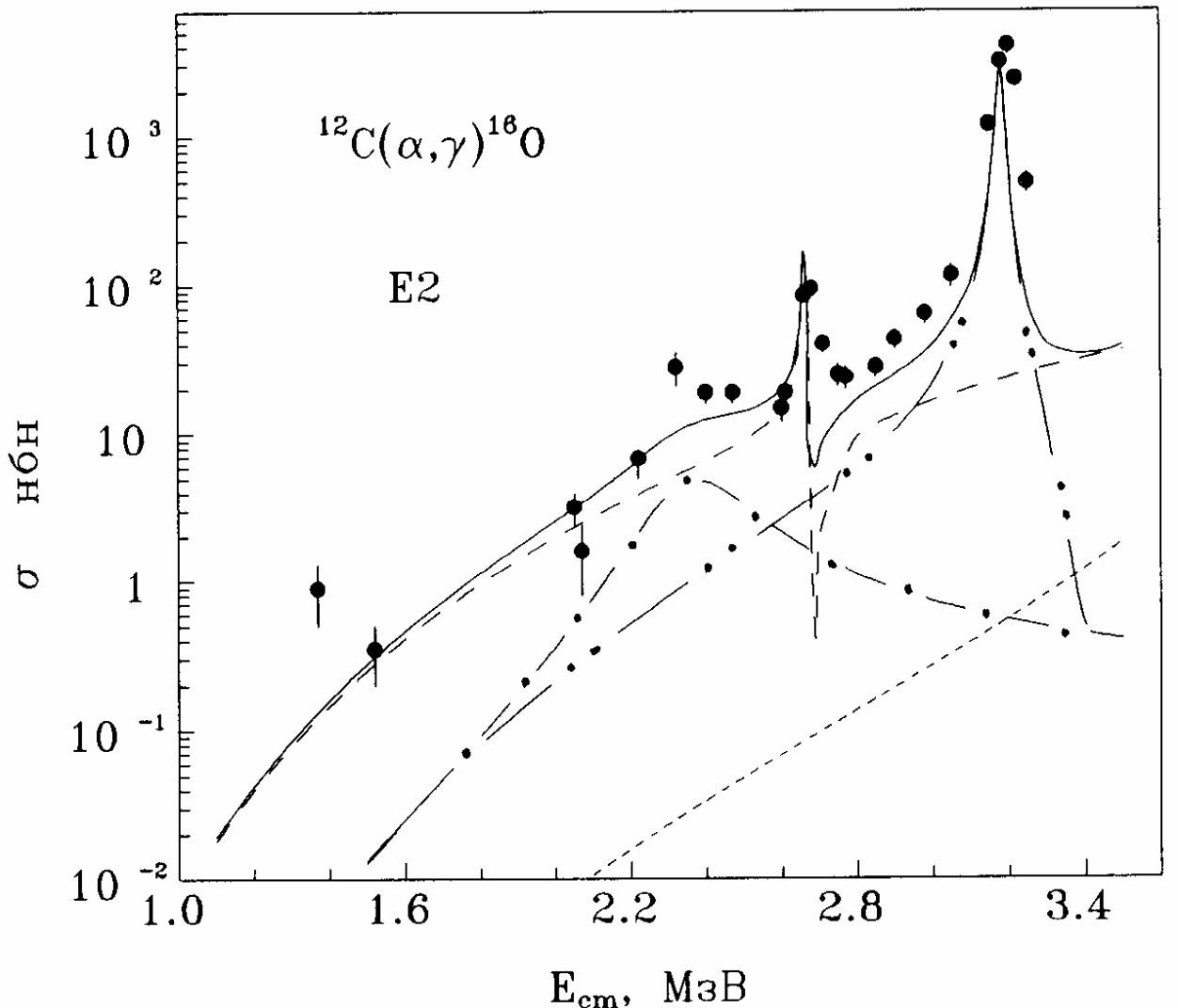

Рис. 5.17б. Полные сечения $^4He^{12}C$ захвата на $2^+$ состояние ядра $^{16}O$. Кривые - расчеты для потенциалов из табл.5.4 [33]. Эксперимент из работы [39].

Тем самым, видно, что используемые потенциалы в $^4He^{12}C$ канале ядра $^{16}O$, согласованные с фазами рассеяния для непрерывного спектра, энергиями и вероятностями радиационных переходов для связанных состояний, вполне позволяют передать полные сечения радиационного захвата на $2^+$ уровень только на основе E2 переходов.

Сечения захвата на основное состояние описываются только в области $2^+$ резонанса. При других энергиях основной вклад в сечения дает, по-видимому, E1 переход $P \rightarrow 1S$, отсутствующий в рассматриваемой кластерной модели.

---

1. Petitjan C., Brown L., Seylyr R.G. - Nucl. Phys., 1969, v.A129, p.209.

2. Hofman H., Mertelmeier T. - Nucl. Phys., 1983, v.A410, p.208.

3. Fujiwara Y. , Tang Y.C. - Phys. Rev., 1983. v.C23, p.2457; Phys. Lett., 1983, v.B131, p.261.

4. Ling Y. - Phys. Ener. Fort. Phys. Nucl., 1984, v.8, p.227.

5. Fujiwara Y., Tang Y.C. - Phys. Rev., 1984, v.C29, p.2025.

6. Fujiwara Y., Tang Y.C. - Phys. Rev., 1985, v.C31. p.342; Fujiwara Y., Tang Y.C. - Phys. Rev., 1983, v.C23, p.2457.

7. Дубовиченко С.Б., Джазаиров - Кахраманов А.В., Сахарук А.А. - ЯФ. 1993, т.56, №.8, с.90.

8. Van Niftric G.J.C., Brockman K.W., Van Oers W.T.H. - Nucl. Phys., 1971, v.A174, p.173; Hausser O. et al. - Nucl. Phys., 1973, v.A212, p.613; Green S. et al. - Phys. Rev., 1971, v.A4, p.251; Sundholm D. et al. - Chem. Phys. Lett., 1984, v.112, p.1; Vermeer W. et al. - Austr. J. Phys., 1984, v.37, p.273; Phys. Lett., 1984, v.B138, p.365; Weller A. et al. - Phys. Rev. Lett., 1985, v.55, p.480; Rand R., Frosch R., Yearian M.R. - Phys. Rev., 1966, v.144, p.859; Bamberger A. et al. - Nucl. Phys., 1972, v.194, p.193; De Vries H. et al. - Atom Data and Nucl. Data Tables., 1987, v.36, p.495; Suelzle L.R., Yearian M.R., Crannell H. - Phys. Rev., 1967, v.162, p.992.

9. Abbandanno U et al. - Nuovo Cim., 1970, v.A66, p.139.

10. Knox H.D., Lane R.O. - Nucl. Phys. 1983, v.A403, p.205.

11. Rosario - Garcia E., Benenson R.E. - Nucl. Phys., 1977, v.A275, p.453.

12. Mercker F. et al. - Nucl. Phys., 1972, v.A182, p.428.

13. Freyere C. et al. - Phys. Rev., 1954, v.93, p.1086.

14. Afzal S.A., Ahmad A.A.Z., Ali S. - Prepr. Int. Cent. Theor. Phys. Int. Atom. Energy. Agency., ¹ 211, 1968; Rev. Mod. Phys., 1969, v.41, p.247.

15. Baz A.I., Goldberg V.Z., Gridnev K.A., Semjonov V.M., Hefter E.F. - Z. Phys., 1977, v.A280, p.171.

16. Ali S., Bodmer A. - Nucl. Phys., 1966, v.80, p.99.

17. Marguez L. - Phys. Rev., 1983, v.C28, p.2525.

18. Rahman M., Ali S., Afzal S.A. - Lett. Nuovo Cim., 1973, v.6, p.107.

19. Schmid E., Saito S., Fiedeldey H. - Z. Phys., 1983, v.A306, p.37.

20. Дубовиченко С.Б., Джазаиров-Кахраманов А.В. - ЯФ, 1992, т.55, № 11, с.2918; 1993, т.56. № 2, с.87; ЯФ, 1994, т.57, № 5, с.590.

21. Ajzenberg-Selove F. - Nucl. Phys., 1979, v.A320, p.1.

22. Ling Y. et al. - Phys. Ener. Fort. Phys. Nucl., 1985, v.9, p.236.

23. LeMere M., Tang Y.C. - Nucl. Phys., 1980, v.A339, p.43; LeMere M., Tang Y.C., Kanada H. - Phys. Rev., 1982, v.C24, p.2902.

24. Brown L., Steiner E., Arnold L.G., Seyler R.G. - Nucl. Phys., 1973, v.A206, p.353.

25. Stowe H., Zahn W. - Nucl. Phys., 1977, v.A286, p.89.

26. Fujiwara Y. , Tang Y.C. - Phys. Lett., 1989, v.B222, p.311.

27. Stowe H., Zahn W. - Nucl. Phys., 1977, v.A289, p.317.

28. Fujiwara Y., Tang Y.C. - Phys. Rev., 1990, v.C41, p.28.

29. Дубовиченко С.Б. - ЯФ, 1997, т.60, № 2, с.254.

30. Неудачин В.Г., Смирнов Ю.Ф. - Нуклонные ассоциации в легких ядрах. М., Наука, 1969., 414с.; Немец О.Ф., Неудачин В.Г., Рудчик А.Т., Смирнов Ю.Ф., Чувильский Ю.М. - Нуклонные ассоциации в атомных ядрах и ядерные реакции многонуклонных передач. Киев, Наукова Думка, 1988, 488с.

31. Itzykson C. - Rev. Mod. Phys., 1966, v.38, p.95.

32. Berman B.L., Fultz S. - Rev. Mod. Phys., 1975. v.47, p.713.

33. Дубовиченко С.Б. - ЯФ, 1996, т.59, № 3, с.447.

34. Jones C.M., Phillips G.C., Harris R.W., Beckner E.H. - Nucl. Phys., 1962, v.37, p.1.

35. Langanke K. - Nucl. Phys., 1985, v.A439, p.384; Langanke K., Koonin S.E. - 1983, v.A410, p.334.

36. Ajzenberg-Selove F. - Nucl. Phys., 1982, v.A375, p.1.

37. Crannell H. - Phys. Rev., 1966, v.148, p.1107.

38. McDonald L., Uberall H., Numrich S. - Nucl. Phys., 1970, v.A147, p.541; Sick I., Mc carthy J.S. - Nucl. Phys., 1970, v.A150, p.631.

39. Kettner K.U., Becker H.W., Buchmann L., Gorres J., Krawinkel H., Rolfs C., Schmalbrock P., Trautvetter H.P., Vlieks A. - Z. Phys. 1982. v.A308. p.73.

# 6. ЯДЕРНЫЕ СИСТЕМЫ С ТЕНЗОРНЫМИ СИЛАМИ

В этой главе мы рассмотрим результаты, которые были получены для двух ядерных систем, а именно, нуклон - нуклонной и кластерной $^4He^2H$ системы, когда в потенциале взаимодействия присутствуют тензорные силы и имеются запрещенные состояния. Но вначале остановимся на некоторых вариантах чисто центральных нуклон - нуклонных взаимодействий с запрещенными состояниями.

## 6.1. Центральные нуклон - нуклонные взаимодействия

Давно известно, что потенциальное описание кластер - кластерного рассеяния может основываться на глубоких взаимодействиях с запрещенными состояниями [1]. Благодаря присутствию таких состояний, фазы рассеяния подчиняются обобщенной теореме Левинсона и оказываются положительными при всех энергиях, стремясь к нулю в борновской области больших энергий. Такой подход, как было показано в предыдущих главах, позволяет построить целый ряд потенциалов для кластерных систем типа $^4He^2H$, $^4He^3H$, $^3He^3H$ [2]. Они правильно описывают не только фазы упругого рассеяния при низких энергиях, но и основные характеристики легких ядер $^6Li$ и $^7Li$, включая фотоядерные реакции в соответствующих кластерных моделях.

Вполне успешное совместное описание фаз рассеяния и некоторых характеристик основных состояний таких ядер оказывается возможным, не только благодаря большой степени их кластеризации в двухчастичные каналы, но и вследствие того, что все состояния оказываются чистыми по орбитальным схемам Юнга. В состояниях рассеяния и основном состоянии ядра разрешена только одна определенная орбитальная схема.

В более легких кластерных системах типа $N^2H$, $N^3H$, $^2H^2H$, где в непрерывном спектре разрешены две возможные орбитальные симметрии, в то время, как основному состоянию сопоставляется только одна схема Юнга [3,4], необходимо выделение чистой фазы рассеяния по которой параметризуется потенциал взаимодействия, применимый для расчетов характеристик основных состояний ядер $^3He$ и $^4He$ в кластерных моделях [3].

Потенциалы с запрещенными состояниями использовались и для описания характеристик нуклон - нуклонных взаимодействий.

По - видимому, впервые концепция запрещенных состояний была применена к нуклон - нуклонным взаимодействиям около двадцати лет назад в работах [5], где предложены синглетные нуклон - нуклонные потенциалы типа Юкавы, позволяющие, в общем, правильно описать S, D и G фазы рассеяния при энергиях до 300 МэВ.

В последствии были параметризованы центральные синглетные и триплетные гауссовы взаимодействия и показано, что триплетные потенциалы с запрещенными состояниями можно одновременно согласовать с фазами рассеяния до 200-300 МэВ, низкоэнергетическими характеристиками и некоторыми свойствами дейтрона [6].

Дальнейшее развитие эти результаты получили в работах [7], где предложены гауссовы синглетные и триплетные потенциалы с тензорной компонентой и учетом потенциала однопионного обмена (OPEP), позволяющие получить хорошее описание фаз нуклон - нуклонных взаимодействий и характеристик дейтрона. Волновые функции  дейтрона и рассеяния в таких потенциалах имеют узел на расстояниях порядка 0.56 Фм, практически неподвижный при изменении энергии.

В работах [8] предложены центральные гауссовы взаимодействия, правильно описывающие общий ход фаз рассеяния при энергиях до 3 ГэВ и дано подробное обоснование глубоких нуклон - нуклонных потенциалов с точки зрения кварковой структуры нуклонов. Приводится общее качественное сравнение результатов с некоторыми кварковыми моделями, в которых получается узловой вид волновых функций.

Как показано в [5-8], потенциалы с запрещенными состояниями позволяют существенно сократить число варьируемых, подгоночных параметров. Так, триплетный потенциал из работ [7] имеет только три параметра, в то время, как взаимодействия с кором, например, триплетный потенциал Рейда [9], содержит около десятка подгоночных величин.

В приведенных выше работах, в основном, использовался гауссовый тип взаимодействий. Поэтому представляет интерес рассмотрение некоторых других видов центральных потенциалов, например, экспоненциальных или Юкава, чтобы сравнить между собой результаты расчетов различных характеристик рассеяния и связанного состояния дейтрона и сопоставить их с результатами для известных потенциалов с кором.

Если рассматривать дейтрон, как шести - кварковую систему, то его основному состоянию соответствует орбитальная симметрия {42}, совместимая с L=0,2, а связанное состояние с орбитальной симметрией {6} оказывается запрещенным. Запрещено также состояние с симметрией {51} в нечетных волнах [7]. Это значит, что потенциал триплетного S взаимодействия имеет низколежащее запрещенное состояние, а основное состояние дейтрона соответствует второму связанному состоянию в таком потенциале. Связанный, но запрещенный уровень существует и в P взаимодействии. Поэтому триплетные S фазы рассеяния, в соответствии с обобщенной теоремой Левинсона [7], начинаются с $360^o$, а синглетные со $180^o$, имея подъем примерно до $220^o$. Триплетные и синглетные P фазы также должны начинаться со $180^o$. В D волне присутствует одно связанное разрешенное состояние, которое вместе с S волной определяет основное состояние дейтрона.

При расчетах характеристик нуклон - нуклонной системы, в случае центральных сил, можно использовать обычное радиальное уравнение Шредингера в виде (1.3.16) с асимптотикой волновых функций рассеяния, выраженной через функции кулоновского типа (1.2.7). Асимптотика волновой функции для связанных состояний при L=0 представляется в обычном виде, который определен первым уравнением в выражениях (1.5.1). Численная волновая функция сшивается с асимптотикой для связанных состояний и рассеяния на расстояниях порядка 15-20 Фм. Решение задачи рассеяния и связанных состояний выполняется, например, конечно - разностным методом, изложенным в [10] и в первой главе настоящей книги.

В расчетах использовались двухпараметрические потенциалы трех типов [11]:

гауссовый

$$V(r)=-V_g \exp(-\alpha_g r^2) , \qquad (6.1.1)$$

экспоненциальный

$$V(r)=-V_e \exp(-\beta_e r) , \qquad (6.1.2)$$

и Юкавы

$$V(r)=-V_y \exp(-\gamma_y r)/r . \qquad (6.1.3)$$

Экспериментальные данные по фазам нуклон - нуклонного рассеяния взяты из работ [12]. Параметры потенциалов подбирались так, чтобы один набор параметров описывал одну парциальную фазу рассеяния при энергиях до 400-500 МэВ. При нулевой энергии рас-сматривалась длина рассеяния $a_0$ [11]

$$1/a_0 = - \lim_{k \to 0} (k \, Ctg \, \delta_0 ) \qquad (6.1.4)$$

и триплетный эффективный радиус $r_0$ , выраженный через энергию связи дейтрона и длину рассеяния

$$1/a_0 = k_0 - 1/2 \, k_0^2 \, r_0 . \qquad (6.1.5)$$

Параметры, полученных таким образом потенциалов, и энергии запрещенных состояний приведены в табл.6.1 [11]. Для гауссовых S взаимодействий найдено два варианта параметров. Первый из них получен на основе фаз при больших энергиях, а второй из описания только низкоэнергетических характеристик рассеяния. Неправильная форма “хвоста” гауссового потенциала, который определяет асимптотическое поведение волновой функции, не позволяет совместить описание фаз при малых и больших энергиях на основе одних параметров. Для триплетной D фазы с J=1 использовались отталкивающие потенциалы, поскольку не учитывается тензорная компонента взаимодействия.

*Таблица 6.1. Параметры пр потенциалов, разного типа и энергии запрещенных состояний [11].*

| $L_J$ | $V_g$, (МэВ) | $\alpha_g$, (Фм$^{-2}$) | $E_{зс}$, (МэВ) | $V_e$, (МэВ) | $\beta_e$, (Фм$^{-1}$) | $E_{зс}$, (МэВ) | $V_y$, (МэВ) | $\gamma_y$, (Фм$^{-1}$) | $E_{зс}$, (МэВ) |
|---|---|---|---|---|---|---|---|---|---|
| $^1S$ | 1128.6 | 1.55 | 448.3 | 2990 | 3.1 | 514.5 | 3141.5 | 3.0 | 951.6 |
| | 1131.5 | 1.55 | 447 | | | | | | |
| $^1P$ | 1050 | 1.3 | 126.7 | 3600 | 3.1 | 78.0 | 625 | 1.4 | 37.9 |
| $^1D$ | 70 | 0.6 | --- | 320 | 2.0 | --- | 200 | 1.5 | --- |
| $^3S$ | 1406.1 | 1.8 | 581.8 | 3094.7 | 3.0 | 581 | 3360.8 | 3.1 | 1197.6 |
| | 949.7 | 1.2 | 395.9 | | | | | | |
| $^3P_0$ | 830 | 0.8 | 160.9 | 2500 | 2.3 | 125.5 | 660 | 1.3 | 82.5 |
| $^3P_1$ | 990 | 1.2 | 125.6 | 4400 | 3.5 | 75.7 | 660 | 1.5 | 38.1 |
| $^3P_2$ | 3200 | 3.0 | 46.5 | 12300 | 5.0 | 686.6 | 3750 | 3.8 | 875.9 |
| $^3D_1$ | -120 | 0.4 | --- | -530 | 1.6 | --- | -360 | 1.1 | --- |
| $^3D_2$ | 63 | 0.3 | --- | 160 | 1.15 | --- | 125 | 0.7 | --- |
| $^3D_3$ | 11 | 0.3 | --- | 55 | 1.5 | --- | 14 | 0.4 | --- |

Решение уравнения Шредингера на связанные состояния для потенциала Юкавы начинается с 0.003 Фм, что, по-видимому, можно рассматривать как введение радиуса обрезания, так как все результаты заметно зависят от этой величины. Для других потенциалов использовался шаг 0.01 Фм с числом шагов конечно - разностной сетки 2500. Изменение шага от 0.01 Фм до 0.005 Фм несущественно сказывается на результатах вычисления фаз рассеяния и изменяет энергию дейтрона не более чем на 0.3%-0.5%. Асимптотическая константа $C_0$ для волновых функций связанных состояний оставалась постоянной в области 6-25 Фм [11].

В табл.6.2, вместе с экспериментальными данными из работ [12,13], приведены характеристики дейтрона и низкоэнергетические параметры, вычисленные с описанными выше потенциалами взаимодействий. В последнем столбце приведены значения R, при которых волновая функция основного состояния имеет узел [11]. В табл.6.3 дано сравнение результатов расчета фаз для потенциала Рейда, приведенных в работе [9], и полученных в [11] на основе используемого здесь конечно - разностного метода.

Результаты расчетов фаз рассеяния в триплетном состоянии представлены на рис.6.1-6.3 (непрерывные линии) для первого варианта гауссового, экспоненциального и юкавского взаимодействий соответственно.

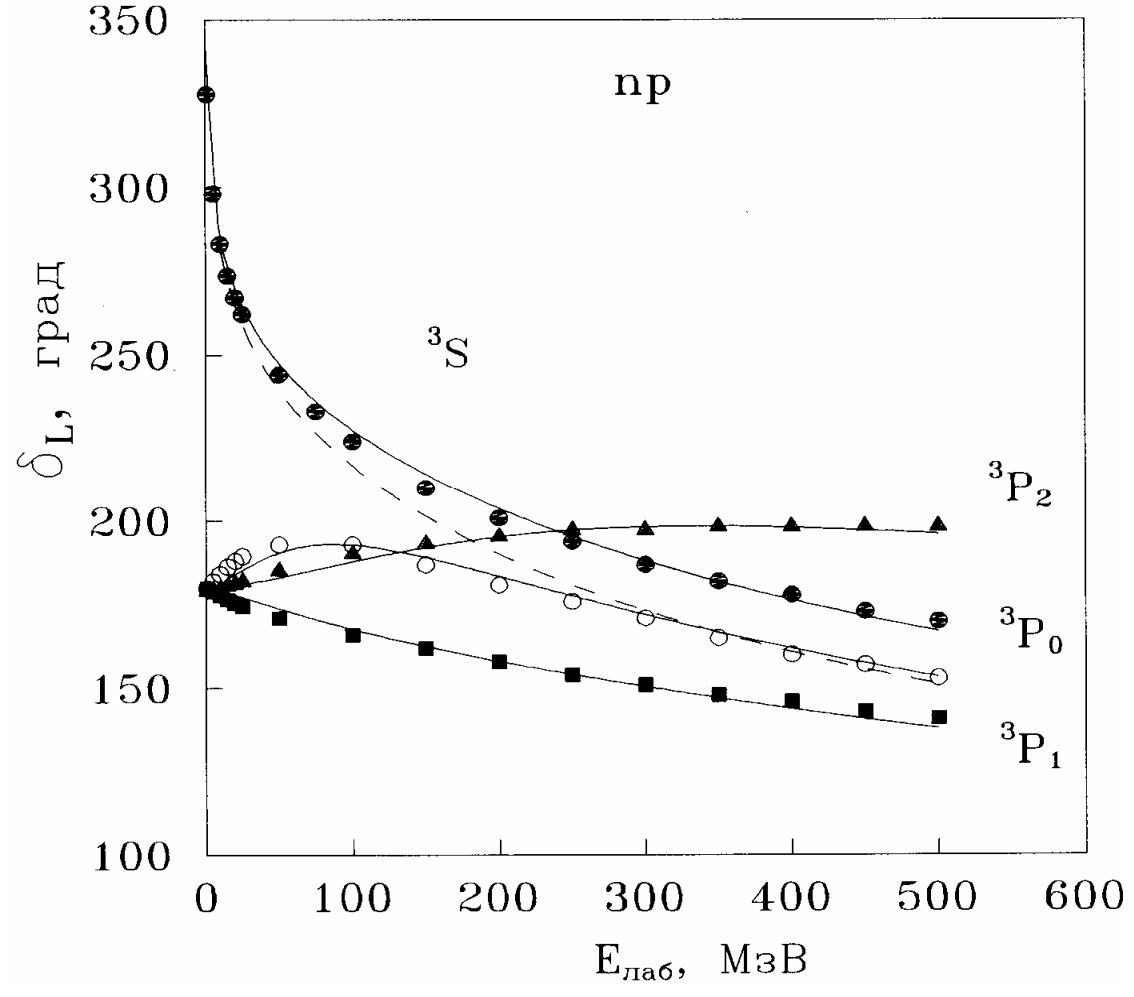

Рис.6.1а. Фазы упругого np рассеяния. Кривые - расчеты для гауссовых потенциалов из табл.6.1. Точки, треугольники, кружки и квадраты - экспериментальные данные из работ [12].

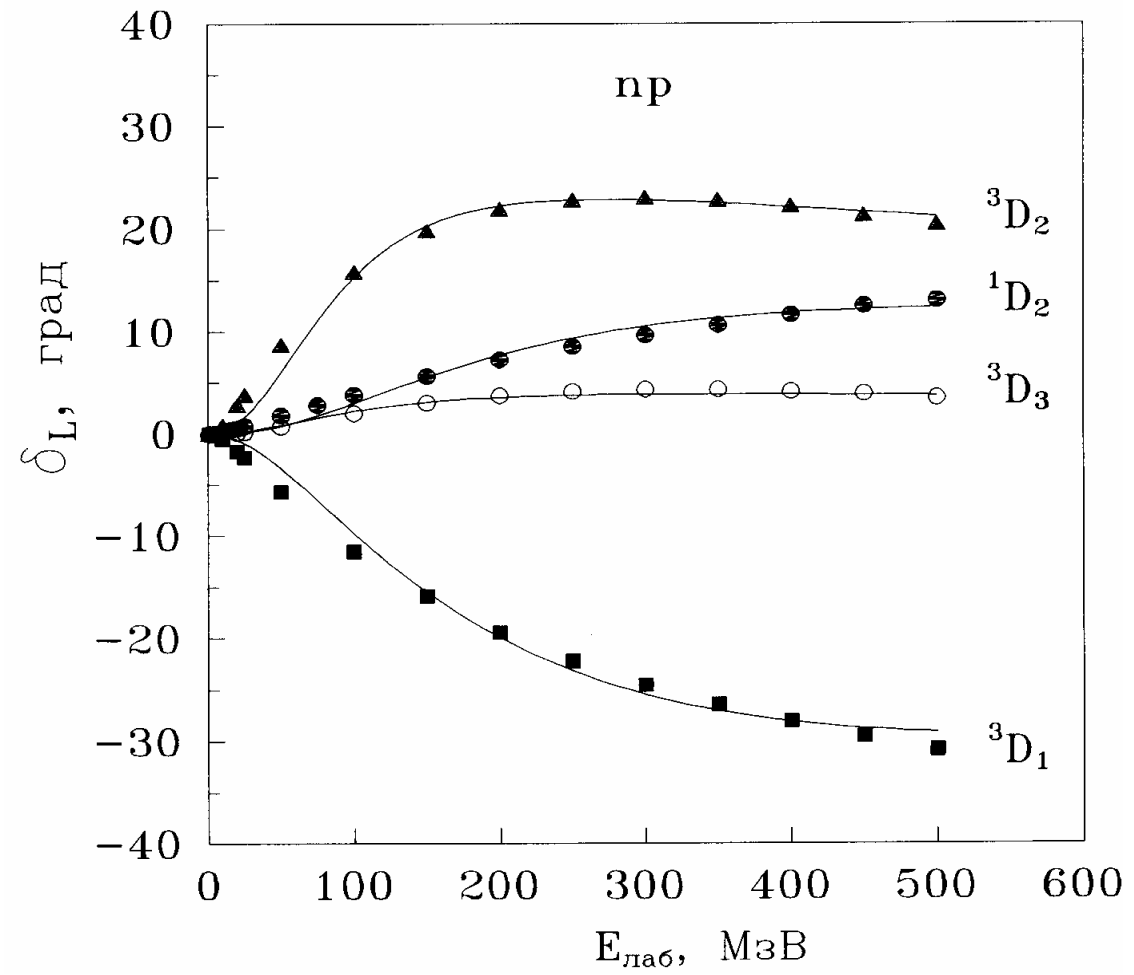

Рис.6.1б. Фазы упругого np рассеяния. Кривые - расчеты для гауссовых потенциалов из табл.6.1. Точки, треугольники, кружки и квадраты - экспериментальные данные из работ [12].

На рис.6.1а штриховой линией показаны фазы второго варианта гауссового S потенциала, который лучше передает низкоэнергетические параметры, но расчетные фазы при больших энергиях идут несколько ниже экспериментальных данных. Штриховыми линиями на рис.6.2 даны триплетные P и синглетная D фазы потенциала Рейда с мягким кором [9].

*Таблица 6.2. Характеристики дейтрона и параметры низкоэнергетического np рассеяния [11].*

| Потенциалы | E, (МэВ) | $R_r$, (Фм) | $R_f$, (Фм) | $C_0$, (Фм$^{1/2}$) | $a_{0t}$, (Фм) | $r_{0t}$, (Фм) | $a_{0S}$, (Фм) | R, (Фм) |
|---|---|---|---|---|---|---|---|---|
| Гаусс.1 | 2.223 | 1.95 | 1.99 | 1.23 | 5.25 | 1.49 | -18.6 | 0.60 |
| Гаусс.2 | 2.223 | 2.02 | 2.06 | 1.29 | 5.41 | 1.75 | -23.1 | 0.71 |
| Экспон. | 2.225 | 2.01 | 2.05 | 1.28 | 5.37 | 1.69 | -24.5 | 0.53 |
| Юкава | 2.223 | 2.00 | 1.73 | 1.27 | 5.33 | 1.64 | -23.6 | 0.51 |
| Рейд [9] | 2.2246 | | | | 5.39 | 1.72 | -17.1 | |
| Экспер. [12,13] | 2.22464 | 1.9660 (68) | | 1.30(2) | 5.414 (5) | 1.750 (5) | -23.719 (13) | |

Синглетные фазы второго гауссового потенциала показаны на рис.6.4а непрерывной линией, штриховой линией даны результаты для экспоненциального взаимодействия, а фазы потенциала Юкавы совпадают с непрерывной линией. Точечной линией показаны фазы для синглетных потенциалов Рейда с мягким кором [9].

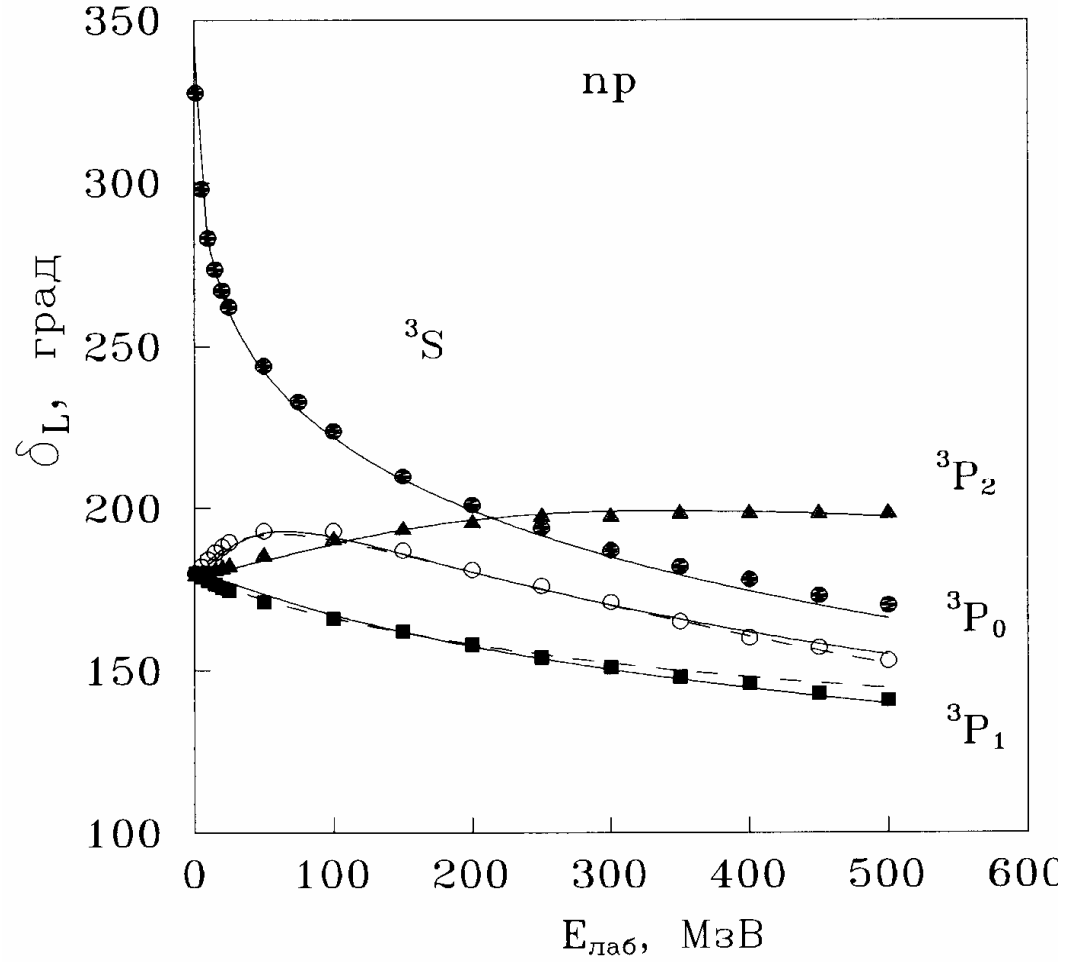

Рис.6.2а. Фазы упругого np рассеяния. Непрерывные кривые - расчеты для экспоненциальных потенциалов из табл.6.1., штриховые линии - расчеты для потенциала Рейда с мягким кором [9].

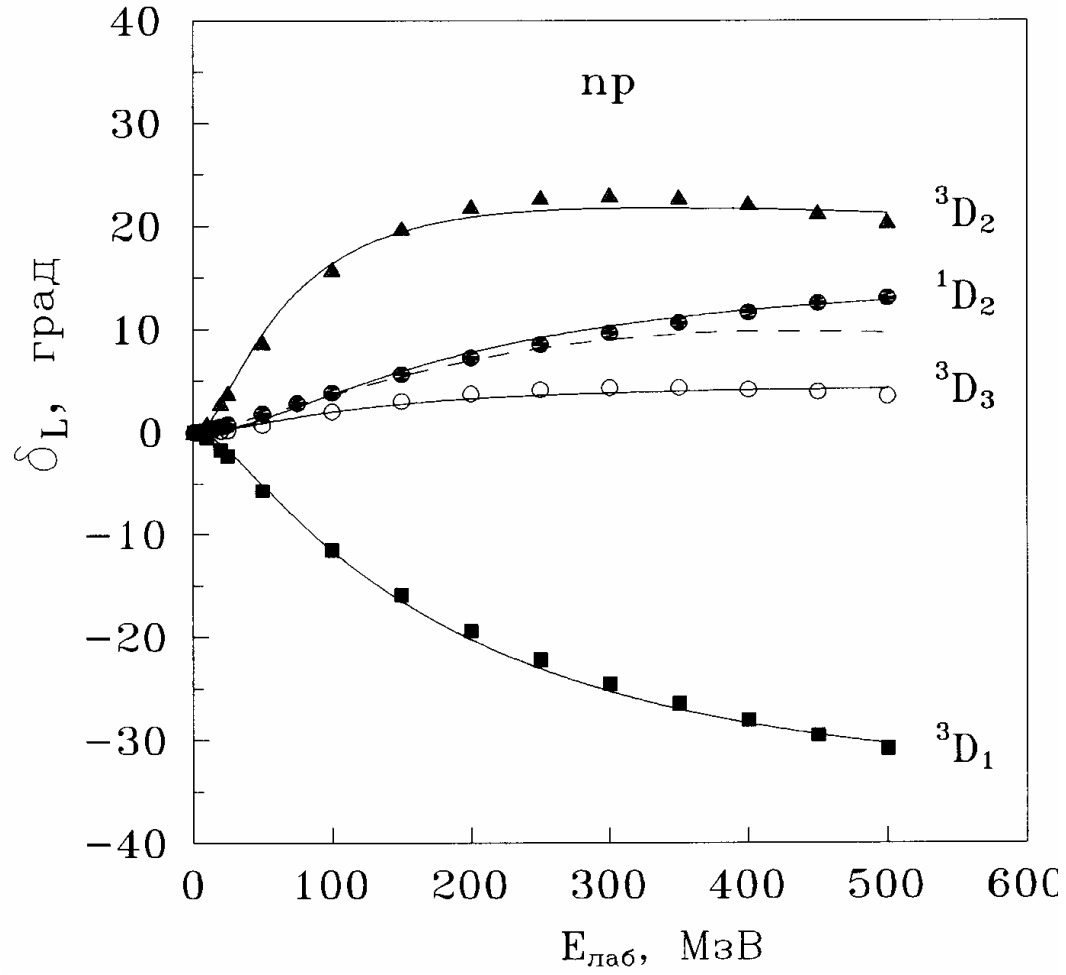

Рис.6.2б. Фазы упругого np рассеяния. Непрерывные кривые - расчеты для экспоненциальных потенциалов из табл.6.1., штриховые линии - расчеты для потенциала Рейда [9].

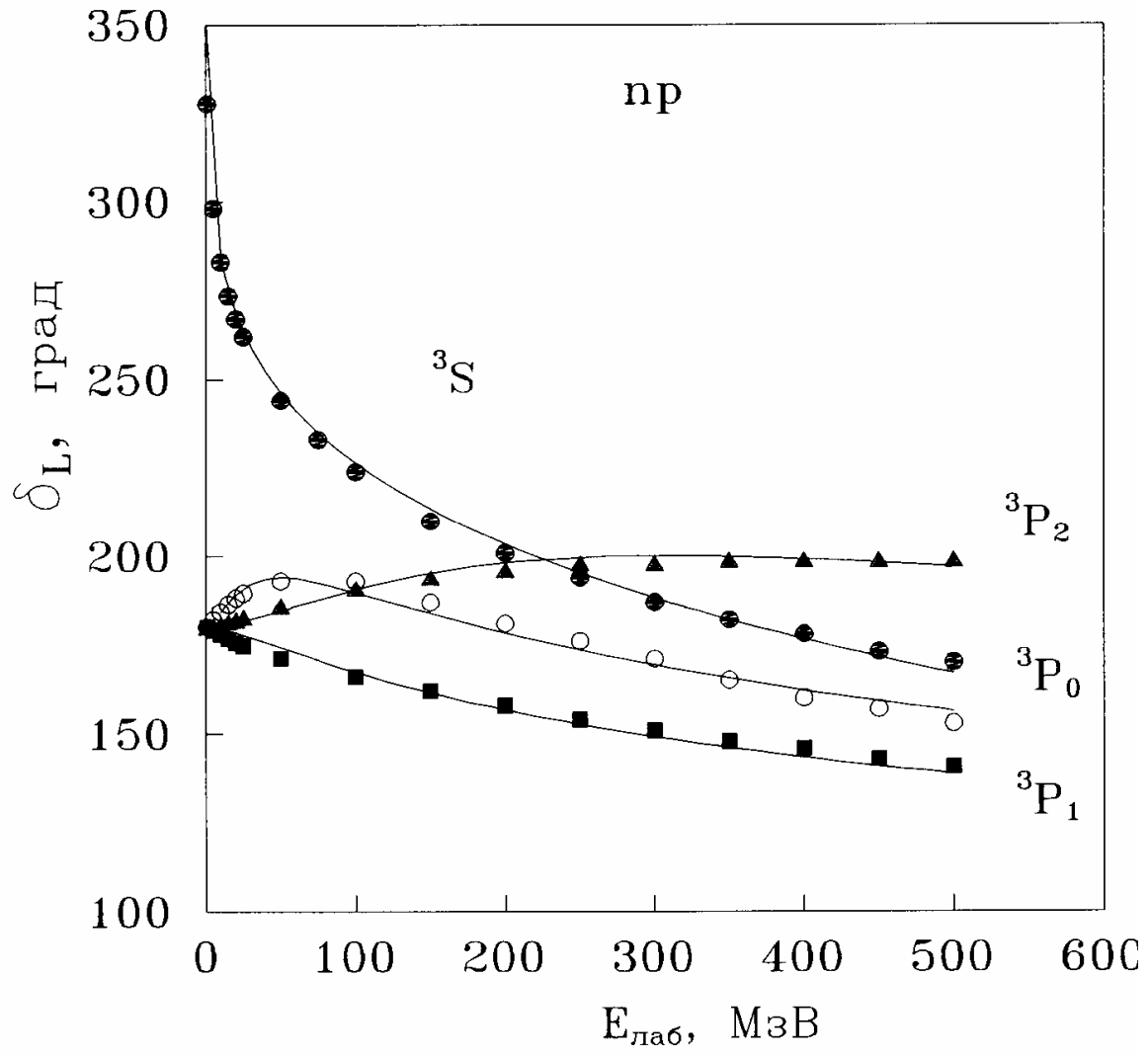

Рис.6.3а. Фазы упругого np рассеяния. Кривые - расчеты для потенциалов Юкавы из табл.6.1. Точки, треугольники, кружки и квадраты - экспериментальные данные из работ [12].

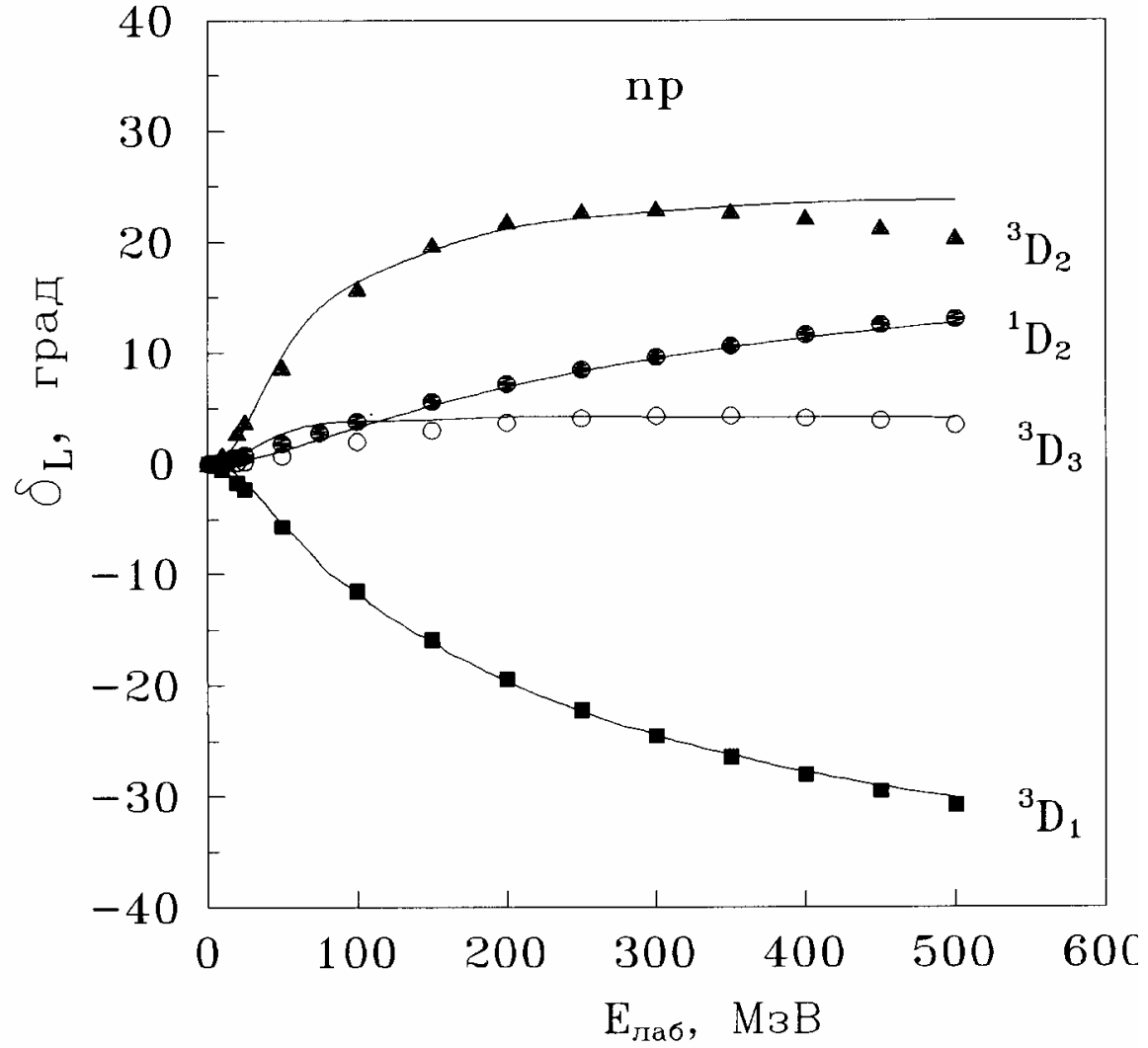

Рис.6.3б. Фазы упругого np рассеяния. Кривые - расчеты для потенциалов Юкавы из табл.6.1. Точки, треугольники, кружки и квадраты - экспериментальные данные из работ [12].

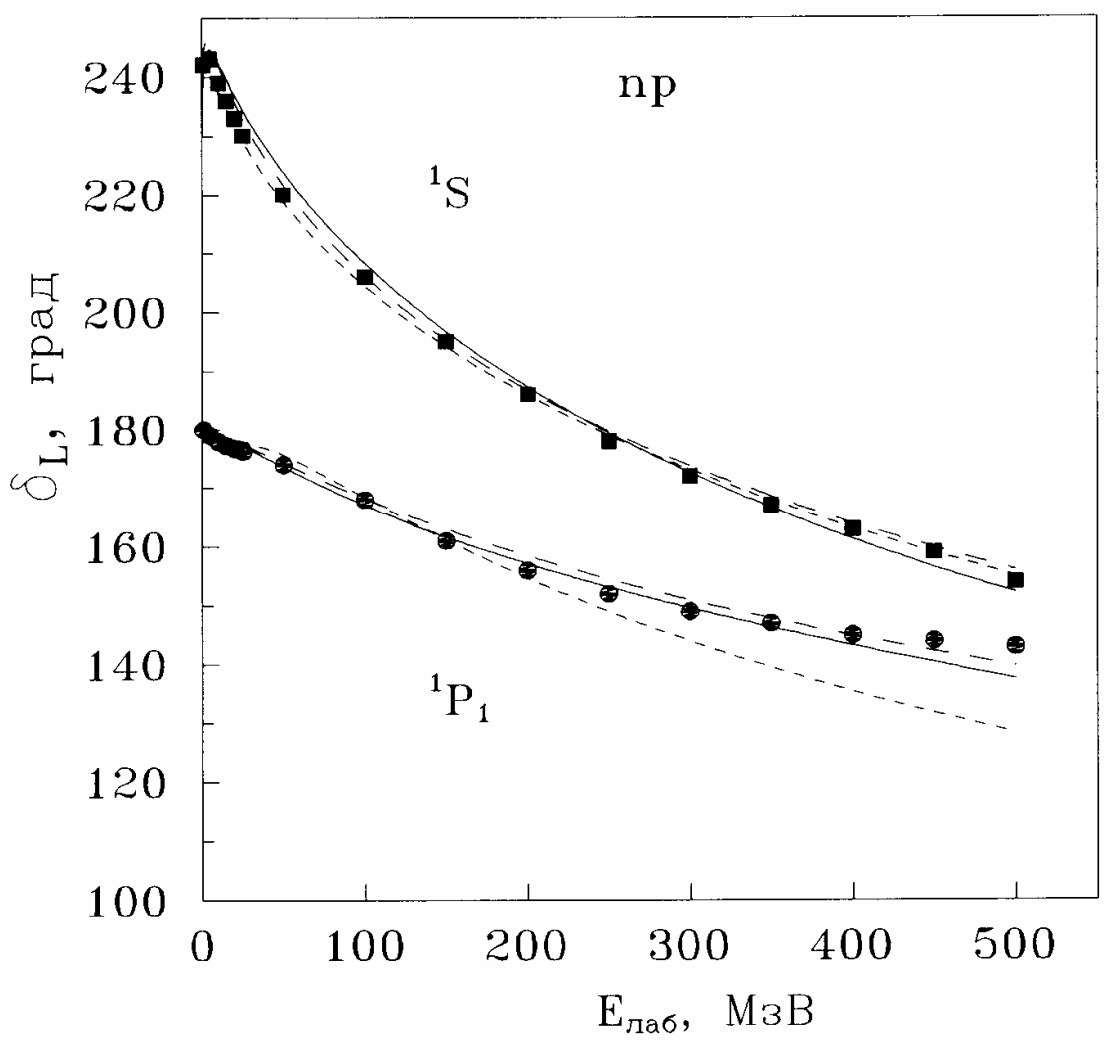

Рис.6.4а. Фазы упругого np рассеяния. Кривые - расчеты для различных потенциалов из табл.6.1., и потенциала Рейда. Точки и квадраты - экспериментальные данные из работ [12].

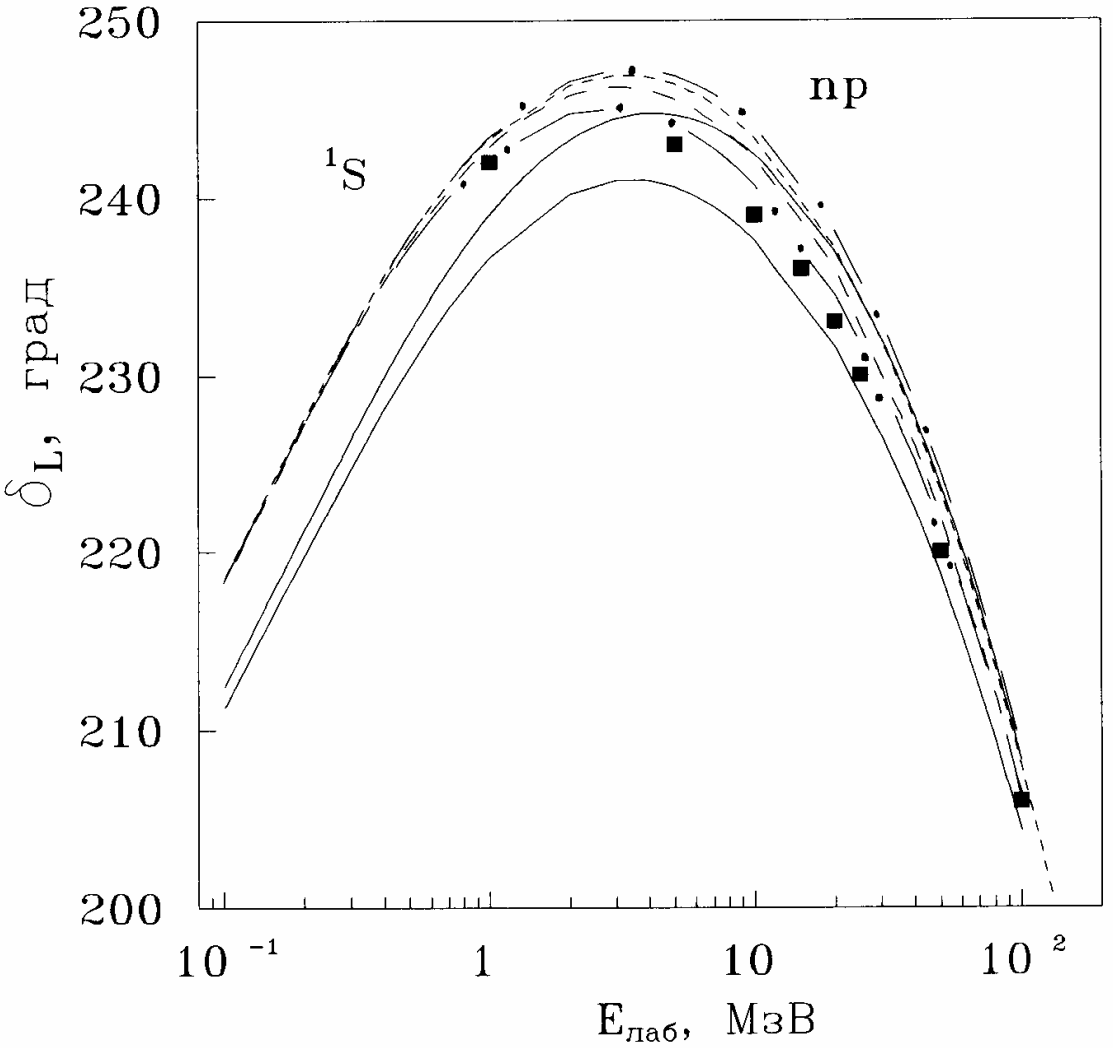

Рис.6.4б. Фазы упругого np рассеяния. Кривые - расчеты для различных потенциалов из табл.6.1., потенциала Рейда и взаимодействия из работ [7]. Точки и квадраты - экспериментальные данные из [12].

На рис.6.4б верхняя непрерывная кривая показывает результаты расчета фаз при малых энергиях для первого синглетного гауссового S потенциала, а штриховая для экспоненциального. Для потенциала Рейда результаты представлены нижней непрерывной линией, потенциала Юкавы фазы даны точечной линией. Штрих - пунктирной линией представлены фазы второго синглетного гауссового потенциала и двойным штрих - пунктиром приведена S фаза гауссового потенциала с OPEP добавкой и обрезанием на малых расстояниях из работ [7]. Фазы этого взаимодействия при энергиях выше 100 МэВ совпадают с непрерывной линией на рис.6.4а. Именно этот потенциал наилучшим образом описывает фазы при энергиях 1-10 МэВ за счет правильного поведения "хвоста" взаимодействия, обусловленного учетом OPEP [11].

Потенциал Рейда и первый гауссовый потенциал несколько занижают синглетную длину рассеяния, что хорошо видно в табл.6.2 и на рис. 6.4б. Рассчитанный формфактор дейтрона для всех рассмотренных типов потенциалов описывает экспериментальные данные работ [14,15] только до 3 $Фм^{-1}$, поскольку здесь не учитывается D - компонента волновой функции. В качестве формфактора протона использовалась гауссова параметризация с $\gamma = 0.0864$ $Фм^2$, а радиус протона принимался равным 0.805 (11) Фм.

При рассмотрении электромагнитных E1 переходов в дейтроне использовалось выражение для полного сечения захвата в виде (1.2.5) и приближенная формула дипольного перехода при фоторазвале [13]

$$\sigma_c\,(E1) = \frac{8\pi}{3}\frac{e^2}{\hbar c}\frac{\hbar^2}{m}\frac{W^{1/2}E^{3/2}}{(E+W)}\left(\frac{1}{1-k_0 r_0}\right), \qquad (6.1.6)$$

где W - энергия связи дейтрона, E - энергия нуклонов в непрерывном спектре.

На рис.6.5 представлены результаты расчетов E1 развала дейтрона. Штриховая линия показывает сечение, полученное из приближенной формулы, непрерывная для второго варианта гауссового потенциала основного состояния и точечная для первого варианта.

Поскольку первый вариант приводит к несколько заниженному эффективному радиусу, сечение развала идет ниже эксперимента, приведенного в работах [16], и приближенных результатов. Для второго гауссового взаимодействия, при относительно верных значениях низкоэнергетических параметров, расчетные сечения вполне передают экспериментальные данные до 70 МэВ.

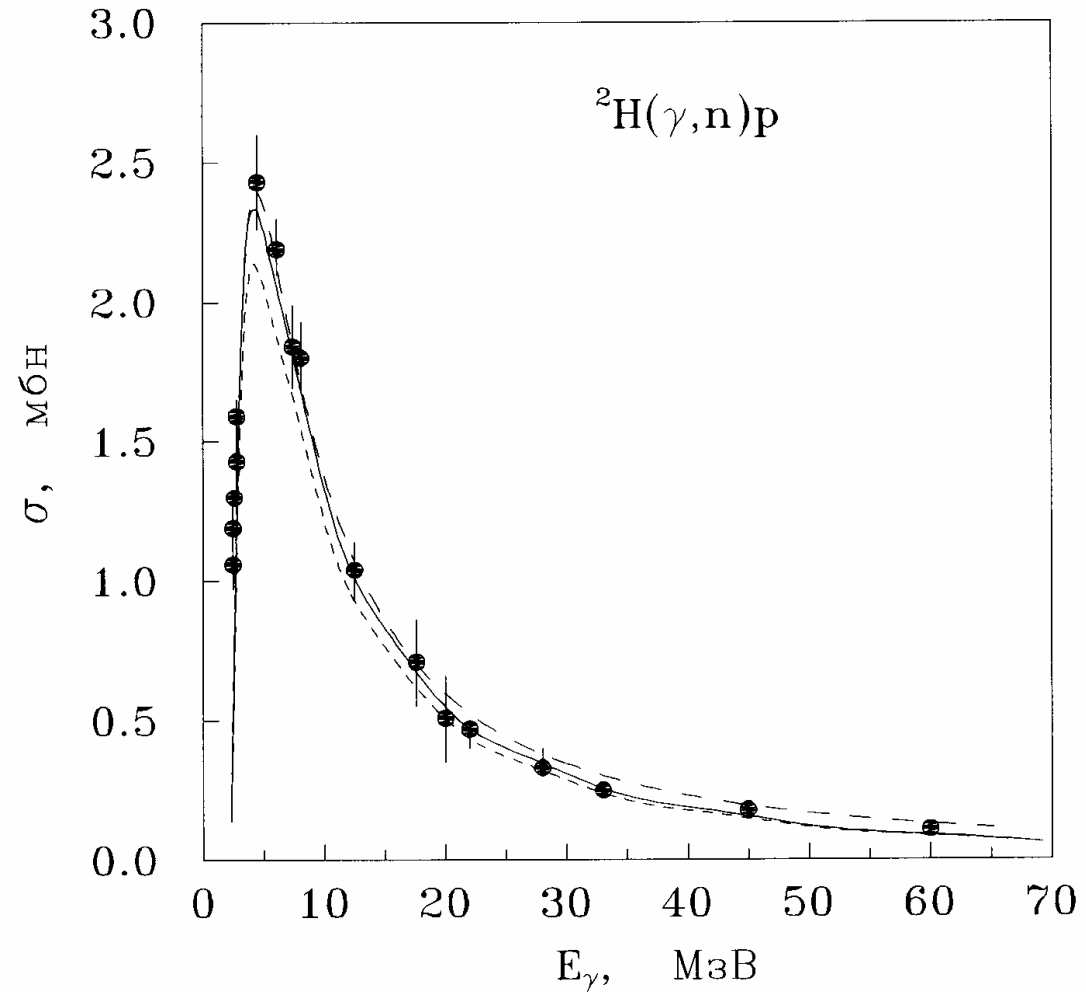

Рис.6.5. Полные сечения фоторазвала дейтрона для потенциалов из табл.6.1. Экспериментальные данные из работ [16].

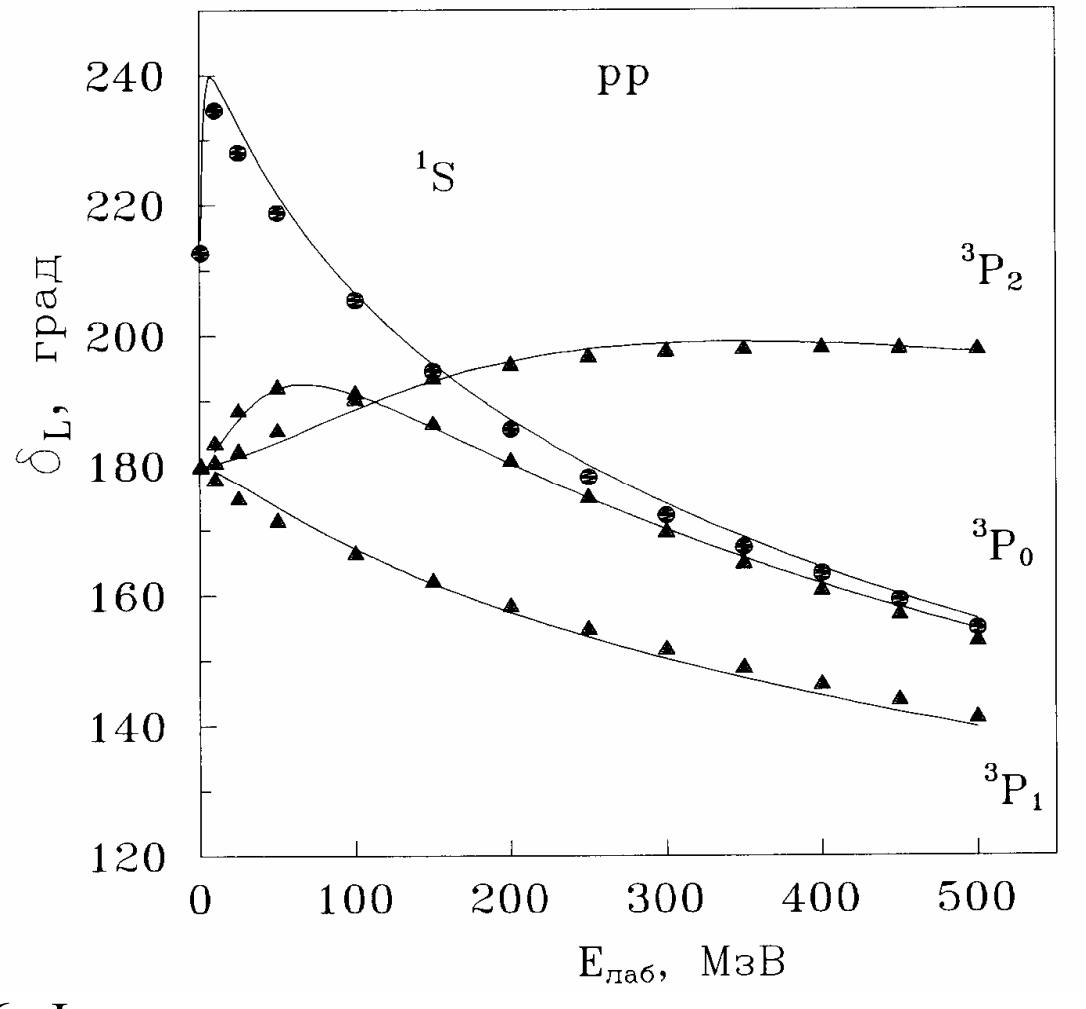

Рис.6.6. Фазы упругого pp рассеяния для экспоненциального потенциала из табл.6.1. Экспериментальные данные взяты из работ [12].

На рис.6.6 даны результаты расчетов фаз pp рассеяния, полученные на основе, приведенных в табл.6.1, параметров экспоненциального потенциала с учетом кулоновского взаимодействия. Видно, что и в этом случае потенциалы позволяют получить хорошее описание экспериментальных данных из работы [12].

*Таблица 6.3. Сравнение пр фаз рассеяния (в радианах) для потенциала Рейда из работы [9] и вычисленных в [11].*

| $L_J$ | E (МэВ) | Расчет из работы [9] | Расчет из работы [11] |
|---|---|---|---|
| $^1S_0$ | 48 | 0.696 | 0.687 |
| | 144 | 0.277 | 0.265 |
| | 208 | 0.093 | 0.082 |
| | 352 | -0.205 | -0.215 |
| $^1P_1$ | 48 | -0.071 | -0.072 |
| | 144 | 0.277 | 0.265 |
| | 208 | -0.456 | -0.458 |
| | 352 | -0.708 | -0.710 |
| $^1D_2$ | 48 | 0.027 | 0.028 |
| | 144 | 0.089 | 0.091 |
| | 208 | 0.123 | 0.125 |
| | 352 | 0.164 | 0.166 |
| $^3P_0$ | 48 | 0.198 | 0.206 |
| | 144 | 0.105 | 0.107 |
| | 208 | -0.012 | -0.010 |
| | 352 | -0.264 | -0.264 |
| $^3P_1$ | 48 | -0.133 | -0.139 |
| | 144 | -0.304 | -0.312 |
| | 208 | -0.386 | -0.393 |
| | 352 | -0.518 | -0.525 |

Таким образом, три рассмотренных типа потенциалов дают практически одинаковое описание экспериментальных фаз рассеяния при всех рассмотренных энергиях. Однако, экспоненциальная форма взаимодействия приводит к несколько лучшему, чем гауссова, описанию экспериментальных данных при малых энергиях.

Оказывается возможным согласовать параметры потенциала с фазами при высоких энергиях и с характеристиками низкоэнергетического рассеяния. Потенциал типа Юкавы, который по форме совпадает с OPEP, позволяет правильно передать S и P фазы, хотя описание D фаз получается несколько хуже.

## 6.2. Нуклон - нуклонные силы с тензорной компонентой

Рассмотрим теперь варианты глубокого феноменологического np взаимодействия экспоненциальной формы с тензорной частью в триплетном спиновом канале и OPEP, который играет основную роль на больших расстояниях. Экспоненциальный потенциал имеет всего два подгоночных параметра $V_0$ и $\alpha$, представляется в виде, аналогичном гауссовому взаимодействию из работ [7]

$$V(r) = V_c(r) + V_т(r)\, S_{12},$$
$$V_c(r) = -V_0 \exp(-\alpha r) + V_{oc}(r) g(r), \qquad (6.2.1)$$
$$V_т(r) = V_{oт}(r) g^3(r).$$

Потенциалы однопионного обмена (OPEP), имеющие вид

$$V_{oc}(r) = -V_1 \exp(-\mu r)/(\mu r), \qquad (6.2.2)$$
$$V_{oт}(r) = -V_1 \left[\, 1 + 3/(\mu r) + 3/(\mu r)^2 \,\right] \exp(-\mu r)/(\mu r)\,,$$

обрезаются на малых расстояниях с помощью множителя $g(r)=1-\exp(-\alpha r)$ [7]. В отличие от гауссового взаимодействия, параметр обрезания OPEP $\alpha$ совпадает с параметром ширины экспоненциального потенциала.

Средняя масса $\pi$ мезонов принята, как в [7,17] $m_\pi$ = 138.03 МэВ, что приводит к $\mu = m_\pi/\hbar c$ = 0.6995 Фм$^{-1}$. Константы $\hbar c$ = 197.327 МэВ Фм и $\hbar^2/m_N$ = 41.47 МэВ Фм$^2$. Если использовать экспериментальное значение константы $\pi NN$ связи $f^2$ = 0.0776(9) [18], то для глубины OPEP получим величину $V_1 = m_\pi f^2$ = 10.71(12) МэВ. Отметим, что в работе [7] использовались несколько иные значения - $V_1$ =10.69 МэВ и $f^2$ = 0.07745.

Используемый потенциал, как и в работе [7], в триплетном и синглетном вариантах имеет глубоколежащее ненаблюдаемое запрещенное состояние, приводящее к узлу в волновых функция дейтрона и процессов рассеяния. Положение узла примерно соответствует расстоянию, на котором потенциалы с кором обычно приводят к “вымиранию” волновой функции относительного движения нуклонов.

Методы решения уравнения Шредингера с тензорными силами изложены в разделе 1.4 первой главы. В результате численных решений можно получить полный вид волновой функции во всей области при $r<R_0$, где радиус сшивки $R_0$ принимается равным 20 Фм. Для численного решения исходного уравнения используется метод Рунге-Кутта с автоматическим выбором шага при заданной точности результатов по фазам и параметру смешивания. При вычислениях эффективных радиусов и длин рассеяния использовались обычные формулы, приведенные, например, в работах [13,16]. Для нахождения волновых функций связанных состояний применялась комбинация численных и вариационных методов, изложенных в последнем параграфе первой главы [17].

Используемые методы расчетов характеристик дейтрона и np системы проверялись на альтернативном потенциале Рейда с мягким кором (RSCA) [9] и Московском потенциале (МП) [7]. В таблице 6.4 приведено сравнение результатов, полученных для этих потенциалов в работах [7] и [9], с вычислениями из работ [17]. Ошибки в расчетных асимптотических константах обусловлены усреднением их значений по интервалу расстояний 7-15 Фм.

*Таблица 6.4. Сравнение характеристик дейтрона и np рассеяния для потенциалов Рейда с мягким кором (RSCA) и МП.*

| Характеристики дейтрона | Расчет из работы [9] (Рейд) | Расчет из работы [17] (Рейд) | Расчет из работы [7] (МП) | Расчет из работы [17] (МП) |
|---|---|---|---|---|
| $E_d$ (МэВ) | 2.22464 | 2.22458 | 2.2246 | 2.22456 |
| $Q_d$ (Фм$^2$) | 0.2762 | 0.276 | 0.2860 | 0.285 |
| $P_D$ (%) | 6.217 | 6.217 | 6.78 | 6.778 |
| $A_S$ | 0.87758 | 0.875(2) | 0.8814 | 0.879(1) |
| $\eta=A_D/A_S$ | 0.02596 | 0.0260(2) | 0.0269 | 0.0270(2) |
| $a_t$ (Фм) | 5.390 | 5.390 | 5.40 | 5.400 |
| $r_t$ (Фм) | 1.720 | 1.723 | 1.73 | 1.728 |
| $a_s$ (Фм) | -17.1 | -17.12 | -23.73 | -23.701 |
| $r_s$ (Фм) | 2.80 | 2.810 | 2.61 | 2.621 |
| $R_d$ (Фм) | 1.956 | 1.951 | 1.961 | 1.954 |

Параметры экспоненциального потенциала подбирались исключительно на основе описания низкоэнергетических характеристик np рассеяния и энергии связи дейтрона, а затем, с полученным потенциалом, рассчитывались высокоэнергетические фазы и параметры смешивания [17], экспериментальные данные, по которым приведены в [12]. Для глубины OPEP $V_1$ использовалась такая же величина, как в работе [7]. В результате было найдено, что описание эксперимента можно получить, если принять $\alpha$= 3.25 Фм$^{-1}$ с глубиной центральной части -1033.885 МэВ.

Расчетные длина рассеяния и эффективный радиус оказались равны: $a_t$ = 5.410 Фм и $r_t$ = 1.741 Фм в хорошем согласии с экспериментом [13,18]. Для энергия связи получена величина -2.2246 МэВ, для квадрупольного момента 0.285 Фм$^2$ и вероятности D состояния 6.78% [17], которые также хорошо согласуются с данными работ [13,18].

Еще в работах [7] для гауссового потенциала отмечалась определенная зависимость параметров $\alpha$ и $\beta$, последний из которых является параметром обрезания OPEP. Фиксируя в определенных пределах один из них, практически всегда можно найти такие значения другого, с которыми удается воспроизвести экспериментальные данные. В результате для параметра обрезания $\beta$ была предложена величина 3.2433 $Фм^{-1}$, очень близкая к полученному здесь значению $\alpha$, поскольку в данном случае $\alpha = \beta$. Тем самым, при использовании экспоненциальной формы взаимодействия оказалось возможным вообще исключить третий параметр и свести потенциал к двухпараметрическому виду [17].

Надо особо отметить, что двухпараметрическая форма позволяет однозначно фиксировать параметры потенциала. Изменение параметра $\alpha$ в любую сторону, от приведенной выше величины, дает завышенные или заниженные значения эффективного радиуса, даже в том случае, если удается правильно передать длину рассеяния. Для правильного и одновременного описания этих двух величин, так чтобы их значения находились в пределах экспериментальных ошибок, параметр $\alpha$ для триплетного потенциала должен находиться в интервале 3.2-3.25 $Фм^{-1}$.

Для синглетного потенциала этот интервал составляет 3.2-3.3 $Фм^{-1}$. Поскольку ошибки для синглетного эффективного радиуса сравнительно велики, то в общем можно найти вариант синглетных и триплетных потенциалов с одинаковым параметром ширины $\alpha$, имеющим значение в интервале 3.22-3.23 $Фм^{-1}$.

Если использовать в качестве $V_1$ величину 10.71 МэВ, то для параметров триплетного потенциала можно найти: $\alpha$= 3.20 $Фм^{-1}$ с глубиной центральной части 1062.81 МэВ.

Для синглетного потенциала получены следующие параметры: $\alpha$ =3.30 $Фм^{-1}$ и $V_0$=3170.72 МэВ, которые приводят к эффективному радиусу 2.732 Фм, в хорошем согласии с экспериментальными данными работы [18]. Если согласовывать синглетный потенциал с данными из [13], то надо принять параметры: $\alpha$= 3.23 $Фм^{-1}$ и $V_0$=3037.4 МэВ, которые приводят к эффективному радиусу 2.768 Фм и длине рассеяния, равной -23.721 Фм.

Характеристики дейтрона и NN рассеяния, вычисленные с этими потенциалами, приведены в табл. 6.5, а фазы рассеяния и параметр смешивания показаны на рис.6.7 точечной линией. Синглетные фазы рассеяния приведены на рис.6.8. непрерывной линией. Штри-ховыми линиями на рис.6.7 и 6.8 даны фазы потенциала из работ [7], которые на рис.6.7а практически совпадают с точечной кривой. На рис.6.9 приведены волновая функция рассеяния для первого варианта синглетного взаимодействия для разных энергий.

*Таблица 6.5. Характеристики дейтрона и np рассеяния для экспоненциального потенциала и модифицированного варианта Московского взаимодействия (МП).*

| Характеристики дейтрона | Экспоненциальный потенциала | Расчет для модифицированного МП | Эксперимент [18] | Эксперимент [13] |
|---|---|---|---|---|
| $E_d$ (МэВ) | 2.2246 | 2.22453 | 2.224579(9) | 2.22452(20) |
| $Q_d$ (Фм$^2$) | 0.286 | 0.286 | 0.2859(3) | 0.286 |
| $P_D$ (%) | 6.736 | 6.776 | | |
| $A_S$ | 0.881(1) | 0.879(1) | 0.8802(20) | |
| $\eta=A_D/A_S$ | 0.0269(1) | 0.0269(1) | 0.0271(4) | |
| $a_t$ (Фм) | 5.418 | 5.401 | 5.419(7) | 5.414(5) |
| $r_t$ (Фм) | 1.752 | 1.729 | 1.754(8) | 1.750(5) |
| $a_s$ (Фм) | -23.716 | | -23.715(15) | -23.719(13) |
| $r_s$ (Фм) | 2.732 | | 2.73(3) | 2.76(5) |
| $R_d$ (Фм) | 1.961 | 1.956 | 1.9560 (68) | |

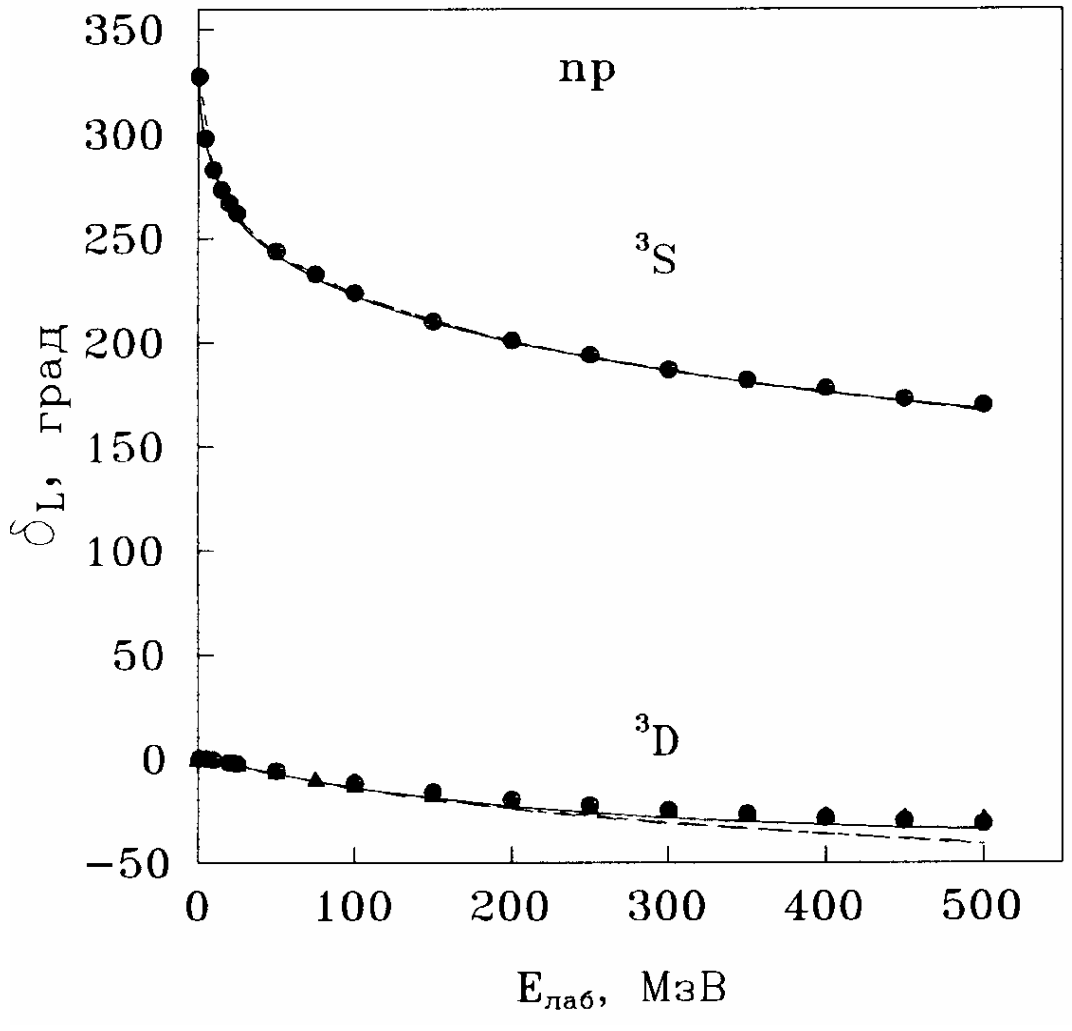

Рис.6.7а. Триплетные фазы np рассеяния. Кривые - расчеты с разными потенциалами. Эксперимент из работы [12].

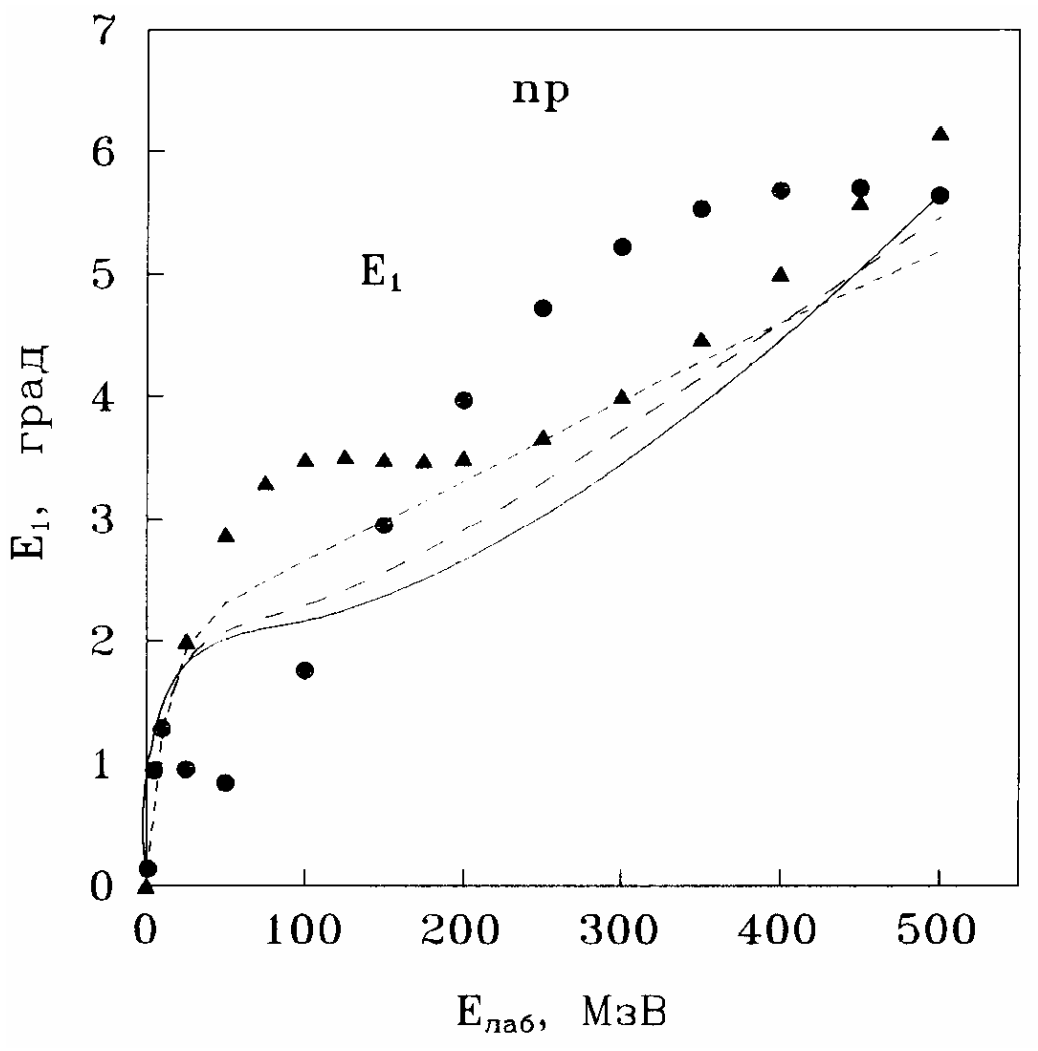

Рис.6.7б. Параметры смешивания для триплетных фаз рассеяния. Кривые - расчеты с разными потенциалами. Эксперимент из работы [12].

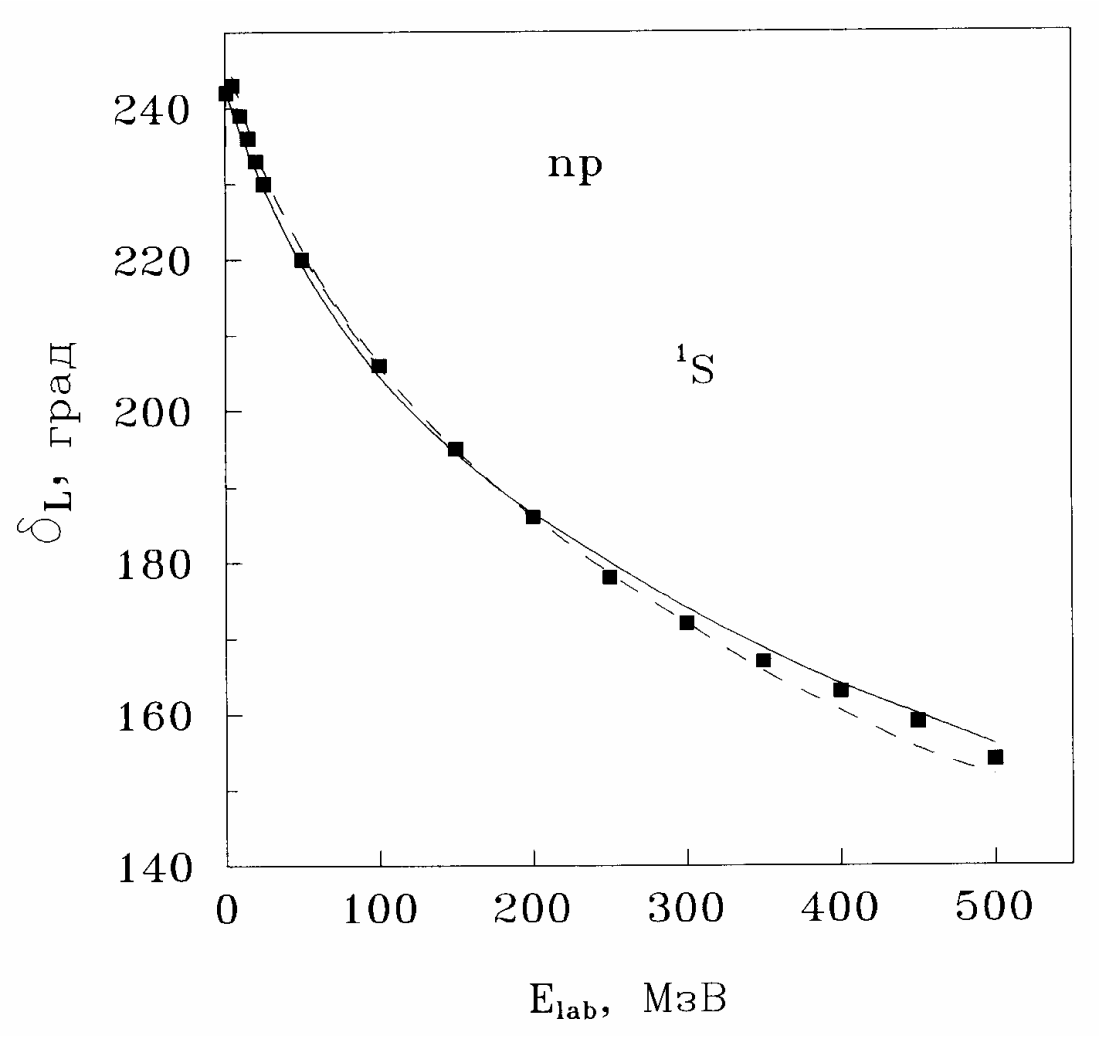

Рис.6.8а. Синглетные фазы np рассеяния. Кривые - расчеты с разными потенциалами. Эксперимент из работы [12].

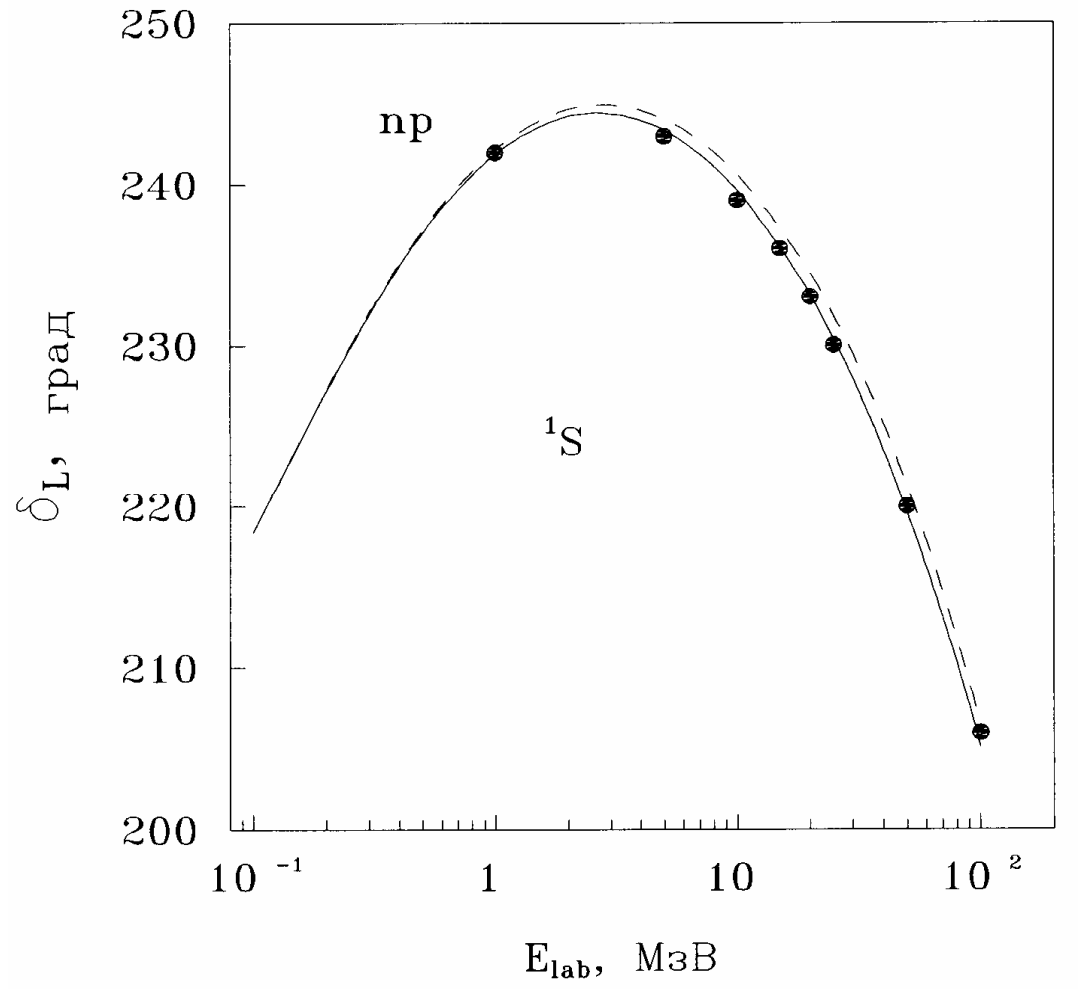

Рис.6.8б. Синглетные фазы np рассеяния при малых энергиях. Кривые - расчеты с разными потенциалами. Экспериментальные данные из работы [12].

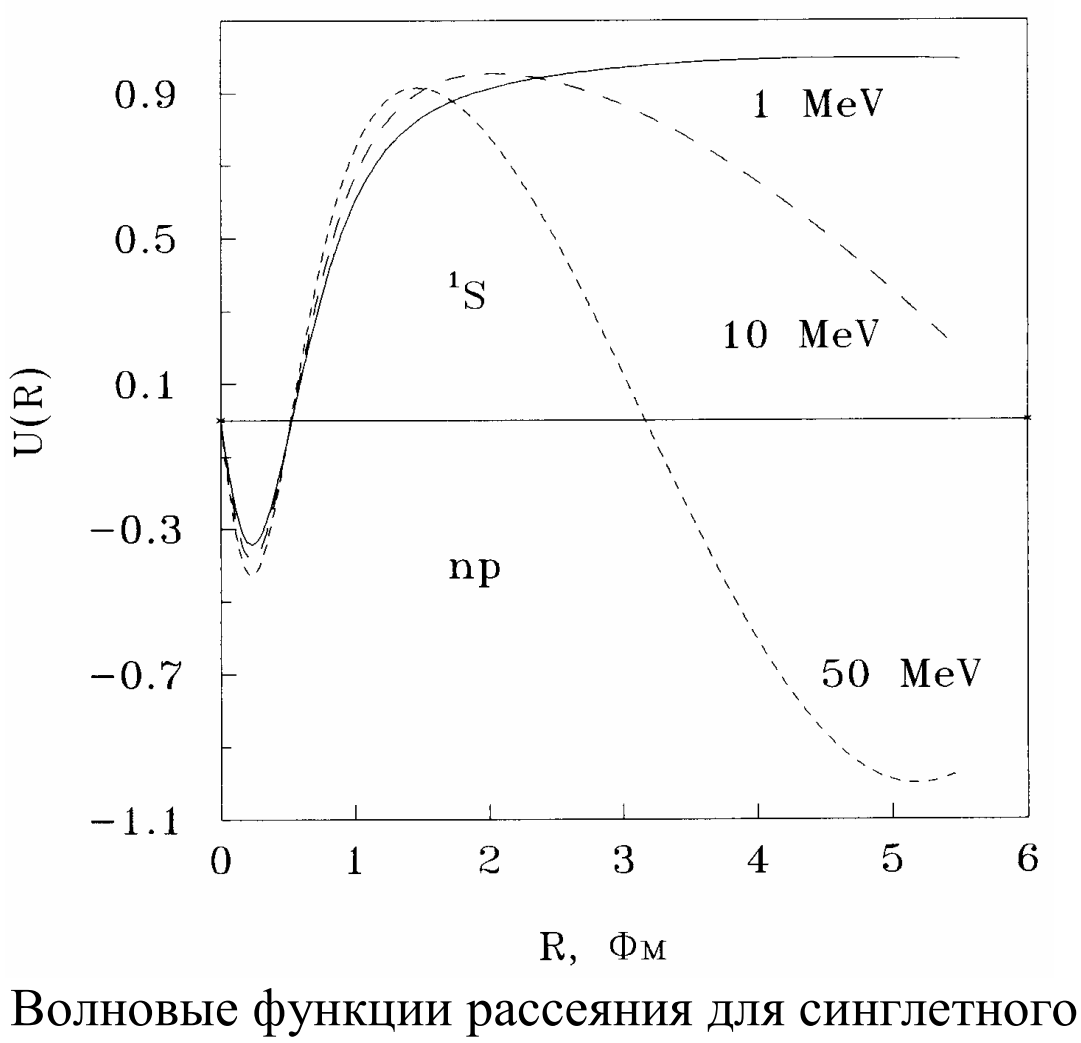

Рис.6.9. Волновые функции рассеяния для синглетного экспоненциального потенциала при разных энергиях.

Непрерывными линиями на рис.6.10 показаны формфакторы дейтрона, полученные для приведенного выше потенциала с $V_1$=10.71 МэВ. Точечными линиями показан формфактор для потен-циала Рейда [9], а штриховой линией даны результаты для московского потенциала [7]. Экспериментальные данные по формфактору приведены в работах [15].

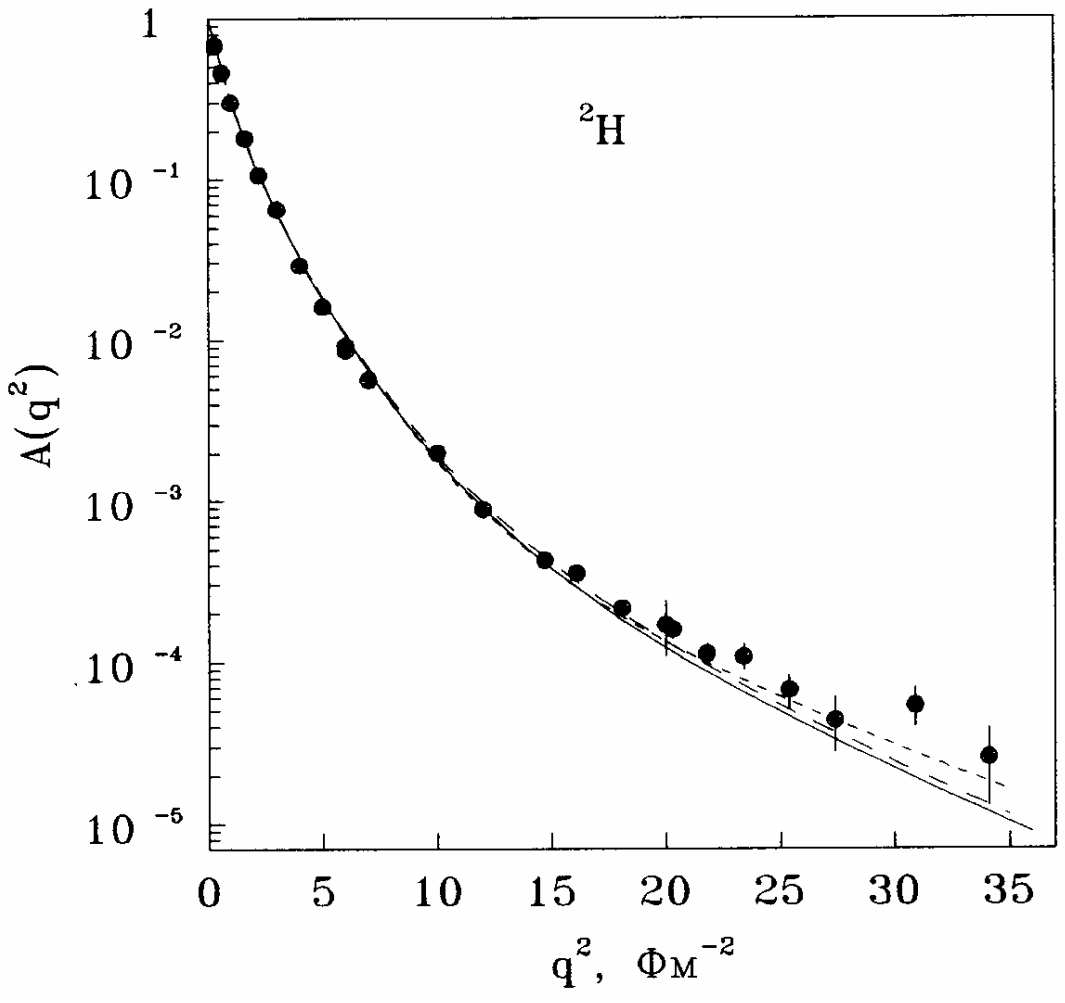

Рис.6.10а. Формфактор дейтрона A(q) для различных потенциалов. Эксперимент по формфактору приведен в [14,15].

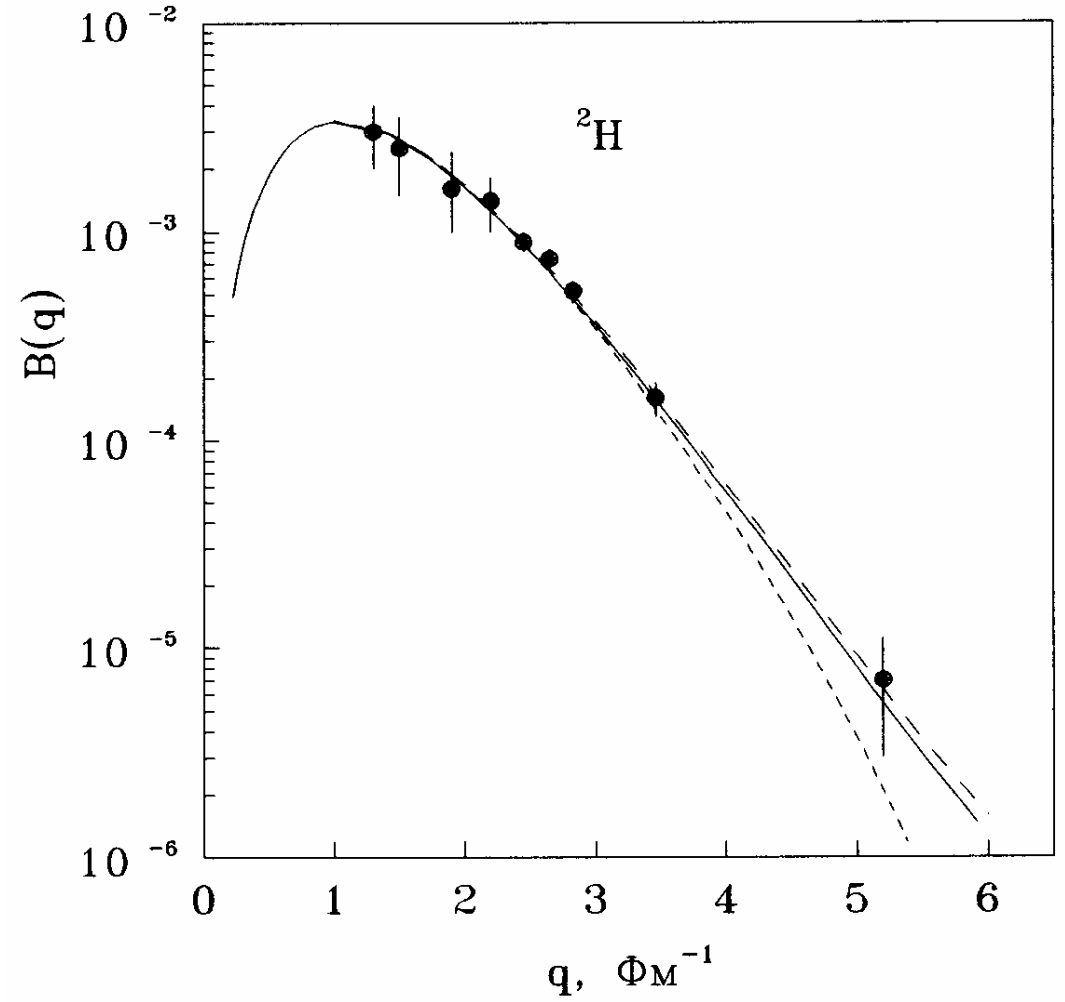

Рис.6.10б. Формфактор дейтрона B(q) для различных потенциалов. Эксперимент по формфактору приведен в [14,15].

## 6.3. Взаимозависимость параметров потенциала

Поскольку параметры триплетного потенциала фиксируются вполне однозначно, то можно попытаться найти взаимосвязь между ними. В частности, положим $V_0 = AV_1$ и $\alpha = B\mu$ , где A и B неизвестные безразмерные параметры, а $V_1$ = 10.71 МэВ и $\mu$ = 0.6995 $Фм^{-1}$.

При использовании такой формы записи параметров удается найти, что параметр B, определяющий радиус потенциала, может быть представлен в виде $B = A^{1/3}$. В таком случае величина параметра глубины A должна быть равна 97.765, что эквивалентно $V_0$=1047.06 МэВ и $\alpha$=3.2224 $Фм^{-1}$. Этот потенциал приводит к длине рассеяния 5.415 Фм и эффективному радиусу 1.748 Фм, а его фазы практически не отличается от результатов для предыдущего варианта взаимодействия.

Для энергии дейтрона получается величина -2.2245 МэВ, для квадрупольного момента 0.285 $Фм^2$, для вероятности D состояния 6.76%. Поскольку константа πNN связи известна с некоторой экспериментальной ошибкой, то и параметр A оказывается заключен примерно в интервале 96-99. А если учесть, что $a_t$ и $r_t$ также имеют определенные экспериментальные ошибки, то этот интервал может быть несколько расширен.

Очевидно, что использование такой взаимосвязи констант всего лишь удачное предположение, не имеющее какого - либо микроскопического обоснования. Тем не менее, используемая форма потенциала позволяет получить, в пределах экспериментальных ошибок, вполне однозначное соответствие между параметрами $V_0$ и $\alpha$, а, значит, и между A и B.

Кроме того, из полученных результатов следует, что можно использовать один и тот же параметр глубины для синглетных и триплетных потенциалов. Но в этом случае для синглетного эффективного радиуса получается величина, находящаяся на верхнем пределе экспериментальных ошибок, а величина триплетного, эффективного радиуса оказывается на нижней границе.

Допуская, что определенная функциональная зависимость между параметрами потенциалов существует всегда, не зависимо от их конкретной формы, можно рассмотреть модификацию триплетного взаимодействия гауссового типа из работы [7]

$V(r)=V_c(r)+V_т(r)\, S_{12}$,
$V_c(r) = -V_0\exp(-\alpha r^2)+V_{oc}(r)g(r)$,
$V_т(r) = V_{oт}(r)g^3(r)$,

с однопионными потенциалами, приведенными в (6.2.2), и обрезанием вида g(r)=1-exp(-βr). Для параметров потенциала в работе [7] были предложены следующие значения: $V_0$ = 349.023 МэВ, α=1.6163 $Фм^{-2}$, β=3.2433 $Фм^{-1}$.

Если пытаться найти функциональную зависимость между параметрами аналогично экспоненциальному взаимодействию, то можно положить $V_0 = AV_1$, $\alpha = B\mu^2$ и $\beta = C\mu$, где A, B и C неизвестные безразмерные параметры, а $V_1$ = 10.69 МэВ и μ = 0.6995 $Фм^{-1}$. При использовании такой формы записи параметров удается найти, что параметры B и C, определяющие радиус потенциала и фактор обрезания OPEP могут быть представлены в виде $B = (1/3A)^{1/2}$ и $C = (2/3A)^{1/2}$. При этом величина параметра глубины A должна быть равна 32.2875, что эквивалентно $V_0$ = 345.1534 МэВ, и α=1.6052 $Фм^{-2}$ и β=3.2453 $Фм^{-1}$. Полученные, таким образом, параметры потенциала мало отличаются от параметров, приведенных выше. Из - за существующих экспериментальных ошибок константы πNN связи, величина параметра A оказывается заключенной примерно в интервале значений 31-33.

Результаты расчетов характеристик дейтрона и np рассеяния для такого модифицированного потенциала приведены в табл.6.5. Фазы и параметры смешивания, волновые функции и формфактор практически совпадают с результатами для исходного потенциала из работ [7], показанными на рис.6.7 штриховыми линиями.

Тем самым видно, что модифицированный вариант гауссового потенциала приводит к таким же результатам, что и взаимодействие с исходными параметрами. Однако, в данном случае мы имеем только одну подгоночную величину A, а остальные связаны с ней определенной эмпирической зависимостью. Причем, значение единственного варьируемого параметра, по экспериментальным данным, определяется вполне однозначно.

## 6.4. NN потенциал без узла в D волне

Все рассмотренные выше NN потенциалы (экспоненциальный и гауссовый) приводят к узлу в S и D компонентах волновой функции дейтрона, что указывает на присутствие запрещенного уровня в D волне. Однако это не согласуется с приведенной выше классификацией, поскольку схемы {6} и {42} допускают только одно связанное D состояние, которое соответствует основному состоянию дейтрона.

Рассмотрим другой вариант глубокого np взаимодействия экспоненциальной формы с тензорной частью и OPEP, который не име-ет узла в D волне, и не приводит к лишнему запрещенному состоянию. Потенциал имеет вид (6.2.1), аналогичный, приведенному в работе [17], однако фактор обрезания в тензорной части стоит не в третьей, а в 20 степени, что позволяет избежать появления узла в D компоненте волновой функции дейтрона.

Параметры потенциала по-прежнему подбираются исходя из описания низкоэнергетических характеристик рассеяния и энергии связи дейтрона, затем с ними рассчитываются высокоэнергетические фазы рассеяния и параметры смешивания, а также другие свойства дейтрона, подробное описание которых вместе с экспериментальными данными можно найти, например, в обзорных работах [18].

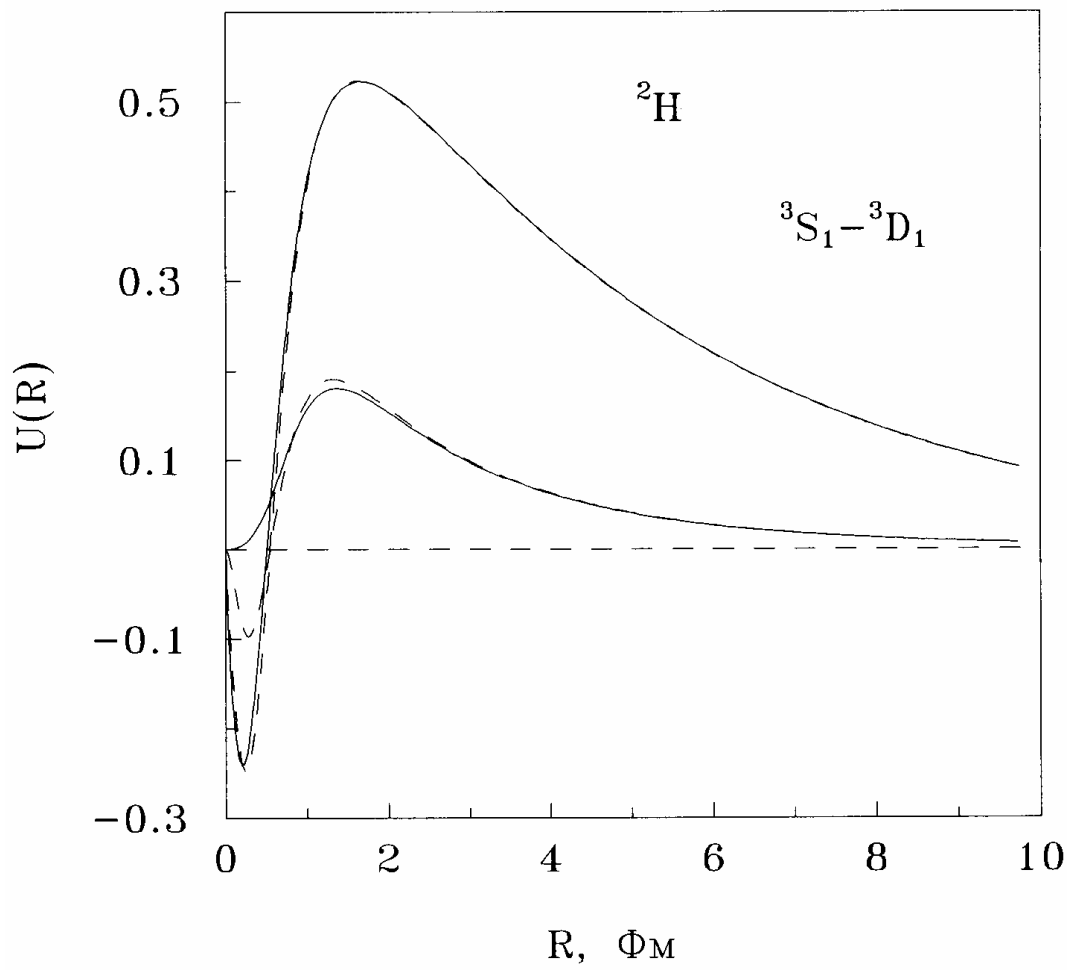

Рис.6.11. Волновые функции связанного триплетного состояния дейтрона для гауссового [7] и экспоненциального np потенциалов.

В результате найдено, что правильное описание эксперимента можно получить, если принять для триплетного потенциала $\alpha$ = 3.9 Фм$^{-1}$ с глубиной центральной части 3652.87 МэВ [19]. Результаты расчетов различных характеристик дейтрона и np рассеяния даны в табл. 6.6, а фазы рассеяния показаны на рис.6.7, непрерывными линиями.

Волновая функция в S волне имеет узел при 0.5 Фм, а функция в D волне безузловая, как показано на рис.6.11 непрерывной линией. Пунктиром приводятся результаты для гауссового потенциала с двумя узлами из работ [7]. Формфакторы дейтрона, вычисленные с этими волновыми функциями, практически не отличаются от результатов, показанных на рис.6.10 непрерывной линии. На рис.6.12 показаны импульсные распределения нуклонов в дейтроне для потенциала Рейда - точечная линия, приведенного выше экспоненциального взаимодействия - непрерывная линия и московского потенциала - штриховая линия.

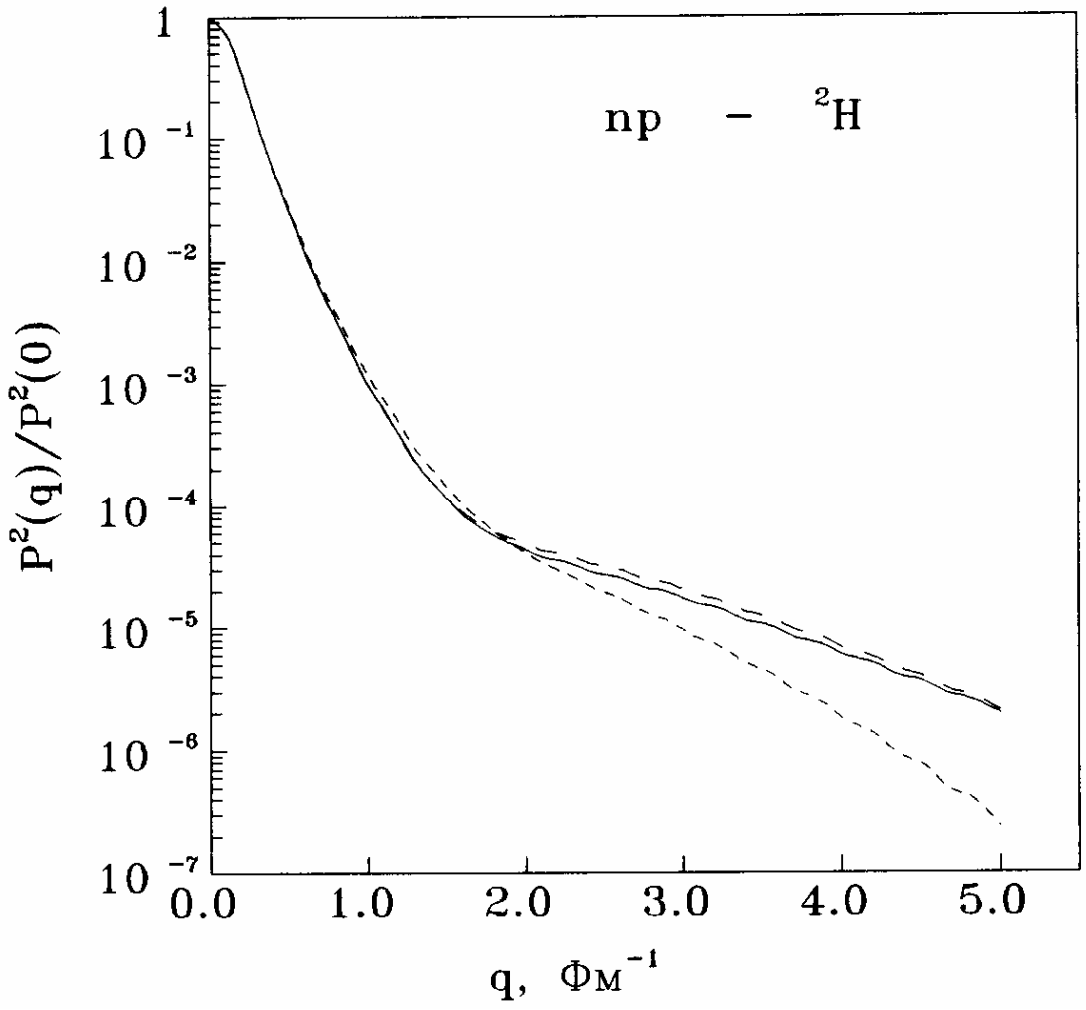

Рис.6.12. Импульсные распределения нуклонов в дейтроне для различных нуклон - нуклонных потенциалов.

*Таблица. 6.6. Сравнение характеристик дейтрона и np рассеяния, вычисленных для экспоненциального потенциала без узла в D волне с экспериментальными данными.*

| Характеристики дейтрона | Экспоненциальный потенциал | Эксперимент из работ [18] |
|---|---|---|
| $E_d$ (МэВ) | 2.2246 | 2.224579(9) |
| $Q_d$ (Фм$^2$) | 0.283 | 0.2859(3) |
| $P_D$ (%) | 6.22 | |
| $A_S$ | 0.881(1) | 0.8802(20) |
| $\eta=A_D/A_S$ | 0.0265(2) | 0.0271(4) |
| $a_t$ (Фм) | 5.417 | 5.419(7) |
| $r_t$ (Фм) | 1.751 | 1.754(8) |
| $a_s$ (Фм) | -23.716 | -23.715(15) |
| $r_s$ (Фм) | 2.732 | 2.73(3) |
| $R_d$ (Фм) | 1.960 | 1.9560 (68) |

Таким образом, видно, что все рассмотренные характеристики np системы могут быть успешно описаны на основе двухпараметрического, глубокого потенциала с запрещенным связанным состоянием в S волне, который не имеет запрещенного D уровня.

## 6.5. Характеристики дейтрона для Нимегенских потенциалов

Сравнительно недавно, было предложено несколько новых вариантов феноменологических нуклон - нуклонных потенциалов [20] с отталкивающим кором, параметризованных на основе фазового анализа, выполненного Нимегенской группой [21]. В настоящее время, варианты потенциалов Nijm.-1, Nijm.-2 и Reid-93 являются одними из лучших, из всех предложенных до сих пор NN взаимодействий, поскольку обеспечивает $\chi^2$ в области энергий до 350 МэВ около 1.03.

Например, классический вариант Рейда 68г. [9] или Парижский 80г. [22] потенциалы приводят к $\chi^2$ около 2-3 в интервале энергий до 300 МэВ. Примерно такую же величину $\chi^2$ = 1.9 дает четвертый вариант Нимегенского потенциала Nijm.-93, но при энергиях до 350 МэВ [20,23]. В работе [23] было показано, что только Аргонский потенциал 84г. [24] дает сравнительно малую величину $\chi^2$ = 3.3 в области до 350 МэВ. Другие потенциалы такие, как Хамада-Джонсон-62, Рейд-68, Урбана-81, Бонн-89 и другие приводят в этой энергетической области к большим значениям $\chi^2$, поскольку параметризованы на основе более узкого энергетического интервала.

Фазовый анализ [21] выполнен в сравнительно ограниченной области энергий 0-350 МэВ, но в целом хорошо согласуется с другими известными результатами [25], где фазовый анализ проводился в области до 1600 МэВ и [26] для области 0-2500 МэВ. На основе сделанного фазового анализа [21] Нимегенской группой были уточнены некоторые характеристики дейтрона и низкоэнергетического np рассеяния [27].

Следует отметить, что для параметризации всех вариантов потенциалов использовалась константа πNN связи $f^2$=0.074, что заметно меньше результатов приведенных в [18] 0.0776(9) и [28] 0.0803(14). Такая величина константы связи скорее согласуется с данными [29], где получено 0.0760(8) и результатами работы [30], где приведена величина 0.0760(2). С полученными в [20] взаимодействиями, был проведен очень подробный анализ многих NN характеристик и свойств дейтрона [21], сделано сравнение результатов с другими известными потенциалами [23]. Выполнены трехтельные расчеты энергии связи ядра $^3$H, которые дают для разных вариантов взаимодействий величину в интервале -(7.6-7.7) МэВ [31] при экспериментальном значении -8.48 МэВ.

Однако, в указанных работах, не рассматривались формфакторы дейтрона. Выполнить такие расчеты можно на основе волновых функций, которые получены в работах [20], а численные результаты приведены в [32]. Но в работе [32] волновые функции в области 0-25 Фм приводятся с переменным шагом, что затрудняет проведение расчетов. Поэтому можно аппроксимировать численную волновую функцию для получения ее значений в произвольных точках.

Для аппроксимации численных волновых функций Нимегенских потенциалов [32] использовалось разложение в ряд по гауссойдам вида

$$R_L(r) = r^L \sum_k C_k \exp(-\alpha_k r^2) \quad , \qquad (6.5.1)$$

где $C_k$ - коэффициенты и $\alpha_k$ - параметры разложения полной радиальной функции, которая для L=0 и 2 представляется в виде $R_0(r) = u(r)/r$, $R_2= w(r)/r$, а u(r) и w(r) решения стандартного радиального уравнения для S и D волн. Суммирование проводилось до N=13, что позволило сравнительно точно аппроксимировать численную функцию в интервале 0-10 Фм. В процессе аппроксимации коэффициенты разложения находились таким образом, чтобы полная функция была нормирована на единицу. Например, для варианта Nijm.-1 нормировка оказалась равна 1.00002, т.е. ошибка нормировки составляет всего 0.002%. При расчетах характеристик дейтрона апроксимационная волновая функция (АВФ) сшивалась на больших расстояниях с асимптотикой вида [20,27]

$$u(r) \to A_s \exp(- r/R) ,$$
$$w(r) \to A_d\{1 + 3R/r + 3(R/r)^2\}\exp(- r/R) .$$

Здесь R = 4.319 Фм и $A_d=\eta A_s$ с асимптотическими константами $A_s$ = 0.8845(8) и $\eta$=0.0253(2) [27]. На расстоянии 9-10 Фм разница между точной волновой функцией, например для варианта Nijm.-1, и ее асимптотикой в S волне составляет величину около $10^{-4}$, для D волны $10^{-3}$. Поэтому на таких расстояниях вполне можно использовать асимптотику волновой функции.

Результаты расчета свойств дейтрона с АВФ и сравнение с точными результатами [20] приведены в Приложении 2 вместе с отно-сительными ошибками, которые показывают отклонение апроксимационной функции от численных результатов [32]. В Приложении 2 даны также вариационные параметры и коэффициенты разложения для всех вариантов Нимегенских потенциалов. Видно хорошее согласие характеристик, полученных с апроксимационными волновыми функциями и результатами работ [20]. Некоторое отличие наблюдается только для варианта Nijm.-1, что можно объяснить более плохим поведением D волны АВФ на больших расстояниях. Причем увеличение в этом случае размерности ряда (6.5.1) до 15 не приводит к заметному улучшению результатов.

На рис.6.13 непрерывной линией показана точная волновая функция дейтрона для варианта Nijm.-1 [20,32], а штриховой линией, которая практически сливается с непрерывной, ее аппроксимация. Точечной кривой приведена асимптотика этой волновой функции. Видно, что уже при 9-10 Фм асимптотические кривые полностью сливаются с точным решением и аппроксимацией, как для S, так и для D волн.

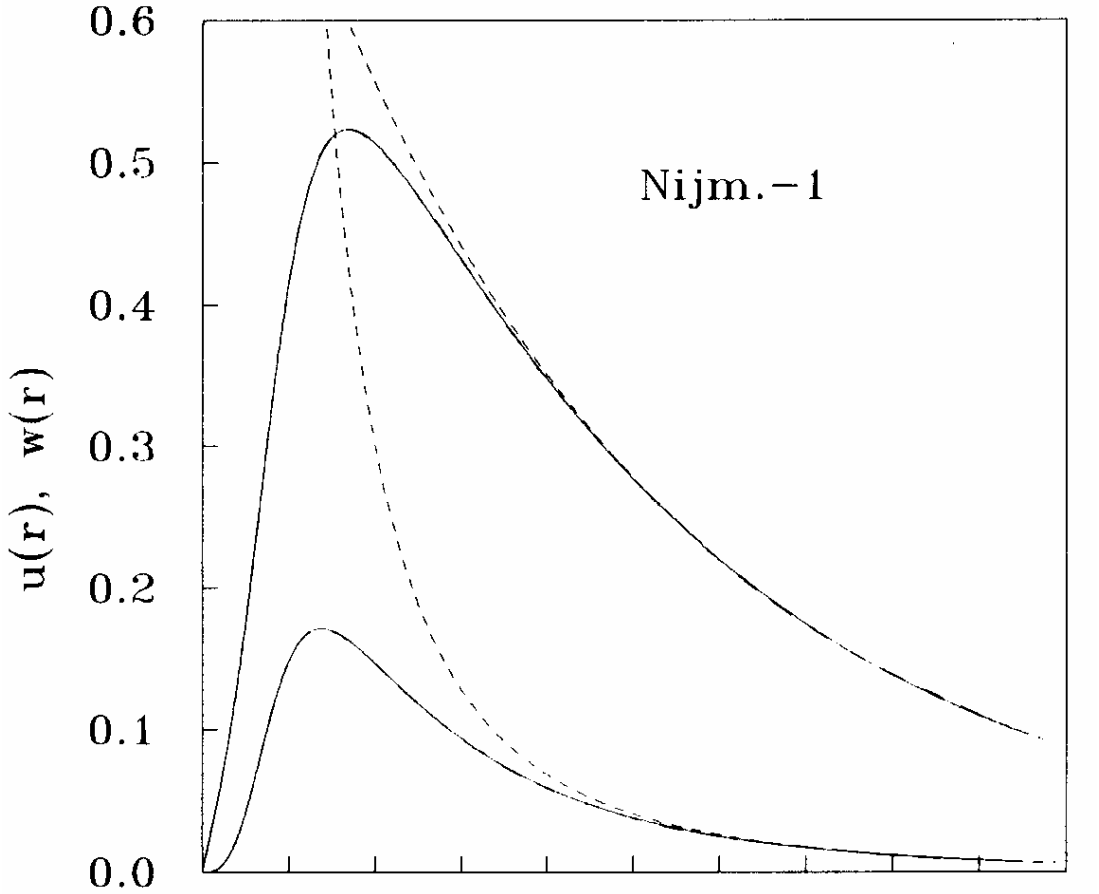

Рис.6.13. Волновые функции для потенциала Nijm.-1 и их асимптотика.

Кроме рассмотренных выше характеристик можно сравнивать и результаты для импульсных распределений, приведенных в работе [32], для различных вариантов потенциала. Вычисленные с АВФ и точные [32] импульсные распределения нормировались на единицу при нулевом переданном импульсе. На рис.6.14 показаны, получен-ные с АВФ импульсные распределения нуклонов в дейтроне для вариантов Nijm.-1 (штрих - пунктирная линия) и Nijm.-93 (штриховая линия) в сравнении с точными результатами [32] - точечная и непрерывная линии соответственно. Видно, что некоторое отличие наблюдается только для потенциала Nijm.-93, а для первого варианта кривые полностью сливаются. Двойным штрих - пунктиром приведены результаты для альтернативного потенциала Рейда с мягким кором [9].

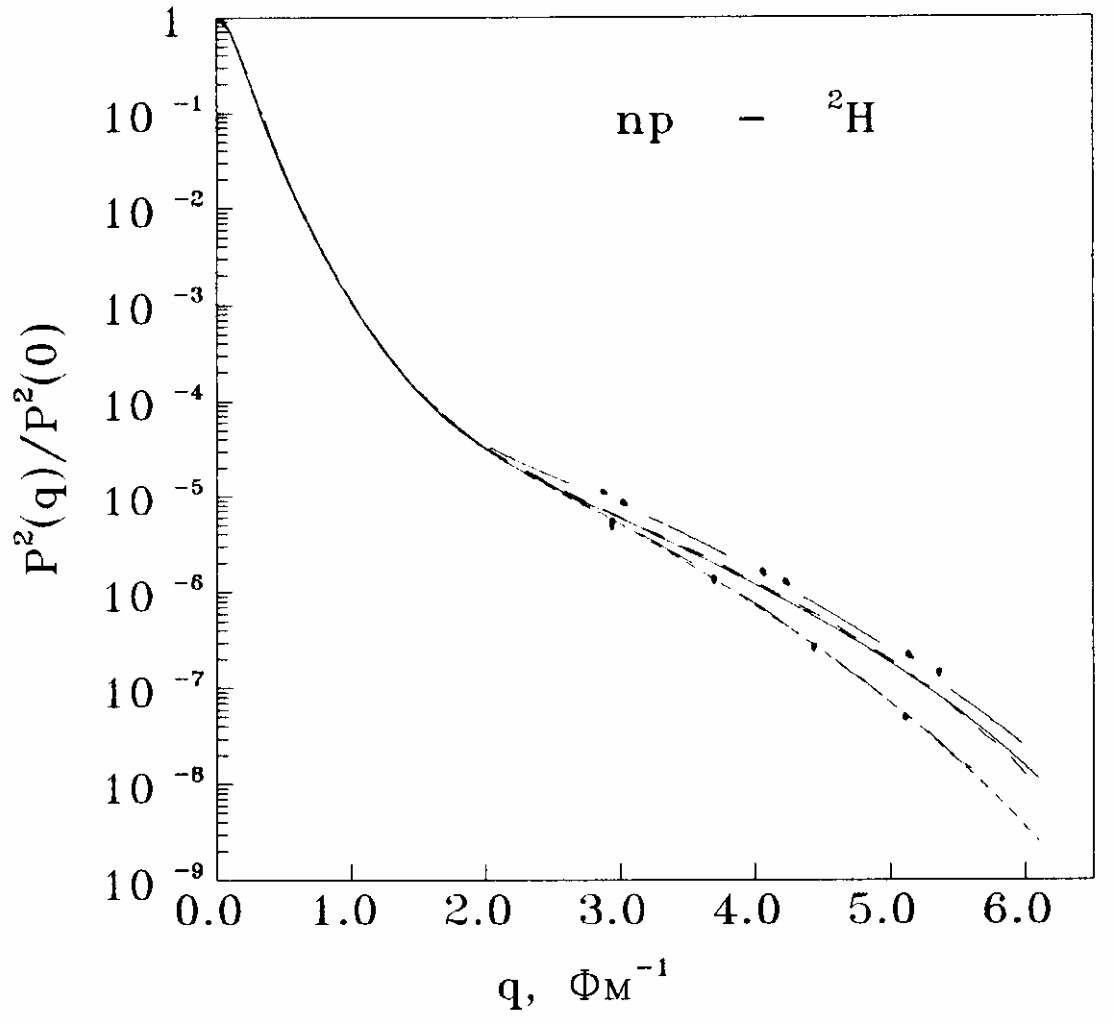

Рис.6.14. Импульсные распределения для разных вариантов Нимегенских потенциалов.

Из приведенных результатов видно, что удается достаточно хорошо согласовать характеристики дейтрона, вычисленные с АВФ и точными численными функциями [32], для всех рассмотренных вариантов потенциалов. Поэтому представляется возможным, использовать, полученные АВФ для расчета формфакторов дейтрона (1.5.9). Результаты расчета формфакторов представлены на рис.6.15 вместе с экспериментальными данными работ [15].

Точечная линия на рис.6.15а представляет результаты для потенциала Рейда-68 [9], а штриховая для варианта Nijm.-1. Результаты для потенциалов Nijm.-2 и Nijm.-93 практически совпадают и показаны непрерывной линий, а для Reid-93 даны штрих - пунктиром, который почти сливается с непрерывной кривой. На рис.6.15б результаты для Reid-93 совпадают с классическим Рейдом-68 и представлены точечной линией. Штриховой линией показан вариант

Nijm.-2, результаты для варианта Nijm.-93 даны непрерывной линией, а для Nijm.-1 - штрих - пунктирной, которая практически совпадает с непрерывной кривой.

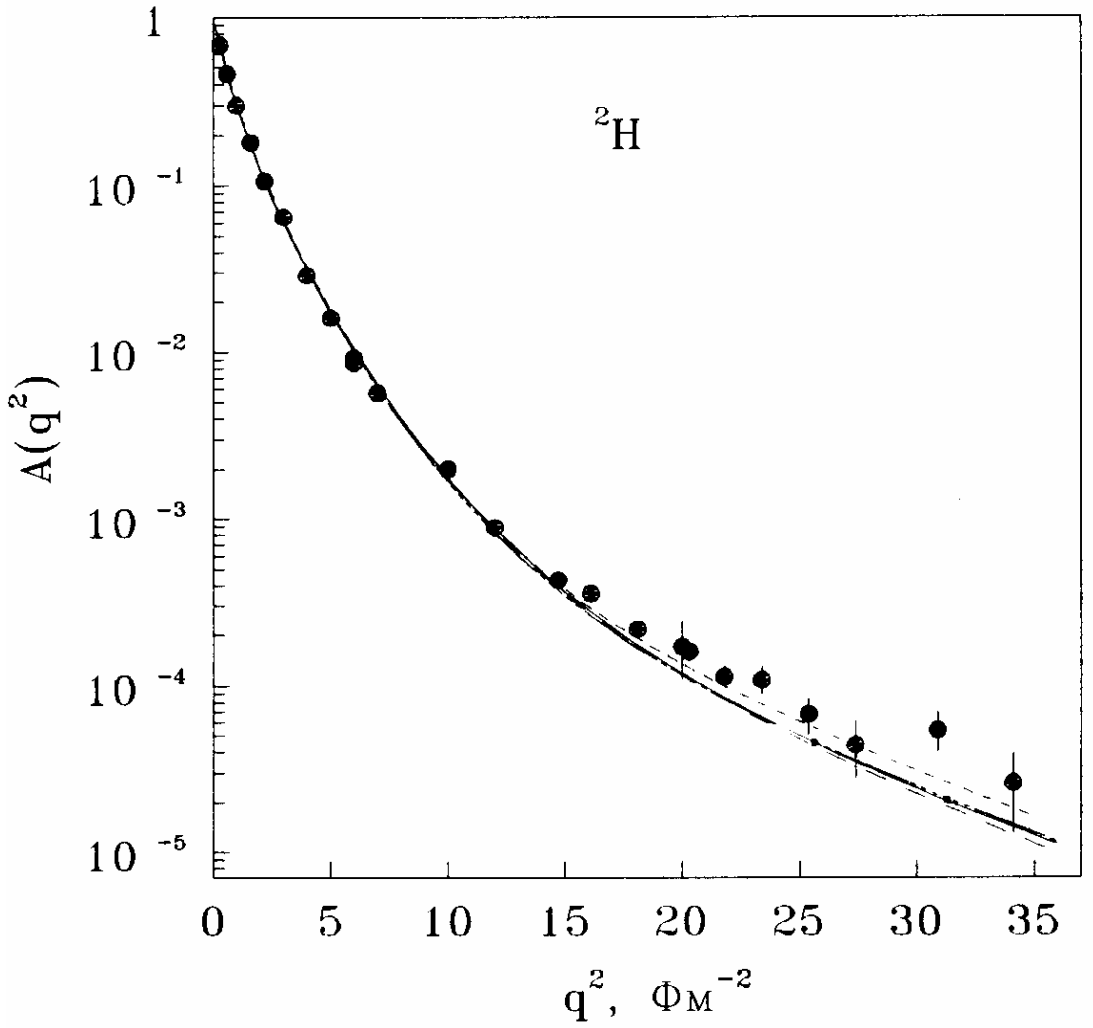

Рис.6.15а. Формфактор A(q) дейтрона для Нимегенских потенциалов.

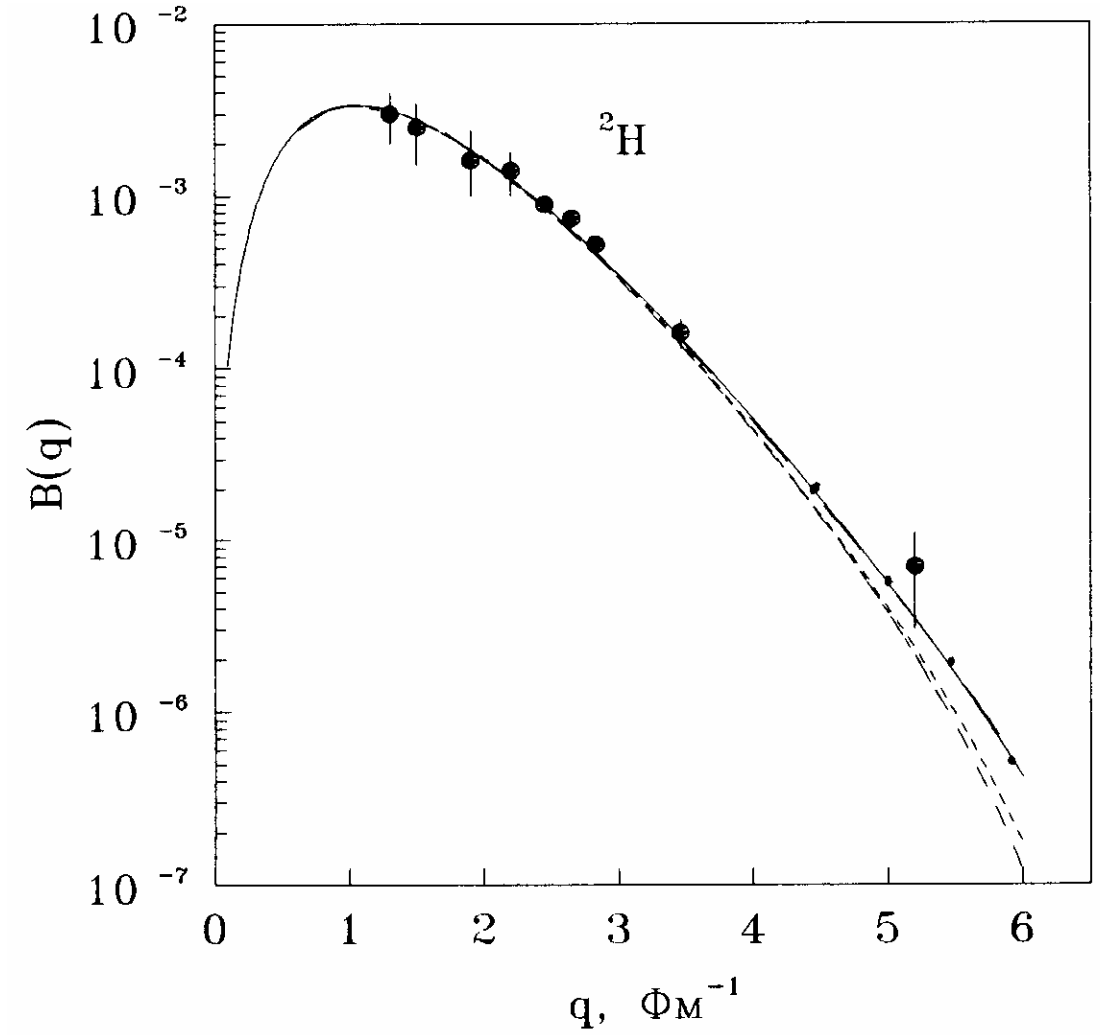

Рис.6.15б. Формфактор B(q) дейтрона для Нимегенских потенциалов.

Нерелятивистские формулы для формфакторов, которые использовались в этих расчетах, применялись и ранее в работах [7,15]. В работе [33] сделано сравнение релятивистских и нерелятивистских методов расчета для Парижского и Аргонского потенциалов и показано, что в области 3.5-4.0 $Фм^{-1}$ эти результаты практически совпадают.

При больших переданных импульсах 5.0-6.0 $Фм^{-1}$ релятивистские эффекты становятся заметными, но существенно не меняют вида формфакторов и не играют доминирующей роли. В работе [34] также рассмотрены релятивистские поправки, и можно считать, что они не дают решающего вклада до 5-6 $Фм^{-1}$, а их величина не превышает имеющихся экспериментальных ошибок.

Таким образом, из приведенных результатов видно, что полученные АВФ вполне способны правильно передать поведение численных функций для всех вариантов Нимегенских потенциалов и позволяют получить хорошее согласие рассмотренных характеристик дейтрона с вычислениями для точных волновых функций. Полученные на основе АВФ формфакторы дейтрона, полностью согласуются с известными экспериментальными данными и мало различаются между собой в рассмотренной области переданных импульсов [35].

## 6.6. Влияние константы πNN связи на описание характеристик дейтрона

Полученный выше глубокий экспоненциальный NN потенциал с узлом в S волне основан на константе πNN связи $f^2 = 0.0776$, найденной из анализа характеристик NN рассеяния в работах [18]. В тоже время, как уже говорилось, имеются и другие результаты, в частности, в работах Нимегенской группы [27] была получена величина $f^2 = 0.074$, а в исследованиях, выполненных группой политехнического института Вирджинии [29] найдено 0.076.

Поэтому представляется интересным выяснить зависимость параметров глубокого потенциала с запрещенными состояниями и качество описания им результатов NN фазового анализа в зависимости от величины константы πNN связи. Будем использовать здесь такой же вид локального NN потенциала, как в (6.2.1) с обрезанием в 20 степени и OPEP вида (6.2.2).

Параметры потенциала для каждого $f^2$ подбирались исходя из описания эффективного радиуса $r_0$, длинны рассеяния $a_0$ и энергии связи $E_d$ дейтрона, а затем с ними рассчитывались фазы np рассеяния до 500 МэВ, параметры смешивания и другие характеристики дейтрона. Недавние данные по NN фазам приведены в работах [26,27], а другие характеристики дейтрона можно найти, например, в работах [18]. Результаты расчетов различных характеристик дейтрона и np рассеяния для полученных вариантов потенциалов и их параметры в зависимости от величины константы $f^2$ даны в Приложении 3 [36]. Видно, что во всех случаях удается хорошо воспроизвести все рассматриваемые характеристики, а величина квадрупольного момента с уменьшением $f^2$ несколько уменьшается, находясь между результатами работ [18] и [27].

Степень фактора обрезания OPEP, равная N=20, выбрана исходя из необходимости, убрать узел в D волне. Малая величина 3-5 не позволяется избавиться от узла [7,17] и только увеличение N до 15 постепенно приводят к его исчезновению. Вообще величина степени обрезания слабо влияет на результаты и может находиться примерно в интервале 15-30. Дальнейшее ее увеличение, приводит к ухудшению результатов по np фазам рассеяния, хотя общая форма волновой функции дейтрона существенно не изменяется.

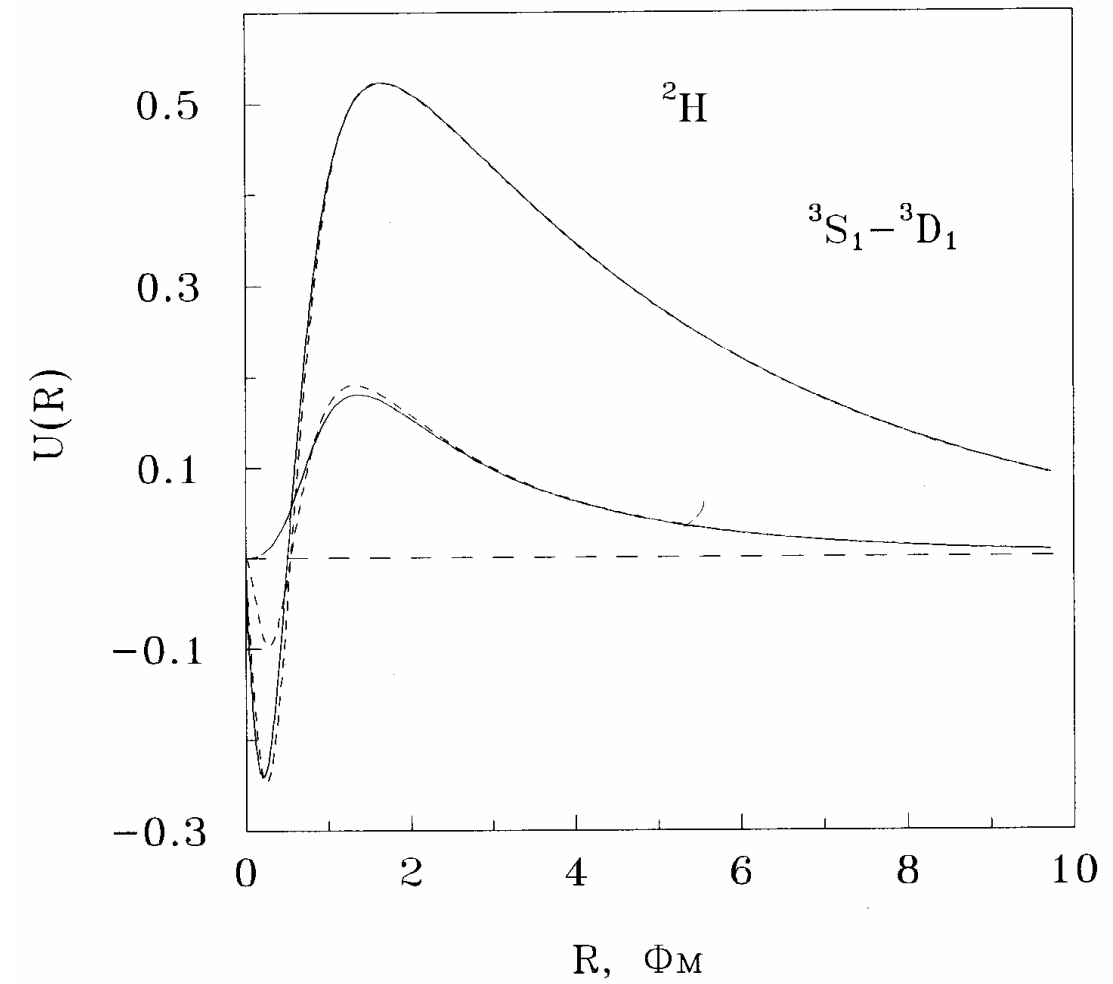

Рис.6.16. Волновая функция дейтрона для разных NN потенциалов.

Волновая функция дейтрона для любого варианта потенциала имеет узел в S - волне примерно в области 0.50-0.52 Фм, а D - волна безузловая, как показано на рис.6.16 для варианта $f^2$=0.074 непрерывной линией. Штриховой линией приведена волновая функция гауссового потенциала из работ [7].

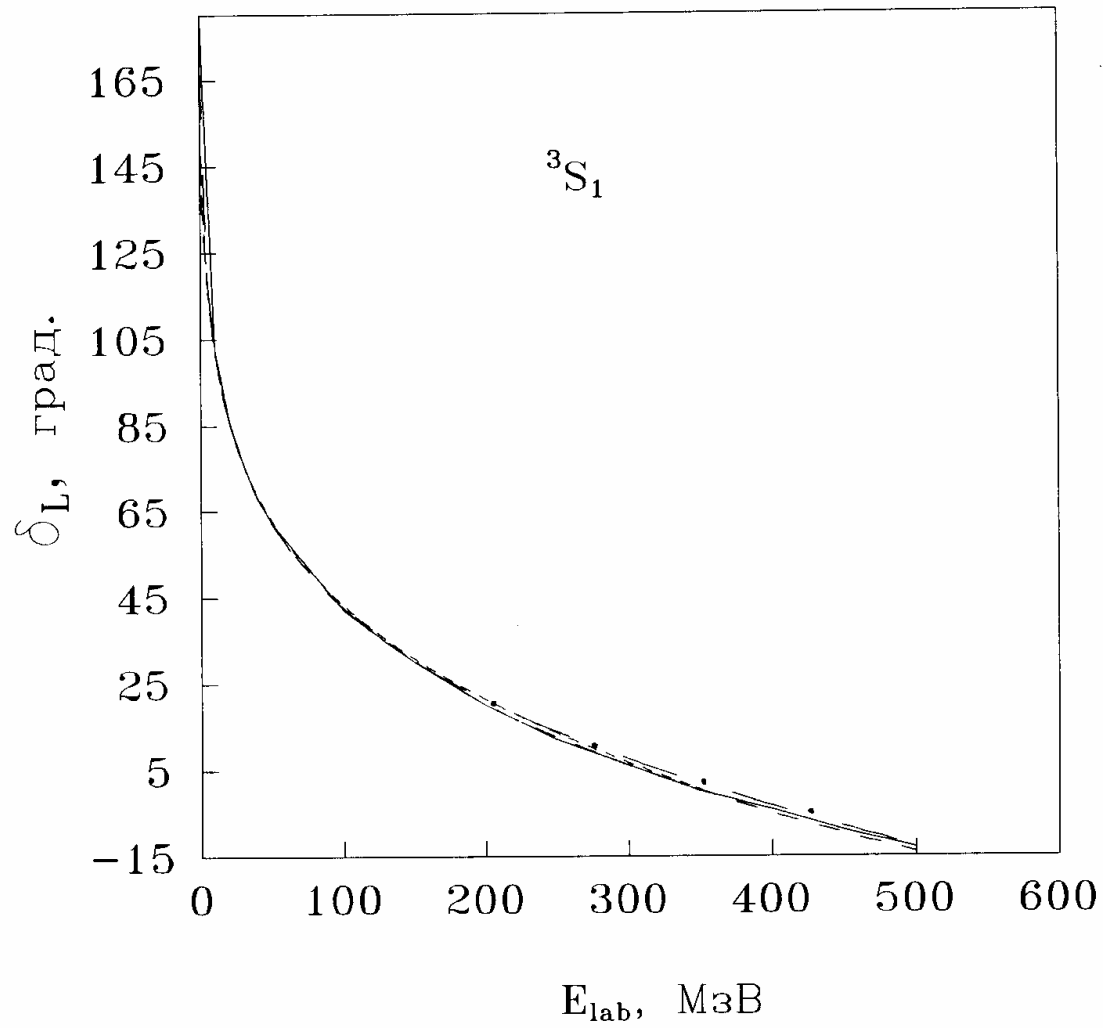

Рис.6.17а. Триплетная S фаза NN рассеяния - непрерывная линия. Штрих - пунктир - результаты [26], а точечная линия - фазы, полученные в [27].

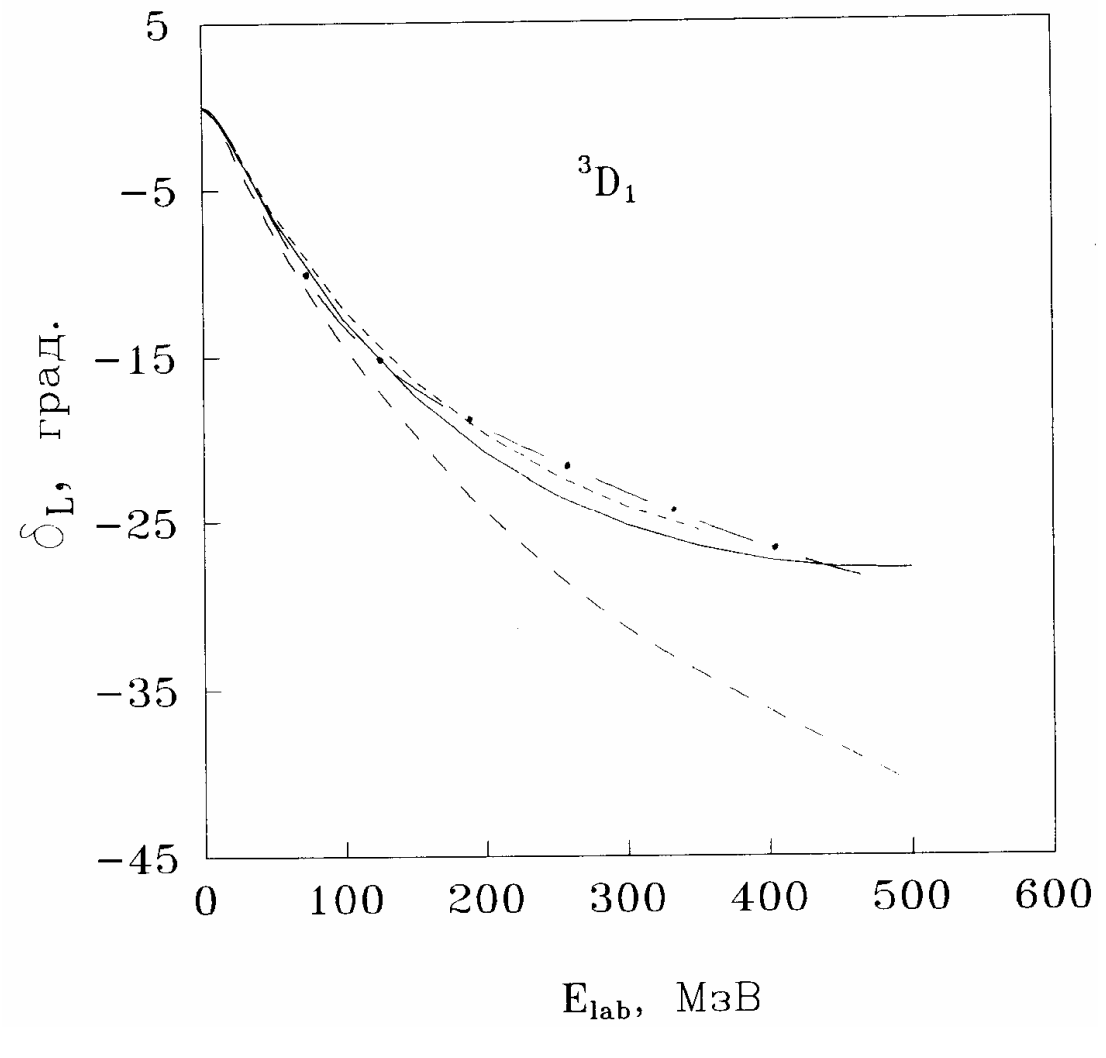

Рис.6.17б. Триплетная D фаза NN рассеяния - непрерывная линия. Штрих - пунктир - результаты [26], а точечная линия - фазы, полученные в [27].

Фазы рассеяния и параметр смешивания, вычисленные с приведенными в Таблице (Приложение 3) параметрами, для варианта $f^2$=0.074 показаны на рис.6.17 непрерывной линией. Штрих - пунктирной линией даны результаты фазового анализа SM97 [26], выполненного группой политехнического института Вирджинии, а точечная линия показывает фазы, полученные Нимегенской группой [27]. Штриховой кривой даны фазы потенциала работ [7].

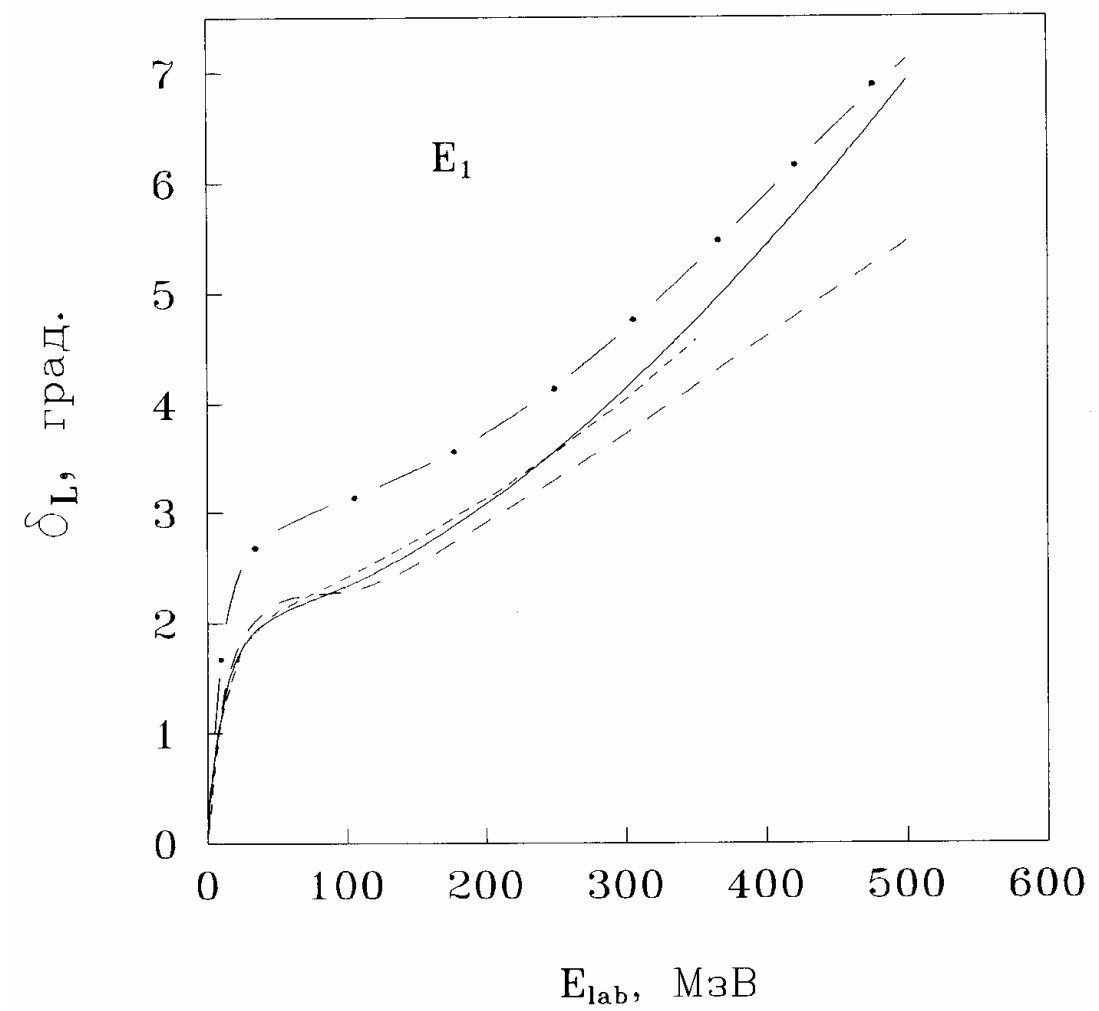

Рис.6.17в. Параметр смешивания для триплетных фаз NN рассеяния - непрерывная линия. Штрих - пунктир - результаты [26], а точечная линия - фазы, полученные в [27].

Качество описания триплетных фаз при $f^2$ больше 0.076, например, для 0.0776-0.0803 [18,28], становится заметно хуже, в частности, D фаза спадает быстрее, чем это следует из результатов фазового анализа. Это хорошо демонстрирует штриховая линия на рис.6.17б для гауссового потенциала с $f^2$ = 0.07745 из работ [7]. С точки зрения лучшего описания D фазы, предпочтительным оказывается значение константы πNN связи, равное 0.074, как было предложено в работах [27].

На рис.6.18 непрерывными линиями приведены формфакторы дейтрона, вычисленные с волновой функцией основного состояния экспоненциального взаимодействия при $f^2$=0.074, а штриховой кривой - результаты для потенциала из работ [7], описанного в разделе 6.2 этой главы. Точечной линией дан формфактор для потенциала Рейда [9], а штрих - пунктирной для взаимодействия Nijm.93 [20].

Экспериментальные данные взяты из работ [15].

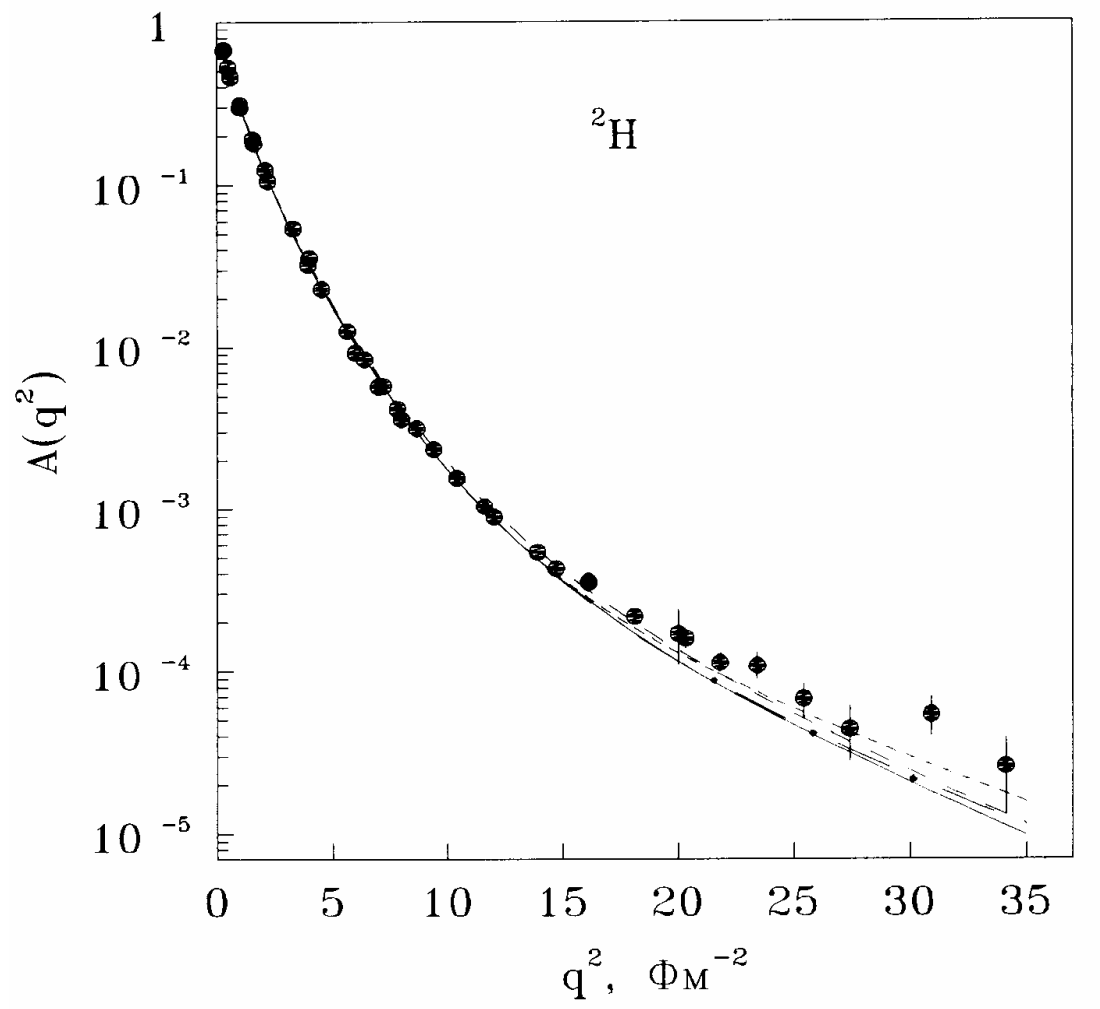

Рис.6.18а. Формфактор A(q) дейтрона для разных NN потенциалов.

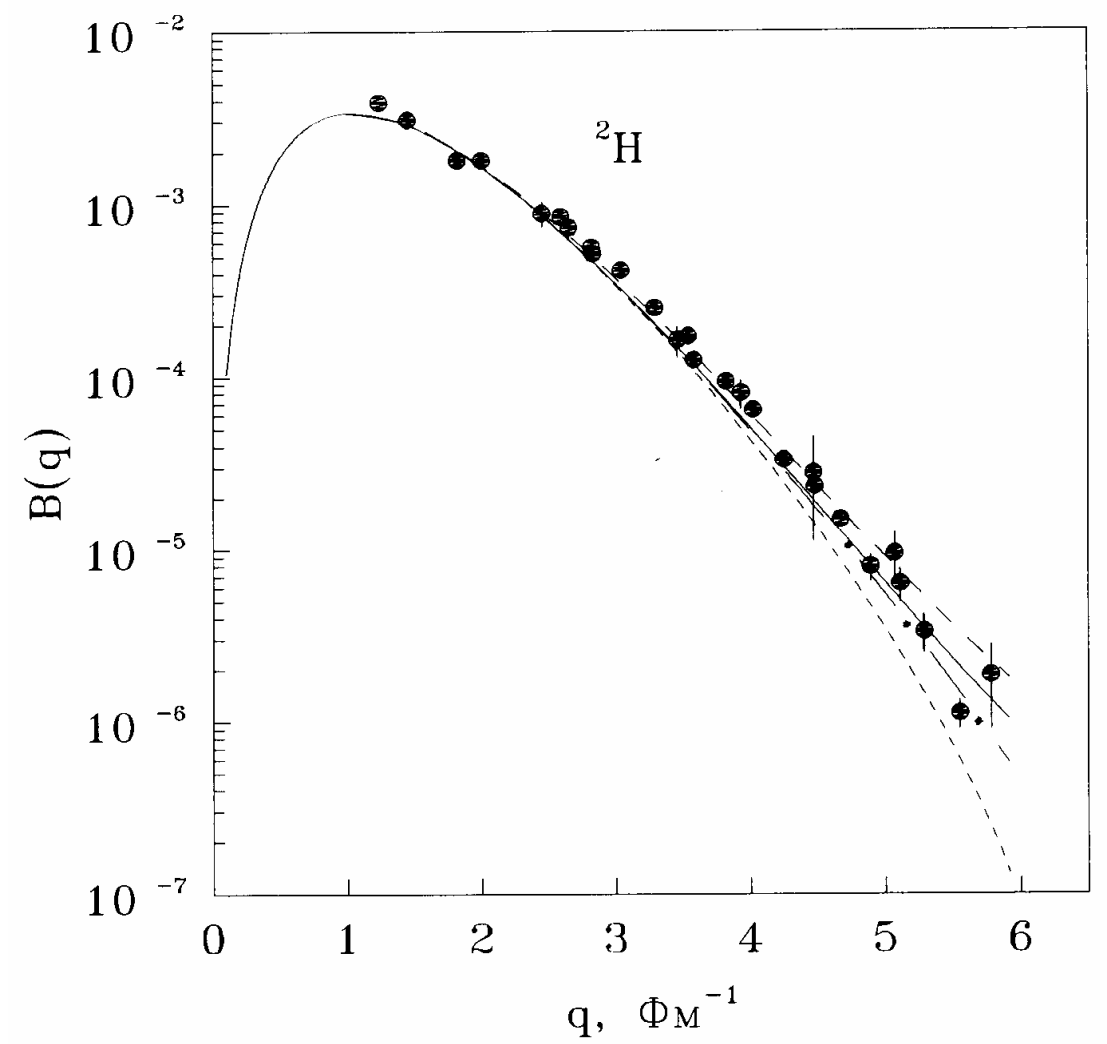

Рис.6.18б. Формфактор B(q) дейтрона для разных NN потенциалов.

В работах [37] было показано, что потенциалы с узлом в D вол-не, полученные в [7], приводят к сильно завышенным сечениям при развале дейтрона протонами и $p^2H$ рассеянии назад при больших переданных импульсах. Такое поведение сечений связывают с относительно большой величиной импульсных распределений для потенциалов, которые приводят к двум узлам в волновой функции, как было получено в [7,17].

Уменьшение величины импульсных распределений при переданных импульсах около 3 $Фм^{-1}$, получающееся для взаимодействия с узлом только в S волне, может привести к улучшению поведения расчетных сечений упругого $p^2H$ рассеяния и развала дейтрона протонами.

С полученными потенциалами рассматривались тензорные и векторные поляризации в упругом $e^2H$ рассеянии, определения которых можно найти, например, в работах [33,38]

$$d\sigma/ d\Omega = S\, (d\sigma/ d\Omega)_{Mott}, \qquad S = A(q) + B(q)\tan^2(\theta_e/2) ,$$

$$A(q) = G_c^2(q)+(8\eta^2/9)G_q^2(q)+(2\eta/3)G_m^2(q) ,$$

$$B(q) = (4\eta/3)(1+\eta)G_m^2(q) ,$$

$$t_{20} = - 1/(\sqrt{2}\, S)\{(8\eta/3)G_cG_q + (8\eta^2/9)G_q^2 + (\eta/3)[1+2(1+\eta)\tan^2(\theta_e/2)]G_m^2\} ,$$

$$t_{21} = 2\eta/(\sqrt{3}\, S)\, [\eta+ \eta^2\sin^2(\theta_e/2)]^{1/2}\, G_mG_q\sec(\theta_e/2) ,$$

$$t_{22} = - \eta/(2\sqrt{3}\, S)G_m^2 ,$$

$$t_{10} = \eta /S \sqrt{\frac{2}{3}}\, [(1+\eta)(1+\eta\, Sin^2(\theta_e /2))]^{1/2}\tan(\theta_e/2)\sec(\theta_e/2)\, G_m^2 ,$$

$$t_{11} = 2/(\sqrt{3}\, S)\, [\eta(1+\eta)]^{1/2}\tan(\theta_e/2)G_m(G_c + (\eta/3)G_q) ,$$

$$G_c (0) = 1 , \quad G_q (0) = M_d^2Q_d = 25.83 , \quad G_m (0) = (M_d/M_p)\mu_d = 1.714 ,$$

где A(q) и B(q) формфакторы дейтрона, $\eta$ - кулоновский параметр, а определения структурных функций $G_c$, $G_q$ и $G_m$ через волновые функции даны в [33,38].

Результаты расчета тензорных и векторных поляризации для варианта потенциала с $f^2$=0.074 показаны на рис.6.19 и 6.20 непрерыв-ной линией, а длинными штрихами даны результаты для потенциала из работ [7]. Штрих - пунктирной кривой показана поляризация для Нимегенского потенциала Nijm.1 [20], точечной для потенциала Рейда [9]. Двойной штрих - пунктир - результаты для Nijm.2 [20] и короткие штрихи для Парижского взаимодействия [22].

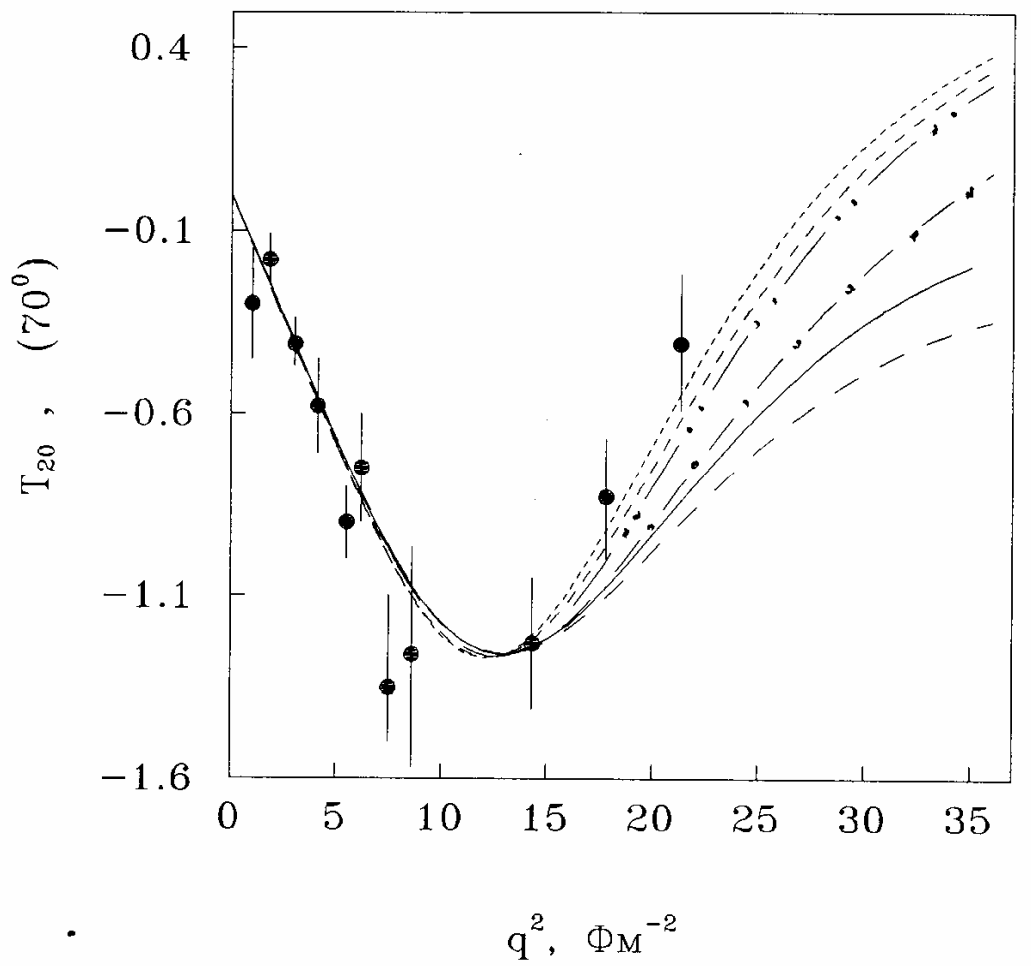

Рис.6.19а. Тензорные поляризации в упругом $e^2H$ рассеянии для различных NN потенциалов.

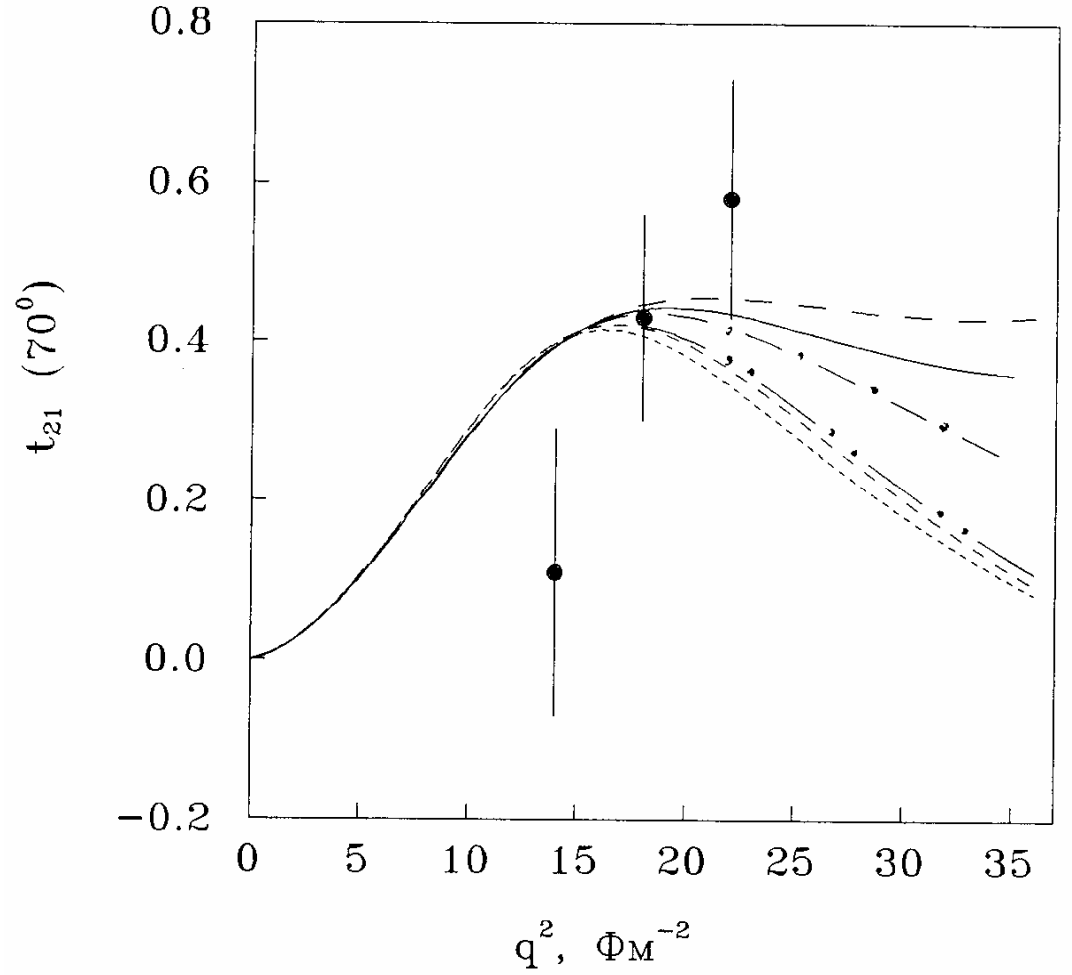

Рис.6.19б. Тензорные поляризации в упругом $e^2H$ рассеянии для различных NN потенциалов.

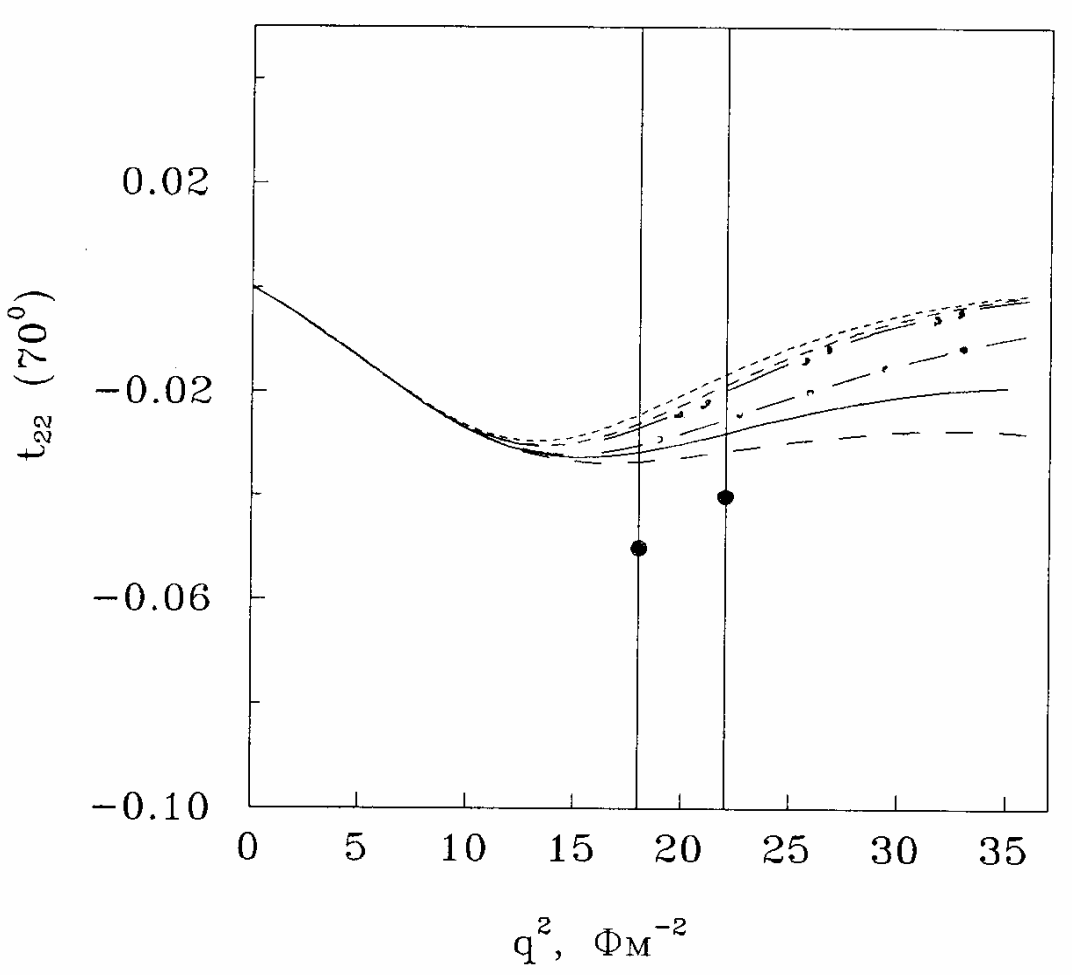

Рис.6.19в. Тензорные поляризации в упругом $e^2H$ рассеянии для различных NN потенциалов.

Точками даны экспериментальные данные по тензорным поляризациям из работ [33,38]. Видно, что при малых переданных импульсах результаты для различных потенциалов практически совпадают. При q>15 Фм$^{-1}$ наблюдается уже заметное различие между результатами для потенциалов с кором и запрещенными состояниями. Однако, отсутствие данных при больших переданных импульсах и большие экспериментальные ошибки не позволяет пока сделать окончательные выводы в пользу какого - либо типа NN взаимодействий.

Но уже сейчас видно, что потенциал из работ [7] приводит к заниженной величине $t_{20}$ при импульсах выше 20 Фм$^{-2}$. Предложенный здесь вариант потенциала при $f^2$=0.074 дает несколько лучшие результаты, хотя и они лежат заметно ниже имеющихся экспериментальных данных. Для других тензорных характеристик $t_{21}$ и $t_{22}$ экспериментальных данных совсем мало, а ошибки измеренных величин слишком велики, чтобы можно было использовать их для выбора определенного типа потенциала. Экспериментальные данные по векторным поляризациям вообще отсутствуют и можно лишь сравнивать результаты расчетов для различных NN потенциалов.

Таким образом, для рассмотренного вида взаимодействия, также как для потенциалов Нимегенской группы [20], предпочтительным оказывается малая величина константы πNN связи, которая позволя-ет лучше передать имеющиеся экспериментальные данные по дейтрону и np рассеянию [38].

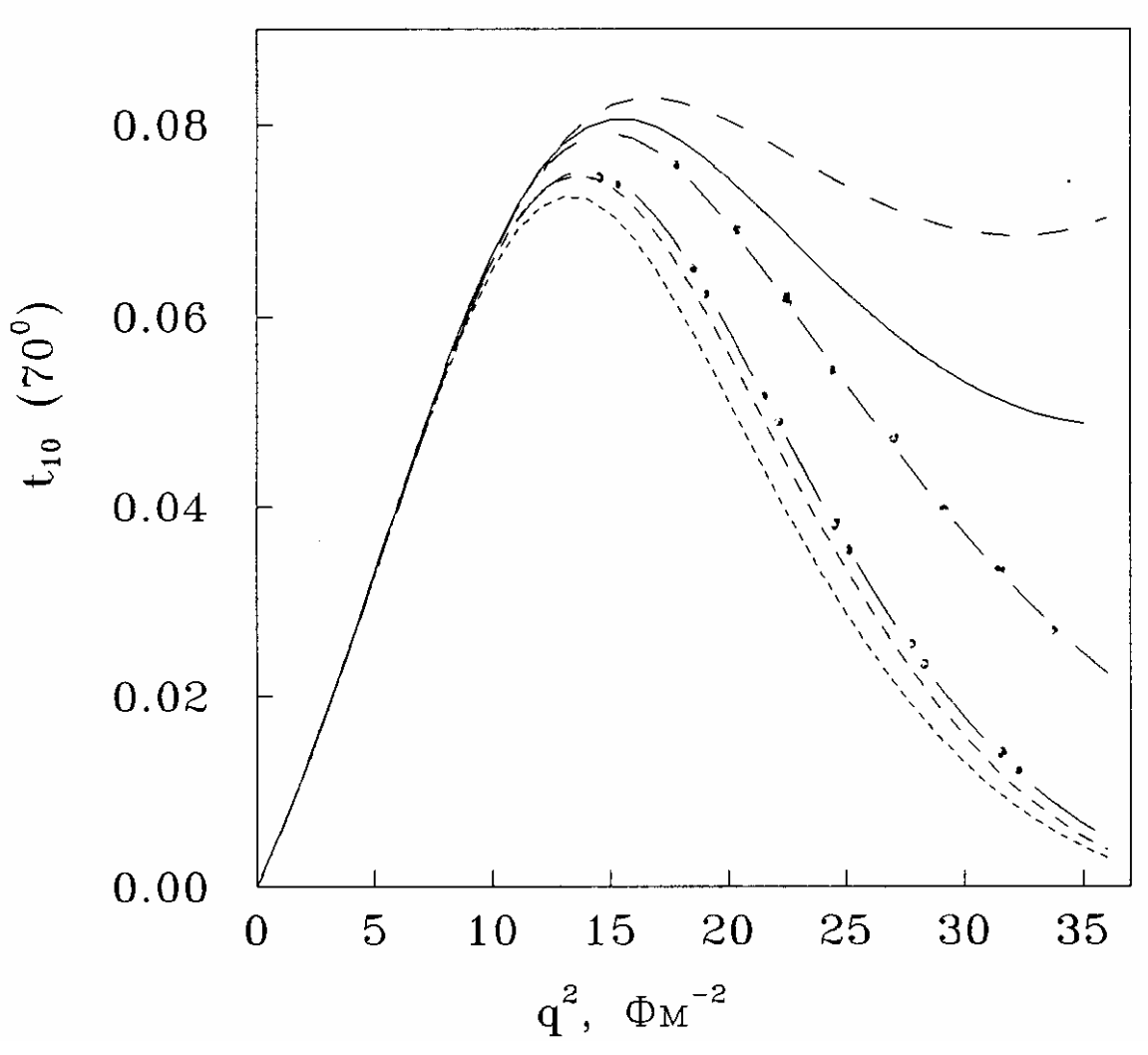

Рис.6.20а. Векторная $t_{10}$ поляризация в упругом $e^2H$ рассеянии для различных NN потенциалов.

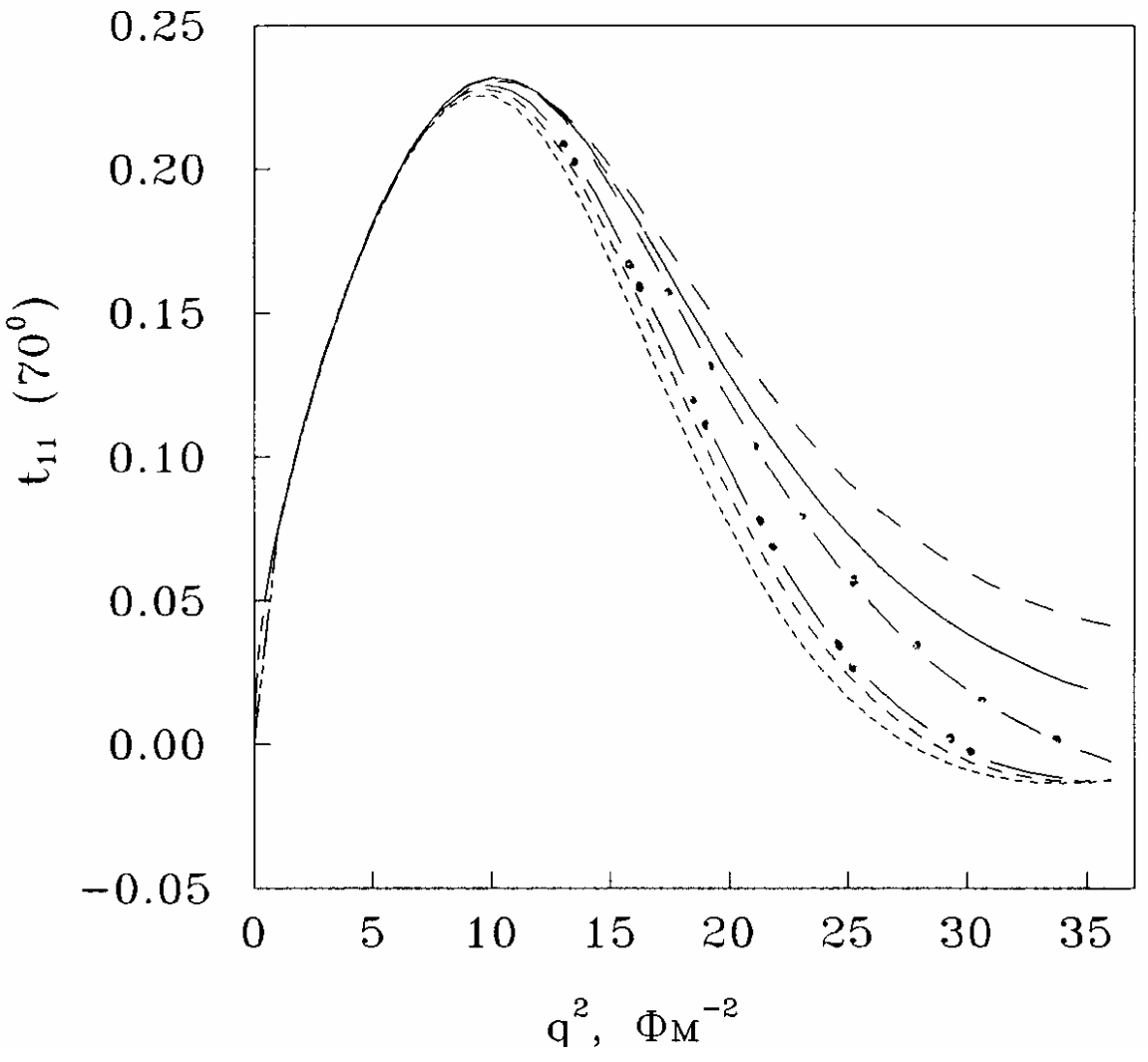

Рис.6.20б. Векторная $t_{11}$ поляризация в упругом $e^2H$ рассеянии для различных NN потенциалов.

## 6.7. Кластерная $^4$He$^2$H система с тензорными силами

Как уже говорилось, в работах [2] были параметризованы межкластерные центральные гауссовы потенциалы взаимодействия, правильно воспроизводящие фазы упругого $^4$He$^2$H рассеяния при низких энергиях, и содержащие запрещенные состояния. Показано, что на основе этих потенциалов в кластерной модели можно воспроизвести основные характеристики связанного состояния ядра $^6$Li, вероятность кластеризации которого в рассматриваемом канале сравнительно высока.

Все состояния такой системы оказываются чистыми по орбитальным схемам Юнга [3] и потенциалы, полученные из фаз рассеяния, можно непосредственно применять для описания характеристик основного состояния ядра. Несмотря на определенные успехи такого подхода, ранее рассматривались только чисто центральные межкластерные взаимодействия. В рамках потенциальной кластерной модели не учитывалась тензорная компонента, которая приводит к появлению D волны в волновой функции связанного состояния и рассеяния, позволяющая рассматривать квадрупольный момент ядра $^6$Li.

Под тензорным потенциалом здесь следует понимать взаимодействие, оператор которого зависит от взаимной ориентации полного спина системы и межкластерного расстояния. Математическая форма записи такого оператора полностью совпадает с оператором двухнуклонной задачи, поэтому и потенциал по аналогии будем называть тензорным.

По - видимому, впервые тензорные потенциалы были использованы для описания $^2$H$^4$He взаимодействия в начале 80 - х годов в работе [39], где предпринята попытка ввести тензорную компоненту в оптический потенциал. Это позволило заметно улучшить качество описания дифференциальных сечений рассеяния и поляризаций.

В работе [40] на основе “фолдинг модели” выполнены расчеты сечений и поляризаций и учет тензорной компоненты потенциала позволил улучшить их описание.

В дальнейшем такой подход был использован в работе [41], где “сверткой” нуклон - нуклонных потенциалов получены $^2$H$^4$He взаимодействия с тензорной компонентой. Показано, что в принципе удается описать основные характеристики связанного состояния $^6$Li, включая правильный знак и порядок величины квадрупольного момента. Однако, в работах [39,40] рассматривались только процессы рассеяния кластеров, а в [41] только характеристики связанного состояния ядра $^6$Li без анализа фаз или сечений упругого рассеяния.

Тем не менее, гамильтониан взаимодействия любой системы должен быть единым для процессов рассеяния и связанных состояний кластеров, как это было сделано в работах [2] в случае чисто центральных потенциалов. Поэтому нужно найти такой потенциал $^2$H$^4$He взаимодействия с тензорной компонентой, который позволил бы правильно передать и характеристики связанного состояния $^6$Li в двухкластерной модели, включая квадрупольный момент ядра, и характеристики кластерного рассеяния при низких энергиях.

Правильный знак и величина квадрупольного момента ранее были получены в работах [42], где использовались феноменологические волновые функции ядра в $^2$H$^4$He модели. Этот результат показывает, что в простой двухкластерной системе можно получить хорошее описание квадрупольного момента одновременно с другими характеристиками основного состояния $^6$Li. Правильный знак квадрупольного момента с величиной -0.076 Фм$^2$ получается и в некоторых расчетах по методу резонирующих групп [43]. Однако зарядовый радиус ядра, в этих расчетах, оказывается несколько заниженным 2.3-2.4 Фм.

Исходя из сказанного выше, представляется интересным рассмотреть влияние эффектов, которые дают тензорные взаимодействия в потенциальной двухкластерной $^2$H$^4$He модели ядра $^6$Li с потенциалом вида

$$V(r) = V_c(r) + V_t(r)\, S_{12}, \quad S_{12} = [6(Sn)^2 - 2S^2],$$
$$V_c(r) = -V_0 \exp(-\alpha r^2), \quad V_t(r) = -V_1 \exp(-\beta r^2). \qquad (6.7.1)$$

Здесь S - полный спин системы, n - единичный вектор, совпадающий по направлению с вектором межкластерного расстояния.

Таким образом, будем искать не параметры феноменологических волновых функций, как это сделано в [42], а межкластерных потенциалов, которые позволяют правильно передать все характеристики основного состояния ядра и фазы упругого $^2$H$^4$He рассеяния при низких энергиях. Тем самым, попытаемся описать характеристики непрерывного и дискретного спектра $^2$H$^4$He кластерной системы на основе единого гамильтониана с феноменологическим потенциалом, содержащим тензорную компоненту.

Поскольку основному состоянию $^6$Li [1,2,4] сопоставляется орбитальная схема {42}, то в S состоянии должен быть запрещенный уровень со схемой {6}. В тоже время в D волне запрещенное состоя-ние отсутствует, так как разрешенная схема {42} совместима с орбитальным моментом 2. Это значит, что волновая функция S состояния будет иметь узел, а D волна должна быть безузловой. Такая классификация запрещенных и разрешенных состояний по схемам Юнга, изложенная во второй главе, в целом позволяет определить общий вид волновой функции кластерной системы.

В известных трехтельных расчетах [44,45] воспроизводятся очень многие характеристики ядра $^6$Li. Однако, квадрупольный момент получается положительным, также как положительна величина $\eta_D$=0.018 - 0.055, экспериментальное значение которой 0.005±0.017, приведенное в [46] допускает отрицательные значения.

Асимптотическая константа $C_0$ в различных вариантах трехтельных расчетов [45] находится в области 2.2 - 2.4 при экспериментальном значении 2.15(10) [46]. В работах [44], для $C_0^W$ получена величина 2.71 при нормировке волновой функции на 70-75% вероятность $^2$H$^4$He канала.

В работах [45] для вероятности S состояния $^2$H$^4$He канала найдено 60-65%, причем вероятность D состояния оказывается очень малой 0.025-0.63%. Зарядовый радиус в трехтельных расчетах находится в пределах 2.26-2.43 Фм и несколько меньше экспериментальных величин 2.56(5) и 2.54(6) Фм, приведенных, например, в работах [43].

Для получения отрицательного знака квадрупольного момента необходимо и отрицательное значение $\eta_D$. Возможность отрицательных $\eta_D$ отмечена в работе [47], где приведены новые экспериментальные данные, которые дают величину -(0.01÷0.015), что возможно только при разных знаках в асимптотике S и D частей волновой функции. Именно такой вид волновых функций был получен в работе [41], что позволило получить правильный отрицательный знак квадрупольного момента.

Из всего сказанного выше становится ясно, какой общей формой должна обладать волновая функция, чтобы правильно описывать квадрупольный момент. Тензорная часть потенциала, скорее всего, должна быть достаточно узкой и мелкой, чтобы не появлялся узел в D волне, а центральная часть будет широкой с глубиной, способной обеспечить наличие узла в S волне. В целом тензорное взаимодействие будет относительно слабым (хотя бы потому, что квадрупольный момент ядра имеет малую величину) и поэтому при построении потенциала можно исходить из результатов работы [2], где приведены параметры центрального гауссового потенциала $V_0$=-

76.12 МэВ и $\alpha$=0.2 $Фм^{-2}$.

*Таблица 6.7. Параметры потенциалов $^2H^4He$ взаимодействия с тензорной частью [48].*

| № | $V_0$, (МэВ) | $\alpha$, ($Фм^{-2}$) | $V_1$, (МэВ ) | $\beta$, ($Фм^{-2}$) |
|---|---|---|---|---|
| 1 | -71.979 | 0.2 | -27.0 | 1.12 |
| 2 | -77.106 | 0.22 | -40.0 | 1.6 |

На основе этих представлений были получены два варианта потенциала, параметры которых приведены в табл.6.7. Параметры взаимодействий подгонялись так, чтобы правильно передать энергию связи, квадрупольный момент ядра и фазы упругого рассеяния при низких энергиях [48].

*Таблица 6.8. Сравнение вычисленных в двухкластерной $^2H^4He$ модели и экспериментальных [46,47,49] значений для характеристик связанного состояния ядра $^6Li$ [48].*

| Характеристики ядра $^6$Li | Расчеты для потенциала №1 | Расчеты для потенциала №2 | Экспериментальные данные |
|---|---|---|---|
| $E_{св}$ (МэВ) | -1.4735 | -1.4735 | -1.4735 |
| $R_r$ (Фм) | 2.60 | 2.56 | 2.56(5); |
| $R_f$ (Фм) | 2.53 | 2.50 | 2.54(6) |
| Q ($Фм^2$) | -0.064 | -0.064 | -0.0644(7) |
| $C_0^0$ | 1.9(1) | 1.9(1) | 2.15(10) |
| $\eta_D$ | -0.0115(5) | -0.0120(5) | -0.0125(25); |
| $\eta_{D2}$ | -0.0119(3) | -0.0122(2) | 0.005±0.017 |
| $C_0^{W0}$ | 2.97(3) | 2.85(5) | --- |
| $C_0^W$ | 3.17(3) | 3.03(3) | --- |
| $\mu_d/\mu_0$ | 0.848 | 0.847 | 0.822 |
| $P_D$ (%) | 1.59 | 1.78 | --- |

Результаты расчета характеристик основного состояния $^6$Li для этих взаимодействий приведены в табл.6.8 вместе с экспериментальными данными из [46,47,49].

Видно, что полученные потенциалы вполне позволяют описать рассмотренные экспериментальные результаты, причем параметры центральной части мало отличаются от параметров чисто центрального взаимодействия [2]. Отметим, что если рассматривать только характеристики связанных состояний, то можно найти очень много вариантов параметров. И только рассматривая одновременно фазы рассеяния и свойства ядра $^6$Li, удается практически однозначно фиксировать параметры потенциала. Зарядовые радиусы $R_r$ и $R_f$, вычисленные на основе этих потенциалов, отличаясь на 2÷3%, находятся в интервале экспериментальных ошибок. Асимптотическая константа $C_0^W$ ≈ 3.0÷3.2, усредненная по интервалу 7÷15 Фм, хорошо согласуется с вычислениями в микроскопической модели [50], где получена величина 3.3 и с трехтельными расчетами, если пересчитать приведенную в [44] константу на единичную вероятность $^2$H$^4$He канала - 3.1÷3.2.

Для вероятности D состояния найдена величина, несколько меньшая, чем требуется для правильного описания магнитного момента. Однако, подобная ситуация существует и в классической NN системе. Для получения магнитного момента дейтрона требуется примесь D состояния около 4%, а феноменологические потенциалы дают величину порядка 6÷7% [18].

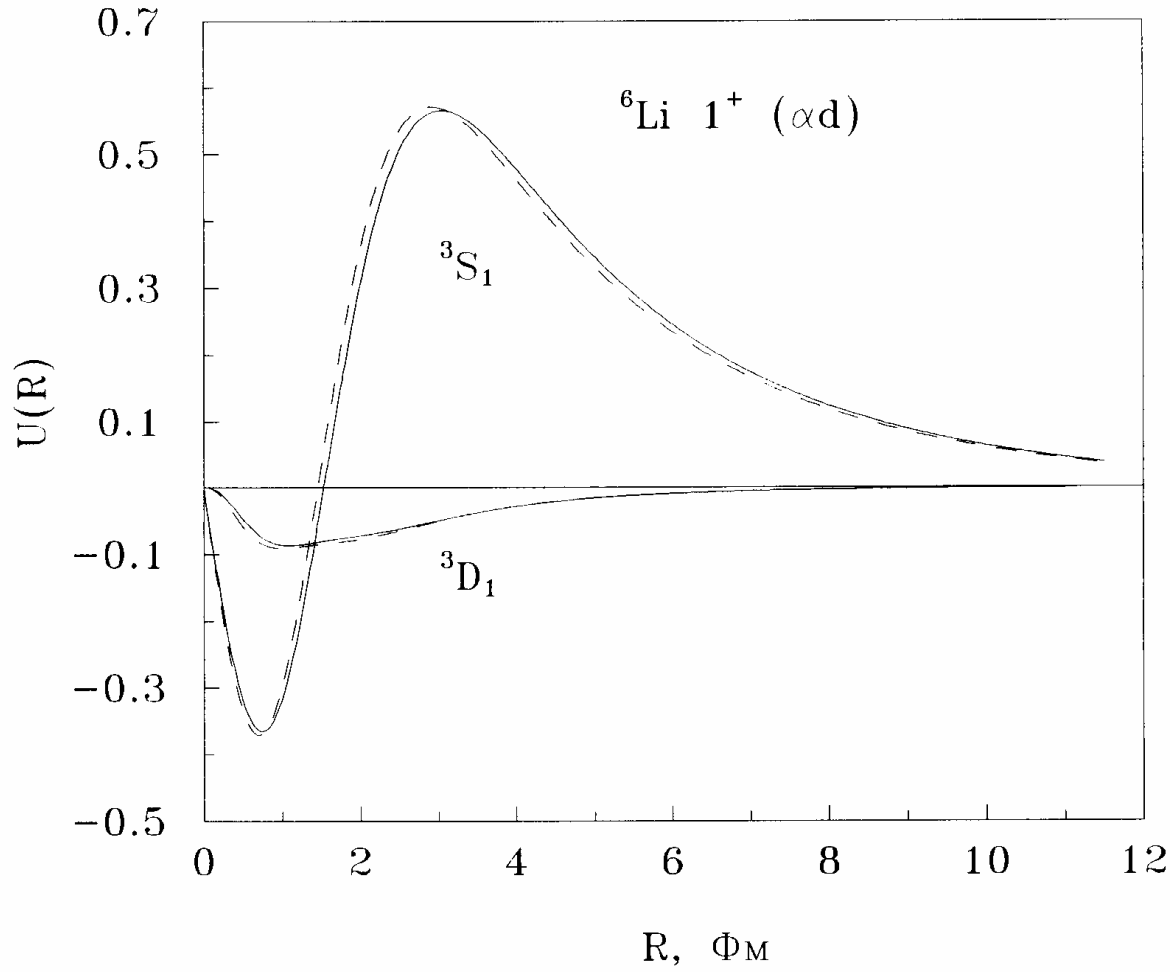

Рис.6.21. Волновые функции связанного состояния ядра $^6$Li в кластерной $^2$H$^4$He модели для потенциалов с тензорной компонентой.

На рис.6.21 показана волновая функция связанного состояния ядра в кластерном $^2$H$^4$He канале. Непрерывной линией даны результаты для первого, а штриховой для второго варианта потенциала.

Видно, что S компонента волновой функции имеет узел на расстояниях порядка 1.6 Фм, а D волна безузловая. Асимптотики S и D функций имеют разные знаки, обеспечивая тем самым отрицательный знак $\eta_D$. Получить такую форму волновых функций оказалось возможным, только при очень узкой тензорной части потенциала. Более широкие потенциалы приводят к смене знака S волновой функции и не позволяют получить правильный знак квадрупольного момента.

На рис.6.22 непрерывной линией показан упругий кулоновский формфактор $^6$Li для первого варианта потенциала и штриховой для второго. Штрих - пунктирной линией приведен вклад C2 компоненты в полный формфактор. Экспериментальные данные взяты из работ [51]. Видно, что вклад C2 формфактора сравнительно мал и существенно не меняет результаты расчетов по сравнению с чисто центральным взаимодействием, которое дает только C0 слагаемое. Точечной линией показаны результаты расчета формфактора в трехтельной модели с полной антисимметризацией волновых функций [52], которые практически полностью описывают экспериментальные данные.

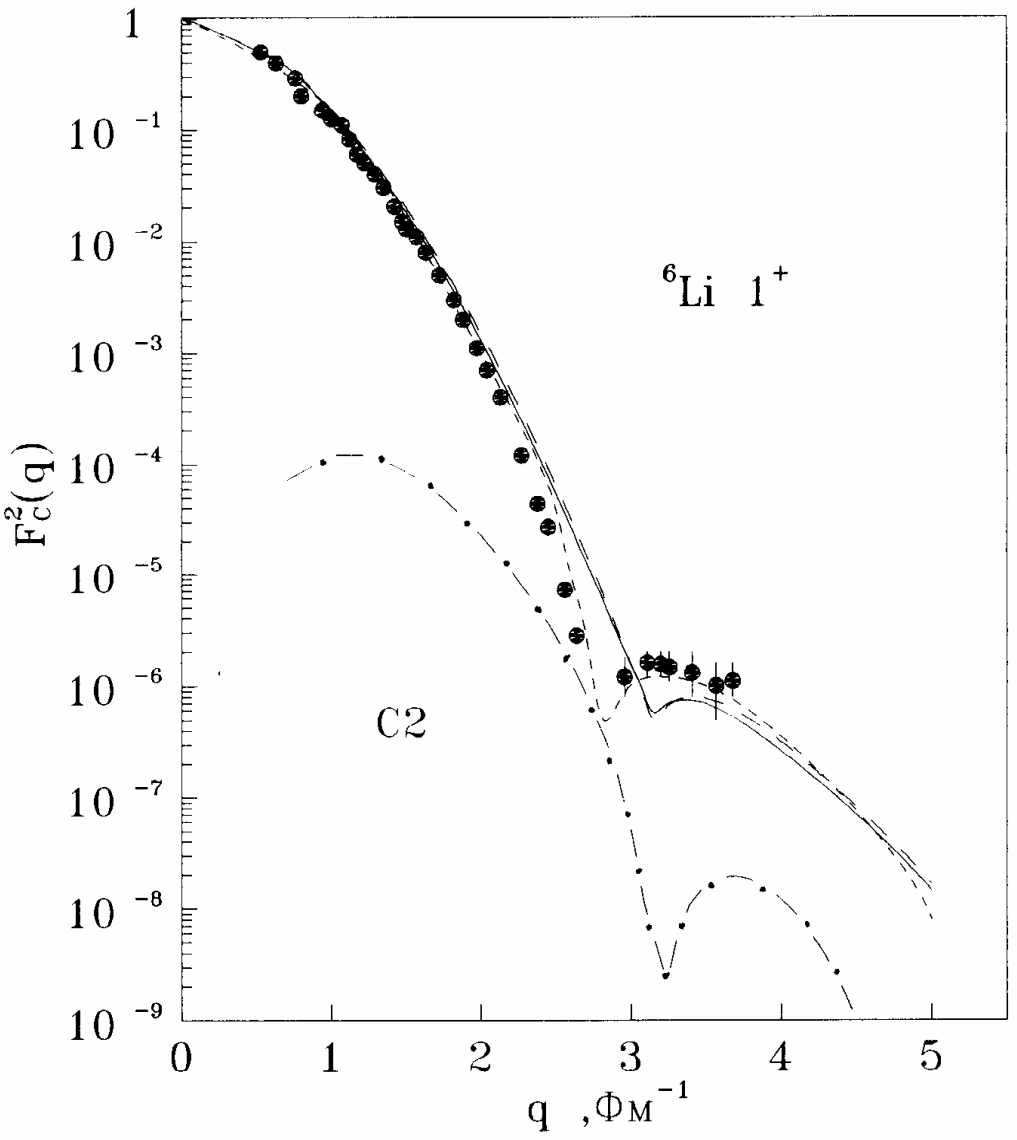

Рис.6.22. Кулоновский формфактор ядра $^6$Li в $^2$H$^4$He модели с тензорными потенциалами. Экспериментальные данные из [51].

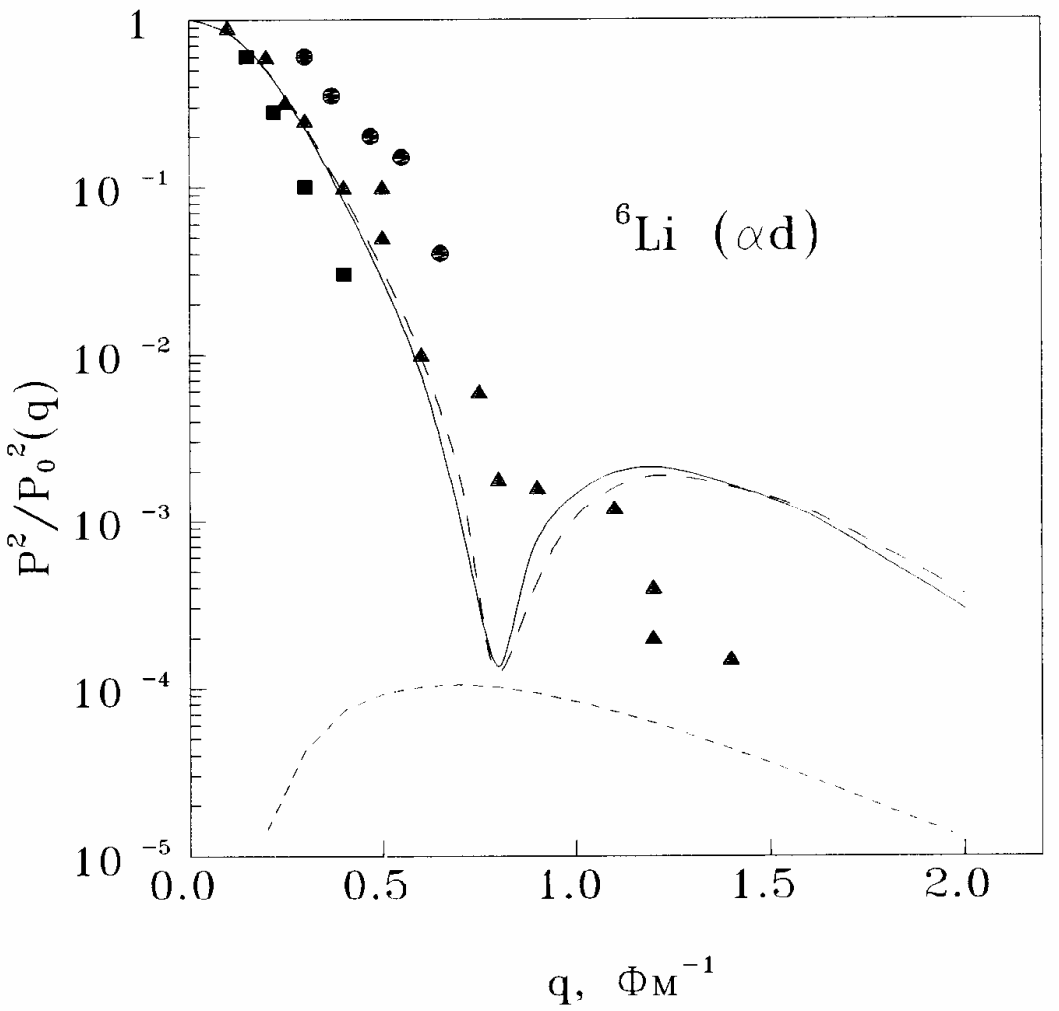

Рис.6.23. Импульсные распределения $^2H^4He$ кластеров в ядре $^6Li$ для тензорных потенциалов. Экспериментальные данные из работ [53].

На рис.6.23 непрерывной линией приведены импульсные распределения $^2H^4He$ кластеров в ядре $^6Li$ для первого и штриховой для второго вариантов взаимодействия, вместе с экспериментальными данными [53]. Точечной линией показан вклад $P_2$ слагаемого, которое несколько сглаживает минимум импульсного распределения в районе 0.7-0.8 Фм$^{-1}$, не приводя к существенным отличиям от результатов для центрального взаимодействия.

На рис.6.24 показаны $^3S_1$ и $^3D_1$ фазы упругого рассеяния и параметр смешивания $\varepsilon_1$ (непрерывная линия для первого и штриховая для второго варианта потенциала) в сравнении с экспериментальными данными [54]. Из рисунка видно, что вполне удается передать поведение экспериментальных фаз упругого рассеяния при малых энергиях.

Отрицательное значение параметра смешивания можно получить только при узком тензорном потенциале. Более широкие потенциалы приводят к смене знака $\varepsilon_1$, и одновременно с ним меняет знак S волновой функции связанного состояния.

Отметим, что существуют и другие данные по фазам рассеяния [55], использующие параметризацию матрицы рассеяния отличную от используемого здесь представления Блатта - Биденхарна.

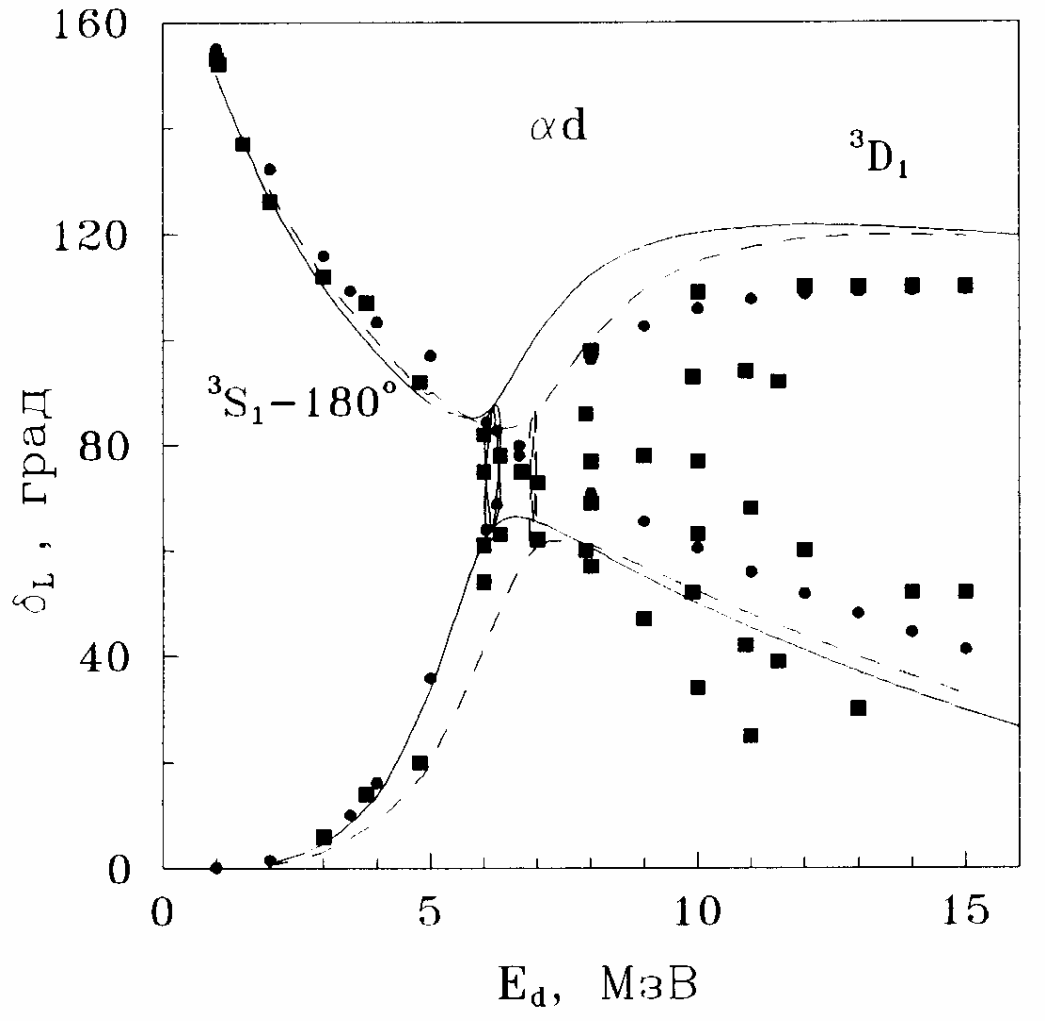

Рис.6.24а. Фазы упругого $^2H^4He$ рассеяния для двух вариантов взаимодействий с тензорной компонентой. Экспериментальные данные [54].

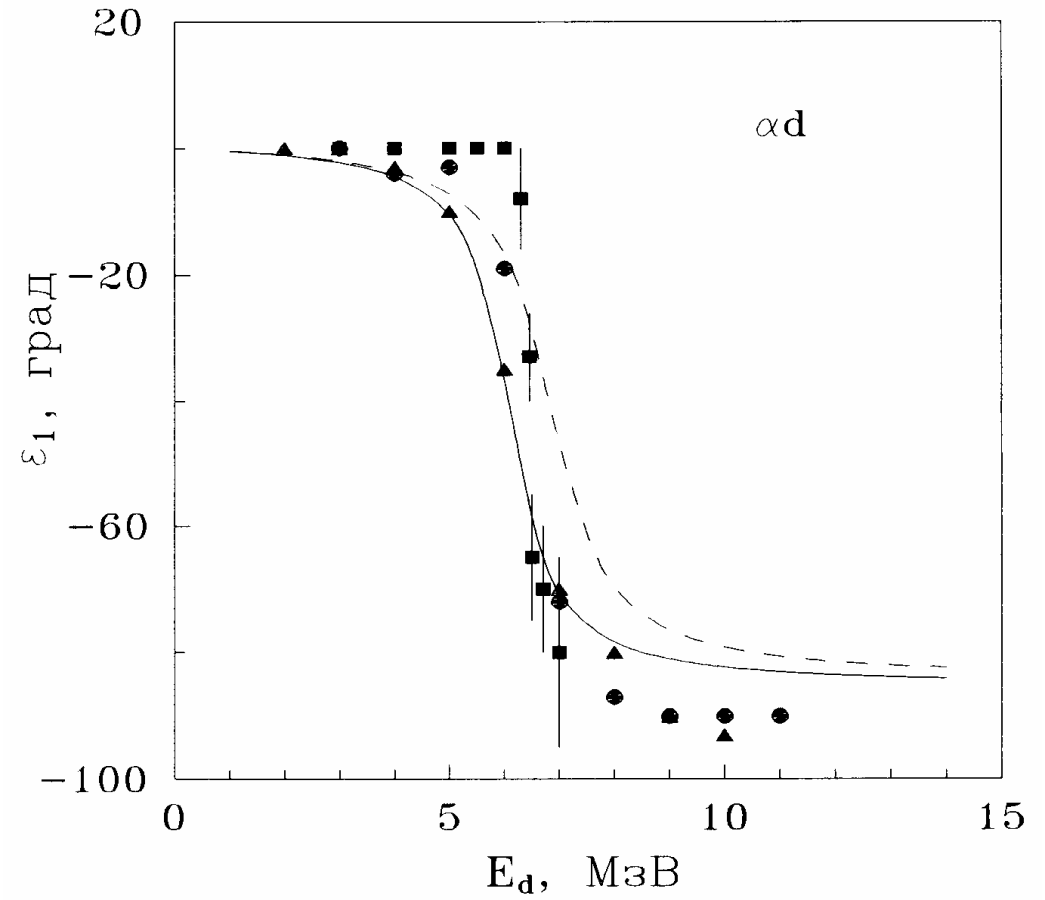

Рис.6.24б. Параметр смешивания фаз упругого $^2H^4He$ рассеяния для двух вариантов взаимодействий с тензорной компонентой. Эксперимент работ [54].

Из полученных результатов видно, что в рамках потенциальной кластерной модели удается получить феноменологические потенциалы $^2H^4He$ взаимодействия с тензорной компонентой, позволяющие правильно передать практически все рассмотренные характеристики связанного состояния ядра $^6Li$ и фазы упругого $^2H^4He$ рассеяния при низких энергиях.

Учет тензорной компоненты мало влияет на большинство характеристик, по сравнению с чисто центральными силами, но позволяет описать квадрупольный момент ядра. Несколько заниженная величина формфактора при больших переданных импульсах, по - видимому, может быть объяснена отсутствием в настоящих расчетах учета обменных эффектов, как это сделано в [52] на основе трехтельной модели.

---

1. Неудачин В.Г., Смирнов Ю.Ф. - Современные проблемы оптики и ядерной физики. Киев. 1974, с.225; ЭЧАЯ, 1979, т.10, с.1236.

2. Дубовиченко С.Б., Джазаиров - Кахраманов А.В. - ЯФ, 1993, т.56, № 2, с.87; ЯФ, 1994, т.57, № 5, с.590; ЯФ, 1995, т.58, с.635; с.852; Dubovichenko S.B., Dzhazairov-Kakhramanov A.V. - In: Int. Nucl. Phys. Conf., Beijing, China, 21-26 August, 1995, p. 5.6-33; 5.6-34;

3. Neudatchin V.G., Kukulin V.I., Pomerantsev V.N., Sakharuk A.A. - Phys. Rev., 1992, v.C45. p.1512; Неудачин В.Г., Сахарук А.А., Смирнов Ю.Ф. - ЭЧАЯ, 1993, т.23, с.480.

4. Neudatchin V.G., Sakharuk A.A., Dubovichenko S.B. - Few Body Sys., 1995, v18, p.159; Дубовиченко С.Б. - ЯФ, 1995, т.58, с.1253; с.1377; с.1973; Dubovichenko S.B., Dzhazairov - Kakhramanov A.V. - In: Int. Nucl. Phys. Conf., Beijing, China, 21-26 August, 1995, p. 5.6-35; 5.6-36; 5.6-37.

5. Neudatchin V.G. , Obukhovsky I.T., Smirnov Yu.F. - Phys. Lett., 1973, v.B43, p.13; Neudatchin V.G., Obukhovsky I.T., Kukulin V.I., Golovanova N.F. - Phys. Rev., 1975, v.C11, p.128; In: Int. Conf. on Clust. Struct. in Nucl. Phys. Chaster. England., 1985, p.353.

6. Дубовиченко С.Б., Кукулин В.И., Сазонов П.Б. - В Сб.: теория квантовых систем с сильным взаимодействием. Калинин. КГУ, 1983, с.65; Дубовиченко С.Б., Жусупов М.А. - Изв. АН КазССР, сер. физ.-мат., 1982, № 6, с.34; In: Proc. Int. Conf. on Nucl. Phys., Florence, 1983, v.1, p.46.

7. Kukulin V.I., Pomerantsev V.N., Krasnopol'sky V.M., Sazonov P.B. - Phys. Lett., 1984, v.135B, P.20; Krasnopolsky V.M., Kukulin V.I., Pomerantsev V.N., Sazonov P.B. et al. - Phys. Lett., 1985, v.165B, p.7; Краснопольский В.М., Кукулин В.И., Померанцев В.Н. - Изв. АН

СССР, 1987, т.51, с.898; Кукулин В.И., Краснопольский В.М., Померанцев В.Н., Сазонов П.Б. - ЯФ, 1986, т.43, с.559; Kukulin V.I., Pomerantsev V.N., Faessler A., Buchmann A.J., Tursunov E.M. - Phys. Rev., 1998, v.C57, p.535.

8. Дородных Ю.Л., Неудачин В.Г., Юдин Н.П. - ЯФ, 1988, т.48, с.1796; Препринт ИЯИ АН СССР, 1988, П-0595, 11с; Neudatchin V.G., Obukhovsky I.T., Smirnov Yu.F - In: Int. Conf. on Clust. Struct. in Nucl. Phys. Chaster., England, 1985, p.353.

9. Reid R.V. - Ann. Phys., 1968, v.50, p.411.

10. Дубовиченко С.Б., Мажитов М. - Изв. АН КазССР, сер. физ. - мат., 1987, № 4, с.55.

11. Дубовиченко С.Б. - ЯФ, 1997, т.60, с.499; In: Particles and Nuclei, XIV International Conference SEBAF, USA, 22-28 May, 1996, p.609; E-Print Archive LANL: Nucl-th/9803024.

12. Arndt R.A. et al. - Phys. Rev., 1983, v.D28, p.97; Mac Gregor M. et al. - Phys. Rev., 1969, v.182, p.1714; Arndt R.A., Strakovsky I.I., Workman R.L. - Phys. Rev., 1996, v.C53, p.430; Arndt R.A., Strakovsky I.I., Workman R.L. - Phys. Rev., 1995, v.C52, p.2246; Arndt R.A., Strakovsky I.I., Workman R.L., Pavan M.M. - Phys. Rev., 1995, v.C52, p.2120.

13. Браун Д.Е., Джексон А.Д. - Нуклон-нуклонные взаимодействия. Москва, Атомиздат, 1979, 246с. (Brown G.E., Jackson A.D. The nucleon-nucleon interaction., North-Holland Pablishing Company, Amsterdam, 1976).

14. Lomon E. et al. - Phys. Rev., 1974, v.C9, p.1329. Hand L. et al. - Rev. Mod. Phys., 1968, v.35, p.335; Simon G. - Nucl. Phys., 1981, v.A364, p.285; Platner D. - In: Europ. Few Body Probl. Nucl. Part. Phys. Sesimbra., 1980, p.31.

15. Buchanan C.D., Yearian M.R. - Phys. Rev. Lett., 1965, v.15, p.303; Ellias J.I. et al. - Phys. Rev., 1969, v.177, p.2075; Arnold R.G. et al. - Phys. Rev. Lett., 1975, v.35, p.776; Simon G.G., Schmitt C., Walther V.H. - Nucl. Phys., 1981,v.A364, p.285; Cramer R. et al. - Z. Phys., 1985, v.C29, p.513; Platchkov S. et al. - Nucl. Phys., 1990, v.A508, p.343; Auffret S. et al. - Phys. Rev. Lett., 1985, v.54, p.649; Bosted P. et al. - Phys. Rev., 1990, v.C 42, p.38; Benaksas D., Drickley D., Frerejacque D. - Phys. Rev., 1966, v.148, p.1327; Drickey D.J., Hand L.N. - Phys. Rev. Lett., 1962, v.9, p.521; Arnold R.G.et al. - Phys. Rev. Lett., 1975, v.35, p.776.

16. Хюльтен Л., Сугавара М.- В кн. строение атомного ядра. М., ИЛ., 1959, с.9. (In. Structure of atomic nuclei. Ed. Flugge S., Springer - Verlag., Berlin-Gottingen-Heidelberg, 1957).

17. Дубовиченко С.Б. - ЯФ, 1997, т.60, с. 704; In: 12-th International Symposium of High Energy Spin Physics, Amsterdam, 10-14 Sep., 1996, p.250; E-Print Archive LANL: Nucl-th/9803001.

18. Ericson T.E. - Nucl. Phys., 1984, v.A416, p.281; Ericson T.E., Rosa - Costa M. - Nucl. Phys., 1983, v.A405, p.497; Ann. Rev. Nucl. Part. Sci., 1985, v.35, p.271.

19. Дубовиченко С.Б. - В кн.: Ядерная спектроскопия и структура атомного ядра. С. П., 1998, с.115; In: XVI-th European Conference on Few-Body Problems in Physics., 1-6 June, 1998, Autrans, France, p.49; E-Print Archive LANL: Nucl-th/9803005; 9805030.

20. Stoks V.G.J., Klomp B.A.M., Terheggen C.P.F., de Swart J.J. - Phys. Rev., 1984, v.C49. p.2950; de Swart J.J., Klomp B.A.M., Rentmeester M.C.M., Rijken Th.A. - Few Body Sys., Sypl., 1995, v.8, p.437 (E-print Archive LANL: Nucl-th/9406039).

21. Klomp A.M.M., Stoks V.G.J., de Swart J.J. - Phys. Rev., 1991, v.C44, p.R1258 ; Stoks V.G.J., Klomp A.M.M., Rentmeester M.C.M., de Swart J.J. - Phys. Rev., 1993, v.C48, p.792.

22. Lacombe M., et al., - Phys. Rev., 1980, v.C21, p.861.

23. Stoks V., de Swart J.J. - Phys. Rev., 1995, v.C52, p.1698 (E-print Archive LANL: Nucl-th/9411002).

24. Wiringa R.B., Smith R.A., Ainsworth T.L. - Phys. Rev., 1984, v.C29, p.1207.

25. Arndt R.A., Strakovsky I.I, Workman R.L. - Phys. Rev., 1994, v.C50, p.2731.

26. Arndt R.A., Chang-Heon Oh, Strakovsky I.I., Workman R.L., Dohrmann F. - Phys. Rev., 1997, v.C56, p. 3005.

27. de Swart J.J., Terheggen C.P.F., Stoks V.G.J. - In: Proc. 3 rd. Int. Symp., Dubna, Russia, 1995, “Deuteron 95” ; Stoks V.G.J., Timmermans R., de Swart J.J. - Phys. Rev., 1993, v.C47, p.512; v.C48, p.792. (E-print Archive LANL: Nucl-th/9509032).

28. Rahm J., et al. - Phys. Rev., 1998, v.C57, p.1077; G. Hohler // Pion-Nucleon Scattering, Landoldt-Bornstein, 1983, v.I/9b2, Ed. H. Schopper, Springer Verlag.

29. Arndt R.A., Strakovsky I.I., Workman R.L., Pavan M.M., - Phys. Rev., 1995, v.C52, p.2120 (E-print Archive LANL: Nucl-th/9505040; SAID - http://www.phys.vt.edu/~igor/appl.html).

30. Arndt R.A., Strakovsky I.I., Workman R.L. - Phys. Rev., 1995, v.C52, p.2246.

31. Friar J.L., Payne G.L., Stoks V.G.J., de Swart J.J. - Phys. Lett., 1993, v.B311, p.4.

32. Stoks V.G.J., Klomp B.A.M., Terheggen C.P.F., de Swart J.J., Rentmeester M.C.M., Rijken Th.A.- NN-ONLINE-http://nn-online.sci.kun.nl/NN/inde. html.

33. Garson M., et al. - Phys. Rev., 1994, v.C49, p.2516.

34. Phillips D.R., Wallace S.J., Devine N.K. - Phys., Rev. 1998, v.C58, p.2261 (E-print Archive LANL: Nucl-th/9802067).

35. Дубовиченко С.Б. - ЯФ, 2000, т.63, с.804-808.

36. Дубовиченко С.Б., Страковский И.И. - ЯФ, 2000, т.63, с.646-651; Изв. РАН, Сер. физ.-мат., 2001, т.65, с.746-748..

37. Perdrisat C.F., Punjabi V. // Phys.Rev., 1990, v.C42, p.1899; Imambekov O., Uzikov Yu.N., Schevchenko L.V. // Z. Phys., 1989, v.A332, p.349.

38. Eerro-Luzzi M. et al. // Phys. Rev. Lett, 1996, v.77, p.2630; Dmitriev V.F. et al. // Phys. Lett., 1985, v.B157, p.143; Войтцеховский Б.Б. и др. // Письма в ЖЭТФ, 1985, v.43, p.567; Gilman R. et al. // Phys. Rev. Lett., 1990, v.65, p.1733; Schulze M.E. et al. // Phys. Rev. Lett., 1984, v.52, p.597; The I. et al. // Phys. Rev. Lett., 1991, v.67, p.173.

39. Frick R. et al. - Phys. Rev. Lett., 1980, v.44, p.14.

40. Nishioka H., Tostevin J.A., Johnson R.C. - Phys. Lett., 1983, v.124B, p.17.

41. Merchant A.C. , Rowley N. - Phys. Lett., 1985, v. B 150, p.35.

42. Il-Tong Cheon - Phys. Lett., 1969, v.30B, p.81; 1971, v.B35, p.276; 1984, v.144B, p.312; Phys. Rev., 1971, v.C3, p.1023; Prog. Theor. Phys., 1970, v.44, p.549.

43. Mertelmeir T., Hofman H.M. - Nucl. Phys., 1986, v.A459, p.387; Kanada H. et al. - Nucl. Phys., 1982, v.A389, p.285.

44. Kukulin V.I., Krasnopol'sky V.M., Voronchev V.T., Sazonov P.B. - Nucl. Phys., 1984, v.A417, p.128; 1986, v.A453, p.365; Kukulin V.I., Voronchev V.T., Kaipov T.D., Eramzhyan R.A. - Nucl. Phys., 1990, v.A517, p.221.

45. Rai M., Lehman D.R., Ghovanlou A. - Phys. Lett., 1975, v.B59, p.327; Lehman D.R., Rajan M. - Phys. Rev., 1982, v.C25, p.2743; Eskandarian A., Lehman D.R., Park W.C. - Phys. Rev., 1988, v.C38, p.2341.

46. Platner G.R. - In: Europ. Few Body Probl. Nucl. Part. Phys., Sesimbra, Portugal, 1980, p.31; Platner G.R., Bornard M., Alder K. - Phys.

Lett., 1976, v. 61B, p.21; Bornard M., Platner G.R., Viollier R.D., Alder K. - Nucl. Phys., 1978, v.A294, p.492.

47. Lehman D.R. - In: 7th - Int. Conf. on Polar. Phen. in Nucl. Phys., Paris, France, 1990.

48. Kukulin V.I., Pomerantsev V.N., Cooper S.G., Dubovichenko S.B. - Phys. Rev., 1998, v. C57, p. 2462; Дубовиченко С.Б. - ЯФ, 1998, т.61, с.210; В кн.: Ядерная спектроскопия и структура атомного ядра. 1998, С. Петербург, с.116; In: XVI-th European Conference on Few-Body Problems in Physics., 1-6 June, 1998, Autrans, France, p. 48.

49. Ajzenberg-Selove F. - Nucl. Phys., 1979, v.A320, p.1.

50. Lovas R.G. et al. - Nucl. Phys., 1987, v.A474, p.451.

51. Lichtenstadt J. et al. - Phys. Lett., 1989, v.B219, p.394; 1990, v.B244, p.173.

52. Eramzhyan R.A., Ryzhikh G.G., Kukulin V.I., Tchu-vil'skymYu.M. - Phys. Lett., 1989, v.B228, p.1.

53. Roos P., Goldberg D.A., Chant N.S., Woody R. - Nucl. Phys., 1976, v.A257, p.317; Watson J. et al. - Nucl. Phys., 1971, v.A172, p.513; Alder J. et al. - Phys. Rev., 1972, v.C6, p.1; Kitching p. et al. - Phys. Rev., 1975, v.C11, p.420; Dollhopf W. et al. - Phys. Lett., 1975, v.58B, p. 425; Albreht D. et al. - Nucl. Phys., 1980, v.A338, p.477; Ent R. et al. - Phys. Rev. Lett., 1986, v.57, p.2367.

54. Bruno M., Cannata F., D'Agostino M., Maroni C., Massa I. - Nuovo Cim., 1982, v.A68, p.35; McIntair L., Haeberli W. - Nucl. Phys., 1967, v.A91, p.382; Keller L.G., Haeberli W. - Nucl. Phys., 1970, v.A156, p.465.

55. Jenny B., Gruebler W., Konig V., Schmelzbach P.A., Schweizer C. - Nucl. Phys., 1983, v.A397, p.61.

# ЗАКЛЮЧЕНИЕ

В использованных кластерных $^4He^2H$ и $^4He^3H$ моделях, удается хорошо воспроизвести не только статические характеристики ядер $^6Li$ и $^7Li$, но и получить вполне приемлемое описание упругих и неупругих кулоновских формфакторов, даже при сравнительно больших переданных импульсах. Параметры межкластерных потенциалов предварительно фиксированы по фазам упругого рассеяния и только при вычислении $3^+$ формфактора $^6Li$ были несколько изменены в $D_3$ волне. На основе тех же потенциалов удается правильно описать полные сечения фотопроцессов, включая астрофизические факторы во всей рассмотренной области энергий.

Определенный успех простой двухкластерной модели может служить очередным подтверждением большой вероятности кластеризации ядер лития в рассмотренные каналы. Для получения согласия результатов с экспериментом, не требуется вводить какие - либо искажения характеристик кластеров и предположение о том, что данным кластерам в ядре можно в целом сопоставлять свойства соответствующих свободных частиц вполне оправдывается. Однако не полное соответствие результатов с экспериментом по формфакторам свидетельствует о приближенном характере модели. Потенциалы межкластерного взаимодействия с запрещенными состояниями позволяют лишь качественно учитывать эффекты, которые определяются полной антисимметризацией волновой функции.

Как мы видели, расчеты, выполненные в $^3He^3H$ модели ядра $^6Li$, позволяют получить вполне разумные результаты для сечений фотопроцессов, однако характеристики ядра описываются сравнительно плохо. Это может служить определенным подтверждением малой вероятности кластеризации основного состояния этого ядра в указанный канал.

Предложенный потенциал $n^5Li$ взаимодействия, основанный на чисто качественных критериях, в принципе, позволяет правильно передать величину фотосечений и их энергозависимость во всей рассматриваемой области энергий, что свидетельствует, по - видимому, о малой чувствительности фотопроцессов в этом канале к виду и форме взаимодействия.

Межкластерные потенциалы, полученные для $^4He^{12}C$ канала ядра $^{16}O$ и согласованные с фазами рассеяния для непрерывного спектра, энергиями и вероятностями радиационных переходов для связанных состояний дают возможность правильно передать полные сечения радиационного захвата на связанный $2^+$ уровень только на основе E2 переходов из различных парциальных волн рассеяния.

Однако согласовать одни и те же потенциалы с энергиями уровней и фазами рассеяния, т.е. получить единые гамильтонианы непрерывного и дискретного спектров, как это было сделано для более легких кластерных систем, все же не удается, что может говорить о некоторой внутренней несогласованности используемой модели и, возможно, не вполне корректном определении количества связанных разрешенных и запрещенных состояний.

Применение одноканальной потенциальной кластерной модели с разделением фаз и потенциалов по схемам Юнга позволяет в целом правильно воспроизвести имеющиеся экспериментальные данные по сечениям фоторазвала в $N^2H$ системе на основе взаимодействий, согласованных с чистыми фазами и характеристиками связанного состояния.

Сравнительно хорошо передается форма полных сечений фоторазвала ядра $^4He$ при рассмотрении E1 переходов с изменением изоспина в $p^3H$ и $n^3He$ каналах. Величина расчетных сечений находится в пределах неоднозначностей различных экспериментальных результатов, а потенциалы основного состояния согласованы с характеристиками ядра и чистыми фазами.

Полученные потенциалы вполне позволяют описать некоторые из экспериментальных данных по полным сечениям фотопроцессов в $^2H^2H$ канале ядра $^4He$. Оказывается возможным избавиться от неоднозначностей при выборе синглетного потенциала основного состояния ядра и согласовать его параметры с чистыми фазами, а так же уточнить форму и структуру синглетного $D_1$ взаимодействия.

Совместное описание характеристик рассеяния, фотореакций и свойств связанных состояний всех легчайших кластерных систем оказывается возможным благодаря разделению фаз и межкластерных взаимодействий по орбитальным симметриям с выделением чистых компонент. Смешанные взаимодействия, в принципе, нельзя согласовать с фазами и свойствами связанных состояний указанных ядер.

Рассмотренные типы центральных нуклон - нуклонных потенциалов приводят к практически одинаковому описанию свойств дейтрона и экспериментальных фаз рассеяния при всех рассмотренных энергиях. Однако, более полное описание характеристик np системы для процессов рассеяния и связанного состояния может быть получено только на основе глубокого с узлами в S и D волнах экспоненциального потенциала с OPEP и тензорной компонентой, параметры которого полностью определены через несколько известных констант нуклон - нуклонного взаимодействия.

Подобные результаты получаются и для другого варианта двухпараметрического, глубокого потенциала, который имеет только одно запрещенное связанное в S волне состояние. Однако, на основе рассмотренных характеристик np системы не удается сделать однозначного выбора между этими двумя типами взаимодействий. Единственное и явное различие заключается в поведении импульсных распределений, которые для потенциала с узлом только в S волне идут примерно в два раза ниже при больших переданных импульсах.

В рамках потенциальной кластерной модели вполне удается найти феноменологические потенциалы $^2H^4He$ взаимодействия с тензорной компонентой, позволяющие правильно передать практически все рассмотренные характеристики связанного состояния ядра $^6Li$ и фазы упругого $^2H^4He$ рассеяния при низких энергиях, включая параметр смешивания. Введение в потенциал тензорной компоненты мало влияет на большинство рассмотренных характеристик по сравнению с чисто центральными силами, но позволяет правильно описать квадрупольный момент ядра.

Во всех случаях и для всех кластерных систем классификация состояний по схемам Юнга позволяет избавиться от известной дискретной неоднозначности параметров межкластерных взаимодействий, которая присутствует в обычной оптической модели, поскольку определенно фиксируется число запрещенных и разрешенных состояний.

Таким образом, мы видели, что используемая двухкластерная модель позволяет получить вполне разумные результаты при рассмотрении самых различных свойств многих легких ядер. Вполне определенный успех такой модели свидетельствует о том, что ее возможности еще не окончательно выяснены и могут оказаться намного шире, чем обычно предполагается.

Конечно, нельзя требовать от нее полного объяснения всех характеристик столь широкого класса ядерных систем. Модель лишь качественно учитывает эффекты, которые могут быть учтены точно, только при полной антисимметризации волновых функций. На данном этапе было важно продемонстрировать общие, принципиальные возможности потенциальной кластерной модели с запрещенными состояниями.

# ПРИЛОЖЕНИЕ 1

## Кулоновские фазы и кулоновские волновые функции

Для практических расчетов характеристик ядерных реакций и процессов рассеяния необходимо знать и, как правило, с высокой точностью, численные значения кулоновских функций и фаз в заданной точке R и широком диапазоне значений кулоновского параметра $\eta_L$. В настоящее время известно достаточно много различных численных методов, применимых для нахождения этих величин, однако, только сравнительно недавно появились достаточно простые и надежные представления для кулоновских функций, а известные способы вычисления кулоновских фаз и сейчас обладают рядом недостатков, так что при их использовании необходимо соблюдать определенную осторожность.

Кулоновские фазы определяются через Г - функцию следующим образом [1]

$$\sigma_L = \arg\{\Gamma(L+1+i\eta)\} \quad , \qquad (П.1)$$

и удовлетворяют рекуррентному процессу

$$\sigma_L = \sigma_{L+1} - \mathrm{Arctg}\left(\frac{\eta}{L+1}\right) , \qquad (П.2)$$

где $\eta = \dfrac{Z_1 Z_2 \mu}{k\hbar^2}$ - кулоновский параметр, $\mu$ - приведенная масса частиц, k - волновое число относительного движения частиц - $k^2 = 2\mu E/\hbar^2$, E - энергия сталкивающихся частиц в центре масс. Откуда сразу можно получить следующее выражение

$$\alpha_L = \sigma_L - \sigma_{L-1} = \sum_{n=1}^{L} \mathrm{Arctg}\left(\frac{\eta}{n}\right), \qquad \alpha_0 = 0 . \qquad (П.3)$$

Наиболее естественное представление для кулоновских фаз получается на основе интегральной формулы для Г - функции [2]

$\sigma_L = \mathrm{arctg}(y/x)$

где

$$y = \int_0^\infty \exp(-t)t^L \mathrm{Sin}(\eta \ln t)dt \quad , \quad x = \int_0^\infty \exp(-t)t^L \mathrm{Cos}(\eta \ln t)dt \ . \qquad (П.4)$$

Однако непосредственное вычисление этих интегралов оказывается достаточно сложной задачей, так как подинтегральные функции являются быстро осциллирующими при $t \to 0$. Поэтому часто используются различного рода приближения и асимптотические разложения, например, такие, как представление фазы при L=0 в виде [3]

$$\sigma_0 = -\eta + \frac{\eta}{2}\ln(\eta^2+16) + \frac{7}{2}\mathrm{arctg}(\eta/4) - [\mathrm{arctg}\eta + \mathrm{arctg}(\eta/2) + \mathrm{arctg}(\eta/3)] -$$

$$-\frac{\eta}{12(\eta^2+16)}[1 + \frac{1}{30}\frac{\eta^2-48}{(\eta^2+16)^2} + \frac{1}{105}\frac{\eta^4-160\eta^2+1280}{(\eta^2+16)^4} + ....],$$

или для L>>1 [1]

$$\sigma_L = \alpha(L+1/2) + \eta(\ln\beta - 1) + \frac{1}{\beta}\left(-\frac{\mathrm{Sin}\alpha}{12} + \frac{\mathrm{Sin}3\alpha}{360\beta^2} - \frac{\mathrm{Sin}5\alpha}{1260\beta^4} + \frac{\mathrm{Sin}7\alpha}{1680\eta^6} - ...\right)$$

$$\alpha = \mathrm{arctg}\left(\frac{\eta}{L+1}\right) \quad , \quad \beta = \sqrt{\eta^2 + (L+1)^2} \ .$$

Используя эти формулы, все остальные фазы определяются из рекуррентных соотношений (П.2). Хотя оба представления обладают высокой скоростью счета на компьютере фазы получаются с некоторой ошибкой оценить которую, можно только сравнив полученный результат с табличными данными или вычислениями по точным формулам. Кроме того последняя формула верна только при L≈100 и рекуррентный процессам вносит дополнительную ошибку в величину фаз. Известны и другие представления, в частности при η>>1 [4]

$$\sigma_0 = \frac{\pi}{4} + \eta(\log\eta - 1) - \sum_{s=1}^{\infty}\frac{B_s}{2s(2s-1)\eta^{2s-1}} \quad , \qquad (П.5)$$

где $B_s$ - числа Бернулли [2]. Однако, подобное разложение хорошо работает только при $\eta \approx 100$. В работах [4] было показано, что можно получить восемь верных знаков с учетом только первого члена суммы только при $\eta = 85$. В области малых $\eta$ ряд сходится плохо и требует, кроме того задания или вычисления чисел Бернулли. В работах [4] имеется и другое определение кулоновских фаз

$$\sigma_L = \eta \Psi(L+1) + \sum_{n=1}^{\infty} \left[ \frac{\eta}{L+n} - \mathrm{arctg}\left( \frac{\eta}{L+n} \right) \right] , \qquad (П.6)$$

которое можно получить из известной формы записи Г функции [2]

$$\Gamma(z)=\Gamma(x+iy)=r\,\exp(i\phi)=r(\mathrm{Cos}\phi+i\mathrm{Sin}\phi), \qquad (П.7)$$

где

$$f = y\Psi(x) + \sum_{n=0}^{\infty} (\mathrm{tg}\omega_n - \omega_n) \quad , \qquad \omega_n = \mathrm{arctg}\left( \frac{y}{x+n} \right) \quad ,$$

и $\Psi(x)$ - логарифмическая производная Г функции [2]

$$\Psi(L+1) = -C + 1 + 1/2 + ..... + 1/L,$$

Здесь C=0.577..... - постоянная Эйлера [2]. Ряд (П.6) будет сходиться тем быстрее, чем меньше $\eta$ и больше L. Эта формула охватывает противоположную представлению (П.5) область и при $1<\eta<100$ оба разложения имеют плохую сходимость. Чтобы оценить остаточный член ряда (П.6) разложим арктангенс в ряд при $\eta/n \ll 1$, что всегда возможно при больших $\eta$. Тогда

$$\sigma_0 = -C\eta + \sum_{n=1}^{\infty} \left( \frac{\eta^3}{3n^3} - \frac{\eta^5}{5n^5} + \frac{\eta^7}{7n^7} - ... \right) . \qquad (П.9)$$

Отсюда видно, что остаток ряда будет иметь порядок величины $\eta^3/n^2$ [5]. Ряд (П.9) при $\eta > 1$ сходится сравнительно плохо, так как для получения, например, относительной точности $10^{-8}$ требуется учитывать десятки тысяч членов ряда. Однако этот ряд допускает существенное улучшение сходимости [6]

$$\sigma_0 = -\eta C + \frac{1}{3}\eta^3 S + \sum_{n=1}^{\infty}\left[\frac{\eta}{n} - \mathrm{arctg}\left(\frac{\eta}{n}\right) - \frac{1}{3}\frac{\eta^3}{n^3}\right] \quad , \qquad (\text{П.10})$$

где $S=\sum_{k=1}^{\infty}\frac{1}{k^3}$ = 1.2020569... . Несложно найти, что остаточный член такого ряда равен $\eta^5/n^4$ и для получения восьми верных знаков требуется учитывать около 100 членов ряда при $\eta\approx1$. В случае $\eta>1$ исходный ряд также допускает улучшение сходимости и преобразуется, например, к виду [6]

$$\sigma_0 = -\eta C + \frac{1}{3}\eta^3 S - \frac{1}{5}\eta^5 D + \sum_{n=1}^{m}\left[\frac{\eta}{n} - \mathrm{arctg}\left(\frac{\eta}{n}\right) - \frac{\eta}{n}\right] - \frac{1}{3}\sum_{n=1}^{m}\left[\frac{\eta}{n}\right]^3 + \frac{1}{5}\sum_{n=1}^{m}\left[\frac{\eta}{n}\right]^5 +$$

$$+ \sum_{n=m+1}^{\infty}\left[\frac{\eta}{n} - \mathrm{arctg}\left(\frac{\eta}{n}\right) - \frac{1}{3}\frac{\eta^3}{n^3} + \frac{1}{5}\frac{\eta^5}{n^5}\right] \quad , \qquad (\text{П.11})$$

где m целая часть $\eta$ и $D=\sum_{k=1}^{\infty}\frac{1}{k^5}$ = 1.036927755... . Первые шесть слагаемых ряда конечны и их вычисление не представляет трудности, а последний сходится очень быстро и имеет остаточный член порядка $\eta^7/n^6$, так что для удовлетворения указанной выше точности требуется учитывать только несколько десятков членов.

Перейдем теперь к рассмотрению кулоновских функций, регулярная $F_L(\eta,\rho)$ и нерегулярная $G_L(\eta,\rho)$ части которых являются линейно независимыми решениями радиального уравнения Шредингера с кулоновским потенциалом [1]

$$\chi_L + \left(1 + \frac{2\eta}{\rho} - \frac{L(L+1)}{\rho^2}\right)\chi_L = 0 \quad , \qquad (\text{П.12})$$

где $\chi_L = F_L(\eta,\rho)$ или $G_L(\eta,\rho)$. Вронскианы этих функций имеют вид

$$W_1 = F_L' G_L - F_L G_L' = 1 \quad , \quad W_2 = F_{L-1} G_L - F_L G_{L-1} = \frac{L}{\sqrt{\eta^2 + L^2}} \quad ,$$

(П.13)

а асимптотика записывается

$$F_L = \mathrm{Sin}(\rho - \eta in 2\rho - \pi L/2 + \sigma_L \;, \quad G_L = \mathrm{Cos}(\rho - \eta in 2\rho - \pi L/2 + \sigma_L \;. \qquad (П.14)$$

Имеется достаточно много методов и приближений для вычисления кулоновских функций [1,3,5,6]. Однако только недавно появилось быстро сходящееся представление, позволяющее получить их значения с высокой точность в широком диапазоне переменных с малыми затратами компьютерного времени [7].

Кулоновские функции представляются в виде цепных дробей вида

$$f_L = F_L^{'} / F_L = b_0 + \cfrac{a_1}{b_1 + \cfrac{a_2}{b_2 + \cfrac{a_3}{b_3 + \ldots}}} \;, \qquad (П.15)$$

где

$$b_0=(L+1)/\rho+\eta/(L+1) \;, \qquad b_n=[2(L+n)+1][(L+n)(L+n+1)+\eta\rho] \;, \qquad (П.16)$$

$$a_1=-\rho[(L+1)^2+\eta^2](L+2)/(L+1) \;, \quad a_n=-\rho^2[(L+n)^2+\eta^2][(L+n)^2-1] \;,$$

и

$$P_L+iQ_L= \frac{G_L^{'} + iF_L^{'}}{G_L + iF_L} = \frac{i}{\rho}\left( b_0 + \cfrac{a_1}{b_1 + \cfrac{a_2}{b_2 + \cfrac{a_3}{b_3 + \ldots}}} \right), \qquad (П.17)$$

где

$$b_0=\rho-\eta \;, \quad b_n=2(b_0+in) \;, \qquad (П.18)$$

$$a_n=-\eta^2+n(n-1)-L(L+1)+i\eta(2n-1) \;.$$

Такой метод расчета, оказывается, применим в области

$\rho \geq \eta + \sqrt{\eta^2 - L(L+1)}$ , и легко позволяет получить высокую точность благодаря быстрой сходимости цепных дробей. Поскольку $\eta$ порядка единицы, а L обычно не более 5, то метод дает хорошие результаты при $\rho > 5$ - 10 Фм. Именно в этой области необходимо знать кулоновские функции при численных расчетах ядерных функций рассеяния и реакций.

Используя (П15, П.17) можно получить связь между кулоновскими функциями и их производными

$$F_L' = f_L F_L \quad ,$$

$$G_L = (F_L' - P_L F_L)/Q_L = (f_L - P_L) F_L / Q_L \quad , \qquad (П.19)$$

$$G_L' = P_L G_L - Q_L F_L = [P_L (f_L - P_L)/Q_L - Q_L] F_L \quad .$$

Таким образом, задавая некоторое значение $F_L$ в точке $\rho$, находим все остальные функции и их производные с точностью до постоянного множителя, который определяется из вронскианов. Вычисления по приведенным формулам кулоновских функций и сравнение их с табличным материалом [8] показывает, что легко получить восемь - девять правильных знаков, если $\rho$ удовлетворяет приведенному выше условию.

---

1. Ходгсон П.Е. - Оптическая модель упругого рассеяния. М., Атомиздат, 1966, 230с. (Hodgson P.E. - The optical vodel of elastic scattering. Clarendon press, Oxford, 1963).

2. Янке Е., Емде Ф., Леш Ф. - Специальные функции. М., Наука, 1968, 344с. (Janke - Emde - Losch. - Tafeln hoherer funktionen., Stuttgard, 1960).

3. Melkanoff M. - Univ. California Pres., Berkley, Los Angeles , 1961, 116p; Lutz H.F., Karvelis M.D. - Nucl. Phys., 1963, v.43, p.31; Melkanoff M. - Meth. Comput. Phys., 1966, v.6, p.1; Gody W.J., Hillstrom K.E. - Meth. Comput., 1970, v.111, p.671; Smith W.R. - Usics Communs., 1969, v.1, p.106.

4. Hull M.H., Breit G. - In: Encyklopedia of Phys., v.XL1/1, Nucl. React., II, Springer - Verlag, 1959, p.408; Abramowitz M. - Tables of Coulomb wave function., v.1, Washington, N.B.S., 1952, 141p.

5. Демидович Б.П., Марон И.Ф. - Основы вычислительной математики., М., Наука, 1966, 664с.

6. Дубовиченко С.Б., Жусупов М.А. - Изв. АН КазССР, сер. физ. - мат., 1981, № 6, с.24.

7. Barret A.R., Feng D.H., Steed J.W. - Comput. Phys. Communs., 1974, v.8, p.377; Barret A.R. - J. Comput. Phys., 1982, v.46, p.171.

8. Абрамовиц М. - Справочник по специальным функциям., М., Наука, 1979, 830с. (Handbook of mathematical functions.. Edit. M. Abramowitz and I. Stegun., NBS., 1964).

# ПРИЛОЖЕНИЕ 2

*Сравнение характеристик дейтрона, вычисленных с точной и апроксимационной волновыми функциями для различных вариантов Нимегенских потенциалов. Здесь $\Delta E(S)$ и $\Delta E(D)$ − средняя относительная ошибка аппроксимации ВФ для S и D волн. Ссылки на экспериментальные данные и результаты вычислений даны на литературу из гл.6.*

| Характеристики дейтрона | Результаты для Nijm-1 из [37] | Результаты для Nijm-1 с АВФ | Результаты для Nijm-2 из [37] | Результаты для Nijm-2 с АВФ | Результаты для Nijm93 из [37] | Результаты для Nijm93 с АВФ | Результаты для Reid93 из [37] | Результаты для Reid93 с АВФ | Эксперимент из [44] |
|---|---|---|---|---|---|---|---|---|---|
| $P_D$ , % | 5.664 | 5.648 | 5.635 | 5.637 | 5.754 | 5.740 | 5.699 | 5.698 | 5.67 |
| $Q_d$ , Фм$^2$ | 0.2719 | 0.2707 | 0.2707 | 0.2708 | 0.2706 | 0.2700 | 0.2703 | 0.2706 | 0.271(1) |
| $R_d$ , Фм | 1.967 | 1.966 | 1.968 | 1.967 | 1.966 | 1.965 | 1.969 | 1.967 | 1.9676(10) |
| $\Delta E(S)$, % | | $2\ 10^{-3}$ | | $4\ 10^{-3}$ | | $2.5\ 10^{-3}$ | | $2.6\ 10^{-3}$ | |
| $\Delta E(D)$, % | | $7\ 10^{-3}$ | | $10^{-2}$ | | $8.0\ 10^{-3}$ | | $1.2\ 10^{-2}$ | |

*Параметры и коэффициенты разложения ВФ в S и D волнах*
*для варианта потенциала Nijm.-1.*

| N | S - волна | | D - волна | |
|---|---|---|---|---|
| | $\alpha_k$ | $C_k$ | $\alpha_k$ | $C_k$ |
| 1 | $2.738730051\ 10^{-3}$ | $6.614956073\ 10^{-4}$ | $1.282555144\ 10^{-2}$ | $2.007853254\ 10^{-5}$ |
| 2 | $1.366334688\ 10^{-2}$ | $1.631205529\ 10^{-2}$ | $4.958672449\ 10^{-2}$ | $1.589707390\ 10^{-4}$ |
| 3 | $2.950269915\ 10^{-2}$ | $8.905930817\ 10^{-2}$ | $9.476378560\ 10^{-2}$ | $3.854176262\ 10^{-3}$ |
| 4 | $5.027672648\ 10^{-2}$ | $-1.408469975\ 10^{-1}$ | $1.483973414\ 10^{-1}$ | $-1.918087341\ 10^{-2}$ |
| 5 | $7.711993903\ 10^{-2}$ | $3.077071309\ 10^{-1}$ | $2.126934528\ 10^{-1}$ | $1.040464938\ 10^{-1}$ |
| 6 | $1.123146415\ 10^{-1}$ | $9.111270308\ 10^{-3}$ | $2.918272913\ 10^{-1}$ | $-2.769045830\ 10^{-1}$ |
| 7 | $1.600000113\ 10^{-1}$ | $-3.891839981\ 10^{-1}$ | $3.930357993\ 10^{-1}$ | $5.651000142\ 10^{-1}$ |
| 8 | $2.279311419\ 10^{-1}$ | $8.382234573\ 10^{-1}$ | $5.293444395\ 10^{-1}$ | $-7.081024051\ 10^{-1}$ |
| 9 | $3.319506347\ 10^{-1}$ | $-5.145812631\ 10^{-1}$ | $7.262902260\ 10^{-1}$ | $7.575676441\ 10^{-1}$ |
| 10 | $5.091819763\ 10^{-1}$ | $3.634730279\ 10^{-1}$ | $1.040969849\ 10^{0}$ | $-4.092908204\ 10^{-1}$ |
| 11 | $8.677174449\ 10^{-1}$ | $6.502671540\ 10^{-2}$ | $1.630128622\ 10^{0}$ | $5.244586468\ 10^{-1}$ |
| 12 | $1.873627305\ 10^{0}$ | $-3.954315782\ 10^{-1}$ | $3.115294456\ 10^{0}$ | $-1.151041761\ 10^{-1}$ |
| 13 | $9.347403526\ 10^{0}$ | $1.930331439\ 10^{-2}$ | $1.204448891\ 10^{1}$ | $-2.547275275\ 10^{-2}$ |

*Параметры и коэффициенты разложения ВФ (1) в S и D волнах для варианта потенциала Nijm.-2.*

| N | S - волна | | D - волна | |
|---|---|---|---|---|
| | $\alpha_k$ | $C_k$ | $\alpha_k$ | $C_k$ |
| 1 | 5.643064622 $10^{-3}$ | 8.409703150 $10^{-3}$ | 8.748187684 $10^{-3}$ | 3.979084431 $10^{-5}$ |
| 2 | 2.576393075 $10^{-2}$ | 5.162835121 $10^{-2}$ | 3.308108822 $10^{-2}$ | -5.403191317 $10^{-4}$ |
| 3 | 5.331781879 $10^{-2}$ | 2.382708341 $10^{-2}$ | 6.255264580 $10^{-2}$ | 6.162323989 $10^{-3}$ |
| 4 | 8.822768182 $10^{-2}$ | 2.229462415 $10^{-1}$ | 9.723804891 $10^{-2}$ | -2.991536632 $10^{-2}$ |
| 5 | 1.321762800 $10^{-1}$ | -5.164530873 $10^{-1}$ | 1.385483891 $10^{-1}$ | 1.049392372 $10^{-1}$ |
| 6 | 1.885450929 $10^{-1}$ | 1.136572838 $10^{0}$ | 1.891129315 $10^{-1}$ | -2.215512991 $10^{-1}$ |
| 7 | 2.634024322 $10^{-1}$ | -8.799225688 $10^{-1}$ | 2.534589767 $10^{-1}$ | 3.310583830 $10^{-1}$ |
| 8 | 3.679800928 $10^{-1}$ | 2.741962671 $10^{-2}$ | 3.396989405 $10^{-1}$ | -2.979401350 $10^{-1}$ |
| 9 | 5.249115825 $10^{-1}$ | 1.174938202 $10^{0}$ | 4.636753500 $10^{-1}$ | 2.086775303 $10^{-1}$ |
| 10 | 7.863841057 $10^{-1}$ | -1.189774513 $10^{0}$ | 6.606617570 $10^{-1}$ | -7.014304399 $10^{-3}$ |
| 11 | 1.301269650 $10^{0}$ | 8.220960498 $10^{-1}$ | 1.026998401 $10^{0}$ | 1.228616163 $10^{-1}$ |
| 12 | 2.692946672 $10^{0}$ | -9.754071832 $10^{-1}$ | 1.941940308 $10^{0}$ | 3.409164250 $10^{-1}$ |
| 13 | 1.229489326 $10^{1}$ | 1.205787435 $10^{-1}$ | 7.343404770 $10^{0}$ | -3.088182807 $10^{-1}$ |

*Параметры и коэффициенты разложения ВФ (1) в S и D волнах для варианта потенциала Nijm.-93.*

| N | S - волна | | D - волна | |
|---|---|---|---|---|
| | $\alpha_k$ | $C_k$ | $\alpha_k$ | $C_k$ |
| 1 | 3.710338380 $10^{-3}$ | 6.383923814 $10^{-3}$ | 1.307721436 $10^{-2}$ | 1.707274350 $10^{-5}$ |
| 2 | 1.751266047 $10^{-2}$ | 9.949852712 $10^{-3}$ | 4.945120960 $10^{-2}$ | 2.338412014 $10^{-4}$ |
| 3 | 3.682383150 $10^{-2}$ | 1.093037128 $10^{-1}$ | 9.350671619 $10^{-2}$ | 3.038547235 $10^{-3}$ |
| 4 | 6.161000952 $10^{-2}$ | -7.773621380 $10^{-2}$ | 1.453561336 $10^{-1}$ | -1.702271774 $10^{-2}$ |
| 5 | 9.312023222 $10^{-2}$ | 9.106899053 $10^{-2}$ | 2.071088254 $10^{-1}$ | 1.087013781 $10^{-1}$ |
| 6 | 1.338701546 $10^{-1}$ | 2.560368478 $10^{-1}$ | 2.826951444 $10^{-1}$ | -3.368652463 $10^{-1}$ |
| 7 | 1.883943826 $10^{-1}$ | -1.601817757 $10^{-1}$ | 3.788827360 $10^{-1}$ | 7.653585672 $10^{-1}$ |
| 8 | 2.651259005 $10^{-1}$ | -2.071751654 $10^{-1}$ | 5.077983737 $10^{-1}$ | -1.099940062 $10^{0}$ |
| 9 | 3.811465502 $10^{-1}$ | 1.013635278 $10^{0}$ | 6.931243539 $10^{-1}$ | 1.239532590 $10^{0}$ |
| 10 | 5.760824680 $10^{-1}$ | -1.000448227 $10^{0}$ | 9.875891209 $10^{-1}$ | -8.485507965 $10^{-1}$ |
| 11 | 9.638445377 $10^{-1}$ | 9.057361484 $10^{-1}$ | 1.535206914 $10^{0}$ | 8.047580719 $10^{-1}$ |
| 12 | 2.026674986 $10^{0}$ | -8.395105600 $10^{-1}$ | 2.902906418 $10^{0}$ | -7.525151968 $10^{-2}$ |
| 13 | 9.565829277 $10^{0}$ | 5.062305927 $10^{-2}$ | 1.097727680 $10^{1}$ | -1.273730397 $10^{-1}$ |

*Параметры и коэффициенты разложения ВФ (1) в S и D волнах для варианта потенциала Reid-93.*

| N | S - волна | | D - волна | |
|---|---|---|---|---|
| | $\alpha_k$ | $C_k$ | $\alpha_k$ | $C_k$ |
| 1 | $4.426570609\ 10^{-3}$ | $7.736701518\ 10^{-3}$ | $1.025847904\ 10^{-2}$ | $-4.320973312\ 10^{-5}$ |
| 2 | $2.089325152\ 10^{-2}$ | $2.241683006\ 10^{-2}$ | $3.752347827\ 10^{-2}$ | $1.156179933\ 10^{-3}$ |
| 3 | $4.393219203\ 10^{-2}$ | $9.419733286\ 10^{-2}$ | $6.983165443\ 10^{-2}$ | $-9.142984636\ 10^{-3}$ |
| 4 | $7.350301743\ 10^{-2}$ | $-2.887684107\ 10^{-2}$ | $1.073626205\ 10^{-1}$ | $5.387797579\ 10^{-2}$ |
| 5 | $1.110958755\ 10^{-1}$ | $1.508325189\ 10^{-1}$ | $1.516261846\ 10^{-1}$ | $-1.789302677\ 10^{-1}$ |
| 6 | $1.597120315\ 10^{-1}$ | $-1.689205170\ 10^{-1}$ | $2.053600550\ 10^{-1}$ | $4.137437940\ 10^{-1}$ |
| 7 | $2.247614563\ 10^{-1}$ | $6.877220869\ 10^{-1}$ | $2.732264400\ 10^{-1}$ | $-6.254537106\ 10^{-1}$ |
| 8 | $3.163050115\ 10^{-1}$ | $-9.494065642\ 10^{-1}$ | $3.635210395\ 10^{-1}$ | $6.933129430\ 10^{-1}$ |
| 9 | $4.547218978\ 10^{-1}$ | $1.158097029\ 10^{0}$ | $4.923470914\ 10^{-1}$ | $-4.936743975\ 10^{-1}$ |
| 10 | $6.872876287\ 10^{-1}$ | $-6.233545542\ 10^{-1}$ | $6.953323483\ 10^{-1}$ | $3.369798362\ 10^{-1}$ |
| 11 | $1.149902105\ 10^{0}$ | $3.788776398\ 10^{-1}$ | $1.069038272\ 10^{0}$ | $1.939450204\ 10^{-2}$ |
| 12 | $2.417898178\ 10^{0}$ | $-6.908371449\ 10^{-1}$ | $1.989494085\ 10^{0}$ | $3.657529354\ 10^{-1}$ |
| 13 | $1.141238785\ 10^{1}$ | $2.619499341\ 10^{-2}$ | $7.277173042\ 10^{0}$ | $1.270186715\ 10^{-2}$ |

# ПРИЛОЖЕНИЕ 3

*Характеристики дейтрона и пр рассеяния для разных вариантов NN потенциалов.*

| Х-ки дейтрона и пр рассеяния | LP1 | LP2 | LP3 | LP4 | LP5 | Nijm.93 [27] | CERN [18] |
|---|---|---|---|---|---|---|---|
| $E_d$ , МэВ | 2.2246 | 2.2246 | 2.2246 | 2.2246 | 2.2246 | 2.224575(9) | 2.224579(9) |
| $Q_d$, Фм$^2$ | 0.271 | 0.274 | 0.279 | 0.285 | 0.290 | 0.271(1) | 0.2859(3) |
| $P_D$, % | 5.62 | 5.75 | 6.00 | 6.23 | 6.56 | 5.67 | --- |
| $A_S$ | 0.884(1) | 0.884(1) | 0.884(1) | 0.884(1) | 0.884(1) | 0.8845(8) | 0.8802(20) |
| $\eta=A_S/A_D$ | 0.0253(1) | 0.0256(1) | 0.0261(1) | 0.0266(1) | 0.0273(1) | 0.0253(2) | 0.0271(4) |
| $a_t$ , Фм | 5.417 | 5.417 | 5.419 | 5.417 | 5.419 | 5.4194(20) | 5.419(7) |
| $r_t$ , Фм | 1.753 | 1.753 | 1.754 | 1.751 | 1.752 | 1.7536(25) | 1.754(8) |
| $R_d$ , Фм | 1.965 | 1.966 | 1.967 | 1.968 | 1.968 | 1.9676(10) | 1.9560(68) |
| $V_1$ , МэВ | 10.076 | 10.214 | 10.490 | 10.710 | 11.080 | --- | 10.71(12) |
| $f^2_{\pi NN}$ | 0.0730 | 0.0740 | 0.0760 | 0.0776 | 0.0803 | 0.074 | 0.0776 |
| $\alpha$ , Фм$^{-1}$ | 3.68 | 3.72 | 3.80 | 3.90 | 4.05 | --- | --- |
| $V_0$ , МэВ | 3495.641 | 3524.172 | 3575.435 | 3652.860 | 3744.980 | --- | --- |

**Дубовиченко Сергей Борисович**

Член - корреспондент Казахстанской Международной
Академии Информатизации,
член Европейского физического общества,
член Нью - Йоркской Академии наук,
Лауреат премии ЛКСМ Казахстана,
Лауреат Международного гранта Сороса,
кандидат физико - математических наук, доцент